\documentclass[final,1p,times]{elsarticle}

\usepackage{txfonts}
\usepackage{slashed,amssymb,amsmath,amsbsy}
\usepackage{bm}
\usepackage{graphicx}
\usepackage[usenames]{color}
\usepackage{url}
\usepackage[bookmarksnumbered=true]{hyperref}
\usepackage{xspace}
\usepackage{longtable}
\usepackage{booktabs}
\usepackage{cleveref}

\newcommand{\settocdepth}[1]{%
	\addtocontents{toc}{\string\c@tocdepth=#1}%
  \setcounter{tocdepth}{#1}%
}

\newcommand{\crefrangeconjunction}{--}

\newcommand{\etal}{\emph{et al.}}
\newcommand{\refcite}[1]{Ref.~\cite{#1}}
\newcommand{\refetal}[1]{\etal{}\ \cite{#1}}
\newcommand{\upto}{\text{-}}

\newcommand{\Pythia}{\textsc{Pythia}}
\newcommand{\Jetset}{\textsc{Jetset}}
\newcommand{\Fritiof}{\textsc{Fritiof}}

\renewcommand{\vec}[1]{\boldsymbol{#1}}
\newcommand{\pr}[1]{{#1}\,'}

\DeclareMathOperator{\RealPart}{\text{Re}}
\DeclareMathOperator{\ImaginaryPart}{\text{Im}}

\newcommand{\res}[3]{#1$_{#2}$(#3)}
\newcommand{\clebsch}[6]{\left\langle  #1\, #2 ; #3\, #4 \vert #5\, #6 \right\rangle^{2}  }
\newcommand{\ii}{\mathrm{i}}
\newcommand{\dd}{\mathrm{d}}
\newcommand{\UNIT}[1]{\ensuremath{\,{\rm #1}}\xspace}

\newcommand{\keV}{\UNIT{keV}}
\newcommand{\MeV}{\UNIT{MeV}}
\newcommand{\GeV}{\UNIT{GeV}}

\newcommand{\MeVc}{\ensuremath{\,{\rm MeV}\!/c}\xspace}
\newcommand{\GeVc}{\ensuremath{\,{\rm GeV}\!/c}\xspace}
\newcommand{\AMeV}{\ensuremath{\,A{\rm MeV}}\xspace}
\newcommand{\AGeV}{\ensuremath{\,A{\rm GeV}}\xspace}

\newcommand{\fm}{\UNIT{fm}}
\newcommand{\fmc}{\ensuremath{\,{\rm fm}/c}\xspace}
\newcommand{\proz}{\UNIT{\%}}

\newcommand{\mb}{\UNIT{mb}}
\newcommand{\mub}{\UNIT{\mu b}}
\newcommand{\GeVminsq}{\mbox{$\GeV^{-2}$}}
\newcommand{\atomfull}[3]{\mbox{${^{#1}_{#2}\text{#3}}$}}
\newcommand{\atom}[2]{\mbox{${^{#1}\text{#2}}$}}
\newcommand{\carbon}    {{\atom{12}{C}}}
\newcommand{\oxygen}    {{\atom{16}{O}}}
\newcommand{\aluminium} {{\atom{27}{Al}}}
\newcommand{\calcium}   {{\atom{40}{Ca}}}
\newcommand{\iron}      {{\atom{56}{Fe}}}
\newcommand{\nickel}    {{\atom{56}{Ni}}}
\newcommand{\cupper}    {{\atom{64}{Cu}}}
\newcommand{\tin}       {{\atom{100}{Sn}}}
\newcommand{\rhodium}   {{\atom{103}{Rh}}}
\newcommand{\xenon}     {{\atom{136}{Xe}}}
\newcommand{\gold}      {{\atom{197}{Au}}}
\newcommand{\lead}      {{\atom{207}{Pb}}}
\newcommand{\leadA}     {{\atom{208}{Pb}}}
\newcommand{\bismuth}   {{\atom{209}{Bi}}}
\newcommand{\uranium}   {{\atom{238}{U}}}
\newcommand{\subQE}{\text{QE}}

\newcommand{\subR}{\text{R}}

\newcommand{\subCC}{\text{CC}}
\newcommand{\subNC}{\text{NC}}
\newcommand{\subEM}{\text{EM}}
\newcommand{\subBG}{\text{BG}}
\newcommand{\subBGpi}{\text{1$\pi$ BG}}
\newcommand{\subtot}{\text{tot}}
\newcommand{\ME}{\mathcal{M}}
\newcommand{\fff}{\mathcal{F}}
\newcommand{\ffc}{\mathcal{C}}
\newcommand{\kz}{k^0}

\newcommand{\kpr}{k'}
\newcommand{\kprz}{{k'}^0}

\newcommand{\abskpr}{|\bvec{k}'|}

\newcommand{\slashp}{\slashed{p}}

\newcommand{\slashq}{\slashed{q}}

\newcommand{\ppr}{p'}

\newcommand{\slashppr}{\slashed{p}'}

\newcommand{\kpi}{k_\pi}
\newcommand{\Mpr}{M'\hspace{.2mm}}

\newcommand{\ml}{m_\ell}

\newcommand{\pT}{p_T}

\newcommand{\op}[1]{\ensuremath{\bm{\mathrm{#1}}}}
\newcommand{\bvec}[1]{\ensuremath{\boldsymbol{#1}}}
\newcommand{\PV}{\mathrm{PV}}
\newcommand{\fslash}{\slashed}
\newcommand{\pb}[2]{\left\{#1,#2 \right\}_{\mathrm{pb}}}

\newcommand{\erw}[1]{\ensuremath { %
     \left \langle {#1} \right \rangle}}
\DeclareMathOperator{\tr}{tr}
\DeclareMathOperator{\Tr}{Tr}
\newcommand{\diag}{\mathrm{diag}}
 \newcommand{\comm}[2]{\ensuremath{ \left[ {#1}, {#2} \right] }}
 \newcommand{\R}{\ensuremath{\mathbb{R}}}

\DeclareMathOperator{\im}{Im}
\DeclareMathOperator{\re}{Re}

\newcommand{\smeson}{$\sigma$-meson }
\newcommand{\pc}{$\pi^{\pm}\pi^0$ }
\newcommand{\pn}{$\pi^0\pi^0$ }

\providecommand{\namecref}[1]{section}
\providecommand{\namecrefs}[1]{sections}

\hypersetup{
  pdfinfo={
    Title={Transport-theoretical Description of Nuclear Reactions},
    Subject={GiBUU},
    Author={The GiBUU Group}
  }
}

\begin{document}

\begin{frontmatter}

\author{O.~Buss}
\author{T.~Gaitanos}
\author{K.~Gallmeister}
\author{H.~van~Hees}
\author{M.~Kaskulov}
\author{O.~Lalakulich}
\author[fn1]{A.~B.~Larionov}
\author{T.~Leitner}
\author{J.~Weil}
\author{U.~Mosel}

\address{Institut f\"ur Theoretische Physik, Universit\"at Giessen,
  Germany} \fntext[fn1]{Also at the National Research Center ``Kurchatov
  Institute'', Moscow, Russia}

\title{Transport-theoretical Description of Nuclear Reactions}

\begin{abstract}
  In this review we first outline the basics of transport theory and its
  recent generalization to off-shell transport.  We then present in some
  detail the main ingredients of any transport method using in
  particular the Giessen Boltzmann-Uehling-Uhlenbeck (GiBUU)
  implementation of this theory as an example. We discuss the potentials
  used, the ground state initialization and the collision term,
  including the in-medium modifications of the latter. The central part
  of this review covers applications of GiBUU to a wide class of
  reactions, starting from pion-induced reactions over proton and
  antiproton reactions on nuclei to heavy-ion collisions (up to about
  $30\AGeV$). A major part concerns also the description of photon-,
  electron- and neutrino-induced reactions (in the energy range from a
  few $100 \MeV$ to a few $100 \GeV$). For this wide class of reactions
  GiBUU gives an excellent description with the same physics input and
  the same code being used. We argue that GiBUU is an indispensable tool
  for any investigation of nuclear reactions in which final-state
  interactions play a role. Studies of pion-nucleus interactions,
  nuclear fragmentation, heavy-ion reactions, hypernucleus formation,
  hadronization, color transparency, electron-nucleus collisions and
  neutrino-nucleus interactions are all possible applications of GiBUU
  and are discussed in this article.
\end{abstract}


\end{frontmatter}

\tableofcontents

\crefname{appendix}{}{}
\Crefname{appendix}{}{}

\section{Introduction and Motivation}
\label{sec:intro}

Any reaction involving nuclear targets poses a challenge to nuclear
theory. Nuclear many-body effects come into play not only in the
reaction mechanism, but also in the reaction amplitude and the
description of final-state interactions. These many-body effects
in general evolve dynamically during the course of a reaction and thus
require a time-dependent framework for their description. For the
experimenters the challenge is to draw conclusions from observed
asymptotic particle yields, spectra and distributions on the
interactions or the state of the matter during early stages of the
collision. A well-known example is that of ultra-relativistic heavy-ion
collisions. There the main aim is to investigate the properties of a
transient state of matter, the quark-gluon plasma, involving quite
different degrees of freedom than those being present both in the
initial and the final states of the reaction, with the help of
asymptotic, `normal' free particles. Another example is given by the
production of hadrons on nuclear targets using high-energy electron beams. Here
the aim is to understand time scales in hadronization and the
search for the phenomenon of color transparency in
hadron- or electron-induced reactions, where the observed final
particle yield must be used to reconstruct the nuclear interactions in
a very early stage of the reaction. This is very much the same as the challenge in
the studies of short-range correlations between nucleons in nuclei which must be inferred
from the final outgoing particles. A final
example is that of neutrino-nucleus reactions, where the nuclear
response must be understood in order to extract neutrino-oscillation
parameters from so-called long-baseline experiments. Common to all these quite
different reactions and physics questions is the need to understand quantitatively the final-state interactions experienced by the outgoing particles. Somewhat simpler are determinations of
total inclusive cross sections. Here final-state interactions enter
only into the first interaction amplitude, but nevertheless they do
determine in this way the total cross sections.

The question then arises, which theoretical framework can be used to
describe all these various reaction types. It is obvious that any
reaction that preserves the phase coherence of the nuclear target can
only be described by a quantum-mechanical reaction theory. An example is
the coherent photoproduction of mesons on nuclear targets. However, as
soon as the final state is not fully specified there are other options
for their description available. The extreme case in this class is the
total inclusive cross section with an elementary projectile without any
information on specific final states of the target. An example is given
by the total photoabsorption cross section on nuclei. Here
quantum-mechanical reaction theory can be used with its standard
approximations, such as the impulse approximation with plane or
distorted waves. The quantum mechanics of the nuclear many-body system
enters here into the wave functions of the initial and final
states. These can be approximated at various degrees of sophistication;
approximations for these states can range from Fermi-gas descriptions
over shell-model wave functions up to very sophisticated results of
state-of-the-art nuclear many-body calculations. The same inclusive
process can also be treated by transport theory by modeling the
collision term for the first, initial interaction by using a plane wave
approximation. Indeed, results of such calculations for fully inclusive
cross sections often show an agreement with data as good as that
achieved in more sophisticated quantum mechanical approximation
schemes. A more complicated case is the investigation of semi-inclusive
reactions, i.e., reactions, where one outgoing particle is measured in
detail, with many other particles and an excited nuclear target being
present. For this reaction type also quantum-mechanical approximations
have been used that describe the final state again by an optical-model
wave function. However, while such an ansatz may be quite appropriate
for the description of total cross sections it must fail in its
description of, e.g., final-state energy distributions of knocked-out
particles. The optical model can describe -- through its imaginary part
of the potential -- the loss of flux of these particles when they
traverse the nuclear target, but it contains no information where the
lost particles go. It can also not describe any side feeding of reaction
channels, where the final observed particle is not the one that was
originally produced, nor can it describe energy distributions in
situations, where the primary particles are being slowed down by
final-state interactions. Semi-inclusive reactions, that in addition
often involve many particles in the final state, thus present a problem
for quantum-mechanical approximations. Here, transport theory is the
method of choice. Such a theory can in principle handle all final-state
channels with many interacting particles present and includes naturally
the processes of rescattering, absorption or side feeding; its
applicability is in general given if the many-body final state evolves
incoherently.

This has been recognized quite early. Precursors of present-day
transport theories were Monte-Carlo (MC) generators written to
describe the final-state interactions in reactions of nucleons with
nuclear targets. While the earliest suggestion of such methods goes
back to Serber~\cite{Serber:1947zza}, the method has then been picked
up by, among others, Metropolis \etal{}~\cite{Metropolis:1958sb},
Bertini~\cite{Bertini:1963zz}, and Cugnon~\cite{Cugnon:1980zz}. In 1983 
Carruthers and Zachariasen gave a very complete description of quantum collision theory
\cite{Carruthers:1982fa}. Later
on, with the advent of heavy-ion reactions, first practical implementations were
developed to describe the dynamical evolution of a colliding
nucleus-nucleus system
\cite{Bertsch:1984gb,Stoecker:1986ci,Bauer:1986zz,Bertsch:1988ik,Danielewicz:1991dh}
while taking into account the hadronic potentials and the equation of
state of nuclear matter within the Boltzmann-Uehling-Uhlenbeck (BUU)
theory.  These codes thus went beyond the simple MC generators used
until then that could not account for potential effects on the
propagation of particles. The GiBUU code, to be discussed later in
this article, has its origins in this era; it started with
applications to heavy-ion reactions (see
\cite{Bauer:1986zz,Cassing:1990dr,Teis:1996kx} and further references
therein). With the availability of ultra-relativistic heavy-ion beams
at the AGS, the CERN SPS, RHIC, and the LHC this field has flourished
ever since. Models such as IQMD \cite{Aichelin:1991xy,Hartnack:1997ez}, relativistic
BUU \cite{Li:1989zza,Blaettel:1993uz}, UrQMD \cite{Bass:1998ca}, RQMD \cite{Santini:2008pk},
HSD \cite{Ehehalt:1996uq} (which grew out of an early predecessor version of GiBUU),
and many other transport models (for a
comprehensive list see Ref.~\cite{Kolomeitsev:2004np}) are still
widely used to extract properties of hot and dense matter from the final
state observables.

On the other hand, little (with protons as projectiles) or no work
involving trans\-port-theo\-re\-ti\-cal methods had been done on the
description of inclusive or semi-inclusive reactions on nuclei with
elementary projectiles, where simple (sometimes oversimplified)
approximations such as the Glauber treatment of final-state
interactions prevailed. The first attempt to use transport-theoretical
methods for the description of elementary processes on nuclei was an
investigation of inclusive pion-nuclear reactions
\cite{Salcedo:1987md} in an MC calculation. A first calculation within
the framework of a BUU theory of pion and $\eta$ photoproduction off
nuclei was performed by the Giessen group in \cite{Hombach:1994gb}
with the same code that had been originally developed to describe
heavy-ion collisions. Since then BUU theory has been used by that
group to analyze a wide class of nuclear reactions involving
elementary projectiles such as hadrons or electrons, photons, and
neutrinos.

The Giessen Boltzmann-Uehling-Uhlenbeck (GiBUU) transport model, which
grew out of these early studies, is a method and simulation code for
hadron-, photon-, electron-, neutrino-, and heavy-ion-induced reactions
on nuclei. It is based on a coupled set of semi-classical kinetic
equations, which describe the dynamics of a hadronic system explicitly
in phase space and in time. The initial state of the hadronic system
either directly corresponds to the experimental conditions
(meson-nucleus, hadron-nucleus, and heavy-ion collisions) or is obtained
via external models (photon-, electron-, and neutrino-nucleus
reactions). The relevant degrees of freedom are mesons and baryons,
which propagate in mean fields and scatter according to cross sections
which are appropriate for the energy range from a few $10\MeV$ to more
than $100\GeV$. In the higher energy regimes the concept of pre-hadronic
interactions is implemented in order to account for color transparency
and formation-time effects.

The GiBUU project is aiming to provide a unified theoretical transport
framework in the MeV and GeV energy regimes for all the reaction types
named above. The basic parameters, e.g., particle properties and
elementary cross sections, assume the same common values for all
reaction types, thus allowing for a validation of the model over a wide
range of phenomena and giving it some predictive power. This is what
makes GiBUU unique among all the other transport codes. The detailed
comparison with many different experimental results also allows one to
get a feeling for the inherent systematic uncertainties of the method,
which are often unknown in other approaches. At the same time, the GiBUU
model contains a number of options like on-shell vs.~off-shell particle
propagation, non-relativistic vs.~relativistic mean fields, various
treatments of the nuclear ground state etc. That also makes GiBUU a
quite flexible theoretical toolkit, allowing to describe a given
reaction on various levels of sophistication.  The code itself is
available under public license from \cite{gibuu}.

The purpose of this article is, on one hand, to set out the general
framework of transport theory and its practical implementation.
In that sense many of the methods and ingredients to
be described can be found also in transport codes for heavy-ion reactions. On the other hand, a
specific implementation of transport theory is used to describe
the power of the method in its application to various physics
questions and to very different reaction types.
Such a implementation is provided by the GiBUU transport code, developed
by the authors mainly to describe interactions of elementary
projectiles with nuclear targets \cite{gibuu}.

In the first part of this paper, in \cref{sec:tranportEquations}, we
briefly describe the theoretical basis of any transport theory, going
back to the seminal work of Kadanoff and Baym \cite{kadanoffBaym} and to
the later derivation of relativistic quantum kinetic equations by
Botermans and Malfliet \cite{Botermans:1990qi} that provided the
theoretical basis for off-shell transport; we will also outline the
approximations that lead to the BUU transport equations that are
actually being used. A particular point here is the discussion of our
treatment of off-shell transport of quasi-particle degrees of freedom
with broad mass spectra, which presents a major development in practical
implementations of transport theory during the last decade. In the
following \cref{sec:BUUingredients} we describe in some detail the
actual ingredients of the GiBUU model. Although similar ingredients can
be found in any transport implementation we felt it necessary to discuss
here as many details as possible in order to make the physics in the
GiBUU model transparent. Only then can meaningful conclusions be drawn
from detailed comparisons with other transport approaches and, more
importantly, from analyses of experimental results.

\Cref{sec:results} then contains the center piece of this article, a
discussion of physics results obtained with GiBUU over a wide range of
nuclear reactions and energies. The reactions discussed here range
from low-energy pion-nucleus interactions over heavy-ion and
antiproton reactions to reactions with electrons and photons on
nuclei. Finally a discussion of the very recent developments of
neutrino-nucleus interactions is given, which is particularly important for neutrino long-baseline
experiments. This broad coverage of very different nuclear
reactions, all with the same physics input and the same code, is what
sets GiBUU apart from other transport codes, mostly limited to
descriptions of heavy-ion reactions. A summary then concludes this
article, to be followed by several appendices that contain detailed
information on the cross sections used, which may be useful also for other
transport models or MC generators, and on the numerical implementation
in GiBUU.

\section{Transport equations}
\label{sec:tranportEquations}

The BUU equation describes the space-time evolution of a many-particle
system under the influence of mean-field potentials and a collision
term, or more precisely, the time evolution of the Wigner transform of
the real-time one-particle Green's function, which is a generalization
of the classical phase-space density. For each particle species, within
the BUU approach one obtains an additional differential equation. All
these equations are coupled through the gain and loss terms, which
represent scattering processes, and by the mean fields, included in the
Hamilton functions. In this section, we sketch the derivation of the
generalized BUU equation from nonequilibrium quantum field theory and
discuss its approximations and solutions.

\subsection{Nonequilibrium quantum field theory}
\label{sec:derivation_kb_eqs}

The appropriate starting point for the derivation of transport equations
from the underlying fundamental many-body-quantum theory is the
Schwinger-Keldysh real-time formalism of quantum field theory. We start
our derivation of a transport equation for broad resonance-like
quasi-particles by a brief review of relativistic off-equilibrium
quantum-field theory in the Schwinger-Keldysh real-time
formalism~\cite{Schwinger:1960qe,Keldysh:1964ud}. We concentrate on the
motion of Dirac particles and derive the transport equations within the
operator formalism. The approach of off-equilibrium many-body theory
within the path-integral formalism can be found, e.g.,
in~\cite{Calzetta:1986cq,Calzetta-Hu:2008}.

\begin{figure}[t]
  \centering
  \includegraphics[width=0.6\linewidth]{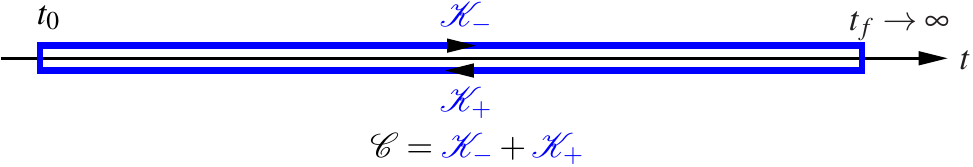}
  \caption[The Schwinger-Keldysh contour]{(Color online) The
    Schwinger-Keldysh contour. Following the convention in~\cite{LP},
    the two branches are denoted with $\mathcal{K}_-$ (chronological
    branch) and $\mathcal{K}_+$ (anti-chronological branch). The
    contour-ordering operator $\mathcal{T}_{\mathcal{C}}$ orders
    operators with time arguments on the contour in the sense of the
    arrows from right to left. For fermionic field operators the symbol
    includes the usual sign change according to the signature of the
    permutation needed to bring the operators in the contour-ordered
    sequence.}
  \label{fig:keldysh-contour}
\end{figure}

We use the interaction picture, splitting the Hamilton operator,
\begin{equation}
  \label{eq:hamiltonian}
  \op{H}=\op{H}_0+\op{H}_i ,
\end{equation}
in a ``free part'', $\op{H}_0$, and an ``interaction part'',
$\op{H}_i$. We assume that the equations of motion for the Dirac-field
operators, $\op{\psi}(t,\bvec{x})$, evolving according to the Hamilton
operator, $\op{H}_0$, can be solved exactly in terms of a time evolution
operator
\begin{equation}
  \begin{split}
    \label{eq:eom-fields}
    \op{\psi}(t,\bvec{x})&=\op{A}(t,t_0) \op{\psi}(t_0,\bvec{x})
    \op{A}^{\dagger}(t,t_0), \\
    \op{A}(t,t_0)&=\mathcal{T}_c \exp \left [\ii \int_{t_0}^{t} \dd t'
      \; \op{H}_0(t') \right ].
  \end{split}
\end{equation}
The operator, $\mathcal{T}_c$, is the usual ``causal time-ordering
operator'', ordering time-dependent (field) operators by increasing
times from right to left. In the case of fermion fields, it also
includes a sign change according to the signature of the permutation
necessary to bring the operators from the original order into the
time-ordered sequence. The time, $t_0$, is the initial time, where the
system is ``prepared'' in terms of a statistical operator, $\op{R}$,
which is a functional of the field operators and, in general, explicitly
time dependent. Its time evolution is given by
\begin{equation}
  \begin{split}
    \label{eq:eom-statop}
    \op{R}[\op{\psi}(t,\bvec{x})] &= \op{C}(t,t_0)
    \op{R}[\op{\psi}(t_0,\bvec{x})] \op{C}^{\dagger}(t,t_0), \\
    \op{C}(t,t_0)&=\mathcal{T}_c \exp \left [-\ii \int_{t_0}^{t} \dd
      t' \; \op{H}_i(t') \right ]
  \end{split}
\end{equation}
and fulfills the equation of motion,
\begin{equation}
  \label{eq:neumann}
  \frac{\dd \op{R}} {\dd t} = \frac{1}{\ii} \comm{\op{H}_i}{\op{R}}.
\end{equation}
Observable quantities are represented by self-adjoint operators,
$\op{O}(t)$. As functions (or functionals) of the field operators their
time evolution is according to $\op{H}_0$ as described in
\cref{eq:eom-fields}. The expectation value of an observable is given by
\begin{equation}
  \label{eq:expt-value.1}
  \erw{\op{O}(t)} = \Tr \{ \op{R}[\op{\psi}(t,\bvec{x})] \op{O}(t) \},
\end{equation}
where the trace has to be taken in the Fock space of the field
operators. Inserting the time evolution for the statistical operator
\cref{eq:eom-statop} one finds
\begin{equation}
  \label{eq:expt-value.2}
  \erw{\op{O}(t)} = \Tr \left [\op{R}_{0} \op{C}^{\dagger}(t,t_0)
    \op{O}(t) \op{C}(t,t_0) \right ].
\end{equation}
Here, we have brought the time-evolution operator $\op{C}(t,t_0)$ to the
right, using the commutativity of operators under the trace operator and
abbreviated the initial state by
$\op{R}_{0}:=\op{R}[\op{\psi}(t_0,\bvec{x})]$. Inserting the explicit
form of the time-evolution operator from \cref{eq:eom-statop} and using
\begin{equation}
  \label{eq:expt-value.3}
  \op{C}^{\dagger}(t,t_0)=\mathcal{T}_a \exp \left [+\ii \int_{t_0}^t \dd
    t' \; \op{H}_i(t') \right ]=\mathcal{T}_a \exp \left [-\ii \int_{t}^{t_0} \dd
    t' \; \op{H}_i(t') \right ],
\end{equation}
where $\mathcal{T}_a$ denotes the ``anti-causal time-ordering
operator'' which orders field-operator products from left to right
with increasing time arguments, leads to the \emph{contour-ordered}
time evolution of expectation values,
\begin{equation}
\begin{split}
  \label{eq:expt-value.4}
  \erw{\op{O}(t)} &= \Tr \left \{ \op{R}_0 \mathcal{T}_{a} \left [\exp
      \left (-\ii \int_{t}^{t_0} \dd t' \; \op{H}_i(t') \right ) \right
    ] \op{O}(t) \mathcal{T}_c \left [\exp \left (-\ii \int_{t_0}^t \dd
        t' \;
        \op{H}_i(t') \right ) \right ] \right \} \\
  & =\Tr \left \{ \op{R}_0 \mathcal{T}_{\mathcal{C}} \left [\exp \left
        (-\ii \int_{\mathcal{C}} \dd t' \; \op{H}_i(t') \right ) \op{O}(t)
    \right ] \right \}.
\end{split}
\end{equation}
In this time integral the time is defined along the
\emph{Schwinger-Keldysh contour} starting at $t_0$, going to a time,
$t_f$, which is larger than any time appearing explicitly in the
considered expression, $t_{\text{max}}$, and then back in opposite
direction to $t_0$ (see \cref{fig:keldysh-contour}). Note that the
opposite sign in \cref{eq:expt-value.3} compared to \cref{eq:eom-statop}
is included by the opposite, i.e., ``anti-chronological'', integration
direction on the lower branch, $\mathcal{K}_+$, of the Schwinger-Keldysh
contour.
The result of the expectation value \cref{eq:expt-value.4} is obviously
independent of the choice of the final time, $t_f$, since the
contributions from the upper and the lower branch of the contour for the
time interval $(t_{\text{max}},t_f)$ cancel each other. For the
following, it is convenient to let $t_f \rightarrow \infty$.

It is clear that this formalism can be extended to contour-ordered
products of an arbitrary number of field operators. E.g., the exact
contour $n$-particle-Green's function is defined by
\begin{equation}
  \begin{split}
    \label{eq:n-point-fct}
    \ii^n S_n^{\mathcal{C}}&(\xi_1,\ldots,\xi_n,\xi_1',\ldots,\xi_n') =
    \erw{\mathcal{T}_{\mathcal{C}} \op{\psi}(\xi_1) \op{\psi}(\xi_2)
      \cdots \op{\psi}(\xi_n) \overline{\op{\psi}}(\xi_n') \cdots
      \overline{\op{\psi}}(\xi_1')} \\
    &= \Tr \left [ \op{R}_0 \mathcal{T}_{\mathcal{C}} \left \{ \exp
        \left [-\ii \int_{\mathcal{C}} \dd t' \; \op{H}_i(t') \right ]
        \op{\psi}(\xi_1) \cdots \op{\psi}(\xi_n)
        \overline{\op{\psi}}(\xi_n') \cdots
        \overline{\op{\psi}}(\xi_1) \right \} \right ],
  \end{split}
\end{equation}
while the corresponding perturbative $n$-particle function is defined by
the leading-order perturbative approximation of the time-evolution
operator,
\begin{equation}
  \begin{split}
    \label{eq:n-point-fct-free}
    \ii^n S_{0n}^{\mathcal{C}}&(\xi_1,\ldots,\xi_n,\xi_1',\ldots,\xi_n')
    = \erw{\mathcal{T}_{\mathcal{C}} \op{\psi}(\xi_1) \op{\psi}(\xi_2)
      \cdots \op{\psi}(\xi_n) \overline{\op{\psi}}(\xi_n') \cdots
      \overline{\op{\psi}}(\xi_1')}_0 \\
    &= \Tr \left [ \op{R}_0 \mathcal{T}_{\mathcal{C}} \left \{
        \op{\psi}(\xi_1) \cdots \op{\psi}(\xi_n)
        \overline{\op{\psi}}(\xi_n') \cdots
        \overline{\op{\psi}}(\xi_1) \right \} \right ].
  \end{split}
\end{equation}
Here, $\xi_j$ includes the contour-time and spatial components of the
four-vectors, $x_j$, as well as the Dirac indices $\alpha_j$, i.e.,
$\op{\psi}(\xi_j) \equiv \op{\psi}_{\alpha_j}(x_j)$. Within this
formalism the perturbative expansion of expectation values can be
established in a way very similar to the usual vacuum-quantum field
theory. The main difference is that time-ordering is replaced by contour
ordering, and that expectation values have to be taken with respect to
the initial Statistical Operator, $\op{R}_0$, and not with respect to
the vacuum state. Also the Wick theorem for expectation values will in
general not be valid in its simple form known from vacuum quantum field
theory, but only for particular forms of the initial Statistical
operator, $\op{R}_0$~\cite{Danielewicz:1982kk}, i.e., if the initial
state of the system is of the form $\op{R}_0=\exp(-\op{A})$, where
$\op{A}$ is an appropriate one-body operator. This means that
many-particle correlations are neglected in the initial state.

However, the full Schwinger-Keldysh formalism cannot be applied for
simulations of reactions. In fact, this formalism contains an infinite
series of self-coupled equations: the one-particle Green's function is
correlated with the two-particle one, the two-particle Green's function
depends on the higher-order one, and so on. In that sense the
Schwinger-Keldysh formalism contains all information on correlations of
the many-body system. Usually one includes these many-body correlations
into a self-energy. The dynamics of the highly correlated system is then
fully characterized by the Dyson equation for the one-particle Green's
function, which contains the self-energy. For practical purposes one
considers a truncation of the Schwinger-Dyson hierarchy on the
two-particle level, and the problem is then reduced to the formulation
of an approximation for the two-particle Green's function. The simplest
ansatz is to neglect correlations completely, i.e., the two-particle
Green's function consists of two uncorrelated one-particle Green's
functions. This approach is well known as the Hartree (or by including
antisymmetrization the Hartree-Fock) method. On this approach
phenomenological models are based, such as the Hartree theory for
quantum hadrodynamics~\cite{Serot:1997xg}. The Dirac-Brueckner (or
Dirac-Brueckner-Hartree-Fock)~\cite{Botermans:1990qi,Horowitz:1986fr}
approach goes beyond the simple Hartree approximation by including a
part of the two-body correlations. In the following we derive a general
form of a transport equation starting from the Dyson equations for the
one-particle Green's functions on the Schwinger-Keldysh contour within
the Dirac-Brueckner approximation.

At first, it is more convenient to split the contour-two-point Green's
functions in a matrix-like fashion~\cite{LP}. The time arguments of the
matrix elements are then defined on $\R$, and the two matrix indices
denote their location at the branch of the Schwinger-Keldysh contour,
leading to the following definitions for the one-particle Green's
functions
\begin{equation}
  \begin{alignedat}{3}
    \label{eq:matrix-gf}
    \ii S_{\alpha \beta}^{c}(x_1,x_2) &=\ii S_{\alpha
      \beta}^{--}(x_1,x_2) &&=&
    &\erw{\mathcal{T}_c \psi_{\alpha}(x_1) \overline{\psi}_{\beta}(x_2)}, \\
    \ii S_{\alpha \beta}^{<}(x_1,x_2) &=\ii S_{\alpha
      \beta}^{-+}(x_1,x_2) &&=&
    -&\erw{\overline{\psi}_{\beta}(x_2) \psi_{\alpha}(x_1)}, \\
    \ii S_{\alpha \beta}^{>}(x_1,x_2) &= \ii S_{\alpha
      \beta}^{+-}(x_1,x_2) &&=&
    &\erw{\psi_{\alpha}(x_1) \overline{\psi}_{\beta}(x_2)}, \\
    \ii S_{\alpha \beta}^{a}(x_1,x_2)&=\ii S_{\alpha
      \beta}^{++}(x_1,x_2) &&=& &\erw{\mathcal{T}_a \psi_{\alpha}(x_1)
      \overline{\psi}_{\beta}(x_2)}.
  \end{alignedat}
\end{equation}
Here, we have written the spinor indices, $\alpha$ and $\beta$,
explicitly. The corresponding Green's function for free particles are
given in the analogous way by using the average $\erw{\cdots}_0$ as
defined in \cref{eq:n-point-fct-free}. As in vacuum-quantum field theory
the perturbative expansion can be cast into a Feynman-diagram formalism,
where expectation values of observables are given by ``closed
diagrams''. In addition to the vacuum-Feynman rules, each vertex carries
a $\pm$-contour index, and for a $+$-vertex an additional factor $-1$
has to be taken into account, which originates from the opposite
integration direction over times on the $+$-branch of the
Schwinger-Keldysh contour. By the same arguments as for the vacuum, the
diagrams can be organized in classes as connected and disconnected
parts, one-particle irreducible (1PI) amputated (or proper vertex
parts), etc. This implies that the same (partial) resummation techniques
as in the vacuum can be formally applied also in the off-equilibrium
many-body case, and the corresponding Schwinger-Dyson equations can be
derived.

For the derivation of transport equations, we introduce the self-energy,
$-\ii \Sigma$, as the amputated 1PI two-point function. Introducing the
matrix notation,
\begin{equation}
  \begin{split}
    \label{eq:dyson.1}
    \hat{S}_{\alpha \beta}(x_1,x_2)&=\begin{pmatrix}
      S_{\alpha \beta}^{--}(x_1,x_2) & S_{\alpha \beta}^{-+}(x_1,x_2) \\
      S_{\alpha \beta}^{+-}(x_1,x_2) & S_{\alpha \beta}^{++}(x_1,x_2)
    \end{pmatrix},\\
    \hat{S}_{0 \alpha \beta}(x_1,x_2)&=\begin{pmatrix}
      S_{0,\alpha\beta}^{--}(x_1,x_2) & S_{0,\alpha\beta}^{-+}(x_1,x_2) \\
      S_{0,\alpha\beta}^{+-}(x_1,x_2) &
      S_{0,\alpha\beta}^{++}(x_1,x_2)
    \end{pmatrix}, \\
    \hat{\Sigma}_{\alpha \beta}(x_1,x_2) &= \begin{pmatrix}
      \Sigma_{\alpha \beta}^{--}(x_1,x_2) & \Sigma_{\alpha \beta}^{-+}(x_1,x_2) \\
      \Sigma_{\alpha \beta}^{+-}(x_1,x_2) & \Sigma_{\alpha
        \beta}^{++}(x_1,x_2)
    \end{pmatrix},
  \end{split}
\end{equation}
for the full and the free two-point Green's function and the
self-energy, the corresponding Dyson equation reads
\begin{equation}
  \label{eq:dyson.2}
  \hat{S}(x_1,x_2)=\hat{S}_{0}(x_1,x_2)+ \hat{S}_{0}(x_1,x_1') \odot
  \hat{\Sigma}(x_1',x_2') \odot \hat{S}(x_2',x_2),
\end{equation}
where $\odot$ stands for the integration over repeated space-time
arguments~\cite{Cassing:1999wx}, including the matrix,
$\hat{\eta}=\diag(1,-1)$, acting on the contour indices, which takes
into account the signs originating from the opposite direction of the
$+$-time branch on the Schwinger-Keldysh contour. Here and in the
following we do not write out the Dirac indices explicitly, if not
needed.

Defining the inverse of the free Green's function as the differential
operator,
\begin{equation}
  \label{eq:dyson.2.2}
  S_{0x}^{-1}=\ii \fslash{\partial}_x-m,
\end{equation}
the equation of motion for the free matrix-Green's function can be
written in the form
\begin{equation}
  \label{eq:dyson.3}
  S_{0x_1}^{-1} \hat{S}_{0}(x_1,x_2) = \hat{\eta} \delta^{(4)}(x_1-x_2).
\end{equation}
Thus, from \cref{eq:dyson.2}, the matrix-Dyson equation can be cast
into the following equation of motion
\begin{equation}
  \label{eq:dyson.4}
  S_{0x_1}^{-1} \hat{S}(x_1,x_2) = \hat{\eta} \delta^{(4)}(x_1-x_2) + \hat{\Sigma}(x_1,y)
  \odot \hat{S}(y,x_2).
\end{equation}
For the following, another set of matrix elements than
\cref{eq:dyson.1} is more convenient,
\begin{equation}
  \begin{split}
    \label{eq:dyson.5}
    S^{\text{ret}}(x_1,x_2) & =S^c(x_1,x_2)-S^{<}(x_1,x_2) = S^{>}(x_1,x_2)-S^a(x_1,x_2), \\
    S^{\text{adv}}(x_1,x_2) &= S^{c}(x_1,x_2) - S^{>}(x_1,x_2) =
    S^{<}(x_1,x_2) -
    S^{a}(x_1,x_2),\\
    \Sigma^{\text{ret}}(x_1,x_2) & =\Sigma^c(x_1,x_2)-\Sigma^{<}(x_1,x_2) = \Sigma^{>}(x_1,x_2)-\Sigma^a(x_1,x_2), \\
    \Sigma^{\text{adv}}(x_1,x_2) &= \Sigma^{c}(x_1,x_2) -
    \Sigma^{>}(x_1,x_2) = \Sigma^{<}(x_1,x_2) - \Sigma^{a}(x_1,x_2).
  \end{split}
\end{equation}
It is easy to prove from the canonical anti-commutation relations for
fermion-field operators that these Green's functions obey the boundary
conditions for the retarded and advanced two-point
functions. Accordingly, from \cref{eq:dyson.4} one finds that the
corresponding equations of motion only involve retarded (advanced)
parts,
\begin{equation}
  \label{eq:dyson.6}
  S_{0x_1}^{-1} S^{\text{ret(adv)}}(x_1,x_2)=\delta^{(4)}(x_1-x_2) +
  \Sigma^{\text{ret(adv)}}(x_1,y) \odot S^{\text{ret(adv)}}(y,x_2).
\end{equation}
For the off-diagonal elements in \cref{eq:dyson.2}, one obtains the
equations,
\begin{equation}
  \label{eq:dyson.7}
  S_{0x_1}^{-1} S^{<}(x_1,x_2) = \Sigma^{\text{ret}}(x_1,y) \odot S^{<}(y,x_2) +
  \Sigma^{<}(x_1,y) \odot S^{\text{adv}}(y,x_2),
\end{equation}
which provide the space-time evolution of the single-particle properties
of the particles in the system encoded in the special propagator,
$S^<$. For example, the following spinor traces at equal space-time
coordinates determine the vector-current and scalar densities,
\begin{equation}
  J^{\mu}(x)=\erw{\overline{\op{\psi}}(x) \gamma^{\mu} \op{\psi}(x)}=-\ii
  \tr [\gamma^\mu S^{<}(x,x)], \quad \rho_S(x)=\erw{\overline{\op{\psi}}(x)
    \op{\psi}(x)}=-\ii \tr [S^{<}(x,x)].
\end{equation}
Later, after Wigner transformation and gradient expansion, one
transforms (\ref{eq:dyson.7}) to generalized transport equations for the
corresponding density distributions in phase-space together with
eqs.~(\ref{eq:dyson.6}) for the retarded or advanced Green's functions,
which determine the spectral properties of the single-particle
states~\cite{kadanoffBaym}.

\subsection{Truncation schemes for the self-energies}

The formula (\ref{eq:dyson.7}) provides an exact equation of motion for
the one-particle Green's function. However, this set of equations
requires the definition of the self-energies in terms of the exact
Green's functions, involving an ``infinite tower'' of coupled
Schwinger-Dyson equations for the complete set of $n$-point functions of
the quantum field theory\footnote{In the context of transport theory
  this is analogous to the well-known
  Boguliubov-Born-Green-Kirkwood-Yvon (BBGKY) hierarchy of
  multi-particle phase-space-distribution functions, which take into
  account multi-particle correlations to an arbitrary order.}. Thus, one
has to choose an approximation scheme in order to cut this hierarchy of
Schwinger-Dyson equations to a finite coupled set of $n$-point Green's
and vertex functions. The usual starting point is perturbation theory,
which however is not well suited for our purpose of off-equilibrium
many-body theory.

The main difficulty lies in the formulation of truncation schemes which
are at the same time ``conserving'' (i.e., obeying the conservation laws
for conserved quantities like total energy, momentum, angular momentum,
and charge-like quantum numbers) and are thermodynamically consistent in
the equilibrium limit, i.e., bulk properties fulfill the usual Maxwell
relations between the corresponding thermodynamical potentials.

A promising starting point for the derivation of transport equations for
particles with large spectral widths is the use of $\Phi$-derivable
approximations, where $\Phi$ is a functional of the mean fields and
exact propagators, defining self-consistent equations of motion for the
mean fields and self-energies by a variational principle, which fulfill
the above mentioned physical constraints. It has been shown by Baym that
the derivability of the equations of motion from a functional of this
kind is not only a sufficient but also a necessary condition for
self-consistent approximations fulfilling all these
properties~\cite{Luttinger:1960ua,Baym:1962sx}. In recent years, the
$\Phi$-derivable-approximation technique has been used to derive
transport equations, e.g.,
in~\cite{Ivanov:1998nv,Ivanov:1999tj,Cassing:1999wx,Knoll:2001jx,Ivanov:2003wa}
which go beyond the quasi-particle approximation, which is discussed in
the next section. An investigation of a corresponding test-particle
description has been given
in~\cite{Leupold:1999ga,Cassing:1999wx,Cassing:2008nn}. A comparison of
the dynamics according to the full Kadanoff-Baym equations with the
corresponding off-shell transport equations has been provided, e.g.,
in~\cite{Berges:2005md,DrJuchem,Cassing:2008nn}.

For $\Phi$-derivable approximations, the self-energy is given by the
variational derivative
\begin{equation}
  \label{eq:os-transp.1}
  \Sigma^{\mathcal{C}}(x_1,x_2) = -\ii \frac{\delta \Phi[S^{\mathcal{C}}]}{\delta
    S^{\mathcal{C}}(x_2,x_1)},
\end{equation}
where the (exact) functional, $\Phi[S]$, is diagrammatically defined as
the sum of all closed Feynman diagrams, where internal lines stand for
full propagators, $S$, which cannot be disintegrated into disconnected
parts by cutting two propagator lines, i.e., the two-particle
irreducible (2PI) closed diagrams\footnote{More formally the $\Phi$
  functional can be defined as the interaction part of an effective
  action with both local and bilocal external
  sources~\cite{Cornwall:1974vz}, which provides the possibility to use
  Feynman-path integral methods to, e.g., derive the equations of motion
  or study the symmetry properties of n-particle Green's
  functions~\cite{vanHees:2002bv}.}. The functional derivative
\cref{eq:os-transp.1} then leads to the skeleton expansion for the
self-energy in terms of full propagators, i.e., as the sum over all
self-energy diagrams with propagator lines that do not contain
self-energy insertions. In this way the $\Phi$-derivable approximation
avoids double counting of diagrams automatically. In equilibrium, for
the solution of the self-consistent Dyson equations for the
self-energies at given temperature and chemical potential(s), at the
same time $\Phi$ becomes part of the grand-canonical potential and thus
leads to definitions for the thermodynamical quantities consistent with
the one-particle dynamics provided by the self-consistent equations of
motion for the self-energies. Of course, for practical purposes, the
$\Phi$ functional has to be truncated in some way, e.g., given by the
number of interaction vertices or loops since otherwise it again leads
to the exact equations of motion, involving the full Schwinger-Dyson
hierarchy of equations for all $n$-point Green's and vertex
functions. Sometimes also partial resummations are used for $\Phi$
functionals, e.g., the ring-summation diagrams to obtain
Landau-Fermi-liquid theory or the Dirac-Brueckner approximation given
below. In this way truncation schemes are defined for which the averaged
Noether currents of underlying symmetries (including energy-momentum
conservation from space-time-translation invariance and
current-conservation laws from (global or local) gauge symmetries)
fulfill the conservation laws exactly. At the same time the
self-energies obey the principle of detailed balance, and the scheme is
thermodynamically consistent\footnote{One should however note that
  approximations of this kind are only partial resummations of the full
  Dyson-Schwinger hierarchy, and the so defined self-consistent
  propagators and self-energies violate Ward-Takahashi identities at the
  order of the expansion parameter used for the
  truncation~\cite{vanHees:2002bv,Arrizabalaga:2002hn}.}.

One example for such a scheme, leading to a transport equation for
spin-symmetric nuclear matter which closely resembles the original
Boltzmann equation, is the Dirac-Brueckner
scheme. Following~\cite{Botermans:1990qi} it can be defined by a
self-consistent set of equations for a ladder-resummed four-nucleon
vertex function, $\Gamma$, and the nucleon self-energy, $\Sigma$, which
itself dresses the nucleon-Green's function, $S$, via the Dyson Equation
by
\begin{alignat}{2}
  \label{eq:transp.1}
  S^{\mathcal{C}}(x_1,x_2) &= S_0^{\mathcal{C}}(x_1,x_2) +
  S^{\mathcal{C}}(x_1,x_1') \odot \Sigma^{\mathcal{C}}(x_1',x_2')
  \odot S_0^{\mathcal{C}}(x_2',x_2), \\
  \label{eq:transp.2}
  \Sigma^{\mathcal{C}}(x_1,x_2) &= -\ii
  [\Gamma^{\mathcal{C}}(x_1,x_1';x_2,x_2') -
  \Gamma^{\mathcal{C}}(x_1',x_1;x_2,x_2')] \odot
  S^{\mathcal{C}}(x_1',x_2'),\\
  \begin{split}
    \Gamma^{\mathcal{C}}(x_1,x_2;x_1',x_2') &=
    U^{\mathcal{C}}(x_1,x_2;x_1',x_2')  \\
    \label{eq:transp.3}
    & \quad + \ii U^{\mathcal{C}}(x_1,x_2;x_1'',x_2'') \odot
    S^{\mathcal{C}}(x_1'',x_1''') S^{\mathcal{C}}(x_2'',x_2''')
    \Gamma(x_1''',x_2''';x_1',x_2').
  \end{split}
\end{alignat}
The contour function, $U^{\mathcal{C}}$, stands for two nucleon-meson
vertices connected by a non-interacting meson propagator. The effect
of meson-mean fields, i.e., the Hartree self-energy diagrams, are not
taken into account explicitly but via one-point contributions to the
self-energy. These contribute to the diagonal elements, $\Sigma^{--}$
and $\Sigma^{++}$, of the contour-matrix self-energy (or equivalently
to the retarded and advanced self-energies) and thus are implicitly
included in the spectral function.

The transport equation for the Green's function, $S^{<}(x_1,x_2)$, which
is known as Kadanoff-Baym equation, is derived from \cref{eq:dyson.7}
and its adjoint equation,
\begin{equation}
  \begin{alignedat}{4}
  \label{eq:transp.4}
  & \ii \left (\fslash{\partial}_{x_1}S^{<}(x_1,x_2) +
  S^{<}(x_1,x_2)\overleftarrow{\fslash{\partial}}_{x_2} \right) &-& [\re
  \Sigma^{\text{ret}}\odot S^{<}](x_1,x_2)
  + [S^{<} \odot \re \Sigma^{\text{ret}}](x_1,x_2) \\
  & \quad &-&[\Sigma^{<}\odot \re S^{\text{ret}}](x_1,x_2)
  + [\re S^{\text{ret}}\odot \Sigma^{<}](x_1,x_2) \\
  = &\frac{1}{2} \biggl\{ [\Sigma^{>} \odot S^{<}](x_1,x_2)+ [S^{<}
    \odot \Sigma^{>}]&(&x_1,x_2) - [\Sigma^{<} \odot S^{>}](x_1,x_2)
    -[S^{>} \odot \Sigma^{<}](x_1,x_2) \biggr\}~.
  \end{alignedat}
\end{equation}
The Kadanoff-Baym equation has to be supplemented by the purely
retarded/advanced Dyson-like equations (\ref{eq:dyson.6}), which, as we
will see below, describe the space-time evolution of the spectral
properties of the particles.

In order to obtain \cref{eq:transp.4}, we have used the relations,
\begin{alignat}{2}
     \Sigma^{\text{ret}}(x_1,x_2) &= \re \Sigma^{\text{ret}}(x_1,x_2)
        + \frac{1}{2} \left(\Sigma^{>}(x_1,x_2)
          -\Sigma^{<}(x_1,x_2)\right) \, , \label{Sigma^ret} \\
     \Sigma^{\text{adv}}(x_1,x_2) &= \re \Sigma^{\text{ret}}(x_1,x_2)
        - \frac{1}{2} \left(\Sigma^{>}(x_1,x_2) -\Sigma^{<}(x_1,x_2)\right)   \label{Sigma^adv}
\end{alignat}
and similar for $S^{\text{ret/adv}}(x_1,x_2)$ with $\re\,
O:=\frac12[O+\gamma^0O^{\dagger}\gamma^0]$ and $\im O:=-\ii(O-\re O)$
for covariant spinor-matrices $O$. In other words, the real and
imaginary parts of Dirac-spinor matrices refer to the corresponding
complex coefficients in the usual decomposition into Lorentz tensors of
this quantity,
\begin{equation}
\begin{split}
\label{dirac-decomp}
&O=O_{\text{S}} + \ii \gamma_5 O_{\text{P}} + \gamma_{\mu}
O_{\text{V}}^{\mu} + \gamma_{\mu} \gamma^5 O_{\text{A}}^{\mu} +
\frac{1}{2} \sigma_{\mu \nu} O_{\text{T}}^{\mu \nu} \quad \text{with} \\
&\gamma^5=\gamma_5=\ii \gamma^0 \gamma^1 \gamma^2 \gamma^3, \quad
\sigma_{\mu \nu}=\frac{\ii}{2} \comm{\gamma_{\mu}}{\gamma_{\nu}}~.
\end{split}
\end{equation}
The self-energy contour-matrix elements are determined from the
Dirac-Brueckner scheme \linebreak \crefrange{eq:transp.1}{eq:transp.3}.

\subsection{Wigner transformation and gradient expansion}
\label{sec:offshelltransport}

To further simplify the task of approximately solving the
self-consistent equations of motion for the Green's function, one aims
at a reduction to a semi-classical transport equation, which admits the
use of numerical approaches like Monte-Carlo test-particle simulations.

The general idea behind such reductions is the assumption that the
many-body system, which is in a state where quantum correlations can be
neglected, on ``macroscopic'' space-time scales behaves like a classical
fluid. Then the Green's functions are rapidly oscillating functions of
the relative coordinates, $\xi=x_1-x_2$ (``fluctuations''), while their
variations in $x=(x_1+x_2)/2$ are slow. Thus, it is convenient to
describe the two-point functions, $F(x_1,x_2)$, by their Wigner
transforms\footnote{The Wigner transforms, of course, inherit the
  Dirac-spinor-index structure from the original two-point functions.},
\begin{equation}
  \label{eq:dyson.8}
  \tilde{F}(x,p)=\int \dd^4 \xi \; \exp(\ii p_{\mu} \xi^{\mu}) F
  \left(x+\frac{\xi}{2},x-\frac{\xi}{2} \right).
\end{equation}
The Wigner transform of convolution integrals can be written in terms of
a Lie-derivative like operator, which resembles a relativistic
generalization of the usual Poisson brackets~\cite{Cassing:1999wx}
\begin{equation}
  \label{eq:dyson.9}
  \diamond(\tilde{F}_1,\tilde{F}_2):=\frac{1}{2} \left (\frac{\partial
      \tilde{F}_1}{\partial p_{\mu}} \frac{\partial \tilde{F}_2}{\partial x^{\mu}} - \frac{\partial
      \tilde{F}_1}{\partial x^{\mu}} \frac{\partial \tilde{F}_2}{\partial
      p_{\mu}} \right)=\frac{1}{2} \pb{\tilde{F}_1}{\tilde{F}_2} .
\end{equation}
Then the Wigner transform of a convolution integral reads
\begin{equation}
  \begin{split}
    \label{eq:dyson.10}
    \widetilde{F_1 \odot F_2}(x,p)&=\int \dd^4 \xi \; F_1
    \left(x+\frac{\xi}{2},y' \right) \odot F_2 \left
      (y',x-\frac{\xi}{2} \right ) \exp(\ii p_{\mu} \xi^{\mu}) \\
    &=\exp \left [\frac{\ii}{2} \left (\partial_p^{(1)}
        \cdot \partial_x^{(2)} - \partial_x^{(1)}
        \cdot \partial_p^{(2)} \right) \right ] \tilde{F}_1
    \tilde{F}_2 =: \exp(\ii \diamond)(\tilde{F_1},\tilde{F_2}),
  \end{split}
\end{equation}
where $\partial_{x}^{(j)}$ and $\partial_{p}^{(j)}$ are the
four-gradient operators with respect to $x$ and $p$, acting on $F_j$ ($j
\in \{1,2 \}$) only. According to the above assumption of a many-body
system close to a ``classical state'', one can neglect all gradients in
the variables, $x$, at second and higher orders.

To derive semi-classical transport equations, the coupled equations
\crefrange{eq:transp.4}{Sigma^adv} are now Wigner transformed. The
convolution integrals are approximated by the first-order gradient
expansion, cf.~\cref{eq:dyson.9,eq:dyson.10}, assuming that the Wigner
function depends only slowly on space-time and momentum
\begin{equation}
  \label{eq:transp.10}
  \widetilde{[F_1 \odot F_2]}_{\alpha \beta}(x,p) \simeq \tilde{F_1}_{\alpha \beta'}
  (x,p) \tilde{F_2}_{\beta' \beta}(x,p) + \frac{\ii}{2}
  \pb{\tilde{F_1}_{\alpha \alpha'}(x,p)}{\tilde{F_2}_{\alpha' \beta}(x,p)}.
\end{equation}
Further, if one is only interested in the spin-averaged behavior, one
can take the spinor trace of the gradient expanded Kadanoff-Baym
\cref{eq:transp.4}. Defining the Wigner form of the vector-current
density by
\begin{equation}
  F^{\mu}_{\mathrm{V}}(x,p)=-\ii\tr[\tilde S^{<}(x,p)\gamma^{\mu}]
\end{equation}
and using the trace's cyclic property one arrives at a generalized
transport equation for Dirac particles,
\begin{equation}
\label{KBxpTr}
\partial_{\mu}F^{\mu}_{\mathrm{V}}(x,p)
-\tr \pb{\re \tilde\Sigma^{\text{ret}}(x,p)}{-\ii\tilde S^{<}(x,p)}
+\tr \pb{\re\tilde S^{\text{ret}}(x,p)}{-\ii \tilde\Sigma^{<}(x,p)}
=C(x,p).
\end{equation}
Here the l.h.s.~describes the non-dissipative and current conserving
part of the transport, the Poisson bracket terms providing the drag- and
back-flow response of the surrounding medium. The dissipative part is
given by the collision term,
\begin{equation}
\label{Cxp}
  C(x,p)=\tr \left [\tilde\Sigma^{<}(x,p)\tilde S^{>}(x,p)
  -\tilde\Sigma^{>}(x,p)\tilde S^{<}(x,p) \right].
\end{equation}

The spectral and spinor properties of the propagator components of
$\tilde{S}$ and thus those of $F_{\mathrm{V}}$ are determined by the
retarded \cref{eq:dyson.6} and can be quite complicated. This makes the evaluation of the
traces appearing in \cref{KBxpTr} quite difficult. A major simplification is thus reached
by invoking the local density approximation in which the particles move locally in a
homogeneous medium, with corresponding self-energies (potentials).
This is in line with the gradient approximation. Furthermore, we now restrict ourselves in
all of the following developments to the transport of particles assuming that the width of the
particles is always much smaller than their mass. Antiparticles can be introduced as a
separate particle species.

The spinor matrix of retarded
and advanced propagators then simplifies to
\begin{equation}
\begin{split}
\label{Sret/adv}
\tilde{S}^{\text{ret/adv}}(x,p)\, \gamma^0 &=\left \{ \gamma^0 \left
    [p_{\mu}\gamma^{\mu}-m
    -\tilde{\Sigma}^{\text{ret/adv}}(x,p) \right ] \right \}^{-1} \\
&=\left [p_0-\tilde{H}_{\text{Dirac}}(x,p) -\ii \gamma^0 \im
  \tilde{\Sigma}^{\text{ret/adv}}(x,p) \right]^{-1},
\end{split}
\end{equation}
here written in the non-covariant, i.e., Hamiltonian form with
$\tilde{H}_{\text{Dirac}}(x,p)$ as the corresponding hermitian Hamilton
function.  The above set, \crefrange{KBxpTr}{Sret/adv}, constitutes the
most general level for our transport description of Dirac particles. The
extension to a multi-particle system is straightforward by either
extending the internal degrees of freedom, e.g., to various flavors such
as isospin, or strangeness, etc. or by adding the equations for further
particles. The self-energies then provide the coupling between the
various species.

If the damping terms, $\im \tilde{\Sigma}^{\text{ret}}$, are small
compared to the mass gap, then $\tilde{S}^{\rm ret/adv}(x,p)\gamma^0$
and consequently $S^{</>}\gamma^0$ become approximately diagonal in the
orthogonal eigen-spinors of $\tilde{H}_{\text{Dirac}}(x,p)$
\begin{equation}
\label{eigenspinor}
[E-\tilde{H}_{\text{Dirac}}(x,E,\bvec{p})]
u^r(x,E,\bvec{p})=0 \quad \text{with} \quad
u^{r\dagger}(x,E,\bvec{p}) u^{r'}(x,E,\bvec{p}) = \kappa \delta^{rr'},\quad r,r' \in \{1,2\}~,
\end{equation}
where the choice of the normalization, $\kappa$, is a matter of
convention (in the case $\kappa \ne 1$, factors $1/\kappa$ have to
appear in expressions explicitly using the spinors defined in
\cref{eigenspinor} in order to restore the normalization). We recall,
that in \cref{eigenspinor} the local homogeneous matter approximation is
implied, i.e., it is assumed that the time and space scales, $\Delta t$
and $\Delta r$, respectively, on which the Hamilton function,
$\tilde{H}_{\text{Dirac}}$, varies are much larger compared to the
corresponding scales of variation of the single-particle wave function,
i.e.,
\begin{equation}
   \Delta t \gg E^{-1}, \quad \Delta r \gg |\bvec{p}|^{-1},
\end{equation}
which is equivalent to the assumptions of the gradient expansion. The
weak $x$-dependence of the eigen-spinors, $u^r$, is therefore neglected
in calculating the gradient terms of the Dirac equation with respect to
that of the quickly oscillating plane-wave factor $\propto \exp(-\ii p
x)$.

In the next section and in \cref{subsec:RMFpot} we will discuss the
quasi-particle case with scalar and vector potentials,
$\tilde\Sigma_{\text{S}}$ and $\tilde\Sigma_{\text{V}}^{\mu}$,
respectively, which are defined from the Lorentz decomposition,
\begin{equation}
   \tilde{\Sigma}^{\text{ret}}(x,p)
 = \tilde\Sigma_{\text{S}}(x,p) - \gamma_\mu \tilde\Sigma_{\text{V}}^{\mu}(x,p). \label{LorDecomp}
\end{equation}
In this case,
\begin{equation}
   \label{H_Dirac}
   \tilde{H}_{\text{Dirac}}(x,p)=p^0-\gamma^0(p^{\ast}_{\mu}\gamma^{\mu}-m^{\ast})
\end{equation}
with
\begin{alignat}{2}
  p^{\ast\mu}(x,p) &:= p^{\mu}+\re\tilde\Sigma_{\text{V}}^{\mu}(x,p)~,  \label{eq:pKin} \\
  m^{\ast}(x,p) &:= m+\re\tilde\Sigma_{\text{S}}(x,p)~  \label{eq:mDirac}
\end{alignat}
being the kinetic four-momentum and the Dirac mass in the medium,
respectively.  Multiplying \cref{eigenspinor} by $\gamma^0$ from the
left and using \cref{H_Dirac,eq:pKin,eq:mDirac}, one obtains the
in-medium Dirac equation,
\begin{equation}
  \label{eq:Dirac-medium}
   [p^{\ast\,\mu}(x,E,\bvec{p})\gamma_\mu - m^{\ast}(x,E,\bvec{p})] u^r(x,p^\ast)=0~.
\end{equation}
From this equation, for the given values of $x$ and $\bvec{p}$, the
energy eigenvalue, $E$, can be found as the solution of the dispersion
relation, $p^{*2}=m^{*2}$, or, for the positive-energy solutions,
\begin{equation}
   \label{os-cond}
   E=E^\ast-\re\tilde\Sigma_{\text{V}}^{0}~~~\text{with}~~~E^\ast:=\sqrt{\bvec{p}^{\,*\,2} + m^{*\,2}}~,
\end{equation}
where the kinetic three-momentum, effective mass and the real part of
vector self-energy are supposed to be calculated at $p^0=E$.
\Cref{os-cond} is convenient, because it provides an analytic solution
of the dispersion relation for the special case of the $p^0$-independent
self-energies $\tilde\Sigma_{\text{S}}(x,\bvec{p})$ and
$\tilde\Sigma_{\text{V}}^{\mu}(x,\bvec{p})$ (e.g., in the case of the
nonlinear Walecka model discussed in \cref{subsec:RMFpot}).

In the energy-spin eigenbasis $\tilde{S}^{\text{ret/adv}}\gamma^0$ are diagonal
and traces over them can easily be performed. Under the assumption of
homogeneity, which is the basis for the gradient expansion, and spin
saturation, other contour components of the fermionic propagator are
related to $\im \tilde{S}^{\rm ret}$ by a multiplicative, scalar
function, $f(x,p)$, as
\begin{alignat}{2}
   \label{SlessSret}
   -\ii \tilde{S}^{<}(x,p) &= -2\, f(x,p)\,
   \im \tilde{S}^{\text{ret}}(x,p), \\
   \label{SgtrSret}
   \ii \tilde{S}^{>}(x,p) &= -2\,[1- f(x,p)]\, \im
   \tilde{S}^{\text{ret}}(x,p),
\end{alignat}
such that the analog of \cref{Sigma^ret} is satisfied. The function,
$f(x,p)$, introduced in \cref{SlessSret,SgtrSret} is a Lorentz
scalar\footnote{For a discussion of this covariance see,
  e.g.,~\cite{DeGroot:1980dk}.}. Later on, in \cref{sect:qp} we will see
that it represents the phase-space distribution of particles and becomes
a local Fermi-Dirac distribution in the case of local equilibration
\cite{Botermans:1990qi}. It also regulates the Pauli principle in the
collision terms (cf.~\cref{sect:qp,sec:collisionTerm}). The contour
components of the propagator are thus related to each other through
\cref{SlessSret,SgtrSret} and the contour relations (\ref{eq:dyson.5})
which also apply for the Wigner transformed propagators.

Thus also the diagonal forms of $S^{</>}\gamma^0$ are
now known, and all traces in the transport equations (\ref{KBxpTr}) and
(\ref{Cxp}) can be evaluated. The most important simplification arises
from the fact that now all components of the vector-current density, the
scalar density as well as the density of hole states can be related to
just one Wigner function,
\begin{equation}
   \label{Fcapital}
   F(x,p) = -2 f(x,p) \tr[\im(\tilde{S}^{\text{ret}}(x,p))\gamma^0]~,
\end{equation}
by interaction-dependent kinematical factors, i.e.,
\begin{alignat}{2}
\label{F.vec}
  F_{\text{V}}^{\mu}(x,p) &= \frac{p^{\ast\,\mu}}{E^\ast} F(x,p), \\
\label{F.scal}
F_{\text{S}}(x,p)
&=-\ii\tr[\tilde{S}^{<}(x,p)]=\frac{m^{\ast}}{E^{\ast}} F(x,p).
\end{alignat}
It is important that the kinetic four-momentum in \cref{F.vec} is taken
on the effective (Dirac) mass shell, i.e., $p^{\ast\,0}=E^{\ast}$.  The
Wigner function, $F(x,p)$, has the Lorentz-transformation properties of
a temporal component of a four vector. It is the generalization of the
phase-space density in the sense of the Wigner representation. With the
gradient expansion as a means of ``coarse graining'' it translates into
a positive semidefinite phase-space distribution of particles, which
then can be simulated with the test-particle ansatz
(cf.~\cref{eq:testparticleansatz}). The corresponding (current)
densities are retrieved upon integrating the above expressions over
four-momenta with the weight, $1/(2\pi)^4$, i.e.,
\begin{alignat}{2}
\label{four-current}
    J^\mu(x) &= \int \frac{\dd^4 p}{(2\pi)^4} \frac{p^{\ast\,\mu}}{E^*}  F(x,p), \\
\label{scalar-density}
    \rho_S(x) &= \int \frac{\dd^4 p}{(2\pi)^4} \frac{m^{\ast}}{E^{\ast}} F(x,p).
\end{alignat}
In the quasi-particle case the separation between particles and
anti-particles is given by the sign of the vector potential (see
\cref{subsec:RMFpot} for details).

It is important to verify that the gradient expansion does not destroy
the consistency and conserving properties of the $\Phi$-derivable
approximations for the Kadanoff-Baym equations. Indeed it has been shown
in~\cite{Knoll:2001jx,Ivanov:2003wa} that the
transport~\crefrange{KBxpTr}{Sret/adv} are conserving for
$\Phi$-derivable truncation schemes and admit the definition of an
effective conserved current for each conserved quantity (energy,
momentum, Noether charges). The effective current is identical with the
expectation values of the corresponding Noether currents of the
underlying symmetries. The $\Phi$-derivable schemes also allow for a
definition of the nonequilibrium entropy for the corresponding transport
equations, including collision terms involving multi-particle processes
beyond the $2 \leftrightarrow 2$ level, obeying the
H-theorem~\cite{Ivanov:1999tj}. In the general case, without assuming
the quasi-particle approximation, one has to take into account the full
\cref{KBxpTr}, including the last term on the left-hand side (l.h.s.).
It turns out that this term describes the dynamics of the finite
spectral width of the particles~\cite{Botermans:1990qi}. It becomes then
particularly important to ensure that particles which are stable in the
vacuum but have a large width in the medium due to collisional
broadening fulfill their proper vacuum mass-shell conditions when they
leave the medium during their dynamical evolution (e.g., to describe the
``freeze-out'' of hadrons in heavy-ion collisions dynamically within the
transport approach).

Further, the first-order gradient expansion has to be performed
consistently. This is not trivial, if the truncated $\Phi$ functional
includes diagrams with more than two vertices, i.e., if the self-energy
diagrams contain internal vertices that are not connected with the
amputated legs. These internal vertices give rise to non-local terms
before the gradient expansion, including complicated memory effects (for
an early discussion of the possible importance of non-Markovian effects
in heavy-ion collisions, see~\cite{Greiner:1994xm}). The first-order
gradient expansion can be described diagrammatically in the Wigner
representation (cf.~\cite{Knoll:2001jx}) and can be applied directly to
the $\Phi$ functional, leading to a local approximation,
$\Phi_{\text{loc}}[x,\tilde{S}]$, where $x$ is an arbitrary reference
point, around which the gradient expansion is performed. It is clear
that then $\Phi_{\text{loc}}$ consists only of loop integrals of exact
propagators (and additional lines, representing space-time and
four-momentum derivatives) to be taken at the reference point, $x$. The
loops in the diagrams, including those containing the additional diagram
elements for the derivatives, then stand for momentum integrals as for
homogeneous systems (e.g., nuclear matter in thermal equilibrium). The
self-consistent equations of motion are then given in terms of the
self-energy defined by this local $\Phi$ functional,
\begin{equation}
  \label{eq:os-transp.2}
  -\ii \tilde{\Sigma}(x,p)=-\frac{\delta \Phi_{\text{loc}}[x,\tilde{S}]}{\delta \tilde{S}(x,p)},
\end{equation}
and the general structure of the transport equation for the Wigner
function (\ref{F.vec}) reads
\begin{equation}
  \label{eq:os-transp.3}
  \mathcal{D} F(x,p) + \tr \pb{\re \tilde{S}^{\text{ret}}(x,p)}{-\ii\tilde{\Sigma}^{<}(x,p)} = C(x,p)~.
\end{equation}
Here we have written
\begin{equation}
  \mathcal{D} F = \pb{p_0 - H}{F} \quad \text{with} \quad H=E^*(x,p)-\re
  \tilde{\Sigma}_V^0(x,p).\label{eq:os-transp.4b}
\end{equation}
The term, $\mathcal{D} F$, contains the first Poisson bracket on the
l.h.s.\ of \cref{KBxpTr}. The collision term, $C$, on the right-hand
side (r.h.s.) of \cref{eq:os-transp.3} is defined above by \cref{Cxp}
and can be, generally, calculated via the diagrammatic rules from
$\Phi_{\text{loc}}$. In a first approximation, one can also directly
express it via the elementary transition probabilities related to the
experimentally measured decay widths of unstable particles and
interaction cross sections in vacuum (see \cref{sec:collisionTerm}
below).

Apart from the definition of the Wigner function, $F$, and the
single-particle Hamilton function, $H$, the transport equation
(\ref{eq:os-transp.3}) has the same form as in the non-relativistic case
\cite{Danielewicz:1982kk,Botermans:1990qi,Ivanov:1999tj,Leupold:1999ga,Knoll:2001jx}.

\paragraph{Boson transport} For relativistic bosons the corresponding
equation has been derived in
refs.~\cite{Mrowczynski:1989bu,Cassing:1999wx,Knoll:2001jx}. For
completeness we give it here,
\begin{equation} \label{bosetransport}
\mathcal{D} F(x,p) + \pb{\re \tilde{D}^{\text{ret}}(x,p)}{\ii \tilde{\Pi}^<(x,p)} = C(x,p) ~,
\end{equation}
with
\begin{alignat}{2}
    \label{Dbose}
\mathcal{D} F &= \pb{p^2 - m^2 - \re \tilde{\Pi}^{\text{ret}}(x,p)}{F(x,p)/2 p^0}~, \\
    \label{Cbose}
C(x,p) &=  \tilde{\Pi}^>(x,p) \tilde{D}^<(x,p) - \tilde{D}^>(x,p) \tilde{\Pi}^<(x,p)~,
\end{alignat}
and $F$ defined by
\begin{equation}
F(x,p) = 2 p^0 \ii  \tilde{D}^<(x,p) ~.
\end{equation}
Here $F$ has the same physical meaning as that in \cref{eq:os-transp.3}
and $D$ is the boson propagator.  Eqs.\ (\ref{bosetransport}),
(\ref{Dbose}) differ from \cref{eq:os-transp.3,eq:os-transp.4b} just by
the typically different appearance of the retarded polarization
function, $\tilde{\Pi}^{\text{ret}}$, for bosons.

\paragraph{Spectral functions}

From the definition of the Green's functions in terms of expectation
values of field operators \cref{eq:matrix-gf} and the Fourier
representation of the Heaviside unit-step function,
\begin{equation}
  \label{eq:dyson.11}
  \Theta(t)=\int \frac{\dd p_0}{2 \pi} \exp(-\ii p_0 t)
  \frac{\ii}{p_0+\ii 0}~,
\end{equation}
it is easy to show that the retarded and advanced Green's functions obey
the dispersion relations,
\begin{equation}
  \label{eq:dyson.12}
  \tilde{S}_{\alpha \beta}^{(\text{ret/adv})}(x,p)= \int \dd p_0'
  \frac{\hat A_{\alpha \beta}(x,p_0',\bvec{p})}{p_0-p_0' \pm \ii 0}~,
\end{equation}
with the spectral functions,
\begin{equation}
  \label{eq:dyson.13}
  \hat A_{\alpha \beta}(x,p)= -\frac{1}{\pi} \im(\tilde{S}^{\text{ret}})_{\alpha \beta}
   =\frac{\ii}{2 \pi} [\tilde{S}_{\alpha \beta}^{>}(x,p)-\tilde{S}_{\alpha \beta}^{<}(x,p)]
    = \frac{\ii}{2 \pi} [\tilde{S}_{\alpha \beta}^{\text{ret}}(x,p)
                       - \tilde{S}_{\alpha \beta}^{\text{adv}}(x,p)]~.
\end{equation}
Integrating this equation over $p_0$, from the canonical equal-time
commutation relations for the Dirac-spinor field operators one obtains
\begin{equation}
  \label{eq:dyson.13b}
  \int_{-\infty}^{\infty} \dd p_0 \; \hat{A}_{\alpha
    \beta}(x,p)=\gamma_{\alpha \beta}^0 \,.
\end{equation}
It is now convenient to introduce the real-valued spectral function by
taking the trace of \cref{eq:dyson.13}. Thus we define a spin-averaged
spectral function,
\begin{equation}
\label{spec-func}
  A(x,p):=\frac{1}{g}\tr[\hat A(x,p)\gamma^0]
         = -\frac{1}{g \pi}\tr[\im(\tilde{S}^{\text{ret}}(x,p))\gamma^0]~,
\end{equation}
where $g=2$ denotes the spin degeneracy. Note that $A(x,p)$ has the
Lorentz-transformation properties of a temporal component of a
four-vector. It can be interpreted as an energy distribution function
because of \cref{eq:dyson.13b} and will be discussed further later in
\cref{sec:selfenergies}. Since we are dealing only with the transport of
particles $A$ receives its strength mainly from positive energy states.

Using the definition of the spectral function (\ref{spec-func}) the
Wigner function \cref{Fcapital} can now be rewritten as
\begin{equation} \label{Ffact}
F(x,p)= 2\pi\, g\,f(x,p)\, A(x,p) ~.
\end{equation}
Here the spectral information contained in $A$ has been separated out of
the generalized phase-space distribution $F$.

\subsection{Off-shell transport}

Particles in the medium have a finite lifetime due to collisional
broadening, even if they are stable in the vacuum. Within the local
density approximation their spectral functions depend on
$\rho(\bvec{x})$ and $p$, i.e., they change over the nuclear volume,
e.g., when going from high- to low-density regions. Therefore, one must
ensure that particles leaving the nucleus have returned to their
vacuum-pole mass (if they are stable in the vacuum like the nucleon) or
to their free spectral function (for baryonic resonances, for example).

This is indeed achieved by the additional term on the l.h.s. of
\cref{eq:os-transp.3} that vanishes for on-shell particles. In
\cref{eq:os-transp.3}, the first term includes the usual drift term,
$\propto \partial H/\partial p^\mu$, as well as the Vlasov term,
$\propto \partial H/\partial x_\mu$, which takes into account mean-field
potentials. The second term is not present in the quasi-particle limit.
However, for broad resonances, this term becomes crucial for the
validity of conservation laws in the off-equilibrium dynamics, providing
the back flow of the conserved quantity from the medium into the channel
described by the particular Green's function under
consideration~\cite{Knoll:2001jx,Ivanov:2003wa,Cassing:2008nn}. Unfortunately,
this back-flow term causes problems in practical calculations since it
cannot be easily interpreted within the test-particle ansatz for Monte
Carlo simulations, because it is not proportional to $F$.

A possible resolution of these problems has been developed by Botermans
and Malfliet~\cite{Botermans:1990qi}. In local thermal equilibrium,
i.e., for the solution of the transport equation, for which the
collision term vanishes identically, i.e., if $\tilde{\Sigma}^<
\tilde{S}^{>}=\tilde{S}^{<} \tilde{\Sigma}^{>}$, we have
\begin{equation}
  \begin{split}
    \label{eq:os-transp.6}
    \tilde{\Sigma}_{\text{eq}}^{<}(x,p) &= \ii \Gamma_{\text{eq}}(x,p) f_{\text{eq}}(x,p), \\
    \tilde{\Sigma}_{\text{eq}}^{>}(x,p) &= -\ii \Gamma_{\text{eq}}(x,p)
    [1-f_{\text{eq}}(x,p)]
  \end{split}
\end{equation}
with
\begin{equation}
  \label{eq:os-transp.7}
  \Gamma(x,p)=-2 \im \tilde{\Sigma}^{\text{ret}}(x,p).
\end{equation}
In the off-equilibrium case one can show that in the decomposition,
\begin{equation}
  \label{eq:os-transp.8}
  \tilde{\Sigma}^{<}(x,p)=\ii \Gamma(x,p) f(x,p) + \xi(x,p),
\end{equation}
the contribution of $\xi$ to the back-flow term on the l.h.s. of
\cref{eq:os-transp.3} is of higher order in the gradient expansion,
i.e., it consists of second-order derivatives with respect to the
space-time components, $x$, or of products of first-order
derivatives. Thus, it is consistent with the first-order gradient
expansion to neglect $\xi$ in \cref{eq:os-transp.8} and to approximate
\cref{eq:os-transp.3} by
\begin{equation}
  \label{eq:os-transp.9}
  \mathcal{D} F(x,p) - \tr \pb{\Gamma f}{\re
    S^{\text{ret}}(x,p)} = C(x,p)~.
\end{equation}
Here the back-flow term becomes proportional to $F$ (cf.~\cref{Ffact})
and can be treated within the test-particle
method. Equation (\ref{eq:os-transp.9}) is known as the Botermans-Malfliet form of
the kinetic equation. As the Kadanoff-Baym form it admits the definition
of an effective conserved current, which however deviates from the
corresponding Noether-current expression by higher-order gradient terms,
even if a $\Phi$-derivable approximation is used to define the
self-energies.

\subsubsection{The quasi-particle limit}
\label{sect:qp}

The spectral functions of many particles of interest in nuclear
transport theory are quite narrow. For example, bound nucleons can be
assumed to have spectral functions of negligible width as long as one is
not looking at exclusive events connected with high-momentum tails in
their wave functions.
In this case it is justified to invoke the further simplifications of
the quasi-particle approximation, for which the particles are assumed to
be on the energy shell, i.e., their energy, $E(x,\bvec{p})$, is the
solution of the dispersion relation (\ref{os-cond}).
Then, neglecting again
anti-particles, we can approximate the spectral function (\ref{spec-func})
by a $\delta$-function and introduce the phase-space occupation
probability, $f(x,\bvec{p})$, by the equation,
\begin{equation} \label{f-onshell}
F(x,p) = 2 \pi g \delta[p_0 - E(x,\bvec{p})] f(x,\bvec{p})~,
\end{equation}
where it is assumed that the self-energies do not depend explicitly on
$p^0$ (otherwise so-called $Z$-factors for normalization appear that
contain the derivative of the self-energies with respect to $p^0$).
Using \cref{f-onshell} allows us to rewrite the vector current of
\cref{four-current} in the usual classical form,
\begin{equation}
  \label{eq:transp.26}
  J^{\mu}(x)= g\int \frac{\dd^3
    \bvec{p}}{(2\pi)^3} \frac{p^{*\,{\mu}}}{E^*}\, f(x,\bvec{p})~.
\end{equation}
Since in the quasi-particle approximation $p_0^*=E^*$, this leads to the
interpretation of $g f(x,\bvec{p})/(2\pi)^3$ as the phase-space density
of the nucleon-quasi particles. Also this function is a Lorentz scalar
for (quasi-)particles fulfilling the on-shell condition (\ref{os-cond}),
leading to the current (\ref{eq:transp.26}) as a manifestly covariant
Lorentz vector, as it should be.

Since by assumption the width of the particle, $\Gamma$, is much smaller than its energy
the second term on the l.h.s. of \cref{eq:os-transp.3} can be
neglected since it is proportional to the width of the spectral function. Therefore, inserting
\cref{f-onshell} into the kinetic equation \cref{eq:os-transp.3} leads,
after performing the $p^0$ integration, to the transport equation for
the one-particle phase-space distribution function, $f$,
\begin{equation}
  \begin{split}
    \label{eq:transp.28}
    \Big[ \partial_t + & (\bvec{\nabla}_{\bvec{p}} E_{\bvec{p}}) \cdot
    \bvec{\nabla}_{\bvec{r}} - (\bvec{\nabla}_{\bvec{r}} E_{\bvec{p}})
    \cdot \bvec{\nabla}_{\bvec{p}} \Big ] f(x,\bvec{p}) = \frac{g}{2}
    \int \frac{ \dd^3 \bvec{p}_2 \; \dd^3 \bvec{p}_1' \; \dd^3
      \bvec{p}_2'}{(2 \pi)^9} \frac{m_{\bvec{p}}^* m_{\bvec{p}_2}^*
      m_{\bvec{p}_1'}^*
      m_{\bvec{p}_2'}^*}{E_{\bvec{p}}^* E_{\bvec{p}_2}^* E_{\bvec{p}_1'}^* E_{\bvec{p}_2'}^*} \\
    & \times (2 \pi)^4 \delta^{(3)}(\bvec{p}+\bvec{p}_2 - \bvec{p}_1' -
    \bvec{p}_2') \delta(E_{\bvec{p}}+E_{\bvec{p}_2}-E_{\bvec{p}_1'}-E_{\bvec{p}_2}') \\
    & \times  \overline{|\mathfrak{M}_{p\,p_2 \to p_1'\,p_2'}|^2}\,
    [ f(\bvec{p}_1') f(\bvec{p}_2') \overline{f}(\bvec{p}) \overline{f}(\bvec{p}_2)
    - f(\bvec{p}) f(\bvec{p}_2) \overline{f}(\bvec{p}_1') \overline{f}(\bvec{p}_2')]
  \end{split}
\end{equation}
with $E_{\bvec{p}}:=E(x,\bvec{p})$ and $\overline{f}=1-f$. The collision
term of \cref{eq:transp.28} is written assuming a spin-saturated system
of identical spin-1/2 fermions neglecting the isospin degree of freedom
(cf.~Eq.~(6.109) in ref.~\cite{Botermans:1990qi} for the collision term
for the spin-matrix phase space density). The transition matrix element
squared and averaged over spins of initial particles and summed over
spins of final particles is denoted as $\overline{|\mathfrak{M}_{p\,p_2
    \to p_1'\,p_2'}|^2}$.  The matrix element, $\mathfrak{M}_{if}$, is
taken in the convention of Bjorken and Drell \cite{Bjorken:1979dk} where
the spinors are normalized according to
\begin{equation}
   \label{eq:BD-norm}
   u^{r\dagger}(x,p^\ast) u^{r'}(x,p^\ast)= \frac{E^\ast}{m^\ast} \delta^{rr'}~~~~\text{or}~~~~
   \bar u^{r}(x,p^\ast) u^{r'}(x,p^\ast)= \delta^{rr'}~.
\end{equation}

Equation (\ref{eq:transp.28}) has the typical form of a semi-classical
BUU equation with a drift term containing the non-trivial space-time
dependent quasi-particle dispersion relations. If meson-mean-field
contributions are taken into account, the transport equation has to be
closed with the corresponding mean-field equations of motion, which are
themselves functionally dependent on the phase-space distribution, $f$.

\subsubsection{Spectral functions and self-energies \label{sec:selfenergies} }

In contrast to the case treated in the previous section the transport of
particles with broad spectral width, for example of mesons with short
lifetimes and/or collision-broadened nucleons, requires the solution of
the full transport equations.  In this case the quasiparticle
approximation cannot be used to simplify the equations, but instead the
nontrivial energy-momentum dependence of the spectral function has to be taken into account.

The
spectral function (\ref{spec-func}) becomes for particle (i.e., positive
energy) states (i.e., $|p^{*0}-E^*| \ll E^*$) \cite{Botermans:1990qi}
\begin{equation}
\label{spec-func-approx}
  A(x,p)\approx \frac{1}{\pi} \frac{\Gamma/2}{(p_0^*-E^*)^2 + \Gamma^2/4}~,
\end{equation}
where
\begin{equation}
\label{Gamma}
   \Gamma := -2 \im [(m^*/E^*) \tilde{\Sigma}_{\text{S}}
             + (\bvec{p}^*/E^*) \cdot \tilde{\bvec{\Sigma}}_{\text{V}}
             - \tilde{\Sigma}_{\text{V}0}]
\end{equation}
denotes the total width of a particle in the calculational frame and
$E^\ast$ is given by \cref{os-cond}. The four-vector, $p^\ast$, now also
describes an off-shell particle, i.e., ${p^\ast}^0 \ne E^*$. Thus, within the same approximation, the spectral
function satisfies the normalization condition,
\begin{equation}
   \int\limits_{0}^{+\infty} \dd p^0 A(x,p) \simeq 1 \label{spec_norm}~.
\end{equation}

Most of the early applications of off-shell transport were in
descriptions of broad vector mesons in nuclei. Therefore, the following
form, suitable also for the description of bosons,
\begin{equation}
  \label{A_R}
  \mathcal{A}(x,p)=\frac{1}{\pi} \frac{\sqrt{p^{*2}} \Gamma_\text{med}}{
    [p^{*2} - m^{*2}]^2 + p^{*2} \Gamma_{\text{med}}^2 } \, ,
\end{equation}
is implemented in the GiBUU model.  Here,
$\Gamma_\text{med}=p^{*0}\Gamma/\sqrt{p^{*2}}$ is the particle width in
the rest frame of this particle. This form has been widely used in
hadronic transport theory \cite{Mao:1999cz} and nuclear matter
calculations \cite{Fuchs:2001fp} and is also the one advocated by the
Particle Data Group (PDG) \cite{Hagiwara:2002fs}. Near the quasiparticle
pole, $|p^{*0}-E^*| \ll E^*$, the Breit-Wigner spectral function
(\ref{spec-func-approx}) and the spectral function \cref{A_R} can be
related by
\begin{equation}
  \mathcal{A} \simeq \frac{ A }{2p^{*0}}.    \label{near_peak}
\end{equation}
The spectral function (\ref{A_R}) is normalized according to the
condition
\begin{equation}
  \int_0^{\infty} \dd (p^{*2}) \; \mathcal{A}(x,p) = 1, \label{A_R:norm}
\end{equation}
where the integration is performed at constant kinetic momentum,
$\bvec{p}^*$, cf.~\cref{eq:pKin}.

For simplicity, in this \namecref{sec:selfenergies} and in
\cref{subsec:OSP}, we consider a model without a vector part of the
self-energy, i.e., $p^*=p$. Defining the complex-valued quantity, $\Pi$,
by
\begin{alignat}{2}
  \re \Pi(x,p) &:= m^{*2} - m^2,     \label{Re_Pi} \\
  \label{Im_Pi}
  \im \Pi(x,p) &:=-\sqrt{p^2} \, \Gamma_\text{med},
\end{alignat}
one can rewrite the spectral function (\ref{A_R}) as
\begin{equation}
  \label{eq:spectralfunction_def}
  \mathcal{A}(x,p)=-\frac{1}{\pi}\frac{\im \Pi}{\left[p^{2}-m^2-\re
      \Pi \right]^2+\left(\im \Pi\right)^2}~,
\end{equation}
which formally coincides with a bosonic spectral function with the boson
polarization function, $\Pi$. Assuming the usual analyticity properties
of the function, $\Pi$, in the upper part of the complex $p^0$-plane, we
can express the real part of $\Pi$ via its imaginary part by using a
once-subtracted dispersion relation,
\begin{equation}
  \label{eq:disp}
  \re \Pi(x,p_0,\bvec{p})=\re \Pi(x,E,\bvec{p})+\frac{p_0-E}{\pi}~
  \PV \int^\infty_{-\infty} \dd p_0^\prime \frac{\im
    \Pi(x,p_0^\prime,\bvec{p})}{(p_0^\prime-E)(p_0^\prime-p_0)}~,
\end{equation}
where $E=\sqrt{\bvec{p}^2+m^{*2}}$ is the on-shell single-particle
energy and $\PV$ denotes the Cauchy-principal value.

The total width of a particle in the medium is given by
\begin{equation}
  \label{eq:gamma_med}
  \Gamma_\text{med}(x,p) = \Gamma_\text{decay}(x,p) +
  \gamma \Gamma_\text{coll}(x,p),
\end{equation}
where $\gamma$ denotes the Lorentz-boost factor from the nucleus
rest-frame to the particle rest-frame (details are given in
\refcite{buss_phd}). The decay width, $\Gamma_\text{decay}$, includes
the Pauli blocking (Bose enhancement) factors for the outgoing fermions
(bosons) (c.f.\ the loss term in \cref{C^1_final}). The actual
parameterizations of the vacuum decay widths are described in
\cref{sec:vacDecays_baryon}. If the invariant mass of the particle is
less than the mass of its final decay products, then
$\Gamma_\text{free}=0$. The width of a particle with four-momentum, $p$,
due to collisions with neutrons ($n$) and protons ($p$) in a nucleus can
be evaluated in the low-density approximation as
\begin{equation}
  \label{eq:coll_width}
  \Gamma_\text{coll}(x,p) =\sum_{i=n,p}\int \frac{g \,
    \dd^3\bvec{p}'}{(2\pi)^3} f_i(x,\bvec{p}')
  \sigma_i(p_0,\bvec{p},\bvec{p}') v_\text{rel}(p_0,\bvec{p},\bvec{p}')~.
\end{equation}
Here, $v_\text{rel}$ denotes the relative velocity of the regarded particle
and a nucleon with momentum, $\bvec{p}'$, and $f_i$ are the nucleon phase-space
occupation probabilities (c.f. \cref{f-onshell}).
The total in-medium nucleon-particle interaction cross sections, $\sigma_i$,
in \cref{eq:coll_width} include the Pauli blocking (Bose enhancement)
factors for the outgoing baryons (mesons), in contrast to the cross
sections used in the GiBUU collision term, where these factors are
written separately (cf.~\cref{C^2_final}). Thus, the collisional width,
$\Gamma_\text{coll}$, accounts for additional decay channels of the
particle inside the nucleus.

Let us take the $\Delta$ resonance as an example. Besides the usual
decay, $\Delta \to \pi N$, also the two-body and three-body
interactions, $\Delta N \to \Delta N$, $\Delta N \to N N$ or $\Delta N N
\to N N N$, occur in the nuclear medium. These pure in-medium
interactions add to the total $\Delta$ width. On the other hand, the
partial $\Delta \to \pi N$ contribution is decreased by Pauli blocking
of the outgoing nucleon. Thus, the total in-medium width of the $\Delta$
resonance is determined by a delicate interplay of the new decay
channels and Pauli blocking effects
(cf.~\refcite{Oset:1987re,Ehehalt:1993cx} and also the discussion in
\cref{gibuu_barBar_xsections} before and after \cref{ImSigma_Oset}).

The quantity, $\re \Pi(x,p)$, can now be readily reconstructed from $\im
\Pi(x,p)$ by using the dispersion relation (\ref{eq:disp}). This leads
to a high accuracy in the evaluation of the spectral function (see
Refs.~\cite{buss_phd,Leitner:2008ue} for further details).

\subsubsection{Off-shell propagation: approximations}
\label{subsec:OSP}
The kinetic equation (\ref{eq:os-transp.9}) holds for stable as well as
broad unstable states and thus allows, e.g., to describe the kinetics of
resonances~\cite{Cassing:1999wx,Leupold:1999ga}. For its solution we use
the so-called off-shell potential (OSP) ansatz, a major simplification,
that contains the relevant physics (correct transition from in-medium to
in-vacuum spectral functions). Within this ansatz the effects of the
second Poisson bracket on the l.h.s.~of \cref{eq:os-transp.9} are
absorbed into a modification of the mean-field potential.

In order to outline this method we start our discussion with the
transport equation in which the back-flow term is removed
\begin{equation}
  \label{eq:OSP}
   \pb{p_0 -H(x,p)}{F(x,p)}= C(x,p).
\end{equation}
The l.h.s.~of \cref{eq:OSP} can explicitly be rewritten as
\begin{equation}
  \pb{p_0 -H}{F} =
  \left[\left(1-\frac{\partial H}{\partial p_0}\right) \frac{\partial }{\partial t}
    +\frac{\partial H}{\partial \bvec{p}} \frac{\partial }{\partial \bvec{r}}
    -\frac{\partial H}{\partial \bvec{r}} \frac{\partial }{\partial \bvec{p}}
    +\frac{\partial H}{\partial t} \frac{\partial }{\partial p^0} \right] F(x,p)
  \label{eq:buueq_leftside_expanded}
\end{equation}
with the single-particle Hamilton function,
\begin{equation}
  \label{eq:singlepart_hamiltonian}
  H(x,p)=\sqrt{m^2+\re\Pi(x,p_0,\bvec{p})+\bvec{p}^2} \, ,
\end{equation}
where $m$ is the pole mass of the particle species under
consideration. The self-energy, $\Pi$, contains the effects of the
potential; in the general off-shell case it can be explicitly dependent
on $p_0$.

We note here that while \cref{eq:OSP} is the transport equation for fermions, the analogous
one for bosons (cf.\ \cref{bosetransport})
\begin{equation}
\pb{p^2 - m^2 - \re \tilde{\Pi}(x,p)}{F(x,p) /(2 p^0)} =
\pb{p_0^2 - H^2(x,p)}{F(x,p) /(2 p^0)} = C(x,p)
\end{equation}
reduces to the one for fermions for $|p^{0}-H| \ll H$ with formally the same Hamiltonian $H$,
provided $H$ is not explicitly time-dependent. Thus the following considerations are also directly applicable to bosonic transport.

The generalized BUU \cref{eq:OSP} can be solved numerically by using the
test-particle technique, i.e., the continuous Wigner function is
replaced by an ensemble of test particles represented by
$\delta$-functions,
\begin{equation}
  F(x,p)= \lim_{n(t)\to \infty}\frac{{\left( 2\pi \right) }^4}{N}
  \sum_{j=1}^{n(t)} \delta[\bvec{r}-\bvec{r}_j(t)]
  \delta[\bvec{p}-\bvec{p}_j(t)] \delta[p^0-p^0_j(t)]~,
  \label{eq:testparticleansatz}
\end{equation}
where $n(t)$ denotes the number of test particles at time, $t$, and
$\bvec{r}_j(t)$ and $p_j(t)$ are the coordinates and the four-momenta of
test particle, $j$, at time, $t$. As the phase-space density changes in
time due to both, collisions and the Vlasov dynamics, also the number of
test particles changes throughout the simulation: in the collision term,
test particles are destroyed and new ones are created. At $t=0$ we start
with $n(0)=N \cdot A$ test particles, where $A$ is the number of
physical particles and $N$ is the number of ensembles (test particles
per physical particle). More details about the numerical treatment of
the Vlasov and collision dynamics can be found in \cref{sec:numerics}.

Combining the time derivatives of \cref{eq:testparticleansatz} and
\cref{eq:buueq_leftside_expanded}, we find the equations of motion,
\begin{alignat}{3}
  \frac{\dd \bvec{r}_{j}}{\dd t}=& &&\left(1-\frac{\partial
      H}{\partial p_0}\right)^{-1} \frac{\partial H}{\partial
    \bvec{p}} \label{eq:buu_ham1} \; ,\\
  \frac{\dd \bvec{p}_{j}}{\dd t}=&-&&\left(1-\frac{\partial
      H}{\partial p_0}\right)^{-1} \frac{\partial H}{\partial
    \bvec{r}} \label{eq:buu_ham2} \; ,\\
  \frac{\dd p^0_{j}}{\dd t}=&&&\left(1-\frac{\partial H}{\partial
      p_0}\right)^{-1}\frac{\partial H}{\partial
    t} \label{eq:buu_ham3} \; .
\end{alignat}
Here the first factors on the r.h.s. are just the so-called $Z$-factors
mentioned at the start of \cref{sect:qp}.  If $\partial H /\partial
p_0=0 $, \cref{eq:buu_ham1,eq:buu_ham2} become the usual Hamilton
equations of motion for stable particles. Energy conservation is
enforced by \cref{eq:buu_ham3}, if $\partial H / \partial t
=0$. Numerically, the generalized Hamilton equations of motion
(\ref{eq:buu_ham1})-(\ref{eq:buu_ham3}), are solved with a
predictor-corrector algorithm (see \cref{sec:numerics}).

Since the Poisson-bracket term in \cref{eq:os-transp.9} has been dropped, there is,
so far, nothing in \crefrange{eq:buu_ham1}{eq:buu_ham3} that restores
the proper vacuum properties of a collision-broadened particle when it propagates
out of the nucleus.
The OSP ansatz approximately restores the proper off-shell propagation.
This idea was originally introduced in
\cite{Effenberger:1999uv,effe_phd} in the spirit of an educated guess,
only taking into account the density dependence of the self-energy. An improved version has
been developed in \refcite{buss_phd} with the intent to solve
the non-relativistic test-particle equations of motion of
\cite{Leupold:1999ga}.

To introduce the (relativistic) OSP method we start by defining the
offshellness (or off-shell potential), $\Delta \mu_j^2$, of the
$j^{\text{th}}$ test particle by
\begin{equation}
  p_j^2=m^2+\re\Pi+\Delta \mu_j^2 \;. \label{eq:offshellness}
\end{equation}
Thus, $\sqrt{m^2+\re \Pi}$ corresponds to the in-medium pole mass and,
consequently, $\Delta \mu_j^2$ is a measure of how far the test-particle
invariant mass, $\mu_j=\sqrt{p_j^2}$, is off the pole\footnote{Remember that we
  suppress the particle-species index $i$ while we keep the index $j$
  which denotes the $j^{\text{th}}$ test particle. For brevity, the
  Coulomb potential is dropped in the derivation of the OSP method, but
  it is taken into account in the calculations (see
  \cref{subsec:nonrelPot}).}. Rewriting \cref{eq:offshellness} yields
for the energy of the test particle
\begin{equation}
  p_j^0=\sqrt{m^2+\re\Pi+\Delta\mu_j^2+\bvec{p}^2} \;.
\end{equation}
The effects of the off-shell propagation, contained in the
Poisson-bracket term in \cref{eq:os-transp.3}, are thus absorbed into
the off-shell potential, $\Delta\mu_j^2$.

The OSP ansatz now consists in regulating the offshellness, $\Delta
\mu_j^2$, such that the vacuum behavior is restored when the particles
leave the nucleus. To achieve this, the offshellness can be written as
\begin{equation}
  \Delta \mu_j^2(\bvec{r},p)=\chi_j\, \tilde\Gamma_j(\bvec{r},p),
\end{equation}
where the off-shell parameter, $\chi_j$, is a constant of motion for
each test particle and therefore can be calculated at the time of
creation, $t_0$, as
\begin{equation}
  \chi_j=\frac{\Delta \mu_j^2(\bvec{r}_j(t_0),p_j(t_0))}{\tilde\Gamma_j(\bvec{r}_j(t_0),p_j(t_0))}. \label{eq:offshellpara}
\end{equation}
Here, $\tilde\Gamma_j$ is related to the total width of a particle in
its $\tilde\Gamma_j(\bvec{r},p)=\mu_j\Gamma_{\text{tot}}(\bvec{r},p)$.

As each test particle is defined with its own off-shell parameter, we
require a separate Hamilton function, $H_j$, for each test particle and
replace
\begin{equation}
  H= \sqrt{m^2+\re\Pi+\bvec{p}^2} \to  H_j=\sqrt{m^2+\re\Pi+\Delta \mu_j^2+\bvec{p}^2}.  \label{eq:Hamilton_OSP}
\end{equation}
Each test particle is thus propagated also under the influence of the
off-shell potential, $\Delta \mu_j^2$. Note that this potential, just
like $\chi_j$, is positive for particles above the pole mass
($\mu_j>\sqrt{m^2+\re \Pi}$) and negative for those below. By inserting
this Hamiltonian into the generalized Hamilton equations of motion
(\ref{eq:buu_ham1})-(\ref{eq:buu_ham3}), one directly obtains the
relativistic test-particle equations of motion given in
\cite{Cassing:1999wx}. Lehr \cite{lehr_phd} has shown that for
non-relativistic particles the OSP method is equivalent to the full
solution of the original Kadanoff-Baym equation as outlined in
\cite{Cassing:1999wx,Leupold:1999ga}; for relativistic particles it
approximates the full solution extremely well.

Using the off-shell parameter, the relativistic equations of motion
simplify to
\begin{alignat}{2}
  \dot{\bvec{r}}_j & = \frac{1}{1-C_j}\frac{1}{2E_j}\left(2\bvec{p}_j + \frac{\partial}{\partial\bvec{p}_j}[\re\Pi_j + \chi_j\tilde{\Gamma}_j]\right) \label{eq:eom1} \; ,\\
  \dot{\bvec{p}}_j & = - \frac{1}{1-C_j}\frac{1}{2E_j}\frac{\partial}{\partial\bvec{r}_j} \left[\re\Pi_j + \chi_j\tilde{\Gamma}_j\right] \label{eq:eom2} \; ,\\
  \dot{\chi}_j & = 0 \label{eq:eom4} \\
\text{with} \quad  C_j & = \frac{1}{2E_j}\frac{\partial}{\partial E_j}\left[\re\Pi_j + \chi_j\tilde{\Gamma}_j\right] \label{eq:eom3} \; ,
\end{alignat}
The earlier OSP ansatz, as applied by Effenberger \cite{effe_phd} and
later by Lehr \cite{lehr_phd} and M\"uhlich \cite{Muehlich_phd}, was
much simpler; assuming $\Gamma_{\mathrm{med}} \approx \gamma
\Gamma_{\mathrm{coll}}$ and $\Gamma_{\mathrm{coll}} \propto \rho$ they
chose a linear dependence in density instead of the full in-medium width
in \cref{eq:offshellpara}, which is a considerable simplification from a
numerical point of view. In addition, these authors introduced a
back-coupling term stemming from the influence of the OSP on the
residual nucleus, which restores also the overall energy
conservation. In reactions, where the nucleus stays approximately in its
ground state, the back-coupling term leads only to minor modifications
and can be neglected.

At first sight the implementation of the off-shell equations of motion,
as given above, seems straightforward. In reality, however, a number of
problems appear, which make the implementation of a fully consistent
off-shell transport scheme difficult.

One such problem concerns the momentum dependence of the total
widths. To describe off-shell transport in a consistent manner, one
should use collisional widths in the medium which are compatible with
the collision term used in the transport model. In general such a
collisional width depends on the particle's momentum (relative to the
medium). Also, the equations of motion shown earlier are sensitive to
such momentum dependences, as they contain terms of the form
$\partial\tilde{\Gamma}_j/\partial\bvec{p}_j$. However, these terms can
cause test particles to become superluminous. To show this,
we express the width in terms of variables $(\mu=\sqrt{(p^{0})^{2}-\bvec{p}^2},\bvec{p})$
instead of $(p^0,\bvec{p})$. Then, by combining \cref{eq:eom1,eq:eom3}
and neglecting $\re\Pi_j$ we obtain the following expression
for the test-particle velocity:
\begin{equation}
  \bvec{v}_j = \dot{\bvec{r}}_j = \frac{\bvec{p}_j}{E_j}
  + \left(1-\frac{\chi_j}{2\mu_j}\frac{\partial\tilde{\Gamma}_j(\bvec{r}_j,\mu_j,\bvec{p}_j)}{\partial \mu_j}\right)^{-1}
  \frac{\chi_j}{2E_j}\frac{\displaystyle\partial\tilde{\Gamma}_j(\bvec{r}_j,\mu_j,\bvec{p}_j)}{\partial\bvec{p}_j}
\end{equation}
If the width depends on invariant mass only and not
on three-momentum the velocity simplifies to the classical expression $\bvec{v}_j=\bvec{p}_j/E_j$.
The same always holds true for on-shell particles, which have $\chi_j=0$.
This is expected, since $\re\Pi_j$ is neglected.
However, there is a correction term $\propto \partial\tilde{\Gamma}_j/\partial\bvec{p}_j$
to the classical limit for off-shell particles
if the width depends on three-momentum. In addition, the velocity is not
guaranteed to be smaller than the speed of light any more. This can
become a problem for particles which are already highly
relativistic. For them, even a small contribution from the extra term
can violate the constraint, $v\leq1$. This problem of superluminous
particles has been noted already, e.g., in \cite{Larionov:2002jw} for
pion propagation. Following \cite{effe_phd,lehr_phd,Muehlich_phd}
we circumvent this problem by keeping only the density dependence
of the width and use a momentum-independent value which
fits the full one, obtained from the collision term, on average.

\section{Potentials and collision terms in GiBUU}
\label{sec:BUUingredients}

The relevant degrees of freedom in the GiBUU model are baryons,
mesons, leptons, their anti-particles and the gauge bosons. The
parameters for all hadrons without strangeness and charm are taken
from the $\pi N$ scattering phase-shift analysis of Manley and Saleski
\cite{ManleySaleski}; the parameters for all other particles are taken
from the PDG group \cite{Amsler:2008zzb}.  Besides the nucleon and the
pion, also $\Lambda$, $\Sigma$, $\Xi$, $\Omega$, $\Lambda_C$,
$\Sigma_C$, $\Xi_c$, $\Omega_C$, $\eta$, $J/\Psi$, $K$, $\bar{K}$,
$D$, $\bar{D}$, $D_s^+$ and $D_s^-$ are assumed to be stable because
their lifetimes are much longer than the usual timescales for nuclear
reactions. A complete list of the properties of the implemented
hadrons is given in \cref{sec:particleproperties}.

Neither the leptons nor the electroweak gauge bosons ($\gamma$,
$W^\pm$, $Z^0$) are explicitly propagated in the simulation, i.e., due
to their small couplings it is assumed that they interact only once in
the initial reaction (e.g., $\nu A \to \mu^- X$) or when they are
produced (e.g., $\eta \to \gamma \gamma$) and that they leave the
nucleus undisturbed afterwards.  Gluons are also not propagated, but
one may re-find them in so-called strings or pre-hadrons which carry
information about gluonic excitations (for details
cf.~\cite{Gallmeister:2005ad,Falter:2004uc}).

The following sections describe the potentials used in the drift term of
the BUU equation, the initialization of the nuclear ground state and --
in some detail -- the collision terms.

\subsection{Potentials}
\label{sec:potentials}

The l.h.s.~of the transport equations
\cref{eq:transp.28,eq:os-transp.9,eq:OSP} describes the propagation of
particles under the influence of mean-field potentials of hadronic and
electromagnetic nature. The potentials determine the Hamilton
function, $H$, for the off-shell-transport equations
\cref{eq:os-transp.9,eq:OSP} or, for the quasi-particle approximation
of \cref{eq:transp.28}, the single-particle energy, $E_p$. In most
implementations of BUU potentials are used only for nucleons, but
there are a few investigations of in-medium corrections of hadronic
self-energies \cite{Leupold:2009kz}, mainly for pions and kaons.

In GiBUU, for the nucleons the non-relativistic Skyrme-like and
relativistic mean field \linebreak (RMF) potentials are alternatively
adopted in the quasi-particle approximation. The off-shell transport
equation (\ref{eq:OSP}) is used with the Skyrme-like potentials
only. In this section, we discuss the concrete form of the mean field
potentials used in our model. The applicability of the Skyrme-like or
RMF potentials depends on the physical situation under consideration.

We would like to note here, that the potentials do not only affect the
particle propagation but also the collision term, e.g., via
particle-production thresholds. The hadronic mean-field potentials are
strong and quite sensitive to the local particle-density
variations. Thus their influence on the collision integral via the
energy-momentum conservation and threshold conditions in interparticle
collisions is fully taken into account. However, the relatively weak
and long-ranged Coulomb mean field potential does not vary much within
the range of the strong interaction $\sim 1\fm$. Therefore, we assume
that it does not affect particle reaction rates, i.e., we neglect it
in the collision term.

\subsubsection{Coulomb potential}
\label{subsec:CoulPot}

We take into account the Coulomb force acting on the charged particles
neglecting the Lorentz force as in our applications the latter is much
smaller than the Coulomb force. In particular this holds true for
hadron-induced reactions, where the bulk of the nuclear matter stays
nearly at rest. Therefore, far away from the nucleus the incoming
charged hadron propagates in an almost pure Coulomb field. In
heavy-ion collisions, the current of charged particles, generally,
exists in any frame. However, for low-energy heavy-ion collisions the
Lorentz force is suppressed with respect to the Coulomb force by a
factor of $\sim (v_\text{proj}^{\text{cm}}/c)^2$, where
$v_\text{proj}^{\text{cm}}$ is the projectile velocity in the
center-of-mass frame of the system. For relativistic heavy-ion
collisions, the Coulomb and Lorentz forces may become comparable in
magnitude. But they both are of minor importance for the strongly
violent dynamics of central collisions governed by inelastic
production processes and the build-up of strong hadronic mean fields
(see \cref{sec:heavyIons}).

The Coulomb potential for a given charge distribution,
$\rho_c(\bvec{r})$, is determined by the Poisson equation,
\begin{equation}
  \label{PoissonEq}
  -\nabla^2 \Phi(\bvec{r}) = 4 \pi \rho_c(\bvec{r}),
\end{equation}
Numerically, this equation is solved using the alternating direction
implicit iterative (ADI) \linebreak method of Douglas
\cite{Varga62}. The boundary conditions for the Coulomb potential are
provided by the multipole expansion (cf.~Appendix B of
\refcite{teis_phd}).

\subsubsection{Non-relativistic mean-field potentials}
\label{subsec:nonrelPot}

In the non-relativistic mean-field approach (cf.~\cite{welke,
  Prakash:1996xs,Chabanat:1997qh}) it is convenient to start with the
energy-density functional,
\begin{equation}
  \epsilon(x) = \int \frac{g\, \dd^3 p}{(2\pi)^3}
  \sqrt{m^2+\bvec{p}^2}\,
  [f_n(x,\bvec{p})+f_p(x,\bvec{p})]
  + \epsilon_\text{pot}(x)~, \label{energy_density_NR}
\end{equation}
where the first term represents the bare part of the total energy
density and allows for relativistically correct kinematics, while the
second term, $\varepsilon_\text{pot}$, is the potential part of the
energy density. We apply the Skyrme-like potential-energy density of
Welke \etal{}~\cite{welke}, supplemented with a symmetry-energy term,
\begin{equation}
  \begin{split}
    \label{potential_energy_density_NR}
    \epsilon_\text{pot}(x) =&\mbox{ } \frac{A}{2}\,\frac{\rho(x)^2}{\rho_0}
    + \frac{B}{\gamma+1}\, \frac{\rho(x)^{\gamma+1}}{\rho_0^\gamma} +
    \frac{C}{\rho_0} \sum_{i=n,p} \sum_{j=n,p} \int \frac{g \dd^3
      p_1}{(2\pi)^3} \int \frac{g \dd^3 p_2}{(2\pi)^3}
    \frac{ f_i(x,\bvec{p}_1) f_j(x,\bvec{p}_2) }{ 1 + (\bvec{p}_1-\bvec{p}_2)^2/\Lambda^2 }  \\
    &\mbox{ }+ d_{\rm symm}\frac{(\rho_p(x)-\rho_n(x))^2}{2\rho_0}~.
  \end{split}
\end{equation}
Here and in the following, $\rho_0=0.168\fm^{-3}$ is the nuclear
saturation density.

The single-particle Hamilton function, $H_i(x,\bvec{p}),~(i=n,p)$ ---
which governs the propagation of particles --- is given by the
functional derivative of the total energy, $E$, with respect to the
phase-space occupation numbers,
\begin{equation}
  \begin{split}
    \delta E &=\mbox{ } \sum_{i=n,p} \int \frac{g \dd^3 r \dd^3 p}{(2\pi)^3}
    H_i(x,\bvec{p})\, \delta f_i(x,\bvec{p})~,   \\
    E &= \mbox{ } \int \dd^3 r\, \epsilon(x)~.  \label{E}
  \end{split}
\end{equation}
Thus, we have
\begin{equation}
  H_i(x,\bvec{p}) = \sqrt{m^2+\bvec{p}^2} + U_i(x,\bvec{p}) \label{H_i}
\end{equation}
with the single-particle potential,
\begin{equation}
  \label{U_i}
  U_i(x,\bvec{p}) = U_N(x,\bvec{p}) + d_{\rm symm}\frac{\rho_p(x)-\rho_n(x)}{\rho_0}\tau^3_i~,
\end{equation}
where $\tau^3_p=1$ and $\tau^3_n=-1$. The isospin-averaged nucleon
potential is
\begin{equation}
  \begin{split}
    \label{U_N}
    U_N(x,\bvec{p}) &= \mbox{ } \frac{U_n+U_p}{2} \\
    &= \mbox{ } A \frac{\rho(x)}{\rho_0} +
    B\left(\frac{\rho(x)}{\rho_0}\right)^\gamma + \frac{2C}{\rho_0}
    \sum_{i=n,p} \int \frac{g\, \dd^3 p^\prime}{(2\pi)^3}
    \frac{f_i(x,\bvec{p}^\prime)}{1 +
      (\bvec{p}-\bvec{p}^\prime)^2/\Lambda^2}.
  \end{split}
\end{equation}
In order to reduce the computation
time for calculating the momentum-dependent part of the potential
\cref{U_N}, we approximate the nucleon phase-space distribution by
a Fermi distribution. This allows to evaluate the momentum
integral on the r.h.s.~of \cref{U_N} as an analytic function of
$|\bvec{p}|$ and of the local baryon density
(see \refcite{welke} for details).
The six free parameters, $A$, $B$, $\gamma$, $C$, $\Lambda$, and
$d_{\rm symm}$, of the nucleon potential \cref{U_i}, are determined
from the following conditions:

\begin{itemize}

\item The momentum-dependent interaction should describe the
  phenomenological real part of a proton-nucleus optical potential and
  also governs the Landau effective mass at the Fermi surface
  (cf.~\cite{Cassing:1990dr,Larionov:2000cu} and refs.~therein)
  \footnote{The Landau effective mass at the Fermi surface and the Dirac
    effective mass of the nucleon are related by
    $m^{*2}_\text{Landau}=p_F^2+m^{*2}_\text{Dirac}$. However, to avoid
    complicated notations, we denote both masses simply as $m^*$ or
    $m^*_N$. If not stated otherwise, the Dirac effective mass is used
    in this work.}
  \begin{equation}
    \label{meff}
    (m_N^*)^{-1} = m_N^{-1} + (p_F^{(0)})^{-1}
    \left ( \frac{\partial U_N}{\partial p} \right)_{p=p_F^{(0)}}~,
  \end{equation}
  which should lie in the range $0.6\,m_N \leq m_N^* \leq m_N$.  These
  conditions constrain the stiffness, $C$, and the range, $\Lambda$.

\item The energy per nucleon in isospin-symmetric nuclear matter has a
  minimum at $\rho=\rho_0$ and a value of $-16\MeV$ at the minimum,
  \begin{equation}
    \left.\frac{\partial \epsilon(\rho,I)/\rho}{\partial
        \rho}\right|_{\rho=\rho_0,~I=0} = 0,
    \quad
    \left.\frac{\epsilon(\rho,I)}{\rho}\right|_{\rho=\rho_0,~I=0} = -16 \MeV,
  \end{equation}
  where $I \equiv (\rho_n-\rho_p)/\rho$ denotes the isospin asymmetry,
  and
  \begin{equation}
    \begin{split}
      \label{edens}
      \epsilon(\rho,I) = &\frac{3}{10m_N} ( p_{F,n}^2 \rho_n +
      p_{F,p}^2 \rho_p ) + \epsilon_\text{pot}(\rho,I)
    \end{split}
  \end{equation}
  is the energy density with the Fermi momenta, $p_{F,i}=(3 \pi^2
  \rho_i)^{1/3}$, $(i=n,p)$.

\item The nuclear-matter incompressibility,
  \begin{equation}
    \label{compr}
    K = 9 \rho_0^2
    \left.\frac{\partial^2 \epsilon(\rho,I)/\rho}{\partial \rho^2}
    \right|_{\rho=\rho_0,I=0} \, ,
  \end{equation}
  is fitted to be in the range $K=200\upto380\MeV$.

\item The potential symmetry-energy coefficient, $d_{\rm symm}$, is
  chosen to reproduce the parameter of the symmetry energy,
  $a_{\mathrm{symm}}=28\MeV$, in the Bethe-Weizs{\"a}cker-mass formula
  using the Migdal relation for symmetric nuclear matter,
  \begin{equation}
    a_{\rm symm} = \frac{p_F^2}{6m_N} + \frac{d_{\rm symm}}{2}~,
    \label{MigdalRelation}
  \end{equation}
  which leads to $d_{\rm symm} \simeq 30\MeV$.

\end{itemize}

Several parameter sets of the nucleon potential \cref{U_i} are
collected in \cref{tab:potential}.
\begin{table}[t]
  \centering
  \begin{tabular}{cccccccc}
    \toprule
    Label & K (MeV) & $m_N^*/m_N$ & A (MeV) & B (MeV) & C (MeV) &
    $\gamma$ & $\Lambda$ (fm$^{-1}$) \\
    \midrule
    SM & 215 & 0.68 & -108.6 & 136.8 & -63.6 & 1.26 & 2.13 \\
    HM & 380 & 0.68 & -10.0  & 38.0  & -63.6 & 2.40 & 2.13 \\
    S  & 215 & 1.00 & -287.0 & 233.7 &  0.0  & 1.23 & -    \\
    H  & 380 & 1.00 & -124.3 & 71.0  &  0.0  & 2.00 & -    \\
    MM & 290 & 0.68 & -29.3  & 57.2  & -63.6 & 1.76 & 2.13 \\
    \bottomrule
  \end{tabular}
  \caption{\label{tab:potential} Parameter sets for the
    non-relativistic mean-field potential \cref{U_i}.
    The meaning of the potential labels is as follows:
    the first, S (soft), H (hard) or M (medium) characterizes
    the incompressibility modulus, $K$, and the second, M, if present,
    indicates the inclusion of the momentum-dependent part
    (the third term in the r.h.s.~of \cref{U_i}) in the potential.}
\end{table}

In a similar way, the single-particle Hamilton functions of other
baryons are calculated, except that for them the symmetry-energy terms
are dropped. The potentials of the $\Delta$-resonances are rescaled by
$U_\Delta=2U_N/3$ \cite{Ericson:1988gk,Peters:1998mb}.  The potentials
of $N^*$-resonances are set equal to the nucleon one, i.e.,
$U_{N^*}=U_N$. The potentials of the $S=-1$ baryons are taken as
$U_Y=U_{Y^*}=2U_N/3$, according to a simple light-constituent-quark
counting picture. At least, for the $\Lambda$-hyperon, this results in
the correct potential depths. However, the empirical $\Sigma$-nucleus
optical potential is strongly repulsive at normal nuclear density,
which is currently a quest for microscopic theoretical models
(cf.~\refcite{Friedman:2007zz} and references therein). The potentials
of the $S < -1$ and $C\neq0$ baryons are neglected, since the
empirical information on them is either absent or quite scarce.

In order to avoid problems with Lorentz invariance, it is very
important to choose the frame, where the mean-field potentials should
be calculated. For this purpose, we have chosen the local rest frame
(LRF) of the nuclear medium, where the spatial components of the
baryon four-current, $j^\mu=(j^0,\bvec{j})$, vanish, i.e., $\bvec{j}
=0$ at the space position of the particle under consideration.
This choice is only possible if the antibaryons --- contributing with negative
sign to the baryon four-current (cf. \cref{j_b} below) --- are not abundant.
In hadron-, photon-, and lepton-induced reactions, the Hamiltonian
propagation of particles is performed in the target rest frame
(computational frame), which practically coincides with the LRF, since
the collective motion of nuclear matter is negligible.  However, even
then, conserving energy in particle-particle collisions requires the
determination of the potentials in the center-of-mass (c.m.) frame of the colliding
particles, where the momenta of the final-state particles are chosen.

The situation becomes even more complex in the case of heavy-ion
collisions, where the nuclear matter develops a space-time dependent
collective-flow field.  Already before the collision, due to the
Lorentz-contraction\footnote{In the case of heavy-ion collisions, the
  computational frame is usually chosen as the center-of-mass frame of
  the col\-li\-ding nuclei.}, the baryon density inside nuclei is
increased by a $\gamma$-factor. Therefore, in order to be able to
apply the potential \cref{U_i}, defined in the LRF only, one first has
to perform a Lorentz transformation of the particle four-momentum to
the LRF. This Lorentz transformation is not trivial since it mixes the
spatial components, $\bvec{p}$, with the time component, $p^0$, of the
particle's four-momentum. The latter, however, is not directly known
since the potential is defined only in the LRF.

By using the Lorentz invariance of the quantity $E^2-\bvec{p}^2$ we
obtain the following equation for the determination of the particle
energy $E$ at the given momentum $\bvec{p}$:
\begin{equation}
  E^2-\bvec{p}^2
  = \Big(\sqrt{m^2+\bvec{p}_\text{LRF}^2}+U_i(x,\bvec{p}_\text{LRF})\Big)^2
  - \bvec{p}_\text{LRF}^2~. \label{eq_LRF}
\end{equation}
Here, the three-momentum in the LRF, $\bvec{p}_\text{LRF}$, is
determined by the Lorentz boost as
\begin{equation}
  \bvec{p}_\text{LRF} = \bvec{p} + \gamma_\text{LRF}\,\bvec\beta_\text{LRF}
  \left(\frac{\gamma_\text{LRF}}{\gamma_\text{LRF}+1}
    (\bvec\beta_\text{LRF}\cdot\bvec{p}) - E\right)~, \label{p_LRF}
\end{equation}
with $\gamma_\text{LRF}=1/\sqrt{1-\bvec\beta_\text{LRF}^2}$ and
$\bvec\beta_\text{LRF} = \bvec{j}_b/j_b^0$, where $\bvec{j}_b^\mu(x)$
is the local baryonic four-current
\begin{equation}
  \label{j_b_NR}
  j_b(x)=\int \frac{\dd^4 p}{(2\pi)^4} (1,\bvec{\nabla}_{\bvec{p}}H(x,p)) F(x,p)~,
\end{equation}
where $F(x,p)$ is the Wigner density of \cref{eq:testparticleansatz}.
One observes, that in any other frame than the local rest frame the
potential gets an explicit $E$-dependence. In the actual GiBUU
implementation, we first try to find a solution of \cref{eq_LRF} for
the energy, $E$, by a Newton root-finding algorithm. If this fails, we
switch to a bisection algorithm.

For the single-particle Hamiltonian, $H(x,p)$, on the l.h.s.~of the
transport equation (\ref{eq:OSP}), i.e., for the Vlasov term, the
Coulomb potential of \cref{PoissonEq} is added to the energy, $E$,
calculated from \cref{eq_LRF}. Thus, $H(x,p) = E +
e(1+\tau^3)\Phi/2$. For the particles produced in a given elementary
collision or resonance decay, the Coulomb part is directly included in
\cref{eq_LRF} by adding it to $U_i$.

\subsubsection{Relativistic mean-field potentials}
\label{subsec:RMFpot}

Another type of baryonic potentials is provided by a Relativistic Mean
Field (RMF) model Lagrangian density
\cite{Blaettel:1993uz,Lalazissis:1996rd,Larionov:2007hy,Gaitanos:2007mm}
\begin{equation}
  \begin{split}
    \label{Lagr}
    \mathcal{L} = & \overline{\psi} [ \gamma_\mu ( i\partial^\mu -
    g_\omega \omega^\mu - g_\rho \bvec\tau \bvec\rho^{\,\mu} -
    \frac{e}{2}
    (1+\tau^3) A^\mu )  -  m_N - g_\sigma \sigma ] \psi \\
    & + \frac{1}{2}\partial_\mu\sigma\partial^\mu\sigma - U(\sigma) -
    \frac{1}{4} \Omega_{\mu\nu} \Omega^{\mu\nu}
    +  \frac{1}{2} m_\omega^2 \omega^2 \\
    & - \frac{1}{4} \bvec{R}_{\mu\nu} \bvec{R}^{\mu\nu} + \frac{1}{2}
    m_\rho^2 \bvec\rho^{\,2} - \frac{1}{16\pi} F_{\mu\nu} F^{\mu\nu}~,
  \end{split}
\end{equation}
where $\psi$ is the nucleon field; $\sigma,~\omega^\mu$ and
$\bvec\rho^{\,\mu}$ are the isoscalar-scalar ($I^G=0^+$, $J^\pi=0^+$),
isoscalar-vector ($I^G=0^-$, $J^\pi=1^-$) and isovector-vector
($I^G=1^+$, $J^\pi=1^-$) meson fields, respectively; $A^\mu$ is the
electromagnetic field. The field-strength four-tensors in \cref{Lagr}
are defined by
\begin{alignat}{2}
  \Omega_{\mu\nu} &= \mbox{ } \partial_\mu \omega_\nu
  - \partial_\nu \omega_\mu~,
  \label{F_munu}\\
  \bvec{R}_{\mu\nu} &= \mbox{ } \partial_\mu \bvec\rho_\nu
  -  \partial_\nu \bvec\rho_\mu~,
  \label{R_munu}\\
  F_{\mu\nu} &= \mbox{ } \partial_\mu A_\nu - \partial_\nu
  A_\mu~.  \label{A_munu}
\end{alignat}
The term,
\begin{equation}
  U(\sigma) = \frac{1}{2} m_\sigma^2 \sigma^2 +  \frac{1}{3} g_2 \sigma^3
  +  \frac{1}{4} g_3
  \sigma^4,  \label{U}
\end{equation}
denotes the self-interactions of the $\sigma$-field.

The meson-nucleon coupling constants, $g_\sigma,~g_\omega$, and
$g_\rho$, the meson mass $m_\sigma$, and the self-interaction
coefficients, $g_2$ and $g_3$, are taken from the non-linear (NL)
Walecka-model-parameter sets NL2 of \refcite{Blaettel:1993uz} and NL3
of \refcite{Lalazissis:1996rd}) and are listed in \cref{tab:RMF}.  The
meson masses $m_\omega$ and $m_\rho$ are fixed to the values $783\MeV$
and $763\MeV$, respectively for both parameterizations. The NL2
parameterization is more appropriate to describe heavy-ion collision
dynamics, while NL3 is mostly tuned to the ground-state properties of
nuclei.

\begin{table}[t]
  \centering
  \begin{tabular}{ccccccccc}
    \toprule
    Label & K (MeV) & $m_N^*/m_N$ & $g_\sigma$ & $g_\omega$ & $g_\rho$ &
    $g_2$ (MeV) & $g_3$ & $m_\sigma$ (MeV) \\
    \midrule
    NL2 & 210 & 0.83 & 8.5     & 7.54    & 0.0   & -50.37  & 6.26    & 508.2 \\
    NL3 & 272 & 0.60 & 10.217  & 12.868  & 4.474 & -10.431 & -28.885 & 550.5 \\
    \bottomrule
  \end{tabular}
  \caption{\label{tab:RMF} Parameter sets for the
    RMF models commonly used and their saturation properties in terms of the
    compression modulus K in units of MeV and the effective mass $m_N^*$ in
    units of the bare nucleon mass $m_N$.}
\end{table}

The Lagrange equations of motion for the mean fields read
\begin{alignat}{2}
  [ \gamma_\mu ( i\partial^\mu - g_\omega \omega^\mu - g_\rho
  \bvec\tau \bvec\rho^{\,\mu} -\frac{e}{2} (1+\tau^3)&A^\mu ) - m_N -
  g_\sigma \sigma ] \psi \ =\
  0~,                          \label{DiracEq}\\
  \partial_\mu\partial^\mu\sigma + \frac{\partial
    U(\sigma)}{\partial\sigma}
  &\ =\  - g_\sigma \rho_S~,
  \label{KGsigma} \\
  (\partial_\mu\partial^\mu + m_\omega^2)\omega^\nu
  &\ =\  g_\omega j_b^\nu~,
  \label{KGomega} \\
  (\partial_\mu\partial^\mu + m_\rho^2)\bvec\rho^{\,\nu}
  &\ =\  g_\rho \bvec{j}_I^{\,\nu}~,
  \label{KGrho} \\
  \partial_\mu\partial^\mu A^\nu &\ =\ 4 \pi e
  j_c^\nu~.  \label{Maxwell}
\end{alignat}
The source terms on the r.h.s.~of the meson-field equations
\crefrange{KGsigma}{Maxwell} are the scalar density,
$\rho_S=\erw{\overline{\psi} \psi}$, the baryon current, $j_b^\nu=
\erw{\overline{\psi} \gamma^\nu \psi}$, the isospin current,
$\bvec{j}_I^{\,\nu}= \erw{\overline{\psi} \gamma^\nu \bvec{\tau}
  \psi}$, and the electromagnetic current, $j_c^\nu = \frac{1}{2} (
j_b^\nu +j_I^{3\,\nu} )$.  Here, $\erw{\cdots}$ denotes the
expectation value, and the meson fields in
\crefrange{KGsigma}{Maxwell} are treated classically. We will neglect
the isospin-mixed nucleon states, which results in the conditions,
$j_I^{1\,\nu}=j_I^{2\,\nu}=0$ and $\rho^{1\,\nu}=\rho^{2\,\nu}=0$.

The dispersion relation for a nucleon quasiparticle is obtained from
the plane-wave solution of the Dirac equation (\ref{DiracEq}), $\psi
\propto \exp(-\ii p x)$,
\begin{equation}
  (p^*)^2 - (m^*)^2 = 0~,          \label{dispRel}
\end{equation}
where $p^*=p-V$ is the kinetic four-momentum, and $m^*=m_N+S$ is the
effective mass. The vector and scalar fields are given by
\begin{alignat}{2}
  V &\ =\ g_\omega \omega + g_\rho \tau^3 \rho^3
  + \frac{e}{2} (1+\tau^3) A~, \label{V}\\
  S &\ =\ g_\sigma \sigma~.  \label{S}
\end{alignat}
The dispersion relation \cref{dispRel} can be rewritten as
\begin{equation}
  p^0=\pm\sqrt{(m^*)^2+(\bvec{p}-\bvec{V})^2} + V^0~.  \label{dispRel_expl}
\end{equation}
The nucleon is described by the positive-frequency solution (upper
sign in \cref{dispRel_expl}). Since we treat antinucleons as a separate 
particle species they are also described by the positive sign in \cref{dispRel_expl}. 
However, the sign of the vector potential they experience is opposite to that felt 
by nucleons as can be seen by using the $G$-parity transformation of the nucleon
potentials \cite{GM} \footnote{In the
  antiproton-nucleus reaction simulations (see
  \cref{sec:antiprotonA}), we allow for the deviations from the exact
  $G$-parity symmetry for the sake of a realistic value of an
  antiproton optical potential.}.  Thus, the antinucleon dispersion relation
reads
\begin{equation}
  p^0=\sqrt{(m^*)^2+(\bvec{p}+\bvec{V})^2} - V^0~.  \label{dispRel_antinuc} 
\end{equation}

The source terms in the meson-field equations
\crefrange{KGsigma}{Maxwell} are expressed in terms of the on-shell
particle distribution functions, $f_i(x,\bvec{p})$
(cf.~\cref{eq:transp.28}), by
\begin{alignat}{2}
  \rho_S &= \mbox{ } \frac{g}{(2\pi)^3} \sum_{i=p, n, \bar p, \bar n}
  \int\,\frac{ \dd^3 p }{ p_i^{*\,0}} m^* f_i(x,\bvec{p})~, \label{rho_S} \\
  j_{b}^\mu &= \mbox{ } \frac{g}{(2\pi)^3} \int\, \dd^3 p \left( \sum_{i=p, n}
    \frac{ p_i^{*\mu} }{ p_i^{*\,0} } f_i(x,\bvec{p})
    -\sum_{i=\bar p, \bar n}  \frac{ p_i^{*\mu} }{ p_i^{*\,0} } f_i(x,\bvec{p}) \right)~, \label{j_b} \\
  j_{I}^{3\,\mu} &= \mbox{ } \frac{g}{(2\pi)^3} \sum_{i=p, n, \bar p, \bar n}
  \int\,\frac{ \dd^3 p }{ p_i^{*\,0} } p_i^{*\mu} \tau^3_i
  f_i(x,\bvec{p})~, \label{j_Ij}
\end{alignat}
The kinetic four-momentum is $p_i^*=p_i-V$ for $i=p,n$ and
$p_i^*=p_i+V$ for $i=\bar p,\bar n$, where
$p_i=(p_i^0,\bvec{p})$. Here, $p_i^0$ is the single-particle Hamilton
function dependent on the particle species and its isospin projection.
For completeness, we will also quote a useful formula for the
canonical energy momentum tensor
(cf.~\refcite{BS,Ivanov:1987ee,Blaettel:1993uz}), $T^{\mu\nu}$,
satisfying the continuity equation, \linebreak
$\partial_\nu T^{\mu\nu}=0$:
\begin{equation}
  \begin{split}
    \label{T^munu}
    T^{\mu \nu} =& \mbox{ } \frac{g}{(2\pi)^3} \sum_{i=p, n, \bar p, \bar n}
    \int\,\frac{ \dd^3 p }{ p_i^{*\,0} }
    p_i^{*\,\nu}   p_i^\mu f_i(x,\bvec{p}) \\
    & \mbox{ } + \partial^\mu\sigma \partial^\nu\sigma
    - \partial^\mu\omega^\lambda\partial^\nu\omega_\lambda
    - \partial^\mu\rho^{3\,\lambda}\partial^\nu\rho^3_\lambda
    - \partial^\mu A^\lambda\partial^\nu A_\lambda \\
    & \mbox{ } - g^{\mu \nu} \left(
      \frac{1}{2}\partial_\lambda\sigma\partial^\lambda\sigma -
      U(\sigma) -
      \frac{1}{2} \partial_\lambda\omega_\kappa\partial^\lambda\omega^\kappa
      + \frac{1}{2} m_\omega^2 \omega^2 \right. \\
    & \mbox{ }\left.  -
      \frac{1}{2} \partial_\lambda\rho^3_\kappa\partial^\lambda\rho^{3,\kappa}
      + \frac{1}{2} m_\rho^2 (\rho^3)^2 - \frac{1}{2} \partial_\lambda
      A_\kappa \partial^\lambda A^\kappa \right)~.
  \end{split}
\end{equation}
In practice, \cref{T^munu} is used to extract the equation of state of
the nuclear matter (cf.~\cref{EoS}). It is also useful for the analysis
of the collective velocity profiles \cite{Larionov:2008wy} and the
approach of thermodynamical equilibrium of collective nuclear motions.

As a technical approximation, time derivatives of the meson fields are
neglected in \crefrange{KGsigma}{Maxwell}, while spatial derivatives
are kept. The resulting equations are solved on the spatial grid by
using the alternating direction implicit iterative (ADI) method of
Douglas \cite{Varga62}.  Neglecting the time derivatives also leads to
some modifications in the expression for the conserved energy-momentum
tensor, $T^{\mu \nu}$ (see \refcite{Larionov:2008wy} for details). The
spatial components of the electromagnetic potential, $A^\mu$, are also
neglected. Thus, in fact, the Maxwell equation (\ref{Maxwell}) is
reduced to the Poisson equation (\ref{PoissonEq}) with $\Phi \equiv
A^0$ and $\rho_c \equiv e j_c^0$.

Let us briefly discuss the applicability of these approximations.  
A thorough consideration can be found in
Refs.~\cite{Weber:1990qd,Blaettel:1993uz,Gaitanos:2010fd} (see also
discussion in \cref{subsec:CoulPot} on the applicability of the
Poisson equation).  In fact, including the space and time derivatives
of mesonic fields would lead to small-amplitude oscillations of
these fields with a frequency $\omega=\sqrt{m^2+\bvec{k}^2}$, where $m$
is the meson mass and $\bvec{k}$ is the wave number of a field
perturbation. If the process under consideration is slow with respect
to the period of these oscillations, $2\pi/\omega\simeq2.5(1.5)\fmc$
for $\sigma$-($\omega$-)field in the long-wave limit $k=0$, one can
approximately average-out the time derivatives of the mesonic fields.
The well known examples of such slow processes are the nuclear giant
resonance vibrations or heavy-ion collisions at low and intermediate
energies (below $E_\text{lab}\simeq1\AGeV$).  The classical meson
field radiation and retardation effects are disregarded in the local
density approximation, when all derivatives of the mesonic fields are
neglected in the meson field equations.  On the other hand, in
heavy-ion collisions at relativistic energies, i.e., above
$E_\text{lab}\simeq5\AGeV$, the radiation and retardation effects
might be significant \cite{Weber:1990qd,Blaettel:1993uz}. In the
low-energy nuclear dynamics, the surface effects are important for the
stability of the ground state nuclei. Including the space derivatives
of the meson fields largely improves the description of a nuclear
surface and, therefore, the nuclear ground state stability (see
\refcite{Gaitanos:2010fd} and \cref{sec:groundstate}).

\subsubsection{Comparison of the non-relativistic and relativistic
  mean-field potentials}
\label{subsec:nonrel_vs_rel}

\begin{figure}[t]
  \centering
  \includegraphics[width=0.6\linewidth]{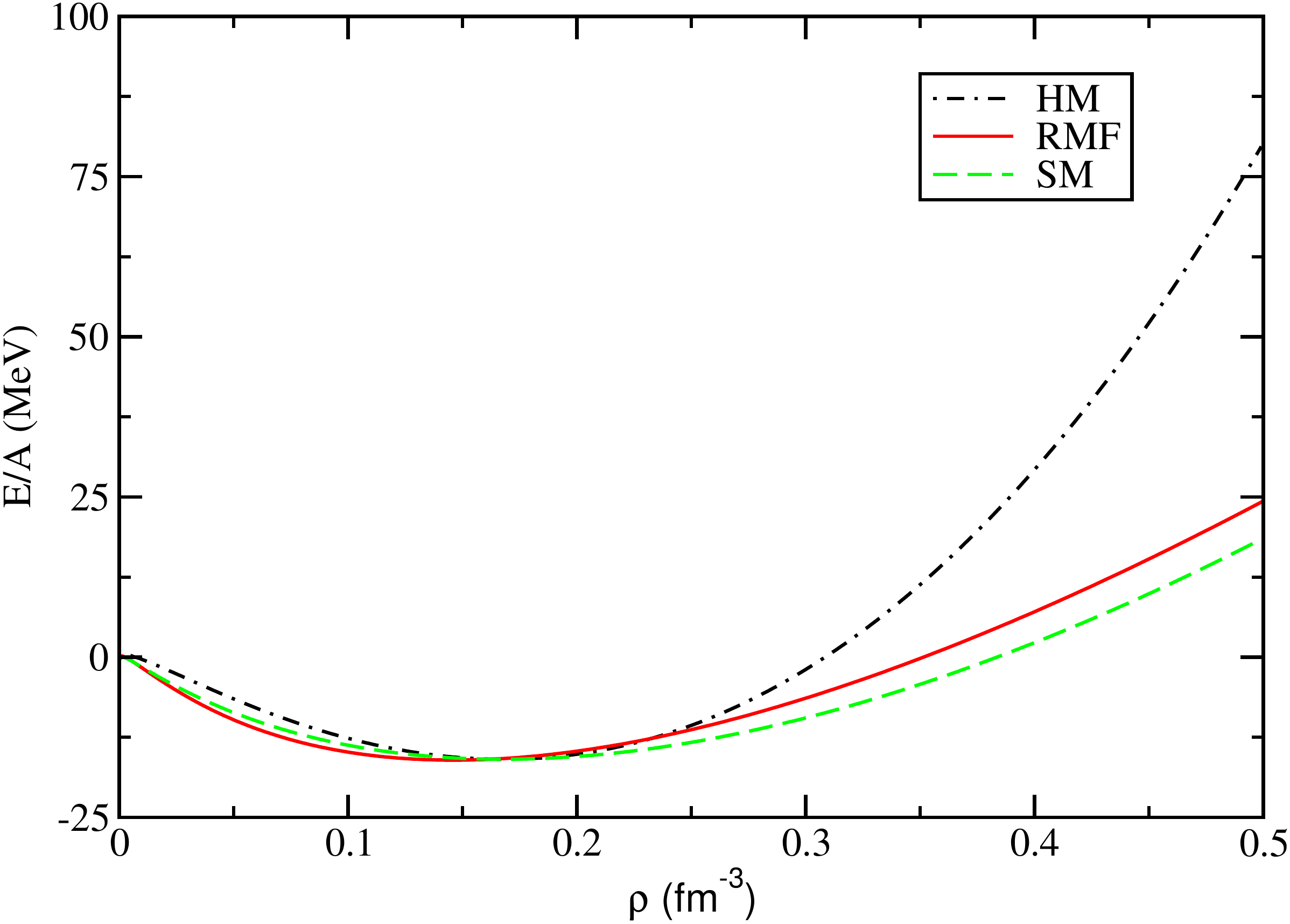}
  \caption{(Color online) \label{EoS}Binding energy per nucleon,
    $E/A$, as function of the baryon density, $\rho$, for different
    mean-field parametrizations used in the GiBUU model: (SM)
    non-relativistic soft momentum-dependent Skyrme, (HM)
    non-relativistic hard momentum-dependent Skyrme, (RMF) non-linear
    Walecka model (NL2 set,~\cite{Blaettel:1993uz}).}
\end{figure}

\begin{figure}[t]
  \centering
  \includegraphics[width=0.6\linewidth]{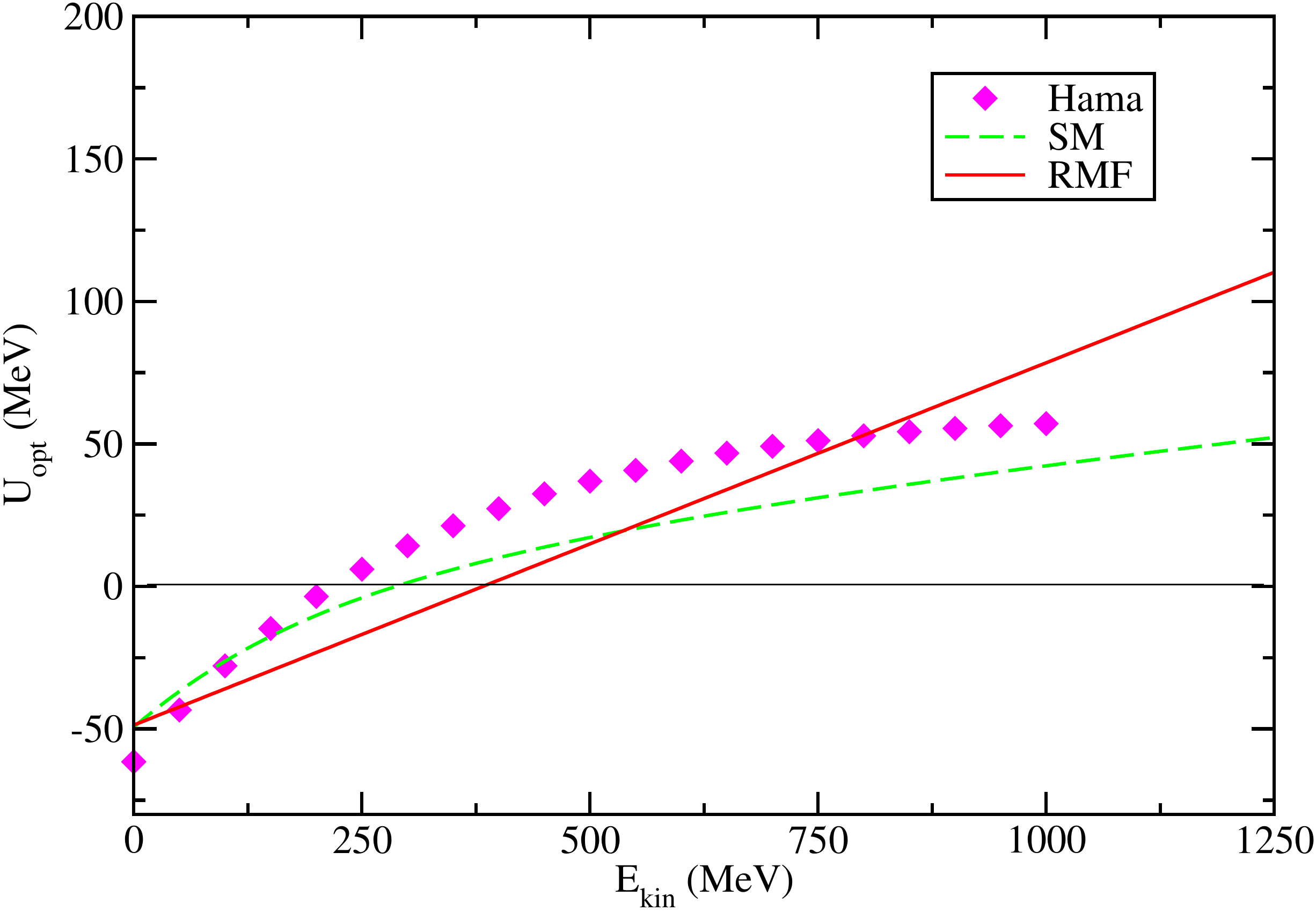}
  \caption{\label{Uopt} (Color online) Kinetic-energy dependence of
    the Schr\"odinger-equivalent optical potential for the Skyrme-like
    interaction (SM) and for the NL2 parameter set of the non-linear
    Walecka model (RMF) for baryon densities at saturation.  These
    potentials are compared with the results of Dirac phenomenology
    \cite{Cooper:1993nx}.}
\end{figure}

The key quantities, which influence the dynamical evolution in
heavy-ion collisions at incident energies up to several GeV per
nucleon, are the equation of state (EoS) and the optical
potential. The study of the nuclear EoS has a long history. It has
been pioneered by the work of Scheid, Ligensa and Greiner
\cite{PhysRevLett.21.1479}, where the compressibility of nuclear
matter has been extracted from light-ion scattering data.  Later, the
extensive studies of different theory groups
(cf.~\refcite{Stoecker:1986ci,Blaettel:1993uz,welke,Cassing:1990dr,%
  Larionov:2000cu,Aichelin:1991xy,Aichelin:1986ss,BertschGupta,%
  Russkikh:1993ct,Maruyama:1993jb,FGW96,Sahu:1998af,Hombach:1998wr,%
  Danielewicz:1999zn,Fuchs:2000kp, Danielewicz:2002pu,Klahn:2006ir})
have been performed to pin down the nuclear EoS from comparison of
microscopic transport simulations with experimental data on collective
flow and particle production.  Taking into account the realistic
momentum dependence of the proton-nucleus optical potential, this
resulted in values of $K \simeq 200\upto270\MeV$, i.e., in a
relatively soft EoS. This range agrees with model analyses of data
on the giant monopole resonance in heavy nuclei
\cite{Blaizot:1980tw,Sharma:2008uy} (see also
\refcite{Gaitanos:2010fd} and refs.~therein).

The EoS of nuclear matter is defined by the density dependence of the
energy per nucleon subtracting the nucleon mass. In the
non-relativistic mean-field model, the EoS is calculated using
\cref{edens}, as $E/A=\epsilon(\rho,I=0)/\rho$.  In the RMF approach
the EoS is determined through the $00$-component of the
energy-momentum tensor, $T^{\mu\nu}$ (see \cref{T^munu}):
$E/A=T^{00}/\rho$.

For a comparison of the energy dependence between the various
mean-field models (Skyrme-like and RMF) we use the
Schr\"{o}dinger-equivalent (SE) optical potential. For the case of the
in-medium interaction of a nucleon with nuclear matter at rest the SE
optical potential is derived from the in-medium dispersion
relation~\cite{Jaminon:1981xg}. In the RMF case, the in-medium
dispersion relation reads $(E-V^0)^{2}-\bvec{p}^2=(m_N+S)^2$, where
the vector ($V$) and scalar ($S$) fields are given by \cref{V,S},
respectively. Here, $E$ denotes the energy of a nucleon in the rest
frame of the nuclear matter at a given baryon density, $\rho$. In the
Skyrme-like potential model, the dispersion relation takes the form
$(E-U_N)^{2}-\bvec{p}^2=m_N^2$, where $U_N$ denotes the
isospin-averaged nucleon potential of \cref{U_N}.

From these dispersion relations one then obtains the in-medium
single-particle energy, $E$, of a nucleon as a function of its
momentum, $\bvec{p}$.  Using energy conservation, i.e., setting the
in-medium single-particle energy, $E$, equal to the asymptotic free
energy, one arrives at the well known form for the SE optical
potential~\cite{Jaminon:1981xg}, which in the RMF case reads
\begin{equation}
  U_{\rm opt} =  \frac{E}{m_{N}} V^0 + S
  + \frac{1}{2m_{N}} \left[ S^{2} - (V^0)^2 \right]
  \:. \label{V_opt}
\end{equation}
In the non-relativistic case, one obtains a similar equation as in
\cref{V_opt} with $S=0$ and $V^0=U_N$.

\Cref{EoS,Uopt} show the EoS and the kinetic-energy, dependence
($E_\text{kin}=E-m_N$) of the optical potential for typical
parametrizations commonly adopted in the calculations, shown later on.
The EoS in the SM and RMF parametrizations show a rather soft behavior
at high densities. For comparison, a hard EoS (HM) is shown too. A
soft density dependence is required for a reliable description of
collective baryon flows and kaon abundances at SIS/GSI energies
(cf.~\cite{Gale:1989dm,Fuchs:2000kp}). Also the energy dependence of
the nuclear mean field is crucial for a consistent characterization of
collective in-plane and out-of-plane transverse flow of matter at
energies around $0.1\upto1\GeV$ per nucleon.

The non-relativistic Skyrme model predicts a saturation of the optical
potential with increasing nucleon energy, which is consistent with the
empirical Dirac phenomenology \cite{Cooper:1993nx}. This originates
from an explicit momentum dependence of the non-relativistic
single-particle Skyrme potential, which is not present in the
relativistic version. In fact, in the latter case the RMF potential
rises linearly with energy. At the energies in the SIS/GSI regime,
i.e., $E_\text{lab}\simeq1\AGeV$, the comparison of the
Schr\"odinger-equivalent potential of the RMF model with the empirical
data \cite{Cooper:1993nx} is quite satisfactory. However, at higher
beam energies the linear rise of the potential leads to a strong
repulsion, which seems to be unphysical and limits the applicability
of this simple version of the RMF model to situations where the
relative motion of the components of the nuclear system is not very
fast.  One example of such situations is the thermalized stage of
heavy-ion collisions. Due to a rather fast equilibration in momentum
space the overlap region of colliding nuclei thermalizes
quickly. Thus, we expect that, except for the relatively short
nonequilibrium stage when the two nuclei penetrate each other,
the RMF model describes the time evolution of the strongly compressed
and heated nuclear system quite reliably.

\subsubsection{Transport equations in the RMF model}
\label{subsec:RMFtransport}

In the case of momentum-independent fields in the spirit of a
Relativistic Mean-Field (RMF) model with a scalar field, $S(x)$, and a
vector field, $V^\mu(x)$, it is convenient to use the distribution
function, $f^*(x,\bvec{p}^*)$, in the kinetic phase space
$(\bvec{r},\bvec{p}^*)$. The kinetic (or effective) four-momentum
$p^{*\mu}$ and the effective mass are defined, respectively, as
(cf.~eqs.~(\ref{eq:pKin}) and (\ref{eq:mDirac}))
\begin{equation}
  p^{*\mu} = p^\mu - V^\mu~.           \label{p^star}
\end{equation}
and
\begin{equation}
  m^* = m + S~,                        \label{m_i^star}
\end{equation}
so that ${p^*}^2=(m^*)^2$ (in-medium on-shell condition).  Now,
$p^{0}=\sqrt{(m^*)^2+(\bvec{p}^*)^2}+V^0$ plays the role of a
single-particle Hamilton function, i.e., $E_{\bvec{p}}\equiv p^{0}$ in
\cref{eq:transp.28}.

The normalization of the distribution function, $f^*(x,\bvec{p}^*)$,
is such that
\begin{equation*}
  f^*(x,\bvec{p}^*) \frac{g \, \dd^3 r \; \dd^3 p^*}{(2\pi)^3}
\end{equation*}
is the number of particles in a kinetic phase-space element, $\dd^3
\bvec{r} \; \dd^3 \bvec{p}^*$. Here, $g$ denotes the spin-degeneracy
factor. Since the Jacobian of the transformation, $(\bvec{r},\bvec{p})
\to (\bvec{r},\bvec{p}^*)$, fulfills
\begin{equation*}
  \det|\partial(\bvec{r},\bvec{p}^*)/\partial(\bvec{r},\bvec{p})|=1\quad,
\end{equation*}
we obtain $f^*(x,\bvec{p}^*)=f(x,\bvec{p})$. With this transformation
of variables, \cref{eq:transp.28} can be expressed as
(cf.~Refs.~\cite{Blaettel:1993uz,Larionov:2007hy})
\begin{equation}
  \label{BUUstar}
  (p^{* 0})^{-1}
  \left[ p^{* \mu} \partial_\mu - ( p_{\mu}^* {\cal F}^{\mu\alpha}
    - m^* \partial^\alpha  m^* )
    \frac{\partial}{\partial p^{* \alpha}} \right] f^*(x,\bvec{p}^*)
  = C(x,p)~,
\end{equation}
where $\alpha=1,2,3$, $\mu=1,2,3,4$, and ${\cal F}^{\mu\nu}
\equiv \partial^\mu V^\nu - \partial^\nu V^\mu$.

For momentum-independent scalar and vector fields, \cref{BUUstar} is
solved by representing the distribution function, $f^*$, in terms of
test particles, i.e., setting
\begin{equation}
  \label{testPart^star}
  f^*(x,\bvec{p}^*) = \frac{ (2\pi)^3 }{ g N }
  \sum_{j=1}^{n(t)} \delta^{(3)}[ \bvec{r} - \bvec{r}_j(t) ]
  \delta^{(3)}[\bvec{p}^* - \bvec{p}_j^*(t) ]~.
\end{equation}
This test-particle representation is similar to that of the more
general off-shell potential ansatz, \cref{eq:testparticleansatz},
except that the energy delta function, $\delta(p^0-p_j^0)$, is
integrated-out. This is because --- in the RMF mode --- we restrict
ourselves to the quasi-particle limit.

The respective equations of motion for the centroids, $\bvec{r}_j$,
and kinetic momenta, $\bvec{p}_j^*$, are obtained by substituting
\cref{testPart^star} into \cref{BUUstar} and setting $C(x,p)=0$, which
leads to
\begin{alignat}{2}
  \dot{\bvec{r}}_j &= \mbox{ } \frac{\bvec{p}_j^*}{p_j^{* \, 0}}~, \label{r_j^dot_RMF} \\
  \dot{p}_j^{* \alpha} &=\mbox{ } -(p_j^{* 0})^{-1} ( p_{j\mu}^* {\cal
    F}^{\mu\alpha} - m^* \partial^\alpha m^*
  )~, \label{p_j^stardot_RMF}
\end{alignat}
where $p_j^{* 0} = \sqrt{(m^*)^2 + (\bvec{p}_j^*)^2}$, and $\alpha =
1,2,3$.

\subsection{Nuclear ground state}
\label{sec:groundstate}

The BUU equation (\ref{eq:transp.28}) is a first-order differential
equation in time. Thus, to solve this equation, the phase-space
distributions of all particle species need to be known at the initial
time $t=0\fmc$, in particular, the initial distribution of the
nucleons within the nucleus (nuclear ground state). As discussed in
the previous section, the BUU equation is treated numerically within
the test-particle method, where the phase-space distribution is
discretized by a finite number of test particles, see
\cref{eq:testparticleansatz,testPart^star} for the non-relativistic
and relativistic cases, respectively. This method leads to Hamilton
equations of motion for the test particles,
\cref{eq:buu_ham1,eq:buu_ham2,eq:buu_ham3,r_j^dot_RMF,p_j^stardot_RMF},
again for the non-relativistic and relativistic cases. In the spirit of
the test-particle ansatz one has to initialize the test particles for
a nucleus at the initial time. Empirical density distributions are
usually adopted to initialize ground-state nuclei in
transport-theoretical simulations, which might not always be
consistent with the energy-density functional used for the propagation
of the system. However, in some particular cases, e.g., low-energy
hadron-induced reaction and fragmentation studies, a very good
stability of the ground-state configurations is required, which is
difficult to reach within the test-particle technique, underlying any
numerical method to solve the BUU equation. Another well known problem
is related to the numerical treatment of Pauli-blocking factors in the
Uehling-Uhlenbeck collision integral. In this Chapter, first we give
an outline of the standard-initialization procedure of nuclear ground
states, before model extensions and results are presented.

\subsubsection{Standard phase-space initialization}
\label{subsec:groundstate-stand}

In order to prepare the phase-space density of the nuclear ground
state, the coordinates of neutrons and protons are sampled according
to empirical density profiles of Woods-Saxon or harmonic oscillator
\cite{DeJager:1974dg} type, for heavy or light nuclei, like \gold{}
and \nickel{} or \carbon{} and \oxygen{}, respectively.

The particle momenta are distributed according to a local
Thomas-Fermi (LTF) approximation,
\begin{equation}
  f_{n,p}(\bvec{r},\bvec{p})=
  \Theta\left[p_{F,n,p}(\bvec{r})-\left|\bvec{p}\right| \right ],
  \label{eq:oneParticleFSD}
\end{equation}
where the momentum distribution is given by an isotropic Fermi sphere
at each point in space with the radius in momentum space determined by
the local Fermi momentum,
\begin{equation}
  p_{F,n,p}(\bvec{r})=[3\pi^2\rho_{n,p}(\bvec{r})]^{1/3}.  \label{eq:fermimom}
\end{equation}
The normalization is chosen such that the proton and neutron densities
--- which serve as an input --- are retrieved by
\begin{equation}
  \rho_{n,p}(\bvec{r})=g~  \int f_{n,p}(\bvec{r},\bvec{p})~ \frac{\dd^3 p}{(2\pi)^3}~.
\end{equation} 
The single-particle phase-space densities,
$f_{n,p}(\bvec{r},\bvec{p})$, are then fully determined, and the
momentum densities read
\begin{equation}
  n_{n,p}(\bvec{p})=g~\int f_{n,p}(\bvec{r},\bvec{p}) \frac{\dd^3 r}{(2\pi)^3}
\end{equation}
with the normalization conditions, $\int \dd^3 p \; n_p(\bvec{p}) =Z$
and $\int \dd^3 p \; n_n(\bvec{p}) =A-Z$.

This standard method provides us with the full phase-space information
at the initial time, before starting the propagation according to the
Hamilton equations of motion.  Smooth distributions in coordinate and
momentum space are achieved by using $\sim10^3$ test particles per
nucleon, which is an important issue for the numerical treatment of
mean-field gradients in the Hamilton equations of motion, for details see 
\cref{sec:Vlasov}.  The smoothness of the test particle distribution
in phase space is also important for numerical evaluation of the
Pauli-blocking factors, $(1-f_{n,p}(\bvec{r},\bvec{p}))$, which enter
the collision term of the BUU equation (see discussion related to
\cref{fig:pauli} and \cref{sec:Pauli}).

\begin{figure}[tbp]
   \centering
  \includegraphics[width=0.6\linewidth]{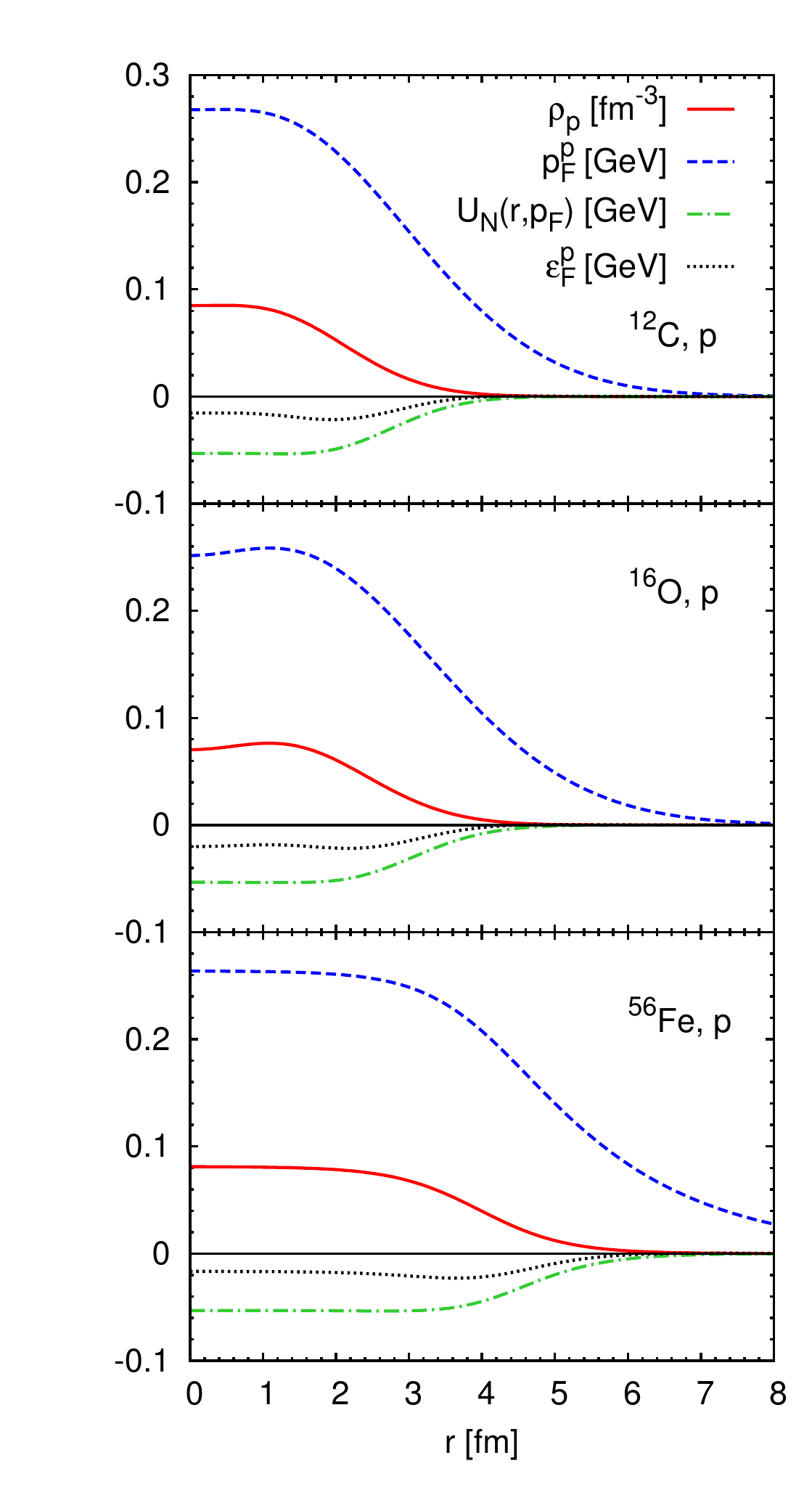}
  \caption{(Color online) The proton density, Fermi momentum,
    mean-field potential and Fermi energy subtracting the nucleon mass
    for $^{12}$C, $^{16}$O and $^{56}$Fe as a function of
    radius.\label{fig:dens_pF_eF_V}}
\end{figure}

In \cref{fig:dens_pF_eF_V} we show the proton density and the Fermi
momentum together with the proton mean-field potential \cref{U_i} and
the resulting Fermi energy
\begin{equation} \label{E_F}
  E_{F,p}=\sqrt{p_{F,p}^2+m_N^2}+U_p(\bvec{r},p_F)
\end{equation}
for \carbon{}, \oxygen{} and \iron{} as a function of the radius at
the initial time $t=0\fmc$.  It is seen in \cref{fig:dens_pF_eF_V}
that the Fermi energy varies rather strongly towards the nuclear
surface and becomes very small there. This leads to problems with
nuclear stability, which are addressed in the next section, where an
improved initialization method is presented.  A similar scheme in the
non-relativistic case has been discussed in
\cite{steinmueller_diplom}.

\subsubsection{Improved phase-space initialization and ground-state
  stability}
\label{subsec:groundstate-improved}

We address the problem of the ground-state stability, which is
discussed in detail in Section 3.2 of \refcite{effe_phd} and in
\refcite{steinmueller_diplom}. The empirical density distributions do
not coincide with the ground-state density distribution corresponding
to the static solution of the Vlasov equation with a local mean field
potential in coordinate space.  The assumed locality of the nucleon
potential \cref{U_i} requires the ground state to be a perfect sphere
of a constant density $\rho_0$. This discrepancy leads to oscillations
of the root-mean-square radius in time (cf.~e.g.~Figs.~3.2 and 3.6 in
\cite{effe_phd}). It has been shown in \refcite{effe_phd} that the
influence on observables in photon-induced reactions is only
minor. Nevertheless, we circumvent this problem by working in the
so-called ``frozen'' approximation, i.e., the test particles which
define the initial particle distribution of the nucleus (the so-called
``real particles'') are not propagated and are not allowed to undergo
any collisions --- they are frozen. Thus, by definition, we obtain a
stable ground state. This treatment is justified by the fact that in
photon- and lepton-induced reactions at around $1\GeV$ beam energy the
nucleus stays close to its ground state, which means that its
phase-space density stays almost constant during the simulation. Only
the reaction products of the lepton-nucleon reactions are propagated
and undergo final-state interactions. These test particles and also
their reaction products are called ``perturbative particles'', and they
do not affect the nucleus phase-space density (see \cref{sec:realPert}
for details).

The ``frozen'' approximation is not applicable to hadron-induced
reactions and heavy-ion collisions at intermediate relativistic
energies. Thus, a good stability of ground-state nuclei has to be
achieved, in particular, when studying reactions with hadron and
heavy-ion beams at low incident energies. We thus have improved the
relativistic transport approach within GiBUU by performing
Thomas-Fermi calculations with the same energy-density functional as
that used in the dynamical evolution. The relativistic Thomas-Fermi
(RTF) equations for a static nucleus with $Z$ protons, $N$ neutrons,
i.e., $A=N+Z$ nucleons, are obtained by applying a variational
principle to the total energy, $E=\int \dd^3 \bvec{r} \;
\epsilon(\rho_p,\rho_n)$, under the constraint of particle number
conservation,
\begin{equation}
  \delta
  \int \dd^3 \bvec{r}
  \left [
    \epsilon(\rho_p,\rho_n)
    -\mu_{p}\rho_{p}(\bvec{r})
    -\mu_{n}\rho_{n}(\bvec{r})
  \right ]
  = 0~.
  \label{var}
\end{equation}
The chemical potentials for protons and neutrons, $\mu_{p,n}$, are
fixed by the conditions
\begin{equation}
  Z = \int \dd^3 \bvec{r} \; \rho_p(\bvec{r}), \quad
  N = \int \dd^3 \bvec{r} \; \rho_n(\bvec{r}) .
  \label{chempot}
\end{equation}

\begin{figure}[t]
  \centering
  \includegraphics[width=0.6\linewidth]{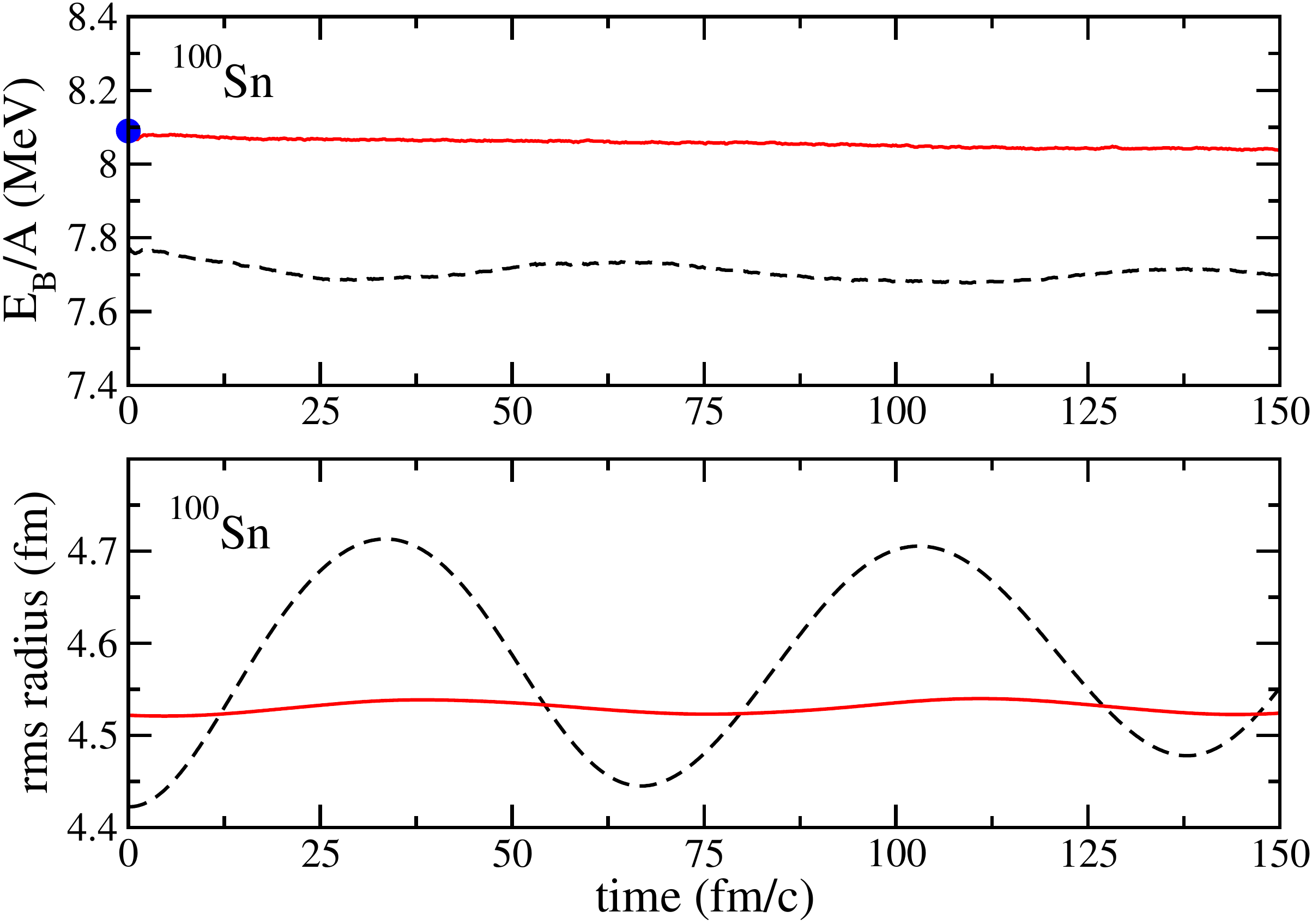}
  \caption{(Color online) \label{BE_Radius}Time evolution of the
    binding energy per nucleon (panel on the top) and root-mean square
    (rms) radius (panel on the bottom) for a ground state \tin{}
    nucleus. Vlasov calculations, using the (dashed) standard
    initialization and the (solid) improved initialization, are
    shown. The filled circle in the top panel at $t=0\fmc$ gives the
    RTF-value of the binding energy.  Taken from
    \cite{Gaitanos:2010fd}.}
\end{figure}

Substituting the energy-density functional, i.e., the $00$-component
of the ener\-gy-mo\-men\-tum tensor \cref{T^munu}, in \cref{var} leads
to the RTF-equations for protons and neutrons,
\begin{equation}
  p_i^0(p_{F,i})=\mu_i , \quad (i=p,n)    \label{RTFeq}
\end{equation}
Here, $p_i^0(p_{F,i})$ are the Fermi energies.  They can be written
explicitly as
\begin{equation}
  \begin{split}
    \label{spEnergies}
    p_p^0(p_{F,p}) & = g_{\omega}\omega^0 + g_{\rho}\rho^{30}
    + e A^0 + E_{F,p}^{*}, \\
    p_n^0(p_{F,n}) & =g_{\omega}\omega^0 - g_{\rho}\rho^{30} +
    E_{F,n}^{*},
  \end{split}
\end{equation}
where $E_{F,p,n}^{*}=\sqrt{p_{F,p,n}^2+(m^*)^2}$. The Fermi energies
\cref{spEnergies} are constants by definition, in contrast to the
Fermi energy defined in \cref{E_F} for the LTF approximation. For a
spherical nucleus, the RTF equations (\ref{RTFeq}) together with the
field equations \crefrange{KGsigma}{Maxwell} completely determine the
radial dependence of the proton and neutron densities and mean fields,
i.e.,~$\rho_{p,n}(r),~\sigma(r),~\omega^0(r)$, $\rho^{30}(r)$, and
$A^0(r)$. In their solution retardation effects are neglected.

\begin{figure}[t]
  \centering
  \includegraphics[width=0.4\linewidth]{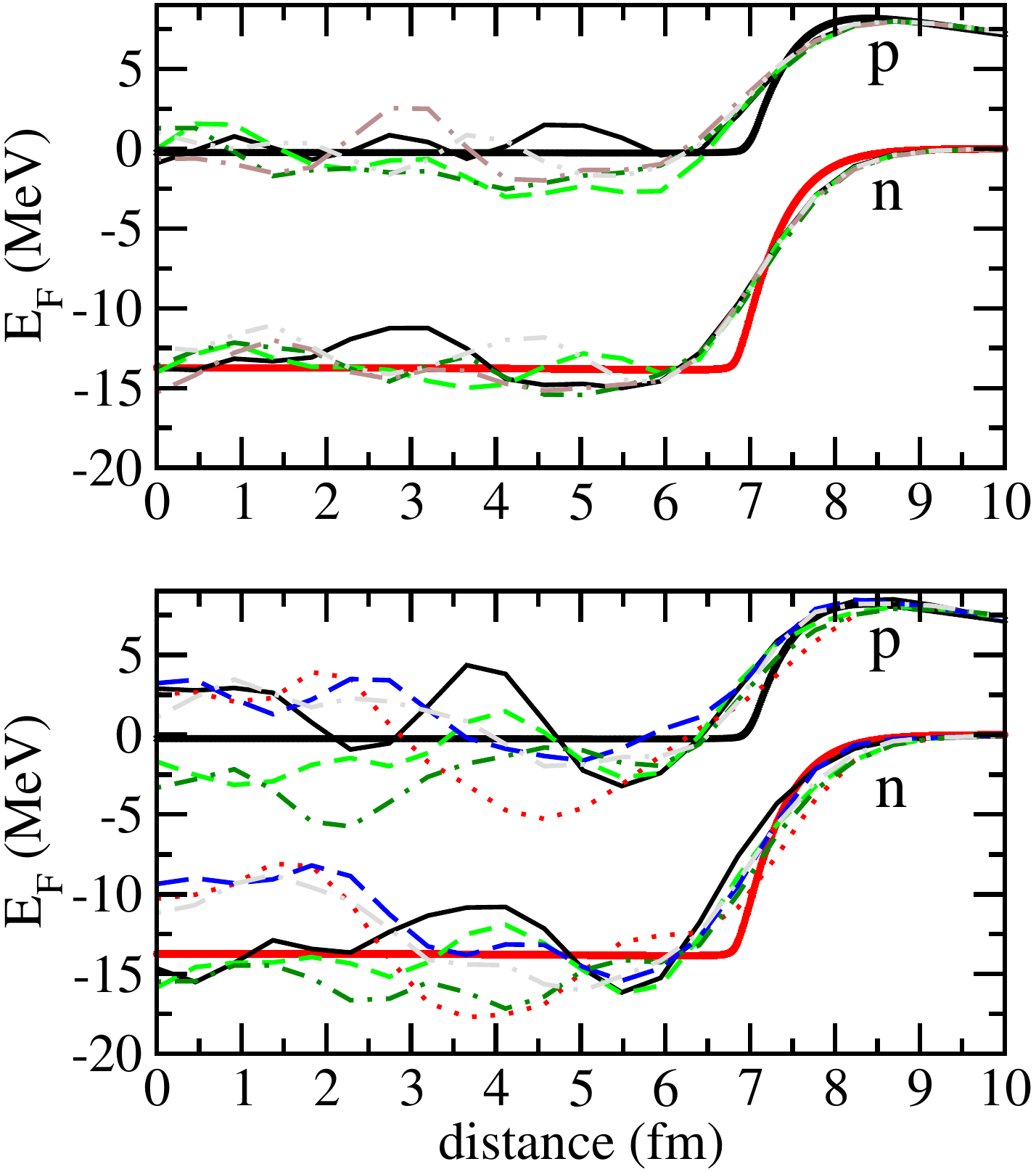}
  \includegraphics[width=0.4\linewidth]{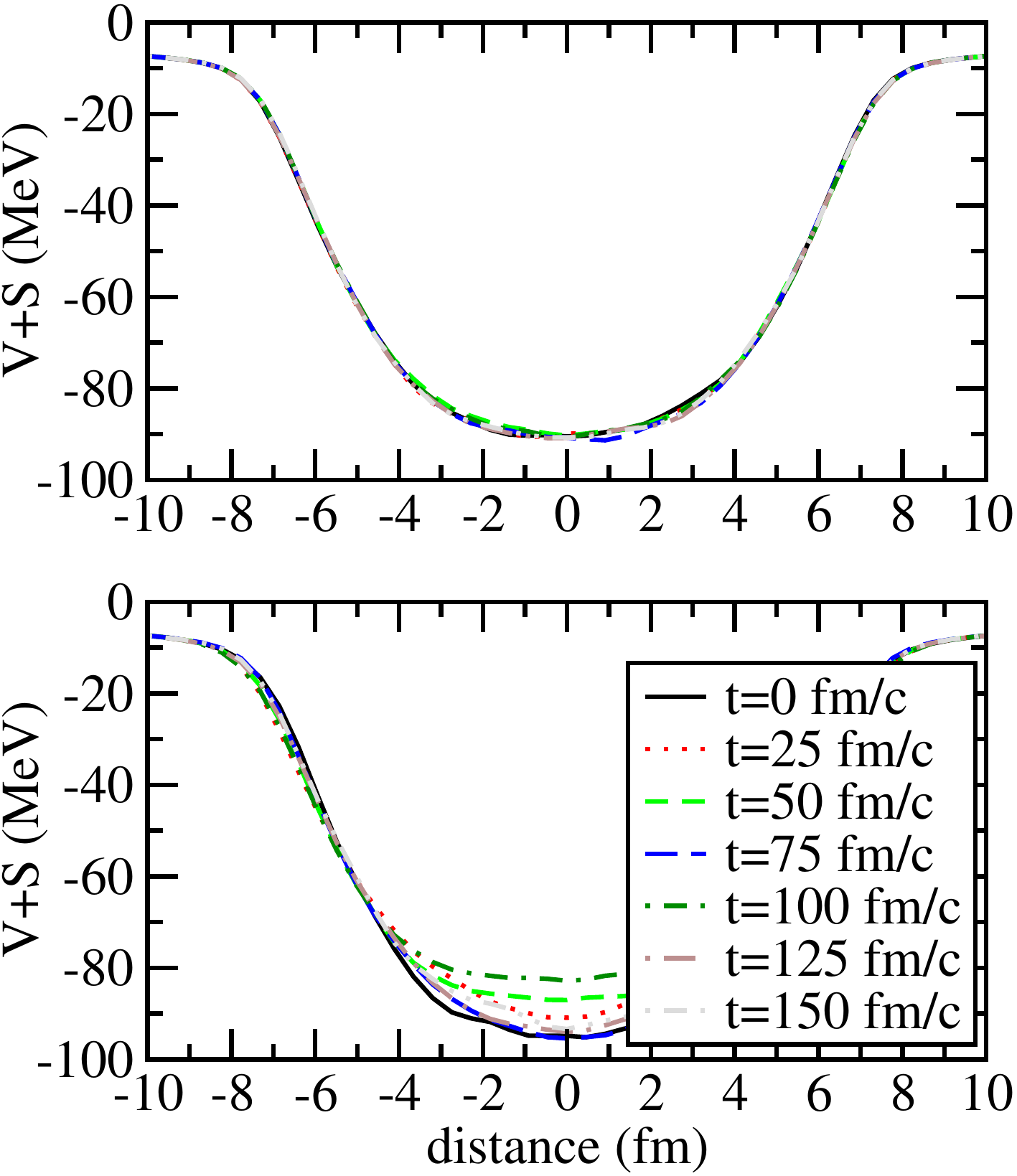}
  \caption{(Color online) \label{RMF_Pot} The Fermi energies for
    protons and neutrons subtracting the nucleon mass,
    $p^{0}_{p,n}-m_N$, with $p^{0}_{p,n}$ given by \cref{spEnergies}
    (left panels) and the nucleon mean-field potential, $V^0+S$, where
    the vector and scalar potentials are defined by \cref{V,S}, (right
    panels) along the $z$-axis passing through the center of the
    \tin{} nucleus. The thick curves in the figures on the left show
    the result of the RTF calculation.  The other curves show the
    Vlasov results at different times (as indicated) using the
    improved initialization (upper panels) and the standard one (lower
    panels).}
\end{figure}

The initialization of neutron and proton densities according to the
Thomas-Fermi calculation largely improves the ground-state stability
in numerical simulations. As an example, \cref{BE_Radius,RMF_Pot} show
the time evolution of the binding energy per nucleon, the rms-radius
and of the relativistic mean-field potential for Vlasov calculations.

The standard-initialization method using the empirical Woods-Saxon
density distribution produces a binding energy smaller by
$0.3\upto0.4\AMeV$ with respect to the RTF value of $E_{B}/A \simeq
8.1\MeV$. This is expected, since the minimum of the total energy is
not reached with the standard initialization. The binding energy
varies with time due to numerical errors in the solution of the
time-evolution \cref{r_j^dot_RMF,p_j^stardot_RMF} and field
\crefrange{KGsigma}{Maxwell}. For the standard initialization, the rms
radius fluctuates quite strongly, even comparable in the amplitude
with the real giant-monopole resonance vibrations. These artificial
temporal oscillations lead also to a significant particle loss with
increasing time, if collisions are included (not shown here). Applying
the improved initialization, in which the same energy-density
functional is used for both the initialization of the nucleus and its
time propagation, stabilizes the nucleus considerably. At $t=0\fmc$
the value of the binding energy per nucleon agrees with the
corresponding RTF value, and both, the binding energy per nucleon and
the rms radius, stay almost constant in time.

In relativistic transport studies the central mean-field potential
consists of the sum of a large negative Lorentz-scalar and a large
positive Lorentz-vector potential. Thus, small spurious density
variations cause strong numerical fluctuations in the mean-field
potential. This is demonstrated in \cref{RMF_Pot} (figures on the
right), where the mean-field potential is displayed as a function of
the coordinate along the central $z$-axis.  The Vlasov calculations
with the standard initialization (panel on the bottom) show large
fluctuations of the order of $10\proz$, while these fluctuations
almost vanish in the calculations using the improved initialization
method. Another good test of the ground state stability within the
modified initialization scheme is the radial dependence of the proton-
and neutron-Fermi energies, which according to the RTF model, see
\cref{spEnergies}, should be constant. Apart from some unavoidable
numerical fluctuations the Fermi energies resulting from the dynamical
Vlasov calculations are rather constant in radius and also in time
within the modified initialization prescription. In particular, they
fit the RTF results and are very stable around the nuclear surface. On
the other hand, the dynamical calculations using the standard
initialization fail to reproduce the RTF calculation and strongly
fluctuate in time.

\begin{figure}[t]
  \centering
  \includegraphics[width=0.6\linewidth]{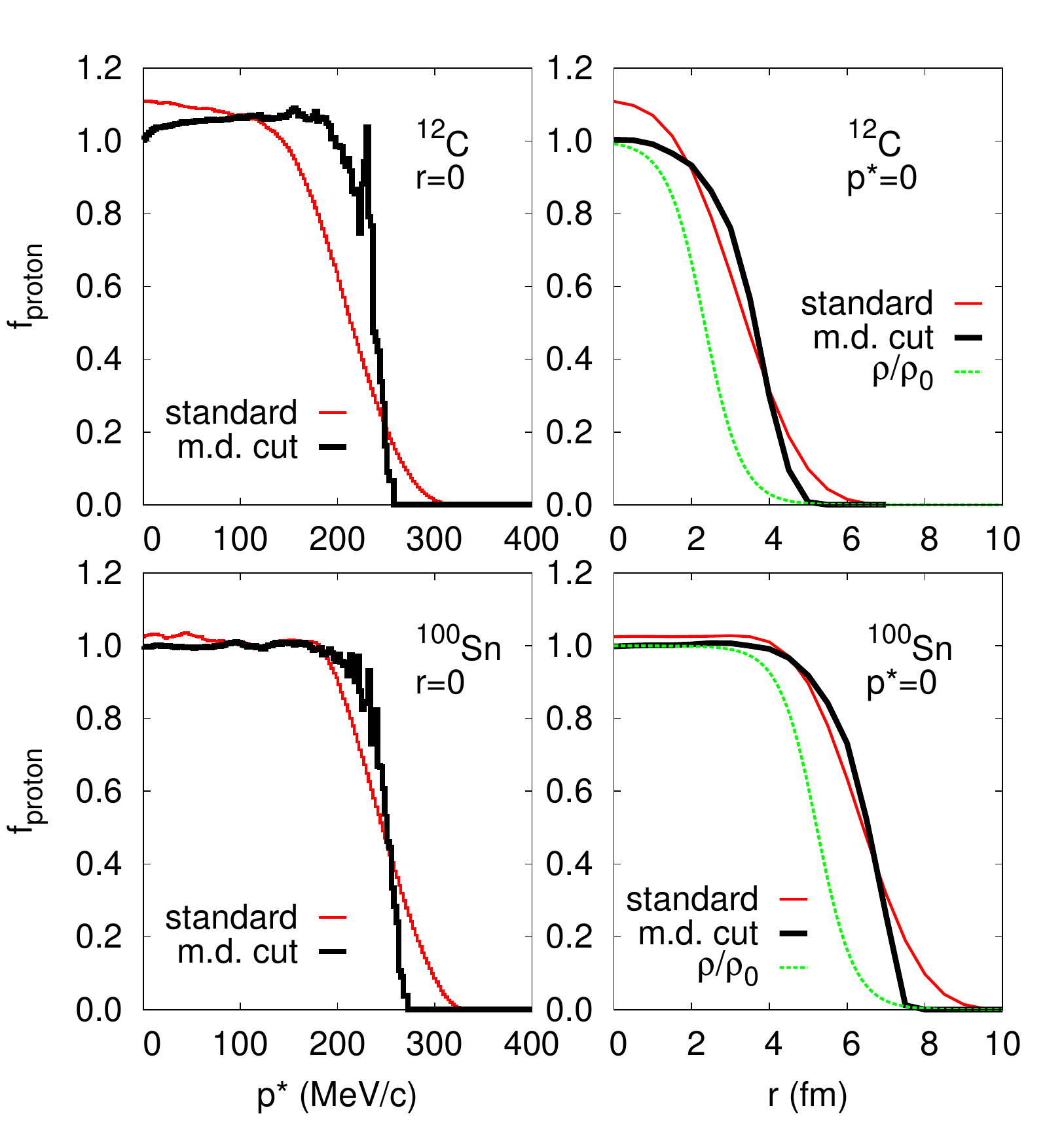}
  \caption{(Color online) \label{fig:pauli}Momentum dependence of the
    proton-occupation number $f_{\rm proton}$ in the center of
    $\atom{12}{C}$ and \tin{} nuclei is shown in the upper and lower
    left panels, respectively.  The radial dependence of $f_{\rm
      proton}$ at zero momentum for \carbon{} and \tin{} is depicted
    in the upper and lower right panels, respectively. The results are
    presented for the standard (thin solid lines) and
    momentum-dependent (thick solid lines) radius, $r_p$ (see \cref{sec:Pauli}).
    The fluctuations of $f_{\rm proton}$ near the Fermi momentum,
    $p_F\simeq250\MeVc$, are due to the finite number of test
    particles per nucleon, which has been set to 10000 in this
    calculation. The nucleon density in units of $\rho_0$ is shown
    additionally by dashed lines in the right panels. We see that at
    the half-central-density radius, the proton occupation number is
    only about 10\proz below unity. Taken from
    \cite{Gaitanos:2010fd}.}
\end{figure}

So far we have discussed the stability of ground state nuclei in
Vlasov simulations, i.e., by neglecting binary collisions. The
frequency of two-body collisions in a Fermi gas depends on the
occupancies of the scattering final states via the hole-distribution
functions $\bar f_i = 1 - f_i$, see the r.h.s.~of the BUU equation
\cref{eq:transp.28}. Due to energy and momentum conservation, no
collisions occur at zero temperature, where
$f_i(\bvec{r},\bvec{p})=\Theta(p_{F,i}(\bvec{r})-|\bvec{p}|)$.  In
test-particle simulations, however, it is impossible to model the
exact $T=0$ Fermi distribution. This causes some spurious two-body
collisions even in the ground-state nucleus. The magnitude of this
effect crucially depends on the numerical treatment of Pauli blocking,
which is described in \cref{sec:Pauli}.

Thus, including two-body collisions requires a careful implementation of
Pauli blocking to prevent the destruction of the ground state. To
demonstrate the accuracy of the test-particle calculations,
\cref{fig:pauli} shows the momentum and radial dependence of the
proton-occupation numbers, which are used in the evaluation of the
Pauli-blocking factors, for \carbon{} and \tin{} nuclei. The
calculations are performed with $N=1000$ test particles per nucleon. The
simulation with the default Pauli-blocking parameters results in a
rather diffuse momentum dependence, especially for the light \carbon{}
nucleus. Using the momentum-dependent radius, $r_p$, and the reduced
width of the Gaussian distribution, largely improves the momentum
dependence of the occupation numbers close to the Fermi surface. The
radial dependence of the occupation numbers also approximates the step
function better for the calculation with the modified Pauli blocking
parameters.

\subsection{Collision term}
\label{sec:collisionTerm}

In general, the collision term, $C(x,p)$, on the r.h.s.~of the kinetic
\cref{eq:os-transp.9,eq:OSP,BUUstar} can be represented as a sum of
one-, two-, three-body, etc.~collisional contributions
(cf.~\cite{LP,Batko:1991xd} and refs.~therein),
\begin{equation}
  C(x,p) = C^{(1)}(x,p) + C^{(2)}(x,p) + C^{(3)}(x,p) + \cdots \; .
  \label{Cdecomp}
\end{equation}
We will start with the general form of the various contributions to
the collision term. At this stage, in order to keep the formulae as
simple and clear as possible, the mean-field effects are ignored.
They will be discussed in \cref{sec:elementaryMedium}.

From now on we will often use the invariant matrix elements $\mathcal{M}_{if}$ in the
PDG convention \cite{Hagiwara:2002fs}, which are related to the used earlier
matrix elements $\mathfrak{M}_{if}$ in the Bjorken and Drell convention \cite{Bjorken:1979dk}
(cf.~\cref{eq:BD-norm}) as
\begin{equation}
   \mathcal{M}_{if}= \mathfrak{M}_{if} \prod\limits_j \sqrt{n_j} \quad \text{with} \quad
    n_j=\begin{cases}
    1 & \text{if $j$ is a boson},\\
    2 m_j & \text{if $j$ is a fermion}~.
  \end{cases}   \label{calM}
\end{equation}
where the product is taken over all incoming and outgoing particles.

The one-body contribution, $C^{(1)}(x,p)$, appears if the considered
particle is an unstable resonance and may decay. For an illustration,
let us consider the one-body collision term for the particle 1
decaying to the particles $1'$ and $2'$, taking into account the
recombination $1'2' \to 1$:
\begin{equation}
  \label{C^1}
  \begin{split}
    C^{(1)}(x,p_1) = &\mbox{ } C^{(1)}_\text{gain}(x,p_1)-
    C^{(1)}_\text{loss}(x,p_1) = \frac{\mathcal{S}_{1'2'}}{2p_1^0g_{1'}g_{2'}}
    \int\,\frac{\dd^4p_{1'}}{(2\pi)^4 2p_{1'}^0}
     \int\,\frac{\dd^4p_{2'}}{(2\pi)^4 2p_{2'}^0} \\
    &\mbox{ } \times (2\pi)^4 \delta^{(4)}(p_1-p_{1'}-p_{2'}) \overline{|\mathcal{M}_{1 \to 1'2'}|^2}
            [F_{1'}(x,p_{1'}) F_{2'}(x,p_{2'})  \overline{F}_1(x,p_1) \\
    &\mbox{ }      - F_1(x,p_1) \overline{F}_{1'}(x,p_{1'}) \overline{F}_{2'}(x,p_{2'}) ]~,
  \end{split}
\end{equation}
where $\overline{|\mathcal{M}_{1 \to 1'2'}|^2}$ is the in-medium matrix
element squared and averaged over the spin states of the initial
particle and summed over the spin states of the final particles;
$F_i(x,p_i)$ ($i=1,1',2'$) is the Wigner function of \cref{Ffact}.  In
\cref{C^1} we have also introduced the auxiliary Wigner function,
\begin{equation}
\label{F.vec-anti}
  \overline{F}(x,p)=
  \begin{cases}
    \ii\tr[\tilde{S}^{>}(x,p)\gamma_0]=2 \pi g A(x,p)[1-f(x,p)] & \text{for fermions,} \\
     2 p^0 \ii  \tilde{D}^>(x,p)=2 \pi g A(x,p)[1+f(x,p)] & \text{for bosons.}
  \end{cases}
\end{equation}
For fermions this Wigner function describes the phase-space density of
hole states and is proportional to the Pauli blocking factor,
$[1-f(x,p)]$. For bosons, the Bose enhancement factor, $[1+f(x,p)]$,
appears instead. Below these factors are denoted as
$\overline{f}(x,p)=1 \mp f(x,p)$ for outgoing fermions (-) or bosons (+).
 To avoid possible misunderstanding, we note that for
the bosons, the spectral function which appears in
\cref{Ffact,F.vec-anti} is defined not by \cref{spec-func} but as
$A(x,p):=2p^0\mathcal{A}(x,p)$ with $\mathcal{A}(x,p)$ being the
spectral function given by \cref{A_R}.

The symmetry factor, $\mathcal{S}_{1'2'}$, in \cref{C^1} takes into
account the possible identity of the final-state particles and is
defined as
\begin{equation}
  \mathcal{S}_{1'2'\dots N'}= \frac{1}{N_a! N_b! \dots}~,   \label{Symmetry_factor_general}
\end{equation}
where $a,b,\ldots$ label the different species of particles appearing in
the final state, $1'2'\ldots N'$, and $N_a$ is the number of particles
of type $a$ among the final-state particles, $1'2'\ldots N'$.

Now we introduce the vacuum decay width of the resonance as
\begin{equation}
  \label{Gamma_decay}
  \begin{split}
    \Gamma_{1 \to 1' 2'}(p_1) =& \mbox{ } \frac{\mathcal{S}_{1'2'}}{2p_1^0}
    \int\,\frac{\dd^4p_{1'}}{(2\pi)^3 2p_{1'}^0}
    \int\,\frac{\dd^4p_{2'}}{(2\pi)^3 2p_{2'}^0}
    (2\pi)^4 \delta^{(4)}(p_1-p_{1'}-p_{2'}) \\
    &\mbox{ } \times \overline{|\mathcal{M}_{1 \to 1'2'}|^2} A_{1'}(p_{1'})
    A_{2'}(p_{2'})~.
  \end{split}
\end{equation}
This expression does not include the Pauli-blocking or Bose-enhancement
factors for the outgoing particles. The spectral functions of the
latter, $A_{1'}(p_{1'})$ and $A_{2'}(p_{2'})$, are also assumed to be
the vacuum ones. In the rest frame of the resonance, after a simple
calculation one can rewrite \cref{Gamma_decay} as
\begin{equation}
  \label{Gamma_decay_rf}
  \begin{split}
    \Gamma_{1 \to 1' 2'}(m_1) &=\mathcal{S}_{1'2'} \int\,\dd m_{1'}^{2} \;
    \mathcal{A}(m_{1'}^{2}) \int\,\dd m_{2'}^{2} \;
    \mathcal{A}(m_{2'}^{2}) \frac{1}{8 \pi m_1^{2}}\,
     \overline{|\mathcal{M}_{1 \to 1'2'}|^2}\,
     q(m_1;m_{1'},m_{2'})~,
  \end{split}
\end{equation}
where $q(m_1;m_{1'},m_{2'})$ is the c.m. momentum of
the decay particles,
\begin{equation}
  q( M; m_1, m_2 )
  := \sqrt { ( M^2 + m_1^2 - m_2^2 )^2 / (4M^2) - m_1^2 }~.
  \label{q}
\end{equation}
In \cref{Gamma_decay_rf}, we have used the spectral functions defined
according to \cref{A_R} and have assumed that they depend on momentum
via the mass, $m^{2}=p^{2}$, only.  Using \cref{Gamma_decay_rf}, we can
rewrite the one-body collision term of \cref{C^1} in the following
simple and practical form:
\begin{equation}
  \label{C^1_final}
  \begin{split}
    C^{(1)}(x,p_1) =&\mbox{ } \mbox{ } \frac{m_1}{p_1^0} \Gamma_{1 \to 1' 2'}(m_1)
    \int\, \frac{\dd \Omega_{\text{cm}}}{4\pi}
    [ f_{1'}(x,p_{1'}) f_{2'}(x,p_{2'}) \overline{F}_1(x,p_1) \\
    &\mbox{ }  - F_1(x,p_1) \overline{f}_{1'}(x,p_{1'}) \overline{f}_{2'}(x,p_{2'}) ]~,
  \end{split}
\end{equation}
where the expression in square brackets is averaged over the direction
of the momentum of particle $1'$ in the rest frame of the decaying
resonance.

The loss term of \cref{C^1_final} is modeled in the following way. For
every resonance test particle of the kind 1, a MC decision is made on
its decay during the time interval, $[t;t+\triangle t]$, according to
the probability,
\begin{equation}
  P_\text{decay}= 1 - \exp\left\{ \frac{-m_1 \, \Gamma_1(m_1) \, \triangle t}{p_1^0} \right\}~,   \label{Pdecay}
\end{equation}
where $\Gamma_1(m_1)$ is the total decay width in the rest frame of a
given resonance test particle obtained by summation over all possible
decay channels,
\begin{equation}
  \Gamma_1(m_1)= \sum_{1',2'}\, \Gamma_{1 \to 1' 2'}(m_1) + \sum_{1',2',3'}\, \Gamma_{1 \to 1' 2' 3'}(m_1)~.  \label{Gamma_1}
\end{equation}
Here, the sums are taken over all two- and (for mesons only)
three-body decay channels. If the resonance decays, its decay channel
is selected by a MC decision with the probability
proportional to the partial width of the given channel. The momenta of
the outgoing particles are sampled isotropically\footnote{Exceptions
  are the decay $\Delta \to \pi N$, where the $\Delta$ has been
  produced in a $\pi N$ collision, and $\rho \to \pi \pi$ for a
  diffractive $\rho$.  In these two cases, the phenomenological
  angular distributions of the decay particles are taken into
  account.} in the rest frame of the resonance.  If one of the decay
particles, e.g., $1'$, is a nucleon, the Pauli-blocking factor in
\cref{C^1_final} is taken into account by accepting the decay event
with the probability, $[1 - f_{1'}(x,\vec{p}_{1'})]$. The
Pauli-blocking factors for other baryons and the Bose-enhancement
factors for mesons are always neglected, since the phase-space
densities of these particles are rather small in all reactions
considered in this review.

The modeling of the gain term in \cref{C^1_final} and also the gain
terms for the two- and three-body processes do not require special
efforts. In fact, the gain terms are \emph{automatically} included in
any test-particle model simulation of the kinetic equations provided
that the final-channel sampling of the corresponding loss terms is
implemented in detail. E.g., the gain term in \cref{C^1_final} is
modeled by the final-state sampling in the simulation of the loss
terms in the kinetic equations for the particles of the kinds $1'$ and
$2'$.

For the processes $1 2 \leftrightarrow 1' 2'$ and $1 2 3
\leftrightarrow 1' 2' 3'$, the two- and, respectively, three-body
collision terms can be written as
\begin{equation}
  \begin{split}
    C^{(2)}(x,p_1) =& \mbox{ } C^{(2)}_\text{gain}(x,p_1)-
    C^{(2)}_\text{loss}(x,p_1) = \frac{\mathcal{S}_{1'2'}}{2p_1^0g_{1'}g_{2'}}
    \int\, \frac{\dd^4p_{2}}{(2\pi)^4 2p_{2}^0}
    \int\, \frac{\dd^4p_{1'}}{(2\pi)^4 2p_{1'}^0} \int\, \frac{\dd^4p_{2'}}{(2\pi)^4 2p_{2'}^0} \\
    & \mbox{ }\times (2\pi)^4 \delta^{(4)}\left(p_1+p_2-p_{1'}-p_{2'}\right)
    \overline{|\mathcal{M}_{12 \to 1'2'}|^2}
           [ F_{1'}(x,p_{1'}) F_{2'}(x,p_{2'}) \overline{F}_1(x,p_1) \\
    & \mbox{ } \times \overline{F}_2(x,p_2) - F_{1}(x,p_{1}) F_{2}(x,p_{2})
                               \overline{F}_{1'}(x,p_{1'}) \overline{F}_{2'}(x,p_{2'}) ]~,
    \label{C^2}
  \end{split}
\end{equation}
\begin{equation}
  \begin{split}
    C^{(3)}(x,p_1) =& \mbox{ } C^{(3)}_\text{gain}(x,p_1)- C^{(3)}_\text{loss}(x,p_1) \\
    =& \mbox{ } \frac{\mathcal{S}_{23}\mathcal{S}_{1'2'3'}}{2p_1^0g_{1'}g_{2'}g_{3'}} \int\,
    \frac{\dd^4p_{2}}{(2\pi)^4 2p_{2}^0} \int\, \frac{
      \dd^4p_{3}}{(2\pi)^4 2p_{3}^0} \int\, \frac{\dd^4p_{1'}}{(2\pi)^4
      2p_{1'}^0} \int\, \frac{\dd^4p_{2'}}{(2\pi)^4 2p_{2'}^0}
    \int\, \frac{\dd^4p_{3'}}{(2\pi)^4 2p_{3'}^0} \\
    & \mbox{ } \times (2\pi)^4
    \delta^{(4)}\left(p_1+p_2+p_3-p_{1'}-p_{2'}-p_{3'}\right)
    \overline{|\mathcal{M}_{123 \to 1'2'3'}|^2} \\
    & \mbox{ } \times [ F_{1'}(x,p_{1'}) F_{2'}(x,p_{2'}) F_{3'}(x,p_{3'})
             \overline{F}_1(x,p_1) \overline{F}_2(x,p_2) \overline{F}_3(x,p_3)     \\
    & \mbox{ } -F_{1}(x,p_{1}) F_{2}(x,p_{2}) F_{3}(x,p_{3})
      \overline{F}_{1'}(x,p_{1'}) \overline{F}_{2'}(x,p_{2'}) \overline{F}_{3'}(x,p_{3'}) ]~.
    \label{C^3}
  \end{split}
\end{equation}

The two-body collision term, $C^{(2)}(x,p_1)$, can be expressed in
terms of the differential scattering cross section $1 2 \to 1'2'$,
\begin{equation}
  \begin{split}
    \dd \sigma_{12 \to 1'2'} = & \mbox{ } (2\pi)^4
    \delta^{(4)}\left(p_1+p_2-p_1'-p_2'\right) \frac{1}{4I_{12}}
    \overline{|\mathcal{M}_{12 \to 1'2'}|^2} \\
    &\mbox{ } \times \mathcal{S}_{1'2'} \frac{\dd^4 p_1'}{(2\pi)^3 2p_{1'}^0}
    \frac{\dd^4 p_2'}{(2\pi)^3 2p_{2'}^0} A_{1'}(x,p_{1'})
    A_{2'}(x,p_{2'})~,
    \label{dsig_12_to_1'2'}
  \end{split}
\end{equation}
where
\begin{equation}
  I_{12} := \sqrt{(p_1p_2)^2 - (m_1 m_2)^2}   \label{I_12_def}
\end{equation}
is the flux factor. Thus, we obtain
\begin{equation}
  \begin{split}
    C^{(2)}(x,p_1) =&\mbox{ } \int\, \frac{\dd^4p_{2}}{(2\pi)^4}
    \int\, \dd \sigma_{12 \to 1'2'} \, v_\text{rel}
                  [ f_{1'}(x,p_{1'}) f_{2'}(x,p_{2'})
                    \overline{F}_1(x,p_1) \overline{F}_2(x,p_2) \\
    &\mbox{ } - F_{1}(x,p_{1}) F_{2}(x,p_{2}) \overline{f}_{1'}(x,p_{1'})
        \overline{f}_{2'}(x,p_{2'}) ]~,
    \label{C^2_final}
  \end{split}
\end{equation}
where
\begin{equation}
  v_\text{rel} := \frac{I_{12}}{p_1^0 p_2^0}  \label{v_rel}
\end{equation}
is the relative velocity of colliding particles\footnote{For
  relativistic particles, $v_\text{rel}=|\bvec{v}_1-\bvec{v}_2|$ only
  if the particles' velocities are collinear. However, for brevity, we
  always call the quantity defined by the r.h.s.~of \cref{v_rel} a
  ``relative velocity''.}. The loss term in \cref{C^2_final} serves as
a basis for various test-particle computational techniques discussed
in \cref{sec:EnsembleTechniques}.

The numerical evaluation of the three-body collision term (\ref{C^3})
is more involved
\cite{effe_phd,Greiner:2000tu,Cassing:2001ds,buss_phd}.  It is usually
done by calculating the three-body collision rate
\begin{equation}
  \Gamma^{(3)}(x,p_1) = C^{(3)}_\text{loss}(x,p_1)/F_1(x,p_1)~.
\end{equation}
Then, $\Gamma^{(3)}$ is treated as a partial width of particle
1. Therefore, the probability for particle 1 to take part in a
three-body collision within the time interval $[t;t+\triangle t]$ is
determined by
\begin{equation}
  P_\text{3b} = 1 - \exp(-\Gamma^{(3)} \, \triangle t)~. \label{P_3b}
\end{equation}
Once it is decided that the particle will participate in a three-body
collision, its collision partners are chosen in some close vicinity,
and the final state is sampled by MC. The sampling of the
final state is done microcanonically, i.e., the probability of a given
momentum configuration is proportional to the corresponding
phase-space volume element.

Below in this \namecref{sec:collisionTerm}, we address the elementary
reaction processes of the GiBUU collision term. First, the vacuum
decay modes of resonances are discussed. Then, we consider reactions
with two and three initial-state particles. Collisions involving more
than three incoming particles are neglected based on the assumption that
the nuclear matter density does not reach too high values.  Such
multi-particle collisions might be important only at very high nuclear
densities ($\sim 10\rho_0$) reached in high-energy heavy-ion
collisions~\cite{Larionov:2007hy}.

\subsubsection{Particle decays}
\label{sec:vacDecays_baryon}

\paragraph{Decays of baryonic resonances}

Our model allows the baryonic resonances to decay only into two-body
final states. Overall, there are 19 different decay channels: $ \pi
N$, $ \eta N$, $ \omega N$, $ K \Lambda$, $ \pi \Delta$, $ \rho N$,
$\sigma N$, $ \pi P_{11}(1440)$, $ \rho \Delta$, $ \pi \Lambda$, $
\overline{K} N $, $ \pi \Sigma $, $ \pi \Sigma(1385)$, $ \eta
\Lambda$, $ \overline{K}^\star N $, $ \pi \Lambda(1520)$, $ \pi \Xi$,
$ \pi \Lambda_c$, $ \pi \Xi_c$. The angular momenta, $L$, for the
final-state particles depend on the resonance and are listed in
\cref{sec:particleproperties}. Following Manley and Saleski, we assume
that the decay width in the rest frame of the resonance is expressed
as (cf.~\cref{Gamma_decay_rf})
\begin{equation}
  \Gamma_{R\to ab}(m)=\Gamma^0_{R\to ab} \frac{\rho_{ab}(m)}{\rho_{ab}(M^0)} ,
  \label{eq:massDependence_gamma}
\end{equation}
where $m=\sqrt{p^\mu p_\mu}$ is the mass of the resonance, $M^0$ its
pole mass and $\Gamma^0_{R\to ab}$ its decay width into a final state
consisting of particles $a$ and $b$ at the pole mass. The function,
$\rho$, is given by
\begin{equation}
  \rho_{ab}(m)=\int \dd p_a^2 \dd p_b^2 \mathcal{A}_a(p_a^2)
  \mathcal{A}_b(p_b^2) \frac{p_{ab}}{m} B^2_{L_{ab}}(p_{ab} R)
  \mathcal{F}^2_{ab}(m) \; .
  \label{eq:baryonwidth}
\end{equation}
In the vacuum, the spectral functions, $\mathcal{A}$, depend only on
the square of the four-momentum. The term $p_{ab}$ denotes the
c.m.~momentum of the final-state products. The Blatt-Weisskopf
functions, $B_{L_{ab}}$, depend on the the angular momentum $L_{ab}$
of the final state particles $a$ and $b$, on the so-called interaction
radius $R=1 \fm$ and on the c.m.~momentum $p_{ab}$.  Compared to
Manley~\cite{ManleySaleski} and the implementation by
Effenberger~\cite{effe_phd}, we modified the large-$m$ behavior by a
cut-off function, $\mathcal{F}_{ab}(m)$. This has become necessary,
because in some channels the width increased too fast with increasing
mass to be used in a dispersion analysis. If the decay channel
includes only stable final-state particles, this modification is not
necessary ($\Rightarrow \mathcal{F}_{ab}(m)=1$); for all other decay
channels we have chosen a form factor according to Post~\cite{postphd}
(Eq.~(3.22) on page 35),
\begin{equation}
  \mathcal{F}_{ab}(m)=\frac{\lambda_{ab}^4+\frac{1}{4}(s_0-M_0^2)^2}{\lambda_{ab}^4+\left(m^2-\frac{1}{2}(s_0+M_0^2)\right)^2} ,
\end{equation}
where $s_0$ is the Mandelstam $s$ for the threshold of the regarded
process. The parameter $\lambda$ has been tuned to pion scattering and
is chosen to be
\begin{equation}
  \lambda=
  \begin{cases}
    0.85 \GeV & \text{for the $\Delta \rho$ channel,} \\
    1.6  \GeV & \text{if there is an unstable meson but no unstable baryon,}\\
    2.0 \GeV & \text{if there is an unstable baryon but no unstable
      meson.}
  \end{cases}
\end{equation}
The $\Delta \rho$ channel plays a special role, since it is the only
channel with two unstable final state particles. The impact of this
cut-off function is discussed in \cite{buss_phd}.

\paragraph{Decays of mesonic resonances}

The mesonic resonances can decay into $16$ different two-body decay
channels ($\pi \pi $, $\pi \rho$, $K \overline{K}$, $K \pi$, $\rho
\gamma$, $\pi \gamma$, $\gamma \gamma$, $D_s^+ \gamma$, $D_s^-
\gamma$, $\pi D_s^+$, $\pi D_s^-$, $\overline{K}\pi$, $D\gamma$,
$\overline{D}\gamma$, $\pi D$, $\pi\overline{D}$) and four distinct
three-body channels ($\pi^0\pi^0\eta$, $\pi^0\pi^-\pi^+$,
$\pi^0\pi^0\pi^0$, $\pi^+\pi^-\eta$ ). For channels with two scalar
final-state mesons with parity $P=-1$ (i.e., the dominant $2\pi$,
$K\overline{K}$, $\pi {K}$, $\pi \overline{K}$ channels), the angular
momentum of the outgoing mesons must be equal to the spin of the parent
resonance. Given this angular momentum, for these channels we can
implement a mass-dependent partial width according to
\cref{eq:massDependence_gamma}. The widths of the other decay channels
are assumed to be mass independent.

\paragraph{Dilepton decay}
In particular, vector mesons and also baryon resonances can decay
into $e^+e^-$ pairs, so-called dileptons. The relevant expressions for
decay widths are given in \cref{sec:dileptons}.

\subsubsection{Two-body collisions}
\label{sec:TwoBody}
\label{matching}

\paragraph{Resonant and non-resonant processes in the low-energy
  region}

For the two-body interactions, we distinguish a \emph{low-energy} and
a \emph{high-energy} region. The low-energy region is dominated by
resonance contributions and small non-resonant background terms.
The resonance model adopted in GiBUU is reliable for baryon-meson
interactions from pion-threshold up to roughly $2.3 \GeV$ center of
mass energy and for meson-meson interactions.  The production cross section
for $a\ b \to R$ is given by the Breit-Wigner formula,
\label{sec:barmes_cross}
\begin{equation}
  \begin{split}
    \sigma_{a\,b \rightarrow R}(s)&=\sum_{f} F_{I}
    \frac{2J_R+1}{\left(2J_a+1\right)\left(2J_b+1\right)}
    \frac{1}{\mathcal{S}_{ab}} \, \frac{4\pi}{p^{2}_{ab}(s)} \,
    \frac{s\, \Gamma_{a b\to R}(s) \, \Gamma_{R \rightarrow
        f}(s)}{\left(s-m_R^2-\RealPart
        \Pi(s)\right)^2+s\Gamma_{\text{tot}}^2(s)}
    \\
    &=F_{I} \frac{2J_R+1}{\left(2J_a+1\right)\left(2J_b+1\right)}
    \frac{1}{\mathcal{S}_{ab}} \, \frac{4\pi}{p^{2}_{ab}(s)} \,
    \frac{s\, \Gamma_{a b\to R}(s) \,
      \Gamma_{tot}(s)}{\left(s-m_R^2-\RealPart\Pi(s)\right)^2+s\Gamma_{tot}^2(s)}
    \label{eq:vacResProd}
  \end{split}
\end{equation}
with
\begin{equation}
  \mathcal{S}_{ab}=\begin{cases}
    1 & \text{if a,b not identical} \\
    \frac{1}{2} & \text{if a,b identical}
    \label{Symmetry_factor}
  \end{cases}
\end{equation}
denoting the symmetry factor of $a$ and $b$. The term $p_{ab}$ denotes
the c.m.~momentum of particles $a$ and $b$, the $J_i$'s define the
total spin of the particles and
\begin{equation}
  F_I=\clebsch{I^a}{I_z^a}{I^b}{I_z^b}{I^R}{I_z^a+I_z^b}
\end{equation}
are the isospin Clebsch-Gordan coefficients squared.  The term,
$\Gamma_{a b\to R}$, denotes the so-called in-width. For stable
particles $a$ and $b$ it is identical to the out-width $\Gamma_{R\to a
  b}$; for unstable particles the final result is given in
\cite{effe_phd} (Eq.~(2.77)).

In this energy region, we have included additional
\emph{non-resonant background processes} to improve the comparison
with experimental data and/or trustworthy model calculations. These
background processes are treated as point-like interactions: While,
e.g., the amplitude of the resonant process $\pi N \to R \to \pi N$ contains a
resonance propagator, the corresponding background process $\pi N
\to \pi N$ is modeled with a point-like four-particle vertex. Such
background processes are defined for several channels
(cf.~\cref{gibuu_barMes_xsections} and \cite[Appendix A]{buss_phd} for
details).

For the baryon-baryon cross sections, matrix elements have been fitted
to $\pi$, $\pi\pi$ and $\rho^0$ and strangeness production up to a
c.m.~energy of ca.~$2.6\GeV$ \cite{Teis:1996kx}. The following
processes are taken into account: $NN \leftrightarrow NN$, $NN
\leftrightarrow N R$, $NN \leftrightarrow \Delta \Delta$, and the
background point-like contribution $NN \to NN \pi$. Details can be
found in \cref{gibuu_barBar_xsections} and \cite[Appendix A]{buss_phd}.

Also the meson-meson interactions are implemented in the model. This is
important, in-particular, for a realistic description of strangeness
production in heavy ion collisions. For details, we refer the
interested reader to \cref{gibuu_mesMes_xsections}.

\paragraph{High-energy processes}

For high energy processes, we use \Pythia{}~\cite{Sjostrand:2006za} as
an event generator. The actual implementation uses version 6.4, while
former code versions used v6.2.\footnote{We have found discrepancies
  between these versions when describing transverse-momentum
  distributions of pions in $p+p$ collisions with beam energies around
  10\GeV{}, indicating that the newer version improves the agreement
  with experiment \cite{Gallmeister:2009ht}.}  The total and elastic
cross sections in the high-energy region are fitted to available data
based on a model using Reggeon and Pomeron exchanges. The model separates
elastic, single-diffractive, double-diffractive and non-diffractive
event topologies. The choice of final state is done with the help of a
fragmentation model implemented in \Pythia{}, i.e., string
fragmentation according to the LUND-String model or cluster collapses
for low mass configurations.

\paragraph{Matching the high energy processes and the resonance
  region}

Altogether, we have two models (\Pythia{} and resonances+non-resonant
background) for two different energy regimes with some overlap region.
Finally, we match both models in this overlap region by introducing a
``transition window'', where we smoothly switch from one prescription
to the other.  We use a c.m.~energy interval of width, $2\Delta$,
around the center point, $\sqrt{s_0}$, in which we mix both types of
events linearly. The probability for a high-energy event is then given
by
\begin{equation}
  p_\text{HiEnergy}(s)=\left\{
    \begin{alignedat}{3}
      &0 \quad&&\text{for} \quad& &\sqrt{s}< \sqrt{s_0}-\Delta\ , \\
      &{\textstyle \frac{\sqrt{s}-(\sqrt{s_0}-\Delta)}{2\Delta}}\quad
      &&\text{for}
      \quad& \sqrt{s_0}-\Delta <&\sqrt{s}< \sqrt{s_0}+\Delta\ ,\\
      &1 \quad&&\text{for} \quad&\sqrt{s_0}+\Delta<&\sqrt{s}\ ,
    \end{alignedat}
  \right.
\end{equation}
where $\sqrt{s}$ is the c.m.~energy, and the probability for a
low-energy event is consequently $1-p_\text{HiEnergy}$. We have chosen
$\Delta=0.2 \GeV$ and $\sqrt{s_0}=2.2 \GeV$ for baryon-meson and
$\sqrt{s_0}=2.6 \GeV$ for baryon-baryon reactions. This procedure
yields a smooth transition also after modifications of individual
cross sections in the low- or high-energy region.

The resulting total and elastic cross section can be found for all
possible incoming particles at the GiBUU homepage \cite{gibuu}.

\paragraph{Detailed balance relations}

The principle of detailed balance states that in the state of thermal
equilibrium the number of collisions $a b \to c d$ is equal to the number
of collisions $c d \to a b$, where the participating particles are characterized
by their momenta only and the spin degrees of freedom are assumed to be
averaged-out (c.f. \refcite{LP}). This is equivalent to the following
relation between the matrix elements
\begin{equation}
  \overline{|\mathcal{M}_{c d \to a b}|^2}
  =\overline{|\mathcal{M}_{a b \to c d}|^2}~,         \label{detBal2}
\end{equation}
where the averaging is done over spin magnetic quantum numbers of
all particles.
If the incoming and outgoing particles are on their mass shell, the
latter equation can directly be transformed to the relation between
the total cross sections of the direct and reversed processes as
\begin{equation}
  \sigma_{c d \to a b}=\sigma_{a b \to c d}
  \left(\frac{p_{ab}}{p_{cd}}\right)^2
  \frac{(2J_a+1)(2J_b+1)}{(2J_c+1)(2J_d+1)}
  \frac{\mathcal{S}_{ab}}{\mathcal{S}_{cd}}~,  \label{detBal}
\end{equation}
where the same notation as in \cref{eq:vacResProd} is used.  Equation
(\ref{detBal}) is often applied to obtain the cross sections, which can
not be obtained empirically, e.g., to determine the $\pi Y \to \overline
K N$ and $\overline K K \to \pi \pi$ cross sections (see
\cref{gibuu_barMes_xsections,gibuu_mesMes_xsections}).

Maintaining detailed balance in the high-energy regime is considerably
more difficult and still presents a problem in many event
generators. This comes about because \Pythia{} can generate, for
example, a typical DIS event, such as $1 + 1 \to n$ with ($n > 2$),
but there is no simple way to describe the time-reversed reaction
where many ($> 2$) hadrons combine to form the simple initial state.
Some progress has only been made for $ 3 \leftrightarrow 2$ and $ 3
\leftrightarrow 3$ reactions
\refcite{effe_phd,Greiner:2000tu,Cassing:2001ds,buss_phd}.

\subsubsection{Three-body collisions}
\label{Sect:3b}
The procedure of the three-body loss term modeling sketched in
beginning of this section requires to know the matrix elements,
$\overline{|\mathfrak{M}_{123 \to X}|^2}$.  Unfortunately, the model independent
information on the three-body matrix elements is restricted by the
processes of the type $1 2 3 \to 1' 2'$, where the detailed balance
relation can be used to extract the matrix element, $\overline{|\mathfrak{M}_{123
    \to 1' 2'}|^2}$, from the cross section of the inverse process,
$1' 2' \to 1 2 3$, (cf.~\refcite{effe_phd}). One example of such a
process is the pion absorption by two nucleons, $N N \pi \to N N$,
implemented in GiBUU.  It is partly due to the two-step mechanism
mediated by the $\Delta$ resonance, $\pi N_1 \to \Delta$ followed by
$\Delta N_2 \to N N$. However, there is also a non-negligible direct
contribution to the two-nucleon pion absorption, which is related to
the background, i.e., non-resonant, part of the $N N \to N N \pi$
cross section important near threshold (cf.~\cref{NN_NNPi}).  The
process, $NN\rightarrow NN \pi$, has been extensively studied in
several experiments over the last twenty
years\cite{Landolt,Andreev:1988tv,Daum:2001yh,Hardie:1997mg,
  Tsuboyama:1988mq,Shimizu:1982dx,Bondar:1995zv}.  We have, therefore,
constructed well defined background cross sections on top of the
resonance contributions for all possible isospin channels.  Another
three-body process implemented in GiBUU is the $\Delta$ resonance
absorption by two nucleons, $NN\Delta \to NNN$.  The rate for the this
process is based on the model of Oset \etal{}~\cite{Oset:1987re}.

A somewhat different, geometrical, approach to many-body collisions
have been developed in \refcite{Batko:1991xd} and optionally
implemented in GiBUU in \refcite{Larionov:2007hy} for the three-body
initial states.  Further details on the three-body processes in GiBUU
can be found in \refcite{effe_phd,Larionov:2007hy,buss_phd}.  The
influence of the three-body collisions on particle production in
heavy-ion collisions is discussed in \cref{sec:heavyIons}.

\subsubsection{Reactions in the medium}
\label{sec:elementaryMedium}
\label{General_aspects_sigmed}
\label{ResProd_inMedium}

Besides the modification of the collision rate via the Pauli-blocking
term, the presence of potentials modifies the in-medium kinematics and
has to be considered in the collision term. After addressing Pauli
blocking, we discuss the general aspects of in-medium cross sections
for the hadron-hadron collisions with arbitrary hadronic final states.
Finally, we consider the resonance processes, where the in-medium
modification of the resonance width has also to be taken into account.

\paragraph{Pauli blocking and Bose enhancement}

For reactions, for which the nucleus stays close to its ground state,
e.g., in low-energy photon and lepton induced events, Pauli blocking
is approximated by the condition that each momentum state below the
Fermi momentum is Pauli blocked. However, in heavy-ion collisions and
reactions induced by pions, protons, leptons, or photons (which
deposit high energies within the target nucleus) one has to simulate
the Pauli blocking dynamically. This requires to determine the nucleon
phase space occupation number at the position and momentum of the
nucleon in the final state of scattering or resonance decay. The
numerical calculation of the nucleon phase space occupation number and
the discussion of its accuracy are given in
\cref{subsec:groundstate-improved} and in \cref{sec:Pauli} 
(see \cref{fig:pauli} and eqs. (\ref{f_i}),(\ref{kappa})).
The Pauli blocking for the other baryons than nucleons as well as the
Bose enhancement for the mesons are always neglected due to presumably
low phase space occupancies of these particles.

\paragraph{General aspects of cross section modifications}

Here, we will discuss the influence of the mean field on the two-body
collision term.  The differential in-medium cross section of the
scattering $12 \to 1'2'$ is defined as (cf.~\cref{dsig_12_to_1'2'} for
the vacuum case)
\begin{equation}
  \begin{split}
    \dd \sigma^*_{12 \to 1'2'} =&\mbox{ }  (2\pi)^4
    \delta^{(4)}\left(p_1+p_2-p_1'-p_2'\right) \frac{n_1^* n_2^*
      n_{1'}^* n_{2'}^*}{4I_{12}^*}
    \overline{|\mathfrak{M}_{12 \to 1'2'}|^2} \\
    &\mbox{ }\times \mathcal{S}_{1'2'} \frac{\dd^4 p_1'}{(2\pi)^3
      2p_{1'}^{*0}} \frac{\dd^4 p_2'}{(2\pi)^3 2p_{2'}^{*0}}
    A_{1'}(x,p_{1'}) A_{2'}(x,p_{2'})~,
    \label{dsig^*_12_to_1'2'}
  \end{split}
\end{equation}
where
\begin{equation}
  I_{12}^* := \sqrt{(p_1^*p_2^*)^2 - (m_1^*m_2^*)^2}   \label{I_12^*_def}
\end{equation}
is the in-medium flux factor, $\overline{|\mathfrak{M}_{12 \to 1'2'}|^2}$ is the
in-medium matrix element squared and averaged over spins of initial
particles and summed over spins of final particles, and
$\mathcal{S}_{1'2'}$ is a symmetry factor for final particles (see
\cref{Symmetry_factor}). Equation (\ref{dsig^*_12_to_1'2'}) explicitly
uses the matrix element in the Bjorken and Drell normalization
\cite{Bjorken:1979dk} with
\begin{equation}
  n_j^*=\begin{cases}
    1 & \text{if $j$ is a boson},\\
    2 m_j^* & \text{if $j$ is a fermion}~.
  \end{cases}
\end{equation}
While the final results of any consistent field-theoretical
calculation should not depend on the choice of the bispinor
normalization, we prefer the normalization of Bjorken and Drell when
discussing the in-medium effects, since, due to the dimensionless Dirac
bispinors ($\bar u u=1$), the matrix elements become less sensitive to the nuclear
medium. This normalization is also the preferred one for the
Dirac-Brueckner calculations of the in-medium cross sections
(cf.~\refcite{TerHaar:1987ce,Fuchs:2001fp}).

This leads to the following expression for the two-body collision term
(cf.~expressions for the on-shell particles with mean fields,
\cref{eq:transp.28}, and for off-shell particles without mean fields,
\cref{C^2_final}):
\begin{equation}
  \begin{split}
    C^{(2)}(x,p_1) =& \mbox{ } \int \frac{\dd^4 p_2}{(2\pi)^4} \int
    \dd\sigma^*_{12 \to 1'2'}\,
    v_\text{rel}^*\,      [ f_{1'}(x,p_{1'}) f_{2'}(x,p_{2'})
                            \overline{F}_1(x,p_1) \overline{F}_2(x,p_2)  \\
    & \mbox{ }       -F_1(x,p_1) F_2(x,p_2) \overline{f}_{1'}(x,p_{1'})
            \overline{f}_{2'}(x,p_{2'}) ]
     =: C^{(2)}_\text{gain} - C^{(2)}_\text{loss}~, \label{C^2_inmed}
  \end{split}
\end{equation}
where
\begin{equation}
  v_\text{rel}^* := \frac{I_{12}^*}{E_1^* E_2^*}  \label{v_rel^*}
\end{equation}
is the in-medium relative velocity of the colliding particles.  In
other words, in order to obtain the in-medium collision term, one
should replace $p \to p^*$ and $m \to m^*$, i.e., the bare vacuum
four-momenta --- by the kinetic ones and the bare vacuum masses --- by
the effective ones, everywhere except for the energy-momentum
conserving $\delta$-function, where the in-medium canonical
four-momenta should appear at the place of the bare vacuum momenta.

In a similar way, the loss terms for more complicated processes like
$12 \to 1'2'\cdots N'$ are given by
\begin{equation}
  C_\text{loss}(x,p_1) = \int\, \frac{\dd^4 p_2}{(2\pi)^4}
  \int \dd \sigma^*_{12 \to 1'2'\dots N'} v_\text{rel}^*
  F_1(x,p_1) F_2(x,p_2) \prod_{i=1'}^{N'} \overline{f}_i(x,p_i)~.   \label{C_loss}
\end{equation}
The differential cross section of the transition $12 \to 1'2'\cdots
N'$ reads
\begin{equation}
  \begin{split}
    \dd\sigma^*_{12 \to 1'2'\dots N'} = & \mbox{ } (2\pi)^4 \delta^{(4)}\left(
      p_1+p_2-\sum_{i=1'}^{N'} p_i \right) \frac{n_1^* n_2^* \prod_{i=1'}^{N'}
      n_i^*}{4I_{12}^*}
    \overline{|\mathfrak{M}_{12 \to 1'2'\dots N'}|^2}      \\
    &\mbox{ } \times \mathcal{S}_{1'2'\dots N'} \prod_{i=1'}^{N'} A_i(p_i)
    \frac{\dd^4 p_i}{(2\pi)^3 2p_i^{*0}}~,
    \label{dsig_12_to_1'2'N'}
  \end{split}
\end{equation}
where we have included the spectral functions (see \cref{spec-func}) for
the general case of broad resonances in the final state.

In practice, however, it is impossible to evaluate the in-medium cross
section (\ref{dsig_12_to_1'2'N'}) for all possible collision types. In
most cases, suitable parametrizations of experimentally measured cross
sections or theoretically predicted cross sections on the basis of
boson-exchange and resonance models
(cf.~\cite{Dmitriev:1986st,Huber:1994ee,Tsushima:1996tv,Tsushima:1998jz})
are used in transport model simulations. All these (semi-) empirical
approaches provide us with the vacuum cross sections,
$\sigma^\text{vac}_{12 \to 1'2'\dots N'}$.  Since the latter depend on
the bare invariant energy $\sqrt{s_\text{free}}$ and possibly other
parameters (such as masses etc.) defined in vacuum, the problem arises
how to use these vacuum cross sections for collisions that take place
in the nuclear medium, in the presence of mean-field potentials acting
on initial and final particles.

We first consider hadron-hadron collisions. Assuming that the
potential energy of incoming resonance and outgoing final states is
exactly the same, one could view the potential as a background field,
which should not affect the reaction rates. For example, consider a
momentum independent potential and an elastic $N_1 N_2 \to N_{1'}
N_{2'}$ scattering process. In this case the potential energy is
exactly conserved during the reaction. One then defines a
potential-corrected, so-called ``free'', c.m. energy
\begin{equation}
  s_\text{free}=(p_\text{1,free}+p_\text{2,free})^2~,
  \label{s_free_nonrel}
\end{equation}
where
\begin{equation}
  p_\text{free}=(\sqrt{m^2+\vec{p}\,^2},\vec{p}) \;
\end{equation}
and uses this to read off the cross section at this energy from its
vacuum parametrization.  This is the prescription used in most
transport codes for treating hadron-hadron collisions.  There are some
intricacies involved in actually determining the quantity
$s_\text{free}$ which are discussed in \cref{sec:sfree}.

The definition (\ref{s_free_nonrel}) relies on the particle's
three-momenta but not on the total in-medium energy of colliding
particles. The latter, however, is important for endothermic threshold
processes, where particles are produced, like, e.g., for meson
production in baryon-baryon collisions. In this case, one can use a
prescription similar to the one applied in
\refcite{Fuchs:2000kp,Fuchs:2005zg} for the treatment of in-medium
thresholds in kaon production processes
\begin{equation}
  \sqrt{s_\text{free}} = \sqrt{s^*} - (m_1^* - m_1) - (m_2^* - m_2)~    \label{s_free_RMF}
\end{equation}
which is used in the RMF calculations (see \cref{sec:sfree} for a
short derivation) with $s^* \equiv (p_1^*+p_2^*)^2$.

For electroweak processes special care has to be taken for treating
the first, initial reaction of the incoming particle (electron,
photon, neutrino) with the nucleon bound in a potential and moving in
the Fermi sea. For such reactions the high-energy cross sections all
depend linearly on the energy of the incoming particle so that any
error in calculating the ``free'' c.m. energy results in a major effect on
the cross section at higher energy. In this case it is most natural to
Lorentz-transform first into the rest frame of the Fermi-moving bound
nucleon. Indeed this is the procedure widely used in the literature
for treating inclusive inelastic scattering cross sections for leptons
with nuclei \cite{Kulagin:2004ie,Kulagin:2007ju}.  Since here the free
(vacuum) cross sections in the rest frame of the Fermi-moving nucleon
depend on the product $m E$ (cf.\ \cref{{DIS_Xsect}}), given by $(s -
{m}^2 + Q^2)/2$, it is suggestive to use the relation
\begin{equation}
  s_\text{free} = s^* + m^2 - (m^*)^2
\end{equation}
where $s^*$ has to be calculated with the properly boosted incoming
energy.

Since transition rates enter the collision terms in the BUU equation
it is reasonable to require that the collision rate of
quasi-particles $(\propto \rho^2\,  \sigma^* v_\text{rel}^*)$ and not the cross section,
is the same as in vacuum \cite{effe_phd}. Therefore, with
non-relativistic potentials, the cross sections are calculated as
\begin{equation}
  \sigma^*_{12 \to 1'2'\dots N'}
  = \frac{v_\text{rel}}{v_\text{rel}^*} \sigma^\text{vac}_{12 \to
    1'2'\dots N'}~,
  \label{sig_12_to_1'2'N'_corr}
\end{equation}
where $v_\text{rel}=|\vec{v}_\text{1,free}-\vec{v}_\text{2,free}|$ is
the relative velocity calculated in the c.m.~frame with vacuum
kinematics, i.e.,
$\vec{v}_\text{free}=\vec{p}_\text{free}/p^0_\text{free}$.  To
estimate the effect of the flux correction factor in
\cref{sig_12_to_1'2'N'_corr}, let us consider the collision between
two nucleons on the Fermi surface. Then, $v_\text{rel}/v_\text{rel}^*
\simeq m_N^*/m_N \simeq 0.7$, where $m_N^*$ is the Landau effective
mass (see \cref{meff,tab:potential}).  So, the flux correction alone
gives an in-medium reduction of the nucleon-nucleon cross section due
to the factor of $\sim m_N^*/m_N$.

One may wish to evaluate the in-medium structure of the differential
cross section (\ref{dsig_12_to_1'2'N'}) in more detail
\cite{Larionov:2007hy,Wagner:2004ee}. It is convenient to perform this
in the RMF model. A reasonable assumption, which has some support from
the correlated basis \cite{Pandharipande:1992zz} and Dirac-Brueckner
\cite{Fuchs:2001fp} calculations of the in-medium elastic $NN \to NN$
scattering cross sections, is to use the vacuum matrix element
$\overline{|\mathfrak{M}_{12 \to 1'2'\dots N'}|^2}$.  Assuming that the particles
in the final state are on their mass shell, this results in the
relation
\begin{equation}
  \sigma^*_{12 \to 1'2'\dots N'}(\sqrt{s^*})
  = \mathcal{F} \sigma^\text{vac}_{12 \to 1'2'\dots N'}(\sqrt{s_\text{free}})       \label{sigMed}
\end{equation}
between the vacuum and in-medium-modified cross sections for the
process, $12 \to 1'2'\dots N'$, where
\begin{equation}
  \mathcal{F} = \frac{n_1^* n_2^* n_{1'}^* n_{2'}^*\dots n_{N'}^*}{n_1 n_2
    n_{1'} n_{2'}\dots n_{N'}}
  \frac{I_{12}}{I_{12}^*}
  \frac{\Phi_{N'}(\sqrt{s^*};m_{1'}^*,m_{2'}^*,\dots,m_{N'}^*)}%
  {\Phi_{N'}(\sqrt{s_\text{free}};m_{1'},m_{2'},\dots,m_{N'})}                         \label{F}
\end{equation}
is the modification factor. Here, we used the $N$-body phase-space
volume (cf. \cref{dPhi_N}),
\begin{equation}
  \Phi_N( M; m_1, m_2, \dots, m_N ) = \int\, \dd\Phi_N(\mathcal{P}; p_1, p_2, \dots, p_N)    \label{Phi_N}
\end{equation}
with the mass-shell constraints $m_i^2 = p_i^2$ ($i=1,2,\dots,N$) and
$M^2 = \mathcal{P}^2$. E.g., the two-body phase-space volume is given
by
\begin{equation}
  \Phi_2(M; m_1, m_2)=\frac{ \pi q( M; m_1, m_2 ) }{ (2\pi)^6 M }~. \label{Phi_2}
\end{equation}
For the valuation of the higher-dimensional phase-space volumes, we
use alternatively either the exact recurrence relations \cite{PDGdata}
or approximate expressions from \cite{Kopylov1962425} accurate at the
level of a few percent. The latter are substantially reducing the
computational time. The vacuum and in-medium flux factors entering
\cref{F} are expressed as
\begin{alignat}{2}
  I_{12}  & =  q( \sqrt{s_\text{free}}; m_1, m_2 ) \sqrt{s_\text{free}}~,   \label{I_12} \\
  I_{12}^* & = q( \sqrt{s^*}; m_1^*, m_2^* )
  \sqrt{s^*}~.  \label{I_12^*}
\end{alignat}
It is also important to note, that, in the derivation of
\cref{sigMed}, we have assumed that the sum of vector fields for
incoming and outgoing particles is the same. This allows us to replace
the canonical four-momenta in the energy-momentum conserving
$\delta$-function by the kinetic ones, since $p_1+p_2-\sum_{i=1'}^{N'}
p_i = p_1^*+p_2^*-\sum_{i=1'}^{N'} p_i^*$.

The in-medium modification of the cross sections according to \cref{sigMed,F}
is most efficient for the baryon-baryon collisions. In this case, the
particles $1$, $2$, $1'$, $2'$ are baryons and the other particles
$3',\dots,N'$, if present, are mesons. Thus, we have
\begin{equation}
  \mathcal{F} \propto \frac{m_1^* m_2^* m_{1'}^* m_{2'}^*}{m_1 m_2 m_{1'} m_{2'}}~.  \label{F_barBar}
\end{equation}
Since $m^* < m$, the baryon-baryon cross sections are reduced in the
nuclear medium according to this simple picture. This effect is
especially strong at high baryon densities reached in heavy-ion
collisions (see \cref{sec:heavyIons}). We also expect on the basis of
\cref{sigMed,F}, that the in-medium modifications of the baryon-meson
and meson-meson cross sections should be less pronounced due to the
smaller powers of the ratio $m^*/m$.

In the limit of low-energy elastic scattering, $NN \to NN$, the
modification factor, $\mathcal{F}$, can be easily calculated. The in-medium
correction factor of \cref{F} can be simplified by using
\cref{Phi_2,I_12,I_12^*},
\begin{equation}
  \mathcal{F}=\left(\frac{m_N^*}{m_N}\right)^4 \frac{s_\text{free}}{s^*}~.   \label{F_NN_LE}
\end{equation}
Then, since $s_\text{free} \simeq (2m_N)^2$ and $s^* \simeq
(2m_N^*)^2$ at low collision energies, one has $\mathcal{F} \simeq
\left(\frac{m_N^*}{m_N}\right)^2$
(cf.~\cite{Pandharipande:1992zz,Fuchs:2001fp,Persram:2001dg,Li:2005jy}).

Once the partial cross sections of all final states are calculated for
a given two-body collision, a particular final state $f$ is selected
by MC sampling according to the probability,
\begin{equation}
  P_f = \frac{\sigma_f}{\sum_{f'} \sigma_{f'}}~.   \label{P_f}
\end{equation}
The MC decision for the momenta and, for the broad particles,
for the masses of the final state particles is performed according to
the algorithm presented in \cref{sec:finalStateDecisions}.

\paragraph{Resonance production}

The resonance cross sections deserve special consideration, since it
has a structure which allows us to take into account the in-medium
modifications of the resonance mass and width.

As follows from the general cross-section formula (\ref{dsig_12_to_1'2'N'}),
the production cross section for a resonance, $R$, in a collision of two particles,
$a$ and $b$, can be written as
\begin{equation}
  \sigma_{ab \to R} = \frac{\pi n_a n_b n_R}{2I_{ab}}
  \overline{|\mathfrak{M}_{ab \to R}|^2} \mathcal{A}_R(p_R)~.          \label{sig_ab_to_R}
\end{equation}
The spectral function of the resonance can be, in general, the
in-medium one (cf.~\cref{eq:spectralfunction_def}). All the other
quantities on the r.h.s.~of \cref{sig_ab_to_R} are taken always at
their vacuum values, for simplicity reasons\footnote{Since, in our
  model, a resonance can be formed only in baryon-meson or in
  meson-meson collisions, the in-medium modifications are not expected
  to be strong anyway (see discussion after \cref{F_barBar}).}.  Due
to detailed balance, the same matrix element determines also the
vacuum-decay width in the rest frame of the resonance,
\begin{equation}
  \Gamma_{R \to ab}^\text{vac}(\mu)=\frac{n_a n_b n_R}{8 \pi \mu^2}
  \frac{(2J_a+1)(2J_b+1)}{2J_R+1}
  \overline{|\mathfrak{M}_{ab \to R}|^2}
  p_{ab}(\mu) \mathcal{S}_{ab}~,            \label{Gamma_R_to_ab}
\end{equation}
where $\mu$ is the vacuum mass of the resonance $R$, and
$p_{ab}(\mu):=q(\mu;m_a,m_b)$ is the c.m.~momentum of particles $a$
and $b$, cf.~\cref{q}.

For the calculation with non-relativistic potentials, the value of
$\mu$ is determined from the known in-medium canonical four-momenta of
the colliding particles, $a$ and $b$, by imposing the conservation of
energy and momentum,
\begin{alignat}{2}
  \vec{p}_a+\vec{p}_b &= \vec{p}_R,                         \label{p_R} \\
  E_a+E_b &= E_R = \sqrt{\vec{p}^2_R+\mu^2} + U_R(x,\vec{p}_R), \label{E_R}
\end{alignat}
where all energies and three-momenta are calculated in the LRF.
Since energy and momentum of the resonance are fixed by conservation
laws, the solution of \cref{E_R} with respect to $\mu$ can be easily
done and {\it does not} require iterations, in
contrast to the general case of the particle energy calculation (see
\cref{subsec:nonrelPot}). By expressing the squared matrix element in
\cref{sig_ab_to_R} via the decay width \cref{Gamma_R_to_ab}, one
obtains
\begin{equation}
  \sigma_{ab \to R} = \frac{2J_R+1}{(2J_a+1)(2J_b+1)\mathcal{S}_{ab}}
  \frac{4\pi^2}{p_{ab}^2(\mu)}
  \mu \Gamma_{R \to ab}^\text{vac}(\mu) \mathcal{A}_R(p_R)~,  \label{sig_ab_to_R_1}
\end{equation}
or, by using \cref{eq:spectralfunction_def} for the spectral function,
\begin{equation}
  \sigma_{ab \to R} = \frac{2J_R+1}{(2J_a+1)(2J_b+1)\mathcal{S}_{ab}}
  \frac{4\pi}{p_{ab}^2(\mu)}
  \frac{\mu \Gamma_{R \to ab}^\text{vac}(\mu) (-\ImaginaryPart\Pi(p_R))}%
  {(p_R^2 - m_R^2 - \RealPart\Pi(p_R))^2 + (\ImaginaryPart\Pi(p_R))^2}~.
  \label{sig_ab_to_R_2}
\end{equation}

In the case of the RMF-mode, the calculations of the
resonance-production cross section, $\sigma_{ab \to R}$, is somewhat
modified: The real part of the resonance self-energy $\RealPart\Pi$ is
neglected. The invariant energy, $\sqrt{s}$, which fully determines
the cross section in the vacuum given by the Breit-Wigner formula
(\ref{eq:vacResProd}), is replaced by the ``free'' invariant energy
(\ref{s_free_RMF}). The flux correction is dropped. Overall, this
simulates the usual vacuum Breit-Wigner formula for the
resonance cross section in the simplest possible way.

\paragraph{Resonance decays}

Due to detailed balance, the partial decay widths of the resonance in
its rest frame, $\Gamma_{R \to cd}$, and the resonance cross section
$\sigma_{cd \to R}$ are not independent. Therefore, the partial decay
widths of resonances are calculated consistently within the
assumptions made in the previous paragraph. Thus, we set
\begin{equation}
  \Gamma_{R \to cd}(p_R)=
  \begin{cases}
    \Gamma_{R \to cd}^\text{vac}(\mu) & \text{if there exists a
      solution
      to the equation, $p_R=p_c+p_d$}, \\
    0 & \text{otherwise,}
  \end{cases}
\end{equation}
where $\mu$ is the vacuum mass of the resonance determined at its
production time in a collision $ab \to R$ according to
\cref{p_R,E_R}. Within this assumption, we do not include
modifications of the final-state phase space, except that we reject
decay events, where we can not fulfill energy and momentum
conservation given the initial four-momentum, $p_R$. E.g., if $\mu >
m_c + m_d$, then $\Gamma_{R \to cd}^\text{vac}(\mu) > 0$.  However if
simultaneously $\sqrt{p_R^2} < m_c^* +m_d^*$, where $m_c^*$ and
$m_d^*$ are the effective masses of particles $c$ and $d$ with zero
momenta in the resonance rest frame, then the event must be
rejected\footnote{In the RMF-mode calculations, in the latter
  condition, we replace $p_R$ by $p_R^*$.}.

\subsection{Hadronization}\label{sec:hadronization}

\begin{figure}[tb]
  \centering
  \includegraphics[width=0.6\linewidth]{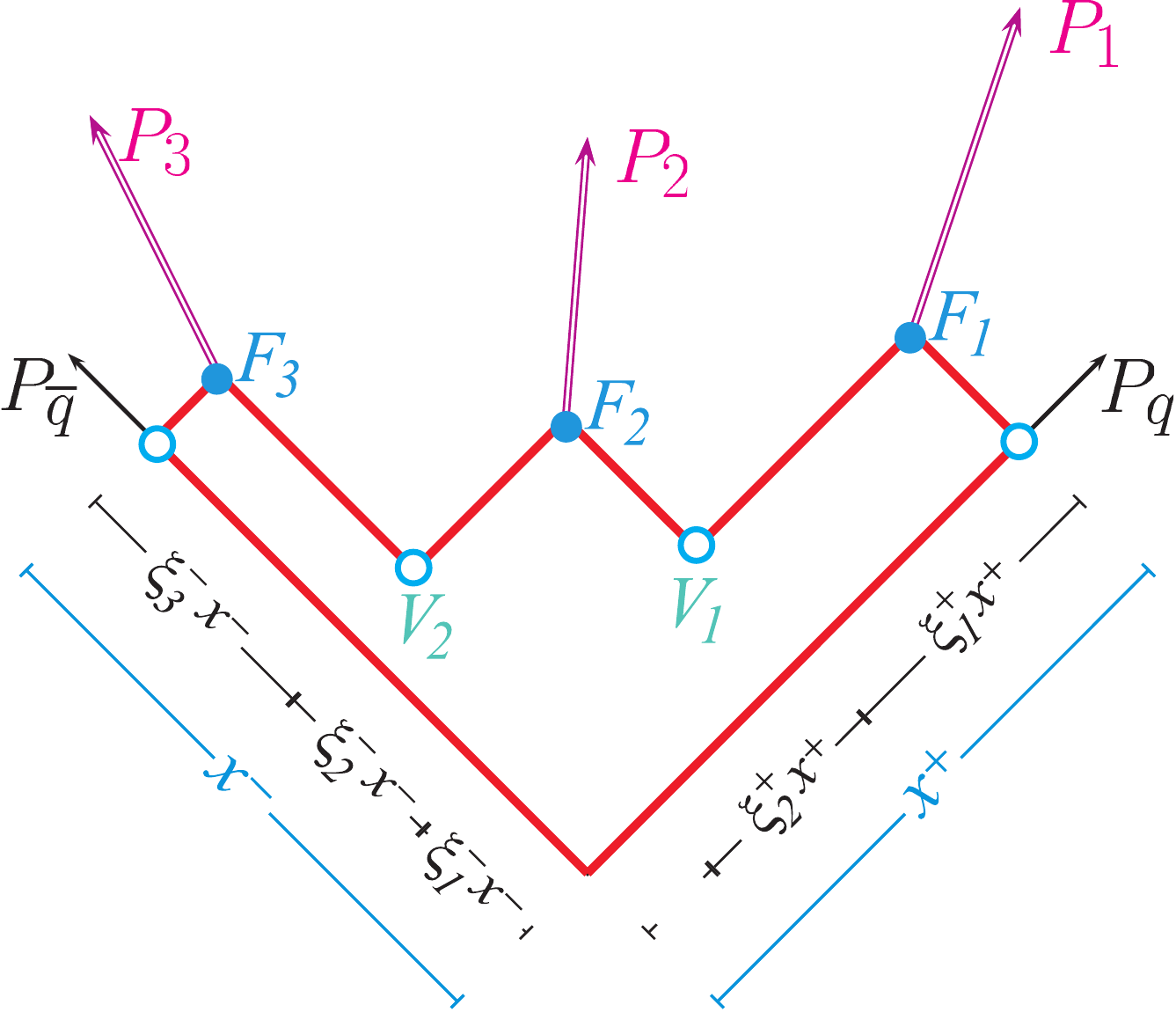}
  \caption
  { (Color online) A sketch of the fragmentation of a one-dimensional
    $q\overline{q}$-string (in the rest frame of the string). The
    $P_x$ denote momenta, while $V_i$ indicate string-breaking
    vertices (production points) and the $F_i$ the yoyo-formation
    points of the hadrons. While in principle all vectors are
    four-dimensional, in this sketch the horizontal axis is the
    $x$-axis and the vertical coordinate represents the time, $t$.
    Fractions of the light cone coordinates $x^\pm$ are also
    indicated. Taken from \cite{Gallmeister:2005ad}.} 
  \label{fig:hadronization_qqString}
\end{figure}

As already explained in \cref{sec:TwoBody},
in the case of high-energy collisions the final state is not known a
priori. It is given via the Lund string fragmentation as provided by
the \Jetset{} part of the \Pythia{} package \cite{Sjostrand:2006za}.
Here one or more strings are formed by the quarks, antiquarks and
diquarks produced by the hard interactions on the partonic level.
These strings fragment by creating new quark--antiquark pairs in
between. From these, the resulting mesons and baryons are built up.
In \cref{fig:hadronization_qqString}
we sketch this mechanism for the case of a simple 1+1-dimensional
quark--antiquarks string.

As can be seen in this figure, we have to
distinguish two classes of particles: Resulting directly from the
fragmentation, any meson or baryon may consist of 0, 1, 2 or even 3
(``leading'') partons, which build up the initial string configurations.
Particles with 0 leading partons are called ``secondary'' or
``non-leading'' particles.

As elaborated in \cite{Gallmeister:2005ad}, in every event during the
MC calculations and for each final particle we extract three
4D-points in \Pythia{}: 
First, two string breaking points correspond usually to two production
points. The meeting point of the
quark/antiquark lines starting at these two string breaking points is
then identified with 
the hadron-formation point. We label the production points by $P_1$
and $P_2$ and the formation point by $F$; the corresponding times are
$t_{P_1}$, $t_{P_2}$ and $t_F$.
In \cref{fig:hadronization_qqString} we illustrate this for a simple
$(1+1)$ case: ``$V_1$'' and ``$V_2$'' correspond to
the two production points of particle 2, while ``$F_2$'' indicates its
formation point.
In the following we will always
identify the ``production time'' of a particle with the ``first'' string
breaking, i.e.~$t_P = \min(t_{P_1},t_{P_2})$.

Particles with 0 leading partons, i.e.~the ``non-leading'' particles,
all have non-vanishing production times, while ``leading'' particles
have at least one parton line directly connected with the hard
interaction point and also have at least one production time which is
zero in all frames.

We now assume, that particles before their \textit{production} may not
interact, while they interact with their full cross section after their
\textit{formation}. Particles before formation we call
``pre--hadrons''. We have shown in \refcite{Gallmeister:2007an} that
only a linear increase of the cross section between these two times is
compatible with data. In that sense the formation time could also be
viewed as an expansion time during which the prehadronic system expands
to its physical ground-state radius, with a correspondingly larger cross
section.  This then makes it possible to investigate color transparency
(CT) effects within GiBUU.

\section{Application to Nuclear Reactions}
\label{sec:results}
In this section we discuss applications of GiBUU to various quite
different reaction types of present interest. We start out with
reactions involving hadrons, i.e.\ pions, protons, antiprotons and
heavy-ions, as projectiles. In a second subsection we discuss the
applications of GiBUU to electron scattering on nuclei, covering both
quasi-elastic scattering and pion production. We also discuss here
studies of hadronization in high-energy electron-nucleus
collisions. This subsection is followed by a discussion of
meson-production on nuclei with real photons. At the end we cover a
relatively new topic, the investigation of neutrino-nucleus interactions
the understanding of which is necessary for extracting information from
neutrino long-baseline experiments.

\subsection{Hadronic Reactions}

The interaction of pions and nucleons is a crucial cornerstone of
every hadronic transport approach. Both particle species are most
abundant in all reaction types and, therefore, very important within
the coupled-channel calculations. To benchmark our description of
pion-nucleon interactions we thus discuss first pion-induced reactions
which set the stage for the discussion of pion production in
heavy-ion, antiproton, electron-, neutrino- and photon-induced
processes.

\subsubsection{Low-energy pions}
\label{sec:pionA_low}
\label{sec:DCX}

In this Section we try to answer the key question how far down in pion
energy a transport model such as GiBUU is applicable since at low
energies the de Broglie wavelength become large and the semi-classical
treatment should start to break down.

\begin{figure}[t]
  \centering
  \includegraphics[height=0.28\linewidth]{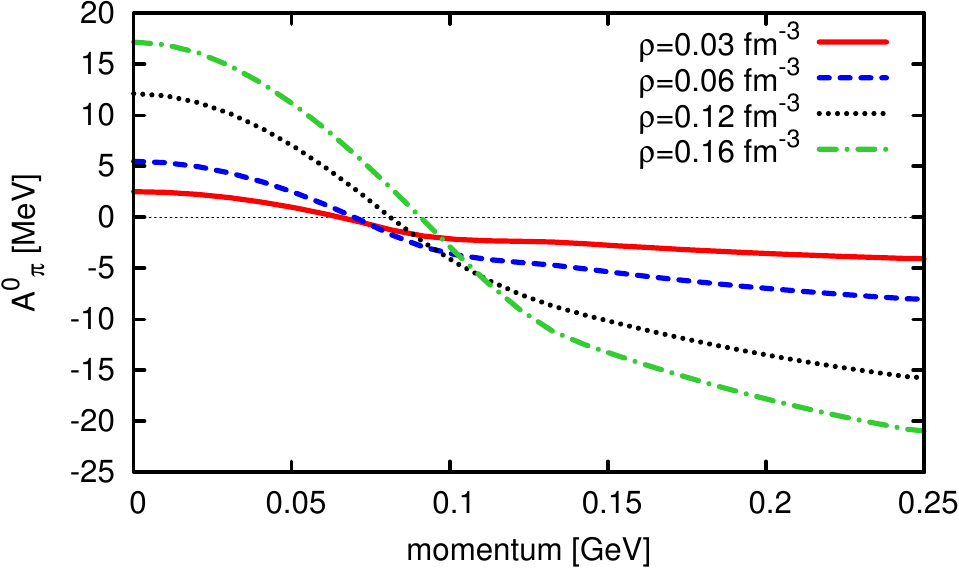}
  \includegraphics[height=0.28\linewidth]{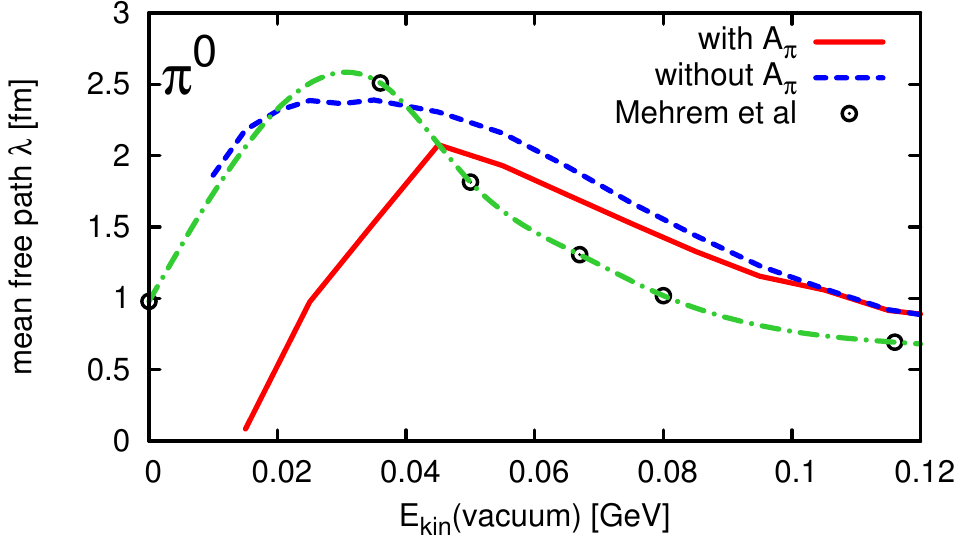}
  \caption[Pion potential and mean free path of a neutral pion]{(Color
    online) Left panel: Hadronic potential, $A_\pi^0$, of the pion as
    a function of pion momentum for symmetric-nuclear-matter density
    at different densities. Right panel: Mean-free path of a neutral
    pion in symmetric nuclear matter at $\rho_0=0.168 \fm^{-3}$ with
    (solid line) and without (dashed line) hadronic potential for the
    pion versus pion kinetic energy. The quan\-tum-me\-cha\-ni\-cal
    calculation by Mehrem \etal{}~\cite{MehremRadi} is shown as
    circles linked by a dashed-dotted line. Source: Taken from \cite{buss_phd}.}
  \label{MeanFreePlot}
\end{figure}

To describe reactions such as close-to-threshold-$\pi\pi$ production
in photon-induced reactions \cite{Messch,Bloch:2007ka}, we aim to a
description of low-energetic pions with kinetic energies down to $30
\MeV$. Already in earlier works of Salcedo
\etal{}~\cite{osetSimulation}, with a simulation of pion propagation
in nuclear matter, and of Engel \etal{}~\cite{Engel:1993jh}, with a
precursor of our present simulation, pions with kinetic energies of
$85\upto300\MeV$ have been investigated in transport models. As
motivated above, we now investigate even less energetic
pions. Therefore, we carefully account for Coulomb corrections in the
initial channel of $\pi$-induced processes and improve the description
of the threshold behavior of the cross sections as compared to earlier
implementations.  Additionally, a momentum-dependent hadronic pion
potential, $A_\pi^0$, shown in the left panel of \cref{MeanFreePlot}
has been
implemented~\cite{Buss:2006yk,buss:bormio06,Buss:2006vh,Buss:2007sa}.
This potential shows a repulsive nature at low momenta, which is due
to the S-wave interactions of pions and nucleons, whereas the P-wave
$\Delta$-hole excitations lead to an attractive contribution, which
dominates the higher momentum.  With Coulomb and hadronic potentials,
the single-particle-Hamilton function for the pion becomes
\begin{equation}
  H_\pi= \sqrt{\bvec{p}_\pi^2 + m_\pi^2} + A_\pi^0 + V_C \ .
\end{equation}

\paragraph{Pion mean free path}

The effect of including the hadronic potential becomes visible in the
mean-free path of the pion in nuclear matter. A proper discussion of
the mean-free path is obviously important in the analysis of
experiments with final-state pions that are produced inside the
nuclear medium.  After having extracted the width of the pion in the
nuclear-matter-rest frame directly from our numerical simulation, the
mean free path is obtained by $\lambda=v/\Gamma$.  Therefore one must
consider the modifications of both the decay width, $\Gamma$, and the
velocity, $v$, due to the potentials.

The velocity of the pions in nuclear matter (in the classically
allowed region $E_\pi-m_\pi=E_{\mathrm{kin}}>A_\pi^0 + V_C$) is given
by Hamilton's equation,
\begin{equation}
  v_i=\frac{\partial H_\pi}{\partial p_{\pi}^{i}}
  =\frac{p_\pi{}_{i}}{\sqrt{\bvec{p}_\pi^{2}+m_\pi^{2}}}+\frac{\partial
    A_{0}^{\pi}}{\partial p_{\pi}^{i}} \ .
  \label{veloEQ}
\end{equation}
At low pion-kinetic energies this velocity, as compared to the vacuum
one, decreases sharply with decreasing kinetic energies, because the
second term in \cref{veloEQ} is always negative. This, in turn,
results in the sharp decrease of the mean-free path, as compared to
the simulation without hadronic potentials.  This is shown in
\cref{MeanFreePlot}, where the mean-free path is plotted as a function
of the pion momentum.  At larger values of the pion momentum the
effects of width and velocity just compensate each other.

As a benchmark for our model, the result of Mehrem
\refetal{MehremRadi} obtained within the quan\-tum-me\-cha\-ni\-cal
framework, solving the full dispersion relation, is also shown in
\cref{MeanFreePlot}. As already mentioned, including the hadronic
potential in our model considerably decreases the mean free path at
low pion momenta. At even lower momenta, where the hadronic potential
becomes repulsive with $A_\pi^0 + V_C > E_{\mathrm{kin}}$ the
semi-classical model breaks down, while quantum mechanical
calculations allow for tunneling, i.e., propagation into such
classically forbidden regions. Indeed, the comparison with the results
of Mehrem in \cref{MeanFreePlot} shows that the semiclassical
transport theory works rather well down to kinetic energies of about
30 MeV, but break down for lower ones.

Our study of the density dependence of the mean-free
path~\cite{Buss:2006vh} has shown, that it is highly nonlinear at low
energies. This non-linearity is generated by the $NN\pi\rightarrow NN$
process, which to first order depends quadratically on the density,
and by the implicit density dependence in the medium modifications. We
conclude, that the naive low-density approximation is qualitatively
and quantitatively not reliable in the energy regime of
$E_{\mathrm{kin}} \lesssim 70 \MeV$, where multi-body collisions,
potential effects, and Pauli-blocking are important.

Since the mean free path is not directly observable, it is ultimately
an open question whether a transport description gives a reasonable
mean-free path for the pion. This can only be answered by
experiment. A test of our model assumptions will, therefore, be the
absorption cross sections which we address next.

\paragraph{Pion absorption and quasi-elastic scattering}

Low-energy pion scattering experiments have been conducted extensively
with elementary targets
(e.g.~\cite{Carter:1971tj,Davidson:1972ky,Kriss:1999cv,Sadler:2004yq}).
However, there exist only a few data points for pions scattering off
complex nuclei
\cite{Carroll:1976hj,Clough:1974qt,Wilkin:1973xd,ashery,Friedman:1991it,nakai,byfield}.
Our results on reaction and charge-exchange cross sections are
presented in \cite{Buss:2006vh,buss_phd}. They show a good agreement
with the few data points existing for {\carbon} and {\bismuth}
\cite{ashery,Friedman:1991it}.

\begin{figure}[p]
  \centering
  \includegraphics[width=0.9\linewidth]{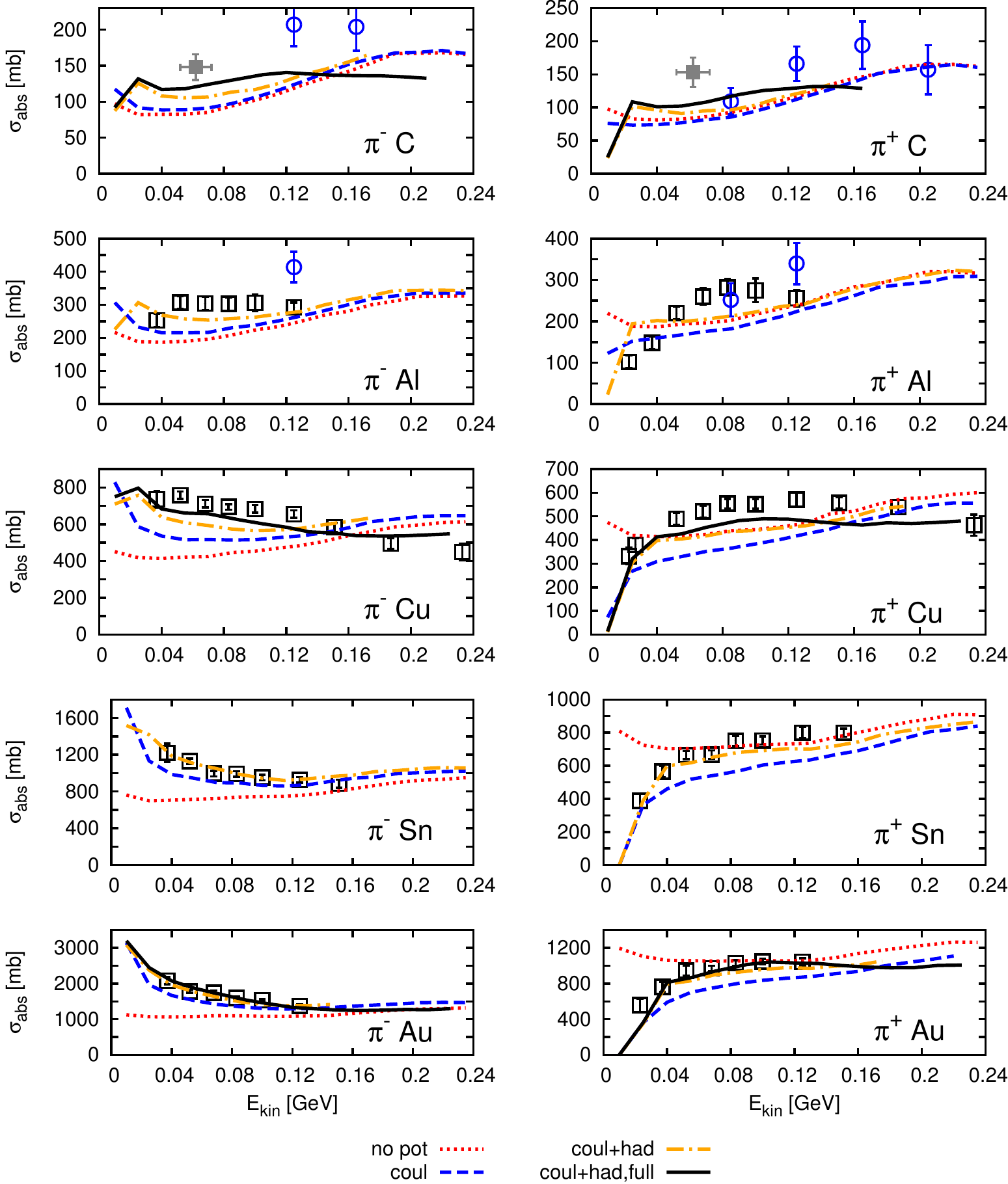}
  \caption[Pion absorption on nuclei]{(Color online) Pion absorption
    on nuclei as a function of pion kinetic energy, depending on the
    choice of potentials for the pion. The data points are taken from
    \cite{ashery} (open circles), \cite{nakai} (open squares) and
    \cite{byfield} (full squares). The dotted line represents a
    calculation without Coulomb potential and hadronic potential for
    the pion such that the pion feels no potentials, all other
    particle species are propagated under consideration of all
    potentials. All other lines are obtained with the Coulomb
    potential for the pion included. The dashed-dotted line
    additionally includes a hadronic potential for the pion. Finally,
    the solid line has been obtained using all potentials and, in
    contrast to the other results, in a full-ensemble calculation.
    Source: Taken from \cite{buss_phd}.}
  \label{absPlot}
\end{figure}

Also for pion absorption, shown in \cref{absPlot}, we achieve good
agreement with existing data. Comparing the curves in \cref{absPlot}
obtained without any potential to those with the Coulomb potential
included, we see that the Coulomb potential alone has only a small
influence for light nuclei, but is very important at low energies for
heavy nuclei. Its long range leads to a sizable deformation of the
trajectories already long before the pions reach the nucleus.  When
one includes the hadronic potential for the pion, another overall
effect sets in. Once the pion enters the nucleus it is affected by the
short-range hadronic potential, which amounts to $-40 \MeV$ at high
momenta and to $+20 \MeV$ at low momenta, as well as the Coulomb
potential which amounts to roughly $\pm 10 \MeV$ in a medium-size
nucleus, and to roughly $\pm 20 \MeV$ in the case of a
\lead~nucleus. At very low energies the two potentials nearly
compensate for the negative pion, while they add up to a strongly
repulsive potential in the case of a $\pi^{+}$.

All results discussed above have been obtained using the
parallel-ensemble scheme (see \cref{sec:EnsembleTechniques}).  In
\cref{absPlot}, we observe deviations between the results from the
superior local-ensemble scheme (box size=$0.5 \fm^3$) compared to
those from the parallel ensemble scheme.  The local ensemble scheme
produces less absorption at higher energies and more absorption at
lower energies. The overall agreement with the data is slightly
improved.

As an overall conclusion, we find that it is critical to include
Coulomb corrections. On top, the absorption cross sections are
sensitive to the hadronic potential of the pion, in particular to the
real part of the self-energy in the medium. As we have already seen in
\cref{MeanFreePlot}, the mean-free path is quite insensitive to the
hadronic potential except at very low energies. We thus conclude that
the modification of the trajectories of the pion is the main effect of
the hadronic potential. In its repulsive regime the hadronic potential
pushes the pion outwards, and the overall path of the pion inside the
nucleus becomes shorter. The probability of absorption is therefore
decreased. The attractive behavior at larger energies causes the
opposite effect.

The overall agreement to data is satisfactory in spite of some
discrepancies, especially for the \carbon~and
\aluminium~nuclei. Considering the fact that the pions have very large
wave lengths at such low energies, the success of the semi-classical
BUU model is quite astonishing. Due to the large wave length one
expects also many-body correlations and quantum interference effects
to be important. Many-body effects are partially included via the mean
fields acting on pions and baryons and the modification of the
$\Delta$ width. Besides this we included only $1\leftrightarrow2$,
$2\leftrightarrow2$ and $2\leftrightarrow3$ body processes in the
collision term. We take the success as an evidence that no higher
order correlations are necessary to describe pion absorption.

\paragraph{Double charge exchange}

Pionic double charge-exchange (DCX) in $\pi A$ scattering is a very
interesting reaction.  The fact that DCX requires at least two
nucleons makes it a very sensitive benchmark for pion rescattering and
absorption. This reaction has received a considerable attention in the
past (see for instance \cite{LAMPF} and references therein). The
mechanism of two sequential single-charge exchanges has traditionally
succeeded to explain the main features of this
reaction~\cite{Becker:1970tk,Gibbs:1977yz} at low energies, although
the contribution of the $A(\pi,\pi\pi)X$ reaction becomes
progressively important as the energy
increases~\cite{Vicente:1988iv,Alqadi:2001pe}. At higher ($\sim1\GeV$)
energies, the sequential mechanism becomes insufficient to account for
the reaction cross
section~\cite{Abramov:2002nz,Krutenkova:2005nh}. Extensive
experimental studies performed at LAMPF has lead to high precision
data for double-differential cross sections on
\atomfull{4}{3}{He}~\cite{Yuly:1997ja} and heavier nuclei (\oxygen,
\calcium, \leadA)~\cite{Wood:1992bi} for kinetic energies of the
incoming pions of $E_\text{kin}=120\upto270 \MeV$.

H\"ufner and Thies~\cite{huefnerThies} have explored for the first
time the applicability of the Boltzmann equation in $\pi N$ collisions
and achieved qualitative agreement with data on single- and
double-charge exchange using some simplifying assumptions of averaged
cross sections and averaged potentials. The work by Vicente
\etal{}~\cite{Vicente:1988iv} is based on the cascade model described
in~\cite{osetSimulation}. There a microscopic model for $\pi N$
scattering has been used as input for the pion-reaction rates in the
simulation. In \refcite{Vicente:1988iv}, pion DCX off \oxygen~and
\calcium~nuclei was explored and fair quantitative agreement with the
data was achieved.

In our work \cite{Buss:2006yk,buss_phd} we explore DCX on heavier
nuclei, comparing with the data measured by Wood
\etal{}~\cite{Wood:1992bi}. We also address the scaling of the total
cross section discussed by Gram \etal{}~\cite{Gram:1989qh}.  To focus
only on single-pion rescattering, we consider incoming pion energies
below $E_\text{kin}=180\MeV$; above this energy, $2\pi$ production
becomes prominent and DCX does not happen necessarily in a two-step
process anymore.  Due to the small mean free path of the incoming
pions, the process is mostly sensitive to the surface of the nucleus.
Therefore, we will discuss and compare two widely used numerical
schemes for the solution of the Boltzmann equation: the
\textit{parallel-ensemble method} employed in the BUU
models~\cite{AB85,BBCM86,BertschGupta,Cassing:1990dr} and in the
Vlasov-Uehling-Uhlenbeck model~\cite{MS85}; and the
\textit{full-ensemble method} used in the Landau-Vlasov~\cite{Gre87},
Boltzmann-Nordheim-Vlasov~\cite{BBD89,BGM94} and Relativistic
BUU~\cite{FGW96,GFW05} models, see \cref{sec:EnsembleTechniques}.
Both schemes are based on the test-particle representation of the
single-particle phase-space density, but they differ in the locality
of the scattering processes. In the following, we first compare both
schemes and, thereafter, point out the impact of neutron
skins. Finally, the transport results are confronted with the
experimental data obtained at LAMPF by Wood
\etal{}~\cite{Wood:1992bi}.

\begin{figure}[t]
  \centering
  \includegraphics[width=0.9\linewidth]{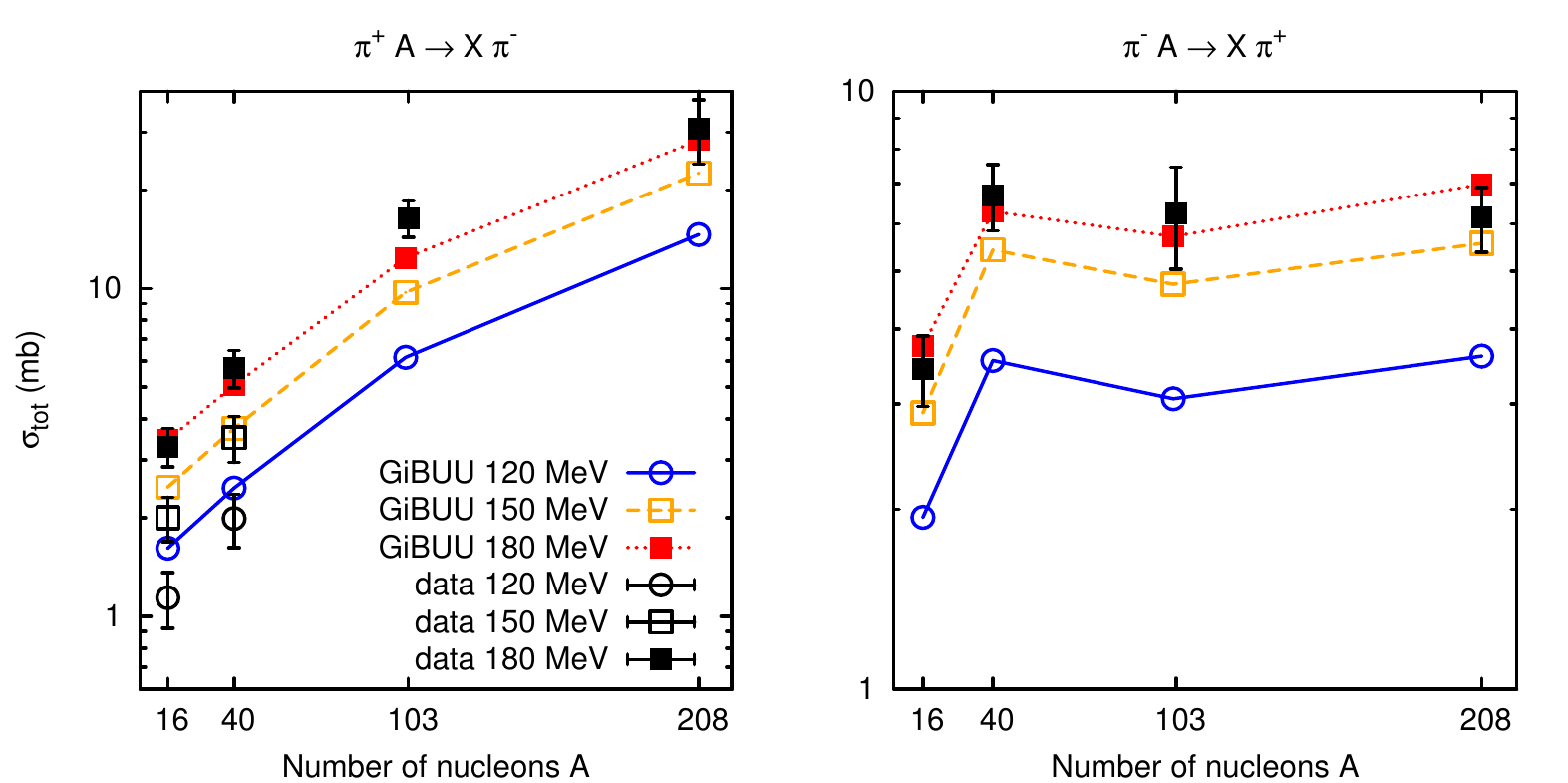}
  \caption[Pion DCX as a function of the nuclear target mass]{(Color
    online) The inclusive double charge exchange total cross section
    as a function of the nuclear-target mass at
    $E_{\mathrm{kin}}=120,150$ and $180 \MeV$. The lines connecting
    our results are meant to guide the eye; the data are taken from
    \cite{Wood:1992bi} (left panel: $E_{\mathrm{kin}}=120 \MeV$ (open
    circles), $150 \MeV$ (open squares) and $180 \MeV$ (full squares),
    right panel: only $180 \MeV$ (full squares)). Source: Taken from
    \cite{buss_phd}.}
  \label{total}
\end{figure}

\begin{figure}[p]
  \centering
  \includegraphics[width=0.8\linewidth]{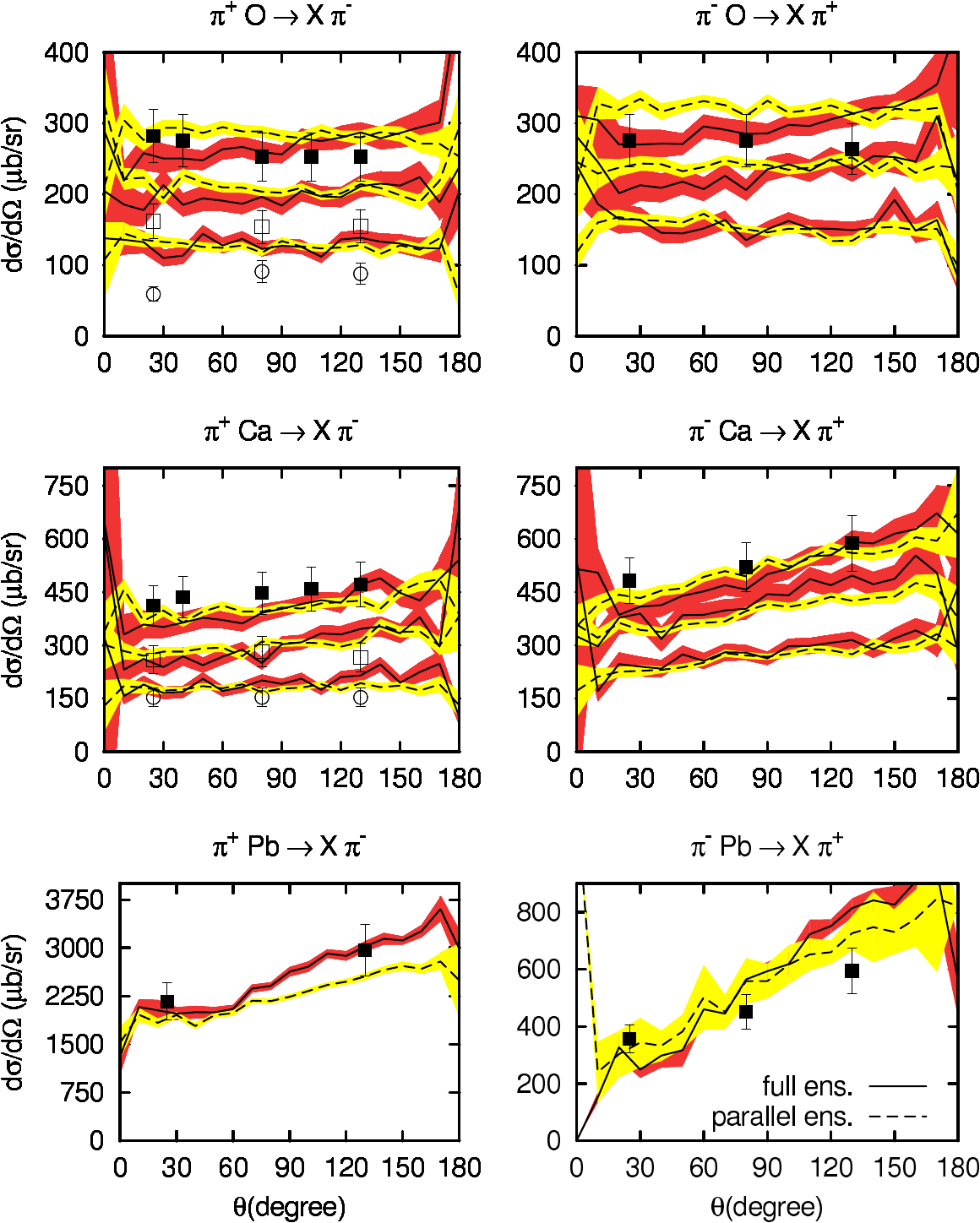}
  \caption[Angular distributions for the double charge exchange
  process $\pi^\pm A\to \pi^\mp X$]{(Color online) Angular
    distributions for the DCX process $\pi^\pm A\to \pi^\mp X$ for the
    incoming pion kinetic energy $E_{\mathrm{kin}}=120,150$ and $180
    \MeV$. The data points are taken from \cite{Wood:1992bi}; only
    systematical errors are shown. The solid lines represent the GiBUU
    results obtained using the full-ensemble scheme, the dashed ones
    the result of the parallel-ensemble scheme. For \leadA~(lower
    panels) we only show the result and the corresponding experimental
    data for $180 \MeV$ kinetic energy of the pion. For \oxygen~(upper
    panels) and \calcium~(middle panels) we show the results for all
    three energies $120 \MeV$ (experimental data: open circles), $150
    \MeV$ (open squares) and $180 \MeV$ (full squares). The lowest
    curves of our results correspond to $120 \MeV$, the highest ones
    to $180 \MeV$. The yellow and red error bands denote the $1\sigma$
    confidence level based upon our statistics. Source: Taken from
    \cite{buss_phd}.}
  \label{parallel}
\end{figure}

In the non-discretized version of the BUU equation, the interactions
are strictly local in space-time. Utilizing the so called
\textit{test-particle ansatz} to solve the problem numerically, this
is no longer the case since usually cross sections in the collision
terms are converted into interaction distances.  The fact that the DCX
reaction depends considerably on the spatial distributions of protons
and neutrons implies that it is also sensitive to the degree of
locality of the scattering processes. Thus we need to elaborate on
this degree of locality of the scattering processes in our
simulation. In \cref{sec:EnsembleTechniques} we have introduced the
concept of the parallel-ensemble approximation, which is expected to
break down for large interaction volumes, $\Delta V_{ij}$. In the
energy regime under consideration, the incoming pions interact
strongly with the nucleons so that the total cross section can reach
more than $200 \mb$. Consequently, the reaction volume, $\Delta
V_{ij}$, exceeds $5 \fm^3$ for a typical time-step size $\Delta
t=0.25\fmc$. The parallel-ensemble approximation is, therefore,
questionable in this energy regime, and thus we evaluate this
approximation scheme by comparison to the full-ensemble method, which
is numerically more time consuming.  In \cite{Buss:2006yk} and in
\cite{buss_phd} the results of both methods were compared. While the
results from both methods are consistent with each other for the
\calcium~nucleus and within the statistical uncertainties also more or
less for \oxygen, some major discrepancy was found for the
$\pi^+\mathrm{Pb} \to\pi^- X$ reaction, in particular at backwards
angles.

The DCX is, due to the small pion-mean-free path in nuclear matter,
very sensitive to the surface properties of the nuclei. In particular,
neutron skins are very interesting because in these skins only $\pi^+$
mesons can undergo charge-exchange reactions. For the positive pions
this causes an enhancement of DCX processes at the surface, so the
pions do not need to penetrate deeply for this reaction.  In
\cite{Buss:2006yk} also the effects of a neutron skin on the DCX cross
sections were explored by comparing results with and without a neutron
skin for the density distribution in \leadA. A significant increase of
about $35\proz$ was found for the reaction $\pi^+ \text{Pb}\to \pi^-
\text{Pb}$ for the density distribution with a neutron skin.

To compare with the data measured at LAMPF by Wood
\etal{}~\cite{Wood:1992bi}, we first discuss the total cross
section. Hereafter, we explore angular distributions and, finally, the
double-differential cross section is addressed as a function of both
angles and energies of the outgoing pions.

In \cref{total} the good quantitative agreement to the total cross
section data at $120$, $150$ and $180 \MeV$ for the \oxygen, \calcium,
\rhodium~and \leadA~nuclei is demonstrated. Only for the
\oxygen~nucleus and the low energy of $120 \MeV$ we find statistically
significant discrepancies. The difference of full- and
parallel-ensemble runs is rather small. Note that GiBUU reproduces the
different $A$ dependences of both $(\pi^+,\pi^-)$ and $(\pi^-,\pi^+)$
reactions. The different $A$ dependences originate from the fact
that, when $A$ increases, the number of neutrons increases with
respect to the number of protons, and this favors the $\pi^+$ induced
reaction over the $\pi^-$ one.

In \cref{parallel} we show $\dd \sigma / \dd \Omega$ for DCX at
$E_{\mathrm{kin}}=120 \MeV$, $150 \MeV$ and $180 \MeV$ on \oxygen,
{\calcium} and {\leadA} as functions of the scattering angle,
$\theta$, in the laboratory frame for both the full- and
parallel-ensemble schemes.  The uncertainties are well under control
except at very small and very large angles, where statistics is very
scarce. Again, there is a very good quantitative agreement for both
\calcium~and \leadA~nuclei. For \oxygen, the data is somewhat
overestimated for low kinetic energies of the pion.

Further comparisons of GiBUU results with DCX data can be found in
\cite{buss_phd,Buss:2006yk}.

\subsubsection{High-energy pion- and proton-induced reactions}
\label{sec:pionA_high}

As already mentioned in \cref{sec:pionA_low}, pion- and proton-induced
reactions are a crucial test for every transport model.

Recently the HARP experiment has published data for $\pi^\pm$
production by proton or pion beams in the momentum range
$3\upto13\GeVc$ impinging on different nuclear targets
\cite{Catanesi:2008zz,Catanesi:2008uv,Catanesi:2005rc}. Here the main
goal is to contribute to the understanding of the neutrino fluxes of
ac\-ce\-le\-ra\-tor-neu\-tri\-no experiments such as K2K, MiniBooNE
and SciBooNE or for a precise calculation of the atmospheric-neutrino
fluxes. Some of the experimental data were also compared to several
generator models used in GEANT4- and MARS-simulation packages
\cite{Catanesi:2008uv}. The overall agreement is reasonable, while for
some models discrepancies up to factors of three are found.
Unfortunately, none of these models is applicable for all energies
considered in the experiment: in the energy region of $5\upto10\GeV$ a
distinction between low energies and high energies has to be
considered, limiting the range of validity of these models. The lack
of high-quality and systematic data concerning hadron-nucleus
collisions in this energy regime has for long been an obstacle for a
serious test of the models. The advent of the HARP experiment has
changed the situation, since it offers charged-pion
double-differential cross sections with a good systematics in angle,
pion momentum, incident energy and target mass. Contrary to other
theoretical frameworks we are able to cover the full energy range of
the HARP experiment.

\begin{figure}[t]
  \centering
  \includegraphics[width=0.6\linewidth]{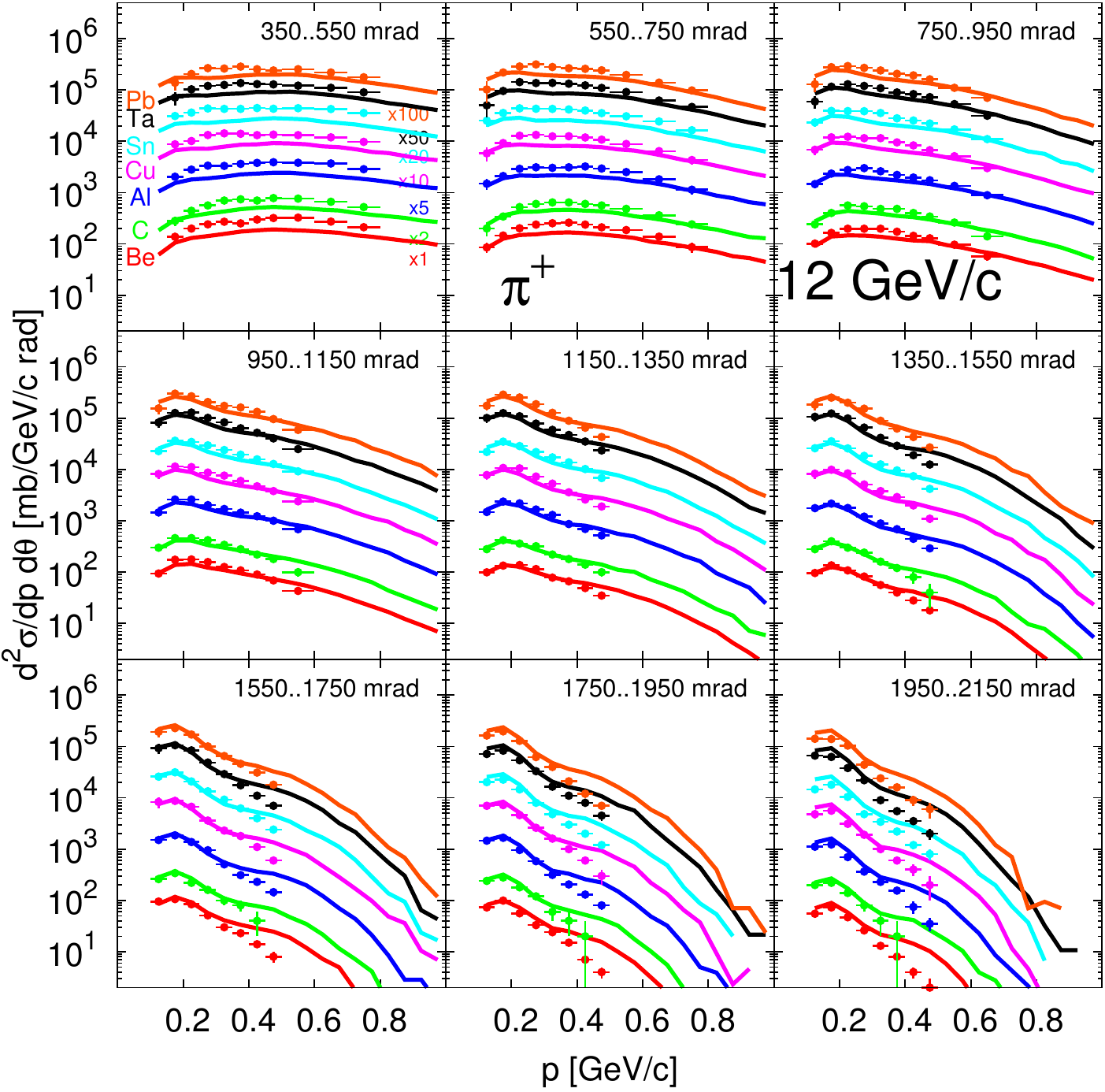}
  \caption{(Color online) Cross section $\dd^2\sigma/\dd p\,\dd\theta$
    for $p+A\to\pi^++X$ with 12\GeVc beam momentum. Experimental data
    are from \cite{Catanesi:2008uv} (HARP large angle analysis),
    curves and data are scaled as indicated.  The targets are
    indicated in the top-left frame. Source: Taken from \cite{Gallmeister:2009ht}.
  }
  \label{fig:HARP_12_p}
\end{figure}

\begin{figure}[t]
  \centering
  \includegraphics[width=0.4\linewidth]{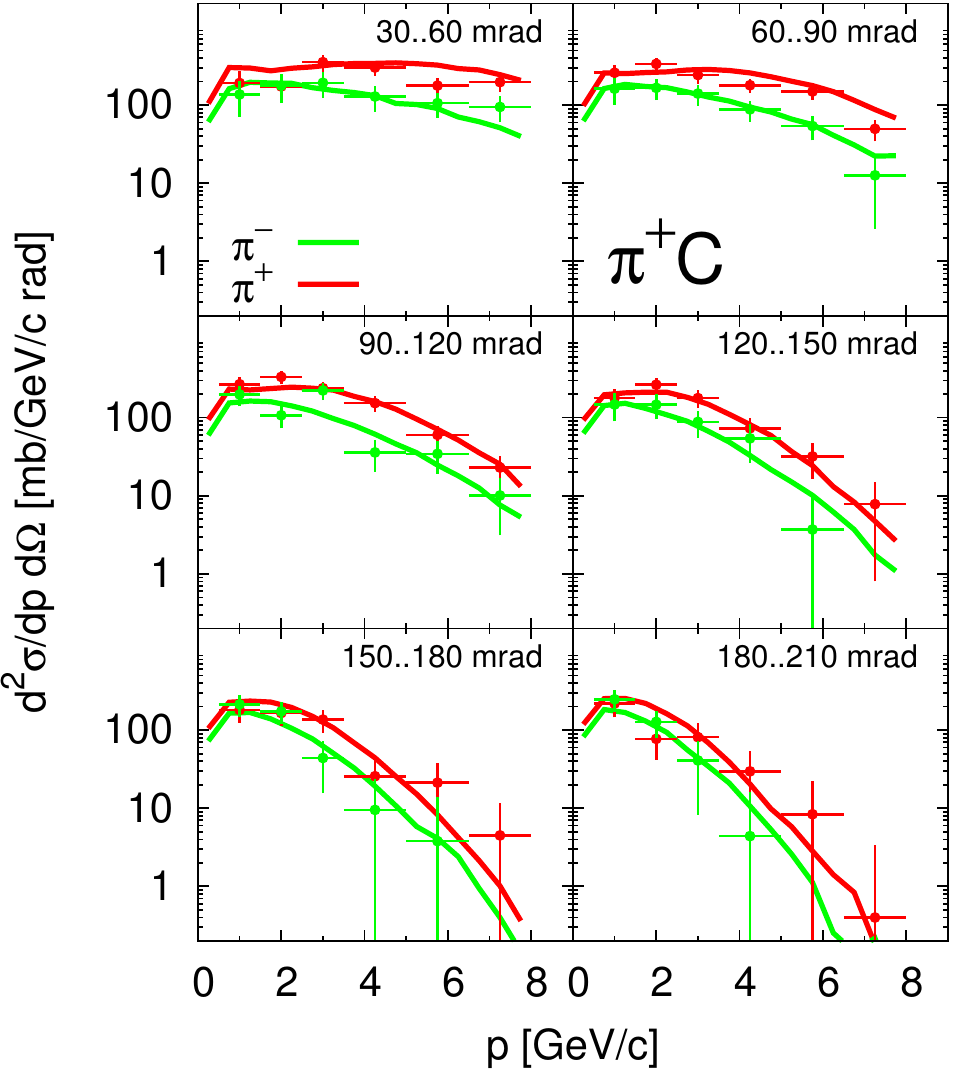}
  \includegraphics[width=0.4\linewidth]{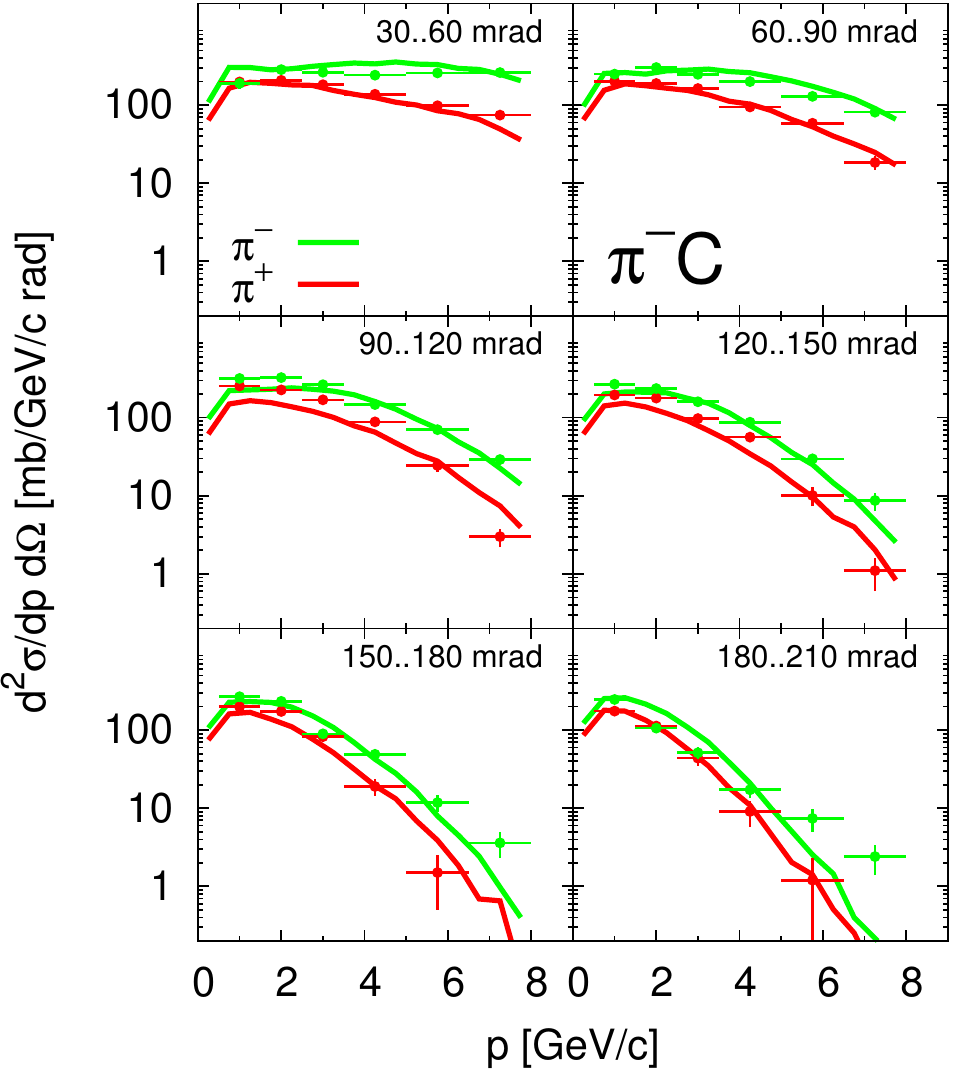}
  \caption{(Color online) Cross section $\dd^2\sigma/\dd p\,\dd\Omega$
    for $\pi^\pm+C\to\pi^\pm+X$ with 12\GeVc beam
    momentum. Experimental data are from \cite{Catanesi:2008zz} (HARP
    small-angle analysis). Source: Taken from \cite{Gallmeister:2009ht}.  }
  \label{fig:HARP_12_pi}
\end{figure}

In \cite{Gallmeister:2009ht} we have shown, that the current
implementation of the proton- and pion-induced collisions in this
energy regime, including all the final-state prescriptions, leads to a
very satisfactory description of the experimental data. In order to
illustrate this excellent agreement with experimental data, we show in
\cref{fig:HARP_12_p} the results for 12\GeV proton-beam energy and the
large-angle analysis and in \cref{fig:HARP_12_pi} the results for
12\GeV pion-beam energy and the small angle analysis.

In Ref.~\cite{Gallmeister:2009ht} we have shown, that the overall
agreement with data is good for all the beam energies $3\upto12\GeV$
for pion and proton beams.\footnote{We note, that after our analysis
  \cite{Gallmeister:2009ht}, more systematic data for the large-angles
  analysis have been published.}  The best description is achieved for
the data with pion beams. The agreement obtained for proton beams is
good over the whole energy-range, except for very forward and very
backward directions.  These deficiencies seem to be due to final-state
interactions (FSI), as a comparison with the corresponding data for
elementary $p + p$ collisions shows. This underlines the need for an
understanding of results on elementary collisions before drawing
conclusions on data taken on nuclei as targets.

In Ref.~\cite{Gallmeister:2009ht}, we have also presented first
theoretical results for the 30\GeV-proton beam in the NA61/SHINE
experiment which aims for a precise determination of the neutrino flux
in the T2K experiment.

\subsubsection{Antiproton-induced reactions}
\label{sec:antiprotonA}

Antiproton-nucleus interactions have been one of the most interesting
fields in ha\-dron-nuc\-leus physics during the last three decades.
The $\overline{p}$ scattering and absorption on nuclei is a direct way
to access the $\overline{p}$-nucleus optical potential. Such
experiments have been performed at LEAR
\cite{Garreta1984266,Garreta:1984rs} and KEK \cite{Nakamura:1984xw} in
the early 80's. Their optical model analysis has led to the conclusion
that the antiproton-nucleus optical potential is moderately attractive
($\text{Re}(V_\text{opt})=-(0\upto70) \MeV$) and strongly absorptive
($\text{Im}(V_\text{opt})=(2\upto3)\text{Re}(V_\text{opt})$).

The low-energy antiprotons colliding with nuclei can be captured in a
Coulomb atomic orbital. The shifts and widths of the
$\overline{p}$-atomic levels caused by strong interactions deliver
important information on the in-medium modifications of the
$\overline{p}$-nucleon scattering amplitude at threshold
\cite{Friedman:2005ad,Friedman:2007zz}.  The possible formation of
deeply bound $\overline{p}$-nuclei predicted by RMF models
\cite{Buervenich:2002ns,Mishustin:2004xa} is expected for the
energetic $\overline{p}$-nucleus collisions.

The antiproton annihilation in a nucleus leads to a large energy
deposition \cite{Rafelski:1979nt,Iljinov1982378,Cahay:1982cp,
  Clover:1982qq}. As a consequence, strongly excited residual nuclei
are formed, which can undergo an explosive breakup to multiply charged
fragments~\cite{Golubeva1988539}.  Another exotic scenario is that the
propagating annihilation fireball absorbing the nucleons on its way
may be transformed into a large quark bag with baryon numbers,
$B\geq1$~\cite{Rafelski:1979nt}. In principle, this can enhance
strangeness production~\cite{Rafelski:1982pu}. However, the
statistical hadronic description can also provide an enhanced
strangeness production for the $B\geq1$-annihilation
channels~\cite{Cugnon:1984zp}. The strangeness enhancement in
$\overline{p}$-annihilations at rest on several nucleons, i.e., for
$B\geq1$-annihilation channels, has been recently found experimentally
in \cite{Bendiscioli:2009zza} which is currently a challenge for
theoretical interpretations.

The ongoing plans for the FAIR project include the antiproton-nucleus
collisions as one of its important research topics. The experiment
PANDA at FAIR~\cite{PANDA} will study, in-particular, the formation of
(double) hypernuclei and $J/\Psi$ production in $\overline{p}$-nucleus
collisions.

In this subsection, we discuss the results of GiBUU calculations of
$\overline{p}$ absorption and of pion production in $\overline{p}$
annihilation on nuclei.  The elementary antibaryon-baryon cross
sections are described in \cref{gibuu_antiBarBar_xsections}.

An important ingredient of the transport simulations of
$\overline{p}$-induced reactions is the antibaryon-mean-field
potential. We describe it on the basis of the RMF model (see
\cref{subsec:RMFpot}). However, since the original non-linear Walecka
model gives a too deep antiproton potential,
$\text{Re}(V_\text{opt})\simeq-660 \MeV$ at normal nuclear density,
some modifications are needed. In the most simple way, this can be
performed by rescaling the antibaryon-meson coupling constants by a
common factor $0 < \xi \leq 1$, as suggested in
\refcite{Mishustin:2004xa}. Thus, we set $g_{\omega \bar B} = -\xi
g_{\omega},~g_{\rho \bar B} = \xi g_{\rho},~g_{\sigma \bar B} = \xi
g_{\sigma}$, where $g_\sigma,~g_\omega$ and $g_\rho$ are the
nucleon-meson coupling constants from the non-linear Walecka
model. The case of $\xi=1$ corresponds to the $G$-parity transformed
nucleon fields or, equivalently, to the antibaryon single-particle
energies derived from the original non-linear Walecka model Lagrangian
density (see \cref{Lagr,dispRel_antinuc}). The vector and scalar
potentials acting on an antibaryon are now redefined as compared to
\cref{V,S} by
\begin{alignat}{2}
  & V_{\bar B} = g_{\omega \bar B} \omega + g_{\rho \bar B} \tau^3
  \rho^3
  + \frac{e}{2} (-1+\tau^3) A~, \label{V_barB}\\
  & S_{\bar B} = g_{\sigma \bar B} \sigma~. \label{S_barB}
\end{alignat}
The antibaryon-single-particle energy reads
\begin{equation}
  \label{dispRel_antinuc_new}
  p^0=\sqrt{(m_{\bar B}^\star)^2+(\bvec{p}-\bvec{V}_{\bar B})^2} + V_{\bar B}^0~,
\end{equation}
where $m_{\bar B}^\star=m_{\bar B} + S_{\bar B}$. Thus, now the
antibaryon single-particle energy has the same form as the nucleon
single-particle energy (\cref{dispRel_expl} with ``+'' sign) with the
nucleon-vector and -scalar potentials \cref{V,S} replaced by the
antibaryon-vector and -scalar potentials \cref{V_barB,S_barB},
respectively.  The antibaryons of the species, $\bar B$, propagate
according to the transport equation (\ref{BUUstar}) with the kinetic
momentum, $p^\star \equiv p_j-V_{\bar B}$, field-strength tensor,
${\cal F}^{\mu\nu} \equiv \partial^\mu V_{\bar B}^\nu - \partial^\nu
V_{\bar B}^\mu$, and effective mass, $m^\star \equiv m_{\bar
  B}^\star$.

In order to conserve energy and momentum, the antibaryon contributions
to the source terms of the field equations for the $\sigma$, $\omega$
and $\rho$ mesons \crefrange{KGsigma}{KGrho} have also been modified
in the case of $\xi \ne 1$. This is especially important for studying
the collective response of a nucleus on an antiproton in its
interior~\cite{Buervenich:2002ns,Mishustin:2004xa,Larionov:2008wy,Larionov:2009er}. For
the full Lagrangian formulation in the case of modified
antibaryon-meson coupling constants we refer the interested reader to
\cite{Mishustin:2004xa,Larionov:2008wy,Larionov:2009er}.  Below, we
will compare our calculations with experimental data on inclusive
observables, such as the $\overline{p}$ absorption cross section, pion
and proton kinetic energy spectra.  Thus, the collective dynamics of
the residual nucleus is not considered here.

\begin{figure}[t]
  \centering
  \includegraphics[width=0.6\linewidth]{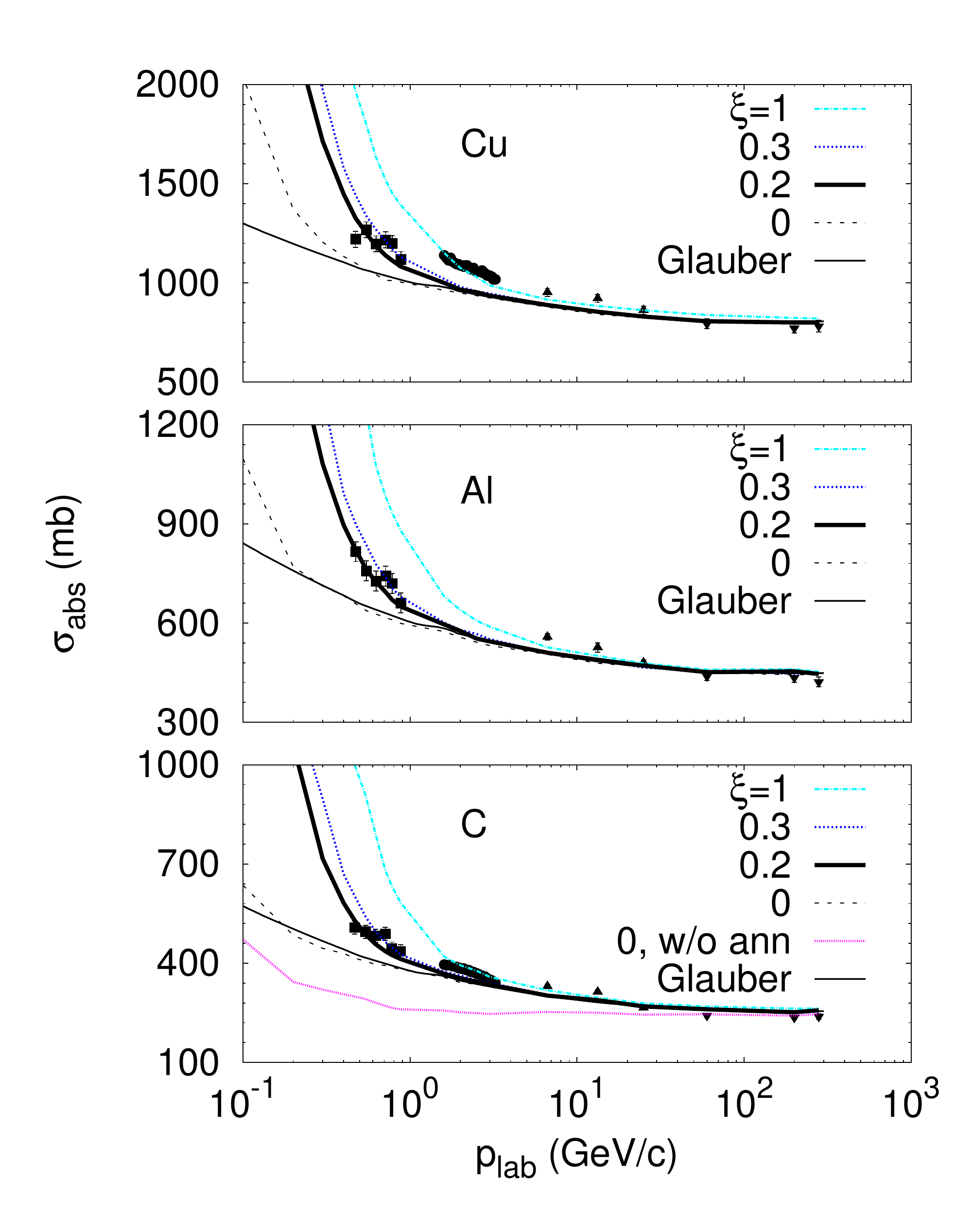}
  \caption{\label{fig:sigabs} (Color online)
    $\overline{p}$-absorption cross section on various nuclei vs the
    beam momentum. The lines marked with the value of a scaling factor
    $\xi$ show the GiBUU results.  Thin solid lines represent a simple
    Glauber-model calculation \cite{Glauber:1970jm,Abrams:1972ab}. For
    the $\overline{p}+\carbon$ system, a calculation with $\xi=0$
    without annihilation is additionally shown by the dotted
    line. Data are from \refcite{Nakamura:1984xw} (filled boxes),
    \refcite{Abrams:1972ab} (filled circles), \refcite{Denisov:1973zv}
    (filled triangles), and \refcite{Carroll:1978hc} (filled
    upside-down triangles). The figure is taken from
    \cite{Larionov:2009tc}.}
\end{figure}

The calculations, presented in this subsection, have been performed in
the full-ensemble technique (see \cref{sec:EnsembleTechniques}).  This
is especially important for the low-energy antiprotons, which have a
big annihilation cross section on nucleons. The full-ensemble
technique helps to better maintain the local structure of the
collision term in this case.  In order to get rid of the spurious
effect of the beam particles' interactions with the secondary produced
particles, which unavoidably appears in the full-ensemble simulations
of the Boltzmann equation, we have turned off the interactions of the
secondaries in the calculations of the absorption cross section.  For
other calculations, this correction has not been done, however.  The
parameters of the non-linear Walecka model have been adopted from the
NL3 set~\cite{Lalazissis:1996rd}, which is well suited to the nuclear
ground state calculations.

\begin{figure}[t]
  \centering
  \includegraphics[width=0.6\linewidth]{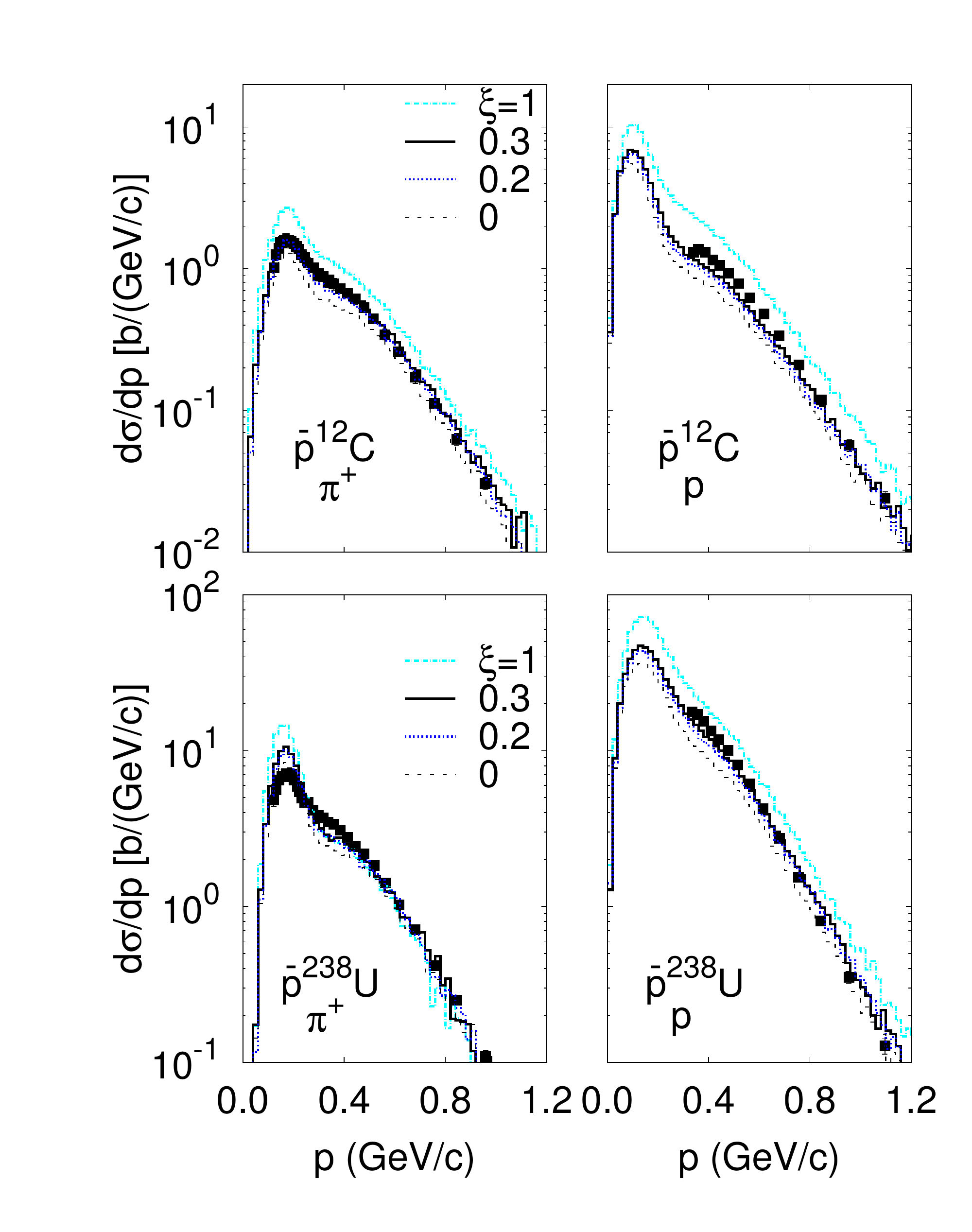}
  \caption{\label{fig:p_spectra}(Color online) The angle-integrated
    $\pi^+$- and proton-laboratory momentum inclusive spectra from
    $\overline{p}$ interactions with \carbon{} and \uranium{} at
    $608\MeVc$. The calculated histograms are labeled with the value
    of the scaling factor, $\xi$. Data are
    from~\cite{Mcgaughey:1986kz}. The figure is taken from
    \cite{Larionov:2009tc}.}
\end{figure}

\Cref{fig:sigabs} shows the $\overline{p}$ absorption cross section on
\carbon, \aluminium{} and \cupper{} as a function of the beam
momentum.  As expected, the calculation without nuclear field acting
on the antiproton ($\xi=0$) is in a good agreement with the
Glauber-model calculations\footnote{It is clear from simple classical
  considerations, that the calculation of the absorption cross section
  based on the Boltzmann equation neglecting the interactions of
  secondaries, exactly reproduces the well known Glauber formula
  (cf.~Eqs.~(6),(7) in \refcite{Abrams:1972ab}). In numerical
  simulations, small differences appear due to finite statistics.},
except for very low beam momenta, where the Coulomb potential causes
the difference. Both, the Glauber-model and GiBUU-($\xi=0$)
calculations describe $\overline{p}$ absorption at high beam momenta
$p_{\rm lab} > 20 \GeVc$ quite well.

However, the experimental data on $\sigma_{\rm abs}$ at low beam
momenta require the introduction of an attractive mean-field potential
for the antiproton in the GiBUU calculations. Indeed, the
antiproton test particles with impact parameters larger than the
radius of a nucleus are subject to two-body scatterings, since their
trajectories are bent by the attractive potential towards the
nucleus. Otherwise, without potential, these particles would propagate
along straight-line trajectories avoiding collisions with
nucleons\footnote{We have observed a similar effect of the attractive
  mean-field potential in pion absorption on nuclei, see
  \cref{absPlot}.}.

\begin{figure}[t]
  \centering
  \includegraphics[width=0.6\linewidth]{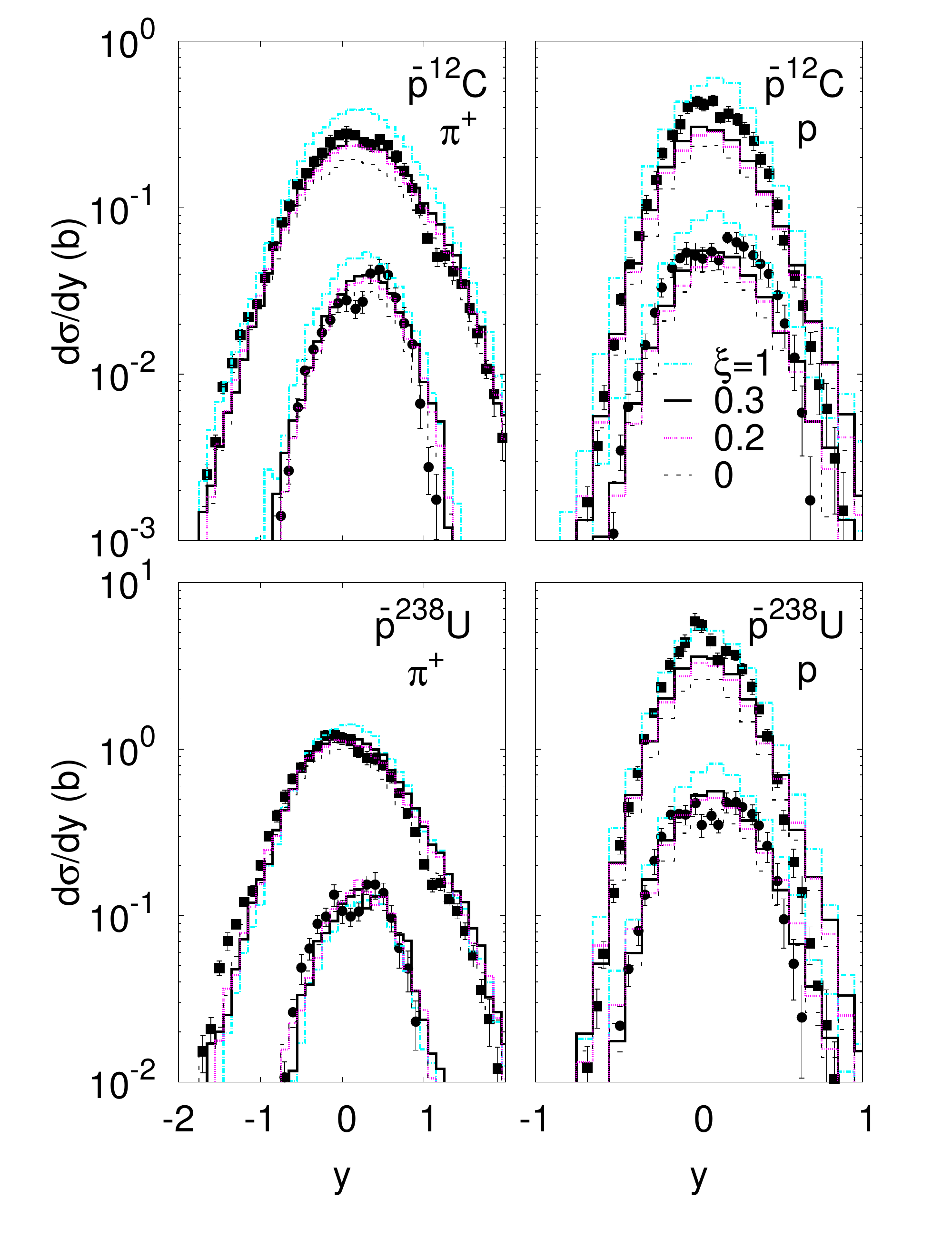}
  \caption{\label{fig:y_spectra} (Color online) $\pi^+$- and
    proton-laboratory rapidity spectra from antiproton interaction
    with \carbon{} and \uranium{} at $608\MeVc$.  The calculated
    histograms are labeled with the value of the scaling factor,
    $\xi$. For the $\pi^+$ (proton) spectra, the upper lines and data
    points correspond to the transverse momenta, $p_T \geq 120~(330)
    \MeVc$, while lower lines and data points are computed for $p_T
    \geq 500(600) \MeVc$.  Experimental data are
    from~\cite{Mcgaughey:1986kz}.}
\end{figure}

The sensitivity of $\overline{p}$-absorption at low beam momenta to
the strength of the attractive potential can be used to determine the
scaling factor, $\xi$, for the antibaryon-meson-coupling
constants. This is illustrated in \cref{fig:sigabs} for several values
of $\xi$. As one can see, $\xi=0.2\upto0.3$ leads to good agreement
with KEK data \cite{Nakamura:1984xw} at $p_\text{lab}=470\upto880
\MeVc$.  If one defines the real part of the antiproton-nucleus
optical potential as a Schr\"odinger-equivalent potential at
$E_\text{lab}=0\GeV$~\cite{Friedman:2005ad},
\begin{equation}
  \mbox{Re}(V_{\rm opt}) = S_{\overline{p}}
  + V_{\overline{p}}^0 + \frac{S_{\overline{p}}^2-(V_{\overline{p}}^0)^2}{2m_N}~,  \label{Re_Vopt}
\end{equation}
the range of $\xi=0.2\upto0.3$ corresponds to $\mbox{Re}(V_{\rm
  opt})=-220$ to $-150 \MeV$ at normal nuclear density.

In \cref{fig:p_spectra,fig:y_spectra}, we show the $\pi^+$- and
proton-momentum and -rapidity spectra from $\overline{p}$ annihilation
on carbon and uranium nuclei at $p_\text{lab}=608 \MeVc$.  One
observes a quite good agreement of the spectral shapes with
experimental data, independent on the choice of the antiproton-mean
field. The absolute yields are best described with $\xi=0.3$, which
corresponds to a strongly attractive antiproton potential,
$\mbox{Re}(V_{\rm opt}) \simeq -200 \MeV$ in the nuclear center. Such
a deep antiproton potential would, in-principle, allow the formation
of a strongly bound compact antiproton-nucleus configurations
\cite{Buervenich:2002ns,Mishustin:2004xa,Larionov:2008wy,Larionov:2009er}.

One observes a peculiar two-slope structure of the pion-momentum
spectra.  The slope change at $p\simeq0.3 \GeVc$ is caused by
pion-nucleon rescattering mediated by the $\Delta(1232)$ resonance.
Higher-momentum pions leave the nucleus practically without
interacting with nucleons. Lower-momentum pions are either absorbed
via $\Delta(1232)$ resonances, $\Delta N \to N N$, or get decelerated
in collisions with nucleons.

The proton-momentum spectrum shows also two slopes. The fast protons
with momenta larger than the Fermi momentum are knocked out from the
nucleus by collisions with energetic pions. The lower-momentum protons
are slowly evaporated from the nucleus after the fast cascading stage
of the reaction.

More discussions on the extraction of the antiproton-nucleus optical
potential from GiBUU calculations and comparisons with other analyses
can be found in \refcite{Larionov:2009tc}. Here, we only mention, that
in an earlier attempt to extract the $\overline{p}$-optical potential
from transport calculations of $\overline{p}$ production from
proton-nucleus and nucleus-nucleus collisions, the authors of \cite{Teis:1994ie} have
found a real part of $-200$ to $-100 \MeV$ consistent with our present
analysis.

\subsubsection{Reactions induced by protons and heavy ions at SIS
  energies}
\label{sec:protonA}

The theoretical description of proton-induced reactions at low to
intermediate beam energies is a powerful tool to test the GiBUU
approach in terms of multifragmentation. A reliable description of the
fragmentation process in low-energy reactions is an important step
before applying the transport model in reactions at higher energies
for the formation of hypernuclei, which is one of the major research
topics at the FAIR facility.

However, the description of fragment formation is a non-trivial task
within BUU in general. This is so because transport models describe
the propagation of one-particle phase-space distributions only and do
not account for the evolution of physical phase-space
fluctuations. The major difficulty here is the implementation of the
real fluctuating part of the collision integral and the reduction of
numerical fluctuations using many test particles per physical nucleon,
which however would require a large amount of computing
resources. Attempts to resolve this issue is still an open
problem~\cite{Colonna:2007hn}.

\begin{figure}[t]
  \centering
  \includegraphics[width=0.6\linewidth]{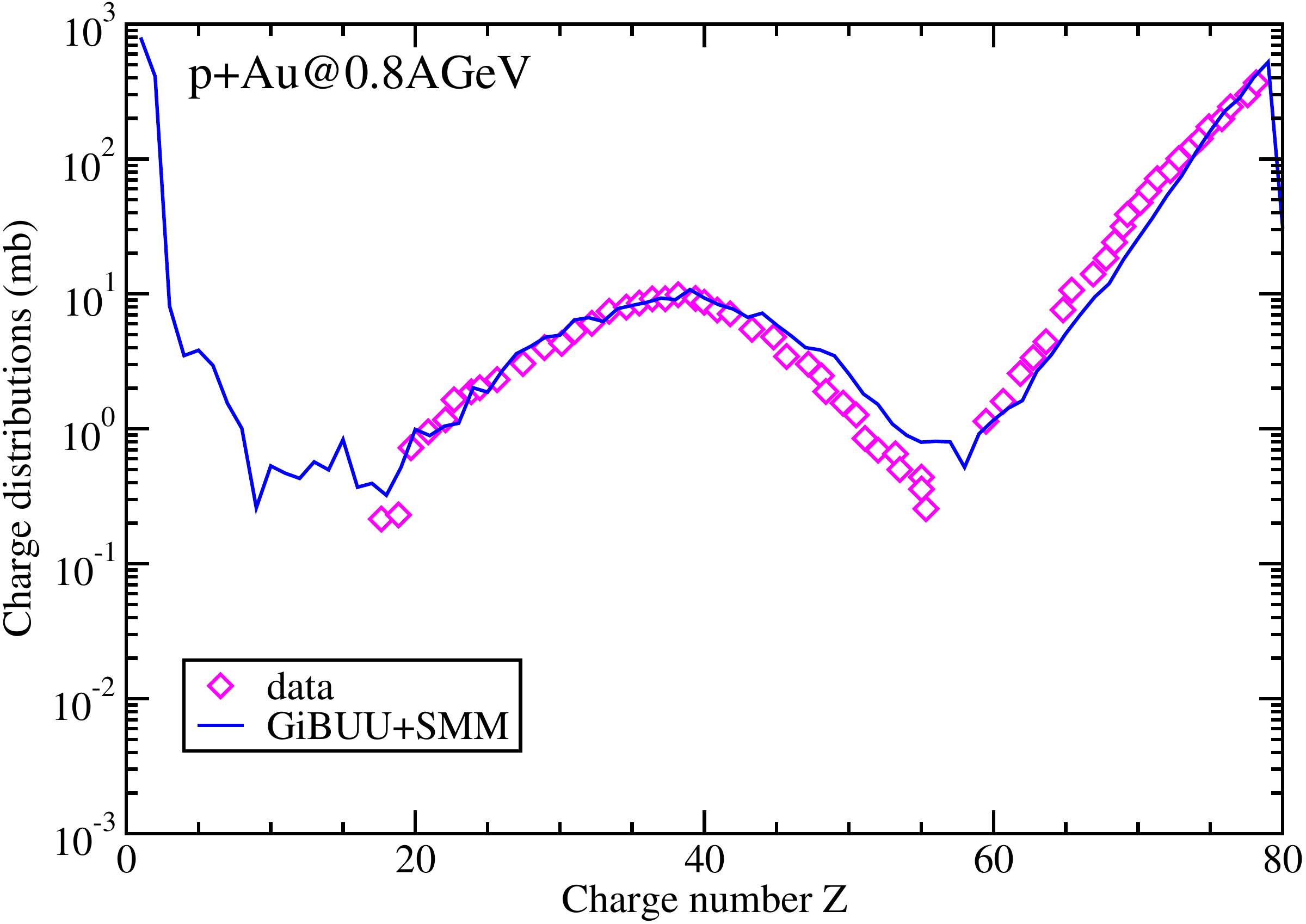}
  \caption{\label{Fig1} (Color online) Charge distributions for
    $p+\gold{}$ reactions at $E_{\text{beam}}=0.8\GeV$ incident
    energy. Theoretical calculations (solid line) within the
    hybrid GiBUU + SMM model are compared with data from
    \protect\cite{Rejmund:2000kn,Benlliure:2000ih} (open diamonds).
    Source: Taken from~\protect\cite{Gaitanos:2007mm}.}
\end{figure}

The standard approach of phenomenological coalescence models for
fragment formation has been found to work surprisingly well in
heavy-ion collisions, as long as one considers only one-body dynamical
observables (see~\cite{Reisdorf:1997fx}).  In particular, coalescence
models are usually applied to the description of central heavy-ion
collisions, in which a prompt dynamical explosion of a fireball-like
system is expected, with the formation of light clusters through
nucleon coalescence. In this dilute matter secondary effects are
negligible.

However, in hadron-induced reactions and in spectators in non-central
heavy-ion collisions the dynamical situation is different. In this
cases compression-expansion effects are only moderate (except in
reactions with antiprotons, see \cref{sec:antiprotonA}), and the
fragmentation process happens over a long time scale (compared to the
short-lived explosive dynamics of fireballs), which is compatible with
a statistical description of the process.

The whole dynamical picture in proton-induced reactions and in
spectators in heavy-ion collisions is therefore modeled by a
combination of dynamical and statistical models.  Usually two types of
microscopic approaches have been applied in proton-induced reactions:
the intranuclear-cascade (INC) model \cite{Boudard:2002yn} and the
quantum molecular dynamics (QMD) prescription
\cite{AbdelWaged:2006ay}, in combination with a statistical
multifragmentation model (SMM) \cite{Bondorf:1995ua}. The SMM model is
based on the assumption of an equilibrated source and treats its decay
statistically. It includes all necessary models of fragment formation,
such as sequential evaporation and fission.  The transition from the
dynamical (BUU) to the statistical (SMM) picture is controlled by the
onset of local equilibration, for details see \cite{Gaitanos:2007mm}.
More information on this field of research can be found in studies of
spallation reaction models where very detailed experimental data are
available~\cite{Mashnik:2010xh}.

As an example for the fragmentation of an excited residual source in
proton-induced reactions within this approach we compare in
\cref{Fig1} our theoretical results for the final charge distribution
to experimental data \cite{Rejmund:2000kn,Benlliure:2000ih} for
$p+\gold{}$ reactions. The theoretical results are in reasonable
agreement both with the \emph{absolute} yields and the shape of the
experimental data.

\begin{figure}[t]
  \centering
  \includegraphics[width=0.6\linewidth]{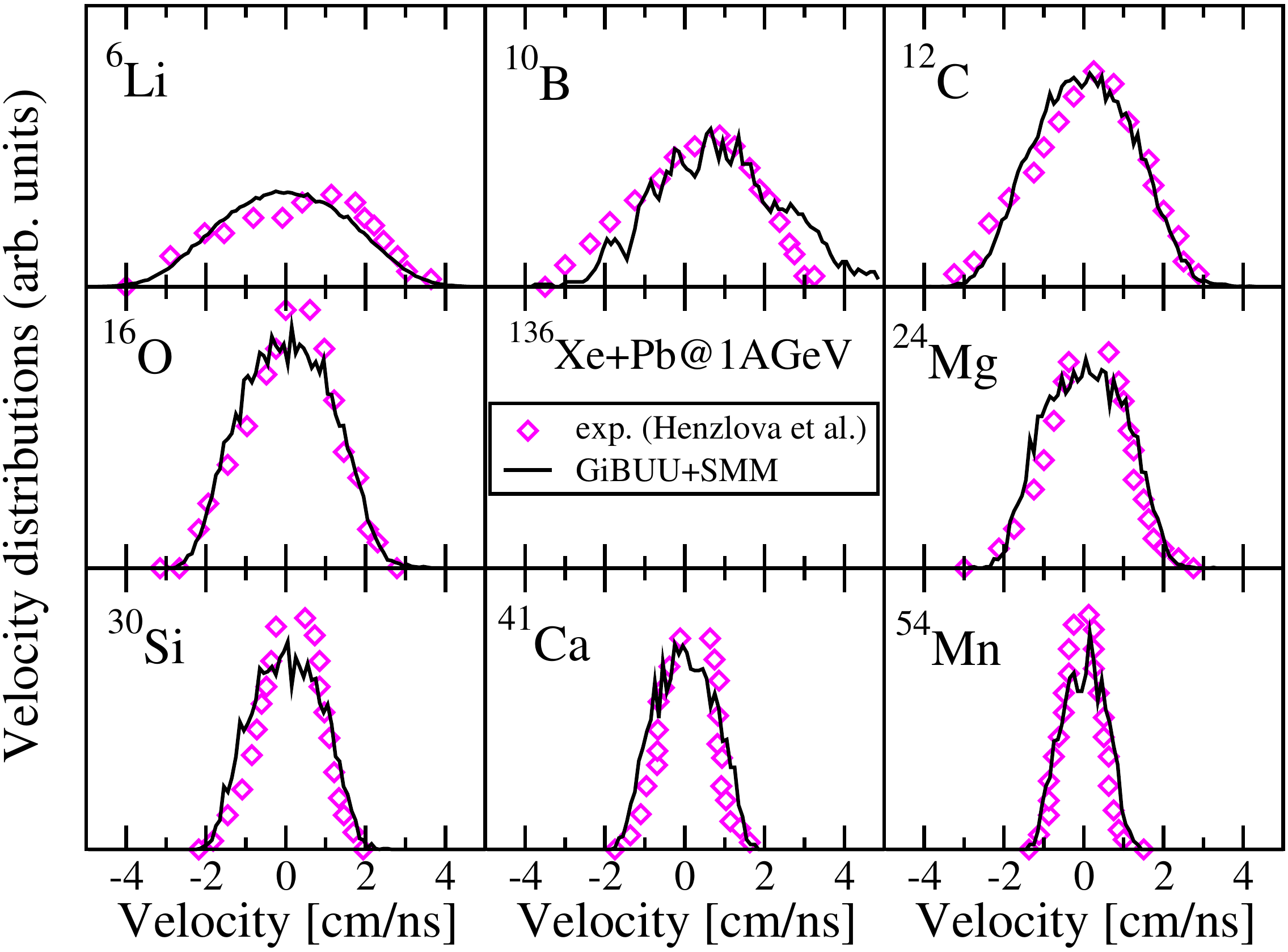}
  \caption{\label{XePb}(Color online) Longitudinal velocity
    distributions in the projectile frame for $\xenon+\lead$
    collisions in $1~\AGeV$ collisions. Theoretical calculations
    (solid curves) are compared with experimental data (open diamonds)
    from \protect\cite{Henzlova:2008zz}. Source: Taken
    from~\protect\cite{Gaitanos:2009at}.}
\end{figure}

In particular, the fragmentation of the residual source is a complex
process involving different mechanisms of dissociation.  According to
the SMM model, heavy nuclei at low excitation energy mainly undergo
evaporation and fission producing the sharp peaks at the very high and
low mass numbers. As the excitation energy (or temperature) approaches
$T\approx 5 \MeV$ the sharp structure degrades due to the onset of
multifragmentation, and at higher excitations, $T=5\upto17 \MeV$, one
expects exponentially decreasing yields with decreasing mass/charge
number. These different mechanisms of dissociation of an excited source
finally lead to a wide distribution in $Z$ as shown in
\cref{Fig1}. Similar results are obtained for the mass distributions,
for the production yields of different isotopes, and also for the
energy spectra of emitted neutrons \cite{Gaitanos:2007mm}.

We have also applied the combined GiBUU+SMM approach to spectator
fragmentation in inter\-me\-di\-ate-energy heavy-ion collisions. A
correct description of velocity distributions of statistically
produced fragments is crucial for their subsequent coalescence with
hyperons. \Cref{XePb} shows fragment-velocity distributions in the
projectile frame in comparison with experimental data taken from
\cite{Henzlova:2008zz}. The velocity distributions are described
rather well on a quantitative level, in particular the width of the
Gaussian-like fragment-velocity distributions is well reproduced.

\begin{figure}[t]
  \centering
  \includegraphics[width=0.6\linewidth]{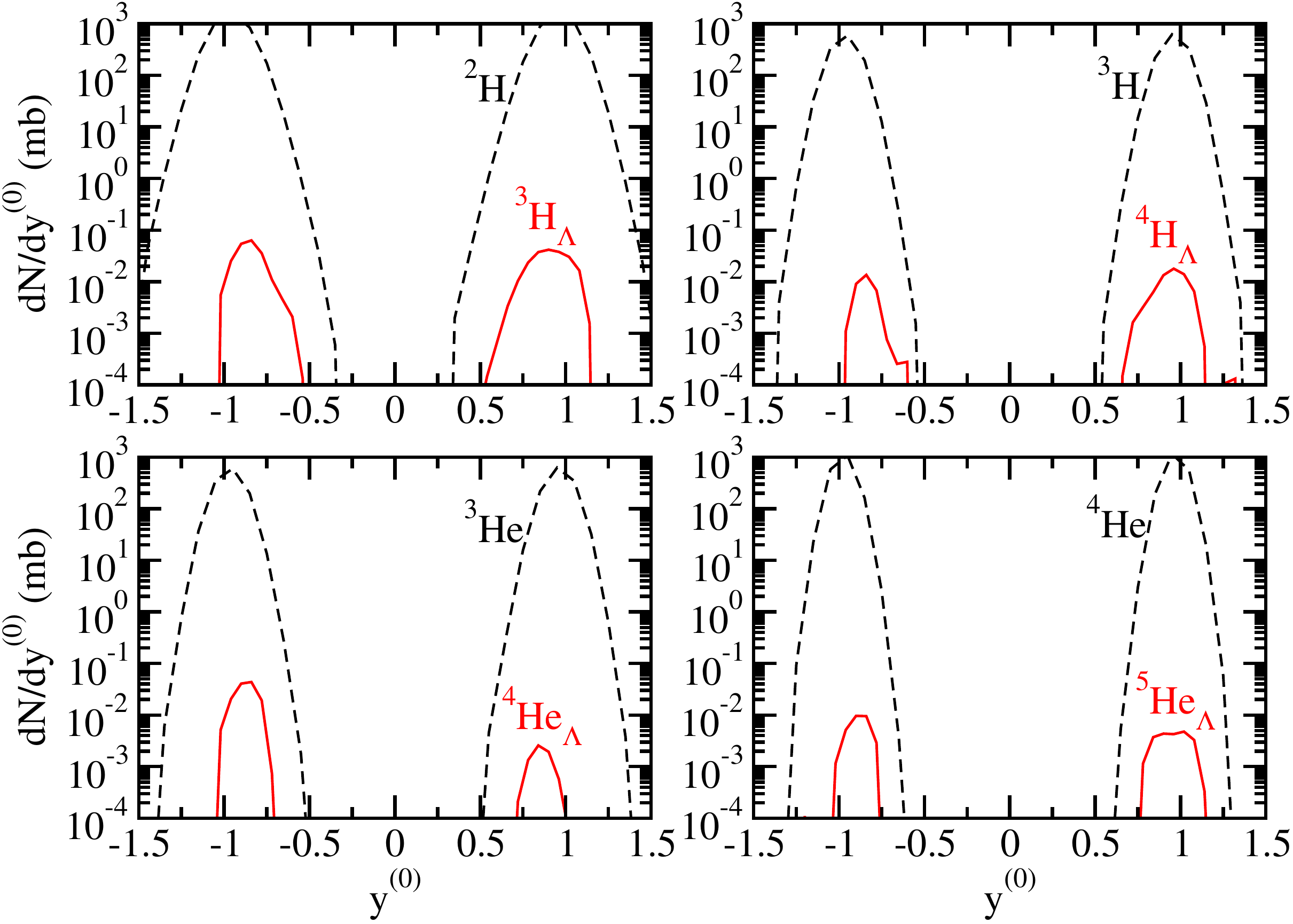}
  \caption{(Color online) \label{hypHI}Rapidity distributions as
    function of rapidity, $y^{(0)}$ (normalized to the projectile
    rapidity in the c.m.~frame) of different particle types (as
    indicated) for \carbon{}+\carbon{} at $2 \AGeV$. Taken
    from~\protect\cite{Gaitanos:2009at}.}
\end{figure}

In general, it turns out that the hybrid model leads to a quite
satisfactory description for the fragmentation of residual nuclei in
proton-induced reactions and for spectator fragmentation in
heavy-ion collisions, which is a non-trivial task in transport-dynamical
approaches. We note again, that the nonequilibrium dynamics has been
treated in a microscopic way using the relativistic coupled-channel
transport approach, which is an important step in extracting the
properties of the excited configuration. Afterwards the excited
fireball-like systems decay statistically according to the SMM model.

An important application of the combined GiBUU+SMM model is the study
of hypernuclei in spectator fragmentation.  The production of
hypernuclei in energetic collisions between light nuclei is one of the
major research topics investigated by the HypHI-collaboration at
GSI. The reason for selecting very light systems is the easier
identification of hypernuclei via the weak decay of a $\Lambda$ hyperon into
a proton and a negative pion. In earlier theoretical studies, see, e.g.,
\cite{Wakai:1992pg}, cross sections of the order of only few microbarn
($\mu \mathrm{b}$) were predicted, which can be easily understood: in
collisions between very light systems, such as \carbon{}+\carbon{} at
$2 \AGeV$, secondary re-scattering events inside the spectator matter,
which are important for producing low energetic hyperons, are rare
processes due to the small interaction volume. The situation is
different in collisions between heavy nuclei due to the high
production rate of strangeness and many secondary scattering
events. The formation of hypernuclei in spectator fragmentation is
performed with a phase-space coalescence prescription between the
hyperons produced dynamically from GiBUU and the fragments generated
from the SMM model.

We have studied the formation of hypernuclei in spectator
fragmentation in $2\AGeV$ $\carbon{}+\carbon{}$ collisions.  Inclusive
rapidity spectra for different light fragments and hyperfragments from
spectator matter are shown in \cref{hypHI}.  The estimated
hyperfragment-production rate is approximately 5 to 6 orders of
magnitude lower than that of fragment production in general. This is
obvious, since the strangeness-production cross sections from
exclusive primary channels, i.e., primary $BB\rightarrow BYK$,
$BB\rightarrow BBK\overline{K}$, and secondary ones ($B\pi\rightarrow
YK$ and $B\overline{K}\rightarrow \pi Y$, where $B$ stands for a
baryon and $Y$ for a hyperon) are very small (orders of few $\mu
\text{b}$) \cite{Tsushima:1998jz}. Among the different processes
contributing to the formation of hypernuclei, $BB\rightarrow BYK$ and
$\Lambda B\rightarrow \Lambda B$ and the secondary one
$B\pi\rightarrow YK$ give the major contribution to the formation of
hypernuclei. Strangeness production channels with three- and four-body
final states are important for the production of low-energy hyperons
that can be captured by spectator matter. The same argument also holds
for secondary re-scattering via elastic hyperon-nucleon and
pion-nucleon processes.

\subsubsection{Dilepton spectra from proton-induced reactions}
\label{sec:protonA_dilep}

While the vacuum properties of most hadrons are known to reasonable
accuracy nowadays, modifications of hadron properties in a strongly
interacting environment have attracted a lot of attention and have been
intensively studied both theoretically and experimentally. These studies
were motivated by the expectation that chiral symmetry may be restored
in a nuclear medium at high temperatures or densities
\cite{Hatsuda:1991ez,Hatsuda:1995dy} and that, as a consequence, the
masses of the light vector mesons should be shifted downwards. In the
vacuum this approximate symmetry is spontaneously broken as visible in
the low-mass part of the hadronic spectrum: chiral partners -- hadronic
states with the same spin but opposite parity -- like the $\rho$ and
$a_1$ meson are different in mass while they should be mass degenerate
if chiral symmetry were not spontaneously broken. It turns out, however,
that a connection between chiral-symmetry restoration and hadronic
in-medium spectral functions is much more involved. QCD sum rules
provide a link between the quark-gluon sector and hadronic descriptions
but do not determine the properties of hadrons in the strongly
interacting medium uniquely. They only provide constraints for hadronic
models, which are still needed for calculating the in-medium self-energies
of hadrons and their spectral functions. Most of these models
predict a sizeable in-medium broadening of the vector mesons, but only
very small mass shifts \cite{Leupold:2009kz}. Experiments looking for
such effects have used various probes (for details see
\cite{Leupold:2009kz,Hayano:2008vn}).

For studying in-medium effects, the more prominent hadronic decay
modes of the vector mesons are unfavorable, since they are affected by
strong final-state interactions with the hadronic medium -- in
contrast to the rare dilepton decay modes, which only feel the
electromagnetic force. Therefore the latter are ideally suited to
carry the in-medium information outside to the detector, undisturbed
by the hadronic medium.

Light vector mesons are particularly suited for these investigations
since -- after production in a nuclear reaction -- they decay in the
nuclear medium with sizable probability because of their short lifetimes
(this is particularly valid for the $\rho$ meson, and to a lesser extent
for the $\omega$ and $\phi$). Experimentally this field has been
addressed in reactions with hadronic and with photon beams (for the
latter see discussion in \cref{sec:omega_photo_nuclei}). Experimental
results are summarized and critically evaluated in recent reviews
\cite{Leupold:2009kz,Hayano:2008vn,Rapp:2009yu,Tserruya:2009zt}. Almost
all experiments report a softening of the spectral functions of the
light vector mesons $\rho$, $\omega$, and $\phi$. Increases in width are
observed depending on the density and temperature of the hadronic
environment. Mass shifts are only reported by the KEK group
\cite{Naruki:2005kd,Muto:2005za}, which studied $\rho$, $\omega$, and
$\phi$ production in proton-nucleus reactions at $12\GeV$.  Also, the
HADES detector at GSI has started an ambitious program for measuring
dilepton spectra from $p+p$, $p+A$ and $A+A$ reactions \cite{:2008yh}.

First transport-theoretical calculations for dilepton production in
heavy-ion reactions were done quite early (see, e.g., the work by Ko
\etal{}~\cite{Li:1994cj} and by the Giessen group using a predecessor
code to GiBUU \cite{Wolf:1991ic}). Here we apply the GiBUU transport
model to the $p+p$ and $p+\text{Nb}$ reactions studied by the HADES
collaboration.  We use GiBUU to generate dilepton events and pass them
through the HADES acceptance filter, in order to compare our
calculations directly to the experimental data measured by HADES.

It is very important to make sure that one understands the elementary
reactions before moving on to heavier systems, which involve effects
of the nuclear medium. Fortunately, HADES has also measured dilepton
spectra from elementary $p+p$ reactions. These provide a base line for
exploring the heavier nuclear systems.

\begin{figure}[t]
  \centering
  \includegraphics[height=0.33\linewidth]{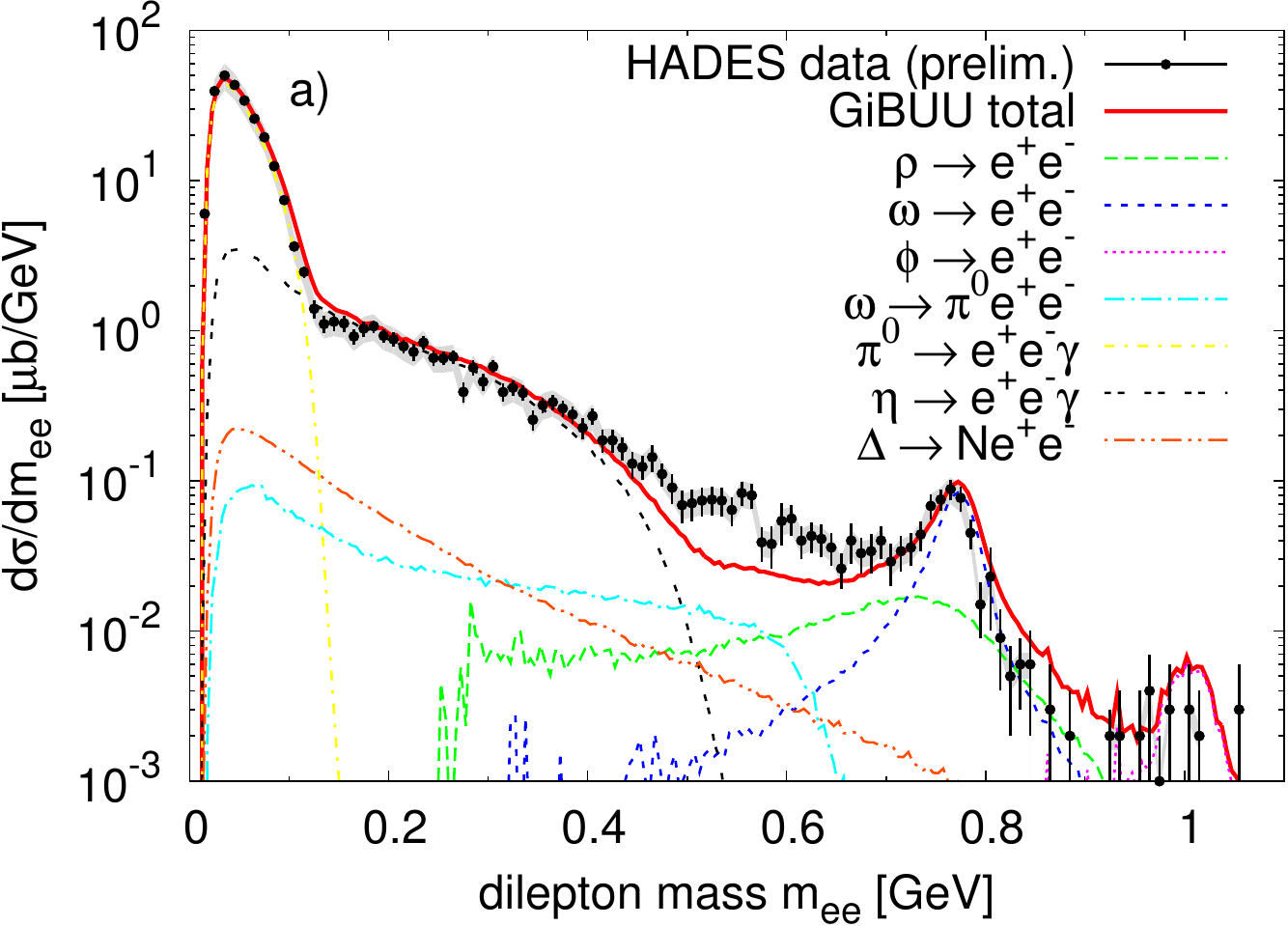}
  \includegraphics[height=0.33\linewidth]{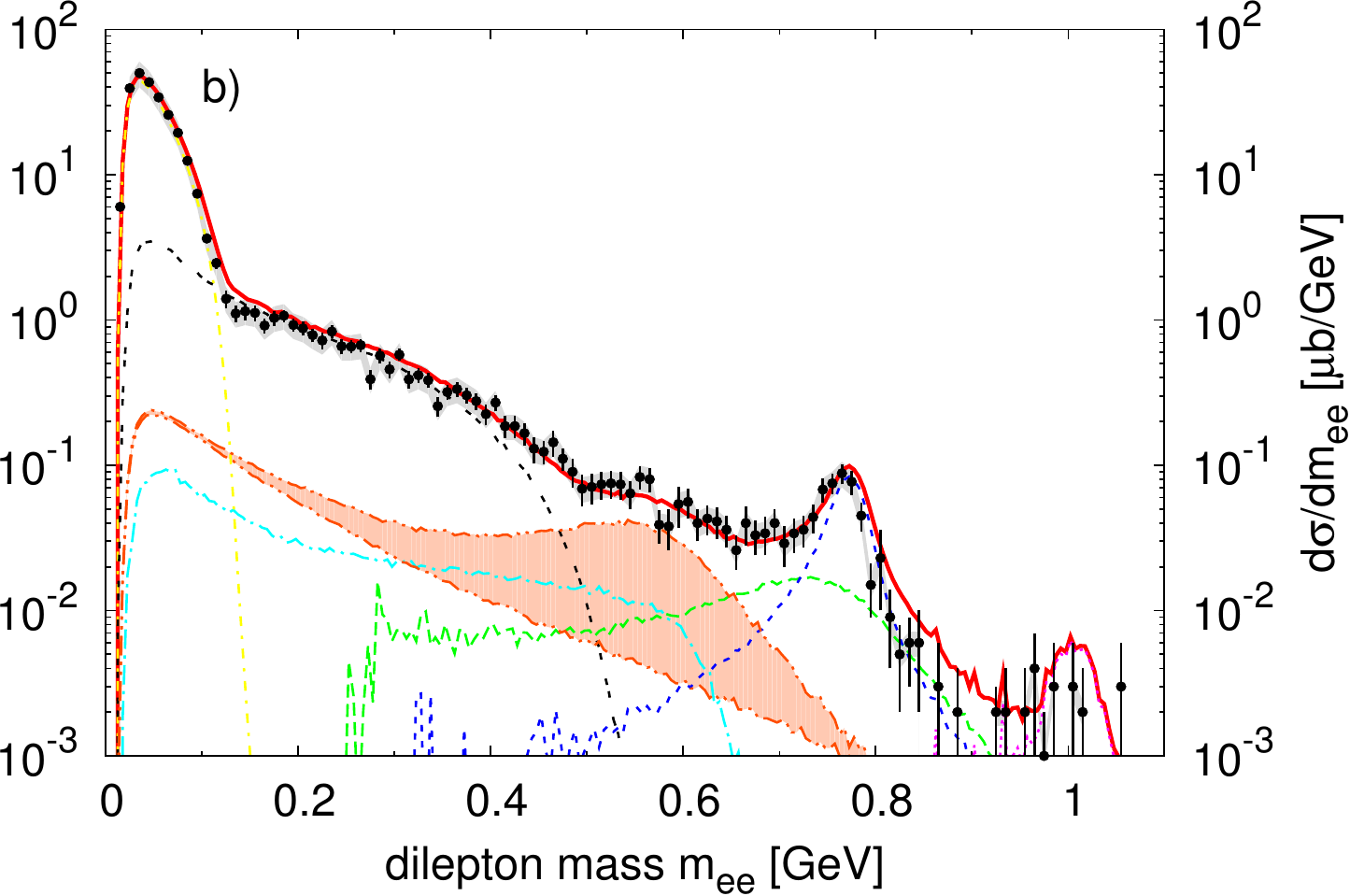}
  \caption{(Color online) Dilepton spectra for $p+p$ collisions at
    $E_{\text{kin}}=3.5\GeV$. a) Mass spectrum without $\Delta$ form
    factor, b) mass spectrum with $\Delta$ form factor. Figure taken
    from \cite{Weil:2011fa}.}
  \label{fig:pp_mass}
\end{figure}

\Cref{fig:pp_mass} shows a comparison plot of a GiBUU simulation to
HADES data \cite{Rustamov:2010zz} for a proton beam of $3.5 \GeV$
kinetic energy impinging on a fixed proton target. This setup
corresponds to a center-of-mass energy of $\sqrt{s}=3.18\GeV$. The
theoretical results have been corrected for the HADES acceptance and
reasonably reproduce the shape of the data over most of the mass
spectrum. In the intermediate mass region around $500\upto600 \MeV$
the inclusion of a proper transition form factor for the
$\Delta$-Dalitz decay \cite{Wan:2005ds} (shaded area) is crucial for
describing the data. Without such a form factor, the calculation
strongly underestimates the experimental data in this region (by at
least a factor of two). The electromagnetic N-$\Delta$ transition form
factor is well-constrained by electron-scattering data in the
space-like region. In the time-like regime of the $\Delta$-Dalitz
decay, however, there is no reliable data available so far. This
missing experimental information on the $\Delta$ form factor
represents one of the largest uncertainties in the dilepton spectra
presented here. Eventually the HADES data might help to shed more
light on this open issue and to provide constraints for form
factor-models like the one of \cite{Wan:2005ds}.

Another channel which could possibly contribute in the
intermediate-mass region is $\eta\rightarrow e^+e^-$. The current
upper limit for the branching ratio of this decay would overshoot the
HADES data by at least a factor of four. However, there is no $\eta$
peak visible in the data, and also the theoretical expectations from
helicity suppression are still orders of magnitude below the current
experimental limit. Therefore, the observed HADES dilepton spectrum at
$3.5 \GeV$ constrains this partial decay width even further.

Within the level of agreement in the elementary $p+p$ collisions shown
in \cref{fig:pp_mass}, we have a good baseline for studying in-medium
effects in $p+\text{Nb}$, although the issue of the $\Delta$-Dalitz
form factor is not completely settled.

\begin{figure}[p]
  \centering
  \includegraphics[width=1.0\linewidth]{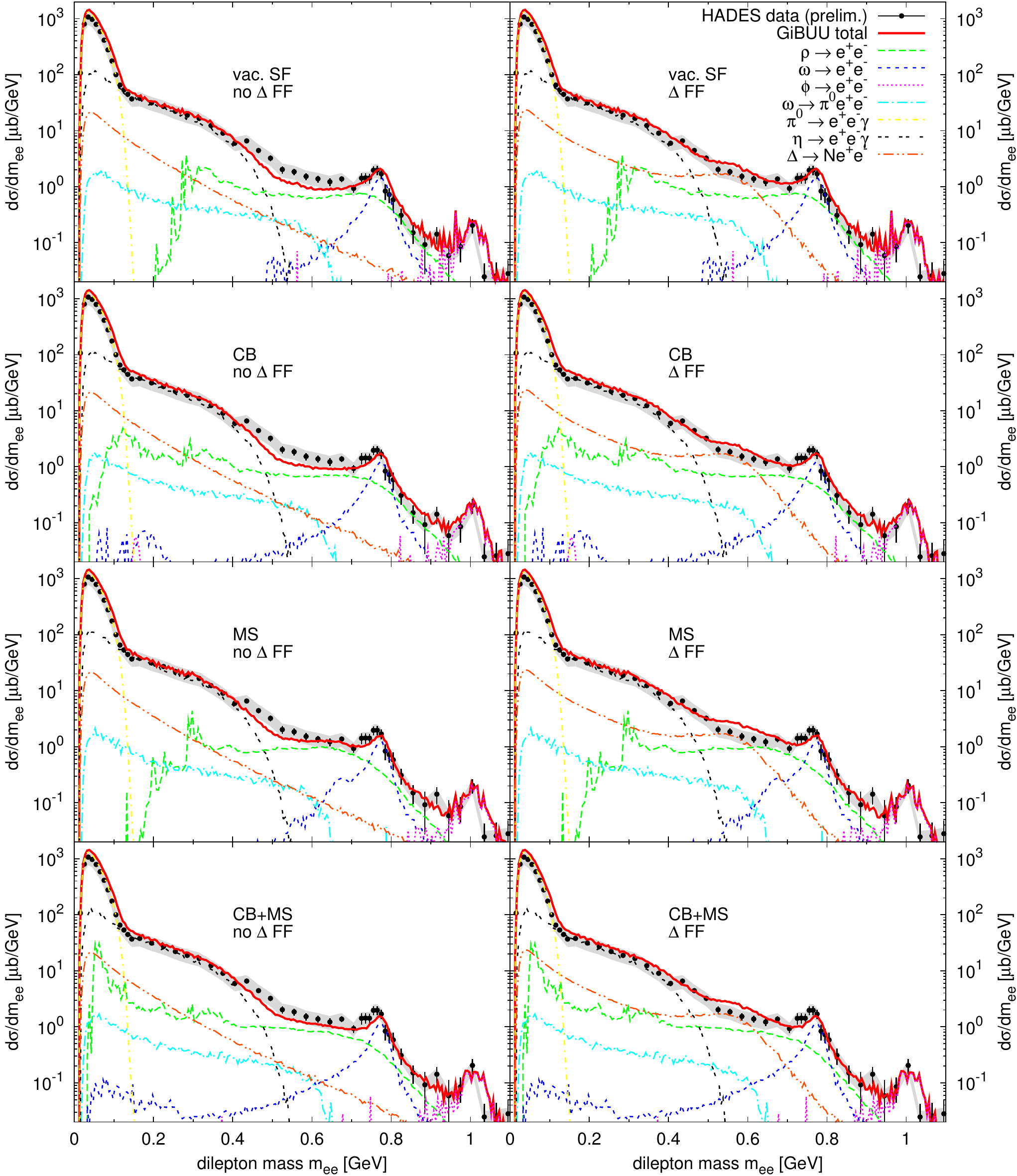}
  \caption{(Color online) Dilepton mass spectra for $p+\text{Nb}$
    collisions at $E_{\text{kin}}=3.5\GeV$. Left column: without
    $\Delta$ form factor. Right column: with $\Delta$ form factor.
    From top to bottom: vacuum spectral functions for the vector
    mesons, collisional broadening, $16\proz$ mass shift, collisional
    broadening plus mass shift. Figure taken from \cite{Weil:2011fa}.}
  \label{fig:pNb_mass}
\end{figure}

In $p+\text{Nb}$ reactions one has to consider some additional
effects, compared to the elementary $p+p$ reactions. First of all, the
primary $p+N$ collisions will be nearly identical, apart from binding
effects and some Fermi smearing, but besides $p+p$ also $p+n$
collisions will play a role.  Furthermore, the produced particles will
undergo final-state interactions within the Nb nucleus, and processes
like meson absorption and regeneration may become important. On
average, the secondary collisions will have lower energies than the
primary $N+N$ collisions, therefore also the low-energy resonance part
of the collision term will be involved. Finally, also the vector-meson
spectral functions may be modified in the nuclear medium.

\Cref{fig:pNb_mass} shows simulated dilepton spectra for $p+\text{Nb}$
collisions at $3.5 \GeV$ in various scenarios, compared to the data
from \cite{Weber:2011zz}.  The overall agreement is not quite as good
as in the $p+p$ case. Already in the pion channel we slightly
overestimate the data. This might have various reasons: too little
absorption, too much secondary-pion production in GiBUU or even a
normalization problem in the data.

According to \cite{weber_phd}, the data have been normalized by
comparing charged-pion spectra measured by HADES in $p+\text{Nb}$ to
those measured by the HARP collaboration. However, the cross sections
obtained by HARP had to be extrapolated to the slightly different beam
energy and nuclear target of HADES. This procedure is responsible for
most of the systematic error of the data (roughly 28\proz), which is
shown as a gray band in the figures.

Another striking feature of the $p+\text{Nb}$ system is that the
simulation gets close to the data in the intermediate mass range, even
without any $\Delta$ form factor. Including the form factor will
slightly overshoot the data. Most of the intermediate-mass gap
observed in $p+p$ is now filled up by low-mass $\rho$ mesons, which
are produced in secondary collisions. Even in $p+p$ collisions, one
might already find a similar effect by describing $\rho$-meson
production via resonance excitation, which could give stronger
contributions in the low-mass part of the $\rho$ spectral function
than \Pythia{}'s string-fragmentation model.

The mass spectrum above $500 \MeV$ can receive further modifications
from the inclusion of in-medium effects in the vector-meson spectral
functions. \Cref{fig:pNb_mass} shows a few typical in-medium
scenarios: The first one includes a collisionally broadened in-medium
width (cf.~\cref{sec:selfenergies}), while the second one assumes a
pole-mass shift according to
\begin{equation}
  m^*(\rho) = m_0\left(1-\alpha\frac{\rho}{\rho_0}\right)
\end{equation}
with a scaling parameter, $\alpha = 16 \proz$.  The third scenario
combines both of these effects. The modifications introduced by these
scenarios are roughly as large as the systematic errors of the
data. This fact, together with the discrepancy in the $\pi^0$ channel
and the uncertainty of the $\Delta$ form factor, presently makes it
impossible to draw any hard conclusions on vector-meson properties in
cold nuclear matter from the HADES data.

In summary, the HADES data from elementary $p+p$ collisions at $3.5
\GeV$ kinetic beam energy can be described very well by the \Pythia{}
event generator with a few adjusted parameters, as employed in the
GiBUU model. It has been found that the intermediate mass region is
sensitive to a VMD-like transition form factor for the $\Delta$ Dalitz
decay. Building on this agreement for the elementary reaction, the
$p+\text{Nb}$ reaction at the same beam energy is reasonably well
described by the GiBUU transport model.

\subsubsection{Heavy-ion collisions at AGS/SPS energies}
\label{sec:heavyIons}

At higher incident energies the meson production starts to dominate
the reaction dynamics.  Heavy-ion collisions (HICs) are very
complicated processes, where all possible hadron-hadron collision
channels can potentially contribute, particularly in the AGS/SPS
energy regime relevant for CBM@FAIR physics. The microscopic transport
description of HICs at these high energies is not only an interesting
physical problem, but also a quality test of practically all parts of
the GiBUU code.

Here we will consider the central $\text{Au}(2\upto20\AGeV)+\text{Au}$
and $\text{Pb}(30\text{ and }40\AGeV)+\text{Pb}$ collisions. The
formation of a highly-compressed baryonic matter ($\rho \sim 10
\rho_0$) is expected in these reactions with a possible
transition to the partonic phase
\cite{Gazdzicki:96,Gorenstein:2003cu,Bratkovskaya:2003ie,Wagner:2004ee,Arsene:2006vf,Larionov:2007hy}.
By observing the deviations of the microscopic transport predictions
with experiment one can, in principle, speculate on the signals of a
deconfinement.

The RMF mode of calculations (see
\cref{subsec:RMFpot,subsec:RMFtransport}) is used in the description
of HICs. The relativistic description of a nuclear mean field is
important since even in the ground state nuclear matter at high
densities the motion of nucleons is relativistic. The NL2 set of
parameters of the RMF model from \refcite{Lang:92} has been applied,
see \cref{tab:RMF} in \cref{subsec:RMFpot}.

\begin{figure}[t]
  \centering
  \includegraphics[width=0.5\linewidth]{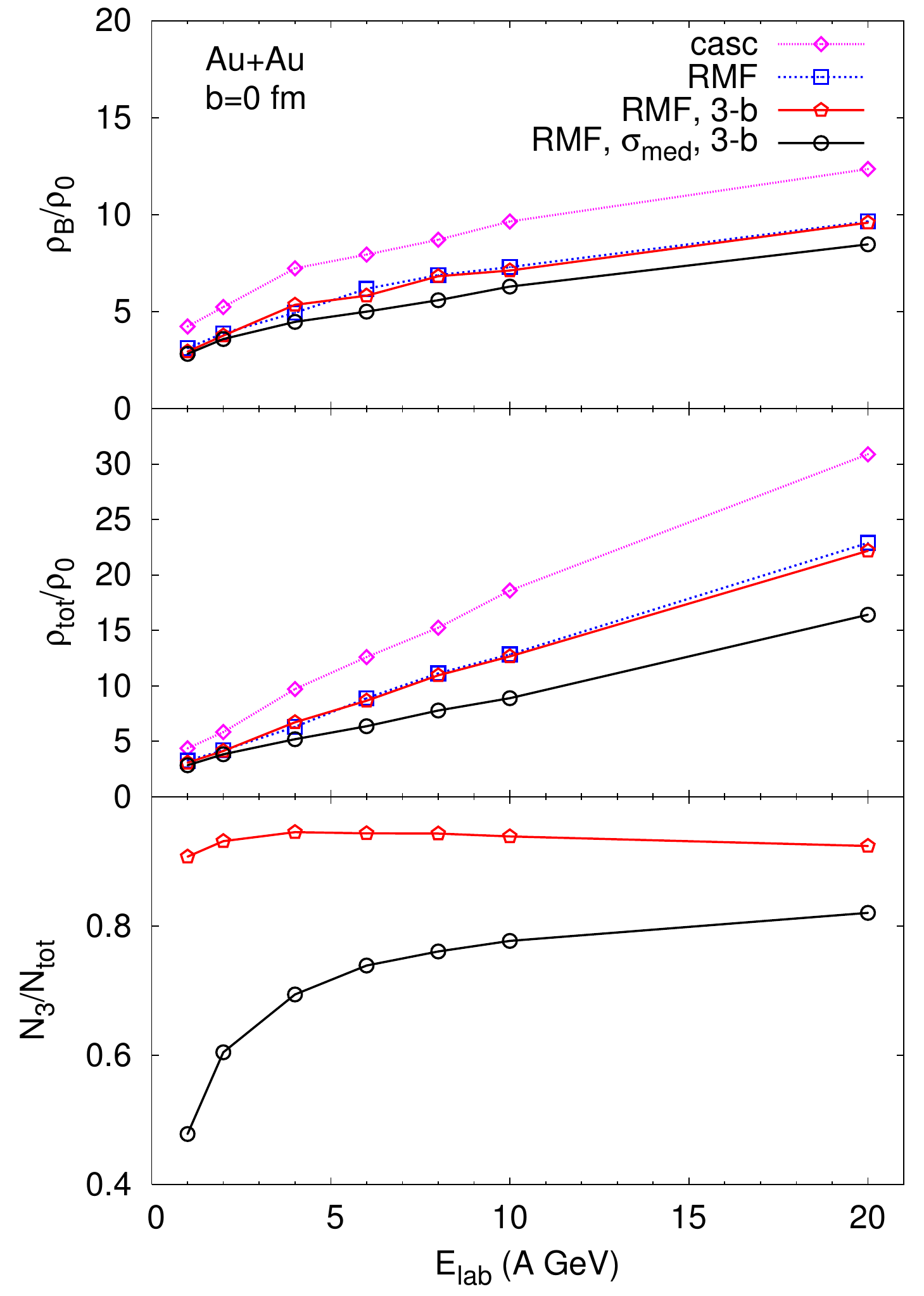}
  \caption{\label{fig:HIC_ratio} (Color online) Top, middle and bottom
    panels, respectively: the maximum central baryon density, the
    maximum central total (i.e.~baryon plus meson) density, and the
    maximum ratio of the three-body collision frequency to the total
    (two- plus three-body) collision frequency reached in central
    Au+Au collisions vs.~the beam energy. The binary-cascade
    calculation (i) is shown by lines with open diamonds. The
    calculation with RMF and vacuum cross sections (ii) is presented
    by lines with open boxes. The results with RMF and three-body
    collisions computed with vacuum cross sections (iii) are plotted
    with lines with open pentagons. Finally, calculations with RMF,
    three-body collisions and in-medium cross sections (iv) are shown
    by lines with open circles.  Figure taken from
    \cite{Larionov:2007hy}.}
\end{figure}

An important feature of relativistic HICs is the Lorentz contraction
of colliding nuclei in the c.m.~frame. In order to resolve
Lorentz-contracted density profiles, in calculating the $\sigma$-field
and the baryon four-current we have used a grid with cell sizes of
$\simeq (1 \times 1 \times 1/\gamma) \fm$. Here, the $z$-axis is the
beam axis, and the $\gamma$-factor is taken in the c.m.~frame of the
colliding nuclei (cf.~\cref{sec:Vlasov}). We have used the
parallel-ensemble technique (cf.~\cref{sec:EnsembleTechniques}) with
$N=200$ test particles per nucleon. For more details concerning the
numerical treatment of the mean-field propagation and high-energy
baryon-baryon and meson-baryon collisions see
Ref.~\cite{Larionov:2007hy,Falter:2004uc}.

The cross sections entering the collision integral are taken from a
fit to data or from simple model calculations tuned to reproduce some
selected data sets on elementary hadron-nucleon collisions. Such a
procedure can only describe vacuum cross sections. In the nuclear
medium, the cross sections are modified, e.g., due to the change of
the energy threshold for particle production by the mean fields. There
is no well established theoretical method to determine the in-medium
cross sections yet. Only for relatively simple processes, like, e.g.,
elastic scattering, $NN \to NN$, \cite{Fuchs:2001fp} or $\Delta(1232)$
resonance excitation, $NN \to N\Delta$,
\cite{TerHaar:1987ce,Larionov:2003av} the theoretical model
calculations of in-medium cross sections are available. These
calculations point toward an in-medium reduction of the baryon-baryon
cross sections at high baryon densities. The same effect is expected
also from a simple screening picture \cite{Danielewicz:1999zn}, since
the geometrical radius of the cross section should not exceed the
inter-particle distance.

\begin{figure}[t]
  \includegraphics[width=0.5\linewidth]{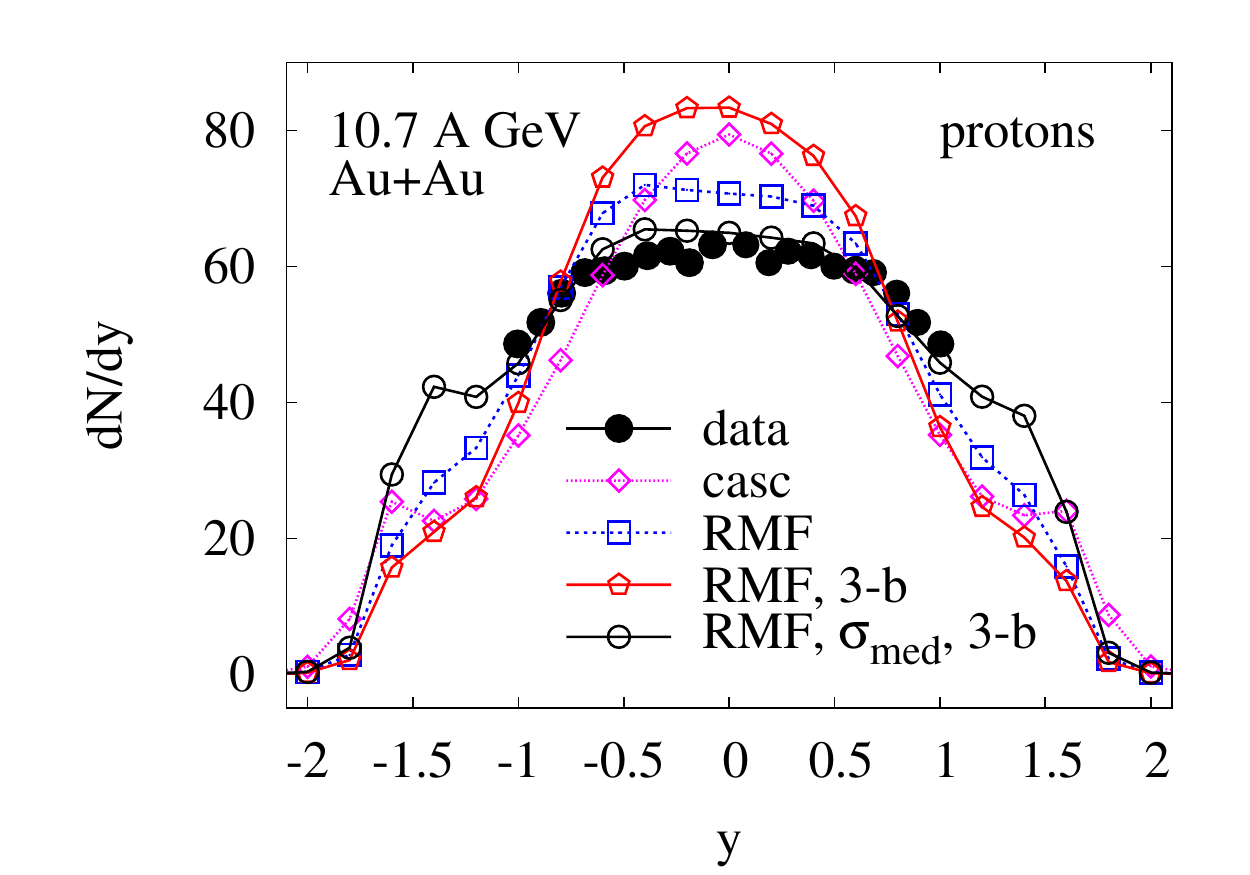}
  \includegraphics[width=0.5\linewidth]{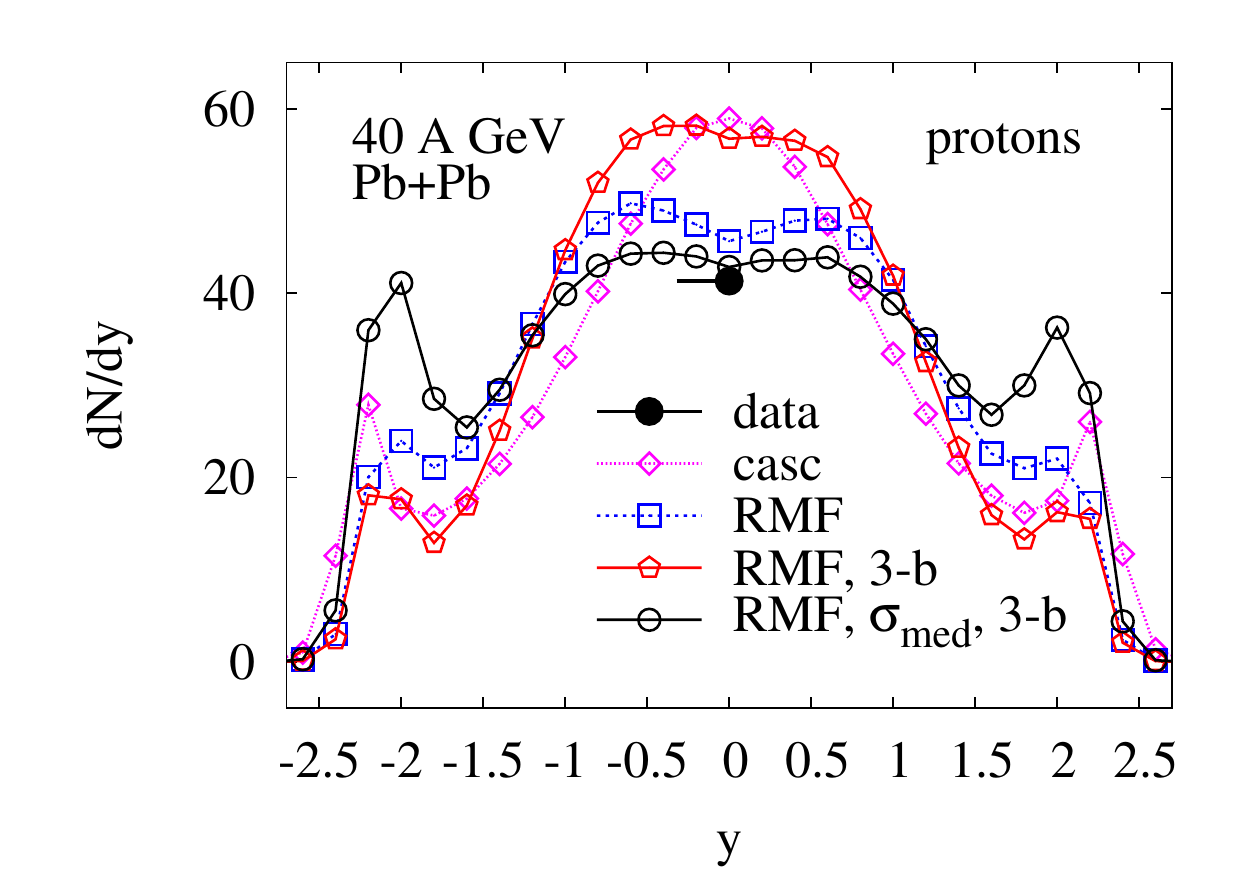}
  \caption{\label{fig:HIC_dNdy} (Color online) Proton rapidity
    distributions for central ($b \leq 3.5 \fm$) Au+Au collisions at
    $10.7\AGeV$ (left panel), and central ($b\leq 4\fm$) Pb+Pb
    collisions at $40\AGeV$ (right panel). The experimental data for
    the Au+Au system are taken from \refcite{Back:2002ic} and
    correspond to the 5\proz most central events. The data for Pb+Pb
    are from \refcite{Anticic:2004yj} (7\proz centrality).  Notations
    are the same as in \cref{fig:HIC_ratio}. Source: Figure taken from
    \cite{Larionov:2007hy}.}
\end{figure}

We have applied two simple models for the in-medium cross sections
described in \cref{General_aspects_sigmed}. In the first scheme, the
``free'' invariant energy $\sqrt{s_\text{free}}$ of the colliding
particles is determined from \cref{s_free_RMF} and used for all
reaction cross sections, $\sigma_{12 \to X}(\sqrt{s_\text{free}})$.
In the second scheme, we apply the in-medium-modified cross sections
of \cref{sigMed} to baryon-baryon collisions. We recall, that these
reduced cross sections are obtained assuming that the matrix elements
of the processes $B_1 B_2 \to B_{1'} B_{2'} M_{3'} M_{4'} ... M_{N'}$
and $B_1 B_2 \to B_{1'} B_{2'}$ are not changed in nuclear medium. The
baryon-meson and meson-meson cross sections are always calculated in
the vacuum. We expect that these cross sections are less subject to
in-medium modifications due to the smaller number of participating
fermions providing the main in-medium reduction effect by powers of
$m^*/m$ in the modification factor (\ref{F}).  For brevity, the first
scheme will be referred to as the calculation with vacuum cross
sections below.  The second scheme will be called the calculation with
in-medium cross sections.

As we will see below, baryon densities $5\upto10\rho_0$ are reached in
central HIC at $E_{\rm lab} =5\upto20\AGeV$ (see also
\cite{Arsene:2006vf} for a comparison of the different transport
models in this energy region). At such high densities, the transition
to a partonic phase is possible, which is under extensive discussion
in the literature
(cf.~\cite{Gazdzicki:96,Gorenstein:2003cu,Bratkovskaya:2003ie,%
  Khvorostukin:2006aw,Arsene:2006vf,Petersen:2008dd,Andronic:2008gu,Satarov:2009zx}).
In our calculations we follow a more conservative approach based on
(pre)hadronic and string degrees of freedom. However, a consistent
description of the dynamics of the high baryon density system requires
some modification of the standard kinetic approach based on two-body
collisions. Indeed, the gas parameter \cite{LP} defined as the number
of particles in the two-body interaction radius can be larger than
unity, $\gamma_{\rm gas} = (\sigma/\pi)^{3/2} \rho_B \simeq 2$. Here
we have taken $\sigma \simeq 40 \mb$ as the asymptotic high-energy
value of the total $pp$-vacuum cross section and $\rho_B=10\rho_0$ as
the maximum density reached in central Au+Au collision at
$20\AGeV$. The Boltzmann equation has a firm theoretical ground only
if $\gamma_{\rm gas} \ll 1$. Therefore, many-body collisions should be
taken into account at high densities. We have included only the
three-body collisions \cite{Larionov:2007hy}, following a simple
geometrical approach of \refcite{Batko:1991xd}.

\begin{figure}[t]
  \centering
  \includegraphics[width=0.6\linewidth]{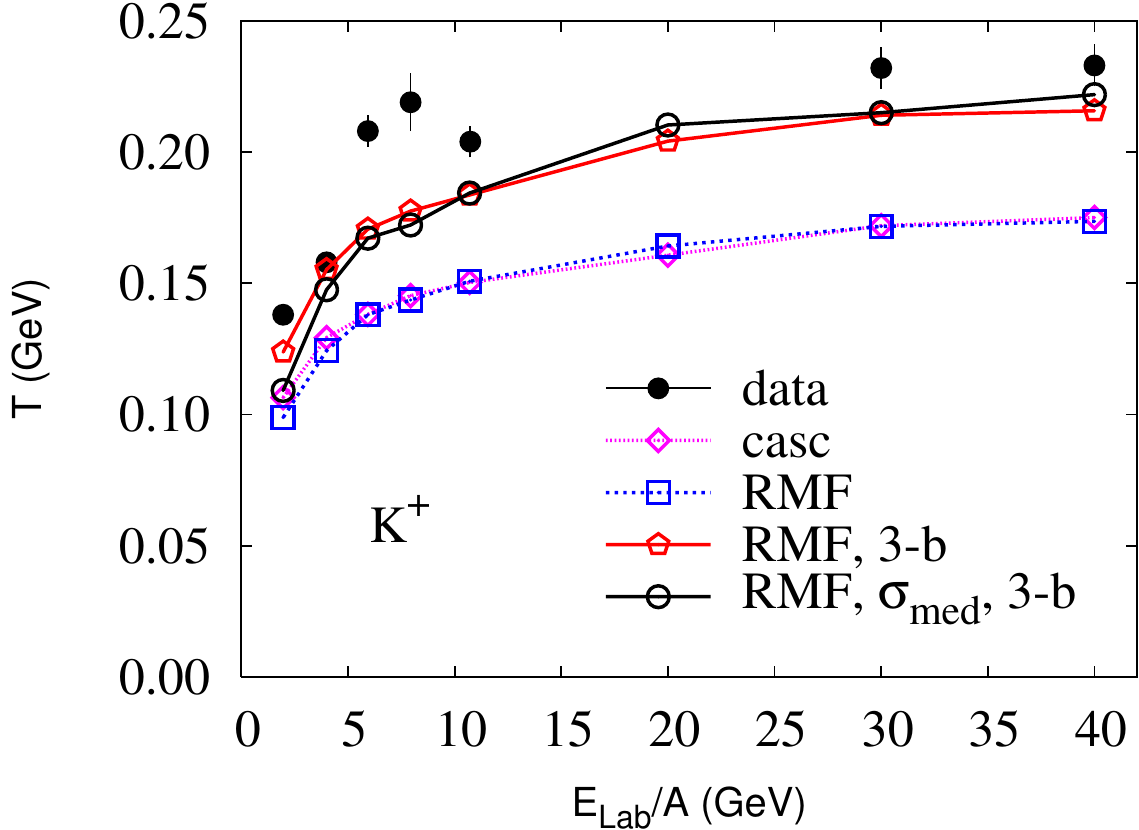}
  \caption{\label{fig:HIC_temp} (Color online) Inverse-slope parameter
    of the $K^+$ transverse mass spectrum at mid rapidity for central
    Au+Au and Pb+Pb collisions as a function of the beam energy. The
    data are taken from
    \cite{Ahle:1999uy,Afanasiev:2002mx,Friese:2003rn}.  Notations are
    the same as in \cref{fig:HIC_ratio}. Source: Figure taken from
    \cite{Larionov:2007hy}.}
\end{figure}

In order to demonstrate the influence of the various physical
ingredients to our model, four types of calculations have been done:
(i) pure binary-cascade calculation without mean field using
vacuum cross sections; (ii) calculation including the RMF, with only
binary collisions using vacuum cross sections; (iii) the same as (ii)
but including three-body collisions; (iv) the same as (iii), but using
the in-medium baryon-baryon cross sections.

\Cref{fig:HIC_ratio} shows the beam-energy dependence of the maximum
central baryon and total densities and of the maximum ratio of the
three-body and the total collision frequency $N_3/N_{\text{tot}}$
reached in central Au+Au collisions. The various schemes lead to an
uncertainty in the ma\-xi\-mum-ba\-ry\-on density of about $30\proz$
and in the maximum total density of about $50\%$. The RMF strongly
reduces the reached maximum-baryon and total densities. This can be
understood as a consequence of the formation of a repulsive vector
field. The three-body collisions do not influence these two
observables. The in-medium cross sections reduce the maximum densities
further. In particular, the total density decreases strongly due to
less pion production. The fraction of three-body collisions is quite
high, about $80\upto90\proz$ of all particle-collision events in a HIC
at $E_{\rm lab}=10\upto20 \AGeV$.  At high beam energies, the
meson-baryon collisions dominate, which are always computed with
vacuum cross sections. Therefore, the sensitivity of the ratio
$N_3/N_{\rm tot}$ to the in-medium baryon-baryon cross sections
becomes weaker with increasing beam energy.

For comparison with experiments we first address the stopping power of
nuclear matter. \Cref{fig:HIC_dNdy} shows the proton-rapidity
distributions in central Au+Au collisions at $10.7\AGeV$
and for the central Pb+Pb collisions at $40\AGeV$. The
cascade calculation produces too much stopping. Including the RMF
reduces the stopping power, leading to closer agreement with the
experimental rapidity spectra. Taking into account three-body
collisions increases the stopping power strongly, which again results
in an overestimation of the mid-rapidity-proton yields. Using
in-medium cross sections reduces the stopping power, leading again to
good agreement with the data. Therefore, a stronger deviation from
thermal equilibrium caused by in-medium reduced cross sections is
compensated by the additional thermalization efficiency due to
three-body collisions.

\begin{figure}[t]
  \centering
  \includegraphics[width=0.8\linewidth]{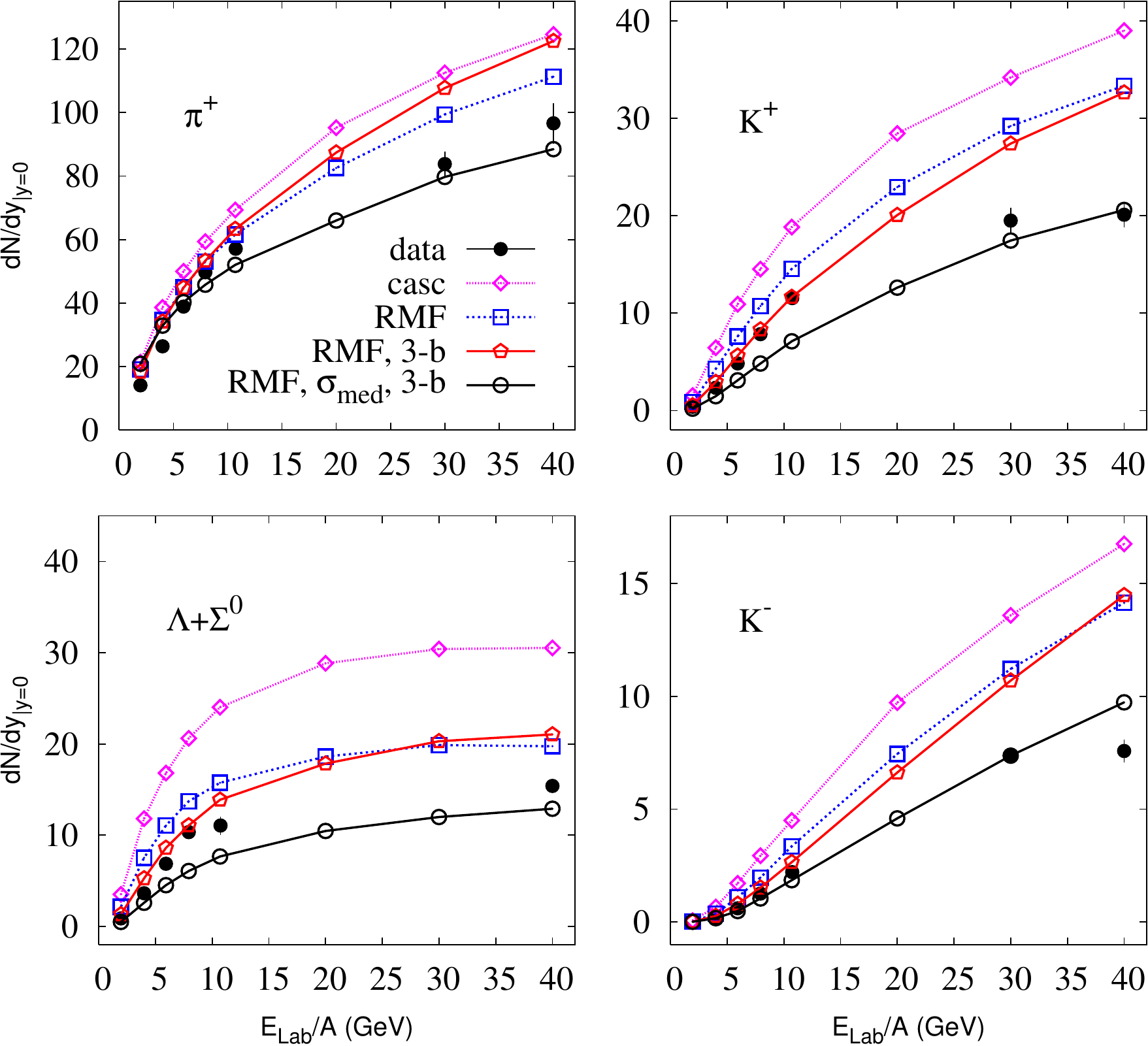}
  \caption{(Color online) The yield of $\pi^+$ (upper left panel),
    $K^+$ (upper right panel), $\Lambda+\Sigma^0$ (lower left panel)
    and $K^-$ (lower right panel) at mid rapidity as function of the
    beam energy for central Au+Au collisions at $E_{lab} \leq 20\AGeV$
    and Pb+Pb at $E_{lab} = 30$ and $40\AGeV$. The data are taken from
    \cite{Ahle:2000wq,Ahle:1999uy,Afanasiev:2002mx,Friese:2003rn,Mischke:2002wt,Mischke:2002ub,%
      Ahmad:1996,Pinkenburg:2002,Antinori:1999hy}.  Notations are the
    same as in \cref{fig:HIC_ratio}. Figure taken from
    \cite{Larionov:2007hy}.}
  \label{fig:HIC_dNdy_midr}
\end{figure}

In \cref{fig:HIC_temp} we present the inverse-slope parameter, $T$
\cite{Ahle:1999uy,Afanasiev:2002mx,Friese:2003rn} of the
$K^+$-transverse mass spectrum at mid rapidity vs.~the beam energy.
Neglecting three-body collisions, we underestimate the inverse-slope
parameter, $T$, by about 30\proz. Including three-body collisions
leads to a fair agreement with the data, except for the points at 5.93
and $7.94 \AGeV$, where we still underestimate the experimental inverse
slope parameter by about 20\proz.

\begin{figure}[t]
  \centering
  \includegraphics[width=0.6\linewidth]{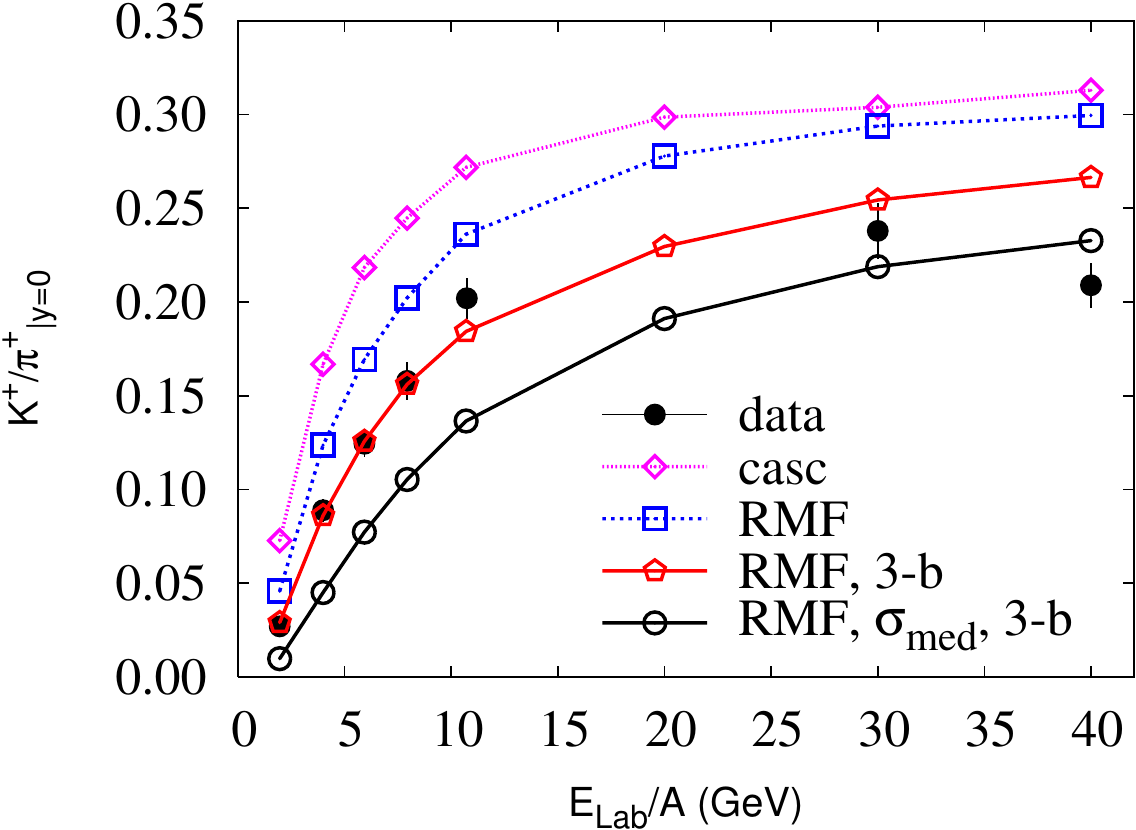}
  \caption{\label{fig:HIC_kp2pip_ratio} (Color online) The yield ratio
    $K^+/\pi^+$ at mid rapidity in central Au+Au and Pb+Pb
    collisions. The data are taken from
    \cite{Ahle:2000wq,Afanasiev:2002mx,Friese:2003rn}. Notations are
    the same as in \cref{fig:HIC_ratio}. Figure taken from
    \cite{Larionov:2007hy}.}
\end{figure}
\Cref{fig:HIC_dNdy_midr} shows the midrapidity yields of $\pi^+$,
$K^+$, $(\Lambda+\Sigma^0)$, and $K^-$ vs.~the beam energy for central
Au+Au collisions at 1.96, 4.00, 5.93, 7.94, $10.7\AGeV$, and $20\AGeV$
and for central Pb+Pb collisions at 30 and $40\AGeV$. The calculations
for the Au+Au system were done in the impact parameter range $b \leq
3.5\fm$ (5\proz of the geometrical cross section,
cf.~\cite{Ahle:2000wq}). For the Pb+Pb system, we have chosen a
slightly larger impact parameter range of $b \leq 4\fm$ (7\proz of the
geometrical cross section, cf.~\cite{Afanasiev:2002mx}).

We observe that the pure cascade calculation overestimates the meson
and hyperon production.  Using the RMF reduces the mid-rapidity-meson
yields by about 15\proz. The mid-rapidity-hyperon yield is reduced
even more, by about 30\proz. This reflects the behavior of the
proton-mid-rapidity yield shown in \cref{fig:HIC_dNdy}, since the mean
field acts on the hyperons too. The inclusion of three-body collisions
affects the mid-rapidity yields of the produced particles rather
weakly. Finally, using the in-medium cross sections reduces the
particle production quite strongly: for mesons by about 30\proz and
for hyperons by about 50\proz. The calculation with the in-medium
cross sections turns out to be in good agreement with the data on pion
and $K^-$ production, while it underestimates the $K^+$ and hyperon
yields below $40\AGeV$.

\Cref{fig:HIC_kp2pip_ratio} shows the $K^+/\pi^+$ ratio at midrapidity
vs.~the beam energy. It is interesting, that the three-body collisions
reduce the ratio quite strongly. This is due to a combination of two
small effects visible in \cref{fig:HIC_dNdy_midr}: increase of the
pion yield and decrease of the kaon yield by the three-body
collisions. The calculation in the RMF mode with vacuum cross sections
and three-body collisions is in the best agreement with the
experimental data below $30\AGeV$. However, we fail to describe the
reduction of the $K^+/\pi^+$ ratio above $30\AGeV$.

To summarize this subsection, the model works quite well for HICs in
the energy region between $2$ and $40\AGeV$. HICs at lower beam
energies remain to be studied in detail within the present
implementation of the model. At higher energies explicit consideration
of partonic degrees of freedom may become important.

\subsection{Electron-induced reactions}
\label{sec:electronA}

In photonuclear reactions, either with electrons or with real photons
in the incoming channel, nearly the entire nucleus is 'illuminated'
during the first interaction\footnote{Here, we disregard the
  phenomenon of shadowing, which may become important for low $Q^2$
  processes at higher energies \cite{Bauer:1977iq}. In
  \refcite{Falter:2002jc, Falter:2002vr} we have developed a method to
  take the shadowing effect into account in BUU calculations by using
  a properly derived profile function for the
  nucleus.}\label{shadowfootnote} This is not so in hadron-induced
reactions, where already a sizeable initial-state interaction
occurs. The photonuclear reactions are, therefore, an ideal testing
ground for theories of in-medium changes of hadronic properties. They
also present an opportunity for a sensitive test of theories and
methods to describe neutrino-nucleus interactions, essential for an
understanding of neutrino long-baseline experiments, that today
exclusively work with nuclear targets (see discussions later in
\cref{sec:nuA}) because photonuclear reactions contain the very same
vector part of the interaction as in the neutrino-induced reactions.

Electron scattering off nuclei in the regime of energy transfer
between 0.1 and $1 \GeV^2$ has been addressed by several experiments
within the last two decades; for a recent review of quasi-elastic
scattering off nuclei cf.~\refcite{Benhar:2006wy}. Comparing the
measured nuclear cross sections to the nucleon cross sections, several
modifications have been observed. First of all, nuclear Fermi motion
leads to a smearing of the peak structures such as in the quasi-elastic
and $\Delta$ regions. Furthermore, one observes a quenching of the
spectral strength around the quasi-elastic peak. In contrast to the
quenching in the peak region, one observes an enhancement in the
so-called dip-region between the quasi-elastic and the $\Delta$
peak. The peak position of the $\Delta$ resonance is found to be both
$A$ and $Q^2$ dependent: it exhibits a shift towards lower invariant
masses for $Q^2 \le 0.1 \GeV^2$ and a shift towards higher invariant
masses for higher $Q^2$ (for more details see the discussions in
\refcite{Leitner:2008ue}).

The basis of nearly all theoretical studies of photonuclear reactions
with nuclei has been the so-called impulse approximation. In this
approximation one assumes that the incoming photon interacts only with
one nucleon at a time; simultaneous interactions with two or more
nucleons are neglected.
This approximation should hold for momentum
transfers large enough that a single nucleon is resolved (i.e.\
roughly for $|\bvec{q}| > 300\MeV$).  Only very few studies have been
undertaken to investigate the influence of 2-nucleon primary
interaction processes (see, e.g.,
\cite{Delorme:1989nh,Dekker:1995zz,Gil:1997bm,Gil:1997jg}, for a
broader discussion of 2N processes see \cite{boffi}).

The influence of nuclear effects on both elastic electron- and
neutrino-scattering cross sections has been investigated by Benhar
\etal{} using an impulse approximation model with realistic spectral
functions obtained from nuclear many-body-theory calculations
\cite{Benhar:2005dj,Nakamura:2007pj}. In particular, they achieve good
agreement with inclusive electron-scattering data in the
quasi-elastic-peak region. Nuclear effects in the quasi-elastic (QE)
region have also been investigated in detail by Nieves \etal{} for
electrons \cite{Gil:1997bm} and neutrinos \cite{Nieves:2005rq} where,
among other nuclear corrections, long range nuclear correlations have
been included. Also this approach describes inclusive electron
scattering data with impressive agreement. A relativistic
Green's-function approach has been applied by Meucci \etal{}~to
inclusive electron \cite{Meucci:2003uy} as well as to inclusive
neutrino nucleus reactions \cite{Meucci:2004ip}. Butkevich
\etal{}~\cite{Butkevich:2007gm} addresses both neutrino and electron
scattering with special emphasis on the impact of different
impulse approximation (IA) schemes: plane wave IA (PWIA) and
relativistic distorted wave IA (RDWIA). Also the Ghent group applies
RPWIA and RDWIA models to neutrino and electron scattering in the QE
region \cite{Martinez:2005xe}; lately they have extended their
framework to pion production \cite{Praet:2008yn}.

In this subsection we will, therefore, compare results of some of
these studies with those obtained within the GiBUU model for
electron-induced reactions on nuclei, starting from low-energy
quasi-elastic scattering over resonance excitations up to the highest
energies available, where one tries to learn something about hadron
formation and attenuation in the medium. We start with a discussion of
electron- and photon-induced processes and then turn to neutrino-induced
reactions. Within our GiBUU framework, we aim at a consistent
treatment of the initial vertex and the final state processes and we
emphasize that these should not be treated separately.

The relevant elementary cross sections are all given in
\cref{sec:elementary_leptonN}.

\subsubsection{Inclusive cross sections}

Within the GiBUU model, which employs a local Thomas-Fermi
distribution for the momentum distribution of nucleons, the inclusive
cross section for scattering off a nucleus is given by
(cf.~\cref{eq:oneParticleFSD})
\begin{equation}
  \label{sigmatot}
  \dd\sigma_{\rm tot}^{\ell A \to \ell' A} = g \int_{\rm nucleus} \dd^3r
  \int \frac{\dd^3p}{(2\pi)^3} \Theta\left(p_F(\bvec{r}) - |\bvec{p}|\right) \; \dd\sigma_{\rm tot}^{\rm med}.
\end{equation}
Here, $g$ is the spin-degeneracy factor. The medium-corrected cross
section, $\dd\sigma_{\rm tot}^{\rm med}$, contains medium-dependent
changes of the elementary cross section such as the spectral function
of the outgoing nucleons, the effects of Pauli-blocking of the final
state and a flux correction factor.

This expression for the total cross section can be rewritten such that
the hole spectral function appears in it. With that aim in mind we
define the hole spectral function by
\begin{equation}
  \label{Phole}
  P(\bvec{p}, E) = \int_{\mathrm{nucleus}} \dd^3r \; \Theta\left [p_F(\bvec{r}) -
    |\bvec{p}|\right] \Theta(E) \delta\left(E - m^* + \sqrt{\bvec{p}\,^2 + {m^*}^2(\bvec{r},\bvec{p})}\right) ~.
\end{equation}
In \cref{Phole}, $m^*$ is the effective mass of the bound nucleon,
which contains the coordinate- and momentum-dependent potential and
$E$ is, as usual, the nuclear separation energy.  \Cref{sigmatot} can
now be rewritten as
\begin{equation}
  \dd\sigma_{\rm tot}^{\ell A \to \ell' A} = g \int \dd E \int \frac{\dd^3p}{(2\pi)^3} P(\bvec{p}, E)\,
  \dd\sigma_{\rm tot}^{\rm med}~.
\end{equation}
Written in this form the cross section has the same appearance as that
in \cite{Benhar:2006wy} or \cite{Nakamura:2002sg}.

\begin{figure}[t]
  \centering
  \includegraphics[width=0.6\linewidth]{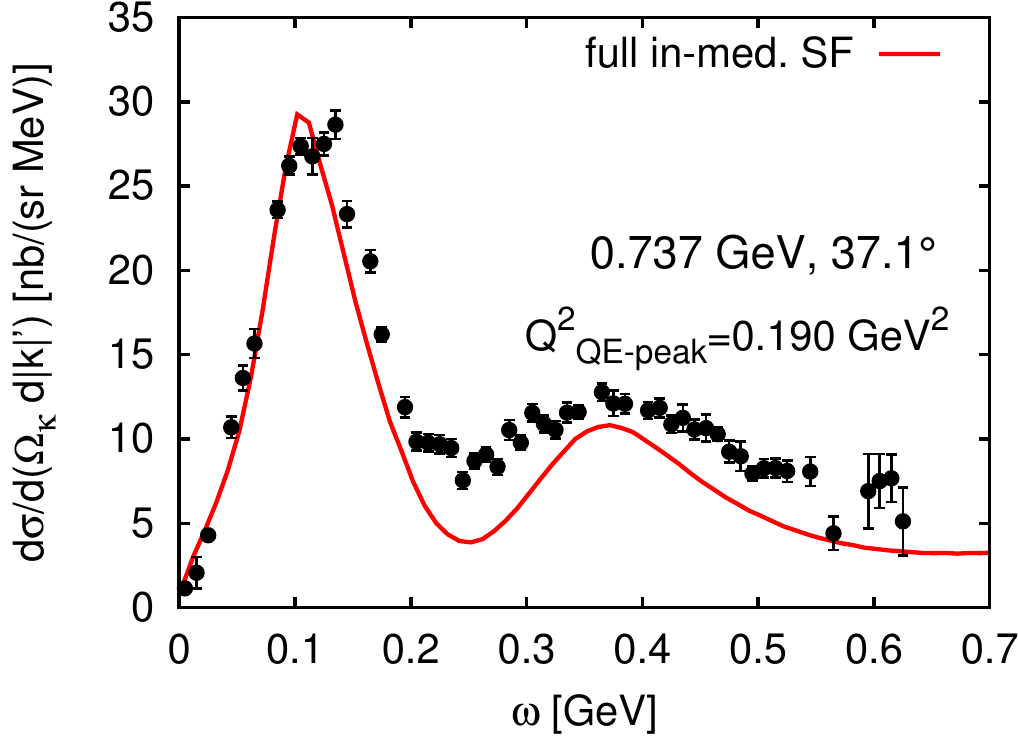}
  \caption{(Color online) Inclusive electron cross section $\dd^2
    \sigma/\dd \omega \dd \Omega_{k'}$ on $\atom{16}{O}$ as a
    function of the energy transfer $\omega$ for a beam energy of
    $0.737 \GeV$ and a scattering angle, $\theta_{k'} =
    37.1\,^{\circ}$, in comparison with the data measured by O'Connell
    \etal{}~\cite{O'Connell:1984nw}. The parameter $Q^2$ denotes the
    four-momentum-squared at the quasi-elastic-peak position (from
    \cite{Leitner:2008ue}).}
  \label{Fig:sigma-e-tot}
\end{figure}

The spectral function in \cref{Phole} contains the real part of the
self-energy through the potential in $m^*$, but it does not account for
the imaginary part. The latter can be implemented by using first
\begin{equation} \label{delta_p2} \delta \left(E - m^* +
    \sqrt{\bvec{p}\,^2 + {m^*}^2(\bvec{r},\bvec{p})}\right) = 2
  {p^*}^0 \delta\left(p^2 - {m^*}^2 \right)
\end{equation}
with the four-vector $ p = ({p^*}^0, \bvec{p})$ and ${p^*}^0 = m^* -
E$. One can then replace the $\delta$-function in \cref{delta_p2} by a
Breit-Wigner distribution of the form (cf.\ \cref{A_R,eq:spectralfunction_def})
\begin{equation}
  \label{spectfunc}
  \mathcal{A}(p) = - \frac{1}{\pi} \frac{\im \Pi(p)}{\left(p^2 -
      m^2 - {\rm Re} \Pi(p)\right)^2 + (\im \Pi(p))^2}~,
\end{equation}
where $\Pi(p)$ denotes the self-energy of the nucleon. The real part
of $\Pi$ contains the effects of the mean field potential (and
possible dispersive corrections)
\begin{equation} {m^*}^2 = (m + U_s)^2 = m^2 - \re \Pi ~,
\end{equation}
the imaginary part, $\im \Pi$, contains the width of the nucleon's
spectral function in medium (see \cref{Im_Pi}). In a low-density
approximation the latter is given by
\begin{equation}
  \Gamma = - \frac{1}{\sqrt{p^2}}\im \Pi(p) = \rho v \sigma_{\rm NN} ~.
\end{equation}
In \cite{Lehr:2000ua,Lehr:2001qy,Froemel:2003dv} we have shown that
transport theory can be used to calculate selfconsistently spectral
functions of nucleons in nuclei which agree very well with those
obtained in state-of-the-art nuclear many-body calculations.

The spectral function in \cref{Phole} differs in two essential aspects
from that of the global Fermi-gas model often used in theoretical
treatments of quasi-elastic scattering off nuclei. While the momentum
distribution of a global Fermi-gas model is constant up to the Fermi
momentum and then drops abruptly to zero, here -- because of the
smooth density distribution -- the momentum distribution has more
strength at lower momenta and drops smoothly to zero for large
$\bvec{p}$ (see Fig.~6 in \cite{Leitner:2008ue}). In addition, the
presence of the $\bvec{r}$-dependent potential, $V$ (implicit in
$m^*$), in \cref{Phole} leads to a smoothing of the distribution in
$E$, which in the global Fermi-gas model exhibits sharp ridges at the
on-shell values of this variable. This $E$ smearing appears even
without using a width (related to the imaginary part of the
self-energy) in the spectral function \cref{spectfunc}. The spectral
function \cref{Phole} is thus much closer to that in realistic models
\cite{Benhar:2005dj} than that of the global Fermi gas.

It is, therefore, not surprising that the spectral function
\cref{Phole}, when used in calculations of the total inclusive
electron cross section on nuclei, leads to very reasonable
results. This is illustrated in \cref{Fig:sigma-e-tot}, where the peak
at the lower energy transfer, $\omega \approx 0.1 \GeV$, is the
quasi-elastic peak.
In general, the agreement with
the data at the quasi-elastic peak is nearly as good as
that of Benhar \etal{}\ \cite{Benhar:2010nx}. The latter is based on a
state-of-the-art nuclear many-body calculation, but does not contain
the $\Delta$ resonance excitation.

\begin{figure}[t]
  \centering
  \includegraphics[width=0.40\linewidth]{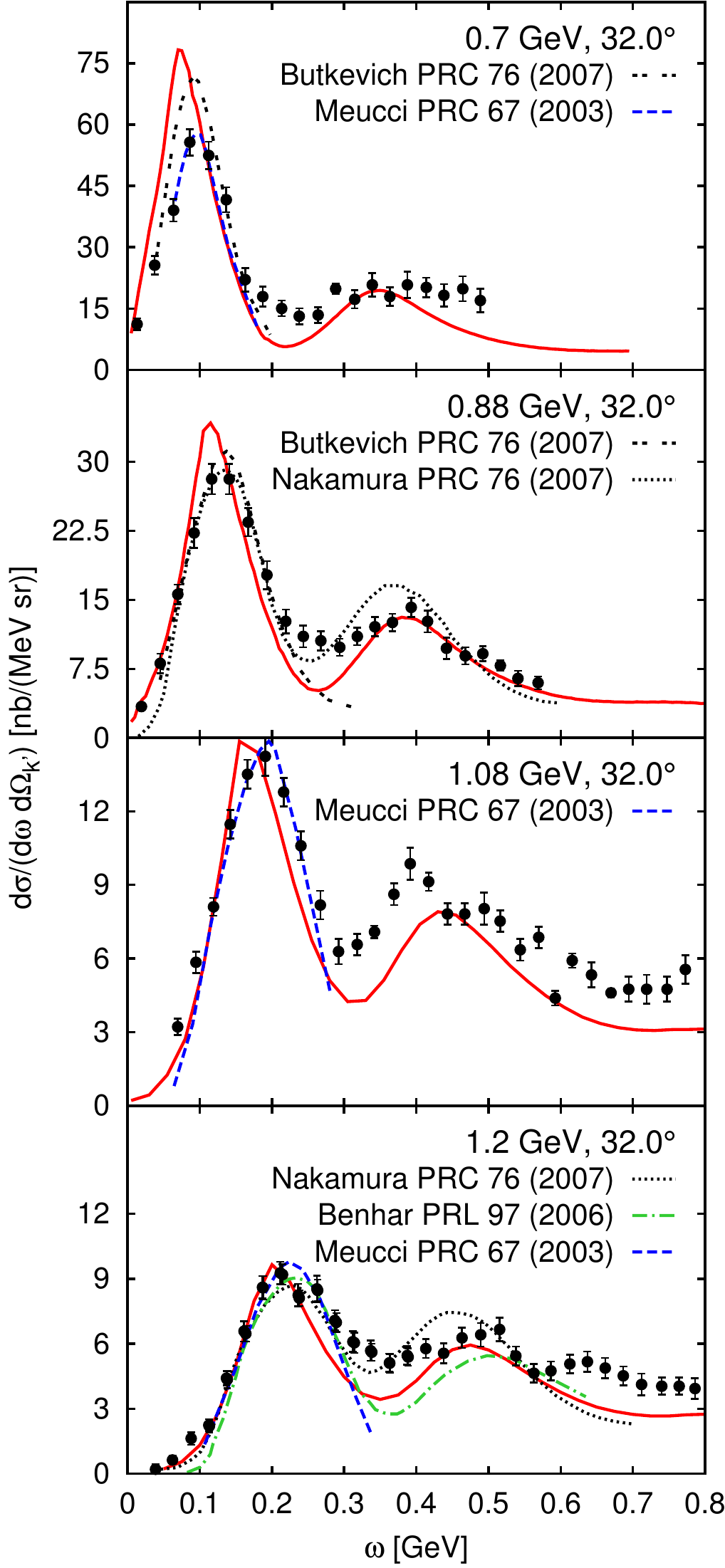}
  \caption{Inclusive electron cross section $\dd^2 \sigma/\dd \omega
    \dd\Omega_{k'}$ on $\atom{16}{O}$ as a function of the energy
    transfer, $\omega$, for four beam energies indicated in the
    figure and a scattering angle $\theta_{k'} = 32\,^{\circ}$. The
    solid lines give the results of the GiBUU calculation, the dashed
    lines are taken from Meucci \etal{}~\cite{Meucci:2004ip}, the
    double-dashed lines give the results of Butkevich
    \etal{}~\cite{Butkevich:2007gm}, the dotted lines those of
    Nakamura \etal{}~\cite{Nakamura:2007pj}, and the dash-dotted line
    shows the result of Benhar \etal{}~\cite{Benhar:2006qv}. The data
    are from Anghinolfi
    \etal{}~\cite{Anghinolfi:1995bz,Anghinolfi:1996vm}.  Taken
    from~\cite{Leitner:2008ue}.}
  \label{Fig:sigma-e-tot-compare}
\end{figure}

In \cref{Fig:sigma-e-tot-compare} we compare the results of the
GiBUU model over a wider range of electron energies with the
results of other theoretical approaches. At the lowest energy, the QE peak
in GiBUU occurs at too low energy transfer with too much
strength, similar to the result obtained by Butkevich
\etal{}~\cite{Butkevich:2007gm}. These deficiencies may be due to the
invalidity of the impulse approximation for the very small momentum
transfer at the quasi-elastic peak. For higher energies the peak
positions and strengths are described quite well by GiBUU. Similar to \cref{Fig:sigma-e-tot}, a
consistent discrepancy shows up at the high-energy side of the QE
peak, where the data are systematically slightly underestimated. This
may be due to either an underestimate of the $\Delta$ strength at this
higher energy, possibly due to coherent excitation, or to some
contribution from more complicated particle-hole configurations than
those contained in the Local Fermi-gas model. The former possibility
indicates that it is indeed essential to describe quasi-elastic
scattering and resonance excitation simultaneously since both
processes overlap in the total inelastic cross section for
intermediate energy transfer.

\subsubsection{Exclusive pion production}

The interactions of high-energy virtual photons with nuclei provide an
important tool to study the early stage of hadronization and
(pre)hadronic FSI at small distances $d \sim 1/\sqrt{Q^2}$. A further
advantage of lepton-induced reactions is that one may vary the energy
$\nu$ and virtuality $Q^2$ of the incident photon independently of
each other. This allows us to study the phenomenon of Color
Transparency (CT), i.e.~the reduced interaction cross section of a
small sized color singlet object produced in processes at high
momentum transfer.

The early onset of CT has been observed at JLAB in semi-exclusive
charged pion electroproduction off nuclei. Particularly for a
theoretical understanding of this kind of experiments the use of BUU
is mandatory since the analysis involves different kinematic and
acceptance cuts. Transport theory can give all particles in the final
state such that the same cuts can be applied also in the calculations.

\begin{figure}[t]
  \centering
  \includegraphics[width=0.6\linewidth]{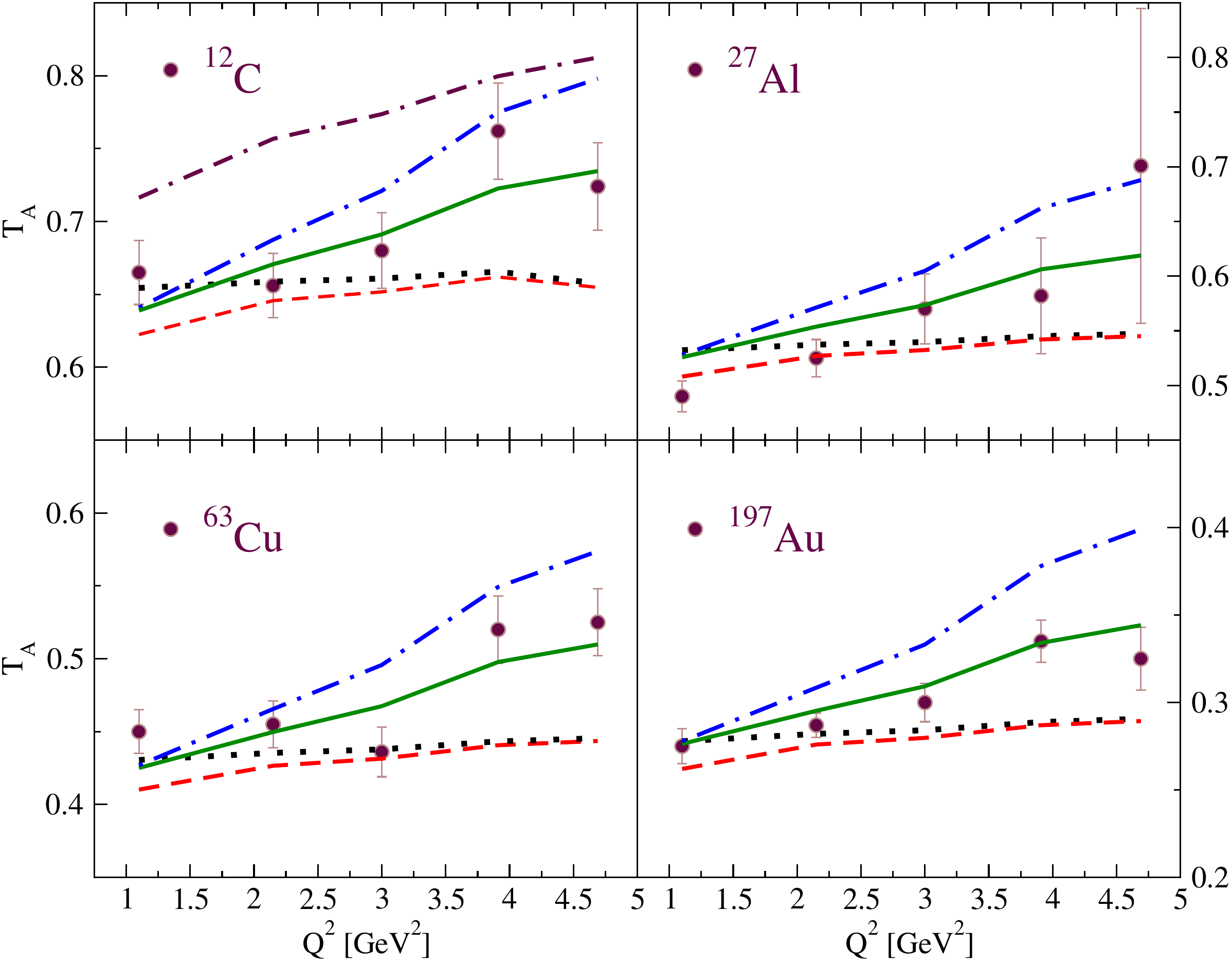}
  \caption{(Color online) Transparency, $T_{A}$, vs. $Q^2$ for
    $\atom{12}{C}$ (left, top panel), $\atom{27}{Al}$ (right, top),
    $\atom{63}{Cu}$ (left, bottom) and $\atom{197}{Au}$ (right,
    bottom).  The dotted curves correspond to FSI with the full
    hadronic cross section, and the dashed curves include the
    shadowing corrections. The dash-dotted curves correspond to the
    in-medium cross sections defined according to the
    Lund-model-formation-time concept, which includes the
    $Q^2$-dependent (pre)hadronic interactions, \cref{eq:scenarioQ},
    for the transverse contribution.  The solid curves describe the
    effect of only time dilatation with a $Q^2$-independent pedestal
    value in the effective cross section. The dash-dash-dotted curve
    in the top-left panel realizes the CT effect both in the
    longitudinal and transverse channels. The experimental data are
    from \refcite{Clasie:2007gqa}. Taken
    from~\cite{Kaskulov:2008ej}. }
  \label{Fig:TQ2new}
\end{figure}

The nuclear transparencies of pions in the reaction $A(e,e'\pi^+)A^*$
off nuclei are presented in \cref{Fig:TQ2new}. The experimental data
are from \refcite{Clasie:2007gqa}. The microscopic input for the
primary interaction of the virtual photon with the nucleon follows the
GiBUU model which describes both the transverse and the longitudinal
cross sections~\cite{Kaskulov:2008xc}.  The coupled--channel GiBUU
transport model has been used to describe the FSI of hadrons in the
nuclear medium. The formation times of (pre)hadrons follow the
time--dependent hadronization pattern as shown
in~\cref{sec:hadronization}.

Our results are consistent with the JLAB data.  The data are well
reproduced if one assumes that point-like configurations are produced
in the regime of hard deep-inelastic scattering (DIS) off partons and
dominate the transverse channel.

\subsubsection{Hadron attenuation}

\begin{figure}[t]
  \centering
  \includegraphics[width=0.8\linewidth]{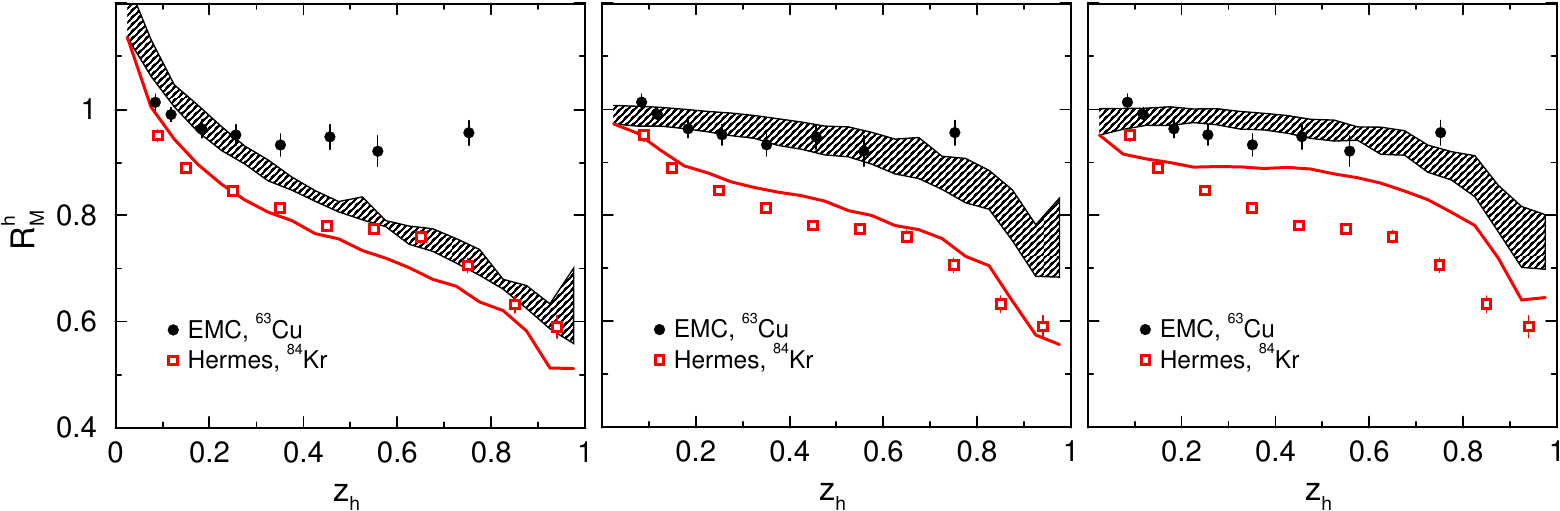}
  \caption{(Color online) Nuclear modification factor for charged
    hadrons.  Experimental data are from HERMES at $27\GeV$
    \cite{Airapetian:2000ks,Airapetian:2003mi,Airapetian:2007vu} and
    EMC at 100 and $280\GeV$ \cite{Ashman:1991cx}. The predictions for
    the two EMC energies are given by the lower and upper bounds of
    the shaded band.  he cross-section-evolution scenarios in the
    calculations are: constant, linear, quadratic (from left to
    right). Taken from \cite{Gallmeister:2007an}.}
  \label{fig:HermesEMC_1}
\end{figure}

In the following discussions we will express the medium modification
of the spectra via the usual nuclear-modification ratio,
\begin{equation}
  R_M^h(\nu, Q^2, z_h,\pT^2,\dots)= \frac{\
    \left[N_h(\nu,Q^2,z_h,\pT^2,\dots)/N_e(\nu,Q^2)\right]_A\ }{\
    \left[N_h(\nu,Q^2,z_h,\pT^2,\dots)/N_e(\nu,Q^2)\right]_{\mathrm{d}}\ },
\end{equation}
where the hadronic spectra on the nucleus (``$A$'') and on deuterium
(``d'') are normalized to the corresponding number of scattered
electrons. As indicated, the nuclear-modification ratio can be
displayed as function of many variables as, e.g., $\nu$, $z_h$,
$\pT^2$, etc. More detailed information would be provided by
multidimensional distributions, which are, however, not yet available
experimentally.

The ``photonic'' parameters of the collisions are given by $\nu$ as
photon energy and by $Q^2$ as the transferred squared four momentum.
The third parameter to fix the lepton/photon kinematics complete is
given by the lepton-beam energy.

The ``hadronic'' variables we focus on are $z_h$ or $\pT^2$. Here
$z_h=E_h/\nu$ stands for the ratio of the energy of the hadron divided
by the energy of the photon, while the squared transverse momentum in
respect to the photon direction is indicated by $\pT^2$.

In \cref{fig:HermesEMC_1} we show the results of our calculations
compared to experimental data
\cite{Airapetian:2000ks,Airapetian:2003mi,Airapetian:2007vu,Ashman:1991cx}
for some different scenarios.  In all scenarios the (pre-)hadronic
cross section is zero before $t_P$ and equals the full hadronic cross
section after $t_F$. The most essential feature of color transparency
-- larger hadrons (smaller $Q^2$) get attenuated more than smaller
ones -- is thus included in all four scenarios.

In the first scenario we assume no time dependence, i.e., the
pre-hadronic cross section is assumed constant,
\begin{equation}
  \sigma^*/\sigma = \text{const} = 0.5,\label{eq:scenarioC}
\end{equation}
where $\sigma$ is the total hadronic cross section. Here, the value
$0.5$ for the constant cross section ratio is chosen, because it leads
to a reasonable description of the HERMES data
\cite{Falter:2004uc}. The following two scenarios are the
``quantum-mechanically inspired'' and the ``naive'' assumptions of
linear or quadratic increase, respectively \cite{Dokshitzer}
\begin{equation}
  \sigma^*(t)/\sigma = \left(\frac{t-t_P}{t_F-t_P}\right)^n,\quad n \in
  [1,2].
  \label{eq:scenarioL}
\end{equation}
All three scenarios for the pre-hadronic interaction mimic color
transparency to some extent, because the interaction rates are reduced
until the formation of the final hadron. The times $t_P$, $t_F$ are as
defined in~\cref{sec:hadronization}.

Finally, we implement the `quantum diffusion' picture of Farrar
\etal{}~\cite{Farrar:1988me} proposed by these authors to describe the
time development of the interactions of a point-like configuration
produced in a hard initial reaction (see also
\cite{Larson:2006ge}). This picture combines the linear increase with
the assumption that the cross section for the \emph{leading} particles
does not start at zero, but at a finite value connected with $Q^2$ of
the initial interaction,
\begin{alignat}{2}
  \sigma^*(t)/\sigma &= X_0
  +(1-X_0)\cdot\left(\frac{t-t_P}{t_F-t_P}\right), &\qquad X_0 &=
  {r_{\rm lead}}\frac{\mathrm{const}}{Q^2},
  \label{eq:scenarioQ}
\end{alignat}
with $r_{\rm lead}$ denoting the ratios of leading partons over the
total number of partons (2 for mesons, 3 for baryons).

The baseline value, $X_0$, is inspired by the coefficient $\langle n^2
k_T^2\rangle/Q^2$ in \refcite{Farrar:1988me}. Our scaling with $r_{\rm
  lead}$ guarantees that summing over all particles in an event, on
average the prefactor becomes unity. The numerical value of the
constant in the numerator of $X_0$ is chosen to be $1 \GeV^2$ for
simplicity, close to the value used in \cite{Farrar:1988me}. This
value is also constrained by the considered $Q^2$ range such that the
pedestal value $X_0 \le 1$ is fulfilled.

\begin{figure}[t]
  \centering
  \includegraphics[width=0.4\linewidth,angle=-90]{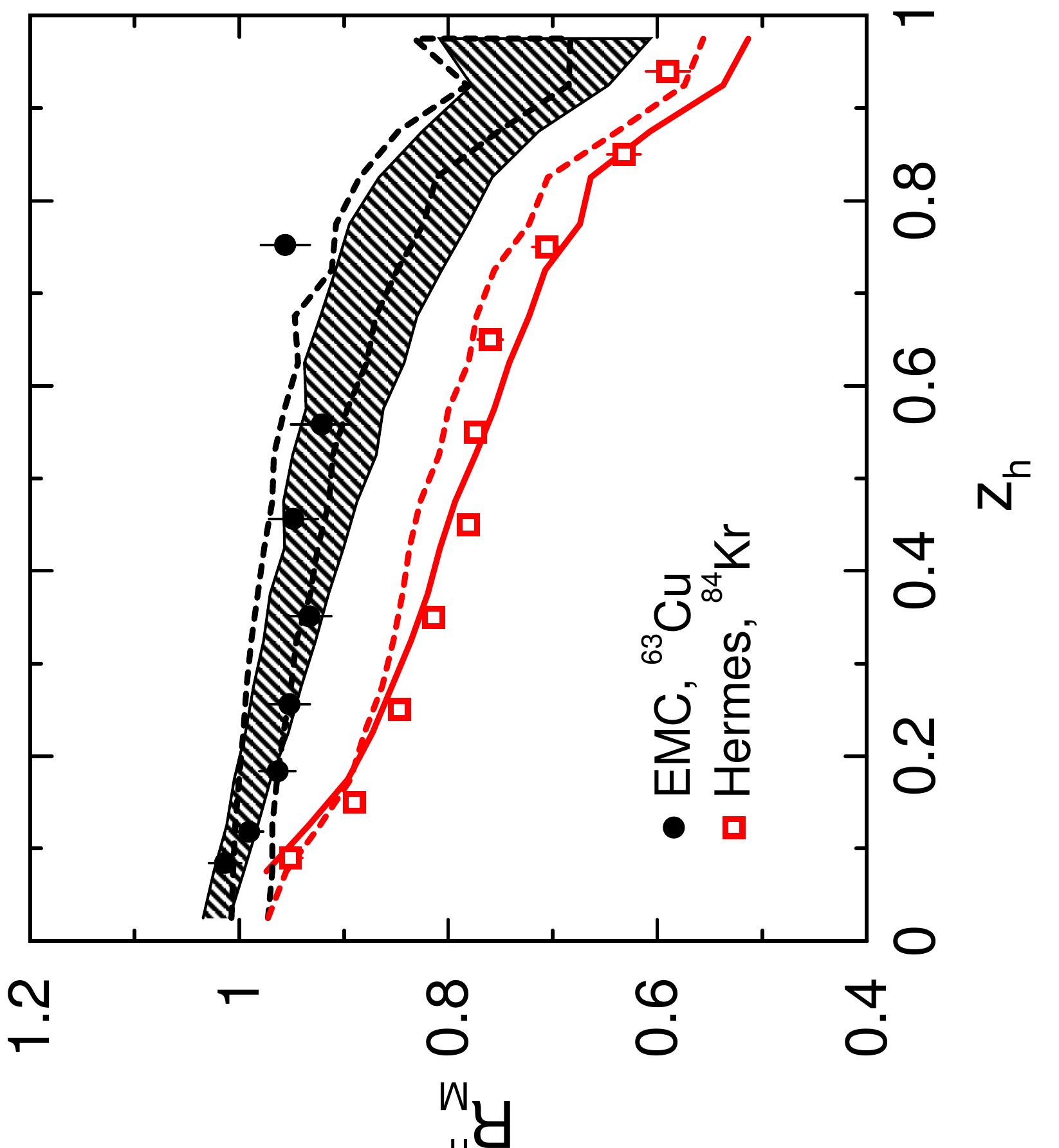}
  \caption{(Color online) Nuclear modification factor for charged
    hadrons as in \cref{fig:HermesEMC_1}. The cross-section-evolution
    scenario in the calculations is according to
    \cref{eq:scenarioQ}. Dashed lines repeat curves from
    \cref{fig:HermesEMC_1} (middle panel). Taken from
    \cite{Gallmeister:2007an}.}
  \label{fig:HermesEMC_2}
\end{figure}

Because it remains unclear, how the (very) different lepton energies
have been considered in the experimental results given in
\cite{Ashman:1991cx} we have performed calculations for the two most
prominent energies of that experiment, i.e., for beam energies of
$100\GeV$ and $280\GeV$. We illustrate the results of our calculations
 by a shaded band in \cref{fig:HermesEMC_1} and \cref{fig:HermesEMC_2}
 that reflects the energy range just mentioned.

\begin{figure}[p]
  \centering
  \includegraphics[width=0.6\linewidth]{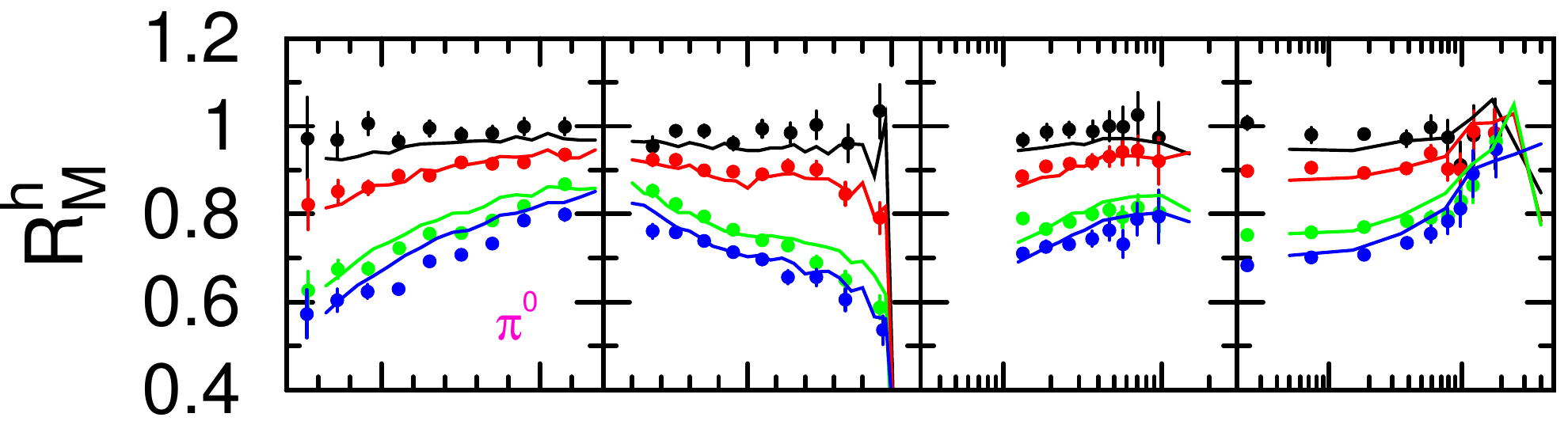}
  \includegraphics[width=0.6\linewidth]{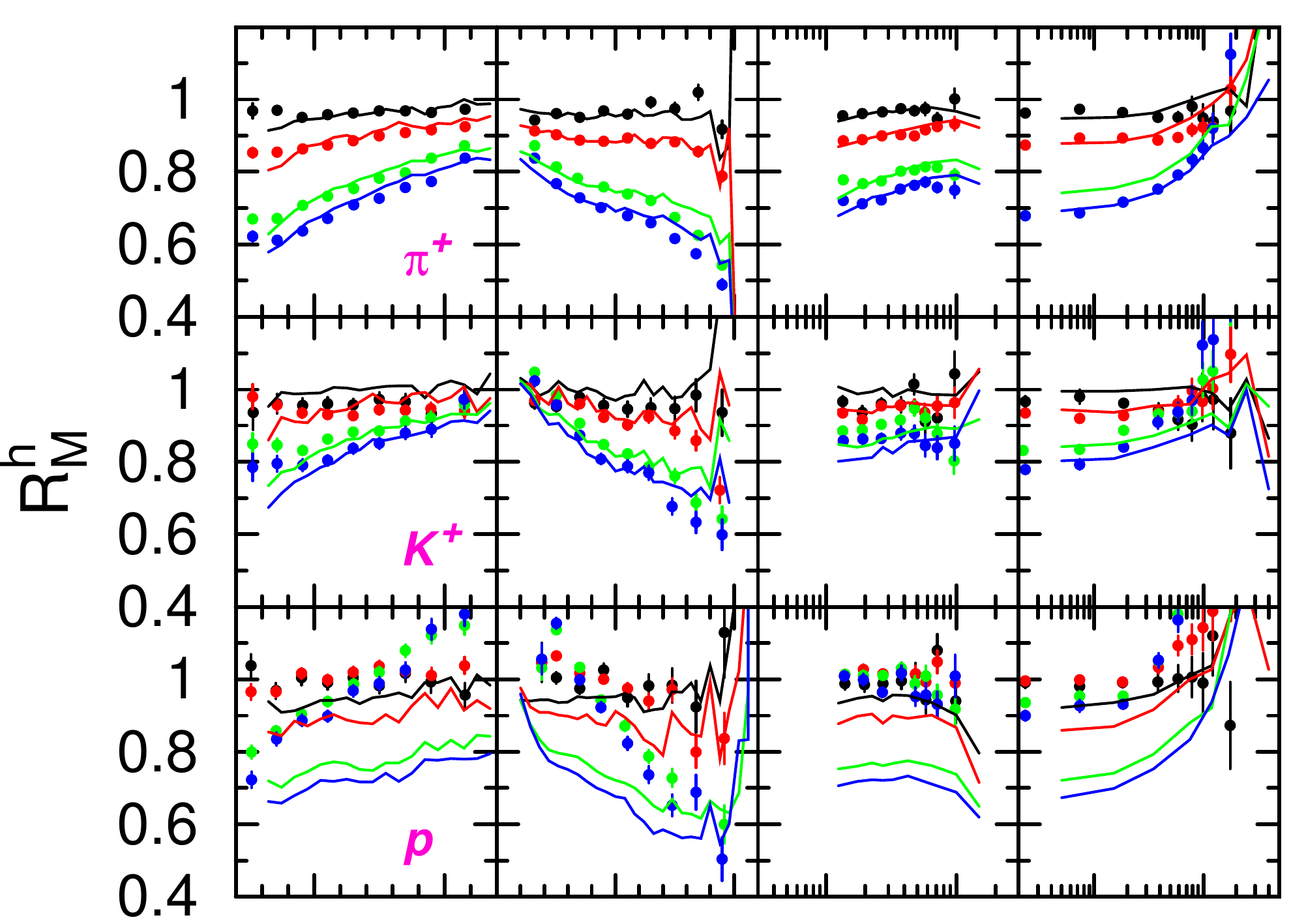}
  \includegraphics[width=0.6\linewidth]{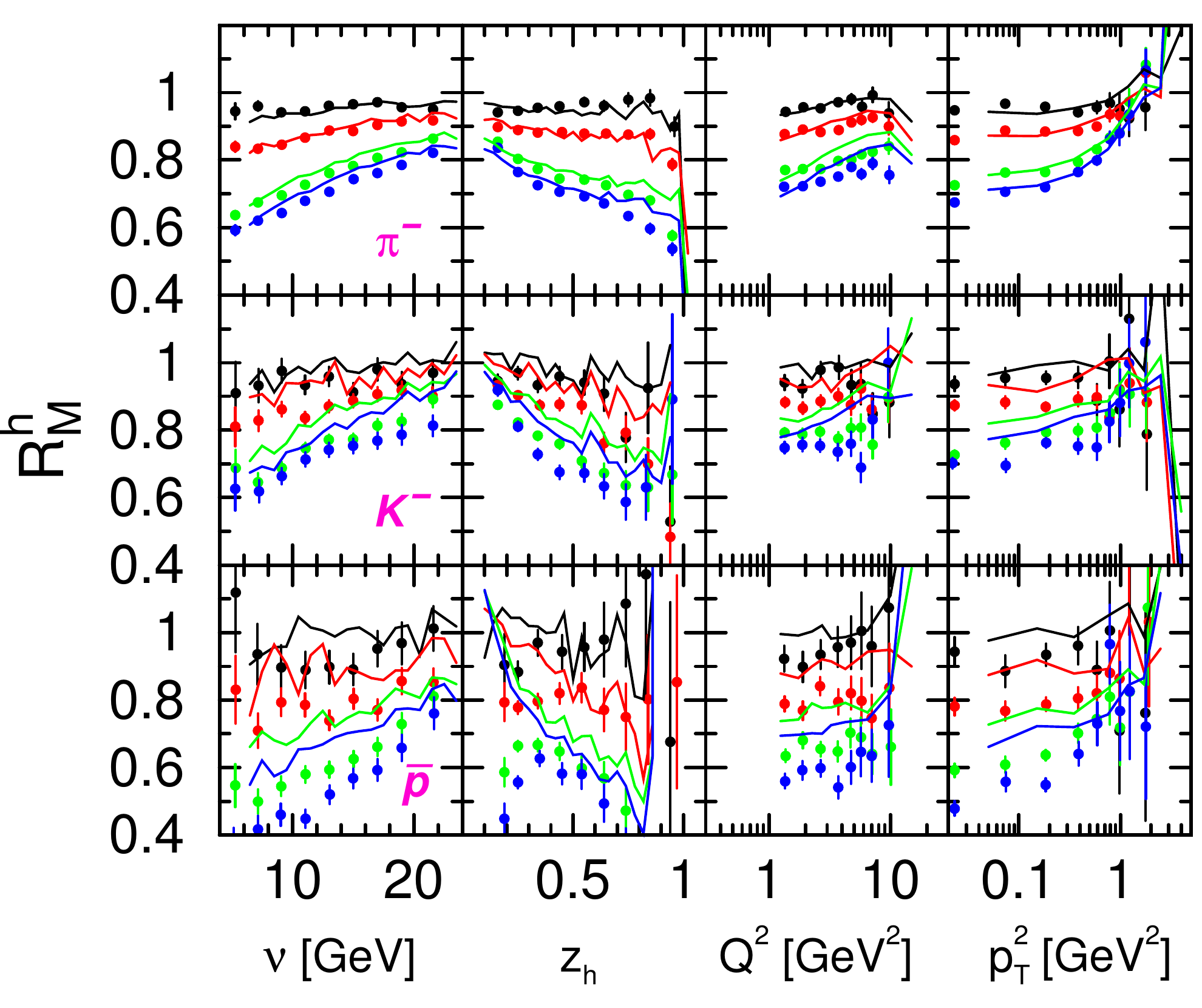}
  \caption{(Color online) Nuclear modification factor for identified
    hadrons for HERMES at $27 \GeV$ with $\atom{4}{He}$,
    $\atom{20}{Ne}$, $\atom{84}{Kr}$, and $\atom{131}{Xe}$-target
    nuclei (top to bottom). Points indicate experimental data
    \cite{Airapetian:2000ks,Airapetian:2003mi,Airapetian:2007vu} while
    the curves represent our calculations with the time-dependence
    scenario cf.~\cref{eq:scenarioQ} with diffractive events
    excluded. Taken from \cite{Gallmeister:2007an}.}
  \label{fig:Hermes27Kr}
\end{figure}

\begin{figure}[t]
  \centering
  \includegraphics[width=0.6\linewidth]{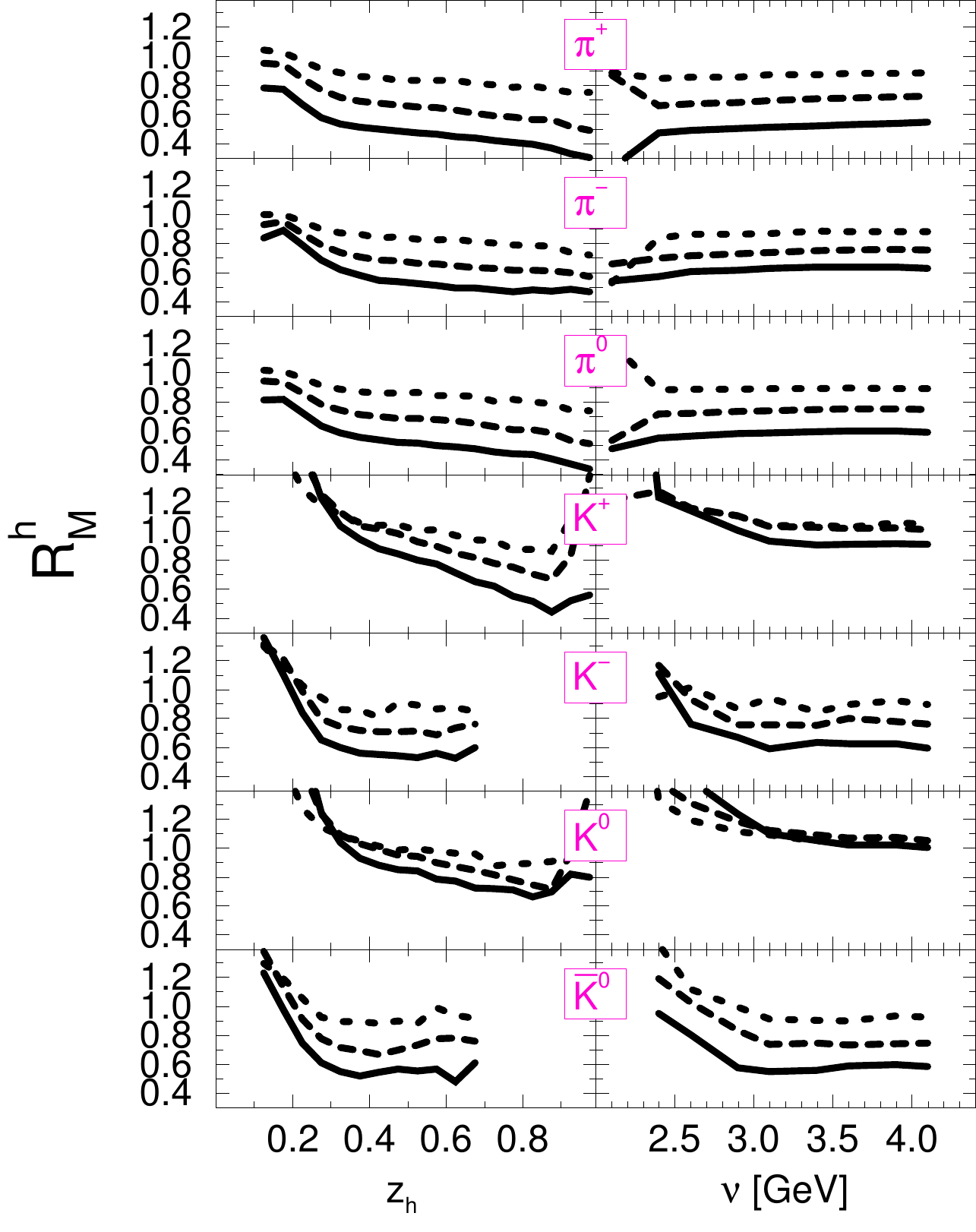}
  \caption{(Color online) Nuclear modification factor of identified
    mesons $\pi^{\pm,0}$ and $K^{\pm,0},\bar K^0$ for JLAB(CLAS) at $5
    \GeV$ with different targets: $\atom{12}{C}$ (dotted),
    $\atom{56}{Fe}$ (dashed), and $\atom{208}{Pb}$ (solid lines).
    Taken from \cite{Gallmeister:2007an}.}
  \label{fig:JLAB5}
\end{figure}

\begin{figure}[t]
  \centering
  \includegraphics[width=0.6\linewidth]{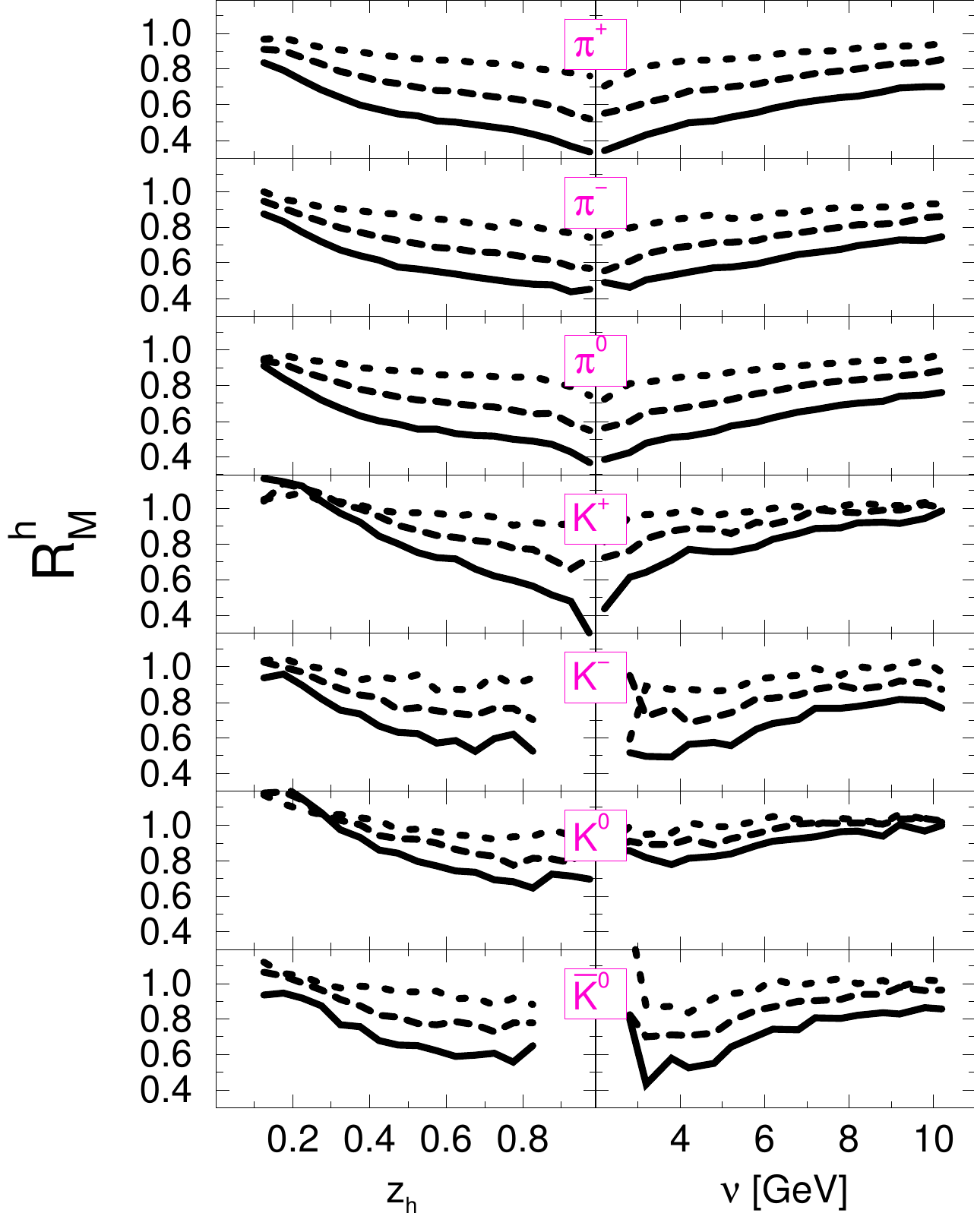}
  \caption{(Color online) The same as \cref{fig:JLAB5}, but for JLAB
    (CLAS) at $12 \GeV$ lepton-beam energy.
    Taken from \cite{Gallmeister:2007an}.}
  \label{fig:JLAB12}
\end{figure}

Assuming a constant cross section (cf.~\cref{fig:HermesEMC_1}), we
obtain a good description of the HERMES results, while the attenuation
for the EMC experiment is severely overestimated
(cf.~\cite{Falter:2004uc}).  Assuming a linear time dependence, both
the HERMES and EMC attenuation are well described.  Going even further
and assuming a quadratic time dependence, leads to a too small
attenuation for both the HERMES and the EMC experiment, with the
discrepancy between theoretical and experimental results being
significant for the HERMES experiment. Only the theoretical scenario
with a cross section evolving linearly in time is able to describe
both data sets at the same time.

\Cref{fig:HermesEMC_2} shows results of our calculations employing the
scenario as given by \cref{eq:scenarioQ}.
Although not very pronounced, the effect of the non-vanishing,
$Q^2$-dependent initial cross section of the leading particles is
visible when comparing \cref{fig:HermesEMC_2} with the middle panel in
\cref{fig:HermesEMC_1}; a slight improvement in the description can be
seen. The observed weak $Q^2$ dependence is in line with experimental
observations of both the HERMES and the EMC experiments
\cite{Airapetian:2000ks,Airapetian:2003mi,Airapetian:2007vu,Ashman:1991cx}.
This scenario (\cref{eq:scenarioQ}) will therefore be used in the following
considerations.

\Cref{fig:Hermes27Kr} shows a comparison of our calculations with the
latest experimental data of the HERMES collaboration with $27\GeV$
beam energy for identified hadrons for the four targets
$\atom{4}{He}$, $\atom{20}{Ne}$, $\atom{84}{Kr}$, and
$\atom{131}{Xe}$.

As expected from \cref{fig:HermesEMC_2}, for the total hadron yield,
the data for pions, which are the most frequently produced hadrons,
are described well by our model. In the large-$z_h$ region charged
pions originate mostly from decays of diffractive $\rho$ mesons. Since
these pions are taken out from the experimental data, we also switch
off diffractive production of $\rho$ mesons in the calculations. While
the description of the data in the strange and anti-baryonic sectors
is also quite good, one still finds the well known discrepancy of data
and calculations for protons: the regions with ``low $z_h$''/``high
$\nu$'' are clearly underestimated in our model. We recall that this
is not a new finding but already known from our previous work
\cite{Falter:2004uc}. The observed discrepancy may reflect a
deficiency in our treatment of final-state interactions at high proton
energies since the (strongly non-perturbative) low-$z_h$ protons arise
mainly from energy-degrading rescattering events. The discrepancy may,
however, also reflect some problems with the treatment of experimental
geometrical acceptance limitations, which affect the data (and are
taken into account in the calculations).

Based on our successful description of the experimental data of the
HERMES collaboration for $27\GeV$ and $12\GeV$ beam energies, we now
make predictions for the meson spectra at the presently available
$5\GeV$ lepton-beam energy and at the future JLAB facility with
$12\GeV$.

We start with a discussion of our results for a 5\GeV{} beam energy in
\cref{fig:JLAB5}.

A comparison of our results \cref{fig:JLAB5} with preliminary
experimental data on the $z_h$ dependence of the $\pi^+$ attenuation
for the three nuclear targets \cite{Hafidi:2006ig} is satisfactory,
both in its magnitude and its target-mass-number dependence.

Contrary to the situation at higher beam energies, feeding effects
leading to attenuation ratios larger than unity at small $z_h$ are
more pronounced and turn out to be an essential feature in this energy
range. For the rarer kaons we stop showing the attenuation at $z_h =
0.7$ because the spectra for $K^-$ drop rapidly at $z_h \simeq
0.7\upto0.8$. On the contrary, the spectra for $K^+$ reach
significantly farther out.  This is a direct consequence of the fact
that contrary to $K^+$ mesons the $K^-$ mesons can only be produced in
the associated strangeness-production mechanism and thus have a higher
threshold than the former. The same holds for $K^0$ and
$\overline{K}^0$, respectively.

At this low energy (and corresponding momentum transfer) the invariant
masses populated in the first interaction are rather small ($\langle
W\rangle=2.2\GeV$) and thus just above the resonance region.  We have
also already noted that at this low energy we have formation times of
only $\simeq 4 \fmc$ at large $z_h$.  Therefore, the interactions of
the formed hadrons are strongly affected by hadronic interactions
while pre-hadronic interactions play only a minor role (at least for
the heavier targets). This can be seen in \cref{fig:JLAB5} in the
different attenuation for $K^+$ and $K^-$, the latter being more
strongly attenuated due to hadronic FSI. We also recall our earlier
finding \cite{Falter:2004uc} that at this low energy also the effects
of Fermi motion are essential and have to be taken into account. The
dynamics in this energy regime is thus more determined by `classical'
meson-nucleon dynamics than by perturbative QCD, which underlies many
of the other theoretical descriptions of the attenuation experiments
\cite{Kopeliovich:2003py,Kopeliovich:2006xy,Wang:2002ri}.

\Cref{fig:JLAB12} shows the calculated results for the multiplicity
ratio of the three pion and four kaon species for the exemplary nuclei
$\atom{12}{C}$, $\atom{56}{Fe}$, and $\atom{208}{Pb}$ with $12 \GeV$
lepton-beam energy as in the future JLAB upgrade.

For all particle species, a strong dependence of the attenuation
ratios on the size of the nucleus is obtained. It is interesting to
observe that at this higher energy the attenuation of $K^+$ and $K^-$
becomes similar at $z_h \approx 0.7$, contrary to the behavior at $5
\GeV$. This reflects the longer formation times at higher energies and
the corresponding predominance of pre-hadronic interactions, which
affect the non-leading $K^-$ only weakly.

\subsection{Photon-induced reactions}\label{sec:photonA}

Photon-induced reactions share with the electron-induced ones the
advantage that there are no initial state interactions, i.e.
photons illuminate the whole nucleus (again disregarding shadowing,
cf.~footnote at the start of \cref{sec:electronA}). GiBUU and its
predecessors have been used to investigate scalar-, pseudoscalar-, and
vector-meson production off nuclei (see
\cite{Effenberger:1996rc,Effenberger:1999jc,Lehr:1999zr,Muhlich:2002tu,Muhlich:2003tj,Krusche:2004uw,Muhlich:2006ps,Mertens:2008np,Wood:2008ee}
and refs. therein). Here we discuss as examples the total
photoabsorption cross sections as well as single- and double-pion
production and omega production.

\subsubsection{Total photoabsorption cross section}

In the total photoabsorption cross section on nucleons, one observes
three major peaks (see \cref{fig:photonBG_2pi} in
\cref{sec:elementary_leptonN_TwoPi}) generated by several overlapping
resonances. The most important ones are the $P_{33}(1232)$,
$S_{11}(1535)$, $D_{13}(1520)$ and $F_{15}(1680)$ resonance states. To
study the properties of these resonance states embedded in nuclear
matter, one has investigated their photon-induced excitation in
nuclei. First experiments using a tagged high-energy photon beam
($E_\gamma=0.3\upto2.6 \GeV$), which offered sufficient energy to excite
the second resonance region, have been performed by the Yerevan
group~\cite{Arakelian:1982ba,Ananikyan:1987}. Following up this
pioneering work, photon absorption on nuclei has been
measured at the Mainz Microton (MAMI)
facility~\cite{Frommhold:1992um,Frommhold:1994zz} with a beam energy of
$E_\gamma=0.05\upto0.8\GeV$, with higher energies of
$E_\gamma=0.2\upto1.2 \GeV$ at the Adone storage ring facility
(Frascati,
Italy)~\cite{Bianchi:1992ze,Bianchi:1993nh,Bianchi:1993mb,Bianchi:1994ax,Bianchi:1995vb},
using the SAPHIR tagged photon beam of $E_{\gamma}=0.5\upto2.67
\GeV$ at ELSA (Bonn, Germany)~\cite{Muccifora:1998ct} and at
Hall B of the Jefferson Laboratory (Newport News,
USA)~\cite{Cetina:2000rw,Cetina:2002mz} with a beam energy of
$E_{\gamma}= 0.17\upto3.84 \GeV$.  In fact, some of the above
experiments have not measured directly the photoabsorption cross
section but only the photofission cross
section~\cite{Frommhold:1992um,Frommhold:1994zz,Bianchi:1993mb,Bianchi:1992ze,Cetina:2000rw,Cetina:2002mz}. Contrary
to earlier assumptions, it has been shown by Cetina
\etal{}~\cite{Cetina:2000rw,Cetina:2002mz} that these two cross sections
need not be identical. Thus, for our analysis we focus on a comparison
with the direct measurements of photon absorption as presented by
Bianchi \etal{}~\cite{Bianchi:1994ax,Bianchi:1995vb} and Muccifora
\etal{}~\cite{Muccifora:1998ct}.

Experimentally, the $\Delta$-resonance region within the nucleus still
exhibits a peak-like structure. However, a slight shift to higher
energies and a broadening is observed, as compared to the vacuum
structure.  Carrasco \etal{}~\cite{Carrasco:1989vq} have shown in a
microscopic model approach, that it is possible to describe the data
in the $\Delta$ resonance region in a very satisfactory manner when
including the $\Delta$ self-energy. This work emphasizes the
importance of many-body absorption channels not included in the
commonly used impulse approximation.

At higher energies the experimental results show no structures in the
second and third resonance region. This is often referred to as the
disappearance of the resonances in the medium and presents a direct
observation of an in-medium effect \cite{Leupold:2009kz}.

\begin{figure}[t]
  \centering
  \includegraphics[width=0.4\linewidth]{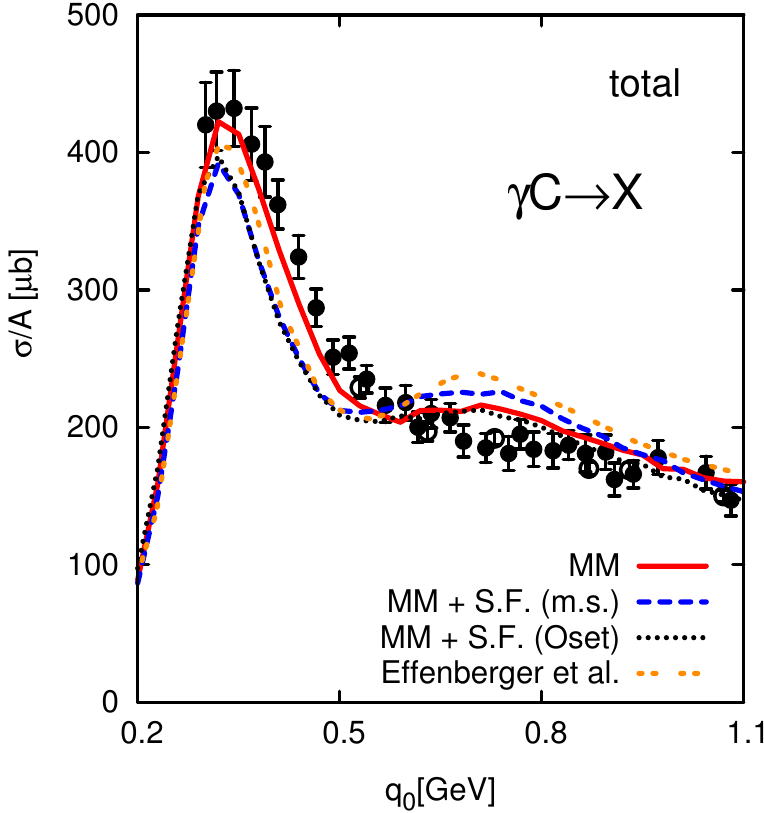}
  \hspace*{1cm}
  \includegraphics[width=0.4\linewidth]{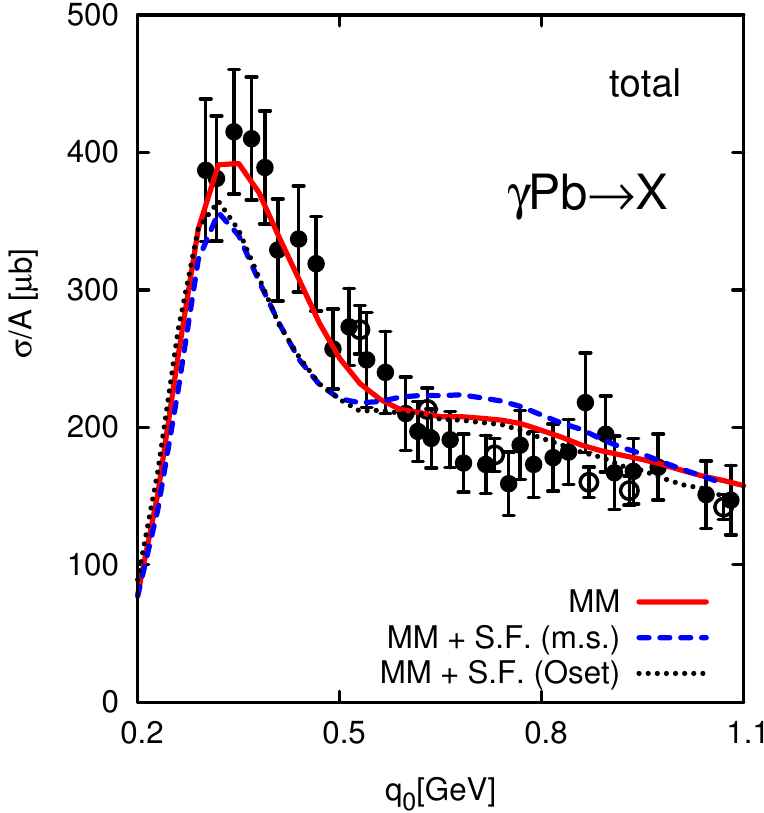}
  \caption[The photoabsorption cross section for \carbon{} and
  \lead{}] {(Color online) The photoabsorption cross section for
    \carbon{} and \lead{}. The (red) solid line denotes the result
    with a momentum-dependent potential (see label ``MM'' in
    \cref{tab:potential}). In addition we show two different
    modifications of the in-medium self-energy: mass shift (m.s.)
    option (blue, dashed), and the result using the Oset option for
    the $\Delta$ (black, dotted). The data are taken from Bianchi
    \etal{}~\cite{Bianchi:1995vb} (full circles) and Muccifora
    \etal{}~\cite{Muccifora:1998ct} (open circles), the error bars
    denote the sum of statistical and systematical errors. Source: 
    Taken from~\cite{buss_phd}.}
  \label{fig:photo_abs_C}
\end{figure}

After the publication of this surprising result in the second resonance
region, various theoretical explanations and analyses have been put
forward. Kondratyuk \etal{}~\cite{Kondratyuk:1993ah} investigated a
baryon-resonance model including the collisional widths of the
resonances as free parameters. The extracted widths turned out to be
extraordinary large (ca.~$320 \MeV$ for the $S_{11}(1535)$,
$D_{13}(1520)$ and $F_{15}(1680)$ resonance), and the whole analysis has
drawn some criticism (cf.~pages 368-369 in \cite{Effenberger:1996im}).
Rapp \etal{}~\cite{Rapp:1997ei,Rapp:1999ej} have applied a vector-meson dominance
(VMD) model to the problem. An energy-independent collisional broadening
of the baryon resonances of $15\MeV$ for the $\Delta$, $250 \MeV$ for
the $D_{13}$ and $50 \MeV$ for all other contributing resonances has
been included by hand. The model reproduces the elementary data on the
proton in the region between the $\Delta$- and the second-resonance
region at the expense of having to use a very soft $\pi$-$NN$ form
factor; it describes the nuclear data in a satisfactory
manner. According to Hirata \etal{}~\cite{Hirata:2001sw}, the
interference patterns among the resonances and the background change
from vacuum to medium, which is the driving force for the disappearance
of the resonances.  Iljinov \etal{}~\cite{Ilinov:1996js} have extended
the Dubna/Moscow INC model, a hadronic transport model, such that it can
be used for high photon energies up to $10 \GeV$. They achieve a good
correspondence with exclusive channels such as single-pion
production. Based on the Dubna/Moscow INC model, the RELDIS
code~\cite{Pshenichnov:2005zz} has achieved good results for
photoabsorption on large nuclei. Both the INC and RELDIS models include
a phenomenological two-body absorption channel on top of single-particle
absorption. Also the LAQGSM model~\cite{Mashnik:2005dm} is based on the
Dubna/Moscow INC model and gives good results for nuclear fissilities.
Deppman \etal{}~\cite{Deppman:2002cg} have successfully applied the
so-called MCMC/MCEF cascade model to evaluate photon-fission cross
sections using the photoabsorption cross section as given
input. Complementary to this first work, Deppman achieved with the CRISP
code~\cite{Deppman:2004vc,Deppman:2006gb}, where the photon absorption
is modeled via a microscopic resonance model, also satisfactory results
for photon absorption.

In view of these numerous attempts to describe the disappearance of
nucleon resonances inside the nuclear medium, Effenberger
\etal{}~\cite{Effenberger:1996im} have attempted to explain this
observation within a precursor of the GiBUU model. Quite a good
description of the $\Delta$ peak has been achieved, although some
strength on the high energy side of the Delta peak is
missing. Furthermore, the model failed to describe the data at higher
energies; the results still showed a prominent structure in the second
and third resonance region.

The present implementation of GiBUU has improved various theoretical
aspects of the calculation (for a detailed discussion see
\refcite{buss_phd}). \Cref{fig:photo_abs_C} shows the results of these
new calculations with a momentum-dependent mean field, as well as
calculations for which the in-medium width and the dispersive
contributions to the real parts have been taken into account (for
details see \refcite{buss_phd}). There is an excellent agreement with
the data when only the momentum dependent potential is switched on.
However, when the improvements to the resonance spectral function are
turned on the agreement gets worse, in particular on the high-energy
side of the $\Delta$ resonance. This is partly due to the lowering of
the peak due to the larger width, but also to an inconsistency; the
increased in-medium width obtained by Oset \etal{}~\cite{Oset:1987re}
contains contributions from 2-body absorption while the calculation
relies on an impulse approximation without such a process.

The results shown in \cref{fig:photo_abs_C} contain the sum of
resonance and background contributions. Note that the background
contributions differ for all calculations, which is caused by the
inclusion of the dispersive parts. These dispersive parts basically
shift strength from the resonances to the background. The figure shows
that the $\Delta$ region is underestimated for both options for the
in-medium width. The situation also does not improve in the second
resonance region, where a strong in-medium broadening, as in the
mass-shift scenario, even leads to an enhancement of the peak
structure due to a shift of $\Delta$ spectral strength towards higher
masses.

For completeness, we show in \cref{fig:photo_abs_C} also the photon
absorption in lead. Compared to the carbon case we do not see a
qualitative change of the picture. We observe the same level of
correspondence with the experimental data as for the carbon case.

\subsubsection{\texorpdfstring{Single $\pi^0$ photoproduction}{Single
    pi0 photoproduction}}
\label{sec:kruschepi_null}

By taking into account rescattering effects leading to a change in the
final-state-particle multiplicities and distributions, the GiBUU
transport model is very well suited for the study of semi-exclusive
reactions. To give a flavor of the GiBUU capabilities, in the next
sections we address $\pi$, $\pi\pi$ and $\omega$
photoproduction. Based upon a precursor version of GiBUU, there also
$\rho$ \cite{:2007mga}, $\phi$ \cite{Muhlich:2002tu,Muhlich:2005kf},
and $\eta$ \cite{Lehr:2003km} production in photon-induced reactions
have been analyzed.

High quality data on $\pi^0$-meson production have been taken by
Krusche~\etal{}~\cite{Krusche:2004zc,Krusche:2004uw} using the TAPS
spectrometer installed at the MAMI facility in Mainz, and on
$\pi^+$-meson production by Fissum~\etal \cite{Fissum:1996fi} at the
Saskatchewan Accelerator Laboratory. The Fissum data are taken in the
very threshold region, where most of the produced pions have energies
below $40 \MeV$ and our FSI model becomes unreliable.

For the $\pi^0$-production experiment performed by Krusche
\etal{}~\cite{Krusche:2004zc,Krusche:2004uw}, the MAMI facility
delivered a photon beam of $200\upto800\MeV$ energy on deuterium,
carbon and lead targets.  In this each produced neutral pion has been
counted as a single event.  Therefore, events with $2\pi^0$ in the
final state are doubly counted. So the total cross section for neutral
pion production is experimentally defined as
\begin{equation}
  \sigma_{\pi^0, \text{ total}}= \sigma_{\pi^0}+2 \sigma_{\pi^0\pi^0}  + \sigma_{\pi^0\pi^+}+ \sigma_{\pi^0\pi^-}+\beta \sigma_{\eta}
\end{equation}
where $\beta\approx 1.2$ is the expected average number of $\pi^0$'s
in the final state of an $\eta$ decay.

\begin{figure}[tb]
  \centering
  \includegraphics[width=0.6\linewidth]{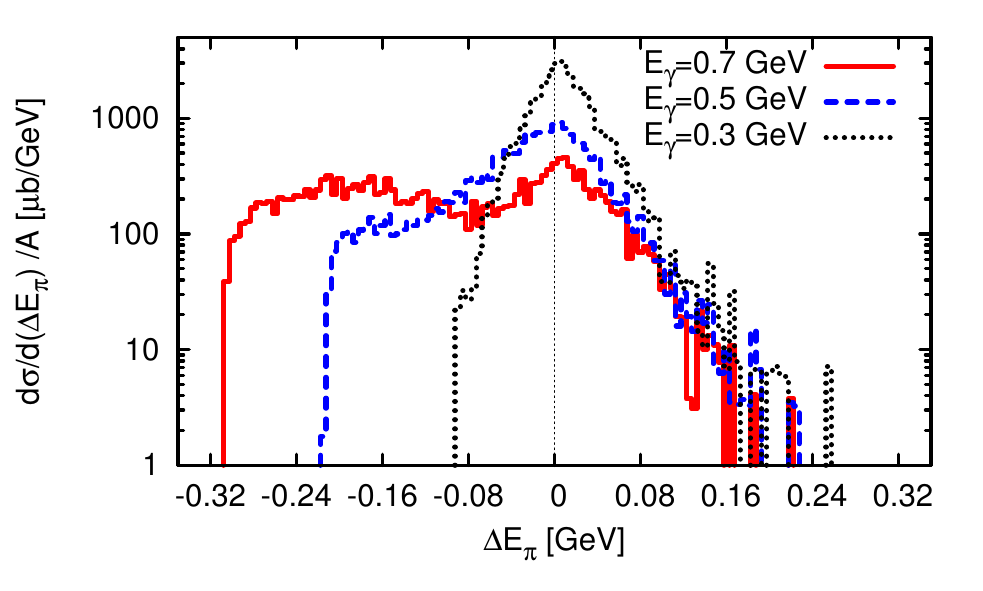}
  \caption[$\dd \sigma_{\pi^0,\text{total}}/\dd(\Delta E_\pi)$ for
  $\gamma \atom{40}{Ca}\to \pi^0 X$]{(Color online) A typical plot for
    $\dd\sigma_{\pi^0,\text{total}}/\dd(\Delta E)$ for a $\gamma
    \atom{40}{Ca} \to \pi^0 X$ reaction at $0.5 \GeV$ and $0.7 \GeV$
    right after production (\emph{i.e., FSI effects are not
      included}). This result is obtained with a momentum-dependent
    potential (label ``MM'' in \cref{tab:potential}) and the
    Oset choice according to \cite{buss_phd} for the self-energies.
    Source: Taken from~\cite{buss_phd}.}
  \label{fig:deltaE_plot}
\end{figure}

In the experimental analysis one tries to differentiate between
quasi-free pion production and in-medium pion production. For the
production of quasi-free pions one assumes that the energy of those is
similar to the energy of a pion produced on a nucleon at rest. This
energy in the $\pi N$ CM frame is given by
\begin{equation}
  E_\pi^\text{free}=\sqrt{|\bvec{p}_\text{CM}|^2+m_\pi^2}
\end{equation}
with
\begin{equation}
  |\bvec{p}_\text{CM}|=
  \sqrt{(s+m_\pi^2-m_N^2)^2/(4s)-m_\pi^2}~,
\end{equation}
where $s=(E_\gamma+ m_N)^2-E_\gamma^2=m_N^2+2m_N E_\gamma$. The
difference of the reconstructed energy in the $\pi N$ CM frame to that
\emph{quasi-free} energy is defined as $\Delta E_\pi$ (cf.~also
\cite{Krusche:2004uw}),
\begin{equation}
  \Delta E_\pi= E_\pi^{\pi N-\text{CM}}-E_\pi^\text{free}(E_\gamma) \; .
\end{equation}

A typical plot of such a distribution without final-state interactions
is shown in \cref{fig:deltaE_plot}. Indeed, one observes a peak at
$\Delta E_\pi\approx 0$, but the peak is not sharp but rather smeared
out due to the potentials and Fermi motion. Additionally, one observes
a broad background at lower $\Delta E_\pi< 0$, which originates from
$\pi\pi$ production. Since this background leaks into the single-$\pi$
peak, it may affect the actual signal.

In the experimental analysis the spectrum has, therefore, been
approximated by a symmetric one, i.e., the left-hand side of the
quasi-free peak is assumed to equal the right-hand side so that
effectively only events with $\Delta E_\pi > 0$ were counted. The
so-called quasi-free cross section, reported by the experiment, is
thus defined as
\begin{equation}
  \sigma_{\pi^0,\text{ quasi-free}}=\frac{2 \times \text{ rate of events
      with } \Delta E_\pi>0}{\text{photon flux} \times \text{density of
      targets} } \; .
\end{equation}

\begin{figure}[tb]
  \centering
  \includegraphics[width=0.8\linewidth]{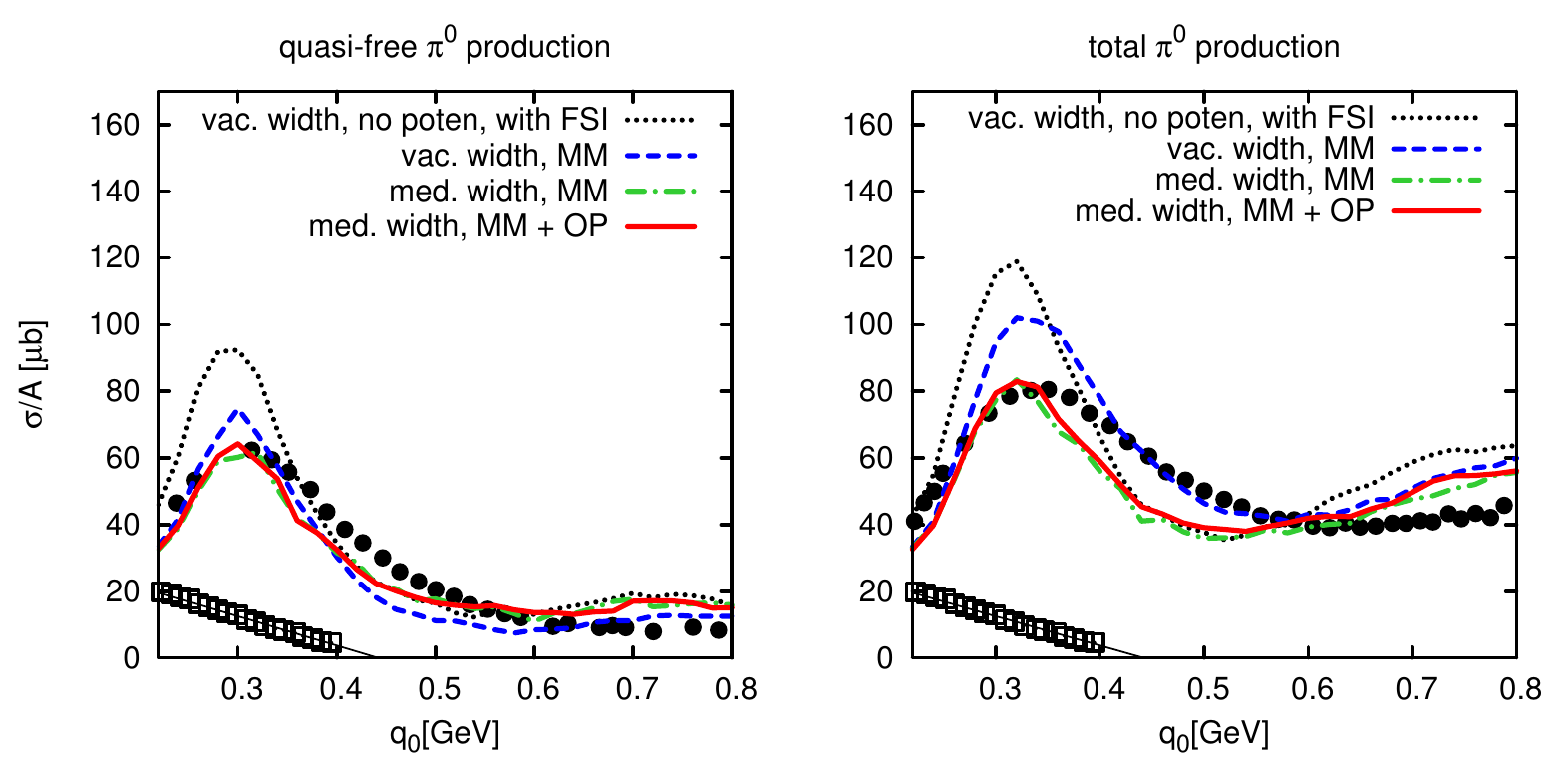}
  \caption[Cross section for $\gamma\atom{40}{Ca}\to \pi^0 X$]{(Color
    online) Quasi-free and total cross sections for the $\gamma
    \atom{40}{Ca}\to \pi^0 X$ reaction: the left panel shows the
    \textit{quasi-free} yield, whereas the right one depicts the total
    yield of $\pi^0$'s. The statistical error bars of the data are
    negligible and have been omitted. The squares show the data for
    the coherent $\pi^0$-production cross
    section~\cite{Krusche:2002iq}, and the circles depict the data for
    the quasi-free, respectively total, cross
    section~\cite{Krusche:2004uw}. The dotted line represents the
    calculation without in-medium modification of widths (i.e., vacuum
    widths) and without potential; the dashed one also uses vacuum
    widths but an in-medium momentum-dependent potential (label ``MM''
    in \cref{tab:potential}); both the dashed-dotted and the dotted
    curves have been obtained by calculations, which include in-medium
    widths and an ``MM'' potential. In the solid curve, the off-shell
    potential (OP) has been additionally included as described in
    \cref{subsec:OSP} (while all other results do not include any
    OP). The thin solid line connecting the coherent data points
    (squares) shows our fit of these data. Source: 
    Taken from~\cite{buss_phd}.}
  \label{fig:Krusche_exclusive}
\end{figure}

\begin{figure}[tb]
  \centering
  \includegraphics[width=0.8\linewidth]{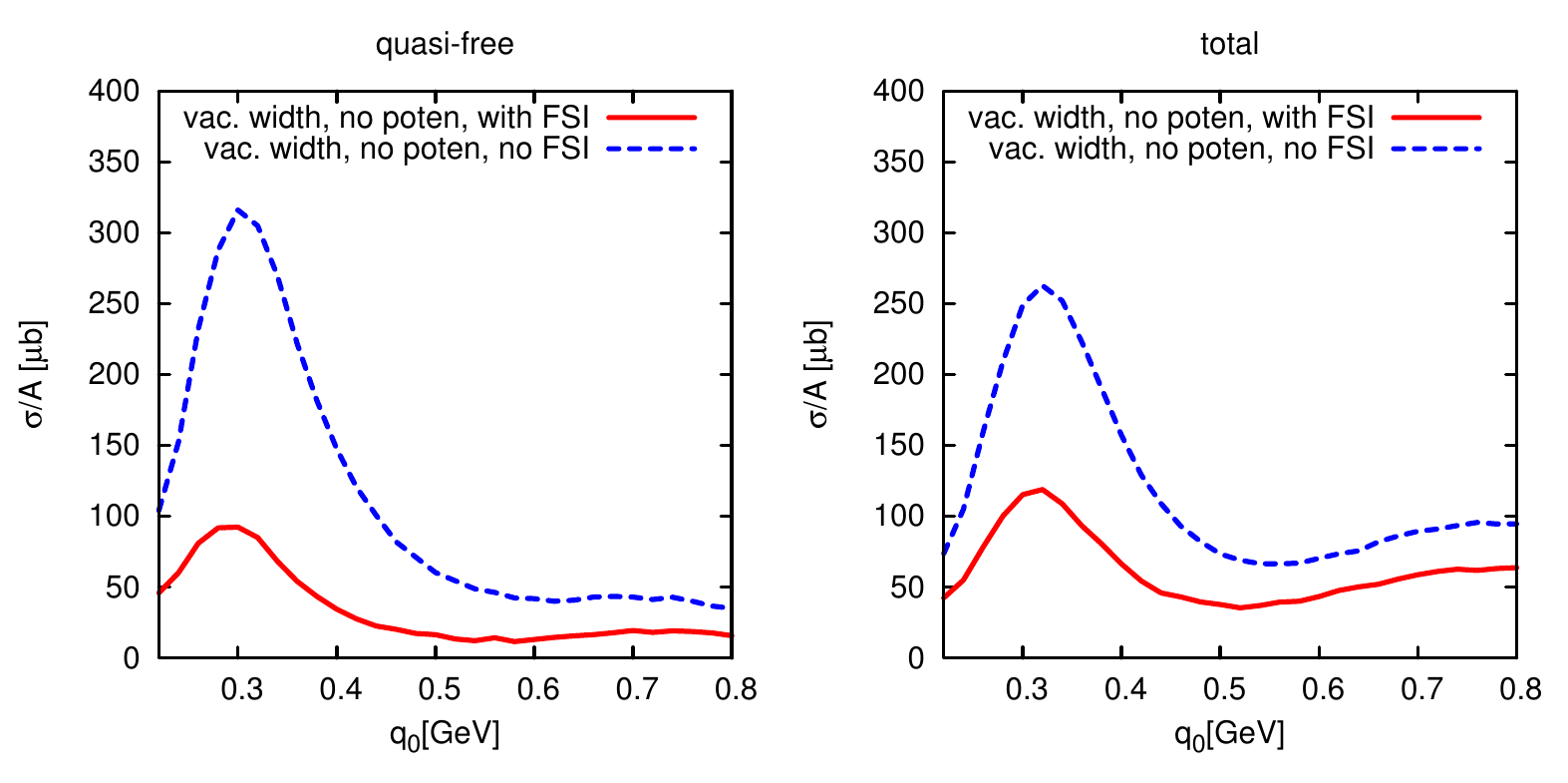}
  \caption[$\gamma\atom{40}{Ca} \to \pi^0 X$: Impact of FSI]{(Color
    online) Quasi-free and total cross sections for the $\gamma
    \atom{40}{Ca}\to \pi^0 X$ reaction with and without final-state
    interactions: the left panel shows the \textit{quasi-free} yield,
    whereas the right one depicts the total yield of $\pi^0$'s. The
    lines represent calculations, which have been performed neglecting
    the in-medium width and potentials. The dashed line does not
    include any FSI, whereas the solid line includes them. Source: 
    Taken  from
    \cite{buss_phd}.}
  \label{fig:Krusche_exclusive_fsi}
\end{figure}

To compare to data, we apply the same counting and cutting scheme also
in our analysis. Note that the sizable coherent
contribution~\cite{Krusche:2002iq} also contributes to
$\sigma_{\pi^0,\text{quasi-free}}$ and $\sigma_{\pi^0, \text{
    total}}$. Since GiBUU only gives the incoherent contribution we
have fitted the data for the coherent process~\cite{Krusche:2002iq} as
shown in \cref{fig:Krusche_exclusive}.

This fit of the coherent data has then been added to our incoherent
result. The resulting curves are shown in
\cref{fig:Krusche_exclusive}, both for the quasi-free cross section,
$\sigma_{\pi^0,\text{ quasi-free}}$, and for the total one,
$\sigma_{\pi^0, \text{ total}}$. When including in-medium width and
momentum-dependent potentials, one retrieves a quite good description
of the quasi-free cross section. The influence of the off-shell
potential is rather minor. Considering the large slowdown of our
simulations due to this off-shell potential, it is favorable, that
neglecting this potential does not change our results in a significant
manner.

For the total cross section the quality of the description data is
similar to that obtained earlier by Lehr \etal{}~\cite{lehr_phd}. One
may speculate whether there is a problem with the $2\pi$ contribution,
which rises continuously from almost 0\proz at $q_0=0.45 \GeV$ to
roughly 30\proz of the $\pi^0$ yield at $q_0=0.8 \GeV$. An additional
in-medium modification of the $\pi\pi$ background could, therefore,
have a major impact on the spectra. One should also note, that
interference effects play a major role in single-$\pi^0$ production at
the energy region of discrepancy (see also \cref{fig:photonBG_2pi}).
In addition, there is the possibility that primary 2N absorption
processes play a role, such as $\gamma N N\to N \Delta$.

In any case, there is major impact of FSI on the spectra. This is
illustrated in \cref{fig:Krusche_exclusive_fsi}, which shows that for
a calculation that does not include any medium modifications of self-energies,
i.e., no potentials or collisional broadening, FSI lead to
an overall reduction of the pion yield\footnote{In \cref{fig:Krusche_exclusive_fsi}, the quasi-free yield
  at low photon energies is higher than the total yield if FSI are not
  included. This is not an inconsistency, but a feature, which in
  implied by the definition of $\sigma_\text{quasi-free}$ in the
  analysis procedure: since the quasi-free peak without FSI treatment
  is slightly shifted towards positive $\Delta E_\pi$ (see
  \cref{fig:deltaE_plot}), we obtain
  $\sigma_\text{quasi-free}=2\int_{0}^{\infty} d\sigma(\Delta
  E_\pi)>\int_{-\infty}^{\infty} d\sigma(\Delta
  E_\pi)=\sigma_\text{tot}$.} by approximately a factor of
$3$. Thus, to achieve a proper agreement
with the data, the accuracy of the FSI must be very high.

\subsubsection{Double pion photoproduction}
\label{sec:pipi_photo_nuclei}

Possible changes of the properties of hadrons when they are embedded
inside the nuclear medium have recently attracted considerable
interest (for two recent reviews see
\refcite{Hayano:2008vn,Leupold:2009kz}). While many studies have
concentrated on the investigation of in-medium properties of vector
mesons (cf.~\cref{sec:omega_photo_nuclei}), there has also been an
interesting prediction of a change of scalar mesons inside nuclei.

In the limit of vanishing quark masses QCD incorporates chiral
symmetry, which is spontaneously broken in vacuum. The order
parameter, the quark condensate, $\langle \bar{q}q\rangle$, of this
symmetry breaking is expected to decrease by about 30\proz already at
normal nuclear-matter
density~\cite{Drukarev:1988kd,Cohen:1991nk,Brockmann:1996iv}. Therefore,
signals for partial chiral-symmetry restoration should be observable
in nuclear-reaction experiments and, in particular, also in
photon-induced processes where initial state interactions are absent.

The modification of the $\sigma$ or $f_0(600)$ meson in the nuclear
medium has been proposed as a signal for partial symmetry
restoration. Theoretical models predict a shift of its spectral strength
to lower masses and narrower widths due to the onset of the
chiral-symmetry restoration~\cite{Bernard:1987im,Hatsuda:1999kd}. The
\smeson is a short-lived state with a width of roughly $600\upto1000
\MeV$~\cite{PDGdata}, decaying predominantly into an $S$-wave $\pi
\pi$-final state. Owing to its short life time, this decay occurs very
close to its production place, i.e., in the medium. If there were no
final-state interactions acting on the pionic decay products, then the
mass of the \smeson could be directly determined measuring the
four-momenta of the pions. Thus experiments have studied the $\pi\pi$
production rate as a function of the total mass of the $\pi\pi$
pair. The major aim is to observe a modification of this signal when
comparing the nuclear production rate to the vacuum rate. Using
different nuclear targets, one probes different effective densities,
which allows for a detailed study of the density dependence of this
production rate. Experiments on $\pi\pi$ production in nuclear matter
have been performed with incident pions by the
CHAOS~collaboration~\cite{Bonutti1,Bonutti2} and with photons by the
TAPS~collaboration~\cite{Messch,Bloch:2007ka,DrGregor}. Both
experiments have shown an accumulation of strength near the $\pi \pi$
threshold in the decay channel of the $\sigma$ in large nuclei. A
possible interpretation of this effect is the in-medium modification of
the $\sigma$ resonance due to partial symmetry restoration.

Roca \etal{}~\cite{Roca:2002vd} have indeed obtained a good
description of these data by assuming pion-pion correlations and using
a straight-line Glauber-like damping factor for pion absorption.

However, the authors of \cite{Muhlich:2004zj} have pointed out the
importance of conventional final-state effects in the analysis of the
TAPS experiment. In this section we improve on our early
calculations using up-to-date input for the elementary rates and using
an improved final-state model. In order to compare to the work by Roca
\etal{}~\cite{Roca:2002vd} the absorption probability for pions is
assumed to be the same as in their paper.

As was already shown in \cref{sec:pionA_low}, the GiBUU transport
model successfully describes pion absorption and rescattering off
complex nuclei. It is, therefore, tempting to apply this model also to
this reaction. The results presented in this Section have been
published
in~\cite{Muhlich:2004zj,AlvarezRuso:2005dh,Buss:2005bm,buss:bormio06,Buss:2006vh}. The
calculations contain all the effects of pion-nucleon-Delta
interactions inside the nucleus, but they do not contain any pion-pion
interaction.

\begin{figure}[t]
  \centering
  \includegraphics[width=0.6\linewidth]{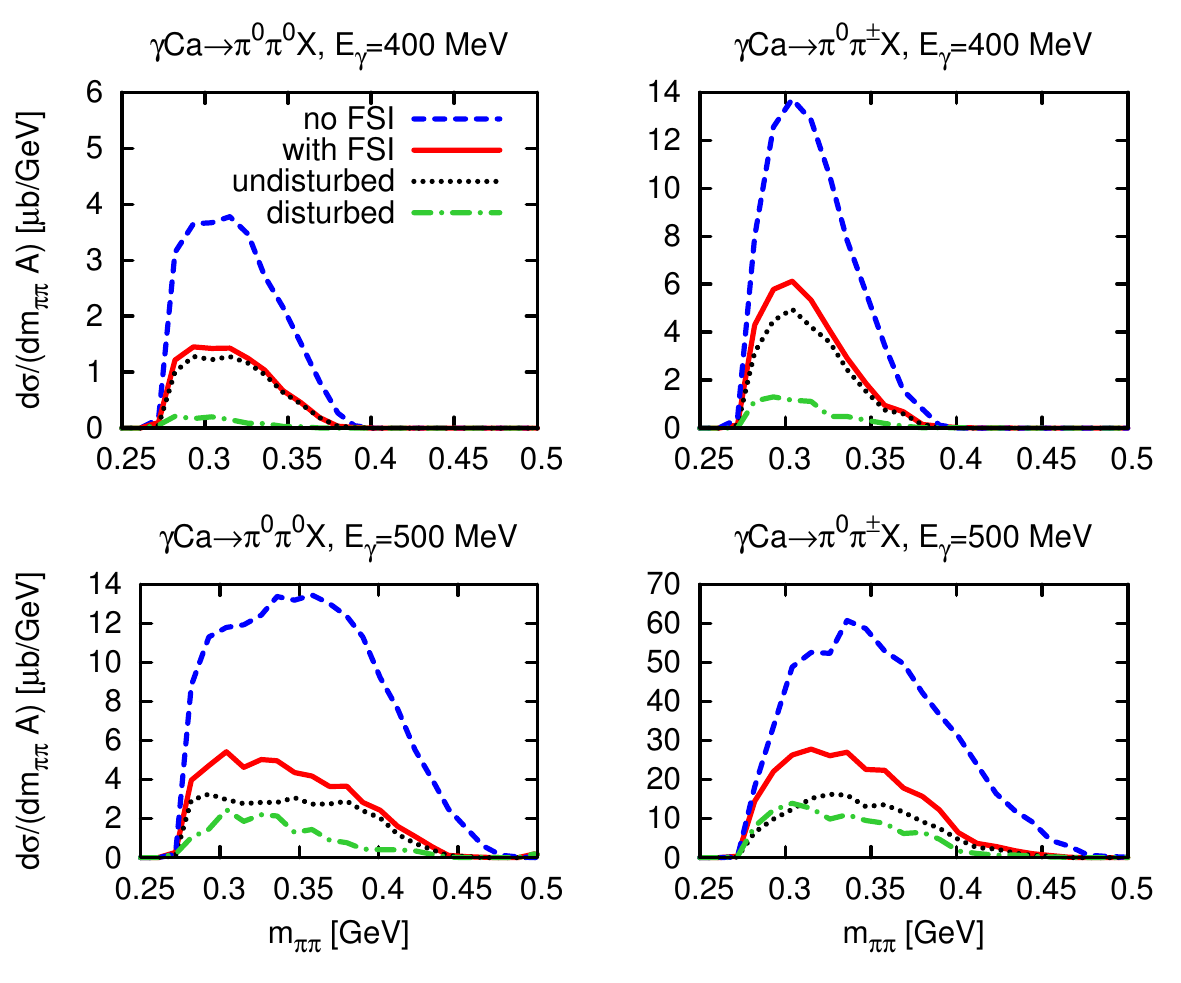}
  \caption[Impact of FSI on \pn and \pc photoproduction]{(Color
    online) The upper figure visualizes the impact of final-state
    rescattering on the mass-differential cross section for $\pi\pi$
    production in $\calcium$. The dashed curve has been obtained
    neglecting FSI, while for the calculations, represented by the
    solid, dotted and dashed-dotted curves, the FSI have been
    included. The solid curve shows the total result including FSI;
    the dotted one represents the ``undisturbed'' contribution of
    pions which left the medium without rescattering; one of the pions
    out of each pair contributing to the dashed-dotted "disturbed"
    contribution underwent at least one scattering event. Source: 
    Taken from~\cite{buss_phd}.}
  \label{fig:400_500_twoPi_coll}
\end{figure}

\begin{figure}[t]
  \centering
  \includegraphics[width=0.6\linewidth]{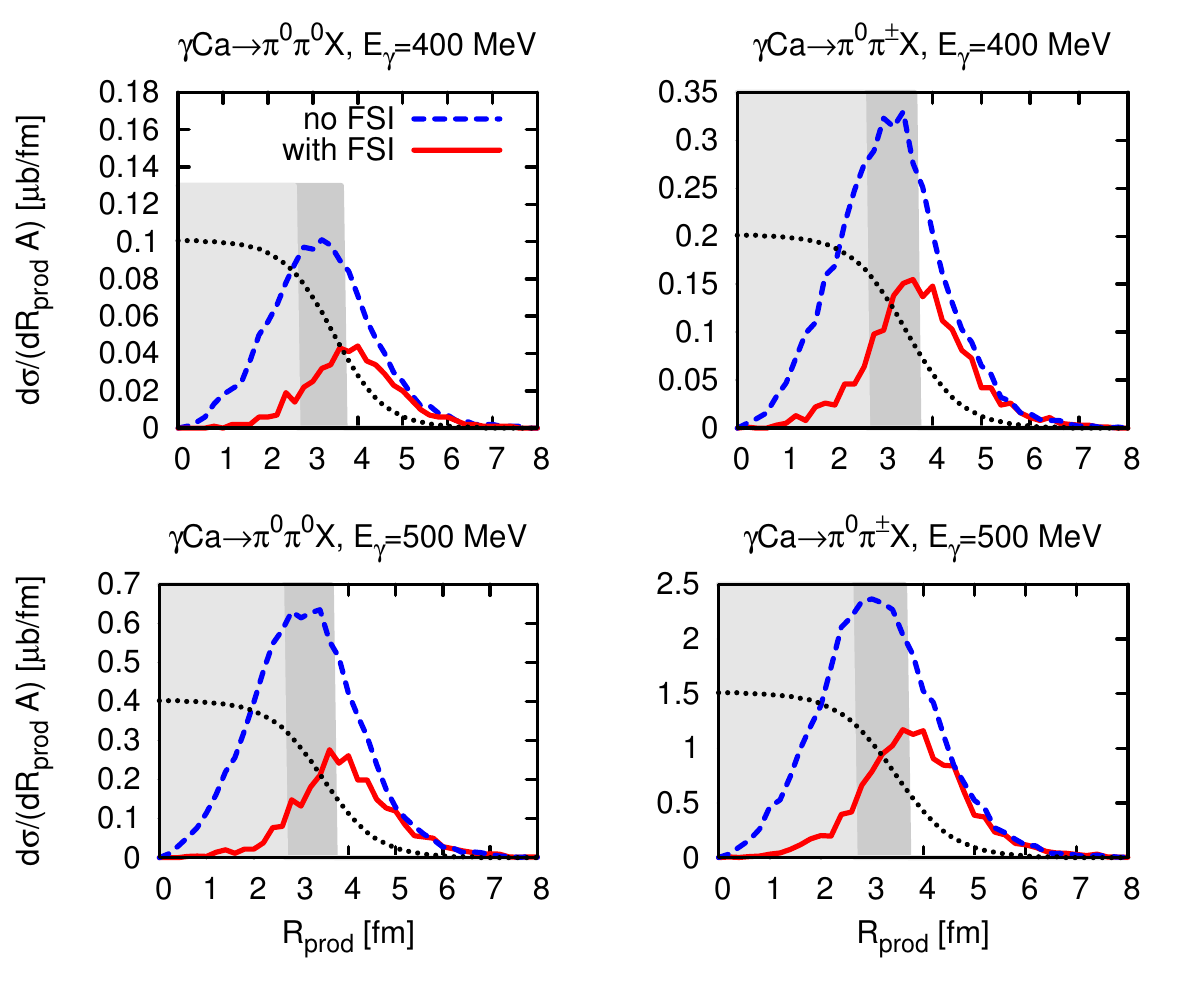}
  \caption[Production places of $\pi\pi$ pairs in \pn and \pc
  photoproduction]{(Color online) Cross section for $\pi\pi$
    production in $\calcium$ as a function of the production radius,
    $R_{\text{prod}}=(x_{\text{prod}}^2+y_{\text{prod}}^2+z_{\text{prod}}^2)^{1/2}$:
    the solid curve depicts the result with FSI, the dashed one does
    not include FSI. Additionally, the dotted line depicts the density
    profile of the $\calcium$ nucleus; the light-gray areas show the
    $R_{\text{prod}}$ region with $\rho\geq 0.15 \fm^{-3}$, the dark
    shaded areas show the region with $0.15 \fm^{-3} \geq\rho\geq
    0.075 \fm^{-3}$. Source: Taken from \cite{buss_phd}.}
  \label{fig:400_500_twoPi_radius}
\end{figure}

As already discussed in \cite{Muhlich:2004zj}, we observe that
absorption, elastic scattering and charge exchange processes cause a
considerable change of the spectra with the peak of the mass
distribution moving to lower masses due to rescattering. This effect
is visualized in \cref{fig:400_500_twoPi_coll}, where we show our
results for $\pi\pi$ production of \calcium{} assuming no FSI and by
including FSI. One observes a reduction of the cross section by a
factor of roughly $2\upto3$ and a shift of the peaks towards lower
masses due to FSI. There are two major effects, which lead to the
modifications: absorption and rescattering. To point out the role of
rescattering, \cref{fig:400_500_twoPi_coll} shows so-called
\textit{disturbed} and \textit{undisturbed} contributions to the cross
section. The undisturbed contribution includes all pion pairs which do
not undergo rescatterings and reach the detector almost undisturbed;
the \textit{disturbed} contribution represents pairs, where at least
one of the two pions scatters with the medium but is not absorbed. The
total cross section is the sum of the disturbed and the undisturbed
contributions. Obviously, the disturbed contribution is shifted
towards lower masses more than the undisturbed one since the pions (on
average) loose energy in a scattering event. At low photon energies
the disturbed contribution is small compared to the undisturbed
one. Here, the energies of the produced pions is small, such that FSI
are dominated by the $NN\pi\to NN$ process and the effect of $N\pi\to
N\pi$ scattering is small. At higher photon energy ($500 \MeV$) also
the average pion energy is higher and, therefore, elastic and
charge-exchange scatterings become more important. Thus the disturbed
and undisturbed contributions are of the same magnitude for
$E_{\gamma}=500\MeV$.

To analyze the possible impact of additional in-medium modifications,
such as e.g.~chiral symmetry restoration, the production points of
those pion pairs which are not absorbed and which are, finally,
observed were studied. \Cref{fig:400_500_twoPi_radius} shows the cross
section for $\pi\pi$ production off {\calcium} at $400 \MeV$ and $500
\MeV$ as function of those production points $R_{\text{prod}}$.
Without FSI the distribution $\dd \sigma_{\pi^0\pi^0}/\dd
R_{\text{prod}}$ is proportional to $\rho(R_{\text{prod}})
R_{\text{prod}}^2$; including FSI the distribution is shifted towards
higher radius and centered around $3.6\fm$ which corresponds to
roughly $\rho=0.075 \fm^{-3}$. Most of the observed signal originates
from low-density regions, which implies that possible in-medium
signals are expected to be rather weak. Also calculations including a
pion potential have been performed; only a minor effect on the
observed pions is found due to the low density at the initial
production point.

\begin{figure}[t]
  \centering
  \includegraphics[width=0.6\linewidth]{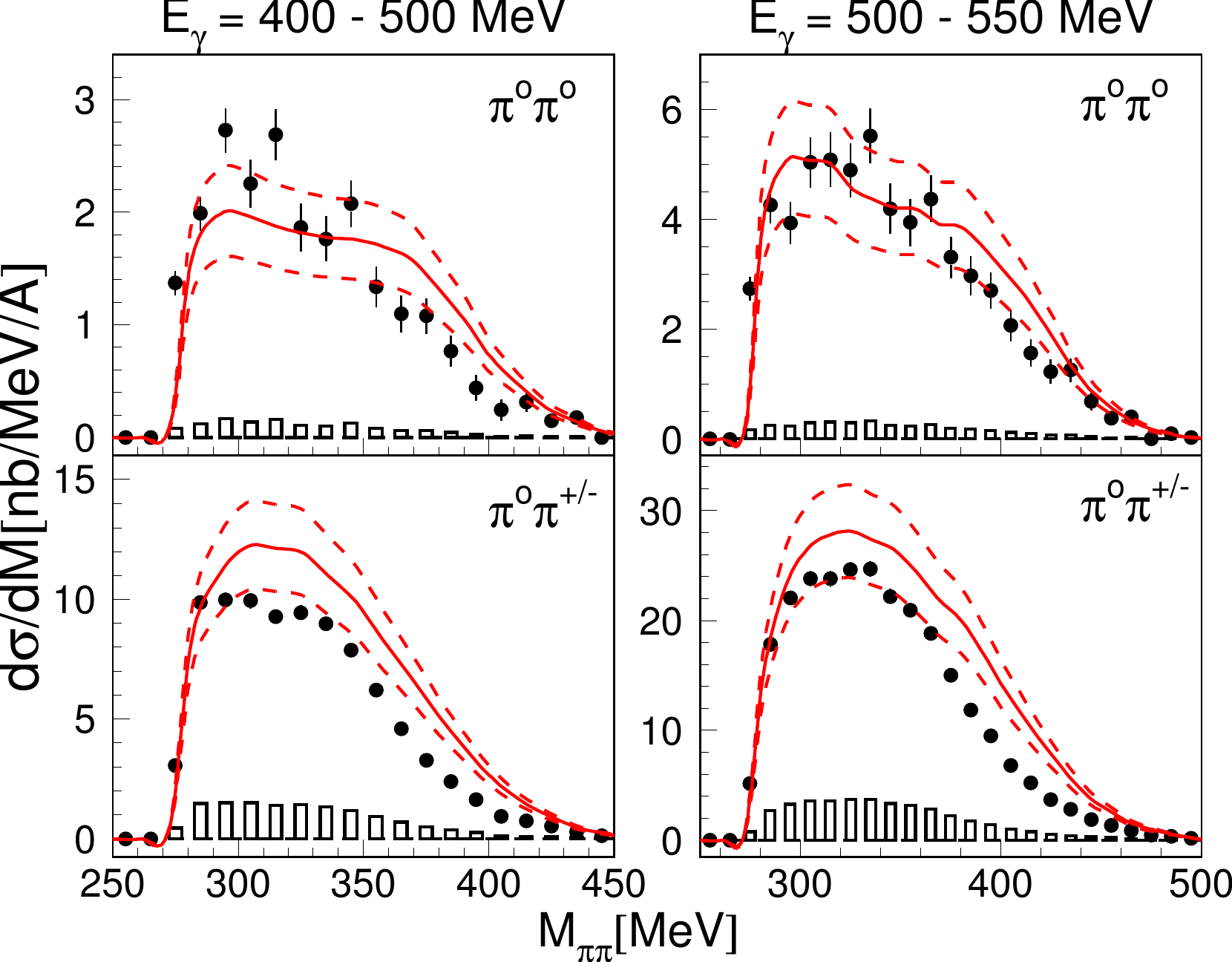}
  \caption{(Color online)Pion-pion invariant mass distributions for
    \calcium{} for the two energy ranges indicated compared to results
    of the GiBUU model in \refcite{Bloch:2007ka}. The bars at the
    bottom represent the systematic uncertainty of the data, the
    dashed lines represent the error band for the GiBUU calculation.
    Source: Taken from~\cite{Bloch:2007ka}.}
  \label{twopiCa}
\end{figure}

The previous discussions showed the strength of a transport model
description of nuclear reactions in that it gives some physical
insight into the 'inner workings' of dynamic many-body processes. The
direct comparison with experimental results then shows how far a model
that contains no $\pi-\pi$ interactions can go. While earlier results
\cite{Muhlich:2004zj} gave a reasonable description of the TAPS data
\cite{Messch}, a later measurement by the same group has resulted in
data that are very well described by the GiBUU model (see
\cref{twopiCa}). The agreement now is excellent.

In summary, final-state interactions of the pions are strong and tend
to shift the maximum of the $\pi\pi$ mass distribution in all channels
towards lower masses. This effect considerably complicates drawing a
link between the experimental data and a possible softening of the
in-medium $I=0$ channel. Any theory aiming to describe the observed
effect on the basis of a partial chiral-symmetry restoration or an
in-medium modification of the $\pi\pi$-production process must include
a state-of-the-art treatment of the final-state effects.

\subsubsection{Photoproduction of vector mesons}
\label{sec:omega_photo_nuclei}

As already discussed in \cref{sec:pipi_photo_nuclei,sec:protonA_dilep}, modifications of hadron
properties in a strongly interacting environment have attracted a lot of
attention and have been intensively studied both theoretically and
experimentally. The dilepton channel is particularly suitable for such
studies since the final state is not affected by FSI. The experiments
with hadronic beams discussed in \cref{sec:protonA_dilep} do suffer from
initial state interactions which may hinder the population of the
higher-density zones in the target. On the contrary, experiments using
photons as incoming particles are nearly ideal for these studies since
they do not involve any initial state interactions. Such an experiment
was first studied with an early version of GiBUU
\cite{Effenberger:1999ay} and has by now been performed with the CLAS
detector at JLAB, where photons with energies of a few \GeV interacted
with nuclei \cite{:2007mga,Wood:2008ee}. The results of this experiment
have been analyzed with GiBUU; a broadening of the $\rho$ meson in line
with theoretical expectations but no mass shift has been observed.  A
recent study on in-medium modifications of the $\rho$ meson in
connection with the CLAS data comes to a similar
conclusion~\cite{Riek:2008ct,Riek:2010gz}.

Another experiment also using photons as incoming particles looks at the
semi-hadronic decay of vector mesons thus suffering from some FSI. Here
the decay $\omega\rightarrow\pi^0\gamma$ is investigated by the CB/TAPS
group in photon-induced reactions at the ELSA accelerator
\cite{Nanova:2010tq}. This experiment has again given a significant
in-medium broadening of the $\omega$ meson \cite{:2008xy} but no mass
shift; the earlier claim of a mass shift of the $\omega$ meson in
photoproduction on Nb \cite{Trnka:2005ey} has not been confirmed in a
reanalysis of the data \cite{Nanova:2010sy}.

In this experiment, incident photon energies covered the range from $900
\upto 2200 \MeV$. Because of the increase of the production cross
section with photon energy most of the observed $\omega$ mesons are
produced with photons of energies larger than $1500 \MeV$. For the
energy range of $1500\upto2200 \MeV$, early transport calculations
\cite{Gallmeister:2007cm,Muehlich_phd} have shown that the $\omega$ line
shape is rather insensitive to different in-medium modification
scenarios like collisional broadening or mass shifts, since most of the
$\omega$ decays occur outside of the nuclear medium, even despite of a
cut on the $\omega$ momentum ($p_{\omega} \leq 500 \MeVc$). Furthermore,
due to the experimentally observed strong absorption of $\omega$ mesons
in the nuclear medium \cite{:2008xy}, $\omega$ mesons produced in the
interior of the nucleus are largely removed by inelastic reactions and
do not reach the detector; information on possible in-medium
modifications thereby is lost. The limited sensitivity of the $\omega$
line shape to in-medium effects has been confirmed experimentally in
\cite{Nanova:2010sy}.

\begin{figure}[t]
  \centering
  \includegraphics[width=0.4\linewidth]{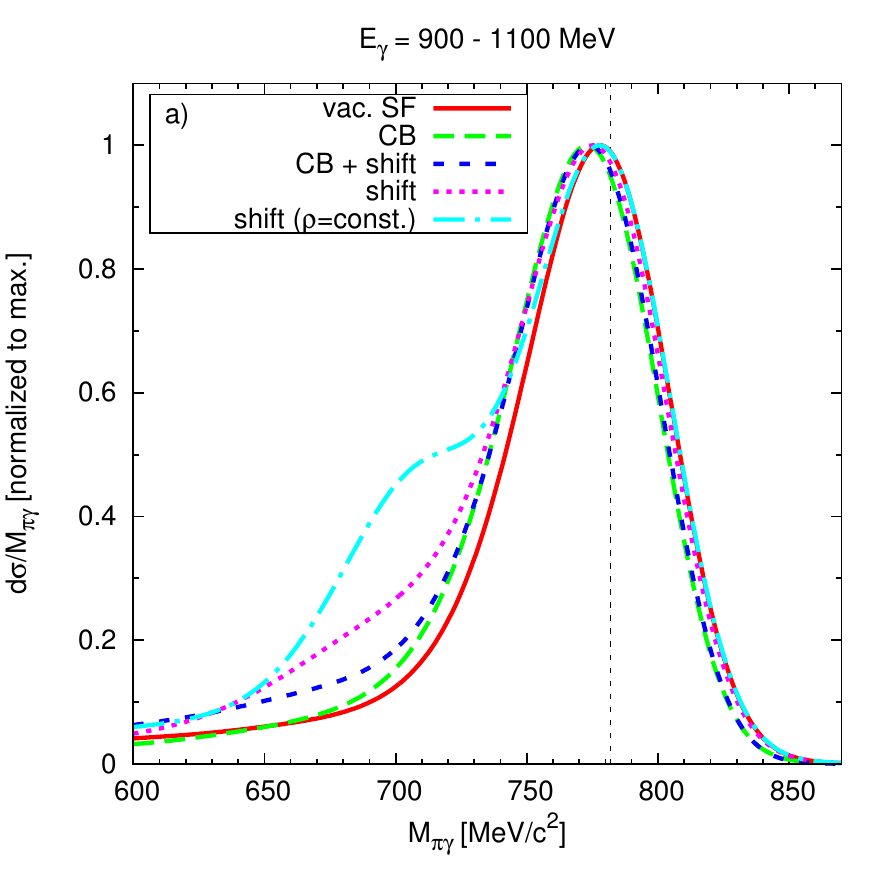}
  \includegraphics[width=0.4\linewidth]{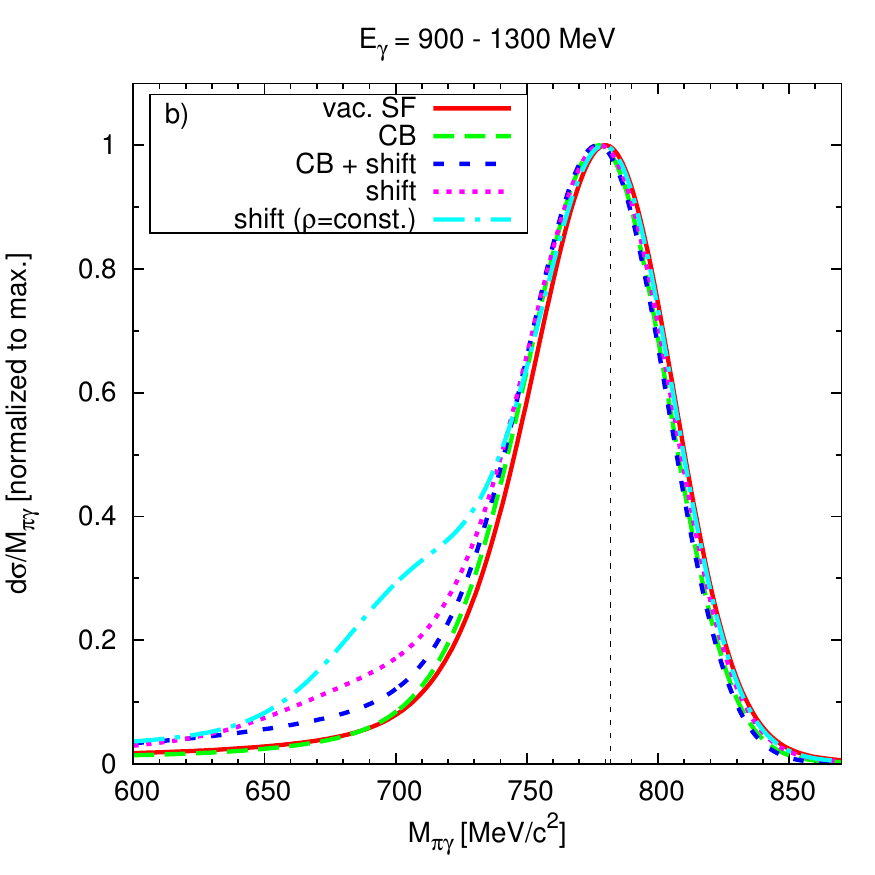}
  \caption{(Color online) $\omega$-meson line shape predicted for a Nb
    target in GiBUU transport-model calculations for different
    in-medium modification scenarios: vacuum spectral function
    (solid), collisional broadening of $\Gamma_{\text{coll}}= 140
    \MeV$ (long dashed), collisional broadening and an attractive mass
    shift of $-16\proz$ at nuclear matter density (short dashed), and
    a mass shift without broadening (dotted). The dash-dotted curve
    shows the results for a constant nuclear density of $\rho = 0.6
    \rho_0$. The signals are folded with the detector response and
    take into account the $1/E_{\gamma}$ weighting of the
    bremsstrahlung spectrum.  a) incident photon energies of
    $900\upto1100 \MeV$; b) incident photon energies of $900\upto1300
    \MeV$. Source: Taken from~\cite{Nanova:2010tq}.}
  \label{fig:omega1}
\end{figure}

It has been argued in \cite{Gallmeister:2007cm} that a search for medium
modifications is much more promising for incident photon energies below
or near the photoproduction threshold (which is at
$E_{\gamma}^{\text{lab}} = 1109 \MeV$ on a free nucleon). New
calculations along these lines illustrate in \cref{fig:omega1} the
expected sensitivity of the $\omega$ signal to various in-medium
changes, such as a mass shift with and without collisional broadening
for two different energy ranges. It is seen that the lower-energy window
indeed leads to a more pronounced -- though not dramatic -- sensitivity
than the higher-energy one. This relatively weak sensitivity is to a
large extent a consequence of the density profile of the nucleus that
spans all densities from 0 to $\rho_0$ and thus smears any
density-dependent signal. Assuming for the sake of the argument a
density profile with a constant density of $0.6 \rho_0$ -- roughly
corresponding to the average density in nuclei -- and a sharp fall off
at the surface the dash-dotted line in \cref{fig:omega1} is obtained;
here, the in-medium signal is significantly stronger. For a realistic
Woods-Saxon nuclear density profile, contributions to the spectral
function from the surface dominate, suppressing contributions from
higher density regions \cite{Mosel:2010mp}.

In both energy windows, a tail towards lower masses is predicted for
the scenario of a dropping $\omega$ mass. This tail is due to $\omega$
mesons which are produced off-shell within the nucleus. In
\cite{Gallmeister:2007cm,Muehlich_phd} an even stronger enhancement in
the low-mass-tail region has been obtained. This calculation used a
phenomenological method for the off-shell propagation, while the
present results are based on the theoretical framework by Leupold
\cite{Leupold:1999ga} and Juchem and Cassing \cite{Cassing:1999wx}. For
details see \cref{subsec:OSP}.

\begin{figure}[t]
  \centering
  \includegraphics[width=0.4\linewidth]{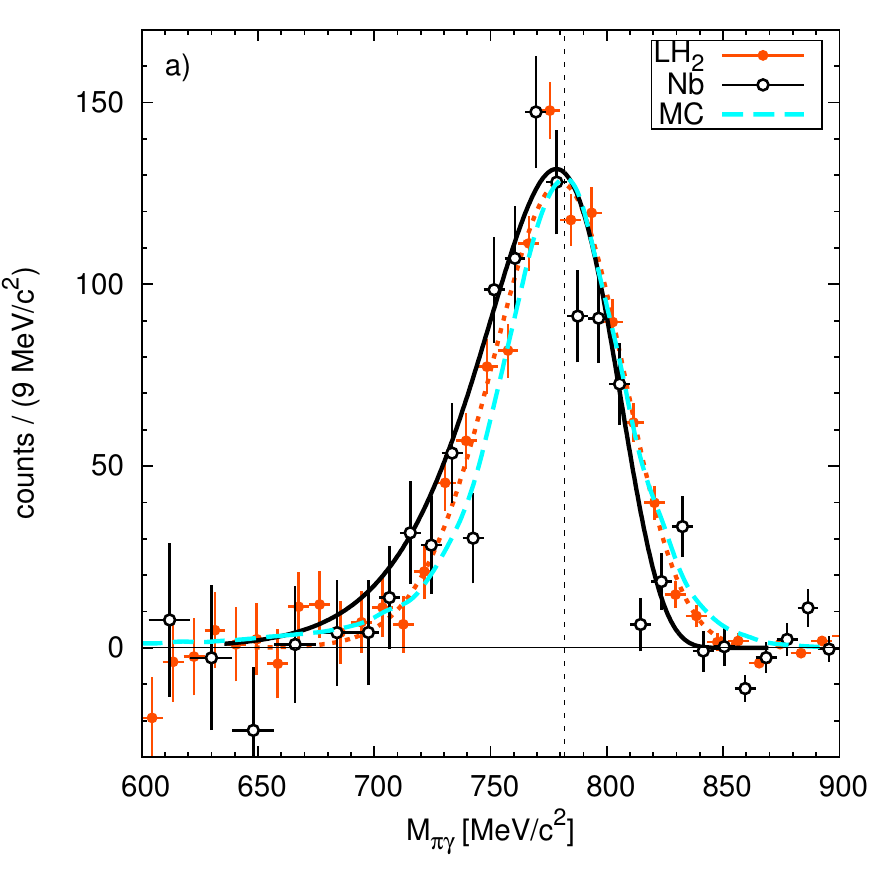}
  \includegraphics[width=0.4\linewidth]{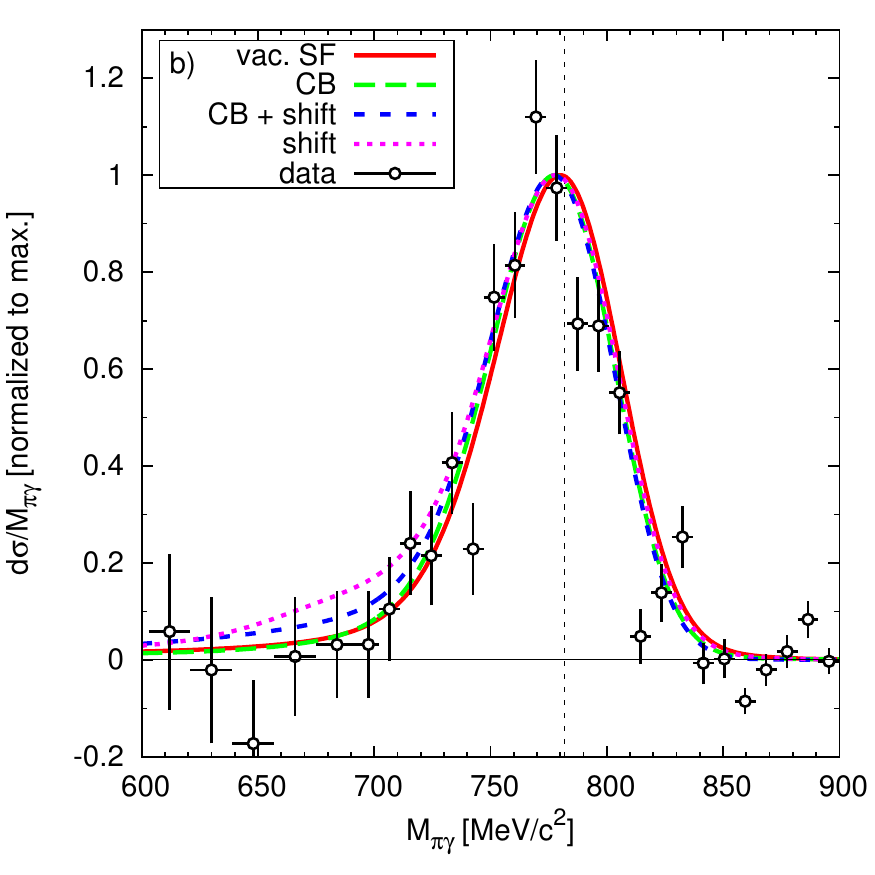}
  \caption{(Color online) (a) $\omega$ signal (solid points) for the
    Nb target and incident photon energies from $900 \upto 1300
    \MeV$. The errors are purely statistical. Systematic errors
    introduced by the background subtraction are of the order of
    $5\proz$.  A fit to the data points is shown in comparison to the
    $\omega$-line shape measured on a $\mathrm{LH}_2$ target and a
    Monte Carlo simulation; (b) $\omega$ signal for the Nb target in
    comparison to recent GiBUU simulations for the following
    scenarios: no medium modification (solid), in-medium broadening of
    $\Gamma_{\text{coll}} = 140 \MeV$ at nuclear saturation density
    (long dashed), an additional mass shift by $-16\proz$ (short
    dashed), mass shift without broadening (dotted) and mass shift
    without broadening assuming a constant nuclear density of
    $\rho=0.6\rho_0$ (dash-dotted). The signals are folded with the
    detector response and take into account a $1/E_{\gamma}$ weighting
    of the bremsstrahlung spectrum. Source: Taken from~\cite{Nanova:2010tq}.}
  \label{fig:omega2}
\end{figure}

Recently, the $\omega$-line shape has also been measured in the
near-threshold range of incident photon energies, i.e., $900 \upto
1300 \MeV$ \cite{Nanova:2010tq}. A tagged bremsstrahlung photon beam
has been impinged on $\mathrm{LH}_2$, C and Nb targets, and photon
triples from $\omega\to\pi^0\gamma$ decays are measured by the
CBELSA/TAPS detector setup at Bonn. After subtraction of the
background from the $\pi^0 \gamma$ spectrum, the $\omega$-line shape
shown in \cref{fig:omega2}~(a) is obtained for the Nb target. The
experimental distribution has been fitted using the Novosibirsk
function \cite{Aubert:2004te}, in order to model the detector response
and the resulting mass resolution.  The resulting fit is compared with
the $\omega$ signal measured on the $\mathrm{LH}_2$ target and with a
MC detector simulation in \cref{fig:omega2}~(a). The
agreement between the $\omega$ signal on the LH${}_2$ target with the
MC simulation demonstrates that the detector response is
under control.  Nevertheless, in view of the systematic and
statistical uncertainties, no significant deviation from the reference
signals is claimed. Higher statistics will be needed to draw any
conclusion. Corresponding data have been taken at MAMI-C using the
Crystal Ball/TAPS set up. The analysis is ongoing.

In \cref{fig:omega2}~(b) the measured $\omega$ signal is compared to
predictions of transport calculations using the GiBUU model for the
same scenarios as in \cref{fig:omega1}. Due to the rather large
statistical errors, the experimental data obviously do not allow to
distinguish between the various theoretical scenarios, in contrast to
initial expectations.

Access to the in-medium spectral function of vector mesons is thus very
limited, mainly due to the dependence of the in-medium properties such
as mass and width on the nuclear density \cite{Mosel:2010mp,Lehr:2001ju}
and the inherent smearing caused by the density profile of nuclei. From the analyses just discussed it
is clear that any extraction of an in-medium effect on spectral
functions of hadrons requires a state-of-the-art treatment of final
state interactions. On the other hand, transparency measurements
\cite{Kaskulov:2006zc,Muhlich:2006ps} can give at least access to the
imaginary part of the in-medium self-energy of the hadron. Another
promising tool could be the measurement of excitation functions
\cite{Muhlich:2003tj}. Such experiments are presently being analyzed.

\subsection{Neutrino-induced reactions}
\label{sec:nuA}

The discovery of neutrino oscillations has renewed the interest in and
the need for a better determination of the neutrino-nucleus cross
sections since nuclear effects are known to be the largest source of
systematic uncertainties in long-baseline neutrino experiments
\cite{NUINT09Talks}.  To increase the neutrino cross section,
experiments use targets with a large atomic mass number, $A$, e.g.,
carbon, oxygen, iron, and lead. This causes a major difficulty:
particles produced in neutrino interactions can reinteract before
leaving the nucleus and can be absorbed, change their kinematics or
even charge before being detected. Nuclear reinteractions limit our
ability to identify the reaction channel, and they change the topology
of the measured hadronic final state. Consequently, the detected rates
on nuclei are changed significantly compared to the ones on free
nucleons.

The oscillation probability depends directly on the neutrino energy:
$\nu_\mu$-disappearance experiments for example search for a
distortion in the neutrino flux in the detector positioned far away
from the source. By comparing the un-oscillated with the oscillated
flux, one gains information about the oscillation probability and with
that about mixing angles and squared-mass differences. However, the
neutrino energy cannot be measured directly but has to be
reconstructed from the final-state particles that are detected. But,
as we have pointed out before, these are affected by in-medium effects
and final-state interactions in the nucleus. Appearance experiments,
on the other hand, search for a specific neutrino flavor in a neutrino
beam of different flavor. The flavor of the neutrino can only be
determined from the charged lepton produced in the
interaction. $\pi^0$-production events in neutral-current reactions
are a source of background in $\nu_e$-appearance searches in a
$\nu_\mu$ beam, because they might be misidentified as charged-current
$(\nu_e,e^-)$ interactions.

To extract the oscillation parameters from the measured particle
yields, the experimental analyses thus have to rely on models for the
neutrino-nucleus interaction that describe many different effects
quite reliably. This is even more important because the neutrino
long-baseline experiments do not work with a fixed neutrino energy,
but instead a broad neutrino spectrum, and thus implicitly sum over
very different reaction types: quasi-elastic (QE), pion production and
DIS. Quasi-elastic scattering alone as well as nucleon knockout is
studied in many different models; only a few study pion
production. The particular strength of GiBUU is that it describes all
these different initial processes and the interactions in their final
particles equally well. In the following we illustrate these points by
discussing some results for neutrino-nucleus cross sections obtained
with the GiBUU model. More detailed discussions can be found in
\cite{Leitner:2006ww,Leitner:2006sp,LeitnerDr}. All the results shown
here rely on the impulse approximation. For true quasi-elastic
scattering this should be an excellent approximation, but for pion
production and total cross sections this may not be sufficient. For a
discussion of the influence of 2N correlations on nuclear
neutrino-induced reactions see
Refs.~\cite{Marteau:1999zp,Martini:2009uj,Martini:2010ex,Nieves:2011pp}.

\subsubsection{Pion production}

Neutrino-induced pion production on nucleons up to neutrino energies
of about $1.5 \GeV$ is dominated by the excitation and subsequent
decay of the $\Delta$ resonance, but, depending on the channel,
non-resonant pion production is not negligible. At higher energies,
higher-mass resonances become increasingly important.

\begin{figure}[t]
  \centering
  \includegraphics[width=0.4\linewidth]{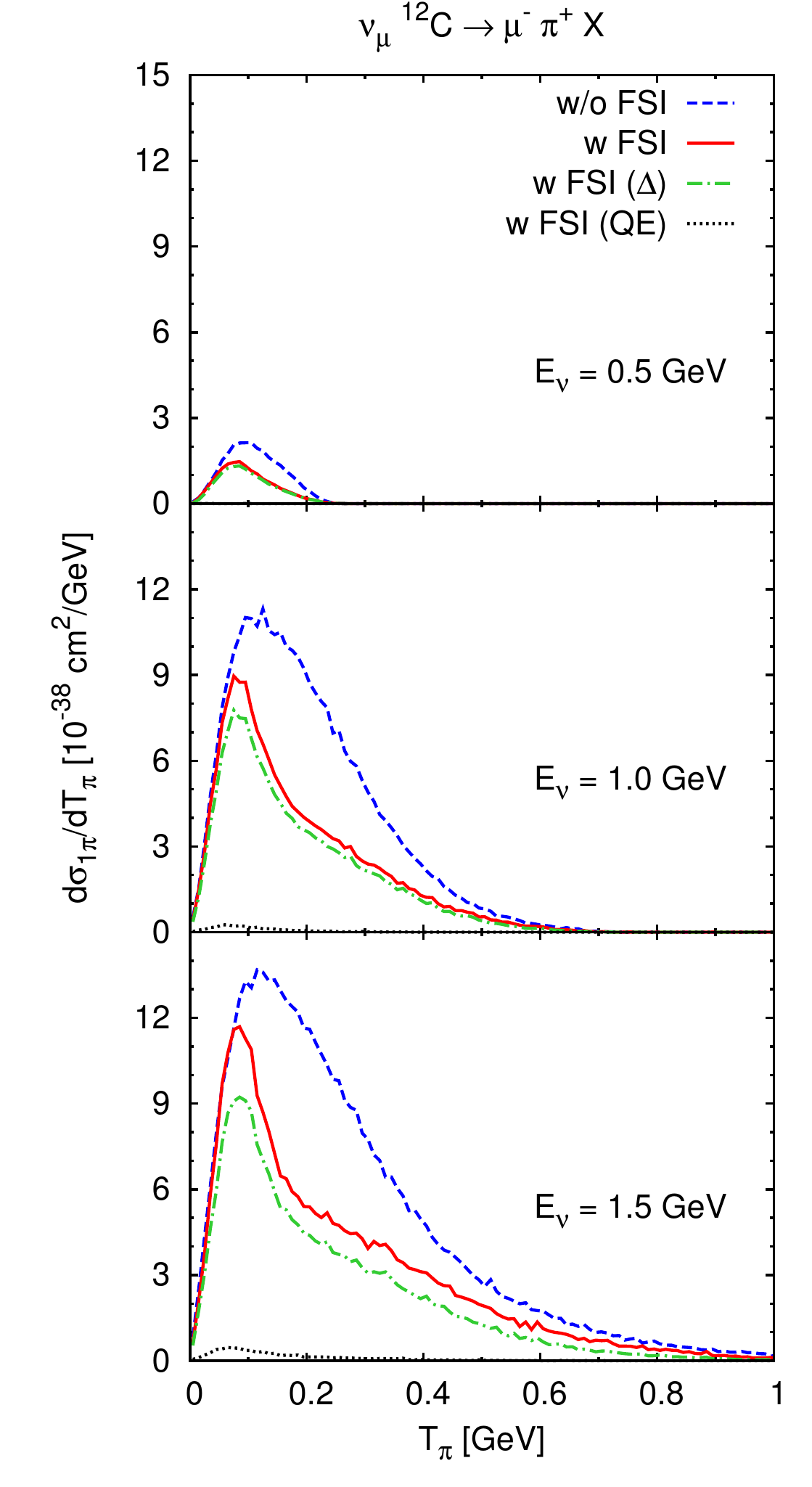}
  \includegraphics[width=0.4\linewidth]{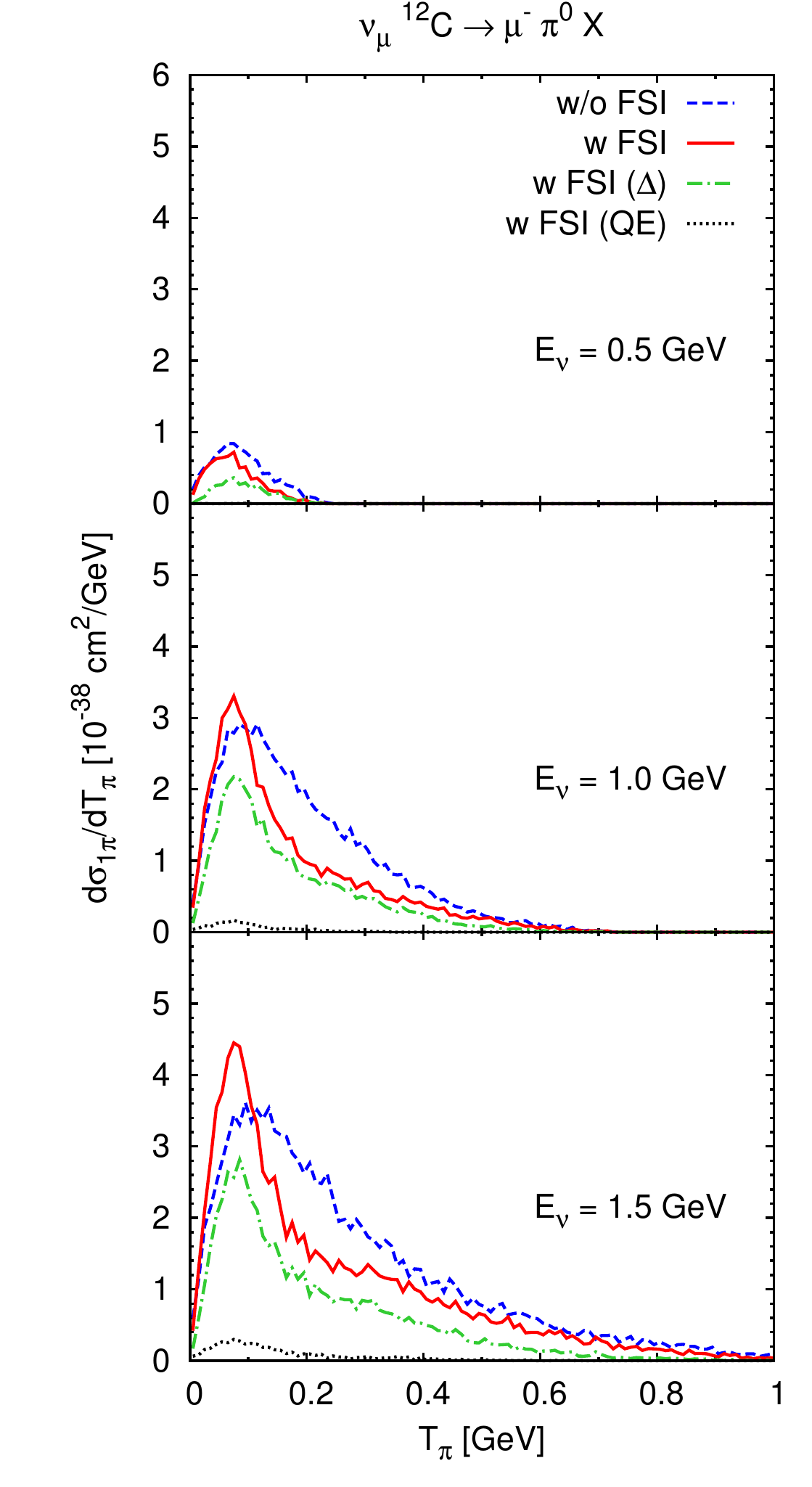}
  
  \caption{(Color online) Kinetic-energy differential cross section
    for CC $1\pi^+$ (left) and $1\pi^0$ (right panels) production on
    $\carbon$ versus the pion kinetic energy $T_{\pi}$ at different
    values of $E_{\nu}$. The dashed lines show the results without FSI
    interactions (only the decay of resonances is possible), the
    results denoted by the solid lines include FSI. Furthermore, the
    origin of the pions is indicated (QE or $\Delta$ excitation). Source: Taken from from \cite{LeitnerDr}.}
    \label{fig:CC_pi_kinetic}
\end{figure}

A realistic treatment of the FSI is an essential ingredient for modeling
pion production off nuclei in a realistic manner. FSI may lead, e.g., to
pionic final-state particles, even though the initial event is
quasi-elastic scattering. As an example, we present the single-pion
(1$\pi$) kinetic-energy distributions.  They are shown in
\cref{fig:CC_pi_kinetic} for $\nu_\mu$ charged current (CC) $\pi^+$ and
$\pi^0$ production (CC $1\pi$) on $\carbon$ target at different values
of $E_{\nu}$.

The maximum of the calculation with final-state interactions peaks at
$0.05 \upto 0.1 \GeV$ in all cases shown in
\cref{fig:CC_pi_kinetic}. This is due to the energy dependence of the
pion absorption. The absorption is higher in the resonance region,
where the pions are mainly absorbed through the reaction $\pi N \to
\Delta$, followed by $\Delta N \to N N$. This strong reduction for
high-energy pions and the corresponding shift of the maximum to lower
energies can be seen by comparing the dashed and the solid
lines. These absorption processes equally affect $\pi^+$ and $\pi^0$
yields. But pions do not only undergo absorption when propagating
through the nucleus. Of particular importance for pions of all
energies is elastic scattering, $\pi N \to \pi N$, which redistributes
the kinetic energy, again shifting the maximum to lower energies.

The contributions from initial $\Delta$ excitation and from initial QE
events are also plotted. $\pi^0$ and $\pi^+$ production through FSI of
QE scattering contributes mostly to the low-energy region of the pion
spectra because of the energy redistribution in the collisions.

The different scale of CC $\pi^+$ and $\pi^0$ is a consequence of
their different production rates in the neutrino-nucleon
reaction. This leads to side feeding from the dominant $\pi^+$ channel
to the $\pi^0$ channel. Pions produced from initial QE events
contribute relatively more to the $\pi^0$ channel. This together with
the side feeding produces the enhancement in the $\pi^0$ cross section
at low kinetic energies compared to the calculation without
final-state interactions for $E_{\nu} \gtrsim 1\GeV$.

\subsubsection{Nucleon knockout}

\begin{figure}[t]
  \centering
  \includegraphics[width=0.4\linewidth]{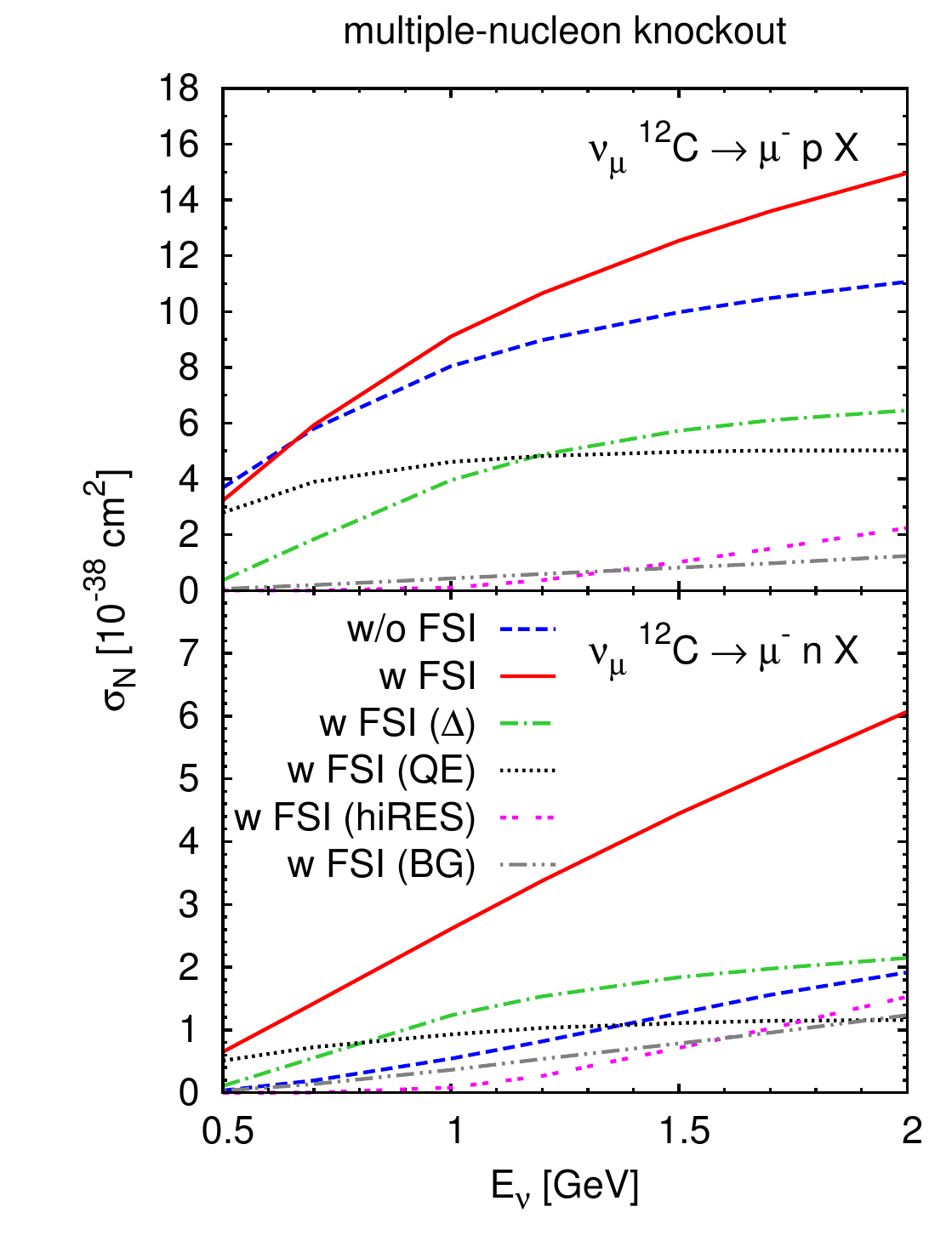}
  \includegraphics[width=0.4\linewidth]{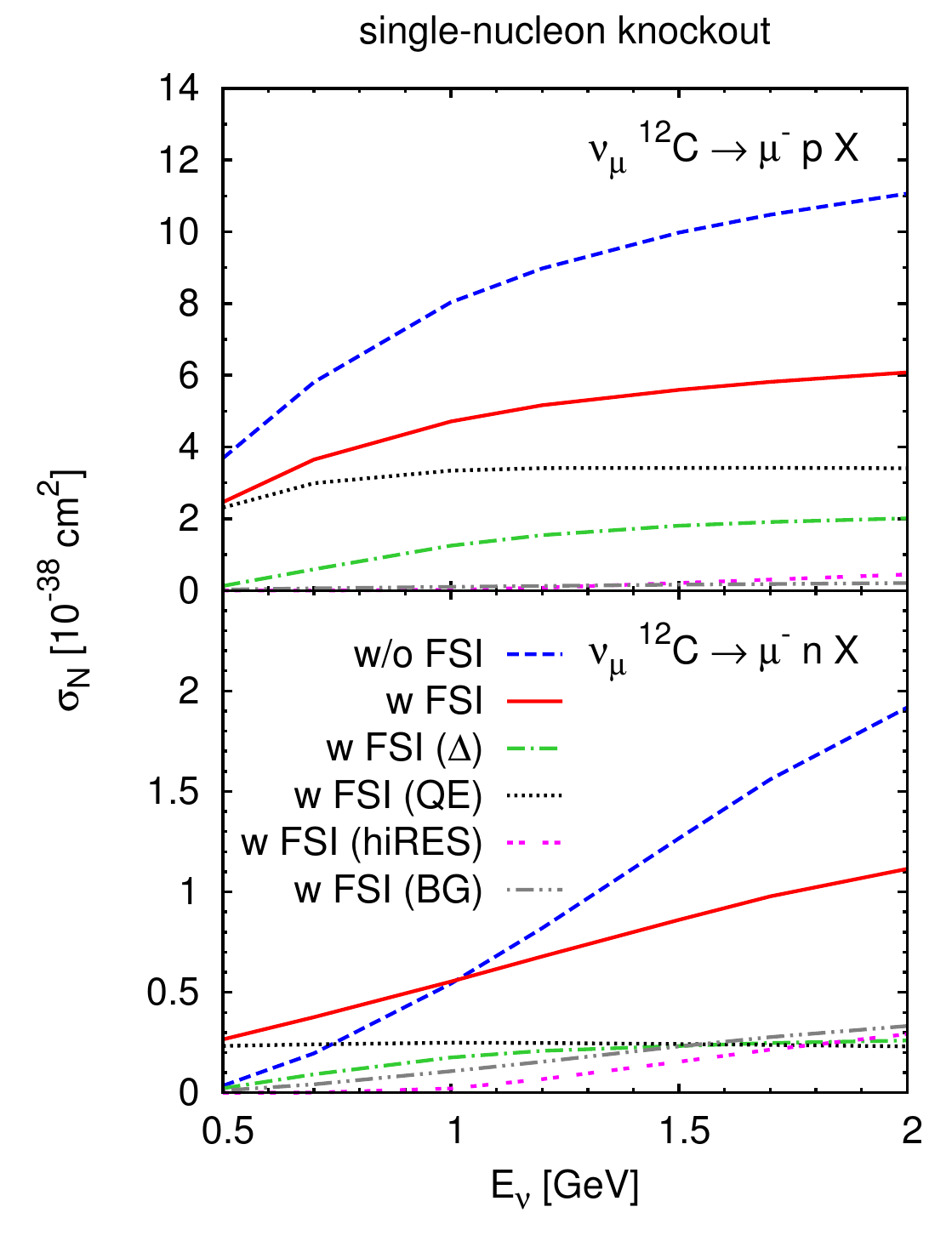}
  
  \caption{(Color online) Total cross section for CC multiple-$p$ (top
    left), $n$ (bottom left) knockout and CC single-$p$ (top right)
    and $n$ (bottom right) knockout on $\carbon$. Multiple-nucleon
    knockout means that the final state may contain any number of
    knocked-out nucleons. Single-nucleon knockout means that the final
    state does not contain any other knocked-out nucleons. The dashed
    lines show the results without FSI (only the decay of resonances
    is possible); the results denoted by the solid lines include FSI.
    Furthermore, the origin of the pions is indicated (QE, $\Delta$
    excitation, excitation of higher resonances, single-pion
    background). Source: Taken from from \cite{LeitnerDr}.}
  \label{fig:CC_nucl_tot}
\end{figure}

The total cross sections for proton and neutron knockout are shown in
\cref{fig:CC_nucl_tot} for multiple-nucleon emission and for
single-nucleon emission for $\carbon$. In the case of multiple-nucleon
knockout, we find, that the result with all
final-state interactions included lies well above the one without FSI
already for the protons, but even more so for the
neutrons. This enhancement is entirely caused by secondary
interactions and cannot be obtained in a Glauber treatment or in any
other quantum-mechanical approach such as the often used relativistic
impulse approximation.

Furthermore, it is indicated in \cref{fig:CC_nucl_tot}, whether the
knocked-out nucleon stems from initial QE scattering or $\Delta$
excitation (the contributions from higher resonances and from the
non-resonant background are also shown). In contrast to the pion case,
both contribute to the total cross section, even though with different
weights depending on the neutrino energy. The phase space for $\Delta$
excitations opens later than for QE; this explains the small
contribution of the $\Delta$ at $E_{\nu}=0.5 \GeV$, which increases
with energy.

Events with multiple nucleons in the final state are disregarded in
the right panels of \cref{fig:CC_nucl_tot}. This leads to a very
different behavior, in particular, the cross section without FSI is in
general above the one with FSI included. Comparing both scenarios
shows, that for single-nucleon knockout the $\Delta$ contribution is
smaller than the QE one while it can be larger for the
multiple-nucleon knockout.  Through processes like $\Delta N \to NN$
and $\Delta \to \pi N$ followed by $\pi N \to \pi N$, the $\Delta$
contributes in large parts to the multiple-nucleon knockout.

One important result of our approach is the finding that already at
$E_{\nu} \approx 1\GeV$ a large part of the ejected nucleons stems
from $\Delta$ excitation and/or other processes different from QE, or,
in other words, QE and non-QE processes are ``mixed'' due to FSI. This
is a unique feature of a coupled-channel approach, such as the GiBUU
model, not present in other models. Except for empirical event
generators, we do not know of any other model for neutrino-induced
nucleon knockout which accounts for QE and non-QE scattering
simultaneously. This mixing, however, has major implications for
neutrino-oscillation experiments since it can lead to
misidentification of events as we shall see in the following.

\subsubsection{Charged-current quasi-elastic identification and energy reconstruction}

\begin{figure}[t]
  \centering
  \includegraphics[width=0.4\linewidth]{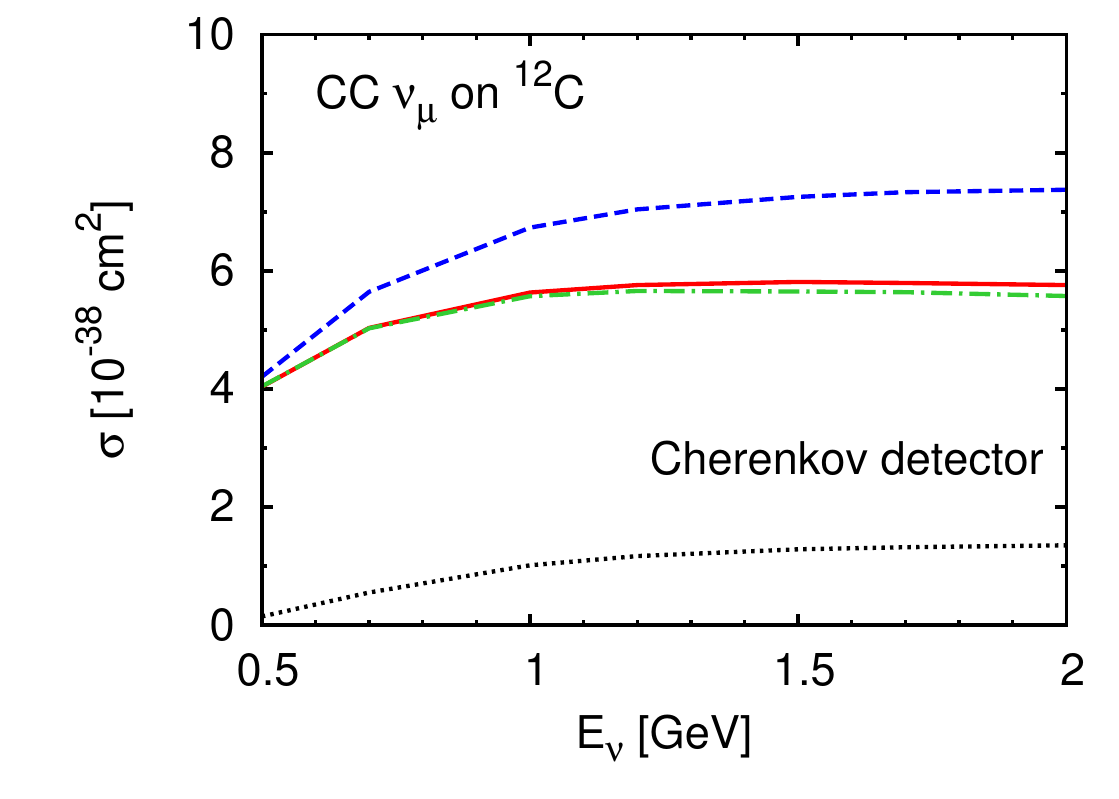}
  \includegraphics[width=0.4\linewidth]{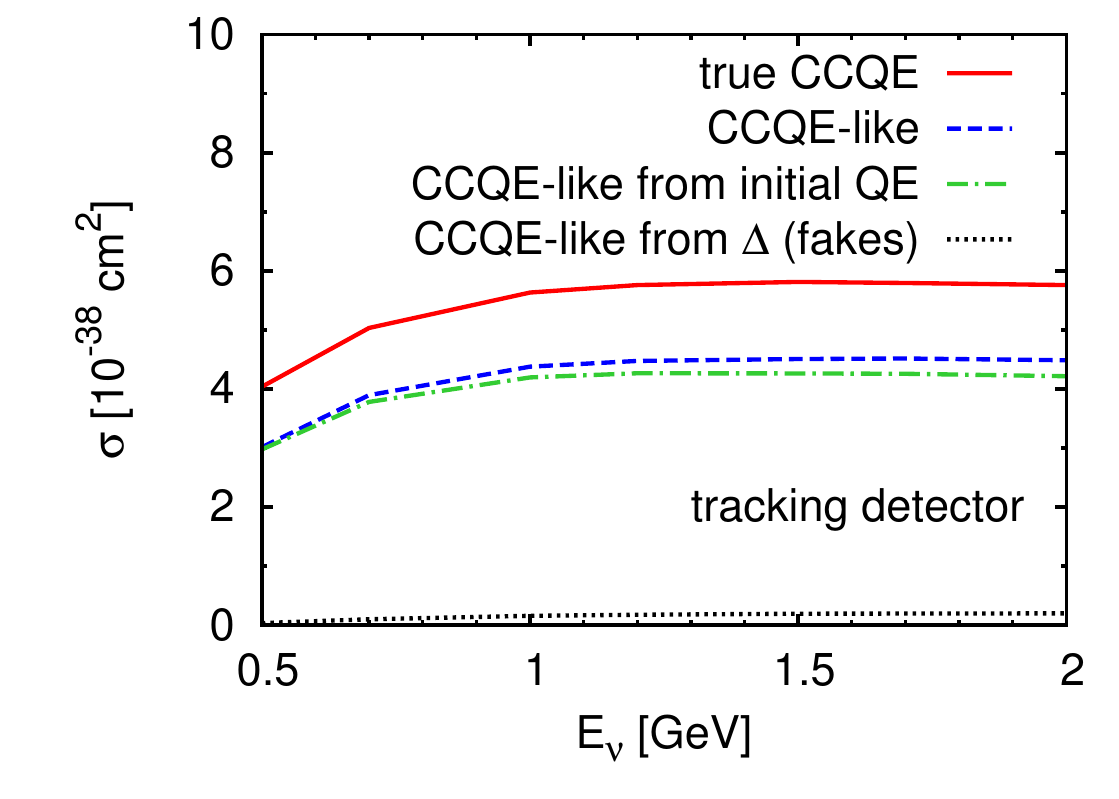}
  \caption{(Color online) Total QE cross section on $\carbon$ (solid
    lines) compared to different methods on how to identify CCQE-like
    events in experiments (dashed lines). The top panel shows the
    method commonly applied in Cherenkov detectors, the lower panel
    the tracking-detector method as described in the text. The
    contributions to the CCQE-like events are also classified
    (CCQE-like from initial QE (dash-dotted), from initial $\Delta$
    (dotted lines)). Experimental detection thresholds are not taken
    into account. Source: Taken from \cite{Leitner:2010kp}.}
  \label{fig:QEmethods}
\end{figure}

The charged-current quasi-elastic (CCQE) reaction, $\nu_\ell n \to
\ell^- p$, being the dominant cross section at low energies, is commonly
used to reconstruct the neutrino energy. In other words, CCQE is the
signal event in present oscillation experiments.

The experimental challenge is to identify \emph{true} CCQE events in
the detector, i.e., muons originating from an initial QE process. To
be more precise: true CCQE corresponds to the inclusive cross
section including all medium effects, or, in other words, the CCQE
cross section before FSI. The difficulty is that the true CCQE events
are masked by FSI in a detector built out of nuclei. The FSI lead to
misidentification of events, e.g., an initial $\Delta$ whose decay
pion is absorbed or which undergoes ``pion-less decay'' contributes to
knock-out nucleons and can thus be counted as CCQE event -- we call
this type of background events ``fake CCQE'' events. We denote every
event which looks like a CCQE event by ``CCQE-like''.

At Cherenkov detectors CCQE-like events are all those, where no pion
is detected, while in tracking detectors CCQE-like events are those,
where a single-proton track is visible and at the same time no pions
are detected.  The two methods are compared in
\cref{fig:QEmethods}. The ``true CCQE'' events are denoted with the
solid lines, the CCQE-like events by the dashed ones. The Cherenkov
detector is able to detect almost all true CCQE (left panel, solid
vs.\ dash-dotted lines agree approximately) but sees also a
considerable amount of ``fake CCQE'' (or ``CC non-QE'') events (left
panel, the dashed line is roughly 20\proz higher than the solid
line). They are caused mainly by initial $\Delta$ excitation as
described in the previous paragraph (absorption of the decay pion or
``pion-less decay''); their contribution to the cross section is given
by the dotted lines. These additional (fake) events have to be removed
from the measured event rates by means of event generators, if one is
interested only in the true QE events. It is obvious that this removal
is the better the more realistic the generator is in handling the
in-medium $\pi$-$N$-$\Delta$ dynamics.

On the contrary, less CCQE-like than true CCQE events are detected
using the method applied in tracking detectors, which trigger both on
pions and protons (right panel, difference between dashed and solid
lines). The final-state interactions of the initial proton lead to
secondary protons, or, via charge exchange, to neutrons which are then
not detected as CCQE-like any more (\emph{single}-proton track). We
find, that at tracking detectors the amount of fake events in the
CCQE-like sample is less than at Cherenkov detectors (dashed and
dash-dotted lines almost agree with each other in the right panel but
not in the left panel). We conclude, that even if the additional cut
on the proton helps to restrict the background, an error of about
20\proz remains, since the measured CCQE cross section underestimates
the true one by that amount. Thus about 20\proz of the total cross
section has to be reconstructed by using event generators. In this
case these generators have to be very realistic in describing the
in-medium nucleon-nucleon interactions. Note that for both detector
types experimental detection thresholds are not yet taken into
account.  Their effect will make the errors even larger
\cite{Leitner:2010kp,Leitner:2010jv}.

The neutrino energy is commonly reconstructed from QE events using a
relation for quasifree scattering on a nucleon at rest,
\begin{equation}
  \label{eq:energy_rec}
  E_\nu^\text{rec} = \frac{2(M_N - E_B)E_\mu - (E_B^2 - 2M_N E_B +
    m_\mu^2)} {2\:[(M_N - E_B) - E_\mu + \abskpr \cos\theta_\mu]},
\end{equation}
with a binding-energy correction of $E_B=34\MeV$
\cite{Aguilar:2007ru}, and the measured muon energy, $E_\mu$, and
scattering angle, $\theta_\mu$.

\begin{figure}[tbp]
  \centering
  \includegraphics[width=0.6\linewidth]{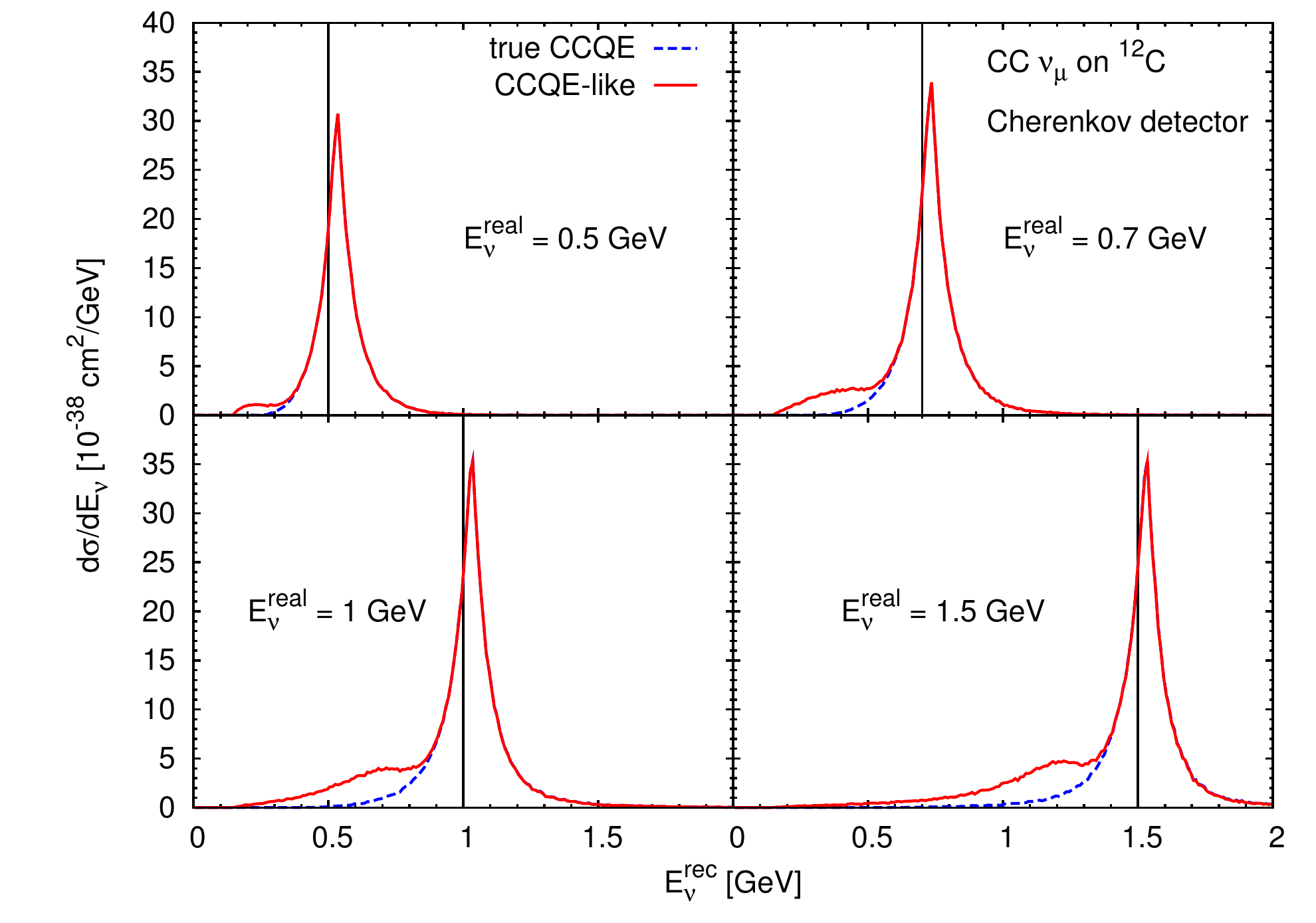}
  \caption{(Color online) Distribution of the reconstructed neutrino
    energy according to \cref{eq:energy_rec} for
    $E_\nu^\text{real}=0.5$, $0.7$, $1.0$, and $1.5\GeV$. The
    reconstructed energy denoted by the dashed lines includes only
    true CCQE events, while the solid lines are obtained by
    reconstructing the energy with CCQE-like events under Cherenkov
    assumptions. Source: Taken from \cite{Leitner:2010kp}.}
  \label{fig:CC_recEnu_Cherenkov}
\end{figure}

In \cref{fig:CC_recEnu_Cherenkov} we plot the distribution of the
reconstructed neutrino energy obtained using \linebreak
\cref{eq:energy_rec} with $E_B=34\MeV$ for four fixed
$E_\nu^\text{real}$ (0.5, 0.7, 1.0, and $1.5\GeV$). The dashed lines
show the true CCQE events only, the solid lines all CCQE-like events
(using the Cherenkov definition but without any threshold cuts). Both
curves show a prominent peak around the real energy, which is slightly
shifted to higher $E_\nu^\text{rec}$. This shift is caused by the
difference between our potential and the specific choice of $E_B$. The
peak has a width of around $100 \MeV$. This broadening is entirely
caused by the Fermi motion of the nucleons -- \cref{eq:energy_rec}
assumes nucleons at rest.

While the distribution of the reconstructed energy for the true CCQE
events is symmetric around the peak, this is not the case for the
CCQE-like distribution. The reconstruction procedure now includes also
CC non-QE events. However, \cref{eq:energy_rec} is entirely based on
the muon kinematics and, in the case of $\Delta$-induced CC non-QE
events, more transferred energy is needed than for true CCQE, thus,
the muon energy is smaller. This lower muon energy then leads to the
second smaller bump at lower reconstructed energies. Thus, the
asymmetry is caused by the CC non-QE events identified as CCQE-like.

We have seen in the previous subsection, that the tracking detector
allows to extract a much cleaner CCQE-like sample than the Cherenkov
detector -- almost no fake, i.e., CC non-QE events spoil the CCQE-like
sample. Consequently, the reconstructed distribution is again
symmetric but at the cost of a lower detection rate (see Fig.~8 in
\cite{Leitner:2010kp}).

\begin{figure}[tbp]
  \centering
  \includegraphics[width=0.6\linewidth]{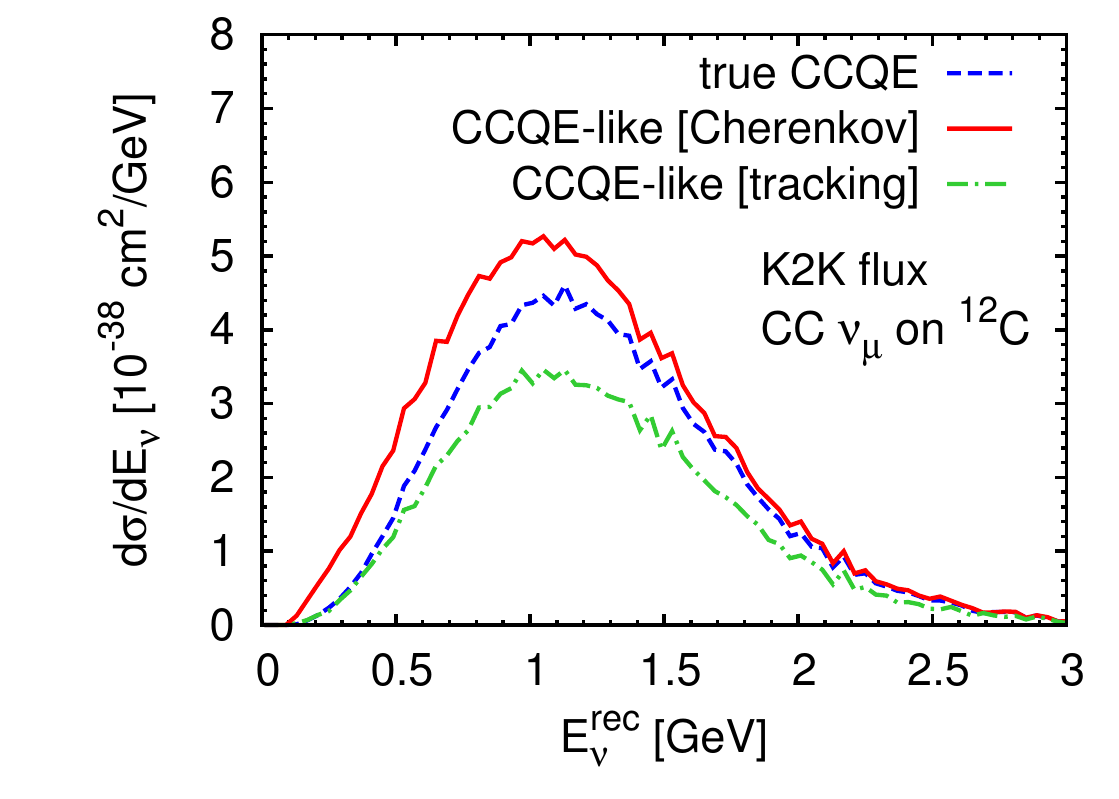}
  \caption{(Color online) Reconstructed energy distribution for the
    K2K flux under different detector
    assumptions. \Cref{eq:energy_rec} is used for the reconstruction,
    but with $E_B$ = 0 (from \cite{Leitner:2010kp}).}
  \label{fig:energyRec_K2Kflux_CC.eps}
\end{figure}

The energy reconstruction influences directly the flux reconstruction
needed for the extraction of neutrino oscillation parameters from
long-baseline experiments which relies on flux com\-pa\-ri\-sons at
the near-side and the far-side detectors. The uncertainties in the
flux related to the energy reconstruction are illustrated in
\cref{fig:energyRec_K2Kflux_CC.eps}.
The figure shows clearly that in particular the Cherenkov-type
detectors lead to a downward shift of the energy distribution which
has to be removed by event generators before a flux comparison can be
done.

To conclude, we have shown that a correct identification of CCQE events
is relevant for the neutrino energy reconstruction and thus for the
oscillation result. A significant part of CC $1\pi^+$ events is detected
as CCQE-like, which is mainly caused by the pion absorption in the
nucleus. This has to be corrected by means of event generators, which is
why the final experimental cross sections contain a significant model
dependence.  Thus, because of the close entanglement of CCQE and CC
$1\pi^+$ on nuclei, both these channels have to be accounted for equally
precise by any model aiming at describing the experimental measurements,
in particular the directly observable rates for nucleon knockout and
$1\pi^+$ production.

In this regard, the GiBUU is a flexible model of nuclear effects and
is thus uniquely suited for a description of broad-band neutrino
experiments. The requirements formulated by Benhar
\etal{}~\cite{Benhar:2010nx} for a new paradigm in neutrino-nucleus
scattering are all met. The GiBUU model can also accommodate the
recently much discussed effects of $2p2h$ initial excitations that are
not contained in the impulse approximation (see \cite{Nieves:2011pp}
and refs. therein). Following the method discussed in \cref{Sect:3b}
for the case of pion-induced reactions such excitations could easily
be implemented.

\section{Summary}
\label{sec:summary}

In this article we have given a comprehensive, detailed discussion of
transport theory and its practical implementation in the GiBUU
transport model. It has been our aim to give all the essential
ingredients of this model so that GiBUU is not a black box to other
theorists or experimenters using it for planning or analyzing an
experiment.

GiBUU is unique in the sense that it uses the same physics input for
very different nuclear reactions and is not a model of many-body
physics optimized for one particular reaction or energy range. For
example, the very same theory and numerical implementation of $\Delta
N \pi$ dynamics is used both for heavy-ion collisions and for
neutrino-induced reactions on nuclei.  As another example, it is able
to give a reasonable description of quasi-elastic scattering off nuclei
as well as of pion production through resonance excitations and is
thus, for example, an ideal tool for an analysis of broad-energy band
neutrino long-baseline experiments that necessarily average over many
reaction types. The primary strength of GiBUU lies in the description
of final-state interactions and here again it can be used both for
heavy-ion and for electroweak reactions. Many important analyses of
fundamental physics questions, such as short-range correlation studies
from $(e,e' p)$ reactions on nuclei, the search for color transparency
in electron- or pion-induced reactions or the investigation of
in-medium changes of hadron-spectral properties depend crucially on a
precise treatment of final-state interactions. The often used Glauber
approximation in its most simple form only describes attenuation of
primarily produced particles on straight-line trajectories. However,
in reality the final, observed particles outside the nucleus, that
reach the detector, may not be the ones that were originally hit by
the incoming probe. They can change their identity, and they can also
change their energy while traveling through the nucleus. All these effects
are not included in the widely used Glauber attenuation
treatment, but are contained in a transport theoretical description.

On the other hand, transport theory also has its limitations. Fully
exclusive or coherent particle production reactions on nuclei can only
be described by quantum-mechanical methods since here the phase
relations between the interaction amplitudes for all the nucleons in
the nuclear target are essential so that transport theory must
fail. However, for inclusive and semi-inclusive reactions on nuclei,
where the phase-coherence between various reaction steps is lost due
to multiple scattering, transport theory is the theory of choice. For
fully inclusive reactions it gives cross sections that are as good as
those obtained from quantum-mechanical methods, such as distorted-wave
approximations or optical-model descriptions. For semi-inclusive
reactions it is, however, superior to the latter ones since it does
not only describe the loss of flux out of a given channel of interest,
but also tells where this flux goes.

In the present article we have first considered the general
theoretical framework starting from nonequilibrium nuclear many-body
dynamics and leading to the off-shell transport equations that are
actually being solved. From this discussion it is clear that even
after so many years the Kadanoff-Baym equations have not been fully
solved, and the back flow term still presents a problem. However, there
have been physics-motivated approaches around this problem, that
lead to the actual transport equations. We have then given all the
details on the mean fields and the collision terms being used in many
of the existing codes. One problem that arises in using self-energies
defined only in the local rest frame, such as those provided by the
Skyrme force, is that of handling the necessary
Lorentz-transformations in a `clean' way. The use of a relativistic
mean-field theory with its well-defined Lorentz-transformation
properties of all involved potentials avoids this problem, but the
presently available relativistic mean-field theory still suffers from the problem of
leading to too repulsive nucleon-nucleus potentials at large
energies. This is clearly a point that deserves more attention in the
future.

We have then presented a broad range of physics illustrations for
results obtained with GiBUU covering a wide range of reaction types
and energies. Common to all of them is that final-state interactions
play a major role. In particular, if one is after physically
interesting and challenging questions such as, e.g.,\ color
transparency or in-medium properties of hadrons, to name just two, one
has to treat the final-state interactions at the same state-of-the-art
level as the actual physics process to arrive at the final
observables. This is still not so in many theoretical studies of such
effects, which employ state-of-the-art QCD-based methods to calculate
the primary physics effects in an idealized environment and then use
much less sophisticated methods to describe the ever-present
final-state interactions. GiBUU provides a method to include the best
presently available treatment of final-state interactions into any
such calculations. We have illustrated that with a broad range of
applications and comparisons to experiment. To name just a few,
examples ranged from antiproton reactions on nuclei over nuclear
fragmentation in heavy-ion collisions to studies of electroweak
interactions with nuclei, including hadronization and the search for
color transparency. This, for example, makes GiBUU a tool well suited for
investigations of $e+A$ reactions at the electron-ion collider which
is now being planned as one of the new machines for the next decade
\cite{EIC}.  The GiBUU method and code has been explained in quite
some detail in this article. The actual code is publicly available
from \cite{gibuu}.

Future developments go into the direction of hybrid codes that contain
both hadronic and partonic degrees of freedom. For ultrarelativistic
heavy-ion reactions this is necessary to pin down observable
consequences of the formation of a quark-gluon plasma phase during the
collision and thus to verify its production. First attempts in this
direction have already been undertaken
\cite{Petersen:2008dd,Lin:2004en,Cassing:2008sv}.  For electroweak
reactions at high energies the observation of produced hadrons can
give valuable information on production, formation and expansion times
of prehadrons or hadrons formed out of the struck quarks inside the
nuclear medium. Thus, both degrees of freedom, hadronic and partonic
ones, have to be described and their time-development has to be
investigated also in the more microscopic reactions. Such work has
only very recently started \cite{Kaskulov:2008ej,Gallmeister:2007an};
transport theory is the only known tool to gain insight into these
questions. We foresee future applications for GiBUU in
antiproton-induced hypernucleus formation at PANDA@FAIR, in
hadronization and color transparency studies at JLAB at 12\GeV and in $e+A$ reactions
at the EIC \cite{EIC}. In addition, the understanding of the detector response in
the new class of long-baseline neutrino experiments will be an
interesting application.

\section*{Acknowledgment}

The authors gratefully acknowledge a very close collaboration with Luis
Alvarez-Ruso (Valencia) on various aspects of resonance properties and
photonuclear and electroweak processes on nuclei. We are also indebted
to Stefan Leupold (Uppsala) for extremely helpful discussions over many
years about various aspects of off-shell transport theory and in-medium
effects. The close contact with Volker Metag (Giessen) on experiments on
photoproduction of mesons and his many suggestions have always been
extremely stimulating. Finally, the authors wish to thank J\"orn Knoll (Darmstadt)
for a very careful reading of the manuscript and many helpful, detailed suggestions for 
presenting the material in this paper.

This work has been  supported by Deutsche Forschungsgemeinschaft, BMBF and the HIC for
FAIR under the LOEWE program. We also gratefully acknowledge support by the Frankfurt Center for
Scientific Computing.

\settocdepth{2} % Keine Unterkapitel von Anhaengen im Inhaltsverzeichnis

\clearpage
\section*{Appendix}
\appendix

In order to make the GiBUU model as transparent as possible and to
give the interested reader and/or practitioner sufficient information
about the 'inner workings' of the model and code we give in the
following extensive paragraphs essential details for the particles
implemented in the actual event description and for the decay rates
and cross sections entering the collision term.  In this way we want
to make sure that GiBUU as a method and code is not a 'black box' to
the user, but that instead the microscopic input is known. We also
give some details about the numerical implementation.

\section{Properties and decay channels of implemented baryons and
  mesons}
\label{sec:particleproperties}

Herein, we list all particles considered in the GiBUU model with their
masses, widths, quantum numbers and decay modes.  Only the strong and,
in some cases, electromagnetic decay modes are taken into account. The
weak decay modes are neglected and the corresponding widths are set to
zero. Also the dilepton decays are usually neglected, but can be
treated perturbatively if required (cf.~\cref{sec:dileptons}).

The properties of the thirty-one non-strange/non-charm resonances
included in the code are listed in \cref{tab:baryon_properties}
together with their decay channels.  These properties are taken from
the partial wave analysis of $\pi N$ amplitudes of
ref.~\cite{ManleySaleski}.  We use the common notation which reads
$l_{ij}$ with the spin $J=\frac{j}{2}$ and isospin $I=\frac{i}{2}$,
while $l$ denotes the relative angular momentum of the $\pi N$ system.
The parity follows from $P=(-1)^{l+1}$.  By default, all the baryonic
resonances listed in \cref{tab:baryon_properties} except for the
$I=\frac{1}{2}$ resonances with one-star (*) rating are allowed to be
created in the meson-baryon collisions.

The properties of the baryons with strangeness and charm quantum
numbers are collected in
\cref{tab:baryon_properties_strange,tab:baryon_properties_strange_charm}.
The masses, widths and quantum numbers are taken from the PDG
analysis \cite{Hagiwara:2002fs}. The most branching
ratios are also adopted from \refcite{Hagiwara:2002fs}. The branching
ratios not given in \cite{Hagiwara:2002fs} (e.g.~$\Sigma^* \pi$ for
$\Lambda(1520)$, $\Lambda(1690)$, $\Sigma(1775)$ and $\Lambda(2100)$)
have been estimated or set to saturate the sum of the branching ratios
to 100\proz{}.

The mesons and their properties are listed in
\cref{tab:meson_properties}. Their quantum numbers and the most
important decay channels with their branching ratios are given
explicitly. The meson properties are defined according to the PDG
analysis \cite{Hagiwara:2002fs}.

\begin{table}[tbp]
\begin{small}
\begin{tabular}{l l c c r r r r r r r r r} 
  \toprule
  && $M_B$&$\Gamma_0$ &  \multicolumn{9}{c} {branching ratio in \%}\\
                    &      &[MeV] & [MeV] & $N\pi$ & $N\eta$ & $N\omega$ & $K\Lambda$ &
  $\Delta\pi$ & $N\rho$ & $N\sigma$ & $N^*\pi$ & $\Delta\rho$\\
  \midrule
  N(938)            &  ****& 938  &  0  &  --- & --- & --- & --- & --- & --- & --- & --- &---\\
  \res{P}{33}{1232} &  ****& 1232 & 118 &  100 & --- & --- & --- & --- & --- & --- & --- &---\\  
  \res{P}{11}{1440} &  ****& 1462 & 391 &  69  & --- & --- & --- &  22 & --- &  9  & --- &---\\  
  \res{S}{11}{1535} &   ***& 1534 & 151 &  51  &  43 & --- & --- & --- &  3  &  1  &  2  &---\\  
  \res{S}{11}{1650} &  ****& 1659 & 173 &  89  &  3  & --- & --- &  2  &  3  &  2  &  1  &---\\  
  \res{S}{11}{2090} &     *& 1928 & 414 &  10  & --- & --- & --- &  6  &  49 &  5  &  30 &---\\  
  \res{D}{13}{1520} &  ****& 1524 & 124 &  59  & --- & --- & --- &  20 &  21 & --- & --- &---\\  
  \res{D}{13}{1700} &     *& 1737 & 249 &  1   & --- & --- & --- &  84 &  13 &  2  & --- &---\\  
  \res{D}{13}{2080} &     *& 1804 & 447 &  23  & --- & --- & --- &  24 &  26 &  27 & --- &---\\  
  \res{D}{15}{1675} &  ****& 1676 & 159 &  47  & --- & --- & --- &  53 & --- & --- & --- &---\\  
  \res{G}{17}{2190} &  ****& 2127 & 547 &  22  & --- &  49 & --- & --- &  29 & --- & --- &---\\  
  \res{P}{11}{1710} &     *& 1717 & 478 &  9   & --- & --- &  37 &  49 &  3  &  2  & --- &---\\  
  \res{P}{11}{2100} &     *& 1885 & 113 &  15  & --- & --- &  2  &  24 &  27 &  32 & --- &---\\  
  \res{P}{13}{1720} &     *& 1717 & 383 &  13  & --- & --- & --- & --- &  87 & --- & --- &---\\  
  \res{P}{13}{1900} &   ***& 1879 & 498 &  26  & --- &  30 & --- & --- &  44 & --- & --- &---\\  
  \res{F}{15}{1680} &  ****& 1684 & 139 &  70  & --- & --- & --- &  11 &  7  &  12 & --- &---\\  
  \res{F}{15}{2000} &     *& 1903 & 494 &  8   & --- & --- & --- &  12 &  75 &  5  & --- &---\\  
  \res{F}{17}{1990} &    **& 2086 & 535 &  6   & 94  & --- & --- & --- & --- & --- & --- &---\\  
  \res{S}{31}{1620} &    **& 1672 & 154 &  9   & --- & --- & --- &  62 &  29 & --- & --- &---\\  
  \res{S}{31}{1900} &   ***& 1920 & 263 &  4   & --- & --- & --- &  16 &  38 & --- &  6  &---\\  
  \res{D}{33}{1700} &     *& 1762 & 599 &  14  & --- & --- & --- &  78 &  8  & --- & --- &---\\  
  \res{D}{33}{1940} &     *& 2057 & 460 &  18  & --- & --- & --- &  47 &  35 & --- & --- &---\\  
  \res{D}{35}{1930} &    **& 1956 & 526 &  18  & --- & --- & --- & --- & --- & --- & --- & 82\\  
  \res{D}{35}{2350} &    **& 2171 & 264 &  2   & --- & --- & --- & --- & --- & --- & --- & 98\\  
  \res{P}{31}{1750} &     *& 1744 & 299 &  8   & --- & --- & --- & --- & --- & --- &  28 & 64\\  
  \res{P}{31}{1910} &  ****& 1882 & 239 &  23  & --- & --- & --- & --- &  10 & --- &  67 &---\\  
  \res{P}{33}{1600} &   ***& 1706 & 430 &  12  & --- & --- & --- &  68 & --- & --- &  20 &---\\  
  \res{P}{33}{1920} &     *& 2014 & 152 &  2   & --- & --- & --- &  83 & --- & --- &  15 &---\\  
  \res{F}{35}{1750} &     *& 1752 & 251 &  2   & --- & --- & --- &  76 &  22 & --- & --- &---\\  
  \res{F}{35}{1905} &   ***& 1881 & 327 &  12  & --- & --- & --- &  1  &  87 & --- & --- &---\\  
  \res{F}{37}{1950} &  ****& 1945 & 300 &  38  & --- & --- & --- &  18 & --- & --- & --- & 44\\  
  \bottomrule
\end{tabular}
\caption{Properties and decay channels for the non-strange/non-charm
    baryons. Given are the rating, the Breit-Wigner mass $M_B$, the width $\Gamma_0$ at
    $M_B$ and the branching ratios into the various decay channels. 
    Adopted from ref. \cite{ManleySaleski}.}
    \label{tab:baryon_properties}
\end{small}
\end{table}

\begin{table}[tbp]
\centering
\begin{tabular}{l l l l r r r r r r r r} 
  \toprule
  & & $M_B$ & $\Gamma_0$ & $J$ & \multicolumn{7}{c} {branching ratio in \%}\\
                  & & [MeV] & [MeV] & & $\Lambda \pi$ & $N \bar K$ & $\Sigma \pi$ & $\Sigma^* \pi$ &
  $\Lambda \eta $ & $N \bar K^*$ & $\Lambda^* \pi $\\
  \midrule
  $\Lambda(1116)$  &  ****& 1116  &  0   & 1/2 &--- &--- &--- &--- & ---&--- &---\\
  $\Sigma(1189)$   &  ****& 1189  &  0   & 1/2 &--- &--- &--- &--- & ---&--- &---\\
  $\Sigma(1385)$   &  ****& 1385  & 36   & 3/2 & 88 &--- & 12 &--- & ---&--- &---\\
  $\Lambda(1405)$  &  ****& 1405  & 50   & 1/2 &--- &--- & 100&--- & ---&--- &---\\
  $\Lambda(1520)$  &  ****& 1520  & 16   & 3/2 &--- & 46 & 43 & 11 & ---&--- &---\\
  $\Lambda(1600)$  &   ***& 1600  & 150  & 1/2 &--- & 35 & 65 &--- & ---&--- &---\\
  $\Sigma(1660)$   &   ***& 1660  & 100  & 1/2 & 40 & 20 & 40 &--- & ---&--- &---\\
  $\Lambda(1670)$  &  ****& 1670  & 35   & 1/2 &--- & 25 & 45 &--- & 30 &--- &---\\
  $\Sigma(1670)$   &  ****& 1670  & 60   & 3/2 & 15 & 15 & 70 &--- &--- &--- &---\\
  $\Lambda(1690)$  &  ****& 1690  & 60   & 3/2 &--- & 25 & 30 & 45 &--- &--- &---\\
  $\Sigma(1750)$   &   ***& 1750  & 90   & 1/2 & 10 & 30 & 60 &--- &--- &--- &---\\
  $\Sigma(1775)$   &  ****& 1775  & 120  & 5/2 & 20 & 45 & 5  & 10 &--- &--- & 20\\
  $\Lambda(1800)$  &   ***& 1800  & 300  & 1/2 &--- & 35 & 35 & 30 &--- &--- &---\\
  $\Lambda(1810)$  &   ***& 1810  & 150  & 1/2 &--- & 35 & 20 &--- &--- & 45 &---\\
  $\Lambda(1820)$  &  ****& 1820  & 80   & 5/2 &--- & 60 & 12 & 28 &--- &--- &---\\
  $\Lambda(1830)$  &  ****& 1830  & 95   & 5/2 &--- & 5  & 60 & 35 &--- &--- &---\\
  $\Lambda(1890)$  &  ****& 1890  & 100  & 3/2 &--- & 30 & 10 & 30 &--- & 30 &---\\
  $\Sigma(1915)$   &  ****& 1915  & 120  & 5/2 & 45 & 10 & 45 &--- &--- &--- &---\\
  $\Sigma(2030)$   &  ****& 2030  & 180  & 7/2 & 25 & 25 & 10 & 15 &--- & 5  & 20\\
  $\Lambda(2100)$  &  ****& 2100  & 200  & 7/2 &--- & 30 & 5  & 45 &--- & 20 &---\\
  $\Lambda(2110)$  &   ***& 2110  & 200  & 5/2 &--- & 15 & 30 &--- &--- & 55 &---\\
  \bottomrule
\end{tabular}
\caption{Properties and decay channels for the baryons with
    strangeness $S=-1$. Given are the rating, the Breit-Wigner mass
    $M_B$, the width $\Gamma_0$ at $M_B$, the spin $J$ and the branching
    ratios into the various decay channels. 
    Based on the PDG review \cite{Hagiwara:2002fs}.}
    \label{tab:baryon_properties_strange}
\end{table}

\begin{table}[tbp]
\centering
\begin{tabular}{l l l l l r r l} 
  \toprule
  & & $M_B$ & $\Gamma_0$       & I(J$^P$)  & S  & C & decay mode  \\
  & & [MeV] & [MeV]            &          &    &   & \\
  \midrule
  $\Xi$        &  ****& 1315  &0         & $\frac{1}{2}({\frac{1}{2}}^+)$  & -2 &0& ---\\
  $\Xi^\star$  &  ****& 1530  & 9.5        & $\frac{1}{2}({\frac{3}{2}}^+)$  & -2 &0& $\Xi\,\pi$ \\
  $\Omega$     &  ****& 1672  &0         & $0({\frac{3}{2}}^+)$            & -3 &0& ---\\
  $\Lambda_c$  &  ****& 2285  &0         & $0({\frac{1}{2}}^+)$            &0 & 1 & ---\\
  $\Sigma_c$   &  ****& 2452  &0         & $1({\frac{1}{2}}^+)$            &0 & 1 & ---\\
  $\Sigma_c^*$ &  ****& 2520  & 15         & $1({\frac{3}{2}}^+)$            &0 & 1 & $\Lambda_c\,\pi$\\
  $\Xi_c$      &  ***& 2466  &0         & $\frac{1}{2}({\frac{1}{2}}^+)$  & -1 & 1 & ---\\
  $\Xi^*_c$    &  ***& 2645  & 4          & $\frac{1}{2}({\frac{3}{2}}^+)$  & -1 & 1 & $\Xi_c\,\pi$\\
  $\Omega_c$   &  ***& 2698  &0         & $0({\frac{1}{2}}^+)$            & -2 & 1 & ---\\
  \bottomrule
\end{tabular}
\caption{Properties and decay channels for the baryons with
    strangeness $S<-1$ and charm $C=1$. Given are the rating, the
    Breit-Wigner mass, $M_B$, the width $\Gamma_0$ at $M_B$, the
    quantum numbers and decay channels. 
    Taken from the PDG review \cite{Hagiwara:2002fs}.}
    \label{tab:baryon_properties_strange_charm}
\end{table}

\begin{table}[tbp]
  \centering
  \begin{tabular}{l c l l l r r l} 
    \toprule
    & $m_m$[MeV] & $\Gamma_0$ [MeV]   & $J$ & $I$ & $S$ & $C$ & decay channels \\
    \midrule
    $\pi$         & 138  &0                 &0&   1 & 0&  0& \\
    $\eta$        & 547  & $1.2\cdot 10^{-3}$ &0&  0& 0&  0& $\gamma \gamma$ (40\%), $\pi^+ \pi^- \pi^0$ (28\%), $3\pi^0$ (32\%) \\
    $\rho$        & 770  & 151                & 1 &   1 & 0&  0& $\pi \pi$\\
    $\sigma$      & 800  & 800                &0&  0& 0&  0& $\pi \pi$\\
    $\omega$      & 782  & 8.4                & 1 &  0& 0&  0& $\pi \pi$ (2\%), $\pi^0 \gamma$ (9\%), $\pi^+ \pi^- \pi^0$ (89\%)\\
    $\eta^\prime$ & 958  & 0.2                &0&  0& 0&  0& $\rho^0 \gamma$ (31\%), $\pi \pi \eta$ (69\%) \\
    $\phi$        & 1020 & 4.4                & 1 &  0& 0&  0& $\rho \pi$ (13\%), $K \bar{K}$ (84\%), $\pi^+ \pi^- \pi^0$ (3\%)\\
    $K$           & 496  &0                 &0& 1/2 &  1 &  0& \\
    $\bar{K}$     & 496  &0                 &0& 1/2 & -1 &  0& \\
    $K^*$         & 892  & 50                 & 1 & 1/2 &  1 &  0& $K \pi$ \\
    $\bar{K}^*$   & 892  & 50                 & 1 & 1/2 & -1 &  0& $\bar{K} \pi$ \\
    $\eta_c$      & 2980 &0                 &0&0  &0 &  0& \\
    $J/\Psi$      & 3097 &0                 & 1 &0  &0 &  0& \\
    $D$           & 1867 &0                 &0& 1/2 &0 &   1 & \\
    $\bar{D}$     & 1867 &0                 &0& 1/2 &0 &  -1 & \\
    $(D^*)^0$     & 2007 & 2                  & 1 & 1/2 &0 &   1 & $D\, \gamma$ (38\%), $D\,\pi$ (62\%)\\
    $(D^*)^+$     & 2007 & $96\times 10^{-3}$ & 1 & 1/2 &0 &   1 & $D\, \gamma$ (2\%), $D\,\pi$ (98\%)\\
    $(\bar{D}^*)^0$ & 2007 & 2                  & 1 & 1/2 &0 &  -1 & $\bar{D}\, \gamma$ (38\%), $\bar{D}\,\pi$ (62\%)\\
    $(\bar{D}^*)^-$ & 2007 & $96\times 10^{-3}$ & 1 & 1/2 &0 &  -1 & $\bar{D}\, \gamma$ (2\%), $\bar{D}\,\pi$ (98\%) \\
    $D_s$         & 1969 &0                 &0&0  & 1  &   1 & \\
    $\bar{D}_s$   & 1969 &0                 &0&0  & -1 &  -1 & \\
    $D_s^*$       & 2112 & 1                  &0&0  & 1  &   1 & $D_s\,\gamma$ (95\%), $D_s\,\pi$ (5\%)\\
    $\bar{D}_s^*$ & 2112 & 1                  &0&0  & -1 &  -1 & $\bar{D}_s\,\gamma$ (95\%), $\bar{D}_s\,\pi$ (5\%)\\
    \bottomrule
  \end{tabular}
\caption[Properties of the light mesons]{Properties and decay channels for the light mesons. 
    Given are the pole mass $m_m$, the width $\Gamma_0$ at $m_m$, spin
    $J$, isospin $I$, strangeness $S$, and the branching ratios into the various decay channels. 
    Based on the PDG review \cite{Hagiwara:2002fs}.}
    \label{tab:meson_properties}
\end{table}

\subsection{Dilepton decays}
\label{sec:dileptons}

In the GiBUU model the following dilepton decay modes are taken into
account:
\begin{itemize}
\item direct decays, as $V \rightarrow e^+e^-$ with $V=\rho^0$,
  $\omega, $ $\phi$, or $\eta \rightarrow e^+e^-$,
\item Dalitz decays, as $P \rightarrow e^+e^-\gamma$ with
  $P=\pi^0,\eta$, $\omega \rightarrow \pi^0e^+e^-$, or $\Delta
  \rightarrow Ne^+e^-$
\end{itemize}
Most of them are treated similarly as in \refcite{effe_phd}. The
leptonic decay widths of the vector mesons are calculated under the
assumption of strict vector-meson dominance (VMD),
\begin{equation*}
  \Gamma_{V\rightarrow e^+e^-}(\mu)=C_V\frac{m_V^4}{\mu^3},
\end{equation*}
with the constants, $C_V$, listed in \cref{tab:vm_dil}.

\begin{table}[bt]
  \begin{center}
    \begin{tabular}{cccccc}
      \toprule
      $V$ & $m_V (\MeV)$ & $\Gamma_{ee} (\keV)$ & $C_V=\Gamma_{ee}/m_V$ \\
      \midrule
      $\rho$   & 775.49   & 7.04 & $9.078\cdot10^{-6}$ \\
      $\omega$ & 782.65   & 0.60 & $7.666\cdot10^{-7}$ \\
      $\phi$   & 1019.455 & 1.27 & $1.246\cdot10^{-6}$\\
      \bottomrule
    \end{tabular}
  \end{center}
  \caption{Dilepton-decay constants for $V\rightarrow e^+e^-$.}
  \label{tab:vm_dil}
\end{table}
While the direct decay of the $\eta$ meson into a $\mu^+\mu^-$ pair
has been observed, for the corresponding $e^+e^-$ decay only an upper
limit of $\mbox{BR}(\eta\rightarrow e^+e^-)<2.7\cdot10^{-5}$ is known
\cite{Berlowski:2008zz}. However, the theoretical expectation from
helicity suppression is still four orders of magnitude lower
\cite{Browder:1997eu}.  The Dalitz decays of the pseudoscalar mesons,
$P=\pi^0,\eta,\eta'$, are treated via the parametrization
\cite{Landsberg:1986fd},
\begin{equation}
  \frac{\dd \Gamma_{P\rightarrow\gamma e^+e^-}}{\dd \mu}=\frac{4\alpha}{3\pi}
  \frac{\Gamma_{P\rightarrow\gamma\gamma}}{\mu}
  \left(1-\frac{\mu^2}{m_P^2}\right)^3 |F_P(\mu)|^2,
\end{equation}
with $\Gamma_{\pi^0\rightarrow\gamma\gamma}=7.8\cdot10^{-6}\MeV$ and
$\Gamma_{\eta\rightarrow\gamma\gamma}=4.6\cdot10^{-4}\MeV$ and the
form factors,
\begin{alignat}{3}
  F_{\pi^0}(\mu) & = 1 + b_{\pi^0}\mu^2, \quad & b_{\pi^0}=5.5\GeV^{-2}, \\
  F_{\eta}(\mu) & = \left(1-\frac{\mu^2}{\Lambda_\eta^2}\right)^{-1},
  \quad & \Lambda_\eta=0.676\GeV.
\end{alignat}
The above value of $\Lambda_\eta$ has been recently determined from
HADES data \cite{spruck_phd}. It should be noted that the form factors
of the $\pi^0$ and $\eta$ Dalitz decays are sufficiently constrained
by data, while the experimental constraints of the $\eta'$ form factor
are much weaker \cite{Landsberg:1986fd}. A VMD form factor for the
$\eta'$ Dalitz decay can be found for example in
\cite{Terschluesen:2010ik}.  The parametrization of the $\omega$
Dalitz decay,
\begin{equation}
  \begin{aligned}
    \frac{\dd \Gamma_{\omega\rightarrow\pi^0e^+e^-}}{\dd \mu} & =
    \frac{2\alpha}{3\pi}\frac{\Gamma_{\omega\rightarrow\pi^0\gamma}}{\mu}
    \left[ \left(1+\frac{\mu^2}{\mu_\omega^2-m_\pi^2}\right)^2
      -\frac{4\mu_\omega^2\mu^2}{(\mu_\omega^2-m_\pi^2)^2}
    \right]^{3/2}
    |F_\omega(\mu)|^2, \\
    |F_\omega(\mu)|^2 & =
    \frac{\Lambda_\omega^4}{(\Lambda_\omega^2-\mu^2)^2+\Lambda_\omega^2\Gamma_\omega^2},
  \end{aligned}
\end{equation}
is adopted from \cite{Bratkovskaya:1996qe} with
$\Gamma_{\omega\rightarrow\pi^0\gamma}=0.703\MeV$,
$\Lambda_\omega=0.65\GeV$ and $\Gamma_\omega=75\MeV$.  We note here
that the form factor of the $\omega$ Dalitz decay is also
well-constrained by data \cite{:2009wb}.  For the $\Delta$-Dalitz
decay we use the parametrization from \refcite{Krivoruchenko:2001hs},
\begin{equation}
  \begin{aligned}
    \frac{\dd\Gamma_{\Delta\rightarrow Ne^+e^-}}{\dd\mu} & = \frac{2\alpha}{3\pi\mu}\Gamma_{\Delta\rightarrow N\gamma^*}, \\
    \Gamma_{\Delta\rightarrow N\gamma^*} & =
    \frac{\alpha}{16}\frac{(m_\Delta+m_N)^2}{m_\Delta^3m_N^2}
    \left[(m_\Delta+m_N)^2-\mu^2\right]^{1/2}
    \left[(m_\Delta-m_N)^2-\mu^2\right]^{3/2} |F_\Delta(\mu)|^2,
  \end{aligned}
\end{equation}
where we neglect the electron mass. The electromagnetic N-$\Delta$
transition form factor $F_\Delta(\mu)$ is an issue of ongoing
debate. Unlike the other semileptonic Dalitz decays, it is poorly
constrained by data. At least at the real-photon point ($\mu=0$) it is
fixed by the decay $\Delta\rightarrow N\gamma$ to
$|F_\Delta(0)|=3.029$, and also in the space-like region this form
factor is well-constrained by electron scattering data on the
nucleon. However, this form factor is basically unknown in the
time-like regime, which is being probed by the $\Delta$ Dalitz decay.

From the theoretical side, many parametrizations are available for the
space-like part, but most of them are not applicable in the time-like
region. One of the few models which take care of the continuation to
the time-like region is the two-component quark model given in
\cite{Wan:2005ds}.

\section{Cross sections}

The cross sections are mostly given by some suitable fits to the
experimental data. In some cases, e.g., for the $NN \to N\Delta$
process, a simple one-boson exchange model is applied. The
meson-baryon collisions are dominated by the intermediate resonance
excitation. At high invariant energies, we rely mostly on the
\Pythia{}~\cite{Sjostrand:2006za} and \Fritiof{}~\cite{Pi:1992ug}
event generators.  In the following, we describe in detail the cross
sections used in the GiBUU collision term.

\subsection{Baryon-baryon cross sections}
\label{gibuu_barBar_xsections}

At $\sqrt{s} > 2.6\GeV$, a baryon-baryon collision is considered as a
highly-energetic one in our model\footnote{See \cref{matching} for
  more precise definition of the low- and high-energy two-body
  collisions.}. In this case, the simulation of elastic collision
events is done according to the cross section parametrization of
\refcite{PhysRevD.50.1173} fitted to the pp data,
\begin{equation}
  \sigma_\text{el} =
  11.9 + 26.9p_\text{lab}^{-1.21} + 0.169\ln^2(p_\text{lab}) - 1.85\ln(p_\text{lab})~,
\end{equation}
where the beam momentum $p_\text{lab}$ is given in $\GeVc$, and the
cross section in mb.  The angular distribution for the elastic
scattering events is chosen as
\begin{equation}
  \label{dsigdt}
  \frac{\dd\sigma_\text{el}}{\dd t} \propto \exp(bt)
\end{equation}
with the energy dependent slope parameter $b=5.0+4s^{0.0808}$ taken
from the \Pythia{} model (see also \refcite{Falter:2004uc}), where $s$
is given in $\GeV{}^2$ and $b$ in $\GeV{}^{-2}$.  The inelastic
collision events are simulated with the help of a \Pythia{} event
generator. The cross section for the inelastic events is given as the
difference $\sigma_\text{tot}-\sigma_\text{el}$ with the total cross
section parametrized in \refcite{PhysRevD.50.1173} to the pp data,
\begin{equation}
  \sigma_\text{tot}= 48.0 + 0.522\ln^2(p_\text{lab})  - 4.51\ln(p_\text{lab})~,
\end{equation}
where $p_\text{lab}$ is in $\GeVc$, and $\sigma_\text{tot}$ in mb.

At the lower invariant energies $\sqrt{s}$, we take into account the
baryon-baryon (BB) elastic scattering or charge exchange (CEX) $BB \to
BB$, resonance (R), and double-$\Delta$ production and absorption in
nucleon-nucleon (NN) collisions $NN \leftrightarrow NR$, $NN
\leftrightarrow \Delta \Delta$, as well as the (direct) pion
production and absorption $NN \leftrightarrow NN\pi$.  Also the
strangeness production in the processes $BB \to K Y B$ and $BB \to N N
K \bar K$ is included in the model. The explicit low-energy BB
collision channels are listed below.

\subsubsection*{\texorpdfstring{${NN \leftrightarrow NN}$}{NN -> NN}}

The cross sections for elastic $pp$ and $np$ scattering at low
energies are based on the parametrizations of Cugnon
\etal{}~\cite{Cugnon:1996kh}. The only difference of our cross
sections compared to those of \refcite{Cugnon:1996kh} is that at the
beam momenta below $0.4\upto0.5\GeVc$ we adopt different functional
form \cite{Gale:87} with numerical parameters refitted to the world
data on $pp$ and $np$ total cross sections \cite{Yao:2006px}. This
results in the following expressions in various beam momentum ranges,
\begin{align}
  \sigma_\text{el}^{pp}&=\left\{
    \begin{alignedat}{3}
      &5.12m_N/(s-4m_N^2)+1.67\quad&&\text{for}\quad&&p_\text{lab} < 0.435\\
      &23.5 + 1000(p_\text{lab}-0.7)^4\quad&&\text{for}\quad&0.435 < &p_\text{lab} < 0.8 \\
      &1250/(p_\text{lab}+50) - 4(p_\text{lab}-1.3)^2\quad&&\text{for}\quad&0.8 < &p_\text{lab} < 2 \\
      &77/(p_\text{lab}+1.5)\quad&&\text{for}\quad&2 < &p_\text{lab} <
      6
    \end{alignedat}
  \right.  \intertext{and} \sigma_\text{el}^{np} &=\left\{
    \begin{alignedat}{3}
      &17.05m_N/(s-4m_N^2)-6.83\quad&&\text{for}\quad& &p_\text{lab} < 0.525 \\
      &33 + 196 |p_\text{lab}-0.95|^{2.5}\quad&&\text{for}\quad&0.525 < &p_\text{lab} < 0.8 \\
      &31/\sqrt{p_\text{lab}}\quad&&\text{for}\quad&0.8 < &p_\text{lab} < 2 \\
      &77/(p_\text{lab}+1.5)\quad&&\text{for}\quad&2 < &p_\text{lab} <
      6
    \end{alignedat}
  \right.~,
\end{align}
where the beam momentum $p_\text{lab}$ is given in $\GeVc$, the
nucleon mass $m_N$ and the invariant energy
$\sqrt{s}=\sqrt{2m_N^2+2m_NE_{lab}}$ (with
$E_{lab}=\sqrt{m_N^2+p_\text{lab}^2}$) in $\GeV$ , and the cross
sections in mb.

The neutron-neutron elastic-scattering cross section is obtained from
the proton-proton one by isospin symmetry,
$\sigma_\text{el}^{nn}=\sigma_\text{el}^{pp}$. The angular dependence
of the $NN$-elastic scattering is chosen according to
\refcite{Cugnon:1996kh}, where accurate and compact parametrizations
of the experimental $pp$ and $np$ angular differential cross sections
are developed.

\subsubsection*{\texorpdfstring{${NN \leftrightarrow N\Delta}$}{NN <->
    NDelta}}
\label{sec:NN_ND_appendix}
The differential cross section for $\Delta$-resonance production in an
$NN$ collision is calculated as a special case of
\cref{twoParticleFinal} as
\begin{equation}
  \frac{\dd\sigma_{NN \to N\Delta} }{\dd\mu_\Delta^2\dd\Omega} =
  \frac{\overline{|\mathcal{M}_{NN \to N\Delta}|^2}}{64\pi^2 s}
  \frac{p_{N\Delta}}{p_{NN}} \mathcal{A}_\Delta(\mu_\Delta^2)~,  \label{dsig_NN_to_ND}
\end{equation}
where $p_{NN}$ and $p_{N\Delta}$ are the initial and final center-mass
(CM) momenta, respectively, and \linebreak
$\mathcal{A}_\Delta(\mu_\Delta^2)$ denotes the spectral function of
the $\Delta$ resonance (see \cref{A_R}). $\overline{|\mathcal{M}_{NN
    \to N\Delta}|^2}$ is the matrix element squared, averaged over
spins of particles in the initial and summed over spins of particles
in the final state. It has been calculated for the channel $pp \to n
\Delta^{++}$ by Dmitriev~\etal{} \cite{Dmitriev:1986st} within the
one-pion exchange model. This model describes the differential $pp \to
n \Delta^{++}$ cross sections at beam momenta $1\upto6\GeVc$ very
accurately \cite{Dmitriev:1986st}.  For other isospin channels, the
cross sections are related to the $pp \to n \Delta^{++}$ cross section
by Clebsch-Gordan coefficients, which gives
\begin{alignat}{2}
  \sigma_{pp \to p\Delta^+}& = \sigma_{pn \to p\Delta^0} = \sigma_{pn
    \to n\Delta^+} =\sigma_{nn \to n\Delta^0}
  = \frac{1}{3} \sigma_{pp \to n\Delta^{++}}~,  \label{isorel1} \\
  \sigma_{nn \to p\Delta^-}& = \sigma_{pp \to
    n\Delta^{++}}~.  \label{isorel2}
\end{alignat}
Using the detailed balance relation \cref{detBal}, the inverse
reaction cross sections are obtained as
\begin{equation}
  \frac{\dd\sigma_{N\Delta \to NN} }{\dd\Omega} =
  \frac{\overline{|\mathcal{M}_{NN \to N\Delta}|^2}}{128\pi^2 s}
  \frac{p_{NN}}{p_{N\Delta}} \mathcal{S}_{NN}~, \label{dsig_ND_to_NN}
\end{equation}
where $\mathcal{S}_{pp}=\mathcal{S}_{nn}=1/2$, $\mathcal{S}_{pn}=1$ is
the symmetry factor for the final-state nucleons.

\subsubsection*{\texorpdfstring{${N \Delta \rightarrow N\Delta}$}{N
    Delta -> NDelta}}
\label{sec:ND_ND_appendix}
The differential cross section of this process can be expressed in a
similar way as for the process, $NN \to N\Delta$, see
\cref{dsig_NN_to_ND},
\begin{equation}
  \frac{\dd\sigma_{N\Delta_i \to N\Delta_f} }{\dd\mu_f^2 d\Omega} =
  \frac{\overline{|\mathcal{M}_{N\Delta_i \to N\Delta_f}|^2}}{64\pi^2 s}
  \frac{p_{N\Delta_f}}{p_{N\Delta_i}} \mathcal{A}_\Delta(\mu_f^2)~,   \label{dsig_ND_to_ND}
\end{equation}
where $p_{N\Delta_i}$ and $p_{N\Delta_f}$ are the initial and final
c.m.~momenta, respectively.  For the matrix element 
$\overline{|\mathcal{M}_{N\Delta_i \to N\Delta_f}|^2}$, we apply the
one-pion exchange model, based on the $NN\pi$ and $\Delta\Delta\pi$
interactions, which results in the following expression (see
e.g.~\cite{effe_dipl,buss_phd} for the details of the derivation),
\begin{equation}
  \begin{split}
    \label{eq:effeNDeltaNDelta}
    \overline{|\mathcal{M}_{N\Delta_i \to N\Delta_f}|^2} = &
    \mathcal{I} \frac{1}{8}
    \left(\frac{f_{NN\pi}f_{\Delta\Delta\pi}}{m_\pi^2}\right)^2
    \frac{F^4(t)}{(t-m_\pi^2)^2}
    \frac{16(\mu_i+\mu_f)^2 m_N^2 t}{9\mu_i^2\mu_f^2} \times \\
    &\times
    (-\mu_i^2+2\mu_i\mu_f-\mu_f^2+t)(\mu_i^4-2\mu_i^3\mu_f+12\mu_i^2\mu_f^2-2\mu_i\mu_f^3+\mu_f^4-2\mu_i^2
    t \\
    &+2\mu_i \mu_f t-2\mu_f^2 t+t^2).
  \end{split}
\end{equation}
Here $F(t)$ is a usual monopole form factor \cite{Dmitriev:1986st}.
The masses of the initial and final state $\Delta$ resonance are
denoted as $\mu_i$ and $\mu_f$, respectively.  The isospin factors,
$\mathcal{I}$, are given in \cref{table:isoNDelta_NDelta}.
\begin{table}
  \centering
  \begin{tabular}{@{}lcl@{}c}
    \toprule
    & &&$\mathcal{I}$	 \\
    \midrule
    $p\Delta^{++} $&$\to$&$ p\Delta^{++}$	&\quad $9/4$\\
    $p\Delta^{+}  $&$\to$&$ n\Delta^{++}$	&\quad $3$\\
    $p\Delta^{+}  $&$\to$&$ p\Delta^{+}$	&\quad $1/4$\\
    $p\Delta^{0}  $&$\to$&$ p\Delta^{0}$	&\quad $1/4$\\
    $p\Delta^{0}  $&$\to$&$ n\Delta^{+}$	&\quad $4$\\
    $p\Delta^{-}  $&$\to$&$ p\Delta^{-}$	&\quad $9/4$ \\
    $p\Delta^{-}  $&$\to$&$ n\Delta^{0}$	&\quad $3$ \\
    \bottomrule
  \end{tabular}
  \caption[Isospin factors for the $N\Delta\to N\Delta$ process.]{Isospin factors for the
    $N\Delta\to N\Delta$ process. For the neutron channels, the isospin factors follow by
    isospin symmetry.}
  \label{table:isoNDelta_NDelta}
\end{table}
The $NN\pi$ coupling constant, $f_{NN\pi}=m_\pi g_A/(2f_\pi)=0.946$,
is used by default in our numerical evaluations of
\cref{eq:effeNDeltaNDelta}.  Here, $g_A=1.267$ is the axial coupling
of the nucleon, and $f_\pi=92.4 \MeV$ is the pion-decay constant
(cf.~\refcite{Pascalutsa:2005nd}). We choose the $\Delta\Delta\pi$
coupling constant, $f_{\Delta\Delta\pi}=9/5 f_{NN\pi}$, motivated by
the large-$N_c$ limit \cite{Pascalutsa:2005nd}. The cutoff parameter,
$\Lambda=0.6 \GeV$, is chosen according to \refcite{Dmitriev:1986st}.

\subsubsection*{\texorpdfstring{In--medium $\Delta$ width based on the
    $NN^{-1}$ and $\Delta N^{-1}$ model}
  {In--medium Delta width based on the particle-hole and Delta-hole
    model}}
\label{appendix:osetWidth}

In our default-model setup, the collisional (or spreading) width of
the $\Delta$ resonance is given by \cref{eq:coll_width} with in-medium
total $\Delta$-nucleon cross sections given by the sum of the
Pauli-blocked vacuum partial $\Delta N \to \Delta N$ and $\Delta N \to
NN$ cross sections calculated within the pion-exchange model as
described above.  The only in-medium effect appears in
\cref{eq:coll_width} due to inclusion of Pauli blocking for the
outgoing nucleon(s). Such a simple semiclassical approximation,
however, completely neglects the in-medium modifications of the
exchange pions and vertex corrections due to the short-range
spin-isospin interactions. It also neglects the contribution of the
$\Delta$ absorption by two nucleons, $\Delta NN \to NNN$.  All these
effects are quite important for a more realistic description of the
processes mediated by $\Delta$-resonance excitation in nuclear matter
as shown within the $NN^{-1}$ and $\Delta N^{-1}$ model by Oset and
Salcedo \cite{Oset:1987re}. Thus, we have optionally included the
description of the collisional width of the $\Delta$ resonance using
the parametrization from \refcite{Oset:1987re} in our model.

In \refcite{Oset:1987re}, a special kinematic situation is considered
when the $\Delta$ is created by a $N\pi$ collision. According to
Eq.~(4.4) of \refcite{Oset:1987re}, the imaginary part of the
collisional contribution to the $\Delta$ self-energy is parametrized
as the sum of the higher order quasi-elastic (Q), two-body (A2), and
three-body (A3) absorption components,
\begin{equation}
  -\ImaginaryPart \Sigma_\Delta
  = C_\text{Q} (\rho/\rho_0)^\alpha + C_\text{A2} (\rho/\rho_0)^\beta
  + C_\text{A3} (\rho/\rho_0)^\gamma~,   \label{ImSigma_Oset}
\end{equation}
where the coefficients, $C_\text{Q}, C_\text{A2}$ and $C_\text{A3}$,
and exponents, $\alpha$, $\beta$, and $\gamma$ are functions of the
incoming pion-kinetic energy, $T_\pi$. Here, the sorting of the
processes is done with respect to the pion interaction with
nucleons. In the quasi-elastic case, the pion reappears in the final
state, which is mainly mediated by the process $\Delta N \to \Delta
N$\footnote{In lowest order, the quasi-elastic pion scattering
  proceeds simply via the decay of the intermediate $\Delta$ resonance
  without any interactions of the latter with the nuclear medium. This
  process is, however, neglected in \cref{ImSigma_Oset}.}. The pion
absorption by two and three nucleons is mediated by the processes,
$\Delta N \to NN$ and $\Delta N N \to NNN$, respectively.

The collisional width of the $\Delta$ resonance can also be decomposed
into the quasi-elastic, two- and three-body absorption-partial widths
as readily follows from \cref{ImSigma_Oset},
\begin{equation}
  \Gamma_\text{coll}=-2\ImaginaryPart \Sigma_\Delta
  = \Gamma_\text{Q} + \Gamma_\text{A2} + \Gamma_\text{A3}~. \label{Gamma_coll_Oset}
\end{equation}
In this expression, all partial widths are given in the $\Delta$-rest
frame and depend on the nucleon density, $\rho$, and on the mass,
$\mu_\Delta=\sqrt{m_\pi^2+m_N^2+2(T_\pi+m_\pi)(\frac{3}{5}E_F+m_N)}$,
of the $\Delta$ resonance produced in the collision of the incoming
pion with kinetic energy, $T_\pi$, with a nucleon in the Fermi sea.
The energy and momentum of the $\Delta$ are correlated since they both
are unique functions of the pion-kinetic energy, $T_\pi$. Thus, the
above formula \cref{Gamma_coll_Oset} is, strictly speaking, applicable
only to the $\Delta$'s created in $\pi N$ collisions. However, this is
the most important $\Delta$-excitation channel for the pion-induced
reactions on nuclei.

In earlier versions of the GiBUU model (cf.~\cite{effe_phd,lehr_phd}),
the absorptional component $\Gamma_\text{A2} + \Gamma_\text{A3}$ of
$\Gamma_\text{coll}$ has been treated by converting the $\Delta$ to
the nucleon or, for the case of perturbative treatment, by just
deleting it with probability, $1-\exp[-\Delta t
(\Gamma_\text{A2}+\Gamma_\text{A3})/\gamma]$, during the time interval
$[t;t+\Delta t]$. Here, $\gamma=p_\Delta^0/\mu_\Delta$ is the Lorentz
factor for the transformation of the widths from the $\Delta$-rest
frame to the laboratory frame, where the nucleus is at rest.  To go
beyond this relatively simple modeling, we now explicitly treat all
three processes, $\Delta N \to \Delta N$, $\Delta N \to NN$, and
$\Delta N N \to NNN$.  This is important, if one is interested, e.g.,
in the spectra of outgoing nucleons or nuclear fragments from
pion-induced reactions.

The (effective) cross sections of the former two processes are
expressed in terms of the corresponding partial widths by formally
using the low-density approximation \cref{eq:coll_width} as follows
\begin{alignat}{2}
  \sigma_{\Delta N \to \Delta N}&= \Gamma_\text{Q}(\rho,\mu_\Delta)/
  \overline{v}_\text{rel}(p_\Delta) \rho,  \label{sig_ND_to_ND_Oset} \\
  \sigma_{\Delta N \to NN}&= \Gamma_\text{A2}(\rho,\mu_\Delta)/
  \overline{v}_\text{rel}(p_\Delta) \rho, \label{sig_ND_to_NN_Oset}
\end{alignat}
where $\overline{v}_\text{rel} (p_\Delta)$ denotes the average
relative velocity of the $\Delta$ and a nucleon from the Fermi sea,
\begin{equation}
  \overline{v}_\text{rel} (p_\Delta)
  = \frac{4}{\rho}\int\limits_{p < p_F} \frac{\dd^3 \bvec{p}}{(2\pi)^3}\,
  \left|\bvec{p}_\Delta/p_\Delta^0 - \bvec{p}/p^0\right|~. \label{v_rel_aver}
\end{equation}
For the derivation of \cref{sig_ND_to_ND_Oset,sig_ND_to_NN_Oset} it is
assumed for simplicity that the cross sections, $\sigma_{\Delta N \to
  \Delta N}, \sigma_{\Delta N \to NN}$, do not depend on the
Fermi-momentum $\bvec{p}$.

The partial width, $\Gamma_\text{A3}$, can be approximately regarded
as a three-body reaction rate, $\Delta N N \to NNN$
\cite{Oset:1987re}.  For the $\Delta$ mass in the pole region, all
partial widths in \cref{Gamma_coll_Oset} are of equal importance:
$\Gamma_\text{Q} \simeq \Gamma_\text{A2} \simeq \Gamma_\text{A3}
\simeq 25 \MeV$ (see Fig.~12 in \refcite{Oset:1987re}). Therefore,
inclusion of the $\Delta$ absorption by two nucleons is the origin of
the main difference between our standard calculations involving vacuum
$\Delta N \to NN$ and $\Delta N \to \Delta N$ cross sections only, and
the modified scheme based on the collisional width
\cref{Gamma_coll_Oset}.

To simulate the process, $\Delta N N \to NNN$, first, it is decided
whether a given $\Delta$ will be absorbed during the time interval
$[t;t+\Delta t]$ according to the probability $1-\exp(-\Delta t
\Gamma_\text{A3}/\gamma)$. Then, the two nucleons are randomly chosen
in the vicinity of this $\Delta$. Finally, the momenta of the outgoing
nucleons are sampled according to the invariant three-body phase
space.

\subsubsection*{\texorpdfstring{${NN \leftrightarrow NR,\, NN
      \leftrightarrow \Delta \Delta,\, NR \leftrightarrow NR'}$}{NN
    <-> NR, NN <-> Delta Delta, NR <-> NR'}}
\label{appendix:NR}
We use the same cross sections as detailed in Appendix A1.2 of
\cite{effe_phd} based on the analysis presented by
Teis~\cite{teis_phd}. Note, that we have chosen the $NN\to
NS_{11}(1535)$ matrix element, $20 \cdot 16\pi \mb\GeV^2$ (in
\refcite{effe_phd}, table A.1, three different values were presented).

\subsubsection*{\texorpdfstring{${NR \leftrightarrow NR}$}{NR <-> NR}}
For all resonances besides the $\Delta$, we assume
\begin{equation}
  \sigma_{NR\to NR}(\sqrt{s})=\sigma_{NN\to NN}(\sqrt{s}-m_R+m_N) \label{eq:NR_massShift_2} \;.
\end{equation}

\subsubsection*{\texorpdfstring{${NN \leftrightarrow NN\pi}$}{NN-> NN
    pion}}
\begin{figure}[t]
  \centering
  \includegraphics[width=0.4\linewidth]{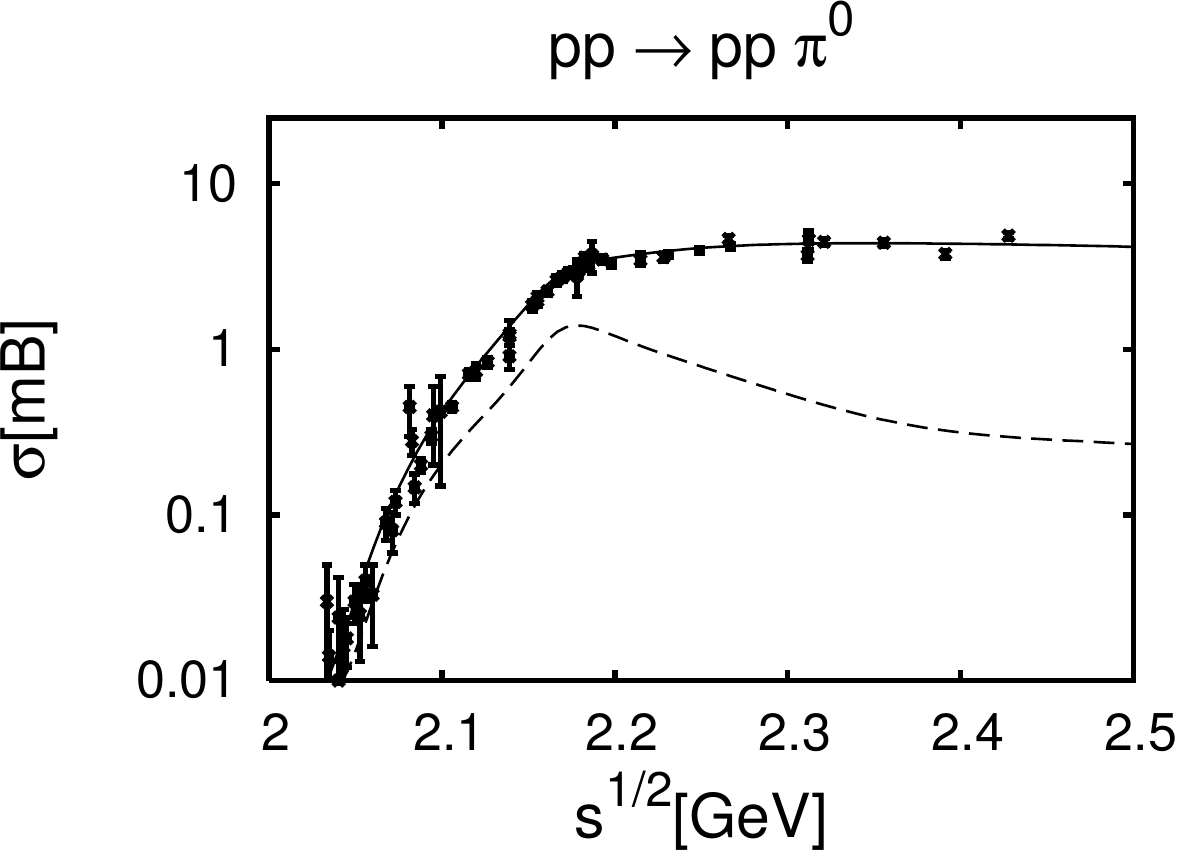}
  \includegraphics[width=0.4\linewidth]{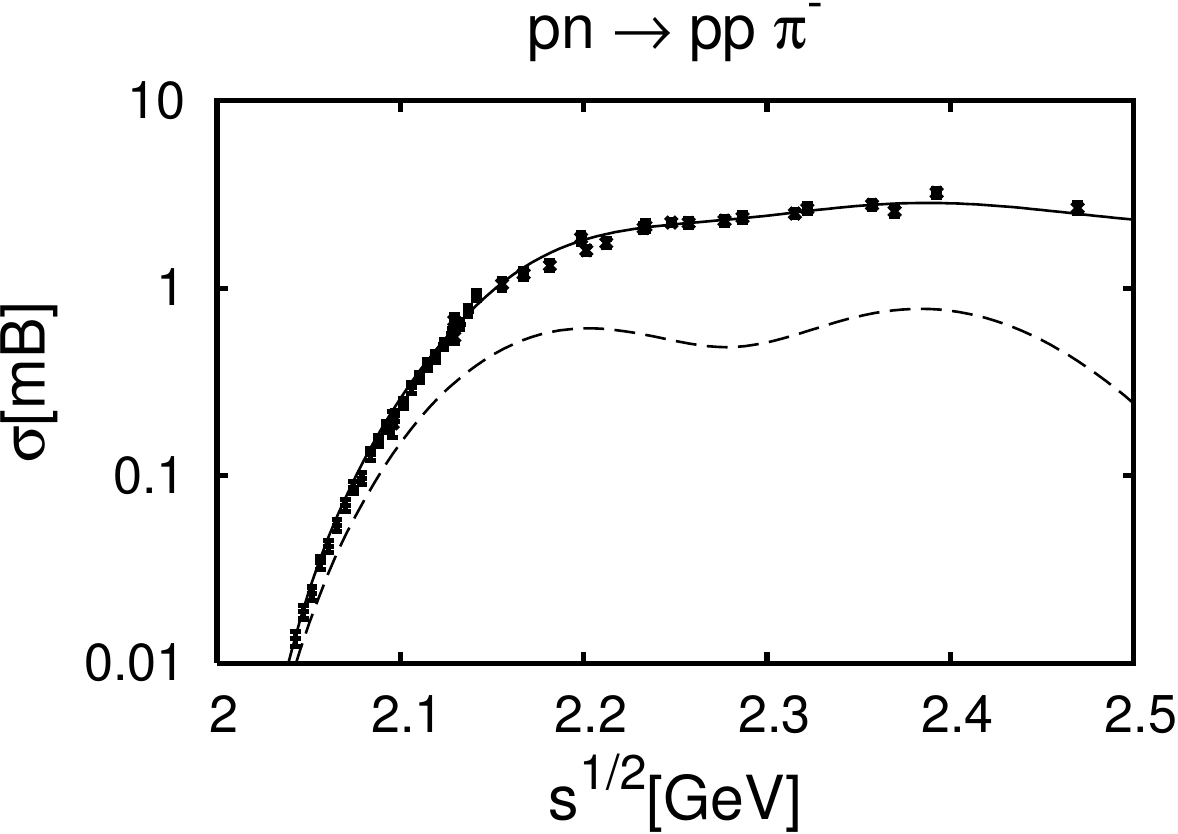}
  \includegraphics[width=0.4\linewidth]{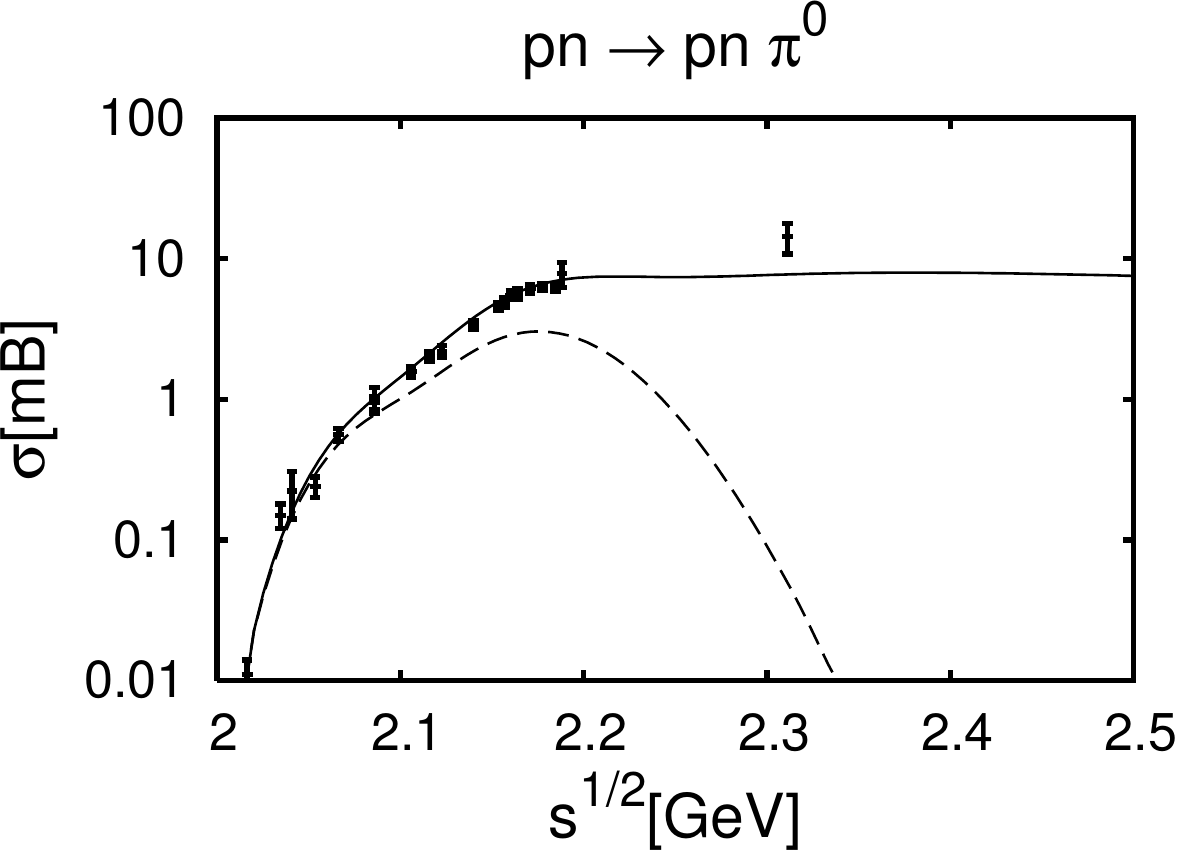}
  \includegraphics[width=0.4\linewidth]{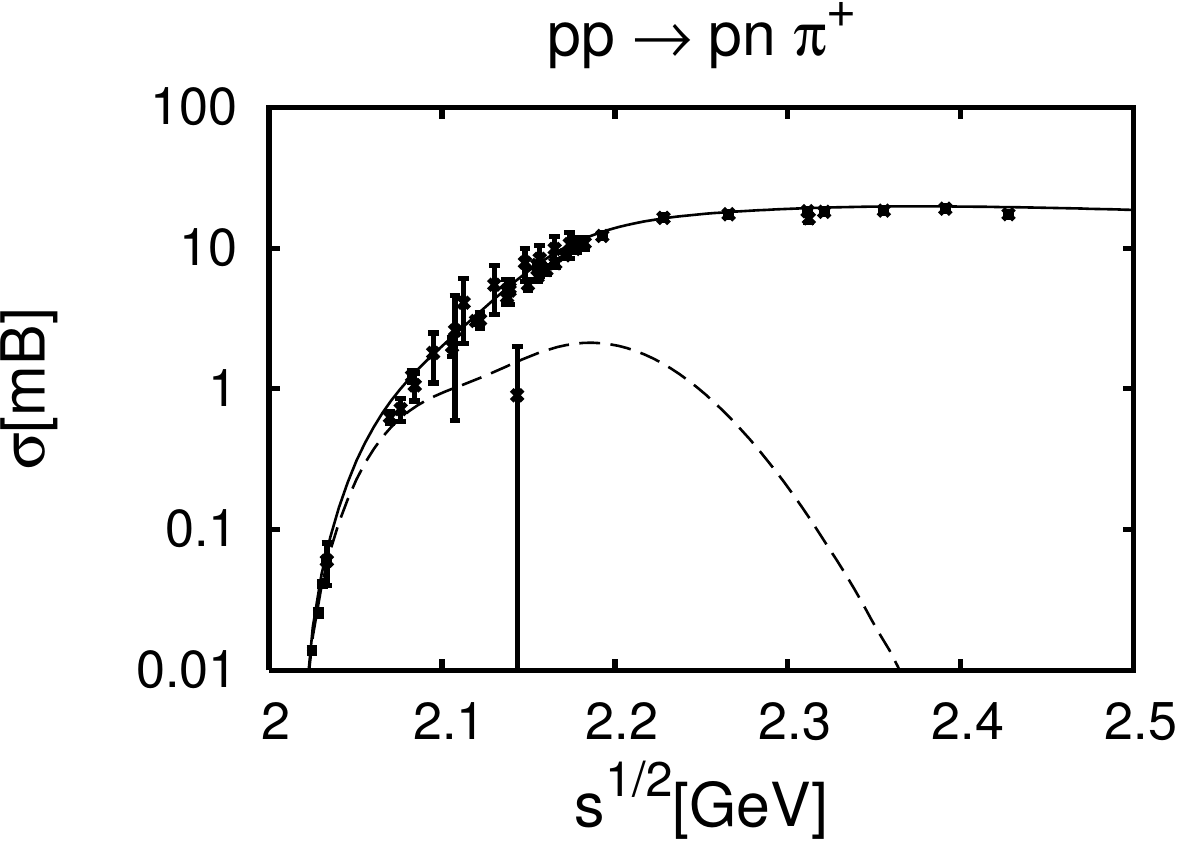}
  \caption[Elementary cross sections for different $NN\rightarrow
  NN\pi$ isospin channels]{Elementary cross sections for different
    $NN\rightarrow NN\pi$-isospin channels. The solid lines show the
    full cross section, whereas the dashed lines represent the
    non-resonant-background contribution. The data are from
    \cite{Landolt,Andreev:1988tv,Daum:2001yh,Hardie:1997mg,Tsuboyama:1988mq,Shimizu:1982dx,Bondar:1995zv}.}
  \label{NN_NNPi}
\end{figure}

For the $NN \leftrightarrow NN\pi$ process, we consider besides
resonance processes, as e.g., $NN\to N\Delta \to NN\pi$, also a
point-like background cross section. In \cref{NN_NNPi} we show the
relevant cross sections.

\subsubsection*{\texorpdfstring{$BB \to BYK$}{BB -> BYK}}
\label{appendix:BYK}

For the kaon-production cross section from baryon-baryon scattering we
use the parametrization adopted by Tsushima
\etal~\cite{Tsushima:1998jz}
\begin{equation}
  \sigma (BB\to BYK) = a\left( \frac{s}{s_{0}}-1\right)^{b}\left( \frac{s_{0}}{s} \right)^{c}~~,
  \label{BB_BYK}
\end{equation}
where $s$ and $s_{0}$ are the squares of the invariant collision
energy and the threshold energy. The parameters, $a$, $b$ and $c$ have
been determined such as to reproduce the calculated energy dependence
of the total cross sections. In the parametrization \cref{BB_BYK} $B$
stands for a baryon ($p$, $n$, or $\Delta$ resonance), $Y$ for a
hyperon ($\Lambda$ or $\Sigma$) and $K$ for a kaon ($K^{+}$ or
$K^{0}$). All isospin channels are taken into account.

\subsection{Antibaryon-baryon cross sections}
\label{gibuu_antiBarBar_xsections}

In this subsection, the beam momentum $p_\text{lab}$ is given in
$\GeVc$ and the cross sections in mb.  For the $\overline p p$-
($\overline n n$-) annihilation cross section we use the following
parametrizations
\begin{equation}
  \label{sig_ann_pbarp}
  \sigma_\text{ann}^{\overline p p} =\left\{
    \begin{alignedat}{3}
      & 51.52p_\text{lab}^{-0.85} + 0.034p_\text{lab}^{-2.94}\quad&&\text{for}\quad&   &p_\text{lab} < 0.51 \\
      & 88.8p_\text{lab}^{-0.4} - 24.2\quad&&\text{for}\quad&0.51 < &p_\text{lab} < 6.34\\
      & 38p_\text{lab}^{-0.5} +
      24p_\text{lab}^{-1.1}\quad&&\text{for}\quad&6.34 < &p_\text{lab}
    \end{alignedat}
  \right.
\end{equation}
At the highest momenta, the formula from \refcite{Cugnon:1989} is
applied in \cref{sig_ann_pbarp}.  The $\overline n p$ ($\overline p
n$) annihilation cross section is significantly lower than the
$\overline p p$ one at low beam momenta
\cite{Armstrong:1987nu,Dover:1992vj} and practically coincide with
$\sigma_\text{ann}^{\overline p p}$ at large beam momenta. We take
this into account as
\begin{equation}
  \sigma_\text{ann}^{\overline n p} =\left\{
    \begin{alignedat}{3}
      & 41.4 + 29/p_\text{lab}\quad&&\text{for}\quad& &p_\text{lab} <0.382 \\
      & \sigma_\text{ann}^{\overline p p}\quad&&\text{for}\quad& 0.382
      < &p_\text{lab}\ ,
    \end{alignedat}
  \right.
\end{equation}
where an expression from \refcite{Armstrong:1987nu} is used at the
lowest momenta.  For the low-energy $\overline p p$ collisions (see
\cref{matching} for the definition of the low- and high-energy
two-body collisions), the elastic $\overline p p$ cross section is
similar to that from \refcite{Cugnon:1989}, but with slightly
readjusted parameters,
\begin{equation}
  \sigma_\text{el}^{\overline p p} = 40p_\text{lab}^{-0.56} + 5.8 \exp[-(p_\text{lab}-1.85)^2]~.
\end{equation}
The charge-exchange cross section, $\sigma_\text{CEX}^{\overline p
  p}$, of the processes, $\overline p p \leftrightarrow \overline n
n$, is adopted from \cite{Cugnon:1989}.  The annihilation and elastic
cross sections for all other possible $\overline B B$ collisions, as
well as the charge-exchange cross sections of the $\overline \Delta N$
and $\overline N \Delta$ collisions (e.g., $\overline\Delta^- n \to
\overline\Delta^{--} p$ and $\overline p \Delta^+ \to \overline n
\Delta^0$) are taken equal to the corresponding cross sections for the
$\overline p p$ collisions with the same relative velocity.

For the low-energy $\overline N N$, $\overline\Delta N$ and $\overline
N \Delta$ collisions, apart from elastic scattering, charge exchange,
and annihilation, we also take into account the processes, $\overline
N N \leftrightarrow \overline N \Delta$ and $\overline N N
\leftrightarrow \overline\Delta N$. The cross sections of the
$\Delta$- ($\overline\Delta$-) resonance production in $\overline N N$
collisions are calculated on the basis of the one-pion exchange model
\cite{Dmitriev:1986st}. For the other low-energy $\overline B B$
collisions, only annihilation and elastic scattering are considered.

\begin{figure}[t]
  \centering
  \includegraphics[width=0.6\linewidth]{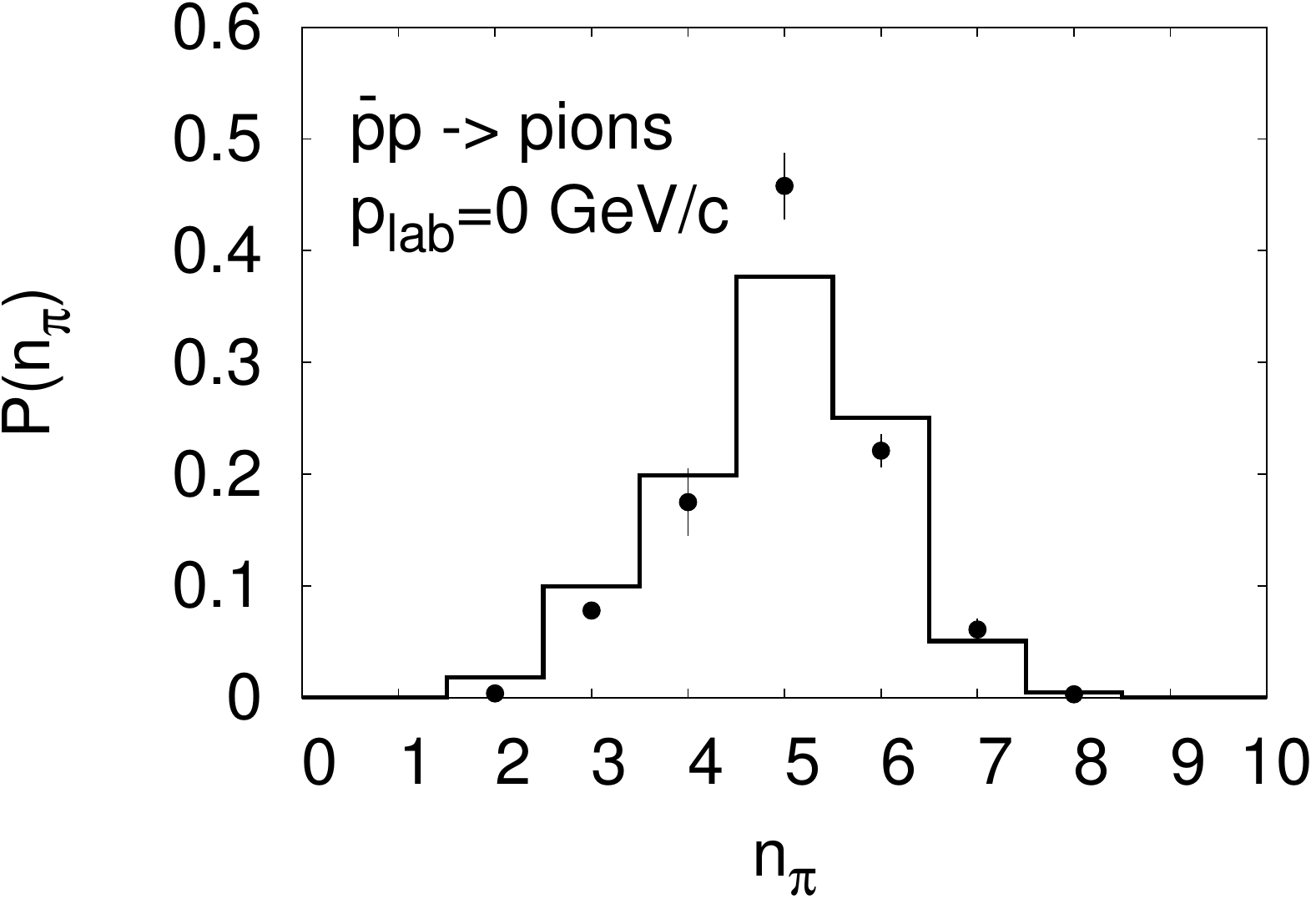}
  \caption{\label{fig:pimul_pbarp} Total pion multiplicity
    distribution for $\overline{p} p$ annihilation at rest (see also
    \cite{Cugnon:1984zp}). Source: The data are from \cite{Ghesquiere:1974}.}
\end{figure}

For the high-energy $\overline B B$ collisions, the annihilation,
elastic scattering, charge exchange (for $\overline N N$,
$\overline\Delta N$ and $\overline N \Delta$ collisions), and
inelastic production processes, $\overline B B \to \overline B B +
\text{mesons}$, are taken into account. The latter process is
simulated via \Fritiof{}~\cite{Pi:1992ug}.  The partial cross section
of the inelastic production is obtained by subtracting the
contributions from annihilation, elastic scattering and charge
exchange from the total cross section.  The total and elastic
$B\overline{B}$ scattering cross sections are parametrized according
to \refcite{PhysRevD.50.1173}:
\begin{equation}
  \begin{split}
    \label{sig_tot_BarBB_HE}
    \sigma_\text{tot}& = 38.4 + 77.6p_\text{lab}^{-0.64} +
    0.26\ln^2(p_\text{lab}) - 1.2\ln(p_\text{lab})~,\\
    \sigma_\text{elast}&= 10.2 + 52.7p_\text{lab}^{-1.16} +
    0.125\ln^2(p_\text{lab}) - 1.28\ln(p_\text{lab})~,
  \end{split}
\end{equation}
where only positive values are allowed.

The angular distribution for the elastic and charge-exchange $\bar p p$
scattering is given by \cref{dsigdt} where the slope parameter (in
$\GeV^{-2}$) is
\begin{equation}
   \label{b_pbarp_EL}
   b = (A+\hbar B/p_{\bar p p})^2/\hbar^2
\end{equation}
for the elastic scattering \cite{Kondratyuk:1986cq} with $A=0.67 \;
\fm$, $B=0.35$ and $p_{\bar p p}$ being the c.m. momentum, and
\begin{equation}
  \label{b_pbarp_CEX}
  b=11\exp(-0.23p_\text{lab})+8p_\text{lab}^{2.2}/(254+p_\text{lab}^{2.2})
\end{equation}
for the charge exchange scattering.

The $\overline{N} N$ annihilation is described with help of a
statistical annihilation model \cite{PshenichnovPhD,Golubeva:1992tr}. In
this model, the probability of a given $\overline{N} N$-annihilation
channel is proportional to the invariant phase-space volume. It is also
proportional to other factors taking into account, in particular, the
$\text{SU}(3)$ symmetry in terms of particle multiplets (for details
see~\cite{PshenichnovPhD,Golubeva:1992tr} and refs.~therein).  The third
isospin components of the final-state particles are randomly sampled
according to the probabilities given by all possible subsequent
couplings to the total isospin, $I$, and its third component, $I_3$, via
Clebsch-Gordan coefficients\footnote{The total isospin, $I$, and its
  projection, $I_3$, are exactly conserved in this model, in contrast to
  the string-based description of $\overline{N} N$ annihilation included
  in the earlier versions of the GiBUU model (cf.~\cite{Larionov:2008wy}
  and refs.~therein).}. Up to six outgoing particles can be produced,
which are various combinations of $\pi, \eta, \omega, \rho, K, \bar K,
K^*$ and $\bar K^*$ mesons. The model is proved to well describe the
$\overline{p} p$ annihilation observables at $1 < p_{\rm lab} < 10
\GeVc$~\cite{PshenichnovPhD}, while at lower beam momenta it fails to
describe the experimental pion-multiplicity distributions.  Therefore,
the original statistical model has been supplemented by empirical
branching ratios for annihilation at rest, which are tabulated in
\cite{PshenichnovPhD,Golubeva:1992tr}. In our calculations, first we
choose randomly whether the empirical branching ratios are used or the
statistical model itself. The probability to select the branching ratios
at rest drops linearly with the invariant energy of the annihilating
pair as
\begin{equation}
  P_\text{at rest} =
  1-\frac{\sqrt{s}-2m_N}{\sqrt{s_\text{max}}-2m_N}~  \label{P_atRest},
\end{equation}
where $\sqrt{s_\text{max}}=2.6\GeV$ is the maximum invariant energy up
to which the annihilation tables at rest still can be selected
(respective beam momentum $p_{\rm lab}=2.5 \GeVc$). (The probability is
set to 0, if the above definition becomes negative.) The momenta of the
outgoing mesons in $\overline{N} N$ annihilation are sampled
microcanonically, i.e., the probability for certain momentum
configuration is proportional to the corresponding phase-space-volume
element, \cref{dPhi_N}.

\begin{figure}[t]
  \centering
  \includegraphics[width=0.6\linewidth]{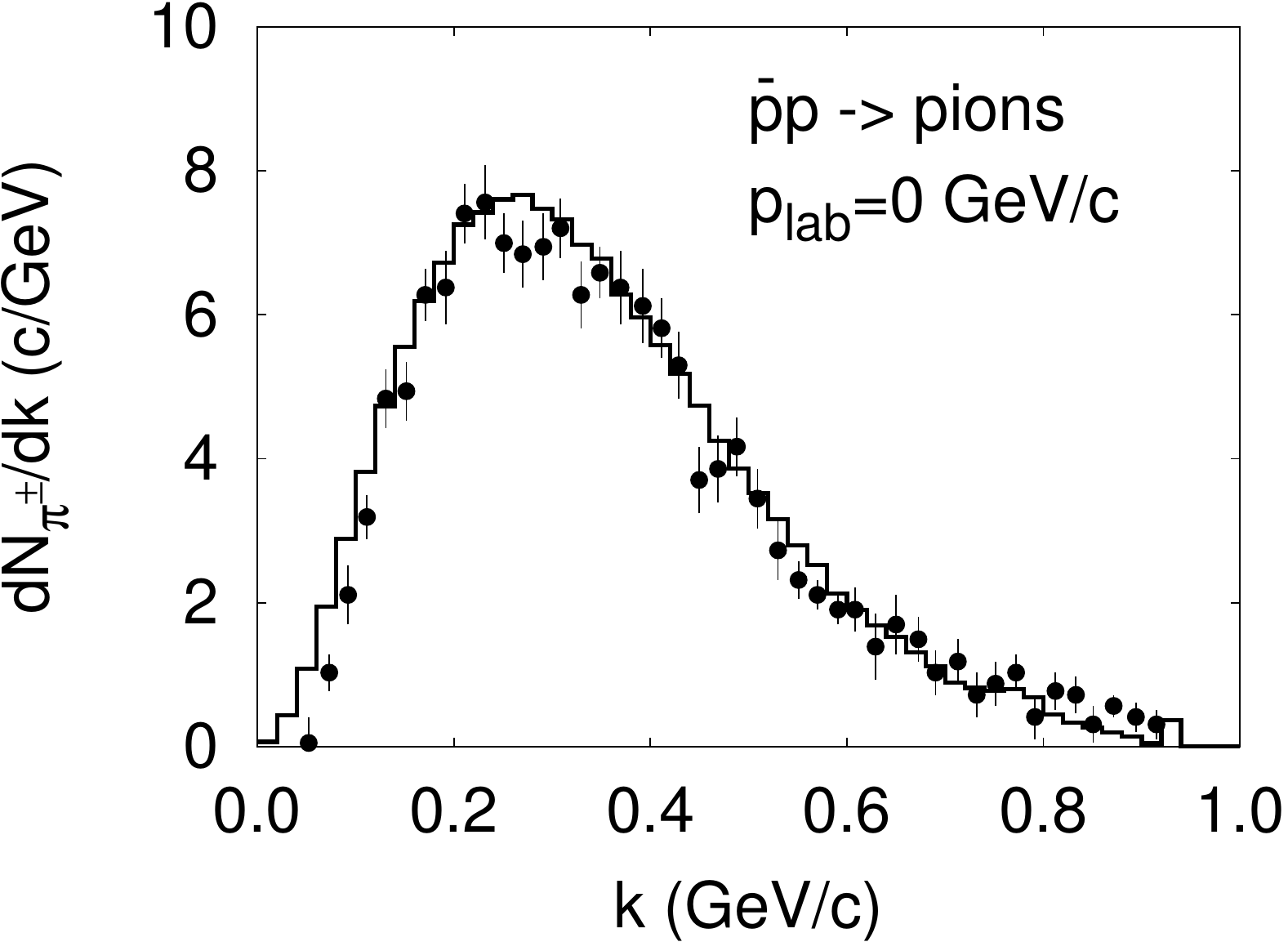}
  \caption{\label{fig:pimom_pbarp} Charged pion momentum distribution
    for $\overline{p} p$ annihilation at rest. Calculations are
    normalized to the number of charged pions per annihilation
    event. Experimental data (in arbitrary units) are taken from the
    review \cite{Dover:1992vj} and rescaled to agree with calculations
    at $k=0.2\GeVc$.}
\end{figure}

\Cref{fig:pimul_pbarp} shows the pion-multiplicity distribution for
$\overline{p} p$ annihilation at rest, calculated taking into account
the decays of $\omega$ and $\rho$ mesons produced in the annihilation
process. The model delivers an excellent description of the experimental
data. Also the momentum spectra of the produced pions in $\overline{p}
p$ annihilation at rest are in a very good agreement with the data, as
as shown in \cref{fig:pimom_pbarp}.

\subsection[Meson baryon cross sections]{Meson-baryon cross sections}
\label{gibuu_barMes_xsections}

\subsubsection[non-resonant backgrounds]{Non-resonant background cross
  sections in the resonance energy region}

Besides the resonance cross sections, non-resonant cross-section
contributions have been implemented in the collision term. The
background cross sections denoted by $\sigma^{\text{BG}}$ are chosen
in such that the elementary cross-section data in the vacuum are
reproduced. Background contributions are instantaneous in space-time,
whereas the resonances propagate along their classical trajectories
until they decay or interact with one or two nucleons in the nuclear
medium.

One may decide not to propagate all know resonances, e.g.~in order to
speed up the simulation, cf.~\cref{sec:particleproperties}. These
resonances are then also not allowed to be produced in the collision
term. We have to compensate for this by introducing additional
background terms to the cross section as a direct interaction.

Additionally, we obtain a background term due to a lack of strength of
the resonance cross sections, which do not describe the full
experimentally observed cross section.

\subsubsection{\texorpdfstring{$\pi N\to X$}{pion N -> X}}

\subsubsection*{\texorpdfstring{$\pi N\to \pi N$}{pion N -> pion N}}
\label{appendix:pionXsections}

\begin{figure}[t]
  \centering
  \includegraphics[width=0.7\linewidth]{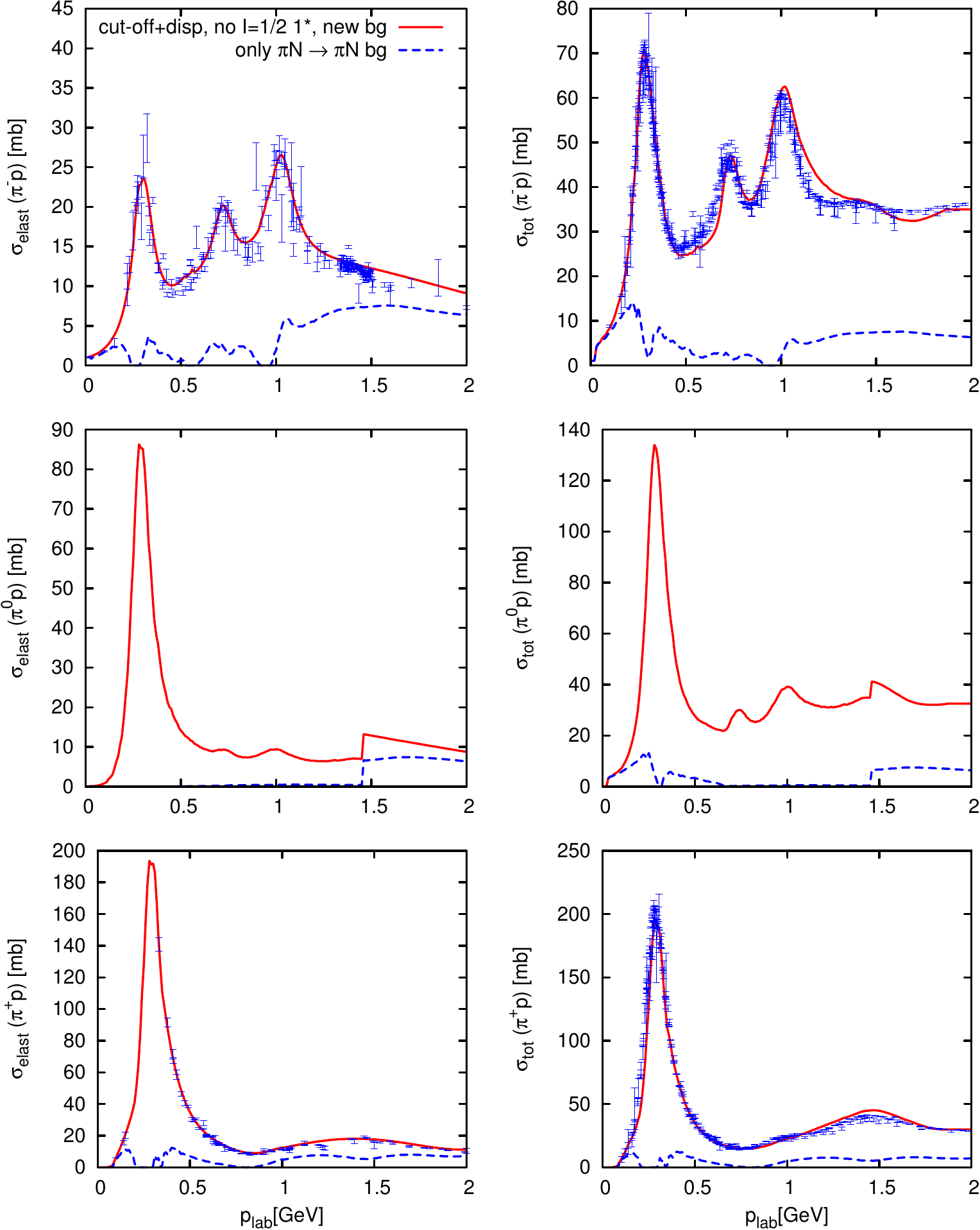}
  \caption[Elastic and total cross sections for the scattering of
  pions and protons]{(Color online) Elastic (left panels) and total
    (right panels) cross sections for the scattering of pions and
    protons. The solid curves show the results with our default
    parameters: all resonances besides the $I=1/2$ 1*-resonances are
    included, the real parts of the self-energies are included in the
    propagators. The dashed curves show the background
    contributions. The data are from \cite{PDGdata}.}
  \label{fig:pionNuc_bg}
\end{figure}

In our model, the cross section for quasi-elastic pion-nucleon
scattering is given by an incoherent sum of background and resonance
contributions,
\begin{equation}
  \label{backDef}
  \sigma_{\pi N \rightarrow \pi N}=\sigma_{\pi N \rightarrow R
    \rightarrow \pi N}+\sigma^{\text{BG}}_{\pi N \rightarrow \pi N} .
\end{equation}
Pion nucleon scattering can be categorized into four different isospin
channels,
\begin{alignat}{2}
  \sigma_{\pi^{-} n \rightarrow \pi^{-} n}&=\sigma_{\pi^{+} p \rightarrow \pi^{+} p} ,\label{c1}\\
  \sigma_{\pi^{-} p \rightarrow \pi^{0} n}&=\sigma_{\pi^{0} n
    \rightarrow \pi^{-} p}
  =\sigma_{\pi^{+} n \rightarrow \pi^{0} p}=\sigma_{\pi^{0} p \rightarrow \pi^{+} n} , \label{c2}\\
  \sigma_{\pi^{-} p \rightarrow \pi^{-}p}&=\sigma_{\pi^{+} n \rightarrow \pi^{+} n} , \label{c3}\\
  \sigma_{\pi^{0} n \rightarrow \pi^{0} n}&=\sigma_{\pi^{0} p
    \rightarrow \pi^{0} p} . \label{c4}
\end{alignat}
The cross sections in the individual channels are either connected by
time-reversal or isospin symmetry. The first channel, \cref{c1}, is a
pure isospin-$I=3/2$-scattering process, whereas the other three
channels are mixtures of $I=1/2$ and $I=3/2$. The cross section for
the $I=3/2$ channel, $\sigma_{\pi \ N \rightarrow \Delta \rightarrow
  \pi \ N}$, is given explicitly in \cite{effe_phd} based on the
resonance analysis by Manley and Saleski \cite{ManleySaleski}.

\begin{figure}[t]
  \centering
  \includegraphics[width=0.8\linewidth]{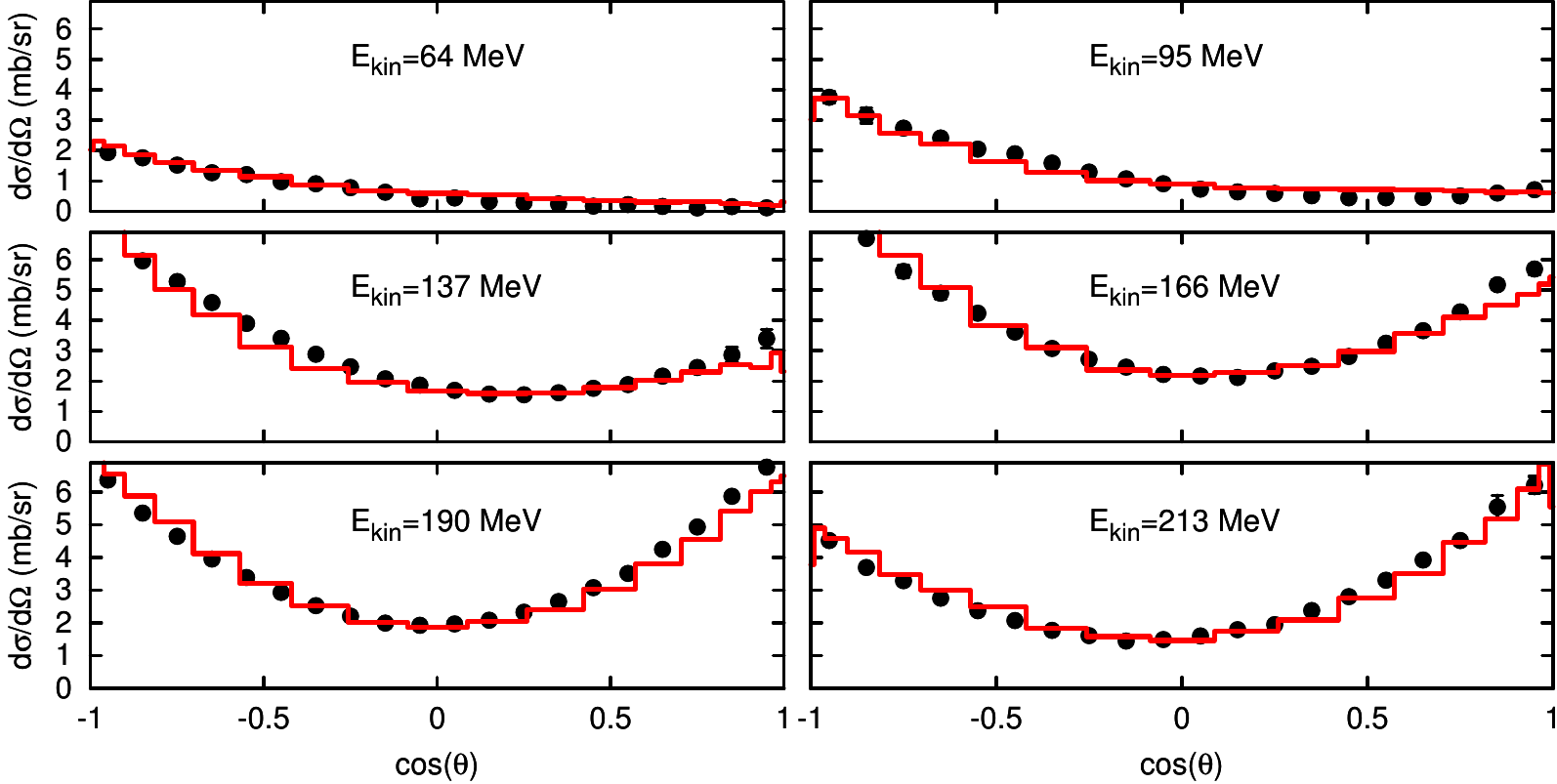}
  \caption[The angular distributions for the CX process, $\pi^- p \to
  \pi^0 n$ ]{(Color online) The angular distributions for the
    charge-exchange process, $\pi^- p \to \pi^0 n$, in the CM frame of
    pion and proton. The plots are labeled with the kinetic energies
    of the pions in the laboratory frame. The data are  from
    \cite{Sadler:2004yq}.}
  \label{fig:ele_pionNucleon}
\end{figure}

There are accurate data sets for the $\pi^{+} p \to \pi^{+} p$,
$\pi^{-} p \to \pi^{-} p$ and $\pi^{-} p \to \pi^{0} n$ for the first,
second and third channel, \cref{c1,c2,c3}, down to very low energies.
Hence, we introduce a background term on top of our resonance
contributions for a better description of these channels. The last
channel, \cref{c4}, is inaccessible for experiment; therefore we can
not introduce any background term. In this approach, we describe in a
satisfying manner all available data, as can be seen in
\cref{fig:pionNuc_bg}.

The cross sections on the neutron follow from isospin symmetry.

We include a realistic angular distribution for the elastic scattering
of the pions \cite{Buss:2006vh}. Due to the $P$-wave nature of the
$\Delta(1232)$ resonance, we assume for $\pi N \to \Delta \to \pi N$
in the resonance rest frame a distribution of the pion scattering
angle, $\theta$,\footnote{The angle, $\theta$, is defined by the
  incoming and outgoing pion momenta. In the simulation, we must store
  for each $\Delta$ produced in a $\pi$N collision the momentum of the
  corresponding pion in the resonance-rest frame.} according to
\begin{equation}
  f^\Delta(s,\theta)=\left(1+3\cos^2(\theta) \right) g(s,\theta) ,
\end{equation}
which is peaked in forward and backward direction. The function,
$g(s,\theta)$, depending on Mandelstam $s$, parametrizes the energy
dependence of the $\pi N$ angular distribution. In a coherent
calculation the angular distribution is generated by interference
effects, which can not be accomplished by our transport model. In our
ansatz we have to split the cross section in an incoherent way to
preserve our semi-classical resonance picture. Therefore we take
\begin{equation}
  g(s,\theta)=\left[\alpha-\cos(\theta) \right]^{\beta \left(m_\Delta-\sqrt{s}\right)/m_\Delta}
\end{equation}
with the $\Delta$-pole mass, $m_\Delta=1.232 \GeV$. For the background
events we assume
\begin{equation}
  f^\text{BG}(s,\theta)=g(s,\theta) \; .
\end{equation}
The constants, $\alpha=1.9$ and $\beta=26.5$, are fitted to the
angular distributions measured in the Crystal
Ball~\cite{Sadler:2004yq} experiment; a comparison of our
parametrization to this data is shown in \cref{fig:ele_pionNucleon}.

\subsubsection*{\texorpdfstring{$\pi N\to \pi\pi N$}{pion N -> pion pion  N}}
\begin{figure}[t]
  \centering
  \includegraphics[width=0.6\linewidth]{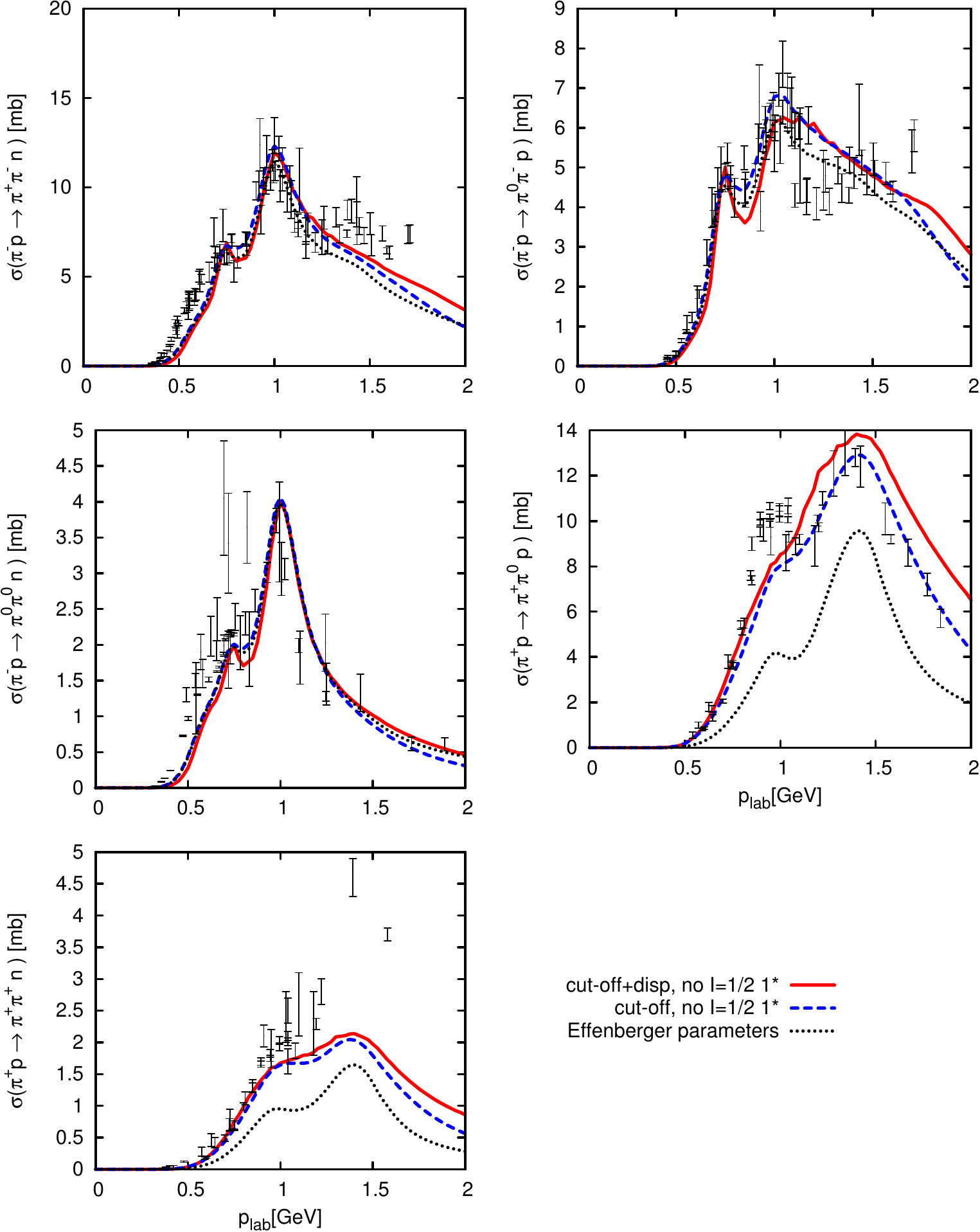}
  \caption[Cross sections for $\pi N\to \pi NN$]{(Color online) Cross
    sections for $\pi^- p\to \pi^0\pi^0 n$, $\pi^- p\to \pi^+\pi^- n$,
    $\pi^- p\to \pi^0\pi^- p X$, $\pi^+ p\to \pi^+\pi^+ n$, $\pi^+
    p\to \pi^+\pi^0 p$. The solid line represents the full model in
    which only those 1*-resonances are not included which have isospin
    1/2. The dashed curve neglects the effects of the dispersion
    relations and the dotted line represents the result according to
    the former choice of parameters~\cite{effe_phd}.}
  \label{fig:pionN_2pionN}
\end{figure}

The $\pi N\to \pi \pi N$ cross section is given by
\begin{equation}
  \sigma_{\pi N\to N \pi\pi}=\sum_R \sigma_{\pi N\to R} \frac{\Gamma_{R\to N \pi \pi}}{\Gamma_\text{tot}} .
\end{equation}
In \cref{fig:pionN_2pionN} we show the model results for all pion
proton channels: $\pi^- p\to \pi^0\pi^0 n$, $\pi^- p\to \pi^+\pi^- n$,
$\pi^- p\to \pi^0\pi^- p X$, $\pi^+ p\to \pi^+\pi^+ n$, $\pi^+ p\to
\pi^+\pi^0 p$.  Obviously, calculations, where we include the
$(I=3/2)$-1*-resonances and exclude only the $(I=1/2)$-1*-resonances,
fit the data better than those, which neglect all 1*-resonance of the
Manley analysis, especially in the $\pi^+ p\to \pi^+\pi^+ n$ and
$\pi^+ p\to \pi^+\pi^0p$ channels.  For the first case we also see,
that there is only a modest impact, whether we include real parts of
the self-energy in the resonance propagators.  We conclude that the
choice to exclude only the $(I=1/2)$-1*-resonances seems to fit the
data better than the former choice to exclude all 1* resonances.

\subsubsection*{\texorpdfstring{$\pi N\to \eta \Delta$}{pion N -> eta Delta}}
The mass differential cross section is given by the following
expression
\begin{equation}
  \frac{\dd\sigma_{\pi N \rightarrow \eta \Delta}}{\dd m_{\Delta}}=\clebsch{1}{I_z^\pi}{\frac{1}{2}}{I_z^N}{\frac{3}{2}}{I_z^\Delta}
  \left|\mathcal{M}_{\pi N \rightarrow \eta \Delta} \right|^2 \frac{p_{\eta\Delta}(m_\Delta)}{s~p_{\pi N}}
  \mathcal{A}_{\Delta}(m_\Delta)~,                                       \label{dsigma_piN_to_etaDelta}
\end{equation}
where the matrix element is assumed to be a constant,
$\left|\mathcal{M}_{\pi N \rightarrow \eta \Delta}
\right|^{2}=7\mb{}\GeV{}^2$. The corresponding total cross section is
given by the integral of the r.h.s.~of \cref{dsigma_piN_to_etaDelta}
over the mass of the $\Delta$ resonance.

\subsubsection*{\texorpdfstring{$\pi N \to \omega N$ and  $\pi N \to \phi N$}{pion N -> omega N}}
The total cross section parametrizations $\sigma_{\pi^- p \rightarrow
  \omega n}$ and $\sigma_{\pi^- p \rightarrow \phi n}$ are adopted
from \refcite{Cugnon:1990xw} and \refcite{Sibirtsev:1996ag},
respectively. In the case of the $\omega$ production, the
contributions from intermediate resonances \res{G}{17}{2190} and
\res{P}{13}{1900} which have significant branching ratios to the
$N\omega$ decay channel (see \cref{tab:baryon_properties}) are
subtracted from the parametrization of the total $\sigma_{\pi^- p
  \rightarrow \omega n}$ cross section such that
\begin{equation}
  \sigma_{\pi^- p \rightarrow \omega n}^{\text{BG}} = \sigma_{\pi^- p \rightarrow \omega n}
  - \sigma_{\pi^- p \rightarrow \omega n}^{\text{Resonances}}~.
\end{equation}
For the $\phi$ production, there is no intermediate resonance
contribution and we set
\begin{equation}
  \sigma_{\pi^- p \rightarrow \phi n}^{\text{BG}} = \sigma_{\pi^- p \rightarrow \phi n}~.
\end{equation}
The background cross sections for other charge channels are obtained
from a simple isospin relation
\begin{equation}
  \sigma_{\pi N_1 \rightarrow M N_2}^{\text{BG}} =
  \clebsch{1}{I_z^\pi}{\frac{1}{2}}{I_z^{N_1}}{\frac{1}{2}}{I_z^{N_2}}
  \frac{3}{2} \sigma_{\pi^- p \rightarrow M n}^{\text{BG}}~,~~~M=\omega,\phi \label{sig_piN_to_phiN}~.
\end{equation}

\subsubsection*{\texorpdfstring{$\pi N \to \omega \pi N$ and $\pi N \to \phi \pi N$}{pion N -> omega pi N}}
We utilize the parametrizations of \refcite{Sibirtsev:1996ag} for the
inclusive cross sections $\sigma_{\pi^+ p \rightarrow \omega X}$ and
$\sigma_{\pi^+ p \rightarrow \phi X}$.  It is assumed, further, that
the inclusive cross sections for the different charge states of
initial particles are the same, i.e.,
\begin{equation}
  \sigma_{\pi N\rightarrow M X}=\sigma_{\pi^+ p \rightarrow M X}~,~~~M=\omega,\phi~.
\end{equation}
Then, the cross sections of the processes $\pi N \to \omega \pi N$ and
$\pi N \to \phi \pi N$, summed over different charge states of the
outgoing pion and nucleon, are obtained by the subtraction
\begin{equation}
  \sigma_{\pi N\rightarrow M \pi N} =  \sigma_{\pi N \rightarrow M X}
  - \sigma_{\pi^- p \rightarrow M n}~,~~~M=\omega,\phi~.
\end{equation}
The probabilities for the different charge states of the outgoing
particles are set equal to each other.

\subsubsection*{\texorpdfstring{$\pi N\to \Sigma K$}{pion N -> Sigma
    Kaon}}

We use a parametrization of \refcite{Tsushima:1996tv} for the cross
sections, $\sigma_{\pi^{+} p \to K^{+} \Sigma^{+}}$, $\sigma_{\pi^{0}
  p \to K^{+} \Sigma^0}$, $\sigma_{\pi^{-} p \to K^{0} \Sigma^0}$, and
$\sigma_{\pi^{-} p \to K^{+} \Sigma^-}$. Assuming isospin symmetry and
isospin-$I=\frac{1}{2}$ dominance in the intermediate resonance states
we obtain $\sigma_{\pi^{0} p \to K^{0} \Sigma^+}= \sigma_{\pi^{-} p
  \to K^{0} \Sigma^0}$.  Channels with an initial neutron are defined
by the isospin reflection symmetry from respective channels with
initial proton.

\subsubsection*{\texorpdfstring{$\pi N\to \Lambda K$}{pion N -> Lambda Kaon}}
For the total cross section $\sigma_{\pi^{-}p\rightarrow \Lambda
  K^{0}}$, we again use the parametrization of
\refcite{Tsushima:1996tv}.  Since a part of this cross section is
caused by intermediate resonances (\res{P}{11}{1710} and
\res{P}{11}{2100}, see \cref{tab:baryon_properties}), we define the
corresponding background cross section by subtracting the resonance
contributions as
\begin{equation}
  \sigma_{\pi^{-}p\rightarrow \Lambda K^{0}}^{\text{BG}}=\sigma_{\pi^{-}p\rightarrow \Lambda K^{0}}
  -\sigma_{\pi^{-}p\rightarrow \Lambda K^{0}}^{\mathrm{Resonances}}~.
\end{equation}
The background cross sections of the other charge channels can be
directly reconstructed by isospin relations since the total isospin of
the final state is fixed to 1/2,
\begin{equation}
  \sigma_{\pi N \rightarrow \Lambda K}^{\text{BG}}
  = \clebsch{1}{I_z^\pi}{\frac{1}{2}}{I_z^N}{\frac{1}{2}}{I_z^K}
  \frac{3}{2} \sigma_{\pi^{-}p\rightarrow \Lambda K^{0}}^{\text{BG}}~.
\end{equation}

\subsubsection*{\texorpdfstring{$\pi N\to K \overline{K} N $}{pion N -> Kaon Antikaon N}}
The cross section $\sigma_{\pi^- p\rightarrow p K^0 \overline{K}^-}$
is parametrized according to \refcite{Sibirtsev:1996rh} as
\begin{equation}
  \sigma_{\pi^- p\rightarrow p K^0 \overline{K}^-}
  = 1.121 \mb \cdot
  \left(1-\frac{s_0}{s}\right)^{1.86} \left(\frac{s_0}{s}\right)^2
  \equiv \frac{1}{2} \sigma_0     \label{sig_sibirtsev}
\end{equation}
with $s_{0}=\left(m_{N}+2m_{K}\right)^2$. The cross sections of the
other charge channels are obtained from the isospin symmetry of the
$K^*$ and $\pi$ exchange diagram (see Fig. 1c in
\refcite{Sibirtsev:1996rh} and \refcite{effe_phd}) which gives:
\begin{equation}
  \begin{alignedat}{4}
    & \sigma_{\pi^- p\rightarrow n K^0 \overline{K}^0}&=\sigma_0,
    \quad && \sigma_{\pi^- p\rightarrow n K^+
      \overline{K}^-}&=\sigma_0, \\
    & \sigma_{\pi^0 p\rightarrow n K^+ \overline{K}^0}&=\sigma_0,
    \quad && \sigma_{\pi^0 p\rightarrow p K^0
      \overline{K}^0}&=\frac{1}{4} \
    \sigma_0, \\
    & \sigma_{\pi^0 p\rightarrow p K^+ \overline{K}^-}&=\frac{1}{4} \
    \sigma_0, \quad && \sigma_{\pi^+ p\rightarrow p K^+
      \overline{K}^0}&=\frac{1}{2} \ \sigma_0~.
  \end{alignedat}
\end{equation}
The $\pi n$ cross sections are directly given by the the isospin
reflection symmetry from the corresponding $\pi p$ cross sections.

\subsubsection*{\texorpdfstring{$\pi N\to \Sigma K\pi,\ \Lambda
    K\pi$}{pion N -> Sigma Kaon pion, Lambda Kaon pion}}
In order to describe the strangeness production plus one pion
channels, we performed a combined fit for all channels to experimental
data \cite{Baldini:1988th} of the form
\begin{align}
  \sigma(s) = a_i\ A
  \left(\frac{s}{s_0}-1\right)^B\left(\frac{s}{s_0}\right)^{-C}
  \qquad,\qquad \sqrt{s_0}=\begin{cases}
    (1.750+0.100)\GeV&\quad\text{for}\quad\Lambda K \pi\\
    (1.823+0.100)\GeV&\quad\text{for}\quad\Sigma K \pi
  \end{cases}
\end{align}
leading to the general constants,
\begin{equation}
  A=86.027\quad,\qquad B=2.197\quad,\qquad C=7.363\quad,
\end{equation}
and the process dependent normalization values, $a_i$, as listed in
\cref{table:fitSigmaKpi}.
\begin{table}
  \begin{center}
    \begin{tabular}[htb]{c@{$\,\to\,$}cl@{$\,\qquad\qquad\,$}c@{$\,\to\,$}cl}
      \toprule
      \multicolumn{2}{c}{Process} & $a_i$ (mb) &
      \multicolumn{2}{c}{Process} & $a_i$ (mb)\\
      \midrule
      $\pi^- p$ & $\Lambda \,K^0\,\pi^0$ & 0.169  & $\pi^+ p$ & $\Lambda \,K^+\,\pi^+$ & 0.217\\
      & $\Lambda \,K^+\,\pi^-$ & 0.140  &           & $\Sigma^0\,K^+\,\pi^+$ & 0.0426 \\
      & $\Sigma^0\,K^0\,\pi^0$ & 0.1    &           & $\Sigma^+\,K^+\,\pi^0$ & 0.126 \\
      & $\Sigma^0\,K^+\,\pi^-$ & 0.0724 &           & $\Sigma^+\,K^0\,\pi^+$ & 0.0887 \\
      & $\Sigma^-\,K^+\,\pi^0$ & 0.0520 & \multicolumn{3}{c}{} \\
      & $\Sigma^-\,K^0\,\pi^+$ & 0.117  & \multicolumn{3}{c}{} \\
      & $\Sigma^+\,K^0\,\pi^-$ & 0.0514 & \multicolumn{3}{c}{} \\
      \bottomrule
    \end{tabular}
    \caption{Normalization factors for the fits to $\pi^\pm p\to
      \Sigma K\pi, \Lambda K \pi$ data.}\label{table:fitSigmaKpi}
  \end{center}
\end{table}
The fit range was $p_{\rm lab}<3.5\GeV$. The additional constant value
of $100\MeV$ added to the threshold energies was introduced to achieve
better numerical convergence.

\subsubsection{\texorpdfstring{$\pi\Delta \to X$}{pion Delta -> X}}
Besides resonance production channels, background contributions for
$\Lambda K$ and $\Sigma K$ production are implemented according to
\cite{Tsushima:1996tv}.

\subsubsection{\texorpdfstring{$\rho N \to X$}{rho N -> X}}
Besides resonance-production channels, a $\pi N$ background is
introduced to absorb missing inelasticities above $\sqrt{s}>1.8\GeV$,
\begin{equation}
  \sigma_{\rho N\to\pi N} = \sigma_\text{tot, data}-\sigma_\text{tot, resonances} \; .
\end{equation}

\subsubsection{\texorpdfstring{$\eta N \to X$}{eta Delta -> X}}
\subsubsection*{\texorpdfstring{$\eta N \to \pi N$}{eta Delta -> pi
    n}}
Additionally to resonance production, a $\pi N$ background is defined
for $\sqrt{s}>2\GeV$,
\begin{equation}
  \sigma_{\eta N\to\pi N} = \sigma_\text{tot, data}-\sigma_\text{tot, resonances} \; .
\end{equation}
In this approach, missing resonance strength is attributed to the $\pi
N$ channel.

\subsubsection{\texorpdfstring{$\eta \Delta \to X$}{eta Delta -> X}}
\subsubsection*{\texorpdfstring{$\eta \Delta \to \pi N$}{eta Delta ->
    pi n}}
By applying the detailed balance relation \cref{detBal} to
\cref{dsigma_piN_to_etaDelta}, we get
\begin{equation}
  \sigma_{\eta \Delta\to\pi N} = \frac{1}{2} \clebsch{1}{I_z^\pi}{\frac{1}{2}}{I_z^N}{\frac{3}{2}}{I_z^\Delta}
  |\mathcal{M}_{\pi N\to\eta\Delta}|^2 \frac{p_{\pi N}}{s~ p_{\eta\Delta}} .
\end{equation}
The factor $1/2$ is due to (2j+1)-terms in the cross sections and
different spins in initial and final state.

\subsubsection{\texorpdfstring{$\omega/\phi N \to X$}{omega/phi N -> X}}
The elastic $\sigma_{MN}^\text{el}$ ($M=\omega,\phi$) and inelastic
$\sigma_{MN}^\text{inel}$ cross sections on nucleon are taken from
\refcite{Lykasov:1998ma} for $\omega$-induced reactions and from
\refcite{Golubeva:1997na} for $\phi$-induced ones. In the case of the
$\omega N$ elastic scattering, the resonance contribution is
subtracted to get the background elastic scattering cross section.
The background cross sections $\sigma_{M N \to \pi N}^{\text{BG}}$ are
obtained by detailed balance relations from the corresponding cross
sections of the inverse reactions. Additionally, the $\pi\pi N$
channel is included as a background in order to describe the remaining
part of inelastic cross sections:
\begin{equation}
  \begin{alignedat}{2}
    \sigma_{\omega N \to \pi\pi N}^{\text{BG}} &= \sigma_{\omega
      N}^\text{inel} - \sigma_{\omega N \to R}^\text{inel}
    - \sigma_{\omega N \to \pi N}^{\text{BG}}~, \\
    \sigma_{\phi N \to \pi\pi N}^{\text{BG}} &= \sigma_{\phi
      N}^\text{inel} - \sigma_{\phi N \to \pi N}^{\text{BG}}~,
  \end{alignedat}
\end{equation}
where $\sigma_{\omega N \to R}^\text{inel} \equiv \sigma_{\omega N \to
  R} - \sigma_{\omega N \to R \to \omega N}$ is the resonance part of
the $\omega N$ inelastic scattering cross section.

\subsubsection{\texorpdfstring{$K N \to X$}{K N -> X}}
The kaon-nucleon interactions are described following Effenberger (see
Appendix A.2.4 of \cite{effe_phd}). The experimental elastic cross
section $K^+p \to K^+ p$ is parametrized as
\begin{equation}
  \sigma_{K^+p \to K^+ p} = \frac{a_0 + a_1p + a_2p^2}{1 + a_3p + a_4p^2}~,
  \label{sig_kp_elast}
\end{equation}
where $p$ is the kaon momentum with respect to the nucleon-rest frame,
$a_0=10.508\mb$, $a_1=-3.716\mb{}/\GeV{}$, $a_2=1.845\mb{}/\GeV{}^2$,
$a_3=-0.764\GeV{}^{-1}$, $a_4=0.508\GeV{}^{-2}$. The $K^+ n$ elastic
and charge-exchange cross sections are assumed each to be 50\proz of
$K^+ p$ elastic cross section,
\begin{equation}
  \sigma_{K^+n \to K^+ n} = \sigma_{K^+n \to K^0 p}
  = \frac{1}{2}\sigma_{K^+p \to K^+ p}~.    \label{sig_kn_elast_cex}
\end{equation}
The inelastic $K^+N$ cross sections are obtained by a spline
interpolation of the experimental data points for the total $K^+N$
cross sections \cite{Landolt} at $p_{\rm lab} < 6\GeVc$ after
subtraction of the elastic and charge-exchange contributions. The
inelastic $K^+N$ cross section is assumed to be entirely composed of
the $K \pi N$-outgoing channels. The charges of the outgoing kaon,
pion and nucleon are randomly selected such that all charge
combinations that are possible by charge conservation are equally
probable. The $K^0N$ cross sections for the different outgoing
channels are equal to the cross sections of the corresponding
isospin-symmetric $K^+N$-collision channels. The kinematics of the
outgoing particles is chosen according to the isotropic angular
distribution for the $KN$-final state or by the three-body phase-space
sampling for the $K \pi N$-final state.

\subsubsection{\texorpdfstring{$\overline K N \to X$}{Kbar N -> X}}
The antikaon-nucleon interaction is governed mostly by the $S=-1$
intermediate resonance excitation (see
\cref{tab:baryon_properties_strange}) according to
\cref{eq:vacResProd}. On the other hand, the resonance contributions
do not completely saturate the measured cross sections. In particular,
at low invariant energies the background terms have to be
included. This is especially important for the charge-neutral initial
channels, $K^- p$ and $\overline{K^0}n$. In this case, the
$\Lambda(1405)$ resonance, whose pole mass is only $30 \MeV$ below the
$\overline K N$ production threshold, determines the low-energy
scattering.  However, in our approach the explicit coupling of the
$\Lambda(1405)$ resonance to the antikaon-nucleon channel is not
included. Thus, the large cross sections of various $K^- p$ scattering
channels at small invariant energies have to be considered as the
non-resonant background.

\begin{figure}[t]
  \centering
  \includegraphics[width=0.6\linewidth]{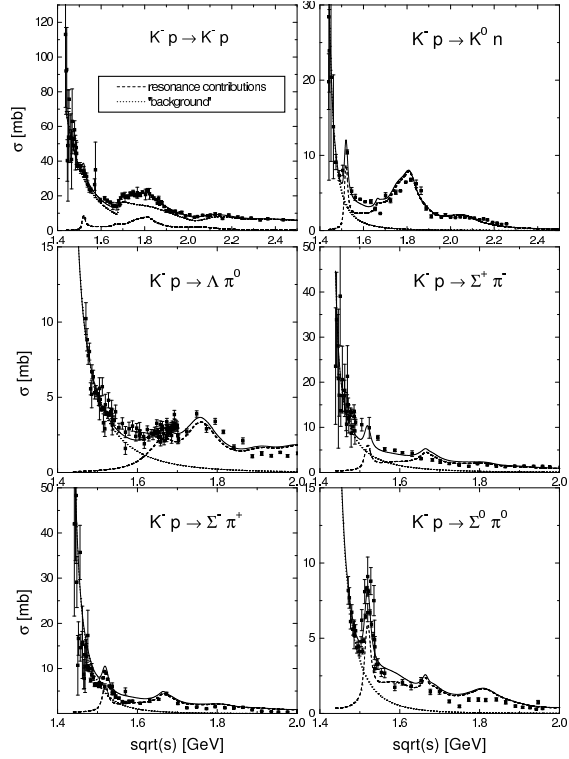}
  \caption{\label{fig:KminProton_part} The cross sections of elastic
    scattering, charge-exchange and hyperon production in $K^- p$
    collisions. The total cross sections are shown as solid lines. The
    dashed lines represent the resonance contributions and the dotted
    lines the non-resonance background. The data are from
    \cite{Baldini:1988th}. Taken from \cite{effe_phd}.}
\end{figure}

The background terms have been parametrized by Effenberger
\cite{effe_phd}. In the case of a two-body outgoing channel
$\alpha\equiv K^- p,~\overline{K^0} n,~\Lambda \pi^0,~\Sigma^+ \pi^-,
~\Sigma^- \pi^+,~\Sigma^0 \pi^0 $,
\begin{equation}
  \sigma_{K^- p \to \alpha}^{\rm bg} = a_0 \frac{p_f}{p_i s}
  \left(\frac{a_1^2}{a_1^2+p_f^2}\right)^{a_2}~,   \label{Kbar2BodyBg}
\end{equation}
is used for the background cross section, where $p_i$ and $p_f$ are
the c.m.~momenta of the incoming and outgoing particles,
respectively. The parameters $a_j$, j=0,1,2, are listed in
\cref{tab:Kbar2BodyBg} for the different outgoing channels $\alpha$.
\begin{table}
  \centering
  \begin{tabular}{cccc}
    \toprule
    channel & $a_0$ [mb GeV$^2$] & $a_1$ [GeV] & $a_2$ \\
    \midrule
    $K^- p$          & 150    & 0.35    & 2 \\
    $\overline{K^0} n$    & 100    & 0.15    & 2 \\
    $\Lambda \pi^0$  & 130    & 0.25    & 3 \\
    $\Sigma^+ \pi^-$ & 600    & 0.1     & 2 \\
    $\Sigma^- \pi^+$ & 5000   & 0.1     & 3 \\
    $\Sigma^0 \pi^0$ & 2500   & 0.1     & 3 \\
    \bottomrule
  \end{tabular}
  \caption{\label{tab:Kbar2BodyBg} Parameters of the non-resonant
    background cross sections for various two-body outgoing channels
    of $K^- p$ scattering.}
\end{table}
For $K^- p \to K^- p$ elastic scattering, the parametrization
\cref{Kbar2BodyBg} is used only for invariant energies $\sqrt{s} <
1.7\GeV$.  At higher$\sqrt{s}$, the spline interpolation of the
experimental data is used to describe the broad maximum in the cross
section at $\sqrt{s}\simeq1.8\GeV$. As can be seen from
\cref{fig:KminProton_part}, the sum of the resonance and background
contributions provide a very good description of experimental data for
the processes $K^- p \to K^- p,~\overline{K}^0 n, ~\Lambda
\pi^0,~\Sigma^+ \pi^-, \Sigma^- \pi^+$ and $\Sigma^0 \pi^0$.

\begin{figure}[t]
  \centering
  \includegraphics[width=0.6\linewidth]{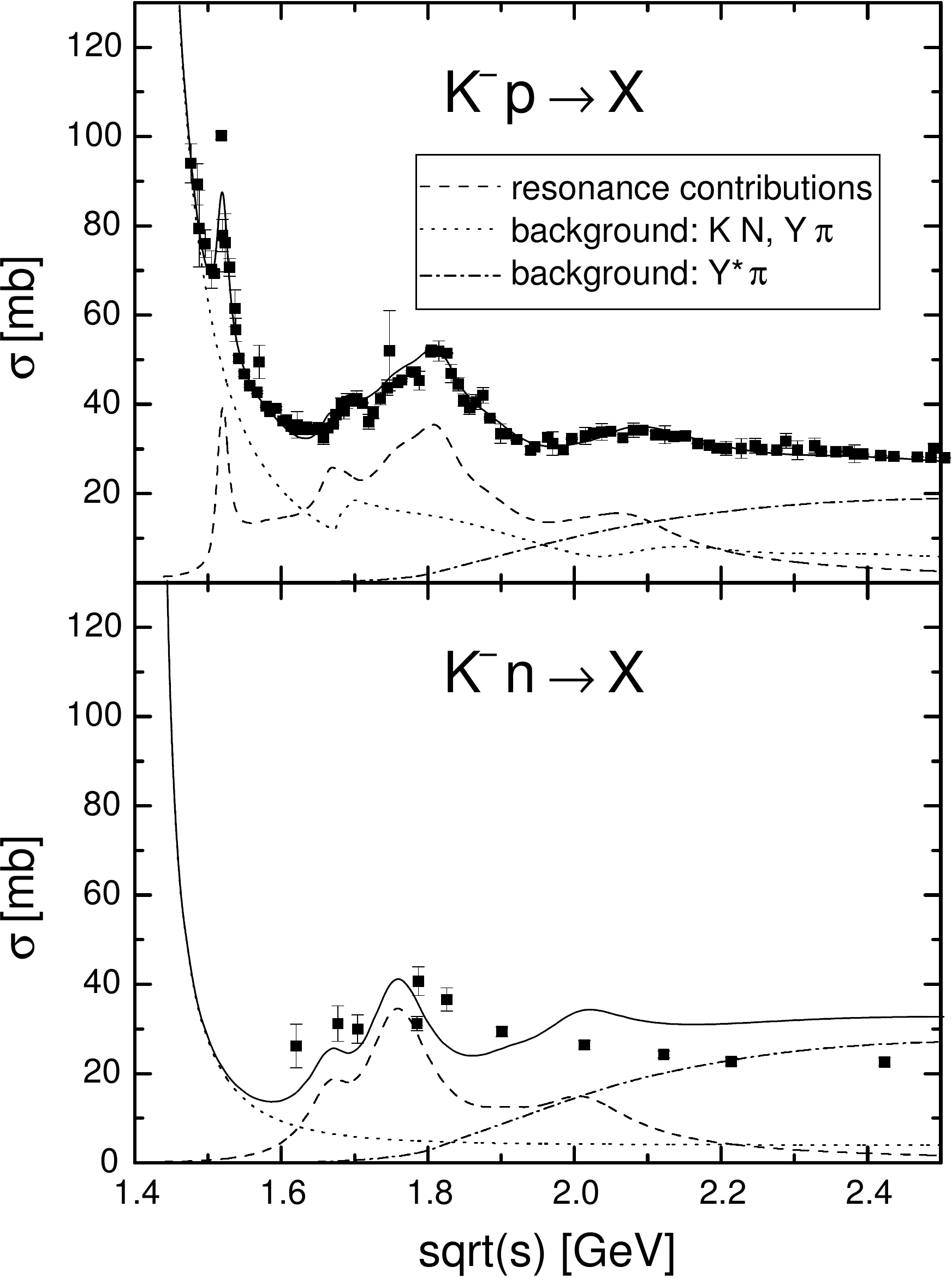}
  \caption{\label{fig:KminNucleon_tot} The total cross sections (solid
    lines) of the $K^- p$ and $K^- n$ collisions with respective
    resonance contributions (dashed lines) and background
    contributions from $\overline K N \to \overline K N, \quad \pi Y$
    (dotted lines) and $\overline K N \to \pi Y^*$ (dash-dotted
    lines). The data are from \cite{Baldini:1988th}. Taken from
    \cite{effe_phd}.}
\end{figure}

In order to describe the total $K^- p$ cross section at invariant
energies $\sqrt{s}$ up to $2.2\GeV$, one needs to take into account
the channels with more than two particles in the final state. To this
aim, the process $\overline K N \to Y^* \pi$ is implemented by
assuming a constant matrix element for all hyperonic resonances,
\begin{equation}
  \sigma_{\overline K N \to Y^* \pi} = C \frac{|{\cal M}|^2}{p_i s}
  \int\limits^{\sqrt{s}-m_\pi}_{m_{\Lambda(\Sigma)}+m_\pi}\,\dd (\mu^2) \,p_f {\cal A}_{Y^*}(\mu)~,
  \label{sig_KbarN_to_YstarPi}
\end{equation}
where the coefficient $C$ is a combination of isospin-Clebsch-Gordan
coefficients of the incoming (1,2) and outgoing (3,4) particles given
by
\begin{equation}
  C= \sum_I\,\clebsch{I^1}{I_z^1}{I^2}{I_z^2}{I}{I_z}\,
  \clebsch{I^3}{I_z^3}{I^4}{I_z^4}{I}{I_z}~.    \label{C}
\end{equation}
The total $K^-p$ cross section is well reproduced (see
\cref{fig:KminNucleon_tot}), if the value of the matrix element
$|{\cal M}|^2=22\mb\GeV^2$ is chosen. Applying
\cref{sig_KbarN_to_YstarPi}, only the hyperon resonances with masses
larger than or equal to $1.6\GeV$ are taken into account. Using the
parametrization \cref{sig_KbarN_to_YstarPi} and the detailed-balance
relation \cref{detBal2}, one obtains the cross section for the inverse
reaction, $\pi Y^* \to N \overline K$,
\begin{equation}
  \sigma_{\pi Y^* \to N \overline K}= C \frac{2}{2J_{Y^*}+1}
  \frac{p_f}{p_i s} |{\cal M}|^2~.
  \label{sig_YstarPi_to_KbarN}
\end{equation}
The $K^- n$ elastic scattering is described in terms of the resonance
contributions and a small constant background cross section
\begin{equation}
  \sigma_{K^- n \to K^- n}^{\rm BG} =4~\mbox{mb}~.
  \label{sig_KminNeutron_el_bg}
\end{equation}
Since the isospin of the final state in the processes $K^- n \to
\Lambda \pi^-$ and $K^- p \to \Lambda \pi^0$ is fixed ($I=1$), we have
the exact relation
\begin{equation}
  \sigma_{K^- n \to \Lambda \pi^-}^{\rm BG}
  =2\sigma_{K^- p \to \Lambda \pi^0}^{\rm BG}.
  \label{sig_KminNeutron_to_LambdaPi}
\end{equation}
For the $\Sigma \pi$-outgoing channel, we assume the background cross
section,
\begin{equation}
  \sigma_{K^- n \to \Sigma^- \pi^0}^{\rm BG}
  =\sigma_{K^- n \to \Sigma^0 \pi^-}^{\rm BG}
  = \frac{1}{2}\left(  \sigma_{K^- p \to \Sigma^- \pi^+}^{\rm BG}
    + \sigma_{K^- p \to \Sigma^+ \pi^-}^{\rm BG}
  \right)~.             \label{sig_KminNeutron_to_SigmaPi}
\end{equation}
The total $K^- n$ cross section is shown in
\cref{fig:KminNucleon_tot}. One observes a fair agreement with the
available experimental data.

The cross sections for $\overline K^0 N$ scattering are obtained
isospin symmetry,
\begin{equation}
  \sigma_{\overline K^0 p} = \sigma_{K^- n}~,
  ~~~\sigma_{\overline K^0 n} = \sigma_{K^- p}~.
  \label{sig_Kbar0Nucleon}
\end{equation}
Finally, the cross sections for the $\pi Y \to \overline K N$
processes, important for heavy ion collisions, are directly obtained
from the detailed-balance formula \cref{detBal}. This results in
values of about $4\mb$ for the isospin-averaged $\pi \Lambda \to
\overline K N$ and $\pi \Sigma \to \overline K N$ cross sections at
$\sqrt{s} < 2\GeV$.

\subsubsection{Meson-Baryon Annihilation Processes}
\label{sec:manni}

It is worthwhile to reemphasize, that the cross sections listed above
are only used for low energies, cf.~\cref{matching}. At higher
energies, the final state decisions are done by event generators like
\Pythia{}. As developed in \refcite{Wagner:2004ee}, we also have to
take into account additionally quark-antiquark annihilation processes
in meson-baryon collisions.

For this we split up the incoming hadrons into their quark content and
check, whether an annihilation between a quark and its antiquark is
possible. If so, we construct a string out of the remaining three
quarks, which has the invariant mass of the incoming colliding hadron
pair. This string decays according the string fragmentation used in
\Pythia{}.

\begin{figure}
  \centering
  \includegraphics[width=0.8\linewidth]{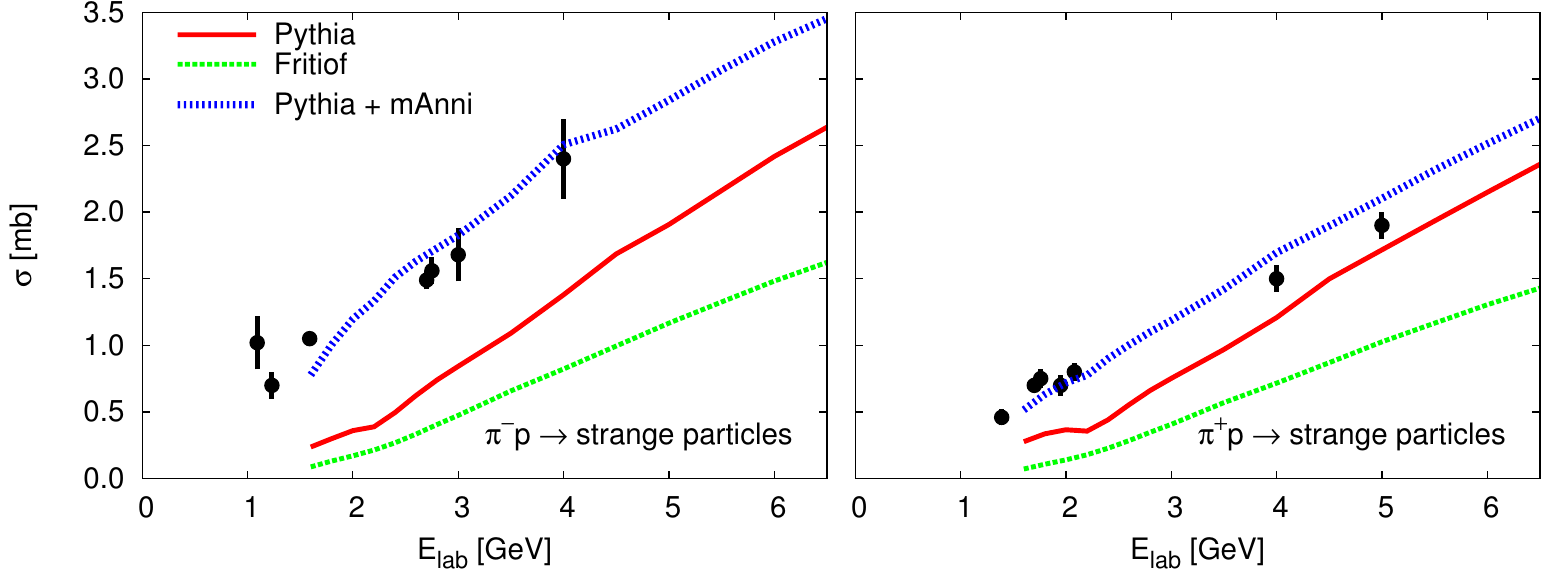}
  \caption{The cross section of production of strange particles in
    $\pi p$ collisions as function of the projectile energy. Only
    the high energy contributions are shown. The event generators
    \Pythia{} (red solid) or \Fritiof{} (green dotted) underestimate
    the experimental data \cite{Baldini:1988th}. The contribution of
    the meson-baryon annihilation process on top of the \Pythia{}
    contribution is shown as the blue dashed curve.}
  \label{fig:manni}
\end{figure}

The probability for these annihilation events is chosen such, that we
we match the strangeness production in $\pi p$ events, as shown in
\cref{fig:manni}. We find
\begin{align} {\rm Prob}({\rm annihilation}) = R_{\rm
    comb}\cdot\left(a-b\sqrt{s}\right) \quad\text{with}\quad a=0.6\ ,\
  b=0.1\GeV^{-1}
\end{align}
as a suitable prescription, which is slightly different from what was
used in \refcite{Wagner:2004ee}. Here $R_{\rm comb}$ is a
combinatorial factor, which is 1, if only one quark-antiquark
combination is possible, and 2, if there are, e.g., two quarks for the
corresponding antiquark (e.g.~for $\pi^+p$, there is only one
$\overline{d}d$ combination, while for $\pi^-p$ one finds two
possibilities for $\overline{u}u$ annihilation).

Comparisons with experimental data for other final state channels, as,
e.g., two pion production, sustain the findings of \cref{fig:manni},
that \Pythia{} plus the annihilation prescription describes best all
the data.

\subsection{Meson-meson cross sections}
\label{gibuu_mesMes_xsections}

In central heavy ion collisions at high energy a hot and dense medium
consisting of baryons and mesons is produced. With increasing beam
energy, the mesonic degrees of freedom become more and more important.
Already at AGS energies ($E_{\rm lab} \simeq 10\AGeV$) the number of
produced pions becomes comparable to the original number of nucleons
in the colliding nuclei. In such a system, the meson-meson (mostly
$\pi\pi$) collisions can not be neglected. The meson-meson collisions
are in particular important for a realistic description of strangeness
production in heavy ion collisions in kinetic models based on hadronic
degrees of freedom \cite{Weber:2002pk,Wagner:2004ee,Tomasik:2005ng}.

The $\pi\pi$-elastic scattering and total cross sections are dominated
by intermediate $\rho$ and $\sigma$ meson excitation via the resonance
mechanism of \cref{eq:vacResProd}. Also other resonance channels of
meson-meson collisions, e.g., $\rho\pi \to \phi$ and $K\pi \to K^*$
have to be considered (see \cref{tab:meson_properties} for all
possible meson-meson collisions leading to a mesonic resonance
excitation included in the model). However, the resonance mechanism
does not account for the process $\pi^+ \pi^- \to K^+ K^-$ which has a
rather large cross section of about $1.5\mb$ near the threshold
\cite{Petersen:1977cs}.

In \refcite{Brown:1991ig}, the cross sections of the processes,
$\pi\pi\to K\overline K$, $\rho\rho\to K\overline K$, $\pi\rho\to K
\overline K^*$, and $\pi\rho\to K^* \overline K$, were calculated on
the basis of a kaon- and $K^*$-exchange model in the impulse
approximation. We use the parametrization of the $\pi\pi\to
K\overline K$ cross section from \refcite{Cassing:1996xx}, which is
based on these calculations,
\begin{equation}
  \sigma_{\pi\pi\to K\overline K} = 6.075 \mb \cdot C
  \left(1-\frac{(2m_K)^2}{s}\right)^{0.76},
\end{equation}
where the isospin-dependent factor $C$ is given by \cref{C}.

For simplicity, \cref{C} is also applied to evaluate the $\rho\rho\to
K\overline K$, $\pi\rho\to K \overline K^*$ and $\pi\rho\to K^*
\overline K$ cross sections since the isospins of the incoming
particles are the same \cite{Wagner:2004ee}. For other
strangeness-production reactions involving two non-strange mesons in
the incoming channel, $\pi\eta \to K\overline K$, $\pi\eta^\prime \to
K\overline K$, $\pi\omega \to K^*\overline K + K\overline K^*$,
$\eta\eta \to K\overline K$, $\rho\eta \to K^*\overline K + K\overline
K^*$, $\eta\omega \to K^*\overline K + K\overline K^*$,
$\eta\eta^\prime \to K\overline K$, $\rho\omega \to K\overline K$,
$\rho\eta^\prime \to K^*\overline K + K\overline K^*$, $\sigma\sigma
\to K\overline K$, $\omega\omega \to K\overline K$, $\omega\eta^\prime
\to K^*\overline K + K\overline K^*$, $\eta^\prime\eta^\prime \to
K\overline K$, a constant cross section of $2 \mb$ is assumed.  The
respective back reactions are included, and their cross sections are
calculated according to the detailed-balance relation
\cref{detBal}. The cross sections of reactions with a vector and a
pseudoscalar meson in the initial channel and outgoing $K \overline
K$, such as, e.g., $\rho \pi \to K \overline K$, are small due to
$P$-wave suppression and are not included directly (see also
discussions in \cite{Brown:1991ig,Tomasik:2005ng}).  However, the
latter reaction proceeds via an intermediate $\phi$ meson. The
contributions from the processes involving a scalar $\sigma$ meson and
a pseudoscalar or vector meson in the initial channel either vanish
due to parity violation (as, e.g., $\sigma\pi \to K \overline K$) or
are small due to the $p$-wave suppression (as, e.g., $\sigma\rho \to
K^*\overline K + K\overline K^*$) and are neglected as well.

\section{\texorpdfstring{Elementary $\ell N$ reactions}{Elementary l-N reactions}}
\label{sec:elementary_leptonN}

Elementary interactions of leptons and photons with nucleons are
treated differently than those of mesons and baryons.  In this section
we explain in detail our implementation of the corresponding cross
sections.

In the region of intermediate lepton beam energies
($E_{\text{beam}}\sim 0.5 \upto 2\GeV$), the cross section contains
contributions from quasi-elastic (QE) scattering ($ \ell N \to \ell'
N'$), resonance (R) excitation ($\ell N \to \ell' R$) and direct,
i.e., non-resonant, single-pion production ($\ell N \to \ell' \pi N'$)
treated as background (1$\pi$ BG).  At higher energies, channels open
up, which can not be described by resonance decays any more.

Thus we assume
\begin{equation}
  \label{eq:generalDecomposition}
  \dd \sigma_\subtot=\dd \sigma_\subQE + \sum_R \dd \sigma_\subR + \dd
  \sigma_\subBGpi  + \dd\sigma_{2\pi\text{BG}} + \dd\sigma_\text{DIS}.
\end{equation}
The dynamics of the interaction is encoded in the absolute value of
the matrix element squared, averaged (summed) over initial (final)
spins,
\begin{equation}
  \overline{|\ME_{\subQE,\subR,\subBG}|^2} = C_{\subEM,\subCC,\subNC}^2 L_{\mu \nu} H^{\mu \nu}_{\subQE,\subR,\subBG} ,
\end{equation}
with $C_{\subEM}=4 \pi \alpha/Q^4$ for electromagnetic (EM)
interactions ($e^- N \rightarrow e^- X $), $C_{\subCC}=G_F \cos
\theta_C /\sqrt{2}$ for charged-current (CC) weak interactions
($\nu_\ell N \rightarrow \ell^- X$) and $C_{\subNC}= G_F /\sqrt{2}$
for neutral-current (NC) ones ($\nu N \rightarrow \nu X$).  Here,
$Q^2$ is the four-momen\-tum transfer; $\alpha \simeq 1/137$ the
fine-structure constant, $G_F = 1.16637 \cdot 10^{-5}\GeVminsq $ the
Fermi constant, and $\cos\theta_C = 0.9745$ the Cabibbo angle. The
leptonic tensor is given by
\begin{equation}
  L_{\mu \nu}= \frac{1+a}{2} \Tr \left[ (\slashed{k} + m_{\ell} )
    \tilde{l}_\mu (\slashed{k}' + m_{\ell'} ) l_\nu \right]
  , \label{eq:leptontensor}
\end{equation}
where $l_\mu=\gamma_{\mu} (1- a \gamma_5)$ and $\tilde{l}_\mu=\gamma_0
l_\mu^\dagger \gamma_0$; $k$ ($k'$) denotes the four-momentum of the
incoming (outgoing) lepton and $m_\ell$ ($m_{\ell'}$) the
corresponding mass. The parameter $a$ depends on the reaction
process: $a=0$ for EM processes and $a=1$ for CC or NC neutrino
scattering.

The hadronic currents in $H^{\mu \nu}_{\subQE,\subR,\subBG}$ have to
be parametrized in terms of form factors and thus depend not only on
the final state but also on the specific process, namely EM, CC, or
NC.

In the following subsections we present a brief summary of our model;
an extended discussion with all details can be found in
\cite{buss_phd,Leitner:2008ue,LeitnerDr}.

\subsection{Quasi-elastic scattering}

The cross section for quasi-elastic scattering, $\ell(k) N(p) \to
\ell'(\kpr) N'(\ppr)$, is given by
\begin{align}
  \frac{\dd\sigma_{\subQE}}{\dd \omega \; \dd \Omega_{\kpr}} =
  \frac{|\bvec{k}'|}{32 \pi^2} \; \frac{ \delta (\ppr^2 - \Mpr^2)
  }{\sqrt{ (k \cdot p)^2- \ml^2 M^2}\ } \; \overline{|\ME_\subQE|^2} \
  , \label{eq:QEdoublediff_xsec}
\end{align}
with $M= \sqrt{p^2}$; $\omega=\kz-\kprz$ is the energy transfer and
$\Omega_{\kpr}$ is the solid angle between incoming and outgoing
leptons.

The hadronic tensor, $H^{\mu \nu}_{\subQE}$, for quasi-elastic
scattering is given by
\begin{align}
  H^{\mu \nu}_{\subQE} = \frac12 \Tr \left[ (\slashp + M )
    \tilde{\Gamma}^{\mu}_{\subQE} (\slashppr + \Mpr )
    \Gamma^{\nu}_{\subQE} \right] , \label{eq:QEhadronictensor}
\end{align}
with
\begin{equation}
  \tilde{\Gamma}^{\mu}_{\subQE}=\gamma_0 {\Gamma^{\mu}_{\subQE}}^{\!\!\!\!\dagger} \gamma_0 .
\end{equation}
$\Gamma^{\mu}_{\subQE}$ has a $V-A$ Dirac structure,
\begin{equation}
  \Gamma^{\mu}_{\subQE}=\mathcal{V}^{\mu}_{\subQE} - \mathcal{A}^{\mu}_{\subQE} , \label{eq:QEcurrent}
\end{equation}
with the vector part,
\begin{equation}
  \mathcal{V}^{\mu}_{\subQE}=\fff_1 \gamma^\mu  + \frac{\fff_2}{2 m_N} \ii \sigma^{\mu\alpha} q_\alpha ,
  \label{eq:QEVECcurrent}
\end{equation}
and the axial-vector part,
\begin{equation}
  -\mathcal{A}^{\mu}_{\subQE}= \fff_A \gamma^\mu \gamma_5 + \frac{\fff_P}{m_N} q^\mu \gamma_5 .
  \label{eq:QEAXcurrent}
\end{equation}
Here, $q_{\mu}=p'_{\mu}-p_{\mu}$; $\fff_i$ ($i=1,2$) stands either for
the CC form factors, $F_i^V$, for the NC form factors,
$\tilde{F}_i^N$, or the EM form factors, $F^N_i$, with $N=p,n$;
$\fff_A$ for the CC form factor, $F_A$, and the NC form factor,
$\tilde{F}_A^N$ (analogous for $\fff_P$). All form factors depend on
$Q^2=-q^2$. Details on the form factors for QE scattering are given in
Chapter~4 of~\refcite{LeitnerDr}.

\subsection{Resonance excitation}
\label{sec:elementary_leptonN_Res}

This section is devoted to the second part in our general
decomposition of the cross section given in
\cref{eq:generalDecomposition}, namely the excitation of resonances
$\sum_R \dd \sigma_R$.

The cross section for resonance excitation $\ell(k) N(p) \to
\ell'(\kpr) R(\ppr)$ is given by
\begin{equation}
  \frac{\dd\sigma_{\subR}}{\dd \omega \; \dd \Omega_{\kpr}} =
  \frac{\abskpr}{32 \pi^2} \; \frac{ \mathcal{A}(\ppr^2) }{\sqrt{ (k
      \cdot p)^2- \ml^2 M^2}\ } \; \overline{|\ME_\subR|^2}
  ; \label{eq:RESdoublediff_xsec}
\end{equation}
with $M= \sqrt{p^2}$ and the resonance-spectral function,
$\mathcal{A}(\ppr^2)$, according to \cref{sec:selfenergies}.

The hadronic tensor for the excitation of a spin-1/2 resonance is
given by
\begin{equation}
  H^{\mu \nu}_{1/2} = \frac12 \Tr \left[ (\slashp + M )
    \tilde{\Gamma}^{\mu}_{1/2} (\slashppr + \Mpr ) \Gamma^{\nu}_{1/2}
  \right],
\end{equation}
with $\Mpr= \sqrt{{\ppr}^2}$ and
\begin{equation}
  \tilde{\Gamma}^{\mu}_{1/2}=\gamma_0 {\Gamma^{\mu}_{1/2}}^{\!\!\!\!\dagger} \gamma^0.
\end{equation}
For states with positive parity (e.g., \res{P}{11}{1440}),
\begin{equation}
  \Gamma^{\mu}_{1/2+}=\mathcal{V}^{\mu}_{1/2} - \mathcal{A}^{\mu}_{1/2} ,
\end{equation}
and for states with negative parity (e.g., \res{S}{11}{1535}),
\begin{equation}
  \Gamma^{\mu}_{1/2-}=\left[ \mathcal{V}^{\mu}_{1/2} - \mathcal{A}^{\mu}_{1/2} \right] \gamma^5 , \label{eq:spinhalfcurrentnegparity}
\end{equation}
where the vector part, $\mathcal{V}^{\mu}_{1/2}$, is given by
\begin{equation}
  \mathcal{V}^{\mu}_{1/2}=\frac{\fff_1}{(2 m_N)^2} \left( Q^2 \gamma^\mu + \slashq q^\mu \right) + \frac{\fff_2}{2 m_N} \ii \sigma^{\mu\alpha} q_\alpha  \label{eq:vectorspinhalfcurrent}
\end{equation}
and the axial part $\mathcal{A}^{\mu}_{1/2}$ by
\begin{equation}
  -\mathcal{A}^{\mu}_{1/2}= \fff_A \gamma^\mu \gamma_5 + \frac{\fff_P}{m_N} q^\mu \gamma^5 . \label{eq:axialspinhalfcurrent}
\end{equation}
As in the QE case, $\fff_i$ ($i=1,2$) stands either for the CC form
factors, $F_i^V$, for the NC form factors, $\tilde{F}_i^N$, or the EM
form factors, $F^N_i$, with $N=p,n$; analogous for $\fff_A$ and
$\fff_P$.

The excitation of a spin-3/2 final state is described within a
Rarita-Schwinger formalism, where the hadronic tensor is given by
\begin{equation}
  H^{\mu \nu}_{3/2}= \frac12 \Tr \left[ (\slashp + M )
    \tilde{\Gamma}^{\alpha \mu}_{3/2} \Lambda_{\alpha \beta}
    \Gamma^{\beta \nu}_{3/2} \right] ,
\end{equation}
with the spin-3/2 projector
\begin{equation}
  \begin{split}
    \Lambda_{\alpha \beta}=& - \left(\slashppr + \Mpr \right) \\
    &\times \left( g_{\alpha \beta} - \frac{2}{3} \frac{\ppr_{\alpha }
        \ppr_{\beta }}{\Mpr^2} + \frac{1}{3} \frac{\ppr_{\alpha }
        \gamma_{\beta} - \ppr_{\beta } \gamma_{\alpha}}{\Mpr} -
      \frac{1}{3} \gamma_{\alpha} \gamma_{\beta} \right),
  \end{split}
\end{equation}
and
\begin{equation}
  \tilde{\Gamma}^{\alpha \mu}_{3/2}=\gamma_0 {\Gamma^{\alpha \mu}_{3/2}}^{\dagger} \gamma_0 .
\end{equation}

For states with positive parity as the \res{P}{33}{1232}, we have
\begin{equation}
  \Gamma^{\alpha \mu }_{3/2+} = \left[ \mathcal{V}^{\alpha \mu }_{3/2} -
    \mathcal{A}^{\alpha \mu }_{3/2}\right] \gamma_{5} ,
\end{equation}
and for the negative-parity ones (e.g., \res{D}{13}{1535}),
\begin{equation}
  \Gamma^{\alpha \mu }_{3/2-} = \mathcal{V}^{\alpha \mu }_{3/2} - \mathcal{A}^{\alpha \mu }_{3/2} .
\end{equation}
In terms of the form factors, the vector part is given by
\begin{equation}
  \begin{split}
    \mathcal{V}^{\alpha \mu }_{3/2} =& \frac{\ffc_3^V}{m_N} (g^{\alpha
      \mu} \slashq - q^{\alpha} \gamma^{\mu})+
    \frac{\ffc_4^V}{m_N^2} (g^{\alpha \mu} q\cdot \ppr - q^{\alpha} {\ppr}^{\mu}) \\
    & + \frac{\ffc_5^V}{m_N^2} (g^{\alpha \mu} q\cdot p - q^{\alpha}
    p^{\mu}) + g^{\alpha \mu}
    \ffc_6^V \label{eq:vectorspinthreehalfcurrent}
  \end{split}
\end{equation}
and the axial part by
\begin{align}
  -\mathcal{A}^{\alpha \mu }_{3/2} =& \left[\frac{\ffc_3^A}{m_N}
    (g^{\alpha \mu} \slashq - q^{\alpha} \gamma^{\mu})+
    \frac{\ffc_4^A}{m_N^2} (g^{\alpha \mu} q\cdot \ppr - q^{\alpha}
    {\ppr}^{\mu}) \right. \nonumber \\ & \left.  + {\ffc_5^A}
    g^{\alpha \mu} + \frac{\ffc_6^A}{m_N^2} q^{\alpha} q^{\mu}\right]
  \gamma_{5} . \label{eq:axialspinthreehalfcurrent}
\end{align}
As before, the calligraphic $\ffc$ stands either for the CC form
factors, $C_i^{V,A}$, $i=3,\ldots,6$, the electromagnetic transition
form factors, $C^N_i$, with $N=p,n$, or the NC form factors,
$\tilde{C}_i^{V,A \; N}$, as detailed in Table 5.5 in
\refcite{LeitnerDr}. Note that current conservation implies $C_6^N=0$.

Any formalism describing resonances with spins greater than 3/2 is
highly complicated \cite{Shklyar:2004ba}. Thus, for simplification, we
assume that all resonances with spin $> 3/2$ can be treated with the
spin-3/2 formalism. This rough approximation introduces only small
errors since the contribution of resonances beyond the $\Delta$ is
small and soon overshadowed by the opening of DIS processes.

The electromagnetic/vector form factors can be fixed using helicity
amplitudes extracted from electron scattering experiments. The
explicit relations between the form factors and the helicity
amplitudes are given in Appendix E of \refcite{LeitnerDr}. We use
these relations to extract the form factors from the results of the
recent MAID2005 analysis
\cite{tiatorprivcomm,MAIDWebsite,Tiator:2006dq,Drechsel:2007if} for
the helicity amplitudes including their $Q^2$ dependences. In this
analysis, 13 resonances with invariant masses of less than $2 \GeV$
are included. They are listed in \cref{table:resPhoto}

Experimental information on the $N-R$ axial form factors is very
limited. In QE case, Goldberger-Treiman relations allowing to express
the nucleon weak axial coupling through the strong pion-nucleon
coupling have been derived \cite{Fogli:1979cz} from the condition of
partially conserved axial current (PCAC).  We follow this approach and
apply PCAC and pion-pole dominance to derive the axial couplings for
the resonances.  The equations obtained in this way are sometimes
called the off-diagonal Goldberger-Treiman relations.  In the case of
the $\Delta$, we use the data to fit the $Q^2$ dependence. For
higher-lying resonances, which have only a small effect on
observables, we simply approximate the $Q^2$ dependence by a dipole
form factor with axial mass $M_A = 1\GeV$. An extended discussion of
axial form factors is given in \cite{LeitnerDr}.

The cross sections for photon-induced reactions are given by the limit
$Q^2\to 0$ of the above expressions,
\begin{equation}
  \begin{split}
    \sigma_{\gamma N \to R}&= \frac{1}{4\sqrt{(q_\alpha p^\alpha)^2} }
    \frac{\dd^4 \pr{p}}{(2\pi)^4} ~2\pi \mathcal{A}_R(\pr{p}^2)
    ~\overline{|\ME_R|^2}~
    (2\pi)^4 \delta^4\left(p+q-\pr{p}\right) \\
    &= \left. \frac{1}{4\sqrt{(q_\alpha p^\alpha)^2} } ~2\pi
      \mathcal{A}_R(\pr{p}^2) ~\overline{|\ME_R|^2}~
    \right|_{p+q=\pr{p}},
  \end{split}
\end{equation}
where $q$ is the four-momentum of the real photon (with $q^2=0$); the
averaged squared matrix element is given by
\begin{equation}
  \overline{|\ME_R|^2}=-\frac{1}{2} H_R^{\mu\nu} g_{\mu\nu}\;.
\end{equation}
As in the lepton case, we use helicity amplitudes extracted from the
MAID analysis \cite{MAIDWebsite} to determine the contributions of
individual resonances and to fix the form factors \cite{buss_phd}.

\begin{table}
  \centering
  \begin{tabular}{lccc}
    \noalign{\vspace{-8pt}}
    \hline
    \toprule
    name            &  spin    & isospin   & parity     \\
    \midrule
    P$_{33}$(1232)  &  3/2 & 3/2 & +     \\
    P$_{11}$(1440)  &  1/2 & 1/2 & +     \\
    D$_{13}$(1520)  & 3/2 & 1/2  & - \\
    S$_{11}$(1535)  & 1/2 & 1/2  & - \\
    S$_{31}$(1620)  & 1/2 & 3/2 & - \\
    S$_{11}$(1650)  & 1/2 & 1/2 & - \\
    D$_{15}$(1675)  & 5/2 & 1/2 & - \\
    F$_{15}$(1680)  & 5/2 & 1/2 & + \\
    D$_{33}$(1700)  & 3/2 & 3/2 & - \\
    P$_{13}$(1720)  & 3/2 & 1/2 & + \\
    F$_{35}$(1905)  & 5/2 & 3/2 & + \\
    P$_{31}$(1910)  & 1/2 & 3/2 & + \\
    F$_{37}$(1950)  & 7/2 & 3/2 & + \\
    \bottomrule
  \end{tabular}
  \caption[Resonances included in the MAID analysis]{Resonances included in the MAID analysis \cite{MAIDWebsite}.}
  \label{table:resPhoto}
\end{table}

\subsection{One-pion background}
\label{sec:elementary_leptonN_OnePi}

This section is devoted to the third term of
\cref{eq:generalDecomposition}, i.e., non-resonant $\pi$ production.
The need for a one-$\pi$ non-resonant background contribution, $\dd
\sigma_\subBGpi$, is justified by the fact, that the pure resonance
contributions underestimate the total one-$\pi$-production cross
section in the isospin $1/2$ channels.

Recent progress has been achieved in models treating the one-$\pi$
background as a sum of Feynman diagrams with a pion and a nucleon in
the final state
\cite{Sato:2003rq,Hernandez:2007qq,Barbero:2008zza}. In this approach
the process of $\Delta$-resonance production with its following decay
is just one of the considered diagrams ($\Delta$ pole), along with six
others (crossed $\Delta$, nucleon pole, crossed nucleon, contact term,
pion pole and pion in flight). These six diagrams introduce new
vertices, which can be described within various models for
nucleon-meson interactions.

As one of the possibilities to account for the background
contribution, in the GiBUU code we adopt the approach of Hernandez
\refetal{Hernandez:2007qq}, where the new vertices are described
within the $\text{SU}(2)$-nonlinear $\sigma$ model. The model is
applied to electron and neutrino scattering on nucleons
\cite{Lalakulich:2010ss}.  The ANL and BNL data for integrated cross
sections and various one-differential distributions
\cite{Radecky:1981fn,Kitagaki:1986ct} are described with an accuracy
which is of the same order of magnitude as the experimental
uncertainties between different data sets.

The advantage of this theoretical approach is that it gives a clear
picture of the underlying processes and their kinematics, and it takes
interference effects into account explicitly. The disadvantage is
related to the fact, that the new vertices mentioned above, as well as
the relative signs of the various amplitudes, are not always
known. In~\cite{Lalakulich:2010ss} it is shown, that the model gives
good results for pion-nucleon invariant masses $W<1.4 \GeV$.  However,
its extension to higher invariant masses, at least to the second
resonance region, is not yet achieved and requires significant
efforts.

Therefore, to obtain the background in the entire resonance region,
that is up to $W<2\GeV$, we have adopted a phenomenological approach,
described in detail in~\cite{buss_phd} and shortly summarized in the
following. This approach starts from the cross sections, since the
amplitudes are not available, and separates them into a background and
a resonance contribution.  We illustrate this first for
electromagnetic processes. We use the measured unpolarized
one-$\pi$-production cross section $\ell(k) N(p) \to \ell'(\kpr)
\pi(\kpi) N(\ppr)$ and subtract the theoretically deduced resonance
contribution.  The rest, which then includes the genuine background,
but also the resonance-background interference terms, we use as
one-$\pi$ background, $\dd \sigma_\subBGpi$.  We thus
obtain~\cite{Leitner:2008ue}
\begin{equation}
  \frac{\dd\sigma_\subBGpi^{\text{V}}}{\dd \omega \dd\Omega_{\kpr} \dd\Omega_{\kpi} }  =
  \frac{\dd\sigma_{N \pi}^{\text{V}}}{\dd \omega \dd\Omega_{\kpr} \dd\Omega_{\kpi} }
  -\sum_{R}\frac{\dd\sigma_{\ell N\to \ell R\to \ell N\pi}^{\text{V}}}{\dd \omega \dd\Omega_{\kpr} \dd\Omega_{\kpi} } .
  \label{eq:piBG}
\end{equation}
The superscript, $V$, denotes here, that in electromagnetic processes
only the vector part of the current contributes.

\begin{figure}[t]
  \centering
  \includegraphics[width=0.6\linewidth]{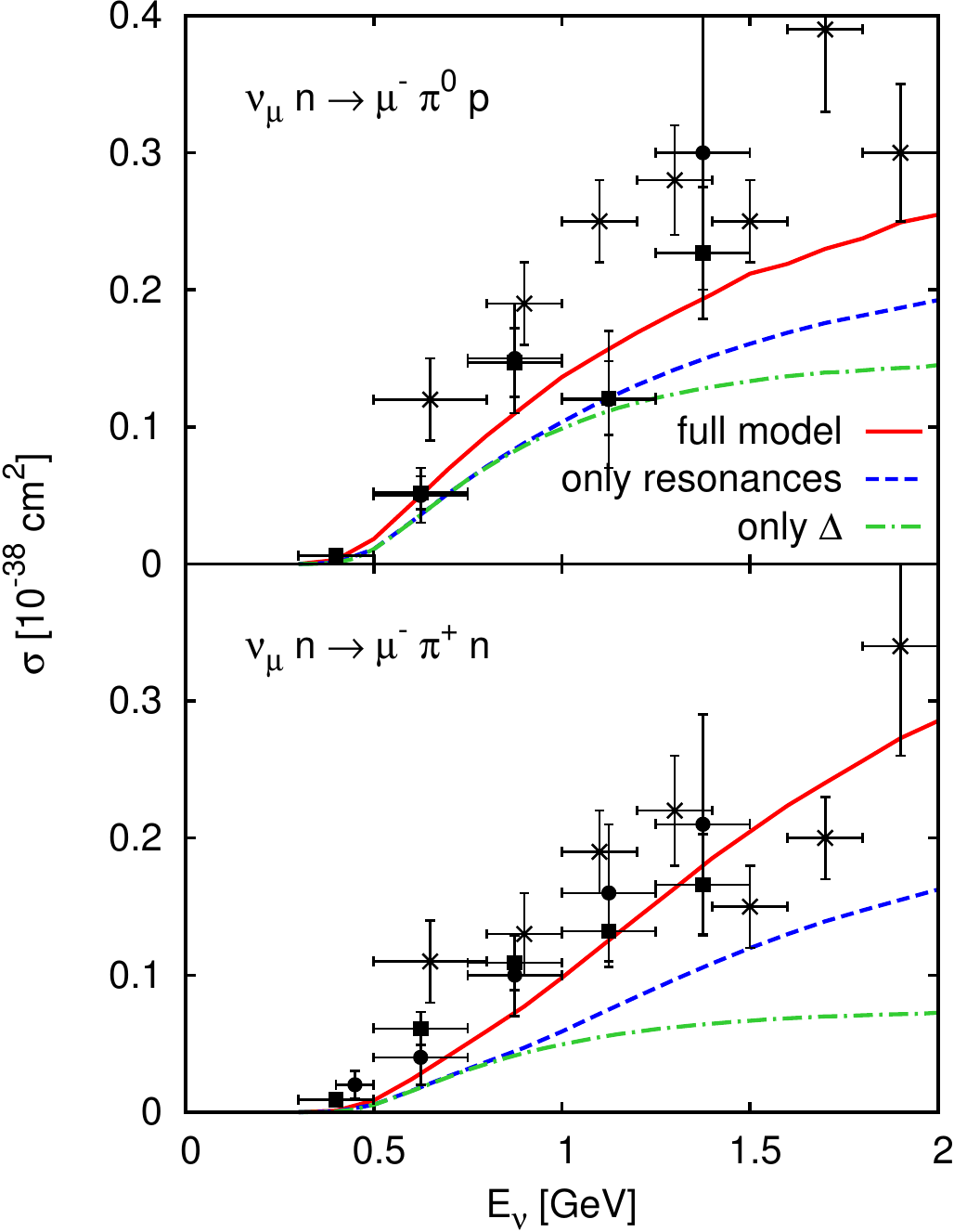}
  \caption{(Color online) Total CC pion-production cross sections for
    the mixed isospin channels as a function of the neutrino energy
    compared to the pion-production data from ANL
    (Refs.~\cite{Barish:1978pj} ($\bullet$) and \cite{Radecky:1981fn}
    ($\blacksquare$)) and BNL (\cite{Kitagaki:1986ct} ($\times$)). The
    solid lines denote our full result including the non-resonant
    background following \cref{eq:non-resBG} with $b^{p \pi^0}=3$ and
    $b^{n \pi^+}=1.5$.  Furthermore, we show the results for pion
    production only through the excitation and the subsequent decay of
    all resonances (dashed lines) or through the $\Delta$ alone
    (dash-dotted lines). No cut on the invariant mass is applied
    (taken from~\cite{Leitner:2008ue}).}
  \label{fig:xsec_pionprod_wBG}
\end{figure}

The first term of the r.h.s.~of \cref{eq:piBG} denotes the measured
one--pion production cross section $\dd\sigma_{N \pi}^{\text{V}}$. For
simplicity we obtain it from the MAID
analysis~\cite{tiatorprivcomm,MAIDWebsite,Tiator:2006dq,Drechsel:2007if}
(the MAID amplitudes are integrated into the GiBUU code).

The second term of the r.h.s.~of \cref{eq:piBG}, the resonance
contribution, is again obtained from the MAID analysis. In order to
get the required angular distribution we assume that the resonances
produced decay isotropically in their rest frame
\begin{equation}
  \frac{\dd \Gamma_{R \to N\pi}}{\dd \Omega_{k_\pi}^{\text{cm}}} = \frac{\Gamma_{R \to N\pi}}{4 \pi} ~.
\end{equation}
Hence,
\begin{equation}
  \begin{split}
    \frac{\dd\sigma_{\ell N\to \ell R\to \ell N\pi}^{\text{V}}}{\dd
      \omega \dd\Omega_{\kpr} \dd\Omega_{\kpi} } & =
    \frac{\dd\sigma_R^\text{V}}{\dd \omega \dd\Omega_{\kpr}}
    \frac{1}{\Gamma_R} \frac{\Gamma_{R\to N\pi}}{4 \pi}
    \frac{\dd \Omega_{k_\pi}^{\text{cm}}}{\dd \Omega_{\kpi}} \\
    & = \frac{\dd\sigma_R^\text{V}}{\dd \omega \dd\Omega_{\kpr}}
    \frac{1}{\Gamma_{R}} \frac{\Gamma_{R\to N\pi}}{4 \pi}
    \frac{\sqrt{\ppr^2} \bvec{k}_{\pi}^2}{|\bvec{k}_{\pi}^{\text{cm}}
      | \left(|\bvec{k}_{\pi} |{\ppr}^0-|\bvec{p}'| \kpi^0
        \cos\theta_\pi\right)} \ ,
  \end{split}
\end{equation}
where $\Gamma_R$ is the total decay width of the resonance,
$\theta_\pi=\measuredangle(\bvec{k}_{\pi},\bvec{p}')$ and the vector
part of the resonance production cross section has been introduced in
the previous section.  The quantity
$\frac{\dd\sigma_\subBGpi^{\text{V}}}{\dd \omega \; \dd\Omega_{\kpr}}$
can now be retrieved by an integration over the solid angle of the
outgoing pion.

The background for neutrino reactions includes vector, axial, and
interference contributions,
\begin{equation}
  \dd \sigma_\subBGpi = \dd \sigma_\subBGpi^{\text{V}} + \dd \sigma_\subBGpi^{\text{A}}+ \dd \sigma_\subBGpi^{\text{V/A}}
  = \dd \sigma_\subBGpi^{\text{V}} + \dd \sigma_\subBGpi^{\text{non-V}} .
\end{equation}
The vector part is fully determined by electron-scattering data as
described above. The axial and the interference terms collected under
the label ``non-V'' are fitted to the available data for neutrino
reactions in the isospin-$1/2$ channels.

We assume that $\dd \sigma_\subBGpi^\text{V}$ and $\dd
\sigma_\subBGpi^\text{non-V}$ have the same functional form, i.e.,
\begin{equation}
  \dd {\sigma_\subBGpi} = \dd \sigma_\subBGpi^\text{V} + \dd
  \sigma_\subBGpi^\text{non-V} = (1+b^{N \pi})\; \dd
  \sigma_\subBGpi^\text{V} ,
  \label{eq:non-resBG}
\end{equation}
where the global factor, $b^{N \pi}$, depends on the process. For CC
reactions, $b^{p \pi^0}=3$ for $\nu n \to \ell^- p \pi^0$, and $b^{n
  \pi^+}=1.5$ for $\nu n \to \ell^- n \pi^+$ give a reasonable
agreement with the ANL data.  This is shown in
\cref{fig:xsec_pionprod_wBG}. The background in the isospin-3/2
channel is neglected.

For NC reactions the data on one-$\pi$ production are scarce. We find
a good agreement with the data already without a non-resonant part. In
view of this, we abstain from fitting the non-vector part to these
data and neglect a NC background. More results on the one-$\pi$
background are presented in \cite{LeitnerDr,Leitner:2008ue}.

\subsection{Two-pion photoproduction}
\label{sec:elementary_leptonN_TwopiXsec}
\label{sec:twoPi_elementary}

\begin{figure}[p]
  \centering
  \includegraphics[width=0.8\linewidth]{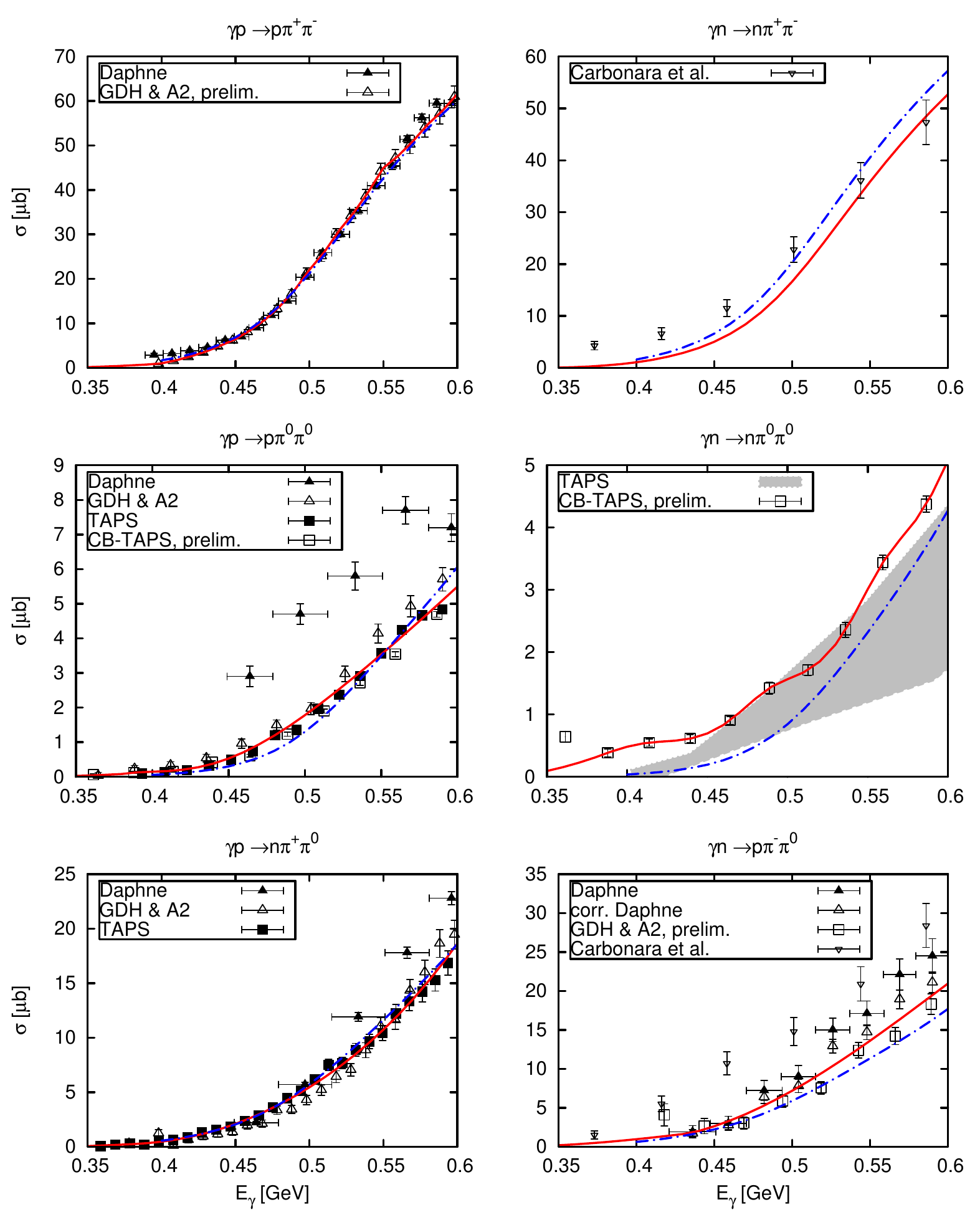}
  \caption[The elementary reaction $\gamma N\to N\pi\pi$]{(Color
    online) Different isospin channels for
    the reaction $\gamma N\to N\pi\pi$. For the error bars we added
    the systematical and statistical uncertainties
    ($\Delta^2=\Delta_\text{stat}^2+\Delta_\text{sys}^2$) as quoted by
    the experiments. The data are taken from
    \cite{Carbonara:1976tg,Zabrodin:1997xd,Braghieri:1994rf,Harter:1997jq,Wolf:2000qt,Kotulla:2003cx,Langgartner:2001sg,Kleber:2000qs,Ahrens:2003na,Ahrens:2005ia};
    preliminary data have been communicated by
    \cite{Jaegele_private,Ahrens:2007zzj,Pedroni_private}. In the panel,
    representing $\gamma n\to n\pi^0 \pi^0$, the shaded area shows the
    cross-section measurement by \cite{Kleber:2000qs}. The solid line
    represents our parametrizations, which we use as input for the
    nuclear targets. The dotted-dashed curves show the result obtained
    with the original model of Nacher
    \etal{}~\cite{Nacher:2000eq,luisPrivate} based on Tejedor
    \etal{}~\cite{GomezTejedor:1994yr,GomezTejedor:1995pe}.  Taken
    from \cite{buss_phd}.}
  \label{fig:2piEle}
\end{figure}

Already in the 1960's and 70's, first extensive photon-induced $\pi\pi$
production experiments on elementary targets have been performed
(cf.~\cite{Hauser1967,:1968ke,Carbonara:1976tg} and references
therein). More recently, the TAPS
\cite{Harter:1997jq,Wolf:2000qt,Kotulla:2003cx,Langgartner:2001sg,Kleber:2000qs},
DAPHNE \cite{Zabrodin:1997xd,Braghieri:1994rf}, GDH
\cite{Ahrens:2003na,Ahrens:2005ia}, and GRAAL
\cite{Assafiri:2003mv}~collaborations have been examining $\pi\pi$
production with high statistical accuracy.  In \cref{fig:2piEle}, the
most recent data sets are plotted as an overview over the relevant
threshold region.  Neutron data (see \cite{buss_phd} for a detailed
description of the experimental situation) have still large
uncertainties, whereas the proton data seem to be reliable after a
period of contradiction between different experiments. The lines
represent our calculations as described below.

For the elementary two-pion-production process on the nucleon we apply
the model of Nacher \etal{}~\cite{Nacher:2000eq}, which is an updated
version of a model developed by Tejedor
\etal{}~\cite{GomezTejedor:1994yr,GomezTejedor:1995pe}.  This model
provides a reliable input for the momentum distributions of the pions
in the elementary process. It is based on a set of tree-level
diagrams, which include the coupling of the nucleons, pions, photons
and baryon resonances ($P_{33}(1232)$, $P_{11}(1440)$, $D_{13}(1520)$,
and $P_{33}(1700)$).

\Cref{fig:2piEle} shows the result of this model for the total cross
sections. Although the agreement of the model to data is quite good we
scale the total cross section to the available data before analyzing the
reaction in complex targets. Hence, for all channels besides $\gamma N
\to N \pi^+\pi^-$ we directly use the data measured by the TAPS, DAPHNE,
and the GDH{\&}A2 collaborations
\cite{Harter:1997jq,Wolf:2000qt,Kotulla:2003cx,Langgartner:2001sg,Kleber:2000qs,Zabrodin:1997xd,Braghieri:1994rf,Ahrens:2003na,Ahrens:2005ia,Ahrens:2007zzj,Pedroni_private,Jaegele_private}
to normalize the calculated cross sections, while we take the decay-mass
and momentum distributions from theory. In the threshold regions of
$\gamma p\to p\pi^0\pi^0$, $\gamma p\to p\pi^-\pi^+$, $\gamma n\to
n\pi^0\pi^0$, and $\gamma n\to p\pi^0\pi^-$, where no data are
available, we estimate the total cross section based on the
three-particle-phase-space structure. We assume the matrix element
$\mathcal{M}$ to be constant in this region and get
\begin{equation}
  \begin{split}
    \label{eq:threshEstimate}
    \sigma_\text{threshold}&= \frac{(2\pi)^4}{4m_N q_0}|\mathcal{M}|^2
    \int \frac{\dd^3 \bvec{k}_1}{2k^0_1(2\pi)^3} \frac{\dd^3
      \bvec{k}_2}{2k^0_2(2\pi)^3} \frac{\dd^3
      \bvec{p}\;^\prime}{2p^\prime_0(2\pi)^3}
    \delta^4\left[q+p-\left( p^\prime+\sum_{i=1}^2 k_i \right ) \right ] \\
    &=\frac{(2\pi)^4}{4m_N q_0}|\mathcal{M}|^2 \times 16\,(2\pi)^7
    \times \int \frac{\dd m^2_{12} \dd m^2_{13}}{s},
  \end{split}
\end{equation}
where $s$ is the Mandelstam $s$ of the process; $\dd m^2_{12}$ and
$\dd m^2_{13}$ are defined by
\begin{equation}
  \begin{split}
    m_{12}&=(k_1+k_2)^2,\\
    m_{13}&=(k_1+\pr{p})^2.
  \end{split}
\end{equation}
The value of $|\mathcal{M}|^2$ is now fixed by the value of the lowest
available data point,
\[
\sigma_\text{threshold}(\text{energy of lowest data
  point})=\sigma(\text{lowest data point}) \; .
\]
\Cref{table:M2Pi} shows a compilation of the invariant matrix elements
$|\mathcal{M}|^2$.
\begin{table}
  \begin{center}
    \begin{tabular}{llll}
      \toprule
      Channel & $E_\gamma$ [MeV]& $\sigma$ [$\mu$b] & $|\mathcal{M}|^2 \times(2\pi)^{11}$ \\
      \midrule
      $\gamma p\to p\pi^0\pi^0$             & 380 &   0.08128   &    4.871  \\
      $\gamma p\to p\pi^-\pi^+$             & 398 &   0.9083  \cite{Ahrens:2007zzj} &    34.74\\
      $\gamma n\to n\pi^0\pi^0$             & 387.9    & 0.3840 \cite{Jaegele_private}   & 18.60 \\
      $\gamma n\to p\pi^0\pi^-$             & 435 &    1.752    &    34.96 \\
      \bottomrule
    \end{tabular}
  \end{center}
  \caption[Threshold estimation for the $\gamma N\to N \pi \pi$ cross
  section]{Parameters for the threshold estimation of the total cross
    section: For each channel, we show the lowest possible energy at which
    we can reliably estimate the cross section from experiment and the
    extracted cross section, $\sigma$. The fourth column shows the
    corresponding value of the threshold-matrix element
    $|\mathcal{M}|^2$.}
  \label{table:M2Pi}
\end{table}

The solid line in the $\gamma n \to n \pi^+\pi^-$ panel in
\cref{fig:2piEle} shows the result of the model by Nacher
\etal{}~\cite{Nacher:2000eq} with a slightly adjusted set of
parameters.  The solid lines in the other panels represent our fits
and threshold estimates. Altogether, these six solid lines represent
our elementary input cross sections.

\subsection{Two-pion background}
\label{sec:elementary_leptonN_TwoPi}

In the previous section we described some involved model, which fits
the photoproduction data very accurately. However, this model is only
applicable to photoproduction and was not extended to nonzero $Q^2$
yet.  In GiBUU, however, we need the description of two-$\pi$
production in electron and neutrino reactions as well.  Keeping also
in mind, that eventually we want to study nuclear reactions, where
resonances modify their properties and propagate out of the nucleus,
we have to split the two-$\pi$ cross section into a resonant and a
background part.

\begin{figure}[t]
  \centering
  \includegraphics[width=0.75\linewidth]{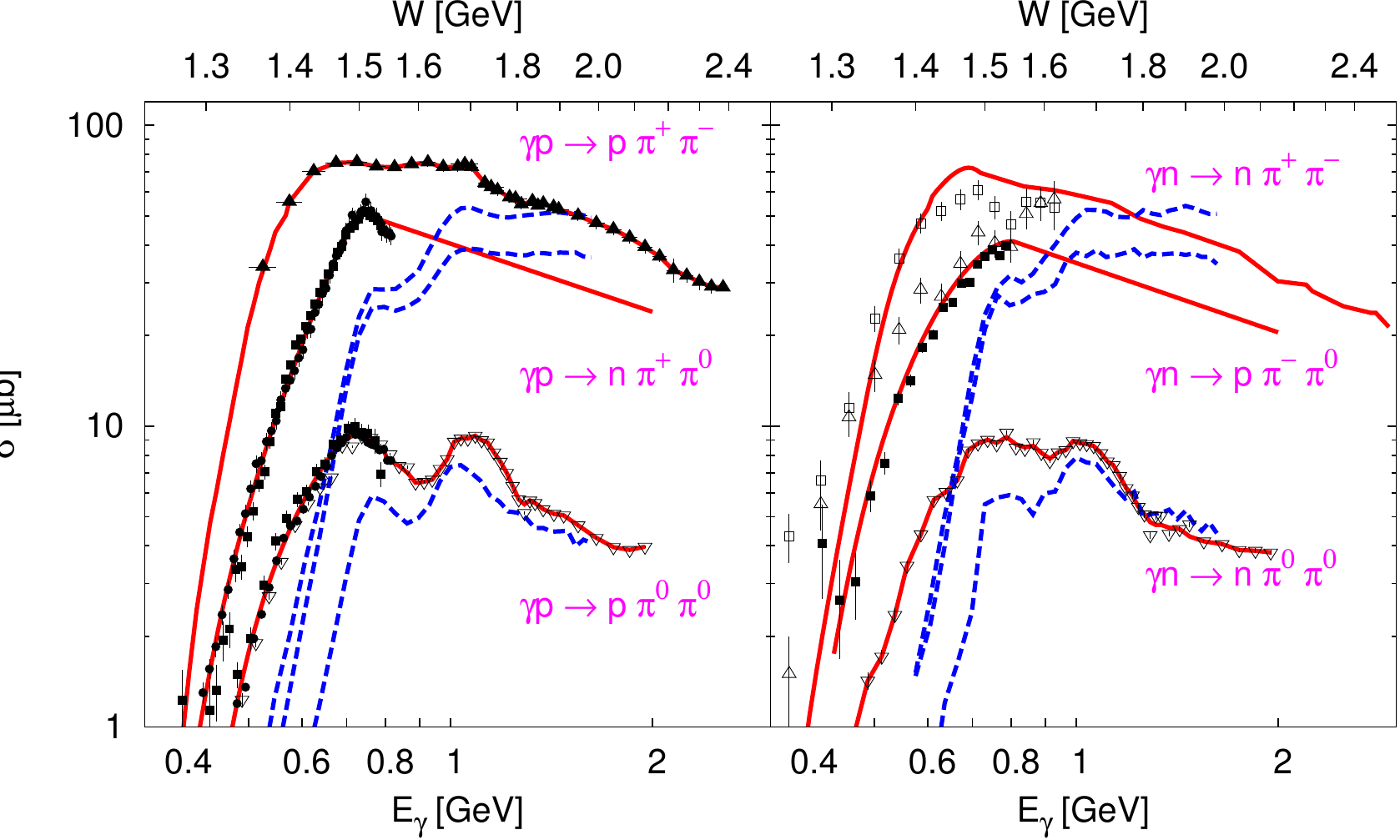}
  \caption{(Color online) Total photoproduction cross section of
    two-pion channels as function of the photon energy (bottom axes)
    or center of momentum energy (top axes), respectively. The solid
    lines indicate the cross-section parametrizations implemented in
    GiBUU intended to describe the experimental data.  The dashed
    lines show only resonance contribution.  Experimental data are
    from Carbonara \etal{} ($\square$,$\triangle$
    \cite{Carbonara:1976tg}), SAPHIR@ELSA ($\blacktriangle$
    \cite{Wu:2005wf}), GDH\&A2@Mainz ($\blacksquare$
    \cite{Ahrens:2003na,Ahrens:2005ia}), TAPS ($\bullet$
    \cite{Langgartner:2001sg,Kotulla:2003cx}), CB-TAPS (preliminary)
    ($\triangledown$ \cite{Jaegele_private}).}
  \label{fig:EM_twoPion_TotalXS}
\end{figure}

For the elementary reaction, the resonances are to be described
according to the general formalism as outlined in
\cref{sec:elementary_leptonN_Res}. Then we can use experimental
two-$\pi$ data to extract the background.

Let us consider EM interactions at low energies, as studied in detail
in \cite{buss_phd}.  In \cref{fig:EM_twoPion_TotalXS}, we show the
resonance contribution to the two-$\pi$ photoproduction cross sections
with various charge combinations in the final state. The data points
are the same that were used in \cref{sec:twoPi_elementary}.  At
$E_\gamma\lesssim 1\GeV$, all curves showing only the resonance
contribution are below the data, because the resonance production with
their following decays opens up only at higher energies.  Thus, we
define the two-$\pi$ background as
\begin{equation}
  \dd\sigma_{2\pi\text{BG}}=\dd\sigma_{2\pi}^{\rm exp} - \sum_R
  \left.\dd\sigma_R\right|_{R\to 2\pi}\quad \text{for} \quad Q^2=0\GeV^2 \ .
\end{equation}
This subtraction must be adjusted for all possible charge channels,
$\pi^+\pi^-$,$\pi^0\pi^0$, and $\pi^0\pi^\pm$, independently and
includes the resonance contributions of the following four channels
\begin{equation}
  \begin{array}{lllllll}
    \gamma N&\to& R &\to& \pi \Delta 	&\to &N'\pi\pi \; ,\\
    \gamma N&\to& R &\to& \pi P_{11}(1440) 	&\to &N'\pi\pi \; ,\\
    \gamma N&\to& R &\to& \rho N 		&\to &N'\pi\pi \; ,\\
    \gamma N&\to& R &\to& \sigma N 		&\to &N'\pi\pi \; .
  \end{array}
\end{equation}
Given $\sigma_{\gamma N\to R}$, the evaluation of $\sigma_{\gamma N\to
  R\to N'\pi\pi}$ for a special $\pi\pi$-charge state involves a
weighting with the partial decay widths for the intermediate $\pi
\Delta$, $\pi P_{11}(1440)$, $\rho N$, and $\sigma N$ channels and the
relevant isospin Clebsch-Gordan factors. The distribution of the
final-state momenta of the $\pi\pi$-background events are assumed to
follow the phase space distribution.

As expected, with increasing photon energy, the background
contribution vanishes.  At $E_\gamma \gtrsim 1\GeV$ the resonance
contribution even overestimates the full model calculations; here we
set the background to zero.

\begin{figure}[t]
  \centering
  \includegraphics[width=0.75\textwidth]{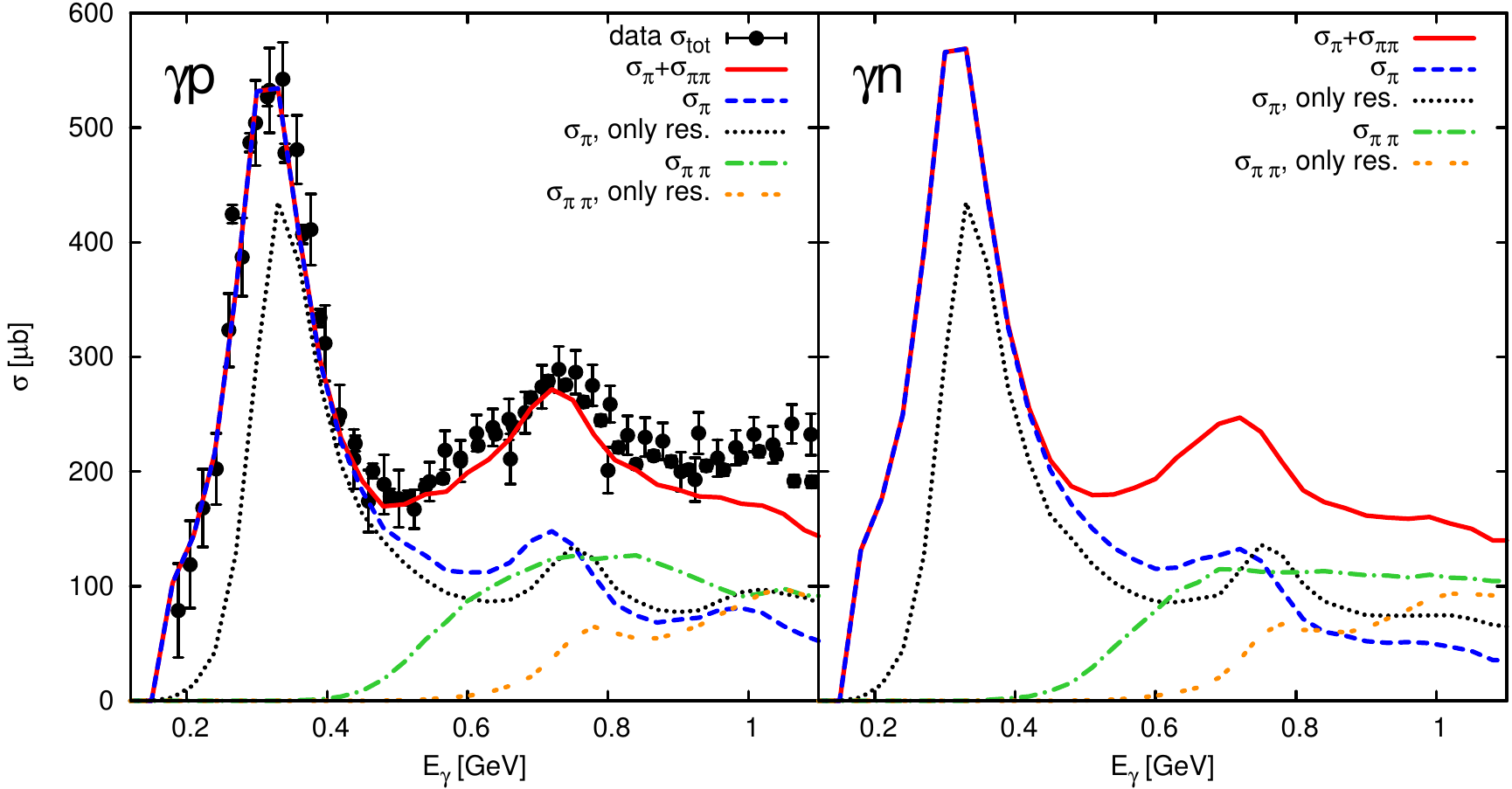}
  \caption[Photon reactions with the proton and the neutron]{(Color
    online) Photon-induced reactions on the proton (left panel) and
    the neutron (right panel) as a function of the incoming photon
    energy. The solid line shows the sum of single-$\pi$ and
    double-$\pi$ production for the full model, the dashed line shows
    the result for single-$\pi$ and the dashed dotted one gives
    $\pi\pi$; dotted and double-dashed curves show the contributions
    of resonances. The data are taken from \cite{Yao:2006px}. Taken
    from \cite{buss_phd}.}
  \label{fig:photonBG_2pi}
\end{figure}

The resulting two-$\pi$ contributions are shown in
\cref{fig:photonBG_2pi} together with the total
pho\-to\-pro\-duc\-tion cross section for both proton and neutron
targets. Obviously, the total cross section is well described for
$E_\gamma<0.9\,\text{GeV}$; the resonance contribution to $\pi\pi$ in
this region is in fact small compared to the total $\pi\pi$-production
cross section, especially at energies below $600 \MeV$.

\begin{figure}[t]
  \centering
  \includegraphics[width=0.4\linewidth]{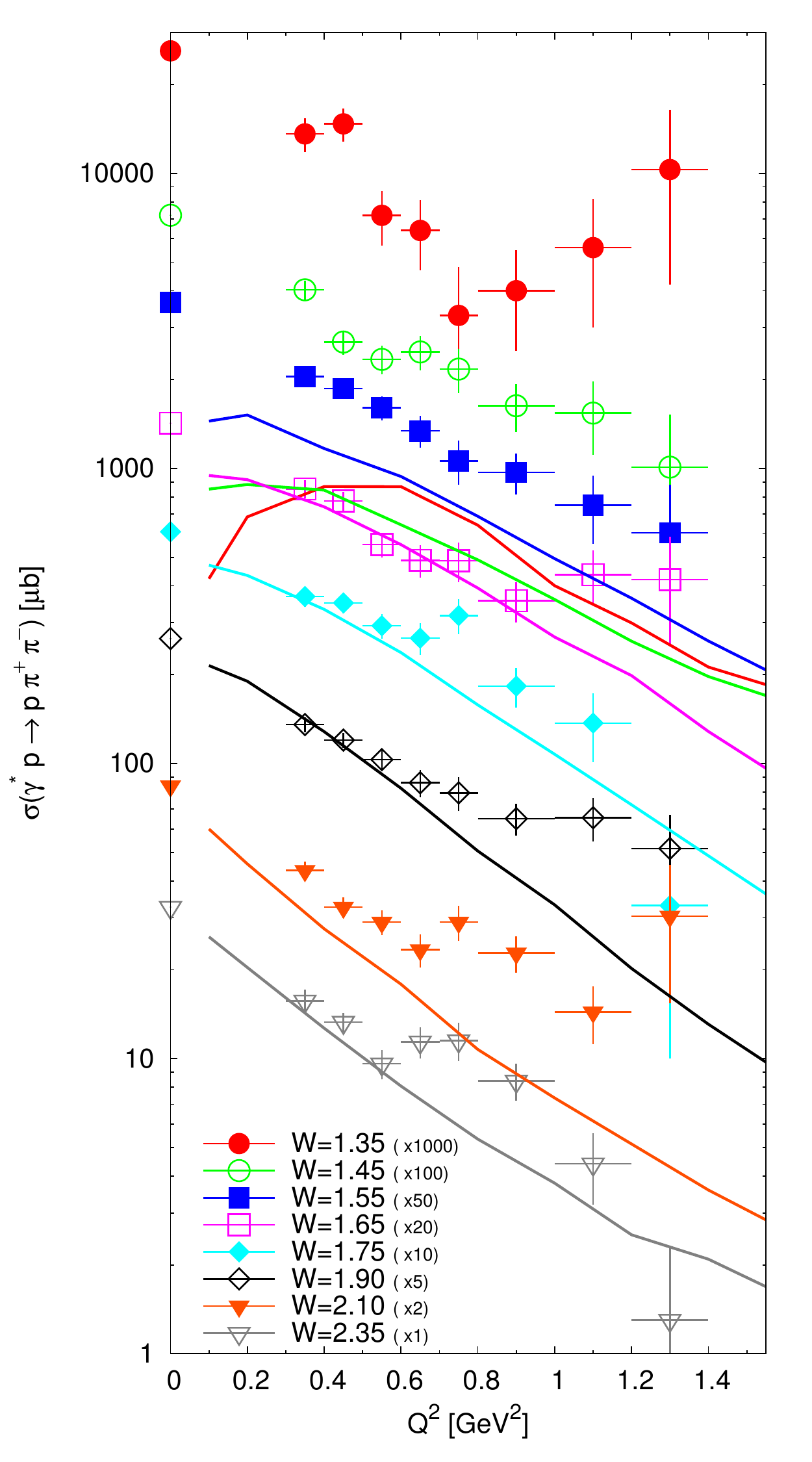}
  \includegraphics[width=0.4\linewidth]{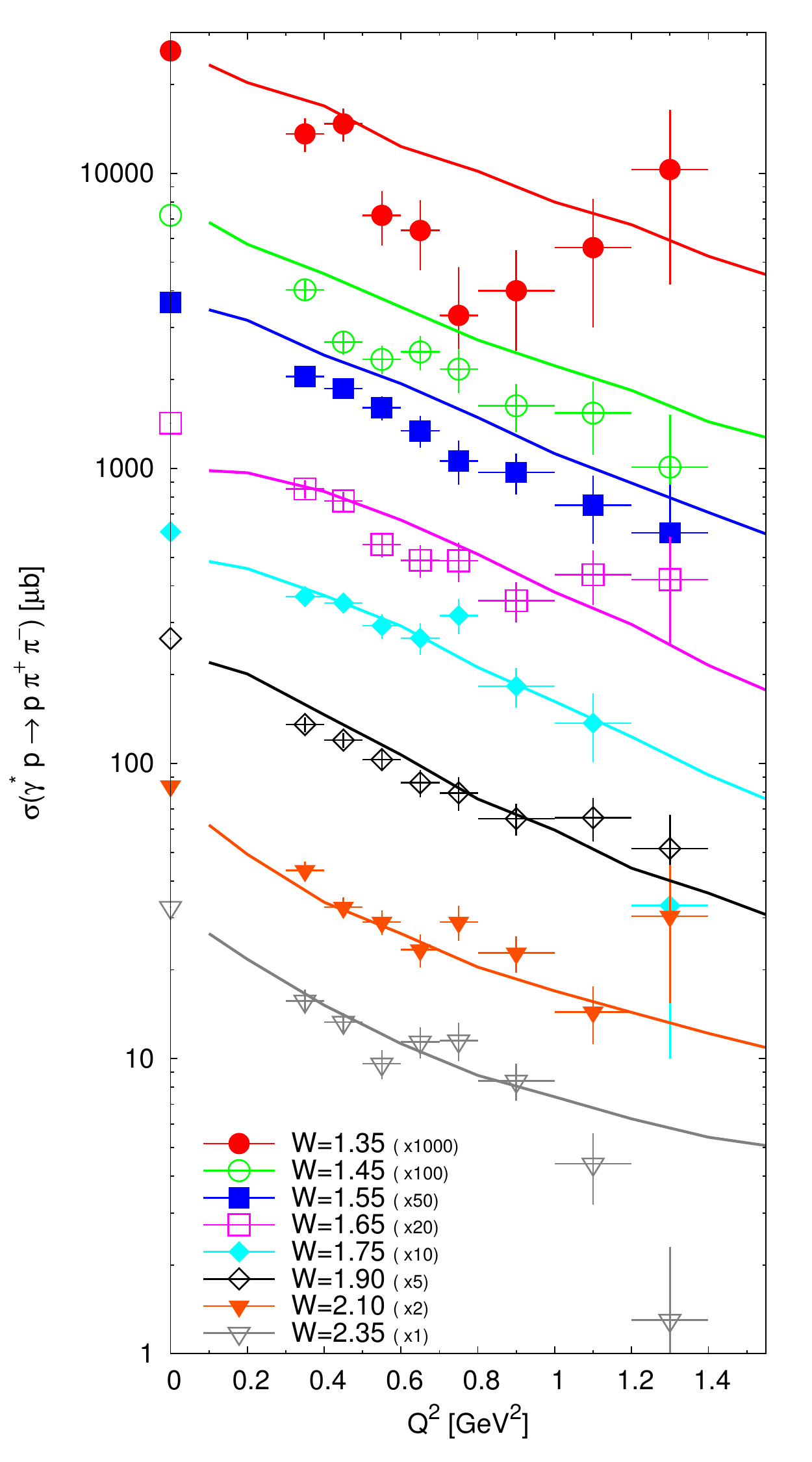}
  \caption{(Color online) Cross section of $\gamma^* p \to p
    \pi^+\pi^-$ for $E_{\rm beam}=7.2\GeV$ as function of
    $Q^2$ for different $W$ bins.
    Both panels repeat the same experimental data
    \cite{Joos:1976nm}, while lines show calculations only within the
    resonance prescription (left) and with inclusion of the two-pion
    background and the DIS contribution (right).
    \label{fig:EM_Joos}}
\end{figure}

In order to generalize this two-pion-background model from real
photons to virtual photons, we assume for simplicity the same
dependence as for the total cross section,
\begin{equation}
  \dd\sigma_{2\pi\text{BG}}(\epsilon,Q^2)=\dd\sigma_{2\pi\text{BG}}\,
  \frac{\dd\sigma_{\rm tot}(\epsilon,Q^2)}{\dd\sigma_{\rm tot}(0,0)}.
\end{equation}
We use the cross-section parametrization of Bosted \etal{}
\cite{Christy:2007ve}.  We show a comparison of the $Q^2$ dependence
of the channel $\gamma^*p\to p\pi^+\pi^-$ with experimental data in
\cref{fig:EM_Joos}.

For neutrino-induced reactions (CC,NC), no experimental data are
available yet to fix the two-pion background. Therefore we neglect it.

\subsection{Deep-inelastic scattering}

It is clear from phase-space considerations, that with increasing
energy the multiplicity in the final state increases. We try to
extrapolate the DIS prescription from the DIS regime ($W\to\infty$,
$Q^2\to\infty$) down to $W\simeq1.6\GeV$.

Conventionally, experimental results for electroproduction reactions
are shown in the form of the cross section for virtual photons
\begin{equation}
  \frac1{\Gamma_t} \frac{\dd \sigma}{\dd \Omega' \dd E'} = \sigma_T + \varepsilon \sigma_L,
  \label{virtual-photon-xsec}
\end{equation}
where $\Gamma_t$ is the flux of the virtual photon field,
\begin{equation*}
  \Gamma_t =\frac{\alpha}{2\pi^2} \frac{E'}{E_e}
  \frac{W^2-m_N^2}{2m_N Q^2} \frac{1}{1-\varepsilon},
\end{equation*}
and $\varepsilon$ is the degree of transverse polarization of the
photon,
\[
\varepsilon=\left[1+2\left(1+\frac{\nu^2}{Q^2}\right)\tan^2\frac{\theta}{2}\right]^{-1}.
\]

\begin{figure}[p]
  \centering
  \includegraphics[width=0.7\linewidth]{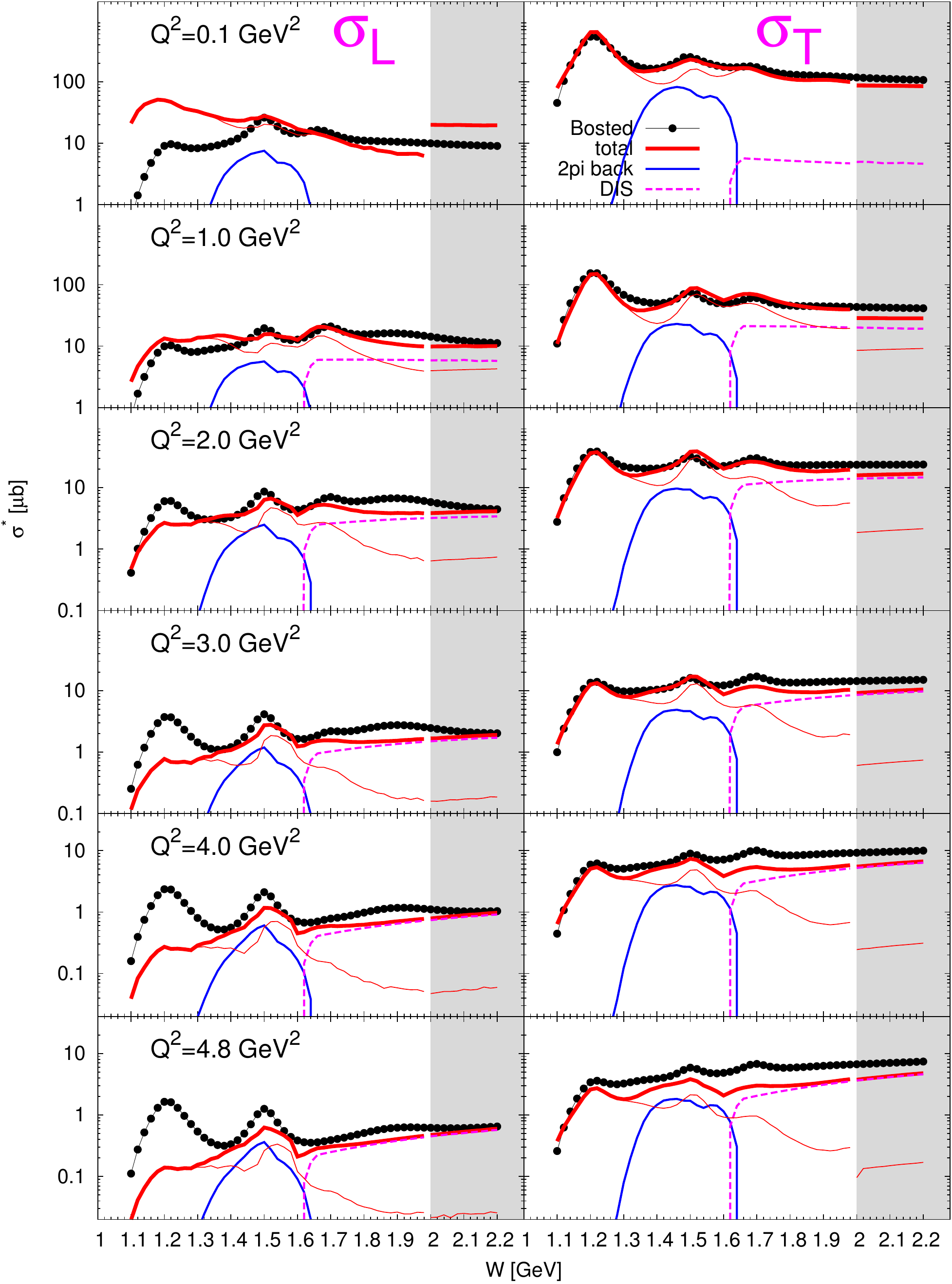}
  \caption{(Color online) Rosenbluth separation done for
    $\epsilon_{1,2}=0.05,0.99$ for $Q^2=0.1\upto4.8\GeV^2$ as function
    of $W$. The points show the data parametrization of Bosted \etal{}
    \cite{Christy:2007ve}. The thick solid line shows the sum of all
    contributions, resonances+1$\pi$BG (thin red), 2$\pi$BG (thin
    blue), and DIS (dashed magenta). The grey-shaded area is the high
    energy region $W>2\GeV$; there we have the VMD contribution
    instead of the resonance contribution (thin red).}
  \label{fig:EM_Rosenbluth2}
\end{figure}

\Cref{fig:EM_Rosenbluth2} compares the model description with the
parametrization of the experimental data by Bosted \etal{}
\cite{Christy:2007ve}. Here we perform a Rosenbluth separation both
for the calculation and the data parametrization, relying on
$\epsilon=0.05$ and $\epsilon=0.99$.

At the high-energy side, we use \Pythia{} \cite{Sjostrand:2006za} for
event generation.  In former code versions, we had not been able to
create events with \Pythia{} down to the energy range of $W=2\GeV$,
since we ran into situations, where \Pythia{} entered infinite loops
in the calculation of diffractive events.  This created a gap between
the \Pythia{} region and the resonance-model region.
Parametrizations of cross sections for some known explicit channels
and the \Fritiof{} model~\cite{Pi:1992ug} for the unknown part had to
be used to bridge the gap in the same spirit as described in
\cref{sec:VecMesProd} for the case of real photons.  In the present
version of GiBUU we have modified the \Pythia{} code such that the
creation of diffractive events in a certain, low-energy parameter
space is forbidden.

Nevertheless, \Pythia{} is just a complex implementation of
leading-order processes, as, e.g., DIS processes. In leading order,
\Pythia{} contains the corresponding standard formulae,
\begin{alignat}{2} \label{DIS_Xsect} \frac{\dd^2\sigma^{(\mu^- p\to
      \mu^-X)}}{\dd x\,\dd y}&=
  2\pi\alpha^2\,\frac{2m_NE}{Q^4}\,\left[1+(1-y)^2\right]
  \,\sum e_i^2\,x\left[q_i(x)+\overline q_i(x)\right], \\
  \frac{\dd^2\sigma^{(\nu_\mu p\to \mu^-X)}}{\dd x\,\dd y}&=
  \frac{G_{\rm F}^2}{2\pi}\,\frac{2m_NE}{(1+Q^2/m_W^2)^2}\,2x\left[d(x)+(1-y)^2\overline u(x)\right],\\
  \frac{\dd^2\sigma^{(\overline\nu_\mu p\to \mu^+X)}}{\dd x\,\dd y}&=
  \frac{G_{\rm
      F}^2}{2\pi}\,\frac{2m_NE}{(1+Q^2/m_W^2)^2}\,2x\left[(1-y)^2u(x)+\overline
    d(x)\right]\quad.
\end{alignat}
with
\begin{equation*}
  x=\frac{Q^2}{2m_N\nu},\quad
  y=\frac\nu{E},\quad
  \nu=E-E',\quad\cos\theta=1-\frac{Q^2}{2\,E\,E'}
\end{equation*}
and $q(x)$ ($\overline{q}(x)$) standing for the parton distribution
function of the quarks (antiquarks) inside the nucleon.

\begin{figure}[t]
  \centering
  \includegraphics[width=0.8\linewidth]{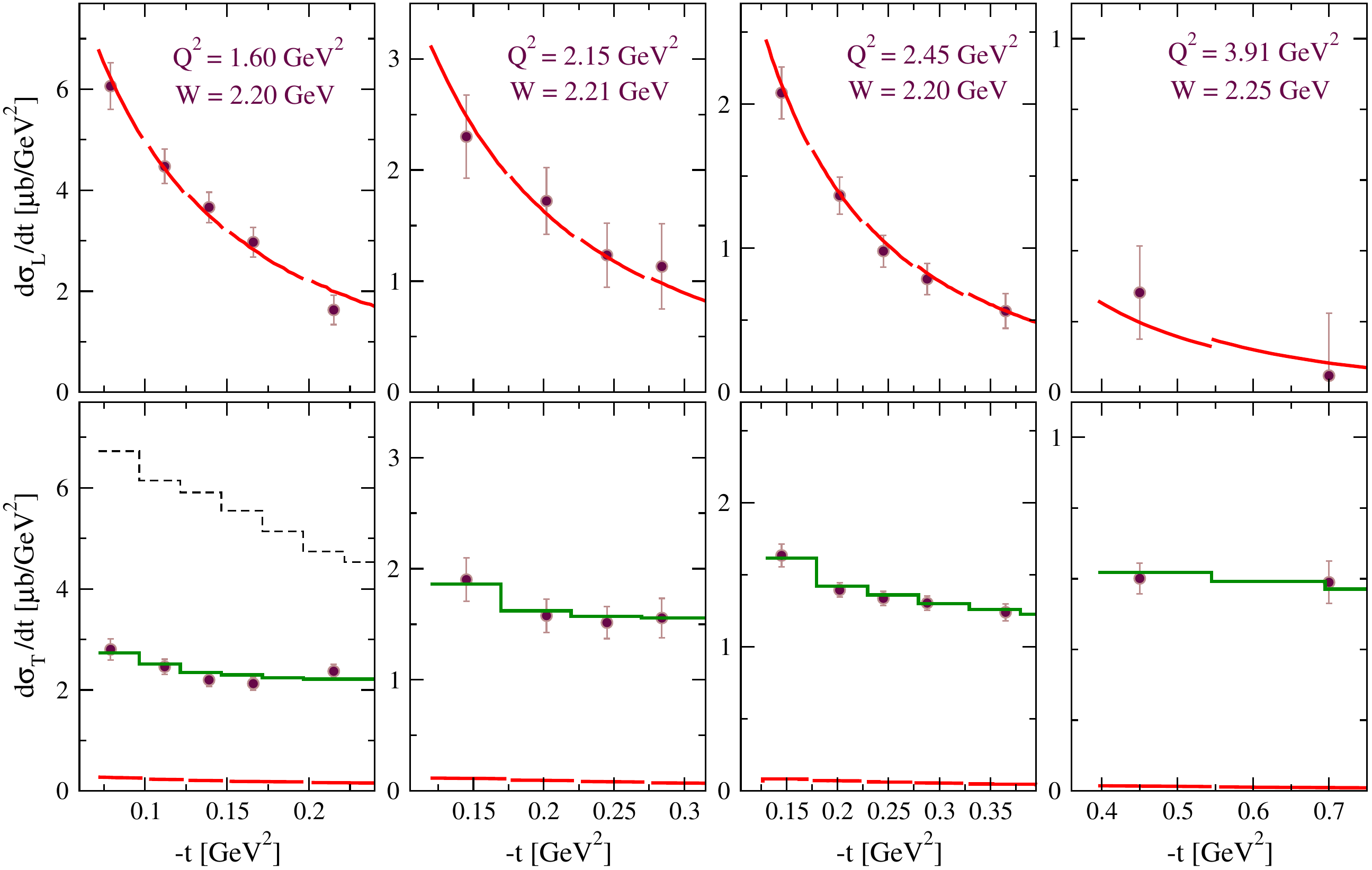}
  \caption{(Color online) The longitudinal ($\dd\sigma_{\rm L}/ \dd
    t$, top panels), and the transverse ($\dd \sigma_{\rm T}/\dd t$,
    bottom panels) differential cross sections of the reaction
    $p(\gamma^*,\pi^+)n$ at average values of $Q^2=1.60~(2.45)
    \GeV^2$ \cite{Horn:2006tm} and $Q^2=2.15~(3.91)\GeV^2$
    \cite{Horn:2007ug}.  The solid curves show the contribution of the
    hadron-exchange model and the histograms the contribution of the
    DIS pions. The discontinuities in the curves result from the
    different values of $Q^2$ and $W$ for the various $-t$ bins.  The
    dashed histogram in the lower left panel shows the contribution of
    the DIS pions for the average transverse momentum of partons
    $\sqrt{\langle k_{\rm t}^2\rangle}=0.4 \GeV$. Taken from
    \refcite{Kaskulov:2008xc}.}
  \label{muratFig1}
\end{figure}

We use the Jacobian determinant,
\begin{equation}
  \dd x\,\dd y =\ \frac{1}{2m_N\nu\,E}\ \dd Q^2\,\dd\nu =
  \frac{E-\nu}{m_N\nu}\ \dd E'\,\dd\cos\theta,
\end{equation}
to transform these analytic expressions to any desired
variables. During the Monte Carlo integration over the results of
\Pythia{} these transformations are straightforward. By using
\Pythia{} we not only obtain the total cross sections, given by
\cref{DIS_Xsect}, but also the complete hadronic final state of the
event. Semi-exclusive events can thus be investigated.

\subsection{Exclusive pion production}

\begin{figure}[t]
  \centering
  \includegraphics[width=0.8\linewidth]{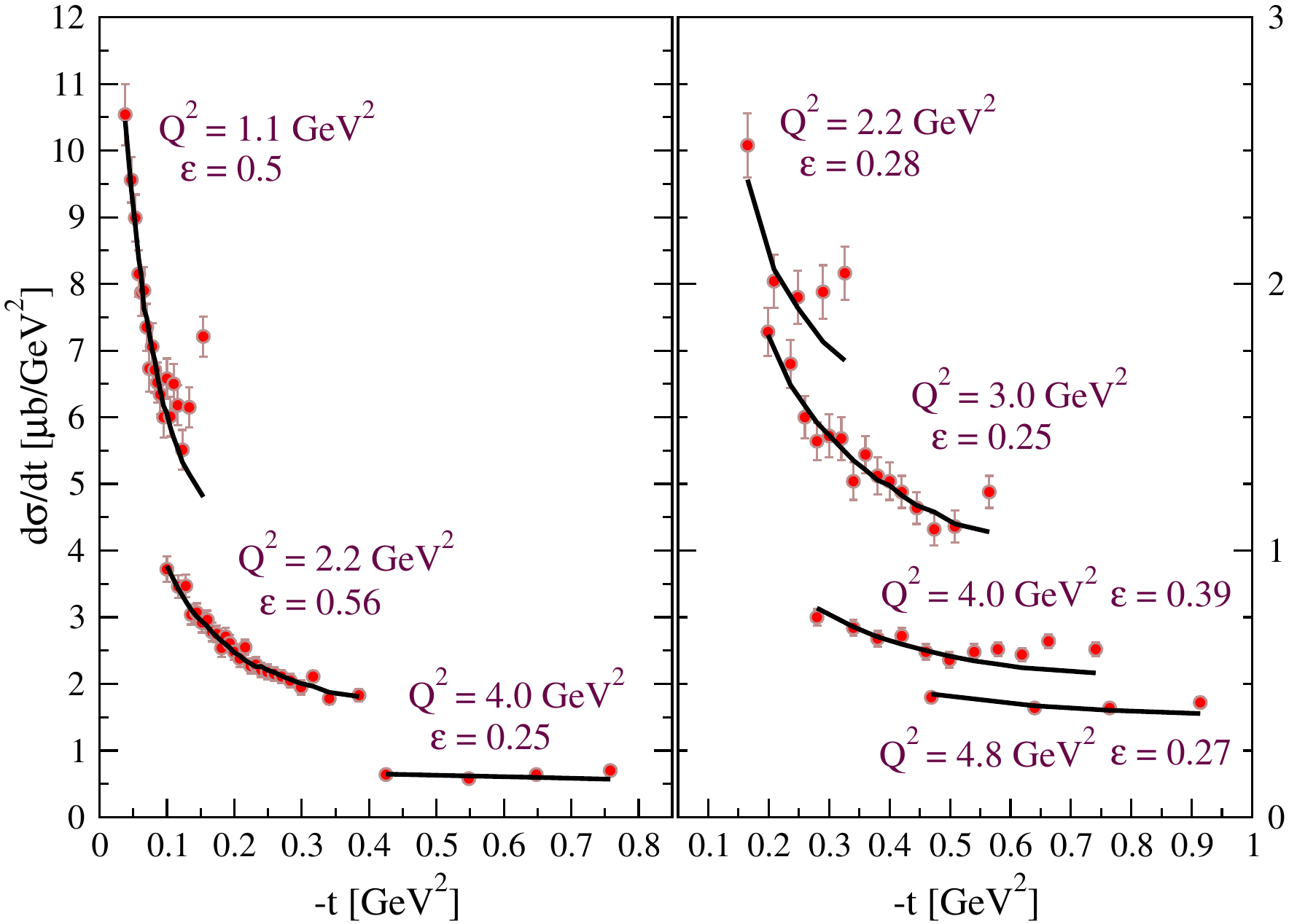}
  \caption{(Color online) Differential cross section $\dd \sigma/\dd t
    = \dd \sigma_{\rm T}/\dd t + \varepsilon \dd \sigma_{\rm L}/\dd t$
    of the reaction, $p(\gamma^*,\pi^+)n$, at JLAB.  The solid curves
    are the model predictions. Taken from \refcite{Kaskulov:2008xc}.}
  \label{muratFig2}
\end{figure}

Until now we have just covered the inclusive particle production as
provided by \Pythia{}. But according to
\cite{Kaskulov:2008xc,Kaskulov:2008ej,Kaskulov:2009gp}, also a
detailed study of exclusive pion production is necessary at high
energies.

The idea proposed
in~\cite{Kaskulov:2008xc,Kaskulov:2008ej,Kaskulov:2009gp} is to treat
exclusive meson production as an exclusive limit of semi-inclusive
DIS. This is in spirit of the exclusive-inclusive connection
\cite{Bjorken:1973gc}. In this case, the reaction mechanism consists
of two components. The first one describes the soft production
mediated by the exchange of meson-Regge trajectories. These peripheral
processes are relevant for the photo production and low-$Q^2$ electro
production. As the second, novel element, at large $Q^2$ the GiBUU
model allows the direct interaction of virtual photons with partons
followed by the fragmentation of excited color string into the
meson-baryon channel. Details can be found in
\refcite{Kaskulov:2008xc,Kaskulov:2008ej,Kaskulov:2009gp}. In
\cref{muratFig1} we show the results for the exclusive production of
charged pions in the kinematics available at JLAB. As one can see, all
longitudinal and transverse cross sections are very well
described. The dashed curves describe the Regge-exchange contributions
and dominate in the longitudinal channel. In the transverse channel
the contributions of mesons are marginal. Here, the direct partonic
interactions dominate the production mechanism. In \cref{muratFig2} we
show the comparison of the model results with new data from JLAB. As
one can see, the model (solid curves) describes the cross section in a
remarkably large $Q^2$ range up to $Q^2\simeq 5 \GeV^2$.

Note that this behavior is rather general and is also observed in
other meson production processes. For instance, the string-breaking
mechanism is effective in the transverse cross section and explains
large cross sections in pseudoscalar $\pi^0(\eta)$ and $\rho^0$
production channels.

\begin{figure}[t]
  \centering
  \includegraphics[width=0.8\linewidth]{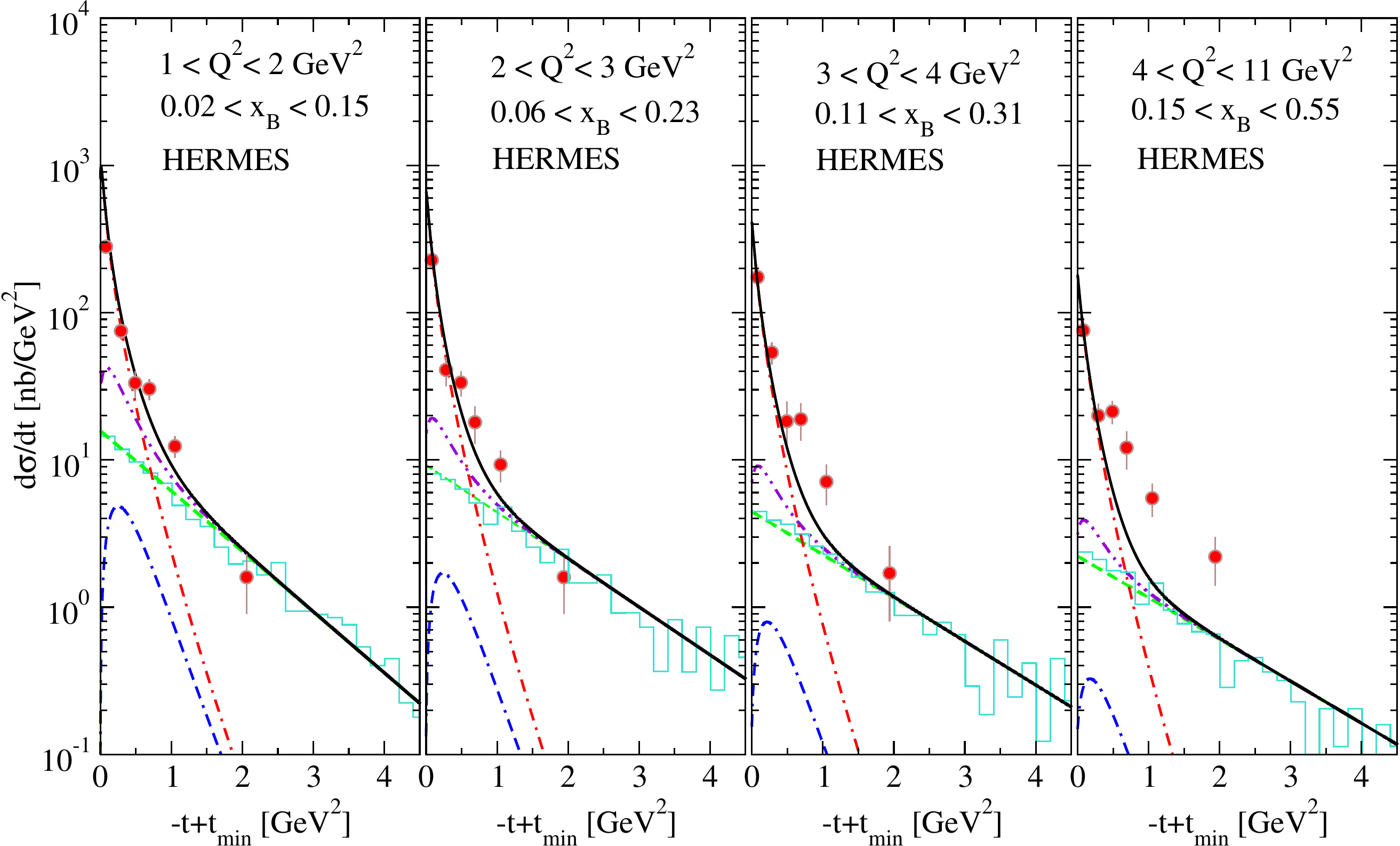}
  \caption{(Color online) The differential cross section, $\dd
    \sigma/\dd t = \dd \sigma_{\rm T}/\dd t + \epsilon \dd \sigma_{\rm
      L}/\dd t$, of the exclusive reaction, $p(\gamma^*,\pi^+)n$, in
    the kinematics of the HERMES experiment. The solid curves are the
    model results. The dash-dotted curves correspond to the exchange
    of $\pi$-Regge trajectory and dash-dash-dotted curves to the
    exchange of $\rho$-Regge trajectory.  The histograms and the
    dashed curves which just fit the histograms describe the partonic
    contributions in line of the DIS mechanism. The dot-dot-dashed
    curves are the contribution of the transverse cross section,
    $\dd\sigma_{\rm T}/\dd t$, to the total unseparated response,
    $\dd\sigma/\dd t$.  Taken from \refcite{Kaskulov:2009gp}. }
  \label{muratFig3}
\end{figure}

In \cref{muratFig3} we show our results for the deep exclusive
production of charged pions in the kinematics of the HERMES
experiment. The two-component model describes again the data very
well. At forward angles the Regge-exchange contributions again
dominate. In the non-forward region the transverse partonic
interactions are large and describe the measured cross sections very
well.

\subsection{Vector-meson production}
\label{sec:VecMesProd}

Exclusive vector-meson production plays a special role in lepton- or
photon-induced reactions. In GiBUU this is usually treated within the
\Pythia{} model, which implements a vector-meson dominance (VMD)
model.

The invariant amplitude for the process $\gamma^*h\to X$ can be
expressed in terms of the on-shell vector-meson-scattering amplitude,
\begin{equation}
  \label{eq:vecmes-ampl}
  {\cal M}_{\gamma^*h\to X}=\sum_V\frac{e}{g_V}\frac{m_V^2}{Q^2+m_V^2}
  {\cal M}_{Vh\to X}\quad.
\end{equation}
Here, $m_V$ denotes the mass of the vector meson, $V$, and $(e/g_V)^2$
gives the probability for the transitions $\gamma^*\to V$. One may use
Pomeron--Reggeon parametrizations for the total hadronic cross
section. Assuming an additive quark model, the total $Vp$ cross
sections can be parametrized as
\begin{alignat}{3}
  \sigma_{\rm tot}^{\rho^0p}&\simeq \sigma_{\rm tot}^{\omega p}&
  &\simeq& \frac12\left(\sigma_{\rm tot}^{\pi^+ p}+\sigma_{\rm
      tot}^{\pi^- p}\right)
  &= 13.63\,s^\epsilon+31.79\,s^{-\eta}\\
  &\sigma_{\rm tot}^{\phi p}&&\simeq\ & \ \sigma_{\rm tot}^{K^+ p} +
  \sigma_{\rm tot}^{K^- p} - \sigma_{\rm tot}^{\pi^+ p} &=
  10.01\,s^\epsilon+1.52\,s^{-\eta}
\end{alignat}
Using \cref{eq:vecmes-ampl} in the optical theorem, one obtains the
`diagonal approximation', i.e., neglecting off-diagonal scattering
$VN\to V'N$ with $V'\ne V$, \footnote{Within this prescription one
  easily may implement medium modifications/shadowing effects by
  modifying the coupling constant, $\frac{e^2}{g_V^2}$, according to
  the spatial coordinate.}
\begin{align}
  \label{eq:vecmes-cross}
  \sigma_{\rm VMD}^{\gamma^*p} = \sum_V \frac{e^2}{g_V^2}\sigma_{\rm
    tot}^{Vp}\quad .
\end{align}
Comparing this with the total $\gamma p$ cross section, one realizes,
that this VMD prescription describes 80\proz of the total interaction
\cite{Falter_Phd}. In addition, the VMD contribution is dominated by
the $\rho^0$ component of the photon. In order to achieve a full
prescription of the photon interaction, one has to take into account
some ``generalized vector meson dominance'' interactions, where the
interaction of the photon is described with some higher excited $q\bar
q$ states, and also direct interactions play a major role
\cite{Falter_Phd}.

The model described above is rather simple and can be used for both
real and virtual photons. Here, we are interested in the case of real
photons. As discussed in \cref{sec:omega_photo_nuclei}, modern
photoproduction experiments allow to investigate the modification of
meson properties in the nuclear medium. Thus, we need a more detailed
description for the elementary input, which can be meaningfully
extended to the in-medium case. So, for photo production we replace
the \Pythia{} treatment with the model described below.

In our model the exclusive cross sections for the photo production of
vector mesons on a nucleon, i.e., $\gamma N \rightarrow V N$ (with
$V=\rho^0,\omega,\phi$), are adjusted to experimental data with the
ansatz \cite{effe_phd}
\begin{equation}
  \sigma_{\gamma N\rightarrow VN}=\frac{1}{p_i s}\int_0^{\mu_{max}}
  \dd (\mu^2) \, \left|\mathcal{M}_V(\sqrt{s}) \right|^2 p_f \mathcal{A}_V(\mu),
  \label{eq:sigma_gammaN_VN}
\end{equation}
where we integrate over the mass, $\mu$, of the vector meson, $V$, up
to a maximum $\mu_{\text{max}}=\sqrt{s}-m_N$.  Here $\sqrt{s}$ is the
total energy available for the reaction.  The center-of-mass momenta
of the initial and final state are denoted by $p_i$ and $p_f$,
respectively:
\begin{align}
  p_i & = \frac1{2\sqrt{s}}\left(s-m_N^2\right) \; ,\\
  p_f & = \frac1{2\sqrt{s}}\sqrt{\left(s-\left(m_N+\mu\right)^2\right)
    \left(s-\left(m_N-\mu\right)^2\right)} \; .
\end{align}

The spectral function of the vector meson is given by
\begin{equation*}
  \mathcal{A}_V(\mu) = \frac{1}{\pi}
  \frac{\mu \Gamma_{V}(\mu)}{(\mu^2-M_V^2)^2 + (\mu\Gamma_{V}(\mu))^2}
  \; .
\end{equation*}
Here, $M_V$ is the pole mass of the vector meson, and $\Gamma_{V}$ its
total width, which for an elementary reaction is just the vacuum decay
width. However, it will include contributions from collisional
broadening, when we describe vector meson production in the medium.

\begin{figure}[t]
  \centering
  \includegraphics[width=0.8\linewidth]{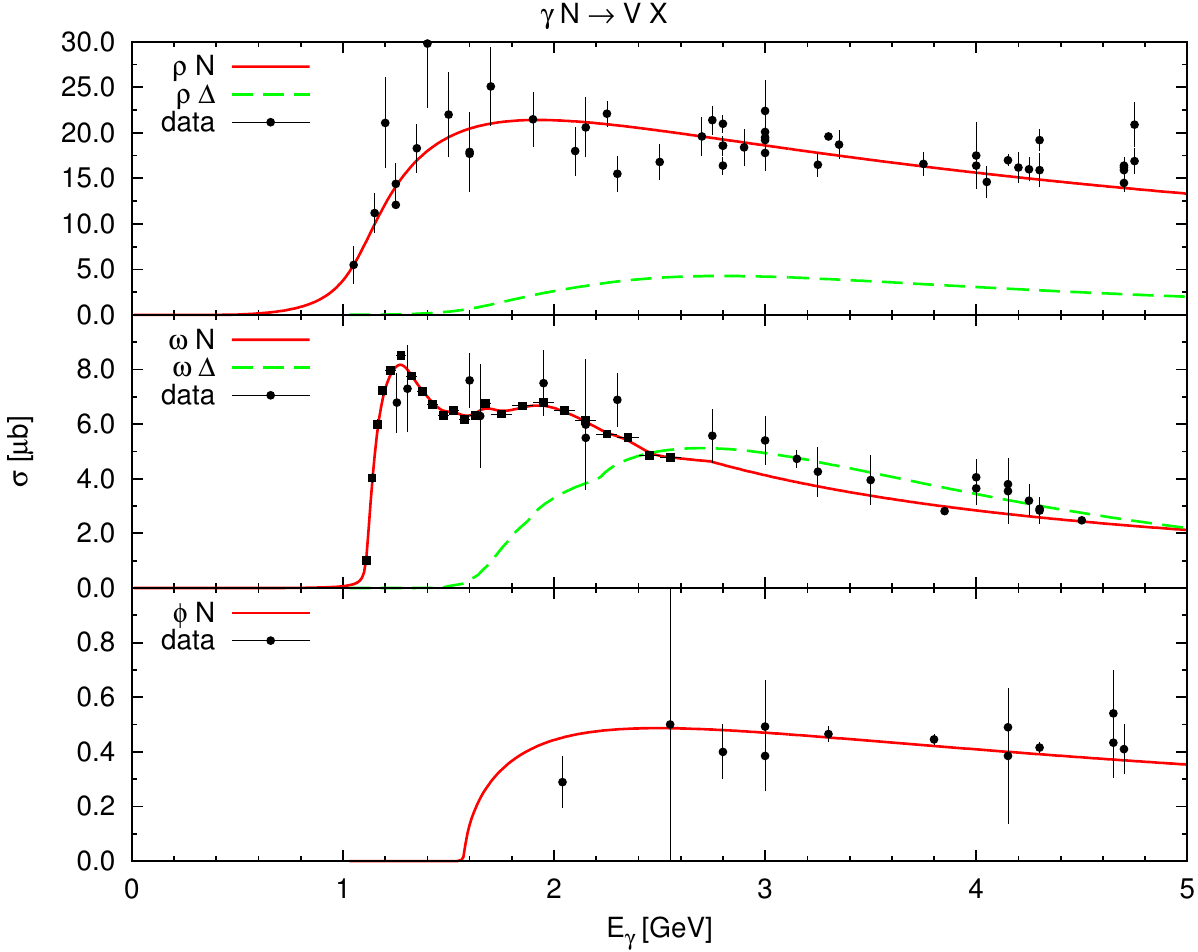}
  \caption{(Color online) Photoproduction cross sections of vector
    mesons. Data taken from \cite{PDGdata} and \cite{Barth:2003kv}.}
  \label{fig:gammaN_VN}
\end{figure}

For the $\rho^0$ and $\phi$ mesons, constant matrix elements
\begin{align}
  |\mathcal{M}_\rho|^2 & = 160\mub\cdot \GeV^2 \; ,\\
  |\mathcal{M}_\phi|^2 & = 4\mub\cdot \GeV^2 \;
\end{align}
are sufficient to obtain a good fit of the data, cf.~\cref{fig:gammaN_VN}.

For the $\omega$ meson we use an $\sqrt{s}$-dependent matrix element.
It can be directly obtained from a spline-fit to the SAPHIR data
\cite{Barth_phd} for $\omega$ photoproduction on a proton.  Using the
fact, that in the vacuum the $\omega$ spectral function is very narrow
\cite{Muehlich_phd} and can be approximated by the Delta-function in
\cref{eq:sigma_gammaN_VN}, we get
\begin{equation*}
  \displaystyle
  \left |\mathcal{M}_\omega(\sqrt{s}) \right|^2 = \frac{p_i \, s\,
    \sigma_{\gamma N\rightarrow\omega N}^{exp}(s)}{\phi_2(s)},
\end{equation*}
where $\phi_2$ stands for the two-body phase-space integral (in the
vacuum).  This matrix element is then used in
\cref{eq:sigma_gammaN_VN} together with the in-medium spectral
function.

In order to describe the photoproduction of mesons with masses below
the pole mass, we follow the idea of \cite{Larionov:2003av} and extend
the matrix element to sub-threshold energies by defining a new
invariant
\begin{equation}
  Q(\mu) = \sqrt{s_0(M_V)} - \sqrt{s_0(\mu)} + \sqrt{s} \equiv \sqrt{s}
  + M_V - \mu.
  \label{Qmu}
\end{equation}
Here $\sqrt{s_0(\mu)}$ is the threshold energy for the production of
$\omega$ mesons of mass $\mu$. With this definition one substitutes
\begin{equation*}
  \mathcal{M}(\sqrt{s}) \rightarrow \mathcal{M}[Q(\mu)]\quad .
\end{equation*}
In the vacuum $Q(\mu)$ is only nonzero in the vicinity of
$Q(\mu)\approx\sqrt{s}$, since the $\omega$ is a very narrow state. At
the same time, \cref{Qmu} provides a reasonable prescription for the
in-medium case (see \cref{sec:sfree} for more details).

For the process $\gamma N\to V\Delta$ we use the following
parametrization of the total cross section \cite{Muhlich:2003tj},
\begin{equation}
  \sigma_{\gamma N\to V\Delta}=\frac{1}{p_i s}\int \dd (\mu_\Delta^2)
  p_f(\mu_\Delta) \mathcal{A}_\Delta(\mu_\Delta)
  \frac{A}{(\sqrt{s}-M)^2+\Gamma^2/4},
  \label{eq:sigma_gammaN_VD}
\end{equation}
where the constants, $A$, $M$, and $\Gamma$, are fitted to the
experimental cross section of \refcite{Barber:1985fr}, yielding
$A=47.3\mub \GeV^4$, $M=2.3 \GeV$ and $\Gamma = 1.8 \GeV$ for the
$\omega$ meson.

For both $\gamma N\to\omega N$ and $\gamma N\to\omega\Delta$, the
angular dependence is usually modeled as an exponential,
\begin{equation}
  \frac{\dd \sigma}{\dd t} \propto \exp(Bt)
\end{equation}
While such a very forward peaked distribution gives a good description
of the data at large photon energies, the experimental distributions
turn out to be flatter at lower photon energies \cite{Barth:2003kv}.  In
\cite{Muehlich_phd} a more complicated tree-level model has been
developed, which includes $s$-, $t$-, and $u$-channel nucleon diagrams
and leads to a good description of $\omega$-photoproduction data at low
energies.  In the GiBUU code one can switch between the two models.

The more inclusive vector-meson-production channels (beyond $\gamma
N\to\omega N$ and $\gamma N\to\omega\Delta$) are described via
\Fritiof{}~\cite{Pi:1992ug}, by assuming strict vector meson dominance
(VMD) and converting the incoming photon into a $\rho^0$.

\section{Numerical realization}\label{sec:numerics}

This appendix introduces the special algorithms used for the GiBUU
model. \Cref{sec:Vlasov} describes the numerical solution of the
Vlasov part of the GiBUU-equation (\ref{eq:transp.28}). Afterwards, in
\cref{sec:realPert}, the concept of perturbative particles is
explained.  The way how the test particles are initialized before the
propagation is described in \cref{sec:Initialization}.
In \cref{sec:AppCollisionTerm} we give some details connected
with the collision term, while in \cref{sec:finalStateDecisions} the
final-state decision algorithm is detailed.

\subsection{Vlasov term: Numerical treatment}
\label{sec:Vlasov}

The propagation of the test particles including mean fields, i.e., the
numerical treatment of the Vlasov equation, is realized in the
framework of the test-particle method, see the general
\cref{eq:testparticleansatz}. In practice, however, the usage of
$\delta$ functions produces numerical noise which has to be
compensated by a large number of test particles per particle. An
alternative and standard representation consists in the use of
Gaussian-like functions in coordinate space with a width, $L$. This
alternative representation leads to smooth densities and fields, even
when using a rather small number of test particles. Furthermore,
heavy-ion collisions are usually simulated in the CM-frame of the two
colliding nuclei with the $z$-axis as the beam direction. In order to
account for the relativistic contraction of the coordinate space in
$z$-direction, the Gaussian distribution is Lorentz contracted along
$z$ and reads
\begin{equation}
  \label{gaussian}
  \rho_i(\bvec{r}) = \frac{\gamma}{(2\pi)^{3/2} L^3}
  \exp\left\{ -\frac{[x-x_i(t)]^2}{2 L^2} - \frac{[y-y_i(t)]^2}{2 L^2}
    -\frac{[z-z_i(t)]^2 \gamma^2}{2 L^2} \right\}~.
\end{equation}
For hadron-induced reactions the simulations are performed in the rest
frame of the target (laboratory system). Thus no Lorentz contraction
is necessary.  To ensure kinetic-energy conservation with an accuracy
of better than 3\proz for the studied reactions, the Hamilton
equations of motion for the test particles are solved by using the
$O(\Delta t^2)$ predictor-corrector method. This numerical method
consists of two steps. First, one predicts in a one-step propagation
the value of the quantity under consideration. This is the predictor
step. This value is then corrected in the secondary corrector step. We
apply this method to the Hamilton equations for the test particles: at
given phase-space coordinates, $\bvec{r}_{i}$, and momenta,
$\bvec{p}_{i}$, for a test particle with label $i$ ($i=1,\cdots,A\cdot N$) at time 
$t$ the predictor step reads
\begin{alignat}{2}
  \bvec{r}_{i}^{\; P} & = \bvec{r}_{i} + \Delta t \frac{\partial
    H(\bvec{r}_{i},\bvec{p}_{i})}{\partial\bvec{p}_{i}},
  \label{pred_r}\\
  \bvec{p}_{i}^{\; P} & = \bvec{p}_{i} - \Delta t \frac{\partial
    H(\bvec{r}_{i},\bvec{p}_{i})}{\partial\bvec{r}_{i}},
  \label{pred_p}
\end{alignat}
where $\Delta t$ is the time step. The single-particle Hamilton
function is given by \cref{eq:Hamilton_OSP} or \cref{dispRel} for the
non-relativistic or relativistic cases, respectively. The predictor
step is performed for all test particles to obtain a predictor value
for the phase-space distribution, $f$. With this $f$ one then
calculates the baryon current, $\bvec{j}_{i}=\bvec{j}(\bvec{r}_{i})$,
and baryon density, $\rho_{i} = \rho(\bvec{r}_{i})$, needed for the
determination of the Hamilton function and its derivatives. The latter
quantities are calculated numerically. These values are then used for
the corrector step, from which one obtains the phase-space
distribution for the next time step,
\begin{alignat}{2}
  \bvec{r}_{i}^{\; C} & = \bvec{r}_{i} + \frac{1}{2}\Delta t \left[
    \frac{\partial H(\bvec{r}_{i},\bvec{p}_{i})}{\partial\bvec{p}_{i}}
    + \frac{\partial H(\bvec{r}_{i}^{\; P},\bvec{p}_{i}^{\;
        P})}{\partial\bvec{p}_{i}^{\; P}} \right],
  \label{corr_r}\\
  \bvec{p}_{i}^{\; C} & = \bvec{p}_{i} - \frac{1}{2}\Delta t \left[
    \frac{\partial H(\bvec{r}_{i},\bvec{p}_{i})}{\partial\bvec{r}_{i}}
    + \frac{\partial H(\bvec{r}_{i}^{\; P},\bvec{p}_{i}^{\;
        P})}{\partial\bvec{r}_{i}^{\; P}} \right].
  \label{corr_p}
\end{alignat}
The technical parameters are therefore the size of shells, $\Delta
\bvec{r}$, the time step, $\Delta t$, and the number of test particles
per nucleon, $N$. The value of the width of the Gaussians in
coordinate space is the same as the step in the spatial grid. Note
also that along the beam-axis the grid is contracted according to the
relativistic $\gamma$-factor.  These parameters are common to all
hadronic processes. Usually the values $\Delta\bvec{r}=0.5\fm$ and
$\Delta t=0.2\fmc$ are used. The number of test particles per nucleon
varies according to the process considered. In heavy-ion collisions
one uses $N=200$, in hadron-induced reactions $N=1000$, in high-energy
photon- and electron-induced reactions $N=200$, in pion-induced ones
$N=500$, and in neutrino-induced ones from $N=500$ to $N=2000$.  Note,
that the various cases listed above can run in parallel mode, which
improves the statistics significantly.

\subsection{Real and perturbative test particles}
\label{sec:realPert}
The test particles mentioned so far can all collide with each
other. This is connected with a large computational effort.

However, there may be physics situations in which only very few
particles are actively involved in a reaction and the remnant nucleus,
involving all the other particles, stays close to its ground state and
acts as a background for the propagation of the active particles. This
may, e.g.\ be the case for low-energy $\pi A$ or $\gamma A$
collisions. In this case the produced particles do not significantly
affect the target configuration. This can also happen in heavy-ion
collisions when rare events are considered \cite{effe_phd}. One
example is the production of $\Lambda$ baryons in $NN$
scattering. This process is so rare that the statistical information
on this channel is very scarce in a conventional run. However, one can
enforce production of an additional $N\Lambda K$ final state in each
$NN$ collision. These particles then are weighted with the production
probability, $\sigma_{N\Lambda K}/\sigma_\text{tot}$. Thus one
achieves large statistics - under the assumption that these rare
events do not lead to an overall change of the particle flow which
acts as a background to their propagation.

In both cases the active particles can scatter with the remnant
without changing it too much. To exploit this situation we introduce
two types of test particles The test particles, which represent the
background nucleons, are called \textit{real} test particles. For the
particles participating actively in a reaction one defines another
type of test particles, which are called \textit{perturbative}. These
perturbative test particles are only allowed to collide with a real
one, but not among themselves. All products of collisions between real
and perturbative testparticles are again represented by perturbative
test particles thus allowing for a build-up of particle multiplicity.
Perturbative particles are neglected in the calculation of the actual
densities which are kept constant.  For example, in $\pi A$ collisions
one may represent the initial-state pion by perturbative test
particles. Thus, the target nucleus is unchanged, and all products of
the collisions of the pion with the target nucleons are assigned to
the perturbative regime.

Since the perturbative particles do not interact with each other and
do not affect the real particles in a reaction one can also split a
perturbative particle into $n$ realizations (several perturbative
particles) during a run. Each realization is given a corresponding
weight, $1/n$. In this way, one simulates $n$ possible final-state
scenarios of the same perturbative particle during one run thus
improving statistics significantly.

As an additional simplification for electron-, photon- and
neutrino-induced reactions one could also freeze the groundstate
configuration by not propagating any real testparticles.

\subsection{Preparation of the initial configuration}
\label{sec:Initialization}

Before starting the time evolution, the initial configuration of test
particles has to be prepared. For different processes we use
different, optimized prescriptions.

\subsubsection{Heavy-Ion and hadron-induced (non perturbative)
  collisions}

The initialization of heavy-ion collisions is straightforward in the
GiBUU model since these event types cannot use the concept of
perturbative test particles. Also $\pi + A$ and $p + A$ reactions can
be run, as an option, in this mode. This is necessary when major
excitation of the target nucleus, or even its break-up, is to be
expected. For the initial configuration in these reactions, real test
particles are distributed according to the phase-space distributions
of the projectile and the target nucleus. The corresponding particles
are separated in space by a given impact parameter in the direction
perpendicular to the beam, i.e.\ in $x$-direction. In $z$-direction,
the beam axis, the separation is adjusted such, that initially there
is no overlap between the two distinct distributions for target and
projectile. For the actual time evolution, the initial velocities of
the test particles are aligned along the $z$-axis. Interactions are
handled via the collision term in the propagation.

Minimum-bias calculations involving an average over all impact
parameters are only possible by performing separate calculations with
fixed values of the impact parameter and averaging the output
afterwards.

\subsubsection{Hadron-induced collisions (perturbative)}

For reactions that can employ the perturbative test particle method
one can simulate impact-parameter averages in
one single run, since the actual impact does not influence the
calculated density distribution.

In this method different weights can be assigned to every test
particle. In the first step, for a given fixed impact parameter, every
real projectile particle is simulated by $n\sim 25$ test particles
(per ensemble), with each one of them being weighted by the factor
$1/n$.

For a correct calculation of the total cross section, the test
particles have to be initialized in a disc in the transverse plane
(displaced by some value in the $z$-direction, see above) with the
radius $b_{\mathrm{max}}$. This value has to be chosen such, that
within the geometrical interpretation of the cross section it is given
by the spatial extension of the target nucleus plus the cross-section
correction factor. Particles outside this radius do not interact with
the nucleus.

The integral $\int_0^{b_{\rm max}} \dd b \, b$ implies that most of
the particles are placed close to the maximal impact parameter,
$b_{\rm max}$. Due to the density-falloff the interaction rate of
these particles is very small. To overcome this problem the individual
weighting of test particles is used, which allows for a drastic
improvement of the numerics. In order to prevent many test particles
passing the nucleus without interaction, we split the initial disc
into an inner disk ($b<b_d$) and an outer ring ($b_d<b<b_{\rm
  max}$). In a first Monte Carlo choice, a test particle is assigned
to the inner disc with probability $p_d\sim0.7$, otherwise it is part
of the outer ring. Then the impact parameter is chosen according to
$\dd b \, b$ between $0$ and $b_d$ or between $b_d$ and $b_{\rm max}$,
respectively.  The weight for every test particle is given by
\begin{equation}
  w_i = \frac{1}{n}\begin{cases}
    \frac{\pi\,b_d^2}{p_d}&\text{for inner disk\ ,}\\[1mm]
    \frac{\pi\,(b_{\rm max}^2-b_d^2)}{1-p_d}&\text{for outer ring\ .}
  \end{cases}
\end{equation}
Interactions are handled via the collision term in the usual
propagation steps.

As an optional additional optimization, an abbreviation of the
propagation of the projectile cloud onto the target nucleus is
implemented. This is possible by first sorting the test particles of
the target nucleus according to their $z$-coordinate. Then for every
projectile test particle, an interaction on every target test particle 
is probed and the loop is stopped, when the collision rate suggests an
interaction to happen. The projectile test particle is taken out of
the particle vector and replaced by the final-state particles of this
interaction.  This optimization also allows for testing some modified
physics: In the case of very high energy proton-induced collisions,
hard probes do not scale with $A^{2/3}$, but instead with $A^1$. This
can be simulated by not quitting the loop over the target test
particles after an interaction, but just continuing. This implies,
that one projectile test particle may interact with more than one
target test particle.

\subsubsection{Lepton-induced collisions (perturbative)}

In the case of collisions, induced by particles which are not directly
propagated in the GiBUU model, as, e.g., electrons, photons, or
neutrinos, initializations as described above are not possible.
Instead, the incoming lepton is split into $A$ test particles in every
ensemble. Now every test particle interacts with one of the target
test particles and produces some final state of particles, which now
may be propagated by the GiBUU model. Only these test particles are
inserted into the particle vector.

Since the final state at the leptonic vertex is not determined by the
input variables, the weight assigned to every particle is given by
$\dd\sigma/\dd E'\,\dd\Omega_k$. For photon-induced reactions, one may
also divide this weight by the flux $f_T$, thus the weight is given by
$\sigma_{\gamma^*}$. These details have to be respected in the
corresponding analysis routines.

\subsection{Collision term}
\label{sec:AppCollisionTerm}

\subsubsection{Ensemble Techniques}
\label{sec:EnsembleTechniques}

To point out the connection between our numerical implementation and
the underlying BUU equation, we focus on the loss term of BUU. We will
not elaborate on the gain term, $I_{\mathrm{gain}}$, which describes
the production of particles. However, its numerical implementation is
analogous to the loss term since both are related by detailed balance.

As shown in \cite{Buss:2006yk}, in terms of the test-particle ansatz
\cref{eq:testparticleansatz} the loss term \cref{C_loss} reads
\begin{equation}
  \begin{split}
    \label{loss2Body}
    &\Delta t \ C_{\mathrm{loss}}(\bvec{r},t,\bvec{p_{A}}) =
    \lim_{N\to\infty} \frac{(2\pi)^4}{gN} \sum_{i=1}^{n(t)} \sum_{\scriptsize\begin{array}{l}  j=1 \\
        j\neq i \end{array}}^{n(t)} \delta(p^0_A-p^0_i)
    \delta(\bvec{p}_A-\bvec{p}_i)  \delta(\bvec{r}-\bvec{r}_i)\\
    &\qquad\qquad\qquad \times \lim_{N\to\infty}
    \frac{1}{\sigma_{ij}}\int \dd\Omega_{\mathrm{CM}} \;
    \mathcal{P}_{a} \mathcal{P}_{b} \frac{\dd\sigma_{ij\to ab}}{\dd
      \Omega_{\mathrm{CM}}}
    \sigma_{ij} \Delta t \, v_{ij} \frac{1}{N} \delta(\bvec{r}-\bvec{r}_j)\\
    &= \lim_{N\to\infty} \frac{(2\pi)^4}{gN} \sum_{i=1}^{n(t)}
    \sum_{\scriptsize\begin{array}{l} j=1 \\ j\neq
        i \end{array}}^{n(t)} \delta(p_A-p_i)
    \delta(\bvec{r}-\bvec{r}_i) \overline{\mathcal{P}_a\mathcal{P}_b}
    \int_{\Delta V_{ij}}\delta(\bvec{r}\; '-\bvec{r}_j) \dd^3r' ,
  \end{split}
\end{equation}
where $\sigma_{ij}$ and $v_{ij}$ are the total interaction cross
section and the relative velocities of the test particles, $i,j$, and
$\mathcal{P}_{a,b}$ are the usual Pauli blocking factors for particles
$a$ and $b$. The volume, $\Delta V_{ij}=\sigma_{ij} \Delta t \, v_{ij}
/N$, is an infinitesimal volume in the vicinity of $\bvec{r}_i$ and
defines the locality of the scattering process of two test
particles. The term,
\begin{equation}
  \overline{\mathcal{P}_a\mathcal{P}_b}
  =\frac{1}{\sigma_{ij}}\int \dd\Omega_{\mathrm{CM}} \; \mathcal{P}_{a}
  \mathcal{P}_{b} \frac{\dd\sigma_{ij\to ab}}{\dd
    \Omega_{\mathrm{CM}}},
\end{equation}
denotes the blocking of the final state averaged over its angular
distribution. We have excluded self interactions -- therefore a test
particle cannot scatter with itself.

The time step, $\Delta t$, is chosen such that the average distance
traveled by the particles during $\Delta t$ is less than their mean
free path. Therefore, $\Delta V_{ij}$ is so small that a particle has
no more than one scattering partner at a given time step.

The kind of simulation for the two-body processes, which we have
described above, is called a \textit{full-ensemble calculation}. There
exists a common simplification to this method: \textit{the
  pa\-ral\-lel-en\-sem\-ble method}~\cite{BertschGupta}. In this
scheme, one performs $\tilde{N}$ calculations in parallel, each of
which includes only one single ensemble ($N=1$). The densities used in
each parallel run are the averaged densities of all $\tilde{N}$
parallel runs. Therefore the propagation part basically stays the
same, whereas the collision term gets very much simplified.

Note that the only justification for this simplification is a great
gain in computation time. In the full-ensemble method, the propagation
part scales according to the number of test particles per nucleon,
$N$, whereas the two-body collision term scales with $N^2$ --
therefore the computation time is $\mathcal{O}(N^2)$. In the
parallel-ensemble method $\tilde{N}$ runs are performed, which results
in $\mathcal{O}(\tilde{N})$ computation time. Thus there is a linear
scaling in a parallel-ensemble run, but a quadratic one in a
full-ensemble run.

The major drawback of the parallel-ensemble scheme is the non-locality
of the collisions. As a showcase, let us consider the pion-nucleon
interaction. There the maximum cross section amounts to roughly $200
\mb$. This leads to a maximal impact parameter of two test particles
of $\sqrt{200 \mb/\pi}\simeq 2.52 \fm$. Hence, the parallel-ensemble
scheme may lead to large non-localities whereas the underlying BUU
collision term is strictly local.

In pioneering works it has been shown by Welke
\etal{}~\cite{PhysRevC.40.2611} and Lang \etal{}~\cite{langBabovsky},
that the parallel-ensemble scheme is a good approximation to the
full-ensemble scheme under the conditions of high-energy heavy-ion
collisions. However, it is still an open question, whether this also
applies to more surface sensitive processes such as, e.g., pion
double-charge exchange in nuclei. We will discuss this problem
together with our numerical results.

The so-called \textit{local-ensemble method}, which has been first
applied to heavy-ion collisions by Lang \etal{}~\cite{langBabovsky},
is a method allowing for a full-ensemble run which is less time
consuming than the standard algorithm. Here one divides position space
into small cubical boxes, $V_i$, with equal volumes, $\Delta V$. The
loss term is then represented by
\begin{equation}
  \begin{split}
    \label{loss}
    \Delta t \ C_{\mathrm{loss}}(\bvec{r},t,\bvec{p_{A}}) &=
    \lim_{N\to\infty} \frac{(2\pi)^4}{gN} \sum_{i=1}^{n(t)}
    \sum_{\scriptsize\begin{array}{l} j=1 \\ j\neq i \end{array}}^{n(t)} \delta(p_A-p_i)  \delta(\bvec{r}-\bvec{r}_i)  \\
    & \qquad\qquad \qquad\qquad \times
    \overline{\mathcal{P}_a\mathcal{P}_b} \Delta V_{ij}
    \frac{1}{\Delta V}\int_{V_A} \dd^3 \bvec{r}' \delta(\bvec{r}\;
    '-\bvec{r}_j) ,
  \end{split}
\end{equation}
where $V_A$ is the box in which the particle A is situated. Now the
whole expression $\overline{\mathcal{P}_a\mathcal{P}_b} \Delta
V_{ij}/\Delta V$ is interpreted as the probability that an interaction
takes place. If there are $n$ test articles within one box, then
there are $n(n-1)/2$ possible scattering events. One now chooses
randomly only $n/2$ collision pairs out of the possible pairs.  To
conserve the overall reaction rate one has to rescale the collision
probability by the factor
\begin{equation}
  \frac{\text{number of possible collisions}}
  {\text{number of collisions}}=\frac{n(n-1)/2}{n/2}.
\end{equation}
Finally, we obtain the probability that a chosen collision takes place
as
\begin{equation}
  P= \frac{n(n-1)/2}{n/2} \overline{\mathcal{P}_a\mathcal{P}_b} \Delta
  V_{ij}   \frac{1}{\Delta V} .
\end{equation}
This method is faster than the original full-ensemble method, since
one can effectively order all test particles into the cells before one
simulates the collision term. Afterwards only such test particles need
to be correlated/compared which are within one cell. In the limit of
$\Delta V\to 0$ and $N\to \infty$ this corresponds to a full-ensemble
calculation. The only drawback is that one has to adjust beforehand
the parameters $\Delta V$ and $\Delta t$ such, that $P\leq1$ for all
boxes.  Therefore, one has to roughly estimate the mean value of $n$
by the nuclear matter density ($\bar{n}\approx \rho \Delta V N$) and
the cross section by some meaningful average value. Larger numbers of
$N$ lead to better estimates of $n$, while low $N$ can lead to huge
fluctuations in $n$. Typical volumes for $N\approx300\upto1500$
are\footnote{The default in our code is chosen to be $(0.5 \fm)^3$. If
  events occur with $P>1$, then the calculation must be repeated with
  a smaller time-step size, $\Delta t$.} $\Delta V=(0.25 \upto 1.0
\fm)^3$.

We have implemented all three algorithms (parallel-, full-, and
local-ensemble method) \linebreak within our model.  The influence on
the results is shown, e.g., in \cref{sec:DCX} for the pionic
double-charge exchange in $\pi A$ scattering.

\subsubsection{Determination of in-medium cross sections}
\label{sec:sfree}

In \cref{sec:elementaryMedium} we have discussed the problem that
arises when using free vacuum cross sections for collisions, where
the colliding particles are inside the nuclear medium and experience
potentials. There we have given three prescriptions for converting the
actual invariant energy, $s^*=(p_1^*+p_2^*)^2$, into a free value at
which the free cross section is read off. We recall, that $p_{1,2}^*$
are the kinetic four-momenta of colliding particles. For the
calculations with nonrelativistic potentials, the kinetic and
canonical four-momenta are the same, $p^*=p$.  However, in the
RMF-mode calculations, they differ by the vector field
(cf.~\cref{p^star}).

All the existing codes, in particular those for heavy-ion reactions,
use a recipe how to achieve this, but this is never spelled out, nor
justified. Since the possible prescriptions used contain some degree
of arbitrariness which may affect the physics result, we discuss this
point here in some more detail.

In order to illustrate the different ways of the calculation of the
``free'' invariant energy, let us consider the simple case of a
momentum-independent scalar field acting on nucleons. This leads to
\begin{equation}
  E=\sqrt{{m_N^*}^2 + \bvec{p}^2}
\end{equation}
for the dispersion relation for the nucleon.

\begin{figure}[t]
  \centering
  \includegraphics[width=0.6\linewidth]{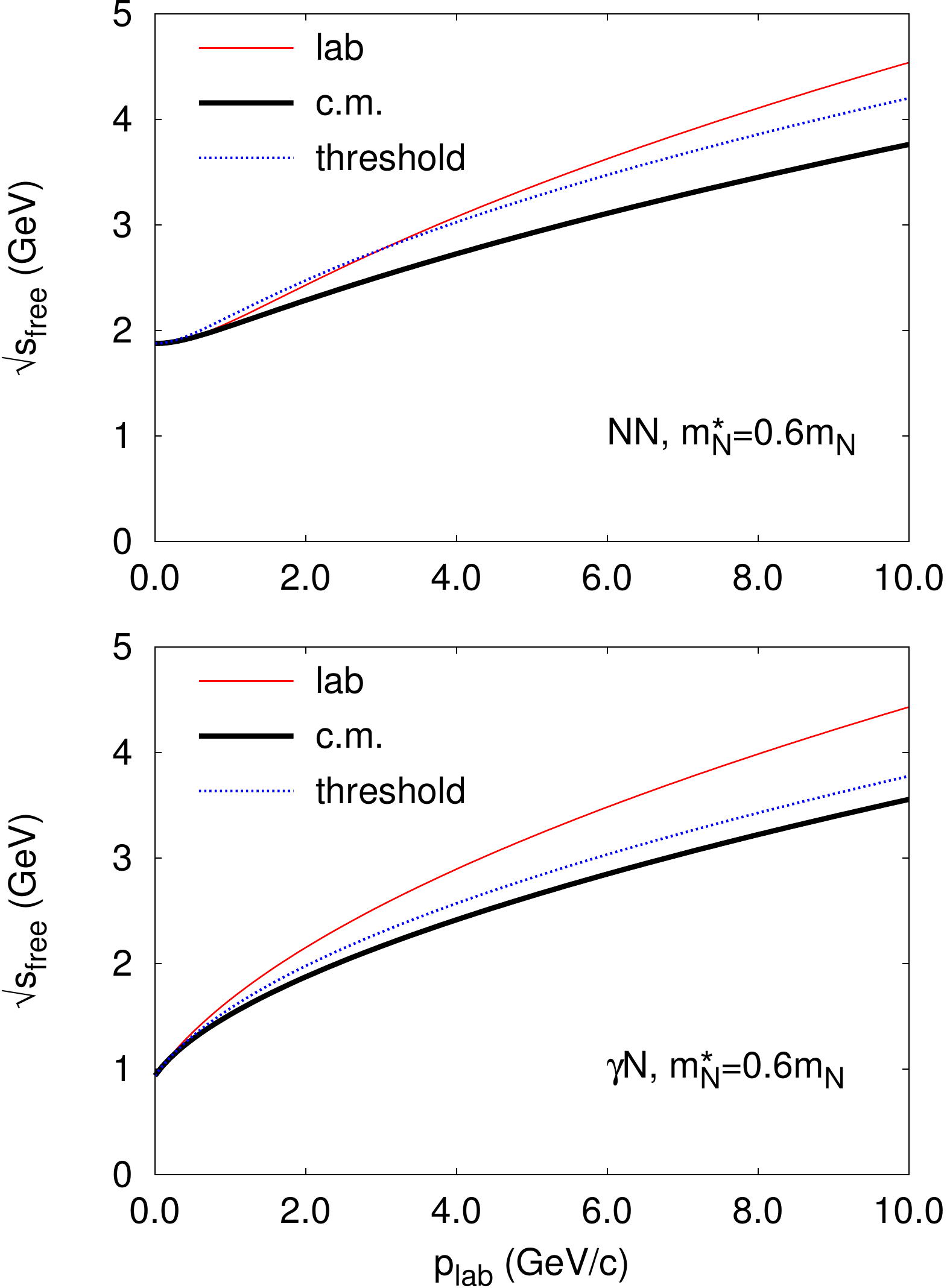}
  \caption{\label{fig:sqrts} (Color online) The ``free'' invariant
    energy calculated from \cref{s_free_nonrel} in the laboratory
    (cf.~\cref{s_free_nonrel_lab}) and c.m.~(cf.~\cref{s_free_nonrel_cm})
    frames as well as from \cref{s_free_RMF} as a function of the
    laboratory momentum for NN (upper panel) and $\gamma$N (lower
    panel) collisions.  The nucleon effective mass is chosen as
    $m_N^*=0.6m_N$ corresponding to the nuclear matter at the
    saturation density. See text for details.}
\end{figure}

As two representative examples, we will discuss NN and $\gamma$N
collisions.  The total invariant in-medium c.m.~energy squared is
calculated as
\begin{equation}
  s^* =
  \begin{cases}
    \Big(\sqrt{{m_N^*}^2+p_\text{lab}^2}+m_N^*\Big)^2 - p_\text{lab}^2
    &~~{\rm for~NN~collision},\\
    (p_\text{lab}+m_N^*)^2 - p_\text{lab}^2 &~~{\rm for}~\gamma{\rm
      N~collision}~.
  \end{cases}
\end{equation}
Applying \cref{s_free_nonrel} in the laboratory frame results in
\begin{equation}
  s_\text{free}\,\big|_\text{lab} =
  \begin{cases}
    \Big(\sqrt{m_N^2+p_\text{lab}^2}+m_N\Big)^2 - p_\text{lab}^2
    &~~{\rm for~NN~collision},\\
    (p_\text{lab}+m_N)^2 - p_\text{lab}^2 &~~{\rm for}~\gamma{\rm
      N~collision}~.
  \end{cases}
  \label{s_free_nonrel_lab}
\end{equation}
Now, if one uses the same relation \cref{s_free_nonrel} in the
c.m.~frame of the colliding particles, one obtains
\begin{equation}
  \sqrt{s_\text{free}}\,\big|_\text{cm} =
  \begin{cases}
    2\sqrt{q^2+m_N^2}  &~~{\rm for~NN~collision},\\
    q+\sqrt{q^2+m_N^2} &~~{\rm for}~\gamma{\rm N~collision}~,
  \end{cases}
  \label{s_free_nonrel_cm}
\end{equation}
where $q=\sqrt{s^*/4-{m_N^*}^2}$ and $q=(s^*-{m_N^*}^2)/2\sqrt{s^*}$
are the c.m.~momenta of the colliding particles for NN and $\gamma$N
collisions, respectively.  Finally, using the relation
\cref{s_free_RMF}, we obtain
\begin{equation}
  \sqrt{s_\text{free}}\,\big|_\text{threshold} =
  \begin{cases}
    \sqrt{s^*} - 2(m_N^*-m_N) &~~{\rm for~NN~collision},\\
    \sqrt{s^*} - (m_N^*-m_N) &~~{\rm for}~\gamma{\rm N~collision}~.
  \end{cases}
\end{equation}
\Cref{fig:sqrts} shows the laboratory momentum dependence of
$\sqrt{s_\text{free}}$ calculated by the different recipes explained
above. One sees that the difference between various prescriptions
grows with beam momentum reaching $\sim 20\proz$ at $10\GeVc$.

In the RMF mode, the ``free'' invariant energy is always calculated
by using \cref{s_free_RMF}. The latter respects the in-medium particle 
production thresholds, as we show below. Let us assume, for simplicity,
that the sum of vector fields is the same for initial and final
particles, i.e., $V_1+V_2=\sum_{i=1'}^{N'} V_i$. Then, the difference
of the canonical four-momenta which appears in the energy-momentum
conserving $\delta$-function of \cref{dsig_12_to_1'2'N'} can be
replaced by the difference of the kinetic four-momenta, and,
therefore,
\begin{equation}
  \dd\sigma^*_{12 \to 1'2'...N'} \propto
  \dd \Phi_N(p_1^* + p_2^* ; p_{1'}^*, p_{2'}^*, \ldots , p_{N'}^*)~,
  \label{dsig_12_to_1'2'N'_propto}
\end{equation}
where we have assumed that the outgoing particles are on their (Dirac)
mass shells, $p_i^{*2}=m_i^{*2}, \quad i=1',...,N'$, and the $N$-body
phase-space-volume element is defined according to \refcite{PDGdata}
as
\begin{equation}
  \begin{split}
    \dd \Phi_N(\mathcal{P}; p_1, p_2, \ldots , p_N) &= \delta^{(4)}(
    \mathcal{P} - p_1 - p_2 - \cdots - p_N ) \\
    &\times \frac{\dd^3 \bvec{p}_1 }{ (2\pi)^3 2p_1^0 } \frac{\dd^3
      \bvec{p}_2 }{ (2\pi)^3 2p_2^0 } \cdots \frac{\dd^3 \bvec{p}_N }{
      (2\pi)^3 2p_N^0 }.  \label{dPhi_N}
  \end{split}
\end{equation}
The in-medium threshold condition follows immediately from
\cref{dsig_12_to_1'2'N'_propto},
\begin{equation}
  Q^* > 0,       \label{Q*}
\end{equation}
with $Q^* = \sqrt{s^*} - \sum_{i=1'}^{N'} m_i^*$ being the in-medium
excess energy and $s^* = (p_1^* + p_2^*)^2$.  Defining now the
``free'' invariant energy as
\begin{equation}
  \sqrt{s_\text{free}} = Q^* + \sum_{i=1'}^{N'} m_i
  = \sqrt{s^*} - \sum_{i=1'}^{N'} (m_i^* - m_i)
  \label{s_free_threshold}
\end{equation}
leads to the evaluation of the vacuum cross section at the in-medium
excess energy, which obviously respects the true in-medium threshold
condition \cref{Q*}.  Unfortunately, the definition
\cref{s_free_threshold} is channel dependent and is hard to apply in
practice. In most cases, the final-state particles are not known until
the actual simulation of a given two-body collision is done. Thus, we
further assume that the sum of scalar fields is the same for initial
and final particles, i.e., $S_1+S_2=\sum_{i=1'}^{N'} S_i$, which
allows us to rewrite \cref{s_free_threshold} as
\begin{equation}
  \label{s_free_RMF2}
  \sqrt{s_\text{free}} = \sqrt{s^*} - (m_1^* - m_1) - (m_2^* - m_2)~.
\end{equation}
We recall here the definition of the Dirac effective mass according to
\cref{m_i^star}.  The expression (\ref{s_free_RMF2}) is used in the
RMF-mode calculations. In the present RMF implementation, the
conditions $V_1+V_2=\sum_{i=1'}^{N'} V_i$ and
$S_1+S_2=\sum_{i=1'}^{N'} S_i$ used in the derivation of
\cref{s_free_RMF2} are fulfilled, since the coupling constants of all
baryons with the meson fields are chosen equal to the nucleon-meson
coupling constants, while the mean-field potentials acting on mesons
are neglected. The only exception is the baryon-antibaryon
annihilation into mesons (see \cref{sec:antiprotonA}).

\subsubsection{Pauli blocking}
\label{sec:Pauli}

The phase-space distribution, $f_i(\bvec{r},\bvec{p})$, which enters
in the Pauli blocking factor, $1-f_i(\bvec{r},\bvec{p})$, is
calculated by counting the number of test particles in the phase-space
volume element composed of small spherical volumes $\Delta V_r$ with
radius $r_r$ centered at $\bvec{r}$ in coordinate space and $\Delta
V_p$ with radius $r_p$ centered at $\bvec{p}$ in momentum space,
\begin{equation}
  f_i(\bvec{r},\bvec{p}) = \sum_{j:~\bvec{p}_j \in \Delta V_p}
  \frac{1}{\kappa(2\pi\sigma^2)^{3/2}}
  \int\limits_{\Delta V_r, |\bvec{r}-\bvec{r}_j| < r_c} \dd^3 \bvec{r}
  \;\exp\left\{ -\frac{(\bvec{r}-\bvec{r}_j)^2}{2\sigma^2}\right\}, \label{f_i}
\end{equation}
where
\begin{equation}
  \kappa = \frac{2\, \Delta V_r\, \Delta V_p\, N}{(2\pi)^3}
  \frac{4\pi}{(2\pi\sigma^2)^{3/2}}
  \int\limits_0^{r_c} \dd r \; r^2 \exp\left\{-\frac{r^2}{2\sigma^2}\right\}
  \label{kappa}
\end{equation}
is a normalization factor. In \cref{f_i}, the sum is taken over all
test particles, $j$, of the type $i=p,n$ whose momenta belong to the
volume $\Delta V_p$. In coordinate space, the test particles are
represented by Gaussians of the width, $\sigma$, cut off at the radial
distance $r_c$, in a similar way as done for the folding of the
density fields with Gaussians, see \cref{sec:Vlasov}. The default
values of parameters are $r_p=80\MeVc$, $r_r=1.86\fm$, $\sigma=1\fm$,
$r_c=2.2\fm$. This set of parameters is a compromise between the
quality of the Pauli blocking in the ground state and the smallness of
statistical fluctuations in the case of simulations with $N\sim200$
test particles per nucleon. Typically, this is sufficient in accuracy
for modeling heavy-ion collisions at beam energies above $\sim
100\AMeV$.

However, for small-amplitude dynamics near the nuclear ground state,
like the giant monopole resonance vibrations studied in
\refcite{Gaitanos:2010fd}, the accuracy provided by \cref{f_i,kappa}
is not sufficient, when the default parameters are used. The main reason
is the constant, i.e., momentum-independent radius, $r_p$, which introduces
a spurious
temperature of the order of several MeV.  To reduce this effect, we
have introduced a position- and momentum-dependent radius of the
momentum-space volume $\Delta V_p$ by
$r_p(\bvec{r},|\bvec{p}|)=\mbox{max}[20\MeVc, p_{F,i}(\bvec{r})
- |\bvec{p}|]$, which provides a sharper Fermi surface. This
allows us to use the reduced parameters also in coordinate space,
$r_r=0.9-1.86\fm$, $\sigma=0.5\fm$, $r_c=1.1\fm$.

\subsection{Final-state decisions for hadron-hadron scattering events}
\label{sec:finalStateDecisions}

The final-state decision is straight forward in the case of resonance
production. If there are two or three particles in the final state,
the treatment is more involved. First, in
\crefrange{1BodyVacuum}{3BodyVacuum}, we discuss the treatment in the
vacuum. Then the medium corrections are discussed.

\subsubsection{Resonance production}
\label{1BodyVacuum}
In the case of resonance production we obtain the final mass of the
resonance by \cref{p_R,E_R}, which completely fixes the kinematics.

\subsubsection{Two-body final states : \texorpdfstring{${X \rightarrow
      c d}$}{X->cd} in the vacuum}
\label{2BodyVacuum}

The general definition of the cross section for $2 \rightarrow n$
reactions is given by (see, e.g., \cite{PDGdata})
\begin{equation}
  \dd\sigma_{a\ b \rightarrow f_1, f_2, f_3, \ldots,f_n}= \dd \Phi_n  \left( 2 \pi
  \right)^4 \mathcal{S}_\text{final} \frac{|\mathcal{M}_{ab \rightarrow
      f_1, \ldots, f_n}|^2}{4I_{ab}},
\end{equation}
where $\dd \Phi_n$ denotes the $n$-particle phase space of the
final-state particles,
\begin{equation}
  \dd \Phi_n = \delta^{(4)} \left (p_a+p_b-\sum_{i=1}^n p_{f_i} \right) \prod_{i=1}^{n}
  \frac{\dd^3 \bvec{p}_i}{(2 \pi)^3 2 E_i},
\end{equation}
$\mathcal{S}_\text{final}$ stands for the symmetry factor of the final
state, and
\begin{equation}
  I_{ab}=\sqrt{\left(p_a \cdot p_b\right)^2-m_a^2 m_b^2}
\end{equation}
represents the invariant flux factor of the particles, $a$ and $b$. In
the center-mass (CM) frame this flux factor reads
\begin{equation}
  I_{ab}=p_{\text{cm}} \sqrt{s} ,
\end{equation}
with the CM momentum, $p_{\text{CM}}$, of the particles, $a$ and
$b$. It can be shown (see, e.g., \cite[cf.~especially Section
4.7]{lehr_phd} \footnote{Note that \cite{lehr_phd} uses a slightly
  different convention for the spectral function which differs by a
  factor $2m$ from the one used in this work.}), that one can express
the cross section for the production of unstable particles $c$ and $d$
in the final state by
\begin{equation}
  \label{twoParticleFinal}
  \frac{\dd \sigma_{a b \rightarrow c d}}{\dd \mu_c\, \dd\mu_d\, \dd
    \Omega}(s)=\frac{1}{64\pi^2\, s} \frac{p_{cd}}{p_{ab}} 2\mu_{c}
  \mathcal{A}_c(\mu_{c},p_c(\mu_c,\Omega)) 2\mu_{d}
  \mathcal{A}_d(\mu_{d},p_d(\mu_d,\Omega)) \left|\mathcal{M}_{a b
      \rightarrow c d } (s)\right|^2 ,
\end{equation}
with $p_{ab}$ and $p_{cd}$ denoting the CM momenta of the $a\,b$ and
the $c\,d$-system, respectively. \Cref{dsig_NN_to_ND,dsig_ND_to_ND}
are specializations of this expression.  Here, one has to assume that
the matrix element depends on $s$ only.  In the vacuum the spectral
functions depend on the squares, $\mu^2=p^{\nu} p_{\nu}$, only, while
in the medium, in general, they are functions of the four momenta.

We want to use a Monte-Carlo method to choose the final state. In
\cref{twoParticleFinal}, we note that the two-particle final state
depends both on the masses of the outgoing particles and their
directions of motion. The Mandelstam variable, $s$, and $p_{ab}$ are
determined by the initial state. Thus we choose the masses, $\mu_c$
and $\mu_d$, at the same time as we choose the angles, $\Omega$.

The transformed random number, $y$,
\begin{equation}
  y(\mu)=2\,\arctan\left[2\,\frac{\mu-M^0}{\Gamma^0}\right],\qquad,\quad
  \dd y= \dd \mu
  \frac{\Gamma_0}{(\mu-M^0)^2+\left(\frac{\Gamma_0}{2}\right)^2}
  \label{eq:randomtransform2D}
\end{equation}
is distributed according to a Cauchy-Lorentz distribution, which leads
to the the function,
\begin{equation}
  \frac{\dd \sigma_{a b \rightarrow c d}}{\dd y_c\, \dd y_d\, \dd
    \Omega}(s)
  \propto
  \frac{1}{s}
  \frac{p_{cd}}{p_{ab}} \frac{\mu_c\mathcal{A}_c(\mu_{c},p_c(\mu_c,\Omega))}{\frac{dy_c}{d\mu_c}}
  \frac{\mu_d\mathcal{A}_d(\mu_{d},p_d(\mu_d,\Omega))}{\frac{dy_d}{d\mu_d}} \left|\mathcal{M}_{a b
      \rightarrow c d } (s)\right|^2,
\end{equation}
smoother than the original function \cref{twoParticleFinal}, which is
advantageous for the application of a rejection method. For $y_{c,d}$
one has to insert the values at the pole position in the vacuum
$M_{c,d}^0$ and $\Gamma_{c,d}^0$ of particle $c$ and $d$.

For the rejection procedure, we choose $y_c$, $y_d$, and $\Omega$
according to a flat distribution. The probability that a random
ensemble, $(y_c,y_d,\Omega)$, will be accepted is then given by
\begin{equation}
  P_\text{accept}(y_c,y_d,\Omega)= \frac{p_{cd}
    \mu_c\mathcal{A}_c(\mu_{c},p_c(\mu_c,\Omega)) \mu_d
    \mathcal{A}_d(\mu_{d},p_d(\mu_d,\Omega)) \frac{\dd \mu_c}{\dd y_c}
    \frac{\dd \mu_d}{\dd y_d}}{\left(p_{cd} \mu_c\mathcal{A}_c \mu_d
      \mathcal{A}_d \frac{d\mu_c}{dy_c}
      \frac{\dd \mu_d}{\dd y_d}\right)_\text{max}} .
\end{equation}
The maximal value in the denominator is actually hard to find. We
parameterize it by
\begin{equation}
  \left(p_{cd} \mu_c\mathcal{A}_c \mu_d\mathcal{A}_d\frac{\dd
      \mu_c}{\dd y_c} \frac{\dd \mu_d}{\dd y_d}\right)_\text{max}=Q_{cd}
  \max\{p^\text{vac}_{cd}\} .
\end{equation}
The dimensionless factor, $Q$, is of the order of $10$ and depends on
the outgoing particles. It has to be readjusted if one introduces
medium effects.\footnote{Recently, we started to implement a method of
  additionally slicing the mass range into smaller regions, where one
  optimizes the rejection method. In addition, the maximal values are
  then precalculated.}

\subsubsection{Three body final states : \texorpdfstring{${X
      \rightarrow c d e}$}{X-> cde} in the vacuum}
\label{3BodyVacuum}

For a three-particle final state one obtains in analogy to
\cref{twoParticleFinal} the complicated structure,
\begin{equation}
  \begin{split}
    \label{threeParticleFinal}
    \frac{\dd \sigma_{a b \rightarrow c d e }}{\dd \mu_c\,
      d\mu_d\,d\mu_e\ \dd |\bvec{p}_{c}|\, \dd \Omega_c \, \dd
      |\bvec{p}_{d}|\, \dd\phi_d}(s)=&\frac{1}{8\left(2\pi\right)^5}
    \frac{1}{p_{ab} \sqrt{s}} \frac{\left|\bvec{p}_{c}\right|
      \left|\bvec{p}_{d}\right|}{E_c E_d} 2\mu_c
    \mathcal{A}_c(\mu_{c},p_c) \\
    & \times 2\mu_d\mathcal{A}_d(\mu_{d},p_d) 2\mu_e
    \mathcal{A}_e(\mu_{e},p_e)\left|\mathcal{M}_{a b \rightarrow c d
        e} (s)\right|^2 \,.
  \end{split}
\end{equation}
Here, $\bvec{p}_{c,d}$ denote the CM momenta of the particles, $c$ and
$d$. The CM momentum of $e$ is given by total momentum conservation,
\begin{equation}
  \bvec{p}_c+\bvec{p}_d+\bvec{p}_e=0 .
\end{equation}
In analogy to the two-particle final state, we apply the variable
transformation \cref{eq:randomtransform2D} to
\cref{threeParticleFinal} and obtain
\begin{equation}
  \begin{split}
    \frac{\dd \sigma_{a b \rightarrow c d e }}{\dd y_c\, \dd y_d\,\dd
      y_e\ \dd|\bvec{p}_{c}|\, \dd\Omega_c \, \dd|\bvec{p}_{d}|\,
      \dd\phi_d}(s) \propto & \frac{1}{p_{ab} \sqrt{s}}
    \frac{\left|\bvec{p}_{c}\right| \left|\bvec{p}_{d}\right|}{E_c
      E_d} \mu_c\mathcal{A}_c(y_{c},p_c)
    \mu_d\mathcal{A}_d(y_{d},p_d) \\
    &\times \, \mu_e\mathcal{A}_e(y_{e},p_e)\left|\mathcal{M}_{a b
        \rightarrow c d e} (s) \right|^2 \, \frac{\dd \mu_c}{\dd y_c}
    \frac{\dd \mu_d}{\dd y_d} \frac{\dd \mu_e}{\dd y_e} .
  \end{split}
\end{equation}
Hence, we need to choose $y_c$, $y_d$, $y_e$, $|\bvec{p}_{c}|$,
$\Omega_c, \phi_d$, and $|\bvec{p}_{d}|$ independently of each
other. The limits for the $y_i$ are given by the smallest and largest
possible masses. The absolute values of the momenta, $|\bvec{p}_{i}|$,
are limited by energy conservation, $|\bvec{p}_{i}|< \sqrt{s}$.  The
value of $\Omega_c$ is determined by choosing a random $\cos \theta
\in[-1,1]$ and $\phi\in[0,2\pi]$.

The Monte Carlo probability to accept an event configuration is given
by
\begin{equation}
  p_\text{accept}= \frac{\mu_c\mu_d\mu_e \frac{\left|\bvec{p}_{c}\right|
      \left|\bvec{p}_{d}\right|}{E_c E_d} \mathcal{A}_c(y_{c},p_c)
    \mathcal{A}_d(y_{d},p_d)  \mathcal{A}_e(y_{e},p_e)
    \frac{\dd \mu_c}{\dd y_c}\frac{\dd \mu_d}{\dd y_d} \frac{\dd
      \mu_e}{\dd y_e}}{m},
\end{equation}
where the factor $m$ in the denominator is chosen such, that it is
larger than the maximum of the numerator.

\subsubsection{Medium Corrections}
\label{sec:finalStateDecision_medium}

In the medium we have more complicated dispersion relations. Therefore
also the phase-space factors differ from the ones used above. Already
in \cite{effe_phd} possibilities to implement the right phase-space
factors for $\Delta N$ and $NN$ scattering have been discussed. Our
treatment does not include such modifications, but preserves the
energy in the medium for all collisions. We use the following
algorithm:
\begin{enumerate}
\item Evaluate $s_\text{vacuum}$.
\item Do the final-state decision with vacuum kinematics assuming
  $s=s_\text{vacuum}$.
\item Correct the final state by scaling the final-state momenta by a
  factor, $x$, in the CM frame.
\end{enumerate}
The last point needs special discussion.  In the CM frame, energy and
momentum conservation in step 2 result in a solution for the momenta
$\bvec{p}_i$, which obeys
\begin{alignat}{3}
  \sum_i \sqrt{\left( \bvec{p}^\text{CM}_i \right)^2 +\left(m_i
    \right)^2} &=\sqrt{s_\text{vacuum}}&\quad&,\qquad& \sum_i
  \bvec{p}^\text{CM}_i &=0.  \intertext{Now we want to define the four
    momenta, $q_i$, in the medium. Momentum and energy conservation
    demand} \sum_i q^0_i \left( \bvec{q}_i \right) &=
  \sqrt{s}&\quad&,\qquad& \sum_i \bvec{q}_i &=0. \label{ECons}
\end{alignat}
The zeroth components of $q_i$ are, due to the potentials, highly
non-trivial functions of the vector components, $\bvec{q}_i$. Hence we
use the following recipe: Using the vacuum result for $\bvec{p}_i$ we
choose
\begin{equation}
  \forall  i: \quad \bvec{q}_i= x \bvec{p}_i,
\end{equation}
where the scaling factor, $x$, is fixed by \cref{ECons}. Since all
momenta are scaled by the same factor, momentum conservation is
fulfilled trivially.

\subsection{Coding}
\label{sec:AppCoding}

The numerical implementation of the GiBUU model is based on
Fortran2003 using modern programming paradigms and a version control
management (Subversion). See the GiBUU website \cite{gibuu} and
Section 3.1 and Appendix C in the dissertation of O.~Buss
\cite{buss_phd} for detailed information on the current code structure
and its programming history. The code runs and has been tested with
numerous compilers on different platforms, details are given on our
website \cite{gibuu}.  At this website, one also finds information
about program options and a documentation of the parameters.  The
GiBUU code is open source under the GPL license \cite{gpl} and can be
downloaded after registration \cite{gibuu}.

\begin{flushleft}

\end{flushleft}

\end{document}